\documentclass[12pt,a4paper,oneside]{report}
\usepackage{geometry}
\usepackage{rotating} 

\usepackage{setspace}
\usepackage{graphicx}
\usepackage{amsmath, amssymb, amsfonts}
\usepackage{hyperref}
\usepackage{url}
\usepackage{array} % (usually already loaded, but safe)
\newcolumntype{Y}{>{\raggedright\arraybackslash}X}

\usepackage[numbers,sort&compress]{natbib}
\usepackage{titlesec}
\usepackage{tocloft}
\usepackage{enumitem}
\usepackage{booktabs}
\usepackage{float}
\usepackage{pdfpages}
\usepackage{alltt}
\usepackage[most]{tcolorbox} 
\usepackage{lipsum}
\usepackage{pdflscape}
\usepackage{lscape}
\usepackage[ruled,vlined]{algorithm2e}
\usepackage{longtable}
\usepackage{tikz}
\usetikzlibrary{arrows.meta,positioning,shapes,calc,fit}
\usepackage{xcolor}
\usepackage[utf8]{inputenc} 
\usepackage{tabularx}
\usepackage{amsthm}
\usepackage{makecell}

\newtheorem{lemma}{Lemma}[chapter]
\newtheorem{proposition}{Proposition}[chapter]
\newtheorem{corollary}{Corollary}[chapter]
\theoremstyle{plain}
\newtheorem{theorem}{Theorem}[chapter] % or [section] depending on your structure
\usepackage{mathtools}
\theoremstyle{remark}
\newtheorem{remark}{Remark}[chapter]
\newtheorem{assumption}[theorem]{Assumption}
\theoremstyle{definition}
\newtheorem{definition}[theorem]{Definition}

\theoremstyle{remark}

\usepackage{etoolbox}
\makeatletter
\pretocmd{\landscape}{\thispagestyle{empty}}{}{}
\makeatother

\definecolor{ImperialBlue}{RGB}{0,33,71}
\definecolor{ImperialGrey}{RGB}{100,100,100}

\newcommand{\VoLL}{\mathrm{VoLL}}
\newcommand{\B}{\mathrm{B}}           % Baseline
\newcommand{\T}{\mathrm{T}}           % Treatment
\newcommand{\DeltaT}{\Delta^{\T}}     % Difference (Treatment vs Baseline)

\newcommand{\Req}{\mathcal{R}}

\newcommand{\AMM}{\text{AMM}}             % Generic AMM
\newcommand{\LMP}{\text{LMP}}             % Generic LMP

\makeatletter
\AtBeginDocument{%
  \@ifpackageloaded{lineno}{\nolinenumbers}{}
  \providecommand\modulolinenumbers[1]{}%
  \providecommand\nolinenumbers{}%
}
\makeatother

\begin{document}

\begin{titlepage}
\thispagestyle{empty}
\sffamily

% ---------------------------------------------------------
% TOP SECTION — LOGO + DEGREE INFO
% ---------------------------------------------------------
\noindent
\begin{minipage}{0.55\textwidth}
    \includegraphics[height=1cm]{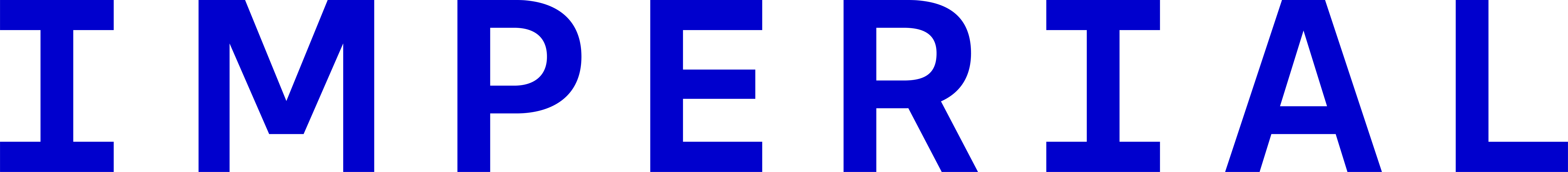}\\[0.25cm]
\end{minipage}
\hfill
\begin{minipage}{0.42\textwidth}
    \raggedleft
    \textcolor{ImperialGrey}{\small Design Engineering Research (PhD)}\\[-0.05cm]
    \textcolor{ImperialGrey}{\small Dyson School of Design Engineering}
\end{minipage}

\vspace{0.6cm}
\noindent\rule{\linewidth}{0.4pt}

% ---------------------------------------------------------
% TITLE BLOCK
% ---------------------------------------------------------
\vspace{1.6cm}

\begin{center}
    {\small\textcolor{ImperialGrey}{Thesis}}\\[0.6cm]

    {\huge\bfseries\textcolor{ImperialBlue}{A Fair, Flexible, Zero-Waste}}\\[0.1cm]
    {\huge\bfseries\textcolor{ImperialBlue}{Digital Electricity Market}}\\[0.7cm]

    {\large\textcolor{ImperialGrey}{A First-Principles Approach }}\\[-0.05cm]
    {\large\textcolor{ImperialGrey}{Combining Automatic Market Making,}}\\[-0.05cm]
    {\large\textcolor{ImperialGrey}{Holarchic Architectures and}}\\[-0.05cm]
    {\large\textcolor{ImperialGrey}{Shapley Theory}}\\[-0.05cm]
\end{center}

% ---------------------------------------------------------
% AUTHOR + CONTACT
% ---------------------------------------------------------
\vspace{1.2cm}

\begin{center}
    {\Large\bfseries Shaun Sweeney, MEng, BSc}\\[0.35cm]

    % LinkedIn row with icon + clean text link
    \raisebox{-0.1cm}{\includegraphics[height=0.55cm]{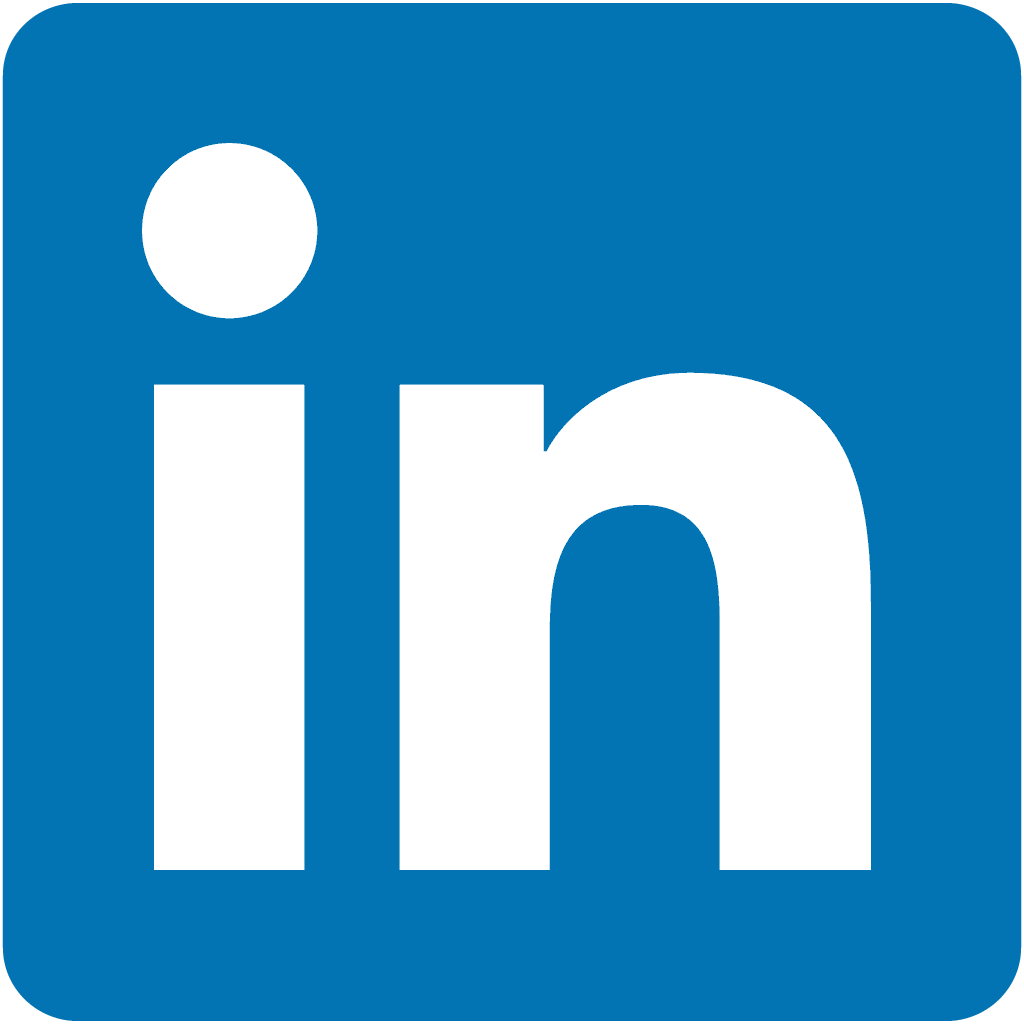}}
    \hspace{0.2cm}
    \href{https://www.linkedin.com/in/shaundsweeney/}{%
      \textcolor{ImperialGrey}{\small linkedin.com/in/shaundsweeney}%
    }\\[0.15cm]

    \textcolor{ImperialGrey}{\small Platform: enleashed.tech}\\[-0.05cm]
    \textcolor{ImperialGrey}{\small Contact: fight@enleashed.tech}
\end{center}

% ---------------------------------------------------------
% SUPERVISORS
% ---------------------------------------------------------
\vspace{1.4cm}

\begin{center}
    \textcolor{ImperialGrey}{\textbf{Supervisors}}\\[0.35cm]
    \textcolor{black}{Prof.\ Robert Shorten \ (Main supervisor)}\\[0.15cm]
    \textcolor{black}{Prof.\ Mark O'Malley \ (Co-supervisor)}
\end{center}

% ---------------------------------------------------------
% BOTTOM — INSTITUTION + YEAR
% ---------------------------------------------------------
\vfill

\begin{center}
    \textcolor{ImperialGrey}{\textbf{Institution}}\\[0.15cm]
    \textcolor{black}{Imperial College London}\\[0.45cm]

    \textcolor{ImperialGrey}{\textbf{Year}}\\[0.15cm]
    \textcolor{black}{2025}
\end{center}

\end{titlepage}

\pagenumbering{roman}
% =========================================================
% ONE-PAGE PROLOGUE
% =========================================================
\cleardoublepage
\thispagestyle{empty}

\vspace*{\fill}

\begin{center}
\textit{\Large Change is possible}
\end{center}

\vspace{1cm}

\begin{center}
\itshape
Change is possible—\\
not by accident,\\
but by choice.\\[1.0em]

We dare to dream\\
because we must,\\
and we dare to act\\
because dreaming alone is not enough.\\[1.0em]

Nothing about us\\
is decided without us;\\
nothing for us\\
is built without our hands.\\[1.0em]

Each step forward counts—\\
run if you can,\\
walk if you must,\\
crawl if you have to,\\
but always move.\\[1.0em]

Freedom begins in the mind,\\
and responsibility gives it form.\\
Character is shaped\\
where limits are accepted\\
and promises are kept.\\[1.0em]

Be the hero of your own story—\\
not alone,\\
but alongside others,\\
lifting as you rise.\\[1.0em]

If we want to,\\
we can.
\end{center}

\vspace{1cm}

\begin{center}
\textit{— Shaun Sweeney}
\end{center}

\vspace*{\fill}

\cleardoublepage

\tableofcontents
\newpage

\chapter*{Abstract}
\addcontentsline{toc}{chapter}{Abstract}

This thesis presents a fundamental rethink of electricity market design at the wholesale
and balancing layers. Rather than treating markets as static spot-clearing mechanisms,
it reframes them as a continuously online, event-driven dynamical control system: a two-
sided marketplace operating directly on grid physics and providing continuous liquidity
across time, space, and system states.

Existing energy-only, capacity-augmented, and zonal market designs are shown to admit
no shock-robust Nash equilibrium under realistic uncertainty, instead relying on price
caps, uplift, and regulatory intervention to preserve solvency and security. In response,
the thesis develops a holarchic Automatic Market Maker (AMM) in which prices are
bounded, exogenous control signals derived from physical tightness rather than emergent
equilibrium outcomes, enabling continuous trade, settlement, and price discovery
without discrete clearing events.

The AMM generalises nodal and zonal pricing through nested scarcity layers, from
node to cluster to zone to region to system, such that participant-facing prices inherit
from the tightest binding constraint. Nodal and zonal pricing therefore emerge as special
cases of a unified scarcity propagation rule.

Beyond pricing, the AMM functions as a scarcity-aware control system and a digitally
enforceable rulebook for fair access and proportional allocation under shortage. Fuel
costs are recovered through pay-as-bid energy dispatch consistent with merit order, while
non-fuel operating and capital costs are allocated according to adequacy, flexibility, and
locational contribution.

Large-scale simulations demonstrate bounded-input bounded-output stability,
controllable procurement costs, zero structural waste, and improved distributional
outcomes. The architecture is climate-aligned and policy-configurable, but requires a
managed transition and new operational tools for system operators and market
participants.

\chapter*{Executive Abstract}
\addcontentsline{toc}{chapter}{Executive Abstract}

Legacy electricity markets were not designed for today’s power systems. As
variable renewables, electrification of heat and transport, constrained
networks, and digitally controllable demand have grown, long-standing
structural weaknesses in market design have become empirically visible:
extreme price volatility, insolvency cascades, rising regulatory intervention,
regressive cost allocation, and weak incentives for flexibility. These outcomes
are increasingly treated as transitional frictions or policy failures. This
thesis argues instead that they are the predictable consequence of legacy market
architectures whose core economic assumptions no longer align with modern grid
physics, balance-sheet risk, or consumer behaviour.

At the root of this misalignment is a category error. Electricity is not a
fungible commodity whose value depends only on aggregate volume. It is a
real-time, spatially constrained, reliability-critical service governed by
network physics and tight balance conditions. As power systems evolve toward
two-way flows and millions of heterogeneous, digitally controllable devices,
electricity markets increasingly resemble dynamical control systems rather than
static equilibrium markets. Prices, allocations, and remuneration must therefore
operate as bounded, state-dependent control signals, not as unconstrained
scarcity outcomes.

Intuitively, the core balancing problem faced by the electricity system at every point in time and space can be understood as a continuously adjusting seesaw. On one side sits demand; on the other, available supply. Both sides carry variable and shifting weights, reflecting network constraints, reliability commitments, and uncertainty over future states. The system admits many possible balance points, but at every instant it must remain balanced. When demand outweighs supply, the control response must increase both the buy price and the offered sell price, attracting additional supply and discouraging excess consumption. When supply exceeds demand, the same mechanism lowers prices on both sides, signalling that additional participation is unnecessary. Crucially, these prices are not equilibrium outcomes but bounded control signals whose role is to restore and maintain balance under physical constraints, both in real time and as projected into the future.

This thesis identifies the structural failure modes common to energy-only,
energy-plus-capacity, and zonal market designs, showing that all three break the
link between cost causation and cost recovery and fail to admit stable,
shock-robust equilibria under realistic uncertainty. As system characteristics
evolve, these designs increasingly rely on ad hoc corrections—price caps,
redispatch, uplift payments, and emergency interventions—rather than coherent
economic signals. The result is a brittle system that struggles to mobilise
flexibility, protect essential demand, or recover fixed costs without political
intervention.

A central claim of this thesis is that \emph{fairness is not optional} in modern
electricity markets, but a structural requirement for stability, investability,
and political durability. In systems characterised by heterogeneous demand,
bounded balance sheets, and physical scarcity, market designs that allocate
costs, access, or risk in ways perceived as arbitrary or exploitative do not
remain economically viable: they invite regulatory override, ad hoc correction,
and ultimately structural fragility. This insight is consistent with findings
from behavioural economics, energy justice, and trust-based participation
literature, which show that acceptance of scarcity, prices, and constraints
depends critically on perceived fairness, transparency, and reciprocity.

Importantly, fairness in this context does not mean equality of prices, payments,
or outcomes. Instead, it is defined axiomatically and operationalised through
four fairness pillars that together govern how costs, access, and risk are
allocated in a physics-constrained system:

\begin{itemize}[leftmargin=*]
  \item \textbf{F1: Fair Rewards} — behaviours that support system reliability,
        such as flexibility provision or congestion relief, should be
        systematically rewarded through lower expected costs or improved
        service outcomes;
  \item \textbf{F2: Fair Service Delivery} — participants who contract for higher
        service or reliability levels should receive those levels in a
        predictable and bounded manner across time and space;
  \item \textbf{F3: Fair Access} — during scarcity, access to energy must not be
        determined solely by willingness-to-pay, but must respect essential
        needs, contractual priorities, and historical contribution;
  \item \textbf{F4: Fair Cost Sharing} — costs should be borne in proportion to
        the burdens imposed and the value derived, rather than through opaque
        cross-subsidies or exposure to coincidental price spikes.
\end{itemize}

Together, these principles ensure that participants who contribute more to
system stability are treated better, those who require essential protection
receive it, and no actor is rewarded or penalised purely by chance, geography,
or financial leverage.

Affordability is therefore not treated as a guaranteed outcome—since it depends
on exogenous factors such as fuel costs and technology trajectories—but as a
probability to be maximised through a zero-waste, incentive-compatible
architecture. Different participants may rationally choose different levels of
reliability, flexibility, and exposure to scarcity, and the role of the market
is not to impose a single notion of fairness, but to implement whichever notion
society selects in a transparent, consistent, and enforceable way.

Crucially, the definition of what constitutes a fair allocation is not assumed
to be fixed over time. As societal priorities evolve—across affordability,
security, decarbonisation, resilience, or investment attractiveness—the same
architectural framework can implement revised fairness parameters without
redesigning the market itself. Policy operates through explicit, tunable controls
embedded in the allocation and remuneration rules, rather than through ex post
intervention. This separation between \emph{architectural structure} and
\emph{policy choice} is what allows the system to remain stable, adaptable, and
politically durable under changing conditions.

This thesis develops an end-to-end re-envisioning of the wholesale
and balancing layers as a digitally regulated, two-way dynamical system,
explicitly grounded in grid physics and capable of scaling to millions of
heterogeneous devices within a holarchic control architecture. Rather than
treating electricity markets as static spot-clearing mechanisms, the proposed
framework treats prices, allocations, and remuneration as state-dependent
control signals that respond to real physical tightness across time, space, and
reliability dimensions.

At the core of the proposed architecture is a holarchic Automatic Market Maker
(AMM) that operates as a network-native scarcity controller, replacing reliance
on extreme spot prices with bounded, interpretable signals linked directly to
system stress. A mathematically defined fairness framework governs both shortage
allocation and generator remuneration, ensuring fair rewards, fair service
delivery, fair access, and fair cost sharing across consumers, suppliers, and
generators. Crucially, the design imposes no constraints on retail business
models: suppliers are free to innovate atop a physically coherent and
economically stable wholesale foundation.

A further conceptual shift is the treatment of prices as \emph{exogenous,
bounded control signals} rather than endogenous equilibrium objects. Unlike
legacy markets, in which prices are expected to simultaneously allocate energy,
signal scarcity, and recover fixed costs, the AMM explicitly computes prices from
the physical state of the system. This inversion replaces price discovery with
price regulation, aligning incentives with grid physics while preserving
competitive participation.

The framework is validated through extensive data-driven simulation using
household-level demand, stylised and GB-scale transmission networks, and a
constrained inter-regional corridor. Rather than fixing procurement cost as an
uncontrolled outcome of scarcity pricing, the proposed architecture makes the
\emph{total cost of procuring the needs bundle an explicit design choice},
bounded between two economically meaningful limits.

The lower bound is a strict cost-recovery level, corresponding to the minimum
revenue required to finance the generator fleet on a regulated, non-subsidised
basis. The upper bound is the maximum aggregate payment that participants are
willing—or politically permitted—to bear, which is not directly observable.
In the experiments, this upper bound is conservatively proxied by the total
revenues distributed under the Baseline LMP design, which are substantially
higher than the cost-recovery requirement for the benchmark network.

Results show that, under the AMM architecture, a wide and non-empty feasible
procurement region exists between these bounds. By calibrating the annual pots,
the regulator can choose where to operate within this region—trading off
investment attractiveness, distributional outcomes, and consumer burden—while
maintaining bounded-input bounded-output stability, smooth scarcity signals, and
fairness-consistent allocation under physical uncertainty. Game-theoretic
analysis further shows that this controlled procurement regime admits
well-defined equilibria, in sharp contrast to the fragility of incumbent
scarcity-priced market forms.

Taken together, the thesis argues that electricity markets must transition from
static spot-pricing paradigms to physically grounded, digitally regulated,
fairness-aware control systems. It shows that such systems can satisfy core
economic properties, respect behavioural constraints, and align incentives with
real grid needs, offering a credible path toward a trustworthy, flexible, and
low-waste electricity system capable of supporting deep electrification.

\chapter*{Technical Abstract}
\addcontentsline{toc}{chapter}{Technical Abstract}

Legacy electricity markets possess \emph{no shock-robust Nash equilibrium}:
under plausible fuel-cost volatility and finite balance sheets, no combination
of retail pricing, hedging, and consumer behaviour can simultaneously satisfy
solvency, affordability, and continuity of essential demand. Even in the
absence of exogenous shocks, the strategic interactions of generators,
retailers, and consumers fail to admit a stable Nash equilibrium that survives
exposure to physical uncertainty and balance-sheet risk. Small perturbations in
prices, demand, or renewable output generate profitable unilateral deviations
that propagate into insolvency cascades, extreme price excursions, or
involuntary curtailment.

Existing market designs fall into three broad families. The first is the
\textbf{energy-only, marginal-cost paradigm}, exemplified by nodal LMP systems,
in which scarcity rents and investment signals are expected to emerge from
unbounded spot prices calibrated by an administratively chosen Value of Lost
Load (VoLL). The second is the \textbf{energy-plus-capacity, heavily regulated
paradigm}, exemplified by the GB model, in which price caps, capacity auctions,
and side-payments are layered atop the energy market to mitigate investment and
affordability failures. A third family—zonal pricing—sits between nodal and
national pricing but inherits the insolvency and volatility dynamics of
energy-only markets while still requiring redispatch and uplift payments.
Lemmas~\ref{lem:risk_volume_instability} and~\ref{lem:price_cap_insolvency} show
that none of these families resolves the structural misalignment between
physical deliverability and financial responsibility.

At a deeper level, all three market families exhibit \textbf{broken links
between cost causation and cost recovery}. Fixed system costs are recovered
through volatile volumetric charges; scarcity-driven adequacy costs are smeared
across consumers regardless of their contribution to peak stress; and
vulnerable households face disproportionate exposure despite exerting limited
control over system risk. As a result, these designs are neither incentive
compatible nor equilibrium stable: those who impose costs are not the same as
those who bear them, leading to systematic unfairness, inefficient investment,
and weak participation incentives. Without trusted and fair participation,
the flexibility required for decarbonisation cannot be mobilised.

At the procurement level, legacy architectures implicitly operate in a
\emph{two-dimensional contract space},
\[
(\text{energy},\ \text{capacity/adequacy}),
\]
while treating service quality, spatiotemporal flexibility, and reliability as
externalities handled through ancillary services, ex-ante tenders, or emergency
interventions. Modern power systems with high renewable penetration and
millions of controllable devices require a \emph{third procurement axis},
\[
(\text{energy},\ \text{capacity/adequacy},\ \text{QoS/flexibility/reliability}),
\]
that is represented natively in the clearing and allocation logic.

This thesis proves that all three incumbent market families are
\textbf{mathematically fragile and physically non-robust}. For energy-only
designs, Lemma~\ref{lem:voll_sensitivity} shows that investment incentives and
scarcity rents are arbitrarily sensitive to the administratively chosen VoLL,
while Lemma~\ref{lem:surplus_welfare_mismatch} demonstrates that surplus-based
welfare maximisation fails to represent social welfare in the presence of
heterogeneous essential needs and income constraints. For energy-plus-capacity
designs, Lemmas~\ref{lem:price_cap_insolvency},
\ref{lem:risk_volume_instability}, and~\ref{lem:affordability_no_cap}, together
with Corollary~\ref{cor:no_free_lunch_retail}, establish that architectures
separating volumetric choice from tail-risk bearing necessarily generate either
insolvency cascades or unaffordable essential bills. These outcomes are not
policy accidents but structural consequences of the designs themselves.

In response, the thesis develops a first-principles redesign of the wholesale
and balancing layers, modelling the power system as a \textbf{digitally
regulated, event-driven cyber--physical control system}. Instead of relying on
ex-post corrective instruments such as price caps, redispatch, or uplift
payments, the proposed architecture embeds physical feasibility, stability,
fairness, and proportional responsibility directly into the clearing law.

A key conceptual shift is the treatment of prices as \emph{exogenous, bounded
control signals} rather than endogenous equilibrium outcomes. In legacy markets,
prices are expected to simultaneously allocate energy, signal scarcity, and
recover fixed costs. This thesis rejects that premise. Prices are instead
computed explicitly from the physical state of the system—tightness,
congestion, and reliability margins—and digitally regulated to ensure bounded
input–bounded output behaviour under uncertainty. This replaces price discovery
as the organising principle of market design with price regulation as a
cyber--physical control mechanism, while preserving decentralised participation.

The thesis makes six principal contributions.

First, it provides a \textbf{physically grounded, operational definition of
fairness} applicable to consumers, suppliers, generators, and system operators,
formalised as four enforceable conditions governing rewards, service delivery,
access under scarcity, and cost sharing.

Second, it introduces a \textbf{holarchic Automatic Market Maker (AMM)} that
functions as a network-native scarcity controller rather than a spot-clearing
auction. Prices become state-aware control signals responding to scarcity,
congestion, inertia, and reserve stress across time, space, and hierarchy.

Third, it develops the \textbf{Fair Play allocation mechanism} for shortage
conditions, which allocates limited energy according to contractual
entitlements, vulnerability, and historical contribution. A convergence result
establishes that long-run delivered service matches contracted QoS levels.

Fourth, it proposes a \textbf{three-dimensional contract framework}—magnitude,
timing sensitivity, and reliability—that realises the missing procurement axis
and links household- and device-level QoS positions to both shortage allocation
and supply-side remuneration.

Fifth, it introduces a \textbf{nested Shapley-value methodology} for generator
remuneration and cost allocation. A structural theorem shows that, under
substitutability and deliverability conditions, the nested allocation is
equivalent to the full generator-level Shapley value, enabling tractable
large-scale evaluation.

Sixth, it presents a \textbf{digitally regulated market architecture and its
experimental evaluation}. Using household-level observational demand data, a
two-region London–Glasgow corridor model, and a stylised GB-scale transmission
network, the thesis demonstrates that the AMM renders the \emph{total cost of
procuring the declared needs bundle an explicit design variable}. Procurement
cost is bounded below by strict generator cost recovery and above by an
affordability proxy conservatively taken as total revenues under the Baseline
LMP design.

Within these bounds, conservative AMM configurations deliver smooth
scarcity-responsive prices, materially fairer distributional outcomes, and
\emph{bounded-input bounded-output stability} under low-inertia and scarcity
conditions. Game-theoretic analysis shows that the AMM admits a well-defined Nash
equilibrium for each physical state, and that the Fair Play–compliant profile
constitutes an \emph{$\varepsilon$-shock-resistant Nash equilibrium} under
bounded perturbations in demand, renewable availability, and network
constraints—unlike the structural fragility observed in incumbent market
designs.

Taken together, these results demonstrate that \textbf{fairness, flexibility,
reliability, and stability can be embedded as programmable primitives within
wholesale electricity market design}, repairing the link between cost causation
and cost recovery and enabling a low-waste, investment-stable, high-electrified
power system.

% =========================================================
\chapter*{Statement of Originality and Copyright}
\phantomsection
\addcontentsline{toc}{chapter}{Statement of Originality and Copyright}
% =========================================================

\section*{Statement of Originality}
\phantomsection
\addcontentsline{toc}{section}{Statement of Originality}

I declare that this thesis is my own work and that it has not been submitted,
either in whole or in part, for the award of any other degree or qualification
at this or any other institution.

All sources of information, data, and ideas that are not my own have been
appropriately acknowledged through citation and reference. Where the work of
others has been used, this has been done in accordance with accepted academic
practice.

\section*{Copyright Declaration}
\addcontentsline{toc}{section}{Copyright Declaration}

The copyright of this thesis rests with the author. Unless otherwise indicated,
its contents are licensed under a Creative Commons
Attribution--NonCommercial 4.0 International Licence (CC BY-NC 4.0).

Under this licence, you may copy and redistribute the material in any medium or
format. You may also remix, transform, and build upon the material. This is
permitted provided that:

\begin{itemize}
    \item appropriate credit is given to the author;
    \item the material is not used for commercial purposes; and
    \item any adaptations are clearly indicated.
\end{itemize}

When reusing or sharing this work, the licence terms must be made clear to
others by naming the licence and providing a link to the licence text.
Where the work has been adapted, you should indicate that changes were made
and describe those changes.

For uses of this work that are not covered by this licence or permitted under
UK Copyright Law, permission must be sought from the copyright holder.

% ---------------------------------------------------------
% ACKNOWLEDGEMENTS
% ---------------------------------------------------------
\chapter*{Acknowledgements}
\addcontentsline{toc}{chapter}{Acknowledgements}

This work stands on the shoulders of giants and would not have been possible
without the support of a large number of people over the years, some of whom
are acknowledged below.

\section{Contributions to the thesis}
The following people directly contributed to the development of the thesis:

\begin{itemize}
    \item \textbf{Prof. Robert (Bob) Shorten} made the following contributions:
    \begin{itemize}
        \item Made Shaun aware of Automatic Market Makers (AMMs).
        \item Encouraged Shaun to explore how aspects of network theory and
        Quality of Service could be applied to electricity markets.
        \item Introduced me to Chris King.
    \end{itemize}

    \item \textbf{Prof. Pierre Pinson} made the following contributions:
    \begin{itemize}
        \item Introduced Shaun to core economic properties: individual
        rationality, incentive compatibility, budget balance, revenue
        adequacy, and economic efficiency. These form a central part of the
        foundation on which the design stands.
        \item Strongly emphasised that the neoclassical economic orthodoxy
        applied to energy should be challenged, and that we can think beyond
        the status quo.
        \item Provided extensive and valuable discussion of concepts and ideas.
    \end{itemize}

    Pierre, Bob, and Shaun independently realised that Automatic Market Makers
    could be formulated as a control system at roughly the same time.

    \item \textbf{Prof. Mark O'Malley} made the following contributions:
    \begin{itemize}
        \item Provided significant reviews of content over the four years.
        \item Highlighted the importance of revenue adequacy, resource
        adequacy, power system reliability, and related concepts.
        \item Emphasised that the work must be compared against Locational
        Marginal Pricing (nodal pricing) and zonal pricing to be taken
        seriously, which now forms part of the core framing.
        \item Provided connections with Julian Leslie, Linda Steg, Muireann
        Lynch, and others.
        \item Identified an error in my original Shapley approach, which had
        incorrectly assumed that transformers have polarity.
    \end{itemize}

    \item \textbf{Prof. Chris King} (Rest in Peace), then at Northeastern
    University, Boston, conducted early work on Automatic Market Makers for
    this application and provided valuable insight into the history of AMMs.
\end{itemize}

\section{Resources}
\begin{itemize}
    \item Moixa: provided one of the datasets.
    \item The Department of Electrical and Electronic Engineering at Imperial
    College London provided me with a MacBook, arranged by Professor Tim
    Green, on which all the work was conducted. When I accidentally spilled a
    pint of water over it while deep in experiments in June 2025, requiring a
    motherboard replacement, Mark O'Malley covered the cost, arranged by
    Güler Eroğlu.
    \item Nick Moult in the Dyson School of Design Engineering was excellent in
    handling all administrative matters.
    \item The IOTA Foundation funded the work.
    \item The Office for National Statistics (ONS) and the UK Statistics
    Authority: the ONS provided training enabling me to become a Digital
    Economy Act (DEA) Accredited Researcher, including completion of Safe
    Researcher training.
\end{itemize}

\section{Special thanks --- An Chlann in \'Eire and further afield}
I am blessed to be extremely close to my incredibly supportive family, and I am
never surprised to discover new cousins and extended family across the world.
To my mum and dad, Elizabeth and Tommy, thank you for always being loving
parents. To my sisters, Claire and Sarah, thank you for being wonderful
sisters. To my godmother and aunty Bried, thank you for always being there when
the rest of the family becomes a little too much.

\section{Special thanks --- Robert Shorten}
Bob, I do not have the words to express my gratitude for all that you have done
for me over the years. Our work on the bike brought us across the world, from
Dublin to Yale to Limerick to London (I never did tell you one of the reasons I
moved to London\ldots). I hugely admire your Trojan work ethic and what you are
building in the Dyson School. I hope we can continue to do exciting work
together for the benefit of people today and those of tomorrow.

\section{Special thanks --- Mark O'Malley}
Mark, thank you for shaping my entire career from Energy Needs Ireland 2013
through to the present day. I do not know if you remember the 21 of us in your
office in UCD, when you told us that ``we would never want any doors opened
again'' due to your contacts. I do not know if you recall that there was one
dissenting voice who asked, ``But how will we know that what we are saying
stands on the strength of our own contributions?'' With this thesis, and the
last twelve years of work since you posed the question ``What is the Smart
Grid?'', I now feel confident that I can answer it. I hope we continue working
together, because there is still a great deal of work to be done.

\section{Special thanks --- David Timoney and everyone who worked on the bike}
Thank you to David Timoney for his kindness, patience, and diligence during my
five years at University College Dublin. Well done for your efforts in trying
to keep up with Bob, both generally and on the bike work in particular. Thank
you also for instigating the Master's Programme in Energy Systems Engineering,
which has been foundational to my career. Further thanks to Giovanni Russo,
Francisco Pilla, Hugo Lhachemi, Rodrigo, and Wynita for their work on the bike.

\section{Special thanks --- Pierre Pinson}
Thank you for your work and kindness during challenging times. I hope that one
day we will work together again.

\section{Special thanks --- All my teachers in Co. Donegal}
I have been blessed with excellent teachers throughout my primary and secondary
education. Special thanks to Kathleen McConigley, Margaret Bonner, Bríd Barr,
Mairéad McDaid, Pádraig Curley, Maureen Curran, Ciarán McLoughlin, Kieran
McTaggart, Andrew Kelly, and Anthony Harkin. In particular, Margaret Bonner,
thank you for stopping me from studying business and redirecting me towards
engineering.

\section{Special thanks --- ReCosting Energy team}
The ReCosting Energy team shone a light on many of the fairness issues in the
energy system, and their work strongly influenced the thinking behind this
thesis.

\section{Special thanks --- The Moixa team}
Thank you for your generosity with time and knowledge over nearly a decade,
and for building something truly special. In particular, thank you to Simon
Daniel and Chris Wright.

\section{Special thanks --- The London--Irish family}
Thank you to Bernie, Jon, Ann Marie, Aunty Una, and the extended Gibbons family
in London for taking me in when I arrived eight years ago with a suitcase, and
for being there ever since.

\section{Special thanks --- Mentors}
I have been fortunate to have excellent mentors throughout my career. I wish
to specifically acknowledge Mr Edward Gunn, Tony Whittle, and Patrick Liddy.

\section{Special thanks --- Louise Spurgeon and Vanessa Maugey}
To my ADHD brain mentors, Louise and Vanessa, who stepped in at exactly the
right times during the PhD to calm me down, keep me fighting, and push me
through the system: I am deeply grateful for your empathy, understanding,
kindness, and laughter.

\section{Special thanks --- Kay Rivers and Paul Lane}
When dark days came in 2022 and I felt the PhD had broken me, I was deeply
grateful to have trained practitioners like Kay and Paul to remind me that
there is a whole world beyond looking inwards.

\section{Special thanks --- Abhinay Muthoo}
Thank you for coming into my life at just the right time and giving me
confidence and motivation. I sincerely hope we will work together in the
future.

\section{Special thanks --- JP McManus}
I was deeply privileged to be awarded the All-Ireland Scholarship in 2012, for
students from disadvantaged socioeconomic backgrounds in Ireland. This
scholarship provided generous funding throughout my first four years at
University College Dublin and enabled me not only to focus fully on my studies
but also to study abroad in Melbourne for six months in 2015. I wear this award
with immense pride and remain profoundly grateful to JP and his family for
their generosity.

\section{Tribute and dedication}
I dedicate any positive contribution of this thesis to those who have gone
before its publication; any negative contributions are entirely my own.
Specifically, my uncles Gerard and Jim Sweeney; my grandparents, particularly
Sally Gibbons, the only one I truly had the chance to know; and my first formal
employer, Sarah McGettigan, who sadly passed away in January 2025. Sarah taught
me the value of a fair day's work for a fair day's pay, peeling potatoes in the
dark and ensuring Northerners got their chips on the 4th of July after a day at
the beach.

Very sadly, a former colleague at Moixa, Toma Moldovan, passed away as COVID was
coming to an end. His final letter can be heard here:
\newline
\href{https://www.mixcloud.com/michaellanigan37/playlists/toma-moldovan/}{https://www.mixcloud.com/michaellanigan37/playlists/toma-moldovan/}

\vspace{10mm}

Rest in power, soldiers. The fight goes on.

\newpage
\pagenumbering{arabic}

\chapter{Introduction}

\section{Motivation}

Contemporary electricity markets are not failing at the margins; they are failing
by design. At a high level, almost all liberalised systems fall into one of three
architectural families:

\begin{enumerate}
    \item \textbf{Energy-only, marginal-cost designs}, exemplified by US-style
    locational marginal pricing (LMP), in which a single energy price (plus
    scarcity adders) is expected to provide both operational and investment
    signals;

    \item \textbf{Energy-plus-capacity, heavily regulated designs}, exemplified
    by the GB model, in which price caps, capacity auctions, Contracts for
    Difference (CfDs), and a dense layer of corrective schemes are added on top
    of an energy market that is known to be insufficient on its own; and

    \item \textbf{Zonal designs}, increasingly proposed in Europe, which
    aggregate nodes into a small number of politically negotiated zones and
    then rely on redispatch and uplift payments to repair the mismatch between
    zonal prices and underlying network physics.
\end{enumerate}

This thesis shows that \emph{all three} design families are structurally
fragile. On the \emph{energy-only} side, Lemma~\ref{lem:voll_sensitivity}
demonstrates that investment signals and scarcity rents in LMP-style designs
depend critically on an administratively chosen Value of Lost Load (VoLL), with
no internal mechanism that pins down a ``correct'' value. Changing VoLL changes
the implied optimal capacity, the present value of scarcity rents, and the
distribution of value between consumers and generators.
Lemma~\ref{lem:surplus_welfare_mismatch} then shows that the standard
surplus-based justification for LMP --- treating consumer plus producer surplus
as a proxy for social welfare --- breaks down once we admit heterogeneous
essential needs, vulnerability, and income constraints. Surplus-maximising
allocations can be systematically misaligned with welfare once we care about
who keeps the lights on, not just aggregate willingness to pay.

On the \emph{energy-plus-capacity} side,
Lemmas~\ref{lem:price_cap_insolvency} and~\ref{lem:risk_volume_instability}
show that any architecture which:
(i) separates volumetric demand choice from tail-risk bearing, and
(ii) constrains retail price responses under volatile input costs, is
\emph{mathematically guaranteed} to generate insolvency cascades or the need for
continual state intervention. Lemma~\ref{lem:affordability_no_cap} and
Corollary~\ref{cor:no_free_lunch_retail} further prove that, in the presence of
unbounded wholesale price shocks, no retail arrangement can simultaneously
guarantee supplier solvency \emph{and} affordability of essential demand. There
is no clever combination of caps, tariffs, and capital buffers that rescues the
current retail architecture from this structural trade-off. Zonal markets,
meanwhile, inherit these insolvency dynamics while still requiring internal
redispatch and uplift payments, failing to resolve the underlying misalignment
between physical deliverability and financial responsibility.

From a game-theoretic perspective, these architectures also lack a
\emph{shock-robust Nash equilibrium}: once fuel-cost volatility, renewable
uncertainty, and finite balance sheets are acknowledged, small perturbations in
physical or financial conditions create profitable unilateral deviations. Any
candidate equilibrium is fragile with respect to shocks, requiring either
ad-hoc interventions or ex-post redistributions to prevent insolvency or
unacceptable price spikes.

At a deeper level, these failures share a common root: the \textbf{link between
who imposes system costs and who pays them is broken}. Fixed system costs (for
example, reserves, stability services, and black-start capability) are often
recovered through volatile volumetric charges. Scarcity-driven capacity costs
are smeared across customers regardless of their contribution to peak stress.
Households with limited agency over their housing, transport, or heating
choices routinely pay a disproportionate share of costs relative to income,
while large, flexible loads can externalise much of the risk they impose. Those
who create the need for capacity and reserves are not the same as those who
bear the bill. This is not just an efficiency problem; it is a fairness problem.

At the procurement level, legacy architectures effectively operate in a
\emph{two-dimensional} space:
\[
(\text{energy},\ \text{capacity/adequacy}),
\]
leaving \emph{quality of service, spatiotemporal flexibility, and reliability}
to be handled by ancillary markets, ex-ante tenders, and emergency measures.
Modern power systems, with high renewable penetration and millions of
controllable devices, require a \emph{third} procurement axis,
\[
(\text{energy},\ \text{capacity/adequacy},\ \text{QoS/flexibility/reliability}),
\]
that is natively represented in the clearing logic rather than bolted on
afterwards. Failing to procure along this third axis is a central reason why
legacy markets misallocate risk, fail to sustain fair participation, and
produce payoff landscapes in which stable, shock-resistant equilibria are hard
to sustain.

A further structural weakness is that existing markets are only loosely coupled
to the \emph{physical laws} that govern electricity systems. In practice,
Kirchhoff’s laws, Ohmic losses, voltage limits, inertia margins and stability
constraints are enforced by system operators and constraint solvers, while the
market layer treats electricity as a scalar commodity priced in discrete
intervals. Physics appears as ex-post redispatch, uplift, balancing and
ancillary-service payments, rather than as the primary object of the pricing
logic itself. As a result, prices often fail to convey the locational, temporal
and stability-related information needed to co-ordinate behaviour in a
high-renewable, low-inertia system.

At the same time, the underlying system is becoming \emph{distributed}. Millions
of devices --- EV chargers, heat pumps, batteries, rooftop PV, flexible
industrial loads --- are connected at the grid edge, equipped with sensors,
communications and controllable inverters. Structurally, the electricity system
increasingly resembles a weighted graph of interacting agents, closer to the
internet than to a classical, centralised utility. Local actions propagate
across a network according to physical flow laws, while congestion, voltage
stress and frequency events emerge endogenously. A market architecture that
clears in rigid time blocks and treats prices as after-the-fact accounting
signals is ill-suited to co-ordinating such a graph-structured,
event-driven system.

In other words, the system fails not because of mismanagement, weak regulation,
or imperfect competition --- but because its underlying design makes stability
and fairness mathematically impossible. When fixed costs are recovered through
volatile marginal prices, and scarcity costs are recovered without regard to
who drives scarcity, both risk and burden are allocated arbitrarily. When
physical constraints and stability margins are only weakly reflected in prices,
actors lack clear, trusted signals about when and where their behaviour matters.
Over time, this undermines trust in the system, weakens willingness to
participate in new programmes, and makes it harder to mobilise the very
flexibility that the energy transition requires. From a strategic viewpoint, it
also means that even if a short-run Nash equilibrium exists in a simplified
model, it is fragile with respect to shocks in demand, renewables, or network
constraints.

This erosion of trust and participation is not an abstract concern. Delivering
a deeply decarbonised, electrified energy system requires millions of households
and businesses to participate actively: shifting EV charging, reshaping heating
demand, allowing devices to be controlled within comfort bounds, and investing
in storage and flexibility. If people experience the system as unpredictable,
opaque, or unfair --- if they see that those who impose costs are not the ones
who pay them --- they are unlikely to enrol their assets, consent to digital
control, or support ambitious climate policy. Fairness is therefore not an
optional ethical add-on; it is a \emph{precondition for participation}.

Participation, in turn, is a precondition for avoiding some of the most harmful
outcomes of climate change and energy poverty. Without flexible, demand-side
participation, decarbonisation must either rely on overbuilt supply and
networks --- which pushes costs up and keeps fuel poverty entrenched --- or
accept higher levels of curtailment and wasted renewable energy. A system that
cannot link cost causation to cost recovery, and cannot allocate risk in ways
that are perceived as legitimate, will struggle to deliver:

\begin{itemize}[leftmargin=*]
  \item a \textbf{zero-waste energy system}, in which available renewable
  energy and flexibility are fully utilised rather than curtailed;

  \item \textbf{low and stable electricity costs}, which are fundamental for
  the large-scale electrification of transport and heating needed to stop
  releasing greenhouse gases into the atmosphere; and

  \item the elimination of \textbf{fuel poverty} as a structural feature of
  the energy system, rather than a by-product to be patched with ad hoc
  subsidies.
\end{itemize}

What was once a theoretical warning is now visibly and empirically true. The
United Kingdom’s electricity market has evolved into a patchwork of corrective
instruments --- price caps, social tariffs, capacity auctions, Contracts for
Difference, bailout mechanisms --- each introduced to compensate for specific
failures in the underlying architecture. Yet, despite these interventions, the
system still generates volatility, insolvency, inequitable cost allocation, and
public distrust, precisely as predicted by the mathematical structure.
Successive layers of corrective instruments have been added to mask specific
failures in the underlying market design rather than to address the systemic
causes. The result is a patchwork of overlapping schemes that is increasingly
decoupled from the physical and economic realities of the electricity system.

Against this backdrop, the central premise of this thesis is that
\textbf{fairness and its delivery must become a core design primitive of the
market architecture}. Electricity markets must be re-specified as
cyber--physical coordination mechanisms, grounded in physics, fairness, and
digital capabilities, rather than as lightly regulated extensions of historical
commodity exchanges. This requires clearing mechanisms that are \emph{event
driven and network native}: they respond directly to physical events (changes
in flows, voltages, inertia margins, reserves) on a graph-structured system,
and use prices as state-aware control signals rather than as ex-post cost
allocation devices. Re-linking cost causation and cost recovery in a way that
is explainable, auditable, and enforceable is essential not only for economic
efficiency, but for restoring trust and participation --- and therefore for
delivering a zero-waste, low-cost, electrified energy system compatible with
rapid decarbonisation. Later chapters show that, under suitable regularity and
incentive conditions, the proposed Automatic Market Maker (AMM) architecture
admits a well-defined Nash equilibrium for each physical state, and that a
Fair Play--compliant strategy profile can be made \emph{shock-resistant} to a
wide range of demand, renewable, and network perturbations.

\section{Objectives}

The thesis pursues eight core objectives:

\begin{itemize}
    \item Develop a physically grounded and operationally meaningful definition of fairness.
    \item Create an asynchronous, event-based clearing mechanism capable of continuous, state-aware operation.
    \item Design a digital regulation architecture consistent with real-time algorithmic governance.
    \item Define a ``zero-waste'' electricity system and develop tools to infer efficiency.
    \item Integrate wholesale, retail, and balancing markets into a coherent unified framework.
    \item Ensure fair compensation to generators using scalable, network-aware Shapley-value principles that overcome classical intractability.
    \item Formulate the AMM--Fair Play system as a game between strategic participants and the mechanism, and establish conditions under which Nash equilibria exist and are locally shock-resistant.
    \item Build a rigorous data and simulation framework to evaluate the resulting system.
\end{itemize}

\section{Research Question}
\label{sec:research_question}

\textbf{How can a national electricity market be redesigned from first principles
to operate fairly, efficiently, and continuously in real time, via event-driven,
state-aware clearing that respects physical constraints, supports two-way power
flows, ensures zero-waste utilisation of system resources, and admits a stable,
shock-resistant equilibrium under realistic uncertainty?}

\section{Scope}

This thesis focuses on the economic and algorithmic design of market structures
and payment flows between consumers, suppliers, and generators. Network charging
mechanisms (DUoS, TNUoS) and infrastructure financing models are out of scope,
except where they provide contextual constraints or interact indirectly with
market operation.

\section{Claimed Contributions}
\label{sec:contributions}

This thesis makes the following original contributions to electricity market
design, cyber--physical systems, and fairness-aware control. Collectively, they
constitute a new architecture for how electricity markets can be operated,
coordinated, and digitally regulated under conditions of high uncertainty,
observability, and participant diversity.

\begin{itemize}

\item \textbf{Physically grounded, operational definition of fairness as a system constraint.}  
The thesis develops a real-time, physically rooted fairness formulation based on
(i) protection of essential needs, (ii) incentive-aligned flexibility rewards,
(iii) fair access rotation and historical equity, and (iv) proportional
responsibility for system stress. Fairness becomes a programmable system constraint,
not an ex post corrective overlay (caps, subsidies, compensation).

\item \textbf{Electricity as a three-dimensional service: Magnitude $\times$ Impact $\times$ Reliability.}  
The work transforms electricity products from simple energy volumes (kWh) to
contracted service bundles characterised by quantity, scarcity timing, and
probability of access under shortage. This 3D product space underpins QoS tiers,
subscription contracts, household classification (P1--P4), and reliability as a
contractible, earned attribute.

\item \textbf{Automatic Market Maker (AMM) as a cyber--physical scarcity controller.}  
The thesis proposes a holarchically organised AMM that synthesises instantaneous,
forecast, and network-based scarcity into time-, space-, and role-specific
price, priority, and access signals. Prices are bounded, monotone in scarcity,
and self-corrective, satisfying BIBO stability. Unlike bid-driven spot markets,
the AMM acts as a digital control layer that regulates scarcity rather than
merely discovering it.

\item \textbf{Voltage as a physical shadow price, AMM price as its digital counterpart.}  
The thesis introduces a new interpretation of measured feeder voltage as a
\emph{physical shadow price} of local supply scarcity (undervoltage) or surplus
(overvoltage). This physical signal is mapped directly into AMM price updates,
creating a \emph{digital shadow price} that activates neighbour-level flexibility
(import, export, charge, discharge), without centralised optimisation. This yields
a stabilising, fairness-aware alternative to Volt/VAR control or OPF-derived DLMPs,
and makes local network physics directly govern digital price behaviour.

\item \textbf{Fair Play: A real-time allocation mechanism for differentiated priority and historical fairness.}  
Fair Play formalises how scarce resources are allocated when fairness, QoS tier,
and flexibility history must be jointly respected. It enables proportional,
non-discriminatory allocation under network and temporal constraints, while
maintaining ex-ante incentive compatibility and avoiding arbitrary rationing.
A law-of-large-numbers style result (the \emph{service-level fairness theorem})
shows that, under Fair Play, long-run delivered service converges to the contracted
share for each QoS tier: premium means premium, basic means basic, in realised
outcomes.

\item \textbf{Dynamic capability bidding for generators and grid-edge devices.}  
Generators, EVs, heat pumps, and storage express their availability as
time-stamped capability profiles—encoding ramp rates, charge/discharge limits,
flexibility windows, minimum runtimes, or notification times—making dispatch and
bidding converge into a single cyber--physical object.

\item \textbf{Nested Shapley value: scalable, role-aware allocation of scarcity rents and reliability value.}  
A hierarchical Shapley method is developed that preserves ranking, respects
physical limits, and enables generator-level remuneration at realistic system
scale. Using network-aware clustering, feasibility pruning, and time-separable
evaluation, it provides physically meaningful Shapley values with tractable
complexity. A structural theorem shows that, under explicit substitutability and
deliverability conditions on the generator clusters, the nested allocation is
\emph{exactly equivalent} to the full generator-level Shapley value; numerical
experiments on an OPF-based network game validate this equivalence and quantify
the computational gains.

\item \textbf{Game-theoretic characterisation and shock-resistant Nash equilibrium.}  
The AMM--Fair Play architecture is formulated as a repeated game between
generators, retailers, and the mechanism. Under mild regularity conditions, the
state-contingent game is shown to admit at least one pure-strategy Nash
equilibrium, and the Fair Play--compliant strategy profile is proved to form an
$\varepsilon$-shock-resistant Nash equilibrium on a neighbourhood of physical
shocks. This provides a formal notion of strategic stability that legacy designs
lack.

\item \textbf{Digitally regulated market architecture with continuous auditability.}  
The thesis designs a governance framework in which compliance, risk allocation,
and policy protections (priority classes, caps, essential guarantees) are
embedded algorithmically within the market engine itself. This enables continuous
audit trails, ex-ante regulatory assurance, and machine-verifiable legitimacy.

\item \textbf{Asynchronous, event-driven clearing and zero-waste inference.}  
An event-based clearing structure replaces periodic auctions, enabling
scarcity-triggered activation of flexibility and live inference of ``zero-waste
conditions'' (unused feasible supply, unused flexibility, missed opportunity).

\item \textbf{Experimental validation under conservative constraints.}  
Validation proceeds in three stages. First, observational UKPN smart meter
traces with EV overlays are used to test the behavioural plausibility of
product differentiation (P1--P4) and to shape the synthetic residential demand
profiles used in later experiments. Second, a stylised London--Glasgow
two-region corridor is used to demonstrate holarchic value propagation,
geographically coherent Shapley allocation, and congestion-informed fairness
behaviour. Third—and only at this stage—the full experimental comparison is
conducted using GB-scale national demand and generation time series, a
clustered transmission network, and synthetic but physically rooted
product-level demand profiles shaped by wind availability. Even with adaptive
features deliberately disabled (fixed subscription settings, static Shapley
weights, no learning or path-dependence), the AMM architecture delivers
tighter and more policy-aligned prices, stronger capacity and bankability
signals, broader participation, smoother volatility, and materially fairer
distributional outcomes than an LMP-style baseline—demonstrating the strength
of the architecture rather than parameter tuning.

\end{itemize}

\section{Thesis Structure}

The thesis proceeds from historical and conceptual background, through philosophy, problem definition, design, implementation, evaluation, and implications:

\begin{itemize}

    \item \textbf{Chapter 1: Introduction}  
    Motivates the problem of electricity market failure, sets the objectives and research
    questions, clarifies the scope and claimed contributions, and provides a high-level
    roadmap of the thesis. It introduces the central reframing of electricity markets as
    continuously online, event-driven systems rather than discrete spot-clearing mechanisms,
    and motivates the need for architectures that provide bounded, continuous price signals
    and system liquidity under physical and economic uncertainty.

    \item \textbf{Chapter 2: Background}  
    Establishes the historical, institutional, and conceptual context of electricity systems and markets. It traces the evolution from early public utilities to liberalised markets, examines energy security, climate policy, digitalisation, and financing arrangements, and highlights the absence of an overall architect for the energy transition.

    \item \textbf{Chapter 3: Literature Review}  
    Reviews core bodies of knowledge across electricity market design, renewable-dominated power systems, cooperative game theory, fairness in energy, local and peer-to-peer markets, digital and event-based control, and broader economic and policy paradigms. It synthesises the main gaps that motivate a new market architecture and positions this thesis within that landscape.

    \item \textbf{Chapter 4: Problem Definition, System Realities, and Solution Concept}  
    Describes the changing nature of the electricity system, misaligned stakeholder incentives, and the physical realities ignored by current market mechanisms. It explains why existing designs cannot scale, articulates the fairness gap, and summarises the problem in a structured way that points toward an event-based, locationally grounded, fairness-aware solution concept.

    \item \textbf{Chapter 5: System Requirements (From First Principles)}  
    Derives system requirements from first principles across physical, economic, digital, behavioural, and fairness domains. It formalises what a viable market–control architecture must satisfy in order to be resilient, fair, financially adequate, digitally enforceable, and implementable in practice.

    \item \textbf{Chapter 6: Design Philosophy and Research Positioning}  
    Sets out the philosophical and conceptual stance of the thesis. It treats fairness as a foundational design driver, reframes electricity as a service rather than a commodity, interprets markets as control systems, and argues for digital regulation, UX, and zero-waste principles as core design levers. It also clarifies the research positioning within engineering, economics, and policy debates.

    \item \textbf{Chapter 7: Methodology}  
    Details the research approach, including the design-science methodology, representation of energy as a contract (magnitude, timing, reliability), data sources and engineering pipeline, modelling of flexible, timing-sensitive device participation, and the validation and evaluation strategy. It also explains how the thesis overcomes Shapley intractability through nested and physically constrained formulations, and maps research questions to methods.
    
    \item \textbf{Chapter 8: Market Designs and Operating Scenarios}  
    Describes the data and physical foundations, and sets out the proposed continuous
    online market instance as a cyber--physical system. It defines the event-driven clearing
    logic, forward and real-time integration, bidding parameters, contract structure,
    cyber--physical synchronisation, and operating regimes (Too Much, Just Enough, Too
    Little). The chapter highlights how continuous pricing and settlement replace discrete
    clearing events, enabling persistent system liquidity across time, space, and operating
    conditions.

    \item \textbf{Chapter 9: Definition of Fairness}  
    Develops a formal fairness framework for electricity markets. It introduces behavioural and theoretical foundations, defines the system model and fairness axioms, derives operational fairness conditions, connects them to existing literature, and proposes consumer and generator-oriented fairness metrics. It also previews the Fair Play mechanism.

    \item \textbf{Chapter 10: The Automatic Market Maker (AMM)}  
    Defines the AMM and its holarchic architecture, explaining how instantaneous, forecast, and network scarcity are integrated into a unified pricing and allocation mechanism. It introduces the interpretation of measured voltage as a \emph{physical shadow price} of local scarcity, and AMM price as its \emph{digital shadow price}, creating a stabilising cyber--physical feedback loop. It analyses the control-theoretic stability of the digital AMM, frames it as a scarcity-control layer, and discusses its interaction with time-coupled requests, subscription products, and digital enforceability.

    \item \textbf{Chapter 11: Mathematical Framework and Implementation}  
    Provides the core mathematical formulation of the proposed architecture. It formalises the fairness mapping and Fair Play allocation mechanism (including the service-level fairness theorem), develops the Shapley-based generator compensation framework (including the nested-equivalence theorem), presents AMM control equations and stability conditions, and introduces dynamic capability profiles, forecasting models, zero-waste efficiency inference, and key analytical properties of the AMM-based market design, including Nash equilibrium existence and local shock-resistance.

    \item \textbf{Chapter 12: Experiment Design}  
    Specifies the experimental programme used to evaluate the architecture. It defines the research questions and hypotheses, treatments and benchmark mechanisms, outcome metrics, scope and conservatism of the design, experimental procedures, inference and decision thresholds, and the pre-analysis plan for the paired simulations.

    \item \textbf{Chapter 13: Results}  
    Reports empirical results for procurement efficiency, price-signal quality, investment adequacy and bankability, participation and competition, revenue sufficiency and risk allocation, and distributional fairness. It compares AMM-based allocation with benchmark mechanisms, includes sensitivity and robustness analysis, and examines generator-level Shapley allocations and fairness metrics. It also presents numerical validation of the service-level fairness result and the nested Shapley equivalence on realistic network instances.

    \item \textbf{Chapter 14: Discussion and Systemic Implications}  
    This chapter synthesises the conceptual and empirical contributions of the thesis.
    It interprets the results of the simulated case studies against the central
    research question and the six headline hypotheses (H1--H6), evaluates their
    robustness under stress-tested conditions, and diagnoses their implications for
    the design of fair, programmable electricity markets operating as continuously
    online, event-driven allocation systems rather than discrete settlement regimes.

    \item \textbf{Chapter 15: Conclusion}  
    Reframes fairness as an operational design rule, summarises how the thesis moves from wholesale markets to holarchic digital clearance, and reflects on evidence of locational, temporal, and reliability value distortion. It distils the main contributions, outlines future research directions, and concludes with a broader vision for fair, digitally regulated, electrified societies.

    \item \textbf{Appendices}  
    Provide detailed dataset documentation, algorithmic and data pipeline descriptions (including Fair Play and synthetic flexible events), extended results and statistical outputs, notation tables, clarification of the experimental AMM configuration, and an epilogue that situates the thesis within wider debates on growth, democracy, finance, and digital governance.

\end{itemize}

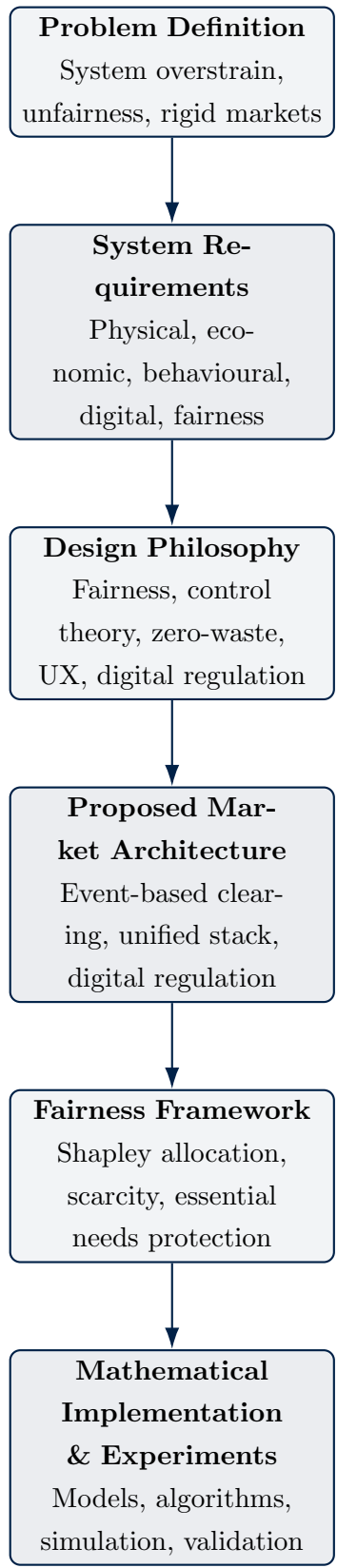
\begin{figure}[H]
\centering
\begin{tikzpicture}[
    node distance=1.2cm and 2.2cm,
    box/.style={
        rectangle,
        rounded corners,
        draw=ImperialBlue,
        thick,
        align=center,
        minimum width=4cm,
        minimum height=1.2cm,
        text width=4.2cm,
        font=\small
    },
    arrow/.style={
        draw=ImperialBlue,
        thick,
        -{Latex[length=3mm, width=2mm]}
    }
]

% Nodes
\node[box, fill=ImperialBlue!5] (problem) { \textbf{Problem Definition} \\ System overstrain, unfairness, rigid markets };
\node[box, fill=ImperialBlue!8, below=of problem] (requirements) { \textbf{System Requirements} \\ Physical, economic, behavioural, digital, fairness };
\node[box, fill=ImperialBlue!5, below=of requirements] (philosophy) { \textbf{Design Philosophy} \\ Fairness, control theory, zero-waste, UX, digital regulation };
\node[box, fill=ImperialBlue!8, below=of philosophy] (architecture) { \textbf{Proposed Market Architecture} \\ Event-based clearing, unified stack, digital regulation };
\node[box, fill=ImperialBlue!5, below=of architecture] (fairness) { \textbf{Fairness Framework} \\ Shapley allocation, scarcity, essential needs protection };
\node[box, fill=ImperialBlue!8, below=of fairness] (implementation) { \textbf{Mathematical Implementation \& Experiments} \\ Models, algorithms, simulation, validation };

% Arrows
\draw[arrow] (problem) -- (requirements);
\draw[arrow] (requirements) -- (philosophy);
\draw[arrow] (philosophy) -- (architecture);
\draw[arrow] (architecture) -- (fairness);
\draw[arrow] (fairness) -- (implementation);

\end{tikzpicture}
\caption{Conceptual flow from problem to implemented solution}
\label{fig:design_flow}
\end{figure}

\chapter{Background}
\label{chap:background}

This chapter provides the historical, economic and institutional context for the
thesis. It traces the evolution of the electricity grid from its early
engineering roots to today’s liberalised, digital and increasingly
decarbonised system; reviews the economic theories that underpin commodity
pricing and electricity market design; summarises key developments in climate
science and the net zero agenda; and examines the fragmented financing and
governance architectures through which the energy transition is currently
being delivered. It then introduces the conceptual tools---automatic market
makers, holarchies, game theory and fairness---that will be formalised and
operationalised in later chapters.

The aim is not to offer a complete history of the electricity sector or
economic thought. Rather, the goal is to identify the specific structural,
behavioural and governance features that motivate the need for a new market
architecture and fairness framework.

% ---------------------------------------------------------
\section{History and Transformational Impacts of the Electricity Grid}
\label{sec:grid_history}

Energy cannot be created or destroyed within a closed system; it can only be
converted from one form to another. Access to a reliable and affordable source
of energy enables people to cook, heat and light their homes, move goods and
people, and run the machinery of modern life. The 19th and 20th centuries were
largely a debate about how to generate and distribute electricity, with key
figures including Faraday, Volta, Edison and Tesla.

The first public supply system powered by a central power station in the UK
was the Holborn Viaduct scheme developed in 1878 by the City of London
Corporation in collaboration with Siemens. It provided street lighting using
arc lamps powered by a dynamo driven by a coal-fuelled steam engine.

In the late 1880s and early 1900s, many independently operating systems
emerged, largely focused on street lighting. These used a variety of primary
energy sources and prime movers: steam engines (coal), waterwheels (hydro),
combustion engines (gas), and early wind technologies such as the Brush
turbine in Cleveland, USA (1888). There was a political and economic push
towards interconnection of these systems to benefit from economies of scale in
larger power plants, promote reliability, and meet increasing demand.

The London Electricity Supply Act of 1908, backed by financial groups and
industrialists, aimed to rationalise electricity supply in London. The London
Power Company was established in 1912 and built large coal-fired power
stations along the Thames including Deptford, Bow and Battersea. To enable
interconnection, it was necessary to standardise the electrical
characteristics of generation and transmission. The chosen standard was a
50\,Hz AC, 132\,kV system. This made it possible to use electricity for a
wider range of activities, including powering tramways and factories.

In 1926, the UK Central Electricity Board was created with the goal of
standardising, centralising and interconnecting electricity supply in
Britain. This led to the ``gridiron'' system across England and Wales, the
first truly national grid anywhere in the world. It reduced generation costs
by around 40\%, enabled coal to be burned at the pithead with electricity
transported to where it was needed, and laid the foundation for later
integration of nuclear and renewable generation.

Initially, municipalities often paid for electricity per ``lamp-hour'', with
energy bills determined by:
\[
  \text{number of lamps installed} \times \text{hours used} \times \text{unit rate}.
\]
Usage could be estimated, or measured using early instruments such as the
Wright demand indicator, which recorded the hours that current flowed. The
development of the Ferranti and Thomas meters allowed electricity to be
measured in ampere-hours or watt-hours, enabling actual consumption to be
billed.

Bills were sent periodically; meter readers visited households and businesses,
read meters and often acted as bill collectors. Prepayment or coin-operated
meters were also widespread. Post-WWII, a programme of rural electrification
brought electricity close to 100\% of households, supplying domestic lighting,
irons, radios, cookers, heaters, fridges and washing machines. Voltage
standards settled at 240\,V single phase for domestic users and 415\,V
three-phase for industry.

Technical developments in the electricity grid continued and its wider impacts
revolutionised virtually every part of life and business, giving countries
with centralised grids competitive economic advantages. Industrial processes
were transformed: electricity replaced waterwheels and steam engines powering
looms and spindles in the textiles industry; electric arc technologies enabled
steel recycling; fertiliser and mechanisation supported increases in
agricultural productivity. Appliance manufacturers scaled to mass-produce
domestic devices, creating new supply chains and new types of jobs, including
electrical engineers and utility workers.

In short, the grid was not merely a technical project; it was a socio-technical
infrastructure that reshaped patterns of work, consumption and everyday life.

% ---------------------------------------------------------
\section{Economics, Commodity Pricing and Market Liberalisation}
\label{sec:econ_commodity}

Economics is a broad school of thought, with the term itself dating back to
Xenophon in Ancient Greece. Microeconomics studies the behaviour of individual
agents and how they interact to allocate scarce resources. It is concerned
with decision making by consumers (demand) and firms (supply), price
formation, resource allocation and efficiency, market structures, and welfare
outcomes (consumer/producer surplus, equity and efficiency).

Commodities are standardised, tradable goods derived from resources. Examples
include wheat, copper, oil and---in some formulations---electricity. There are
differing views within economics on how to price such commodities.

Classical economics (labour theory of value), pioneered by Adam Smith, David
Ricardo and Marx, determines prices based on costs. Neoclassical economics
emerged in the 1870s with Jevons, Walras and Menger, with prices emerging from
supply and demand. In a competitive market, the price of a good is set by the
marginal cost of the last unit produced. Commodity markets for grains existed
on the Chicago Board of Trade from 1848 and for metals on the London Metal
Exchange from 1877.

The theory of marginal pricing says that identical commodities should sell for
the same uniform clearing price. Consumer surplus is the difference between
what consumers are willing to pay and what they actually pay. Producer surplus
is the difference between the market price and producers' marginal cost, with
low-cost producers receiving inframarginal rents. The theory says that social
welfare is maximised when consumer and producer surplus are maximised. At this
point, the market is said to be economically efficient and Pareto efficient:
no individual can be made better off without making someone else worse off.

This framework depends on strong assumptions:
\begin{itemize}
  \item each unit of the commodity is homogeneous;
  \item consumers and producers act rationally to maximise utility or profit;
  \item market participants have perfect or near-perfect information;
  \item no market participant can exert sustained market power;
  \item there are no unpriced externalities (environmental or social);
  \item markets are contestable and entry/exit is relatively frictionless.
\end{itemize}

Bids from producers should represent short-run marginal costs (fuel and
variable operating costs), with fixed costs (capital and overheads) recovered
from inframarginal rents. When market prices are not high enough on average,
this creates the \emph{missing money} problem: there is insufficient revenue
to support investment in capacity that is only needed in peak periods.

\subsection{Keynesian Public Utility Thinking}

Keynesian economics arose in 1936 as a response to the Great Depression.
Keynes argued that neoclassical ideas about self-correcting markets failed to
explain mass unemployment. Markets do not always self-correct; aggregate
demand drives the economy. On commodities, Keynes argued that commodities
essential to welfare (electricity, housing, staple foods) should be shielded
from market volatility. He viewed commodity markets as unstable due to
inelastic supply and demand, and advocated buffer stocks, price supports and
long-term contracts.

Post-war electricity in the UK followed a Keynesian public-utility model. The
1947 Electricity Act created a monopoly structure with generation and
transmission under the British Electricity Authority (later the Central
Electricity Generating Board, CEGB) and 12 Area Boards responsible for
distribution and retail in defined geographic areas. The CEGB was legally
obliged to recover its costs (fuel, operating expenses, capital charges and
investment).

Cost recovery was achieved through the Bulk Supply Tariff (BST), a wholesale
tariff comprising an energy charge (£/kWh) and a capacity charge based on each
Area Board's contribution to system peak demand. Area Boards then designed
retail tariffs to recover distribution and administrative costs. Households
typically saw simple volumetric charges and sometimes fixed daily charges;
industrial customers faced more complex structures including maximum demand
(kW) charges and reactive power penalties.

The BST was uniform across Area Boards, cross-subsidising costs so that a kWh
in rural areas cost the same wholesale as a kWh in urban areas, regardless of
underlying cost differences. Distribution costs were also cross-subsidised so
rural customers were not charged dramatically more than urban customers.
Household tariffs were often held stable, with some cross-subsidy from
industry to households. Electricity was treated as a universal public service,
similar to the Post Office or BT, overseen by the Electricity Council and the
CEGB.

\subsection{Liberalisation and the Neoliberal Turn}

By the 1980s, the political ideology of neoliberalism, grounded in
neoclassical economics, had taken hold. Under this philosophy, the CEGB's
cost-plus, centrally planned, monopolised model was seen as dulling incentives
to cut costs or innovate. There were additional motivations: offering
consumers choice, and shifting risk from taxpayers and bill-payers to private
investors.

The Electricity Act 1989 liberalised the sector. Generation and supply were
opened to competition; ``the wires'' (transmission and distribution) were
recognised as natural monopolies. Nuclear generation remained under state
ownership due to its profitability challenges.

The independent regulator OFFER (later Ofgem) was established. The CEGB was
broken up into generation companies, the National Grid Company (transmission),
and the regional electricity companies (RECs) in distribution and supply.
Wholesale prices were set through the Electricity Pool with marginal pricing.
Consumer choice in supply was gradually introduced through the 1990s.

From a neoclassical perspective, electricity became a \emph{special commodity}:
not storable in bulk, requiring real-time balance, delivered over a physical
network with natural monopoly characteristics. In practice, however, the
liberalised regime ported a commodity market architecture designed for wheat,
oil and copper onto electricity, and assumed that neoclassical marginal
pricing would discipline costs, drive efficient investment and maximise social
welfare.

The rest of this thesis questions whether these assumptions hold in a
decarbonising, digital, capital-intensive system with strong distributional
concerns.

% ---------------------------------------------------------
\section{Operation of the Electricity System and Energy Security}
\label{sec:operation_security}

Alternating current power systems must satisfy fundamental physical conditions
at all times:
\begin{itemize}
  \item supply and demand must balance in real time to maintain frequency
        (50\,Hz in GB);
  \item voltages must remain within acceptable bounds at all points in the
        network;
  \item power flows on each line must remain within thermal and stability
        limits;
  \item protection systems must detect and isolate faults to avoid cascading
        failures and blackouts.
\end{itemize}

The System Operator is responsible for real-time whole-system balancing,
frequency and security. Distribution System Operators (DSOs) manage local
constraints, voltages and connection rights. Reliability concepts such as Loss
of Load Probability (LOLP), resource adequacy and capacity margins emerged
from planning practice long before liberalisation.

Transmission and distribution incur technical losses; energy security is a
function of fuel availability, generating capacity, interconnection,
flexibility resources and effective governance. Liberalised markets, capacity
mechanisms and balancing services were layered on top of this physical
reality. The underlying physics did not change.

Recent geopolitical events and gas price shocks have highlighted how tightly
energy security, affordability and political stability are coupled. When
wholesale price spikes threaten to collapse retail markets, force innovative
suppliers into administration, and require universal subsidies funded through
public borrowing, the underlying system design must be questioned rather than
treated as a fixed background.

\subsection{Inertia, System Operability Tightness and Digital Stability}
\label{sec:inertia_operability}

A crucial but often implicit feature of traditional power systems is
\emph{inertia}: the kinetic energy stored in the rotating masses of
synchronous generators. When supply and demand are not perfectly balanced, the
resulting mismatch is initially absorbed by these rotating machines, causing
frequency to deviate only gradually rather than instantaneously. Inertia
therefore acts as a physical buffer, slowing the rate of change of frequency
and buying operators time to deploy reserves, re-dispatch generation or shed
load in an orderly way.

From an operability perspective, inertia can be viewed as a form of
\emph{system slack}. A high-inertia system is more forgiving: forecasting
errors, sudden plant trips or demand spikes manifest as relatively slow
frequency drifts that can be corrected with conventional tools. A low-inertia
system is far more \emph{tightly coupled}: small imbalances lead to much
faster deviations, narrowing the window within which corrective action must be
taken. In this thesis, we refer to this as \emph{system operability
tightness}: the degree to which the system can tolerate shocks, delays and
errors before violating its physical limits.

Historically, synchronous inertia was an almost accidental by-product of the
generation fleet. Large coal, gas and nuclear plants, directly connected to
the grid, provided substantial rotational mass as part of their basic
engineering design. System operators did not need to procure inertia as a
distinct service; it was simply ``there'' as long as enough synchronous
machines were online to meet demand. The resulting environment was
\emph{inertia-rich and slack}: technical standards, operational procedures and
market designs all evolved under the implicit assumption that frequency
disturbances would unfold on time-scales of seconds rather than milliseconds.

The transition to weather-dependent renewables fundamentally changes this
picture. Modern wind turbines, solar PV and many forms of distributed
generation connect to the grid via power electronics rather than direct
mechanical coupling. Unless explicitly configured to do so, these units
provide little or no natural inertial response. As synchronous plant retires
or runs at low output while inverters carry a larger share of the load, the
system becomes \emph{inertia-scarce}. Disturbances propagate more quickly,
rates of change of frequency increase, and the grid moves into a regime of
much tighter operability.

In response, system operators and technology providers are developing forms of
\emph{synthetic} or \emph{digital} inertia. Batteries, inverter-based
resources, demand response and electric vehicles can be controlled to change
their active power output extremely rapidly in response to measured frequency
or grid conditions. Rather than relying on the passive physics of rotating
mass, the system increasingly depends on \emph{digitally activated,
asynchronous inertia} provided by fast-responding resources distributed
throughout the grid.

This represents a deeper architectural shift. Inertia is no longer an
unpriced, incidental property of a small number of large machines; it becomes
a programmable service, delivered by many small devices coordinated through
signals, contracts and control algorithms. Questions of \emph{who} provides
this stabilising response, \emph{where} it is located, \emph{how} it is
measured and \emph{how} it is paid for are no longer purely technical. They
are design choices in the market and regulatory architecture.

The inertia challenge thus links directly to the broader themes of this
thesis. As the system becomes more tightly coupled and digitally mediated,
operability, fairness and market design cannot be treated as separate domains.
Any credible architecture must:
\begin{itemize}
  \item recognise inertia (and related stability services) as scarce,
        allocatable products rather than background assumptions;
  \item ensure that digitally activated inertia from batteries and other
        fast-responding resources is coordinated in a way that respects
        physical limits on time-scales compatible with modern electronics;
  \item allocate the obligations and rewards for providing stabilising actions
        in a way that is transparent and fair across participants.
\end{itemize}
Later chapters return to these issues when discussing programmable products,
flexibility services and the role of the Automatic Market Maker (AMM) as a
cyber--physical controller.

% ---------------------------------------------------------
\section{Climate Science, Net Zero and Distributional Impacts}
\label{sec:climate_net_zero}

The climate is changing. Industrial and household economic activities in
Western economies throughout the 20th century, relying heavily on fuel
combustion and carbon-emitting energy sources, are with high probability a key
contributor to this. Climate tipping points present such an existential risk
to the planet and civilisation that it is prudent to rapidly decarbonise
economies and to make the case for other countries to do the same.

To lead internationally in making that case, countries in the West that claim
to defend democratic values must demonstrate that climate policy can be
delivered in an economically sustainable way, without impoverishing or placing
undue burden on those in society with the least economic resilience.

To achieve these ideals, we need energy to be affordable for people to meet
their basic needs, aligned with the United Nations Universal Declaration of
Human Rights (1948). Democratic consent and societal buy-in for climate policy
is even more important when climate change is already affecting migration
patterns and creating new categories of climate refugees. Right-minded Western
nations concerned with preserving democratic values should want to lead the
response. Leading such a response will be difficult if current policy unfairly
impoverishes the poorest people in our own societies, justified using
technocratic language that is seldom explained to the public, during a period
where governments are failing to deliver on other basic promises.

The arguments for decarbonising the economy extend beyond climate science.
Decarbonising \emph{supply} using local sources of free (but weather-dependent)
energy such as wind and solar increases national, business and household
energy security and resilience through decentralisation and diversification.
Decarbonising \emph{demand} by phasing out fossil fuels for transport and
heating reduces the very real human health impacts of air pollution and
supports superior technologies such as electric vehicles and heat pumps.

At the same time, a high share of inverter-connected renewables reduces
synchronous rotational inertia on the system, tightening operability and
increasing reliance on digitally activated stability services provided by
batteries, demand response and other fast-responding resources
(Section~\ref{sec:inertia_operability}).

In practice, however, contemporary net zero policies often reveal a gap
between climate objectives and fairness. Grant schemes for heat pumps or home
retrofits, electricity levies that fund renewable subsidies, and carbon
accounting frameworks that outsource embodied emissions to other jurisdictions
can combine to create a pattern in which:
\begin{itemize}
  \item relatively affluent owner-occupiers receive capital subsidies;
  \item running costs remain high due to market design and levy choices;
  \item fuel-poor households in inefficient homes bear a disproportionate
        share of the cost of the transition;
  \item carbon metrics place emphasis on territorial emissions while ignoring
        embodied carbon in imported goods.
\end{itemize}

The Conference of the Parties (COP) process and associated carbon accounting
frameworks (Scope 1, 2, 3 emissions) formalise climate commitments, but are
largely silent on internal distributional questions: who pays, when, for what,
and under which governance structure. These distributional questions are
central to this thesis.

% ---------------------------------------------------------
\section{Digitalisation, Aggregators and Fragmented Governance}
\label{sec:digitalisation_fragmentation}

The development of the electricity grid revolutionised living standards in the
West throughout the 20th century. It supported the spread of democracy by
enabling industrial and social infrastructure, and contributed to periods of
growth that politicians later mis-remember as evidence that GDP growth
automatically delivers improved living standards. In reality, it is the
affordability and accessibility of basic input goods---including energy---that
transforms living standards. GDP is an output metric; growth in GDP does not
automatically correlate with improved quality of life or reduced inequality.

At the same time, we are living through a digital communications revolution.
In 1947, at Bell Labs, Bardeen, Brattain and Shockley built the first
transistor. Digital electronics, wireless communications, the internet, mobile
devices and the Internet of Things (IoT) have transformed virtually every
sector: finance, telecommunications, retail, media, logistics, manufacturing
and more.

The electricity system stands out as one of the last critical infrastructures
that still largely operates using a market design and regulatory mindset
rooted in the 1990s, with analogies to 19th century commodity markets. It is
no longer tenable to pretend that the internet and modern cloud computing have
not been invented.

As digital infrastructure becomes central to system operation, stability
services such as inertia are increasingly delivered through coordinated,
fast-response, inverter-based resources rather than passive synchronous
machines, further tightening system operability
(Section~\ref{sec:inertia_operability}).

\subsection{A Proliferation of Markets and Pseudo-Markets}

The UK energy market today is effectively a collection of at least twelve
different markets and pseudo-markets:
\begin{enumerate}
  \item \textbf{Wholesale market:} A financial market comprised of multiple
        sub-markets (forward, day-ahead, intra-day) with their own rules, in
        which suppliers and large consumers procure energy.
  \item \textbf{Retail market:} The market in which suppliers interact with
        end-users under tariffs, including retail price caps.
  \item \textbf{Carbon market:} Markets for emissions allowances and carbon
        credits.
  \item \textbf{Contracts for Difference (CfD) market:} A scheme that
        guarantees strike prices to low-carbon generators, decoupled from
        instantaneous system value.
  \item \textbf{Capacity market:} A mechanism to pay generators and
        demand-side providers to be available in the future, compensating for
        revenue inadequacy elsewhere.
  \item \textbf{National flexibility market:} Schemes such as the Demand
        Flexibility Service incentivise demand reduction in real time at the
        system level.
  \item \textbf{DSO flexibility markets:} Local markets run by DSOs to manage
        distribution constraints.
  \item \textbf{Meter market:} Markets for meter assets and services (MAPs,
        MOPs, data collectors, data aggregators).
  \item \textbf{Adapter market:} Technical integration markets created to
        connect suppliers to the Data Communications Company (DCC) via
        different intermediaries.
  \item \textbf{Adapter aggregator market:} Additional layers created to cope
        with the adapter market’s fragmentation.
  \item \textbf{Network ownership market:} Financial markets for ownership of
        network companies by international investors with limited direct
        incentive to invest for long-term resilience.
  \item \textbf{Balancing market:} The set of services and arrangements
        through which the System Operator balances the system in real time.
\end{enumerate}

None of these markets is designed as part of a unified architecture. They do
not communicate or coordinate in a principled way; several can be in direct
conflict at the same time. Different organisations are responsible for
different slices. Complexity and opacity have replaced clarity and design.

\subsection{Asset Ownership, Aggregators and Digital Intermediaries}

The transition is also reshaping \textbf{who owns and operates energy assets}.
Previously, generation assets were owned by utilities and financed through
regulated tariffs. Today, millions of distributed assets---EV batteries, heat
pumps, rooftop solar, smart thermostats---are privately owned. As intelligent
devices proliferate, new business models emerge: digital aggregators, virtual
power plants (VPPs), and energy service providers bundle thousands of small
assets and trade their flexibility or capacity in wholesale, balancing, and
local grid markets.

These actors operate at the intersection of finance, software and energy. They
monetise flexibility, arbitrage prices, and provide balancing services without
necessarily owning traditional infrastructure. By capturing value through
automation, optimisation and market visibility, aggregators increasingly
occupy a role that neither legacy utilities nor regulators were designed to
accommodate. Yet their rise has not been matched by a proportional governance
framework, leaving critical questions unanswered:
\begin{itemize}
  \item Who is responsible for data accuracy, control decisions and cyber-risk
        management at scale?
  \item Who arbitrates conflicts when the same device is enrolled across
        incompatible services (for example, a virtual power plant and a
        capacity auction)?
  \item Who guarantees performance of flexibility when aggregated portfolios
        are used for system balancing?
\end{itemize}

\subsection{Universities, Industry and the Acceleration Gap}

Historically, universities acted as primary centres of research and early
innovation. Today, however, the speed of change in digital energy systems
means that many commercial actors---in software, fintech, data analytics, AI
and platform governance---are innovating faster than academia or policy.
Industry-led pilots (flexibility platforms, AI-based load control, EV-grid
integration) routinely move from concept to deployment before universities
have fully developed theoretical frameworks to explain, critique or govern
them.

The effect is an \textit{acceleration gap}: practice outruns theory;
deployment outruns design; financial and digital architectures emerge without
coherent governance oversight. This gap is amplified by slow regulatory
consultation cycles, siloed institutional mandates, and fragmented academic
disciplines. Energy policy, finance theory, behavioural science, computer
science and control engineering each provide only partial views, resulting in
incoherent design principles and fragmented governance.

\subsection{The Absence of an Overall Architect}

Across the modern energy system, no single institution is responsible for
integrating technical, financial, social, digital and behavioural design
dimensions. Regulators govern wires and tariffs; central banks govern
financial risk; data regulators govern privacy; system operators manage
balancing; technology providers build platforms; suppliers manage billing;
aggregators orchestrate flexible assets; and consumers---now active
participants---make autonomous decisions. Yet no actor sees the whole system,
let alone designs it.

This absence of a \textbf{systems architect} leads to:
\begin{itemize}
  \item misaligned incentives between infrastructure investment, price
        formation and consumer participation;
  \item contradictions between policies that seek increased electrification
        while penalising electricity through levies;
  \item digital control systems with no unified cybersecurity, safety or
        interoperability framework;
  \item fragmentation of energy markets with conflicting signals, inconsistent
        rules and no unifying allocation mechanism.
\end{itemize}

As later chapters will argue, it is this lack of coherent architecture---
rather than any single technological gap---that constitutes the defining
challenge of the contemporary energy system.

\subsection{Smart Meter Roll-Out, Data Gaps and Unfairness}
\label{sec:smart_meter_experience}

Digitalisation in the electricity sector has often been framed as a
\emph{technology deployment problem}: roll out smart meters, connect them to
a secure data hub, build apps and time-of-use tariffs, and flexibility will
follow. In practice, the experience of smart meter and advanced metering
roll-outs has exposed deeper architectural and fairness problems.

First, smart meters were sold to the public and policymakers as enablers of:
\begin{itemize}[leftmargin=1.2cm]
  \item near real-time visibility of demand;
  \item cost-reflective time-of-use tariffs;
  \item system-wide load shifting and peak reduction; and
  \item more accurate, automated billing with fewer surprises.
\end{itemize}
In reality, many deployments have delivered:
\begin{itemize}[leftmargin=1.2cm]
  \item data latencies measured in \emph{days or weeks} for settlement
        processes that clear at sub-hourly resolution;
  \item coarse-grained profiles used for wholesale settlement, with high
        resolution data relegated to consumer portals and marketing analytics;
  \item fragmented device standards and firmware that cannot be upgraded
        consistently over the air; and
  \item a persistent gap between the granularity of physical system events
        and the granularity of the data used for pricing and allocation.
\end{itemize}

Second, the incentives created by the current retail-settlement architecture
have produced systematic \emph{unfairness} in where and how digitalisation is
delivered:
\begin{itemize}[leftmargin=1.2cm]
  \item Suppliers have often prioritised smart meter installations where they
        are cheap and convenient (good signal, easy access, high-consumption
        customers), and deprioritised hard-to-reach or low-margin areas.
  \item Households with prepayment meters, multiple occupancy, or complex
        housing arrangements have frequently experienced slower, more
        problematic migrations, or have been exposed to system glitches that
        directly affect their ability to keep the lights on.
  \item Data and control capabilities are unevenly distributed: some customers
        have near real-time in-home displays and app integrations; others
        remain effectively ``offline'' in the informational sense, with
        estimated or profile-based billing.
\end{itemize}

The result is a form of \textbf{digital energy inequality}. Those with better
infrastructure, housing and connectivity are first in line to benefit from
dynamic tariffs, smart appliances and flexibility revenues. Those in
data-poor areas or with weaker digital access remain on coarse tariffs,
vulnerable to bill shocks and with little ability to monetise flexibility.

From a control and market-design perspective, this is more than a social
problem: it is a structural failure. The system invests billions in metering
and communications infrastructure, yet:
\begin{itemize}[leftmargin=1.2cm]
  \item wholesale settlement and balancing still treat large fractions of load
        as opaque or poorly measured;
  \item the operational value of fine-grained data is only weakly connected
        to how suppliers and consumers are actually paid; and
  \item meter and device standards are not systematically tied to
        real-time deliverability, fairness, or system stability.
\end{itemize}

Later chapters argue that this is not an accident but a consequence of the
underlying market architecture. When ex-post settlement risk is not
explicitly priced, and when retail products are defined only in terms of
volume and static tariffs, smart meters become a \emph{compliance cost} rather
than a core instrument of system control and value allocation. This thesis
reverses that logic: it treats granular, trustworthy data as a first-class
input into pricing, fairness and control, and designs the market mechanism so
that suppliers and device providers are financially motivated to deploy,
upgrade and maintain digital infrastructure in a non-discriminatory way.

\subsection{Networks as Graphs, Distributed Optimisation and Algorithmic Limits}
\label{sec:graphs_distributed_control}

Behind the institutional complexity, the electricity system can be viewed as
a \emph{graph}: nodes representing buses, substations, feeders, households or
devices; edges representing lines, transformers or communication links.
Physical constraints (power flows, voltages, thermal limits) and digital
constraints (data paths, latencies, control actions) are both defined on this
graph.

This graph-theoretic view is standard in power systems analysis, but has not
been fully integrated into market design. In particular:
\begin{itemize}[leftmargin=1.2cm]
  \item Optimal power flow (OPF) problems define feasibility regions over
        nodal injections subject to network constraints, but wholesale
        markets often operate on zonal or national abstractions that ignore
        this structure.
  \item Distributed control and optimisation algorithms (consensus methods,
        primal--dual schemes, ADMM, multi-agent reinforcement learning) are
        increasingly used in research prototypes to coordinate assets over
        networks, yet retail and balancing markets still assume centralised,
        batch optimisation with limited real-time feedback.
  \item Communication networks and data platforms introduce their own
        graph structure and bottlenecks, which are rarely modelled explicitly
        when designing tariffs, flexibility services or settlement rules.
\end{itemize}

Digitalisation amplifies these tensions. As millions of devices become
addressable and controllable, the system is no longer a small set of large
plants plus passive demand; it is a \textbf{large-scale, distributed control
problem} on a coupled physical--digital graph. Any realistic market design
must therefore confront:
\begin{itemize}[leftmargin=1.2cm]
  \item algorithmic limitations (computational complexity, scalability,
        convergence under delays and noise);
  \item data limitations (measurement error, cyber-risk, privacy constraints,
        incomplete observability); and
  \item hardware and cryptographic limitations (finite device memory and
        processing power, evolving security standards, including
        post-quantum requirements).
\end{itemize}

From this perspective, the question is not simply whether ``more data'' and
``more optimisation'' are available, but whether the \emph{market mechanism}
is designed to:
\begin{enumerate}[leftmargin=1.2cm]
  \item respect the graph structure of the underlying physical system;
  \item admit decentralised, holarchic implementations in which local
        controllers act on local information while preserving global
        stability; and
  \item expose prices and allocation rules that are computable, auditable and
        robust to model error.
\end{enumerate}

Emerging computing paradigms---including specialised accelerators, quantum
inspired and quantum computing---will likely expand the feasible frontier of
what can be optimised or simulated in real time. However, they do not remove
the need for principled architecture. A poorly designed market that ignores
graph structure, fairness and control constraints will not become fair or
stable simply by running on faster or more exotic hardware.

The thesis therefore takes a complementary stance: it uses graph-theoretic
intuition, distributed optimisation concepts and algorithmic awareness to
inform the design of a holarchic Automatic Market Maker (AMM). The AMM is
crafted so that:
\begin{itemize}[leftmargin=1.2cm]
  \item its pricing and allocation rules can be implemented in a distributed,
        event-based manner over the network graph;
  \item its fairness logic is compatible with local measurements and
        device-level telemetry; and
  \item its computational requirements remain bounded and adaptable as
        digital infrastructure and hardware capabilities evolve, including
        potential future quantum-safe or quantum-assisted implementations.
\end{itemize}

In short, digitalisation, IoT and advanced algorithms are treated not as
decoration on top of a legacy commodity market, but as integral parts of a
control-theoretic market architecture. The following chapters build on this
background, linking these ideas to the problem definition, fairness
framework and AMM design.

% ---------------------------------------------------------
\section{Financing, Levies and the Architecture Gap in the Energy Transition}
\label{sec:financing_levies}

Decarbonising the energy system---by deploying wind, solar, heat pumps,
batteries, EV infrastructure and other low-carbon assets---requires large
upfront capital. Traditional pay-as-you-go electricity tariffs or consumer
payments alone cannot fund the rapid scale-up needed. As a result, a variety
of financing mechanisms have emerged, combining public, private and hybrid
capital. At the same time, many jurisdictions have relied heavily on embedding
transition costs in energy bills via levies and charges. This section reviews
the main financing channels and critically examines why financing via stealth
taxes on energy consumption may conflict with principles of fairness,
democratic consent and behavioural incentives for flexibility.

\subsection{Financing Instruments and Capital Markets}

The required investment in new energy assets increasingly relies on
diversified financing channels, including:
\begin{itemize}
  \item \textbf{Project finance for large-scale renewables:} Offshore wind
        farms and large solar parks are typically financed via non-recourse or
        limited-recourse project finance structures, supported by long-term
        contracts such as Power Purchase Agreements (PPAs) or Contracts for
        Difference (CfDs) that stabilise revenue expectations.
  \item \textbf{Green bonds and sustainability-linked loans:} These instruments
        allow institutional investors to fund clean infrastructure with
        explicit environmental performance targets and constraints.
  \item \textbf{Public--private partnerships and ``green banks'':} Public
        capital can shoulder early-stage technology or policy risk, unlocking
        private capital at scale once risks are better understood.
  \item \textbf{Asset-level and platform finance for distributed assets:}
        Emerging models treat portfolios of EV chargers, heat pumps, rooftop
        solar and batteries as financeable assets, backed by digital meter
        data, performance guarantees and sometimes platform-based cash flows.
\end{itemize}

These instruments demonstrate that large-scale decarbonisation is not
inherently dependent on funding through day-to-day retail tariffs. It requires
a credible, stable policy and market environment in which financial actors can
quantify risks and returns over long timescales.

\subsection{Bill-Based Levies and ``Stealth Taxes''}

Despite the availability of structured finance channels, many governments and
regulators continue to recover a significant share of transition costs through
\emph{levies on household energy bills}. In practice, this equates to
untransparent ``stealth taxes'' funding renewable subsidies, energy efficiency
programmes, social tariffs, and network upgrades.

While sometimes politically convenient, this approach creates several
distortions:
\begin{enumerate}
  \item \textbf{Regressive burden:} Levies are typically applied uniformly per
        unit of electricity, disproportionately impacting lower-income
        households and those in energy-inefficient homes, often pushing them
        deeper into fuel poverty.
  \item \textbf{Electrification penalty:} By making electricity artificially
        expensive relative to fossil fuels, levies discourage adoption of heat
        pumps, electric vehicles and other low-carbon technologies, despite
        these being central to net-zero strategies.
  \item \textbf{Behavioural disincentives:} When levies inflate the fixed
        portion of energy bills independent of real-time system conditions,
        they dampen the effectiveness of dynamic pricing and undermine the
        very flexibility behaviours (load shifting, price-responsive EV
        charging) that smart grid design is meant to encourage.
  \item \textbf{Loss of transparency and democratic legitimacy:} Citizens lack
        clear visibility into what portion of their bill funds energy use and
        what portion subsidises system-wide investment. This weakens public
        trust and erodes democratic accountability over energy policy.
\end{enumerate}

From a market-design perspective, this illustrates a structural confusion: a
failure to distinguish between \textbf{operational pricing} (reflecting
real-time system states) and \textbf{infrastructure financing} (reflecting
long-term capital recovery). Embedding both into a single volumetric charge
leads to inefficiency, inequity and weakened system adaptability.

\subsection{Financing, Behaviour and Fairness}

If the goal is to mobilise flexibility at scale, operational price signals must
be credible, comprehensible and salient. When levies and policy costs dominate
bills, real-time variations associated with flexibility programmes become a
small residual. Households and businesses see high, relatively flat prices
rather than meaningful incentives to adjust behaviour. At the same time, the
distributional pattern of these levies often conflicts with fairness goals,
placing relatively higher burdens on those least able to respond.

A fair financing architecture for the transition would:
\begin{itemize}
  \item rely on capital markets and long-term contracts to fund
        infrastructure, not day-to-day retail levies;
  \item use general taxation or progressive mechanisms for social and equity
        objectives, rather than regressive levies on essential services;
  \item preserve operational price signals for flexibility and efficiency;
  \item make the allocation of costs and benefits transparent and subject to
        democratic scrutiny.
\end{itemize}

These principles align with the fairness definition and market design
objectives developed later in the thesis. The key point here is that financing
design is not neutral; it shapes behaviour, fairness and the feasibility of
any proposed operational market mechanism.

% ---------------------------------------------------------
\section{Conceptual Tools: Automatic Market Makers, Holarchies, Game Theory and Fairness}
\label{sec:conceptual_tools}

The previous sections have described how today’s electricity system combines
complex physics, legacy infrastructure, layered market mechanisms, ambitious
decarbonisation goals and fragmented governance. This section briefly
introduces the conceptual tools that will be used later in the thesis to
design and analyse a new market architecture: automatic market makers,
holarchies, game-theoretic allocation (Shapley values) and fairness.

\subsection{Automatic Market Makers and Holarchies}

An Automatic Market Maker (AMM) is a function that determines prices in a
deterministic way based on an explicit formula. The first widely known AMM was
the Logarithmic Market Scoring Rule (LMSR) developed by Hanson for prediction
markets. In decentralised finance (DeFi), AMMs embedded in smart contracts
provide continuous liquidity without matching buyers and sellers directly.

A \emph{holarchy} (a hierarchy of holons) is a system architecture in which
each entity is simultaneously a whole and a part. The concepts of holon and
holarchy were introduced by Arthur Koestler in \emph{The Ghost in the
Machine}. The Earth can be considered as a holarchy: the planet is made up of
oceans and land; land is made up of countries; countries are made up of
regions and cities; cities are made up of buildings and infrastructure. A
power system can similarly be viewed as a holarchy: transmission systems,
distribution systems, feeders, buildings and devices.

By using an AMM in combination with a holarchy, we can define energy prices at
every point in time and space within a digital marketplace. This architecture
offers a high degree of flexibility and control over pricing design, and can
therefore be used to pursue explicit policy objectives---for example,
encouraging households and businesses to consume or generate electricity at
particular times and locations. Later chapters formalise a specific class of
holarchic AMMs for electricity.

\subsection{Game Theory, Shapley Values and Nash Equilibrium}

Game theory studies strategic interaction between decision-makers, where each
player’s payoff depends not only on their own choices but also on the choices
of others.

Cooperative game theory focuses on how players form binding agreements or
coalitions and how to divide payoffs among them. Tools include the core, the
Shapley value and bargaining solutions. The Shapley value provides a
principled way of allocating the gains (or costs) from cooperation by
attributing to each player their expected marginal contribution across all
possible coalitions.

Non-cooperative game theory studies strategic moves and equilibrium concepts
such as Nash equilibrium. Here, the emphasis is on predicting behaviour when
players cannot commit to binding coalitions.

In this thesis, Shapley values are used as a fairness tool for allocating
value (or cost) among generators and between products such as different
consumer classes. Nash-style equilibrium concepts appear implicitly where
strategic behaviour and incentives are considered. The detailed mathematical
formulation is developed later; the key point here is that game-theoretic
tools provide a language to talk about contribution, responsibility and fair
division.

\subsection{Fairness and Fairness in the Energy Sector}

Fairness is a well-developed concept across multiple domains. In networks and
communications, fairness criteria shape scheduling and congestion control
algorithms. In economics, fairness appears in tax regimes, social welfare
functions and redistribution schemes. In law and public policy, fairness
underpins concepts of equality before the law and non-discrimination.

In the energy sector, fairness is typically invoked in an ad hoc way:
fuel-poverty measures, social tariffs, targeted subsidies, or broad claims
about ``just transitions''. Existing electricity market designs, rooted in
marginal pricing and patchwork regulation, do not provide a physically
grounded, operational definition of fairness that can be embedded in real-time
dispatch and settlement.

This thesis later introduces a specific, physically grounded fairness
definition for electricity markets, based on contribution, responsibility and
reliability received. Here, it is sufficient to note that fairness matters not
only philosophically, but practically: as a prerequisite for political
stability, social cohesion, investment and trust.

% ---------------------------------------------------------
\section{A Proposed Way Forward: A Return to First Principles}
\label{sec:way_forward}

Taken together, the background above motivates a different starting point for
electricity market design:

\begin{itemize}
  \item \textbf{A return to first principles of physics:} Electricity is
        governed by thermodynamics, Maxwell’s equations, Ohm’s law,
        Kirchhoff’s laws and protection constraints. Any market must respect
        these.
  \item \textbf{A re-examination of commodity pricing:} The application of
        19th and early 20th century marginal commodity pricing to electrons in
        a decarbonising, digital system must be scrutinised, including the
        assumptions behind social welfare maximisation.
  \item \textbf{Learning from behavioural science:} Mechanisms must respect
        how humans and organisations actually behave, including issues of
        trust, choice, attention and bounded rationality.
  \item \textbf{An explicit role for fairness:} Fairness must be defined, not
        assumed, and integrated into how costs and value are allocated.
  \item \textbf{Learning from other sectors:} Networks, financial markets and
        digital platforms have all confronted similar scaling, volatility and
        complexity challenges.
  \item \textbf{Digitalisation as an enabler, not an afterthought:} The
        internet and cloud computing can be used to operate markets in ways
        previously impossible, at very low marginal transaction cost.
  \item \textbf{Coherent architecture and governance:} Financing, pricing,
        digital platforms, physical constraints and fairness criteria need to
        be designed as parts of a single socio-techno-economic architecture,
        rather than as disconnected layers.
\end{itemize}

The rest of the thesis builds on this background. The next chapters review
existing literature, articulate the design philosophy and problem definition,
and then propose and evaluate a new architecture for electricity markets that
respects physics, leverages digital technology, and puts fairness at its core.

\chapter{Literature Review}

\section{Introduction}

The rapid transformation of electricity systems---driven by decarbonisation targets, the proliferation of distributed energy resources, and advances in digital technologies---has catalysed a fundamental reconsideration of electricity market design. Classical markets were developed for a world with large fuel-based generators, predictable operational characteristics, and centralised control structures \cite{schweppe1988spot}. Contemporary systems, by contrast, operate with high penetrations of variable renewable energy, rapidly changing net demand, and an increasingly active and heterogeneous consumer base.

This chapter reviews the principal strands of literature relevant to modern electricity market design and situates the thesis within five interconnected domains:

\begin{enumerate}
    \item the evolution of electricity systems under high renewable penetration;
    \item the foundations and shortcomings of classical market design;
    \item fairness, cost allocation, and cooperative game theory in energy applications;
    \item decentralised coordination, local markets, and prosumer participation; and
    \item digitalisation, algorithmic regulation, and event-based computational paradigms.
\end{enumerate}

Together, these literatures highlight a unique research gap: \textbf{the absence of an integrated, event-driven, continuously clearing, fairness-aware electricity market architecture} capable of operating effectively under the conditions expected in future power systems.

% ---------------------------------------------------------
\section{Evolution of Electricity Systems Under Renewable Dominance}

\subsection{Traditional Power System Architecture}

Historically, power systems were designed around large, dispatchable thermal plant operated by vertically integrated monopolies. Planning and operation were dominated by security-constrained economic dispatch and unit commitment, in which a system operator selects an optimal set of generators subject to ramping limits, minimum up- and down-times, and network constraints. The underlying economics are well captured by the marginal-cost paradigm: fuels determine short-run marginal costs, while capital costs are recovered through infra-marginal rents and scarcity prices. 

The formal theory of nodal pricing and optimal dispatch developed by Schweppe et al.\ provided a unifying framework for this architecture, showing that under convexity assumptions, locational marginal prices (LMPs) derived from security-constrained optimisation can support efficient equilibria \cite{schweppe1988spot}. Subsequent work on tracing power flows and assigning network usage costs, such as Bialek's flow-tracing approach \cite{bialek1996tracing}, further embedded the assumption of a relatively small number of large, controllable generators feeding largely passive demand through a meshed transmission network. 

In this traditional setting, uncertainty was treated mainly as a forecast error on demand, with limited temporal coupling beyond unit-commitment constraints. The combination of dispatchable supply, slow structural change, and coarse-grained metering meant that markets (where they existed) could be organised around relatively infrequent, batch-style clearing processes without fundamentally compromising system viability.

\subsection{The Transition to Fuel-Free Systems}

The increasing penetration of non-synchronous renewable generation is reshaping both the operational and economic landscape of power systems. Taylor, Dhople, and Callaway argue that future systems may be fundamentally characterised as ``power systems without fuel'', in which short-run marginal costs approach zero for large fractions of installed capacity and fuel-based unit commitment becomes largely irrelevant \cite{TAYLOR20161322}. In such systems, balancing, price formation, and investment incentives can no longer be understood through the lens of conventional fuel-driven marginal-cost structures.

A substantial literature documents the operational challenges associated with variability, uncertainty, and non-dispatchability. Integration studies emphasise the need for increased flexibility, ramping capability, and reserves as variable renewable energy (VRE) shares grow, together with more frequent cycling and redispatch of the remaining synchronous fleet \cite{milliganIntegrationChallenges2016}. These operational requirements, in turn, affect asset revenues and risk profiles, undermining the conventional assumption that scarcity pricing in energy-only markets will deliver adequate investment signals.

Classical reliability and resource-adequacy theory, as developed by Billinton and Allan, formalises probabilistic indices such as loss of load expectation (LOLE), loss of load probability (LOLP), and expected energy not served (EENS), which underpin traditional planning standards and capacity-requirement definitions \cite{billinton1996reliability, allan2000probabilistic}. This framework has been extended to quantify the capacity value or effective load-carrying capability (ELCC) of variable renewables, using methods surveyed by Milligan and Porter, Keane et al., and Dent et al.\ \cite{milligan2008capacitywind, keane2011capacityvaluewind, dent2010capacityvalue}. In combination, these works provide the probabilistic backdrop for modern discussions of resource adequacy under high VRE shares.

Mays and co-authors highlight how high-renewable systems transform the resource adequacy problem into one of managing correlated weather-driven risk, with energy-constrained storage and flexible demand playing an increasingly central role \cite{maysResourceAdequacy2023}. In this context, temporal interdependence becomes much stronger: system conditions at one time step are heavily influenced by the state of storage, weather patterns, and previous dispatch decisions. The literature thus points to an electricity system whose dynamics are time-coupled, weather-correlated, and increasingly dominated by resources with negligible short-run marginal costs.

\subsection{Demand-Side Flexibility and Distributed Energy Resources}

In parallel with the transformation of the generation mix, the demand side has become more heterogeneous and potentially flexible. Distributed energy resources (DERs)---including rooftop photovoltaics, behind-the-meter batteries, electric vehicles (EVs), smart appliances, and building energy management systems---are now recognised as key actors in system balancing and adequacy. Rather than a single aggregated demand profile, system operators increasingly face millions of devices with device-specific constraints, preferences, and flexibility ranges.

Recent work on prosumer home energy systems and local markets illustrates this shift. K{\"u}hnbach et al.\ investigate electricity trading in local markets from a prosumer perspective, showing how optimised household-level energy management can interact with market signals to provide system services, but also raising questions about participation barriers and distributional effects \cite{KUHNBACH2022122445}. Similar studies of EV charging highlight the coupling between mobility and power systems: charging flexibility is constrained by travel needs, state-of-charge requirements, and user tolerance for delay, yet offers substantial potential for demand-shifting and frequency support when properly coordinated.

This body of work motivates a move from a ``central supply, passive demand'' paradigm to a decentralised, consumer-centric system in which heterogeneous DER capabilities must be integrated into both operational and market design. A key theme is the tension between the technical potential for flexibility and the practical challenges of orchestrating large numbers of small actors with different objectives and constraints.

\subsection{Holonic and Multi-Agent Control Paradigms}

To address the scale and complexity of future systems, researchers have explored distributed control architectures inspired by holonic and multi-agent systems. Holonic approaches conceptualise the power system as a hierarchy of semi-autonomous subsystems (``holons'') that can make local decisions while respecting system-wide constraints. Negeri et al., for example, propose holonic smart grid architectures in which local controllers negotiate with higher-level coordinators to maintain stability and optimise performance across scales \cite{negeriHolonicSmartGrid2012}. Similarly, Howell and colleagues discuss semantic holons as a way to structure interactions between devices, aggregators, and system operators in a modular fashion \cite{howellSemanticHolons2017}.

Multi-agent system (MAS) research extends this perspective, modelling generators, loads, storage units, and aggregators as agents that interact through negotiation, bidding, and contract mechanisms. MAS-based coordination schemes promise resilience and scalability by reducing the reliance on a single central optimiser and allowing local adaptation to changing conditions.

Although this line of work is often framed in control-theoretic rather than market-design language, it provides important conceptual foundations. Holonic and MAS architectures assume continuous, event-driven interactions between agents, and often rely on local observability and algorithmic decision rules rather than periodic, centralised optimisation. In this sense, they are structurally closer to the kind of event-driven, continuously clearing system envisioned in this thesis than traditional unit-commitment-based market architectures.

% ---------------------------------------------------------
\section{Classical and Contemporary Electricity Market Design}

\subsection{Foundations of Market Design}

The foundational theory of electricity market design emerges from the application of marginal-cost pricing and general equilibrium concepts to power systems. In the canonical framework of Schweppe et al., LMPs are derived as the shadow prices of nodal power balance constraints in a security-constrained optimisation problem \cite{schweppe1988spot}. Under standard convexity assumptions, these prices decentralise the optimal dispatch: profit-maximising generators and utility-maximising consumers respond to prices in a way that reproduces the system-optimal solution.

This theoretical foundation underpins energy-only markets with nodal pricing, which have been widely adopted in North America and inform European zonal pricing approaches. Over the past three decades, a rich analytical literature has developed around these structures, examining equilibrium properties, bidding strategies, and the efficiency implications of different congestion-management and settlement arrangements. Key assumptions typically include convex production costs, well-defined balancing markets, and an exogenous, largely inelastic demand profile.

As markets replaced vertically integrated monopolies, strong emphasis was placed on short-term dispatch efficiency and on providing long-term investment signals through spot and forward prices. In this view, market failures could, in principle, be addressed through well-designed pricing rules and competitive entry, with regulatory intervention confined to setting and enforcing market rules.

Cramton provides a comprehensive historical and conceptual assessment of electricity
market design, tracing how liberalised wholesale and balancing markets evolved from
central dispatch to market-based structures, and identifying the core design objectives
of efficiency, reliability, investment adequacy, and mitigation of market power
\cite{cramton2017electricity}. His work underscores that current market architectures
remain rooted in periodic auctions, sequential clearing stages, and static bidding
interfaces—an assumption that becomes increasingly strained in systems with high
renewable penetration, digital control capabilities, and dynamic demand. This reinforces
the central premise that temporal design, not just pricing rules, is a first-order issue
in future market architecture reform.

\subsection{Market Failures: Revenue Adequacy and Missing Money}

Experience with liberalised electricity markets, especially under increasing renewable penetration, has revealed significant limitations of the classical design paradigm. A central issue is the ``missing money'' problem: energy-only markets with price caps and imperfect scarcity pricing often fail to provide sufficient net revenues to support the level and mix of capacity required for reliability. Joskow presents a detailed analysis of capacity payments and their role in imperfect electricity markets, highlighting the tension between reliability standards, price caps designed to protect consumers, and the revenue streams needed to justify investment in peaking and flexible resources \cite{joskow2008capacity}.

Newbery's analysis of electricity market reform in Great Britain further illustrates how interactions between wholesale markets, balancing arrangements, and policy instruments (such as Contracts for Difference and the Capacity Market) can create complex incentive structures that are poorly aligned with long-run decarbonisation and adequacy objectives \cite{NEWBERY2018695}. The erosion of scarcity rents in systems with large shares of low-marginal-cost renewables intensifies these problems: as average prices fall and price volatility increases, merchant investment in flexible capacity becomes more risky and dependent on policy design.

From a resource-adequacy perspective, Cramton and Stoft argue that liberalised markets have converged on a limited set of designs---energy-only with scarcity pricing, capacity payments, and capacity markets---aimed at delivering sufficient firm capacity relative to probabilistic reliability criteria \cite{cramton2006resadequacy}. Yet the effectiveness of these designs depends critically on how well scarcity prices, capacity obligations, and reliability standards are aligned, and on the extent to which weather-correlated renewables and storage reshape the underlying adequacy problem.

Simshauser and others argue that merchant renewable projects, relying on wholesale price signals alone, face substantial revenue risk that can undermine investment and drive demands for additional support mechanisms. Across this literature, a common theme is that energy-only markets, designed for a different technological context, do not naturally deliver adequate and appropriately located capacity under high-renewable, policy-driven transitions.

\subsection{Price Formation Under High Renewables}

The recognition that marginal-cost-based prices may fail to convey appropriate incentives in high-renewable systems has sparked renewed interest in the ``price formation problem''. Eldridge, Knueven, and Mays systematically revisit the theory of uniform pricing in day-ahead markets, arguing that current implementations often blur the distinction between energy and uplift payments, leading to opaque incentives and potential distortions in investment signals \cite{eldridge2023priceformationI, eldridge2023priceformationII}. They emphasise the importance of designing pricing rules that reflect the true marginal cost of serving load while accounting for non-convexities and unit-commitment constraints.

Wang et al.\ revisit the formulation of electricity prices in the presence of low-marginal-cost resources and complex operational constraints, highlighting that commonly used pricing approaches can deviate significantly from theoretically efficient benchmarks \cite{WANG2020117542}. As renewable penetration grows, zero or negative prices become more frequent, not because marginal costs are literally negative, but because policy instruments, network constraints, and inflexible plant interact in ways that decouple spot prices from the underlying scarcity of system services.

A broader literature examines scarcity pricing, uplift mechanisms, and the choice between sequential and simultaneous markets. Many proposals seek to refine existing batch-clearing processes---for example by modifying shortage pricing rules or better integrating reserves into energy markets---but retain the underlying assumption that markets are cleared in discrete time intervals with relatively coarse granularity.

\subsection{Capacity Mechanisms and Insurance Approaches}

In response to revenue adequacy concerns, a wide variety of capacity mechanisms have been introduced. Joskow categorises these into capacity payments, capacity markets, and more complex arrangements such as reliability options, discussing their strengths and weaknesses in different institutional contexts \cite{joskow2008capacity}. Capacity markets, as implemented in Great Britain, parts of the United States, and several European countries, create separate products for capacity, clearing in periodic auctions that are intended to reveal the value of reliability and support investment.

More recently, attention has turned to insurance-style overlays that sit on top of energy-only markets. Billimoria and co-authors propose an insurance-based capacity mechanism in which generators sell reliability contracts that pay out in scarcity conditions, aiming to reconcile energy-only market principles with the need for explicit adequacy instruments \cite{BILLIMORIA2022119356}. Such proposals seek to preserve the informational efficiency of spot markets while providing a more explicit and transparent hedge against reliability shortfalls.

Conejo and colleagues, in their survey of investment and market design under uncertainty, emphasise that all of these mechanisms operate against a backdrop of deep uncertainty about future policy, technology costs, and demand patterns \cite{CONEJO2018520}. This uncertainty complicates the design of capacity mechanisms and raises questions about their robustness as the system moves toward very high shares of renewables and flexible demand.

At the European level, cross-border balancing platforms such as TERRE, MARI, and PICASSO illustrate attempts to harmonise balancing and adequacy across national borders through coupled replacement and balancing-reserve markets \cite{ENTSOE_TERRE_2019, ENTSOE_MARI_2020, ENTSOE_PICASSO_2020, gonzalez2020crossborder, schittekatte2021balancing}. While these initiatives significantly improve operational coordination, they largely retain batch-based auction structures and do not fundamentally alter the underlying market architecture or its treatment of fairness.

\subsection{Lessons for Future Market Redesign}

Drawing together this literature, several themes emerge. First, classical energy-only, marginal-cost-based designs struggle to provide adequate investment signals in systems characterised by low-marginal-cost renewables, strong policy interventions, and correlated weather-driven risks. Second, attempts to patch these markets through capacity mechanisms, scarcity pricing tweaks, and uplift designs often introduce new complexities and may not resolve underlying incentive misalignments. Third, most of the proposed reforms remain rooted in periodic, batch-based market clearing and do not fundamentally question the temporal structure of market operation.

Survey and perspective papers on future electricity markets underline the need to better integrate flexibility, storage, and active consumers into market design, and to align short-run operational signals with long-run decarbonisation objectives. Newbery, Lynch et al., and the ReCosting Energy reports all call for ``whole-system'' approaches that recognise the interactions between wholesale, retail, network, and policy instruments \cite{NEWBERY2018695, LYNCH2021101312, sandysrecosting20}. However, even in these forward-looking works, the core abstractions remain those of periodic markets and ex post settlement. The potential for \emph{continuous}, event-driven markets---in which prices and allocations are updated in real time based on system events and fairness constraints---is largely absent from the mainstream market design literature.

This need for structural reform is reinforced by Honkapuro et al., who systematically
review European electricity market design options and find that the overwhelming
majority of proposals retain periodic auction structures and do not address continuous
clearing or cyber-physical coordination \cite{honkapuro2023systematic}. Their analysis
shows that most reforms merely reconfigure price formation or auxiliary capacity
mechanisms but do not challenge the fundamental batch-based clearing architecture.
This confirms the research gap identified in this thesis: the temporal architecture
of markets remains largely unquestioned.

% ---------------------------------------------------------
\section{Fairness, Cost Allocation, and Cooperative Game Theory}

\subsection{The Role of Fairness in Energy Systems}

Fairness has emerged as a central concern in energy systems, both as a normative objective and as an instrumental factor influencing participation, compliance, and political legitimacy. Distributional outcomes affect who bears the costs of decarbonisation, who benefits from new technologies, and how the burdens and benefits of system operation are perceived across different social groups and regions. 

Granqvist and Grover argue that distributive justice in paying for clean energy infrastructure is critical for maintaining public support and avoiding backlash against climate policies, particularly when the costs are regressive or perceived as unfair \cite{GRANQVIST201687}. Similar concerns are reflected in the energy justice literature, which extends traditional economic efficiency criteria to include considerations such as recognition, procedural justice, and the fair distribution of environmental and economic impacts. In the context of multi-energy buildings and local energy communities, Mohammadi et al.\ highlight the importance of fair cost allocation mechanisms that respect both technical usage and broader notions of energy justice \cite{en16031150}.

At a larger scale, Weissbart shows how different approaches to allocating decarbonisation costs across regions can lead to very different distributional outcomes, with implications for political feasibility and perceptions of fairness \cite{WEISSBART2020104408}. Collectively, these strands of literature underline that fairness is not a secondary concern that can be addressed ex post, but a core design criterion that interacts with investment incentives, participation decisions, and long-term system stability.

\subsection{Cooperative Game Theory Foundations}

Cooperative game theory provides a formal framework for analysing how the costs or benefits of joint actions should be divided among participants. In energy applications, cooperative games are natural whenever agents share infrastructure (such as community energy storage, microgrids, or transmission networks) or undertake joint investments whose benefits depend on group participation. The Shapley value, introduced by Shapley in 1953, is widely regarded as a principled allocation rule, satisfying axioms such as efficiency, symmetry, dummy, and additivity.

In the energy context, the Shapley value has been applied to a range of problems: allocating costs of community storage, sharing the benefits of virtual power plants, and dividing network charges among users. Its appeal lies in its interpretation as the expected marginal contribution of each player to all possible coalitions, which resonates with intuitive notions of ``fair share''. However, exact computation of the Shapley value scales exponentially with the number of players, which poses significant challenges for large-scale energy communities or markets with many participants.

\subsection{Fair Allocations in Energy Markets}

Several concrete applications illustrate how cooperative game theory can support fair allocations in energy settings. Yang, Hu, and Spanos develop a method for optimal sharing and fair cost allocation of community energy storage using the Shapley value, demonstrating how users with different load profiles and contributions to system peaks can be charged in proportion to their marginal impact on storage costs \cite{9440895}. Jafari et al.\ propose a cooperative game-theoretic approach for fair scheduling and cost allocation in multi-owner microgrids, showing that Shapley-based allocations can align individual incentives with system-optimal operation \cite{JAFARI2020115170}.

While allocation rules such as Shapley-based methods offer principled foundations
for ex post revenue allocation, recent work has evaluated how well such mechanisms
align with formal notions of distributive fairness in energy-sharing settings.
Couraud et al.\ analyse energy distribution mechanisms in collective
self-consumption schemes, comparing proportional sharing, marginal-contribution,
and Shapley-based approaches against established fairness axioms
\cite{couraud2025collectivefairness}. They demonstrate that allocation mechanisms
can satisfy efficiency but fail fairness, and vice versa—highlighting the need
for explicit alignment between fairness indicators and allocation logic.

At the level of large energy communities, Alonso-Pedrero and co-authors design scalable strategies for fair investment in shared assets, again using Shapley-inspired principles to divide costs and benefits among participants \cite{ALONSOPEDRERO2024131033}. These studies collectively demonstrate that fairness constraints can be made explicit and analytically tractable, rather than being treated as informal or purely political considerations.

\subsection{Fairness Indicators in Local Electricity Markets}

In parallel with cooperative game-theoretic allocation rules, a growing strand of
literature focuses on \emph{fairness indicators} for local electricity markets.
These indicators aim to quantify how equitably costs and benefits are distributed
among market participants, often drawing on concepts such as energy justice,
income inequality metrics, or proportional sharing. Soares et al.\ review this
emerging field, highlighting the diversity of proposed indicators and the lack
of consensus on which metrics genuinely capture fairness in local energy systems
\cite{SOARES2024123933}.

Dynge and Cali address this gap by explicitly formulating \emph{distributive energy
justice} in the context of local electricity markets and systematically evaluating
how well popular fairness indicators perform relative to that definition
\cite{DYNGE2025125463}. Using simulated local market outcomes based on real
Norwegian household consumption data, they test a suite of indicators that have
been adopted in the LEM literature and examine their behaviour across different
welfare distributions. Their analysis shows that some widely used indicators can
classify clearly unequal outcomes as ``fair'', or conversely penalise outcomes that
are consistent with reasonable justice principles. Dynge and Cali therefore propose
adjustments and further refinements to these indicators, and argue that fairness
metrics should be explicitly aligned with a clear normative definition of justice
before being used to evaluate or compare market designs \cite{DYNGE2025125463}.

Beyond allocation rules, fairness has also been examined in broader demand response
programmes. Saxena et al.\ provide a detailed survey of fairness concepts applied
to DR, distinguishing between envy-freeness, proportionality, max-min fairness,
and regret-based criteria \cite{saxena2021drfairness}. They show that fairness must
be treated as an operational design principle rather than merely an ex post
assessment, and argue for fairness-aware participation and compensation mechanisms
linked directly to system operation.

For this thesis, these contributions are important in two ways. First, they
reinforce the view that fairness must be operationalised through explicit
metrics rather than left as an informal aspiration. Second, they provide a
structured starting point for selecting and adapting fairness indicators for
the empirical ``fairness experiment'' conducted later in the thesis, where
distributional outcomes under different market architectures are compared.

\subsection{Scalability Challenges and Approximations}

Despite their conceptual appeal, cooperative game-theoretic solutions face serious scalability issues. Exact computation of the Shapley value becomes intractable for games with more than a modest number of players. This has motivated the development of approximation methods, such as Monte Carlo sampling and stratified sampling techniques, that estimate Shapley values with controlled error at lower computational cost. Cremers et al.\ propose efficient stratified sampling methods for approximating Shapley values in energy systems, highlighting the trade-off between accuracy and computational effort in realistic applications.

Alonso-Pedrero et al.\ further address scalability by designing allocation schemes that exploit structure in large energy communities, such as clustering participants with similar profiles or leveraging hierarchical decompositions \cite{ALONSOPEDRERO2024131033}. However, these methods typically remain offline: they are applied to historical data over relatively long time horizons in order to compute fair cost allocations or revenue splits after the fact.

\subsection{Gap: Lack of Real-Time Fairness Mechanisms}

Across the fairness and cooperative game theory literature, fairness is overwhelmingly treated as an \emph{ex post} accounting problem. That is, system operation is determined first---through dispatch, market clearing, or optimisation---and fairness comes in later, when revenues or costs are divided among participants according to some allocation rule. While this separation is analytically convenient, it misses an important design opportunity: incorporating fairness directly into the operational decision-making and market-clearing process.

In particular, there is little work on mechanisms that enforce fairness constraints \emph{in real time}, for example by adjusting allocations or prices in response to evolving fairness metrics, or by embedding cooperative-game-inspired rules into continuous market operation. Existing Shapley-based methods also tend to assume fixed coalitions and relatively static participation, which is at odds with the fluid, event-driven nature of future systems in which participants may join, leave, or change behaviour on short time scales.

This thesis addresses this gap by viewing fairness not merely as an accounting exercise but as a constraint and design goal in an event-driven market architecture. The aim is to move from offline, batch allocation of costs to online, continuously updated fairness-aware operation, in which allocation rules and control decisions co-evolve as part of a cyber-physical market system.

% ---------------------------------------------------------
\section{Local Energy Markets, Prosumers, and Distributed Coordination}

As distributed energy resources become more widespread, a growing literature argues that some aspects of coordination should be pushed closer to the edge of the system. Rather than treating households and small businesses as passive consumers, local energy market (LEM) designs and peer-to-peer (P2P) trading schemes seek to activate prosumer flexibility, foster local self-consumption, and reduce network stress. This section reviews key strands of that literature and draws out their limitations from the perspective of system-wide architecture.

\subsection{Local Markets and Community Platforms}

Local energy markets are typically defined as market structures operating at distribution level, in which local participants trade energy and flexibility among themselves, often mediated by a platform or community operator. Soares et al.\ provide a comprehensive review of fairness in local energy systems, classifying LEM designs by their objectives (cost minimisation, self-consumption, emission reduction), coordination mechanisms (central auctioneer, distributed optimisation, P2P), and fairness criteria (envy-freeness, proportionality, Shapley-based allocations) \cite{SOARES2024123933}. They emphasise that fairness and participation incentives are not peripheral concerns but central to the long-term viability of local schemes.

A broader systematic review by Khaskheli et al.\ examines local energy markets
across centralised, distributed and hybrid coordination structures, comparing
auction-based clearing, bilateral peer-to-peer mechanisms, and AMM-derived
liquidity pooling \cite{khaskheli2024lemreview}. While these designs activate
local flexibility, the authors emphasise that current LEMs remain small in scale,
lack interoperability with the system operator, and rarely embed fairness or
multi-layer coordination objectives. This reinforces the structural limitations
highlighted earlier: LEMs are promising, but not yet architecturally integrated
into the wider market system.

Mechanism-design-oriented contributions---such as those of Tsaousoglou et al.\---formalise LEMs as markets in which local aggregators or prosumers submit bids for buying and selling energy, with clearing rules designed to recover network costs, respect voltage and thermal limits, and reward flexibility \cite{tsaousoglouLEM2022}. These models often demonstrate that, under appropriate assumptions, local markets can reduce losses, alleviate congestion, and defer reinforcement by aligning local incentives with system needs.

However, most LEM models operate on relatively short case studies and assume either perfect or highly stylised participation. They rarely address the question of how multiple local markets should interoperate with each other and with wholesale markets in a way that preserves overall system efficiency, nor how to coordinate local clearing with real-time network constraint management at scale.

A recent strand of work goes further by embedding automated market maker protocols
directly into local market clearing, allowing prosumers to trade against liquidity
pools backed by storage assets rather than through double auctions \cite{bevin2023amm}.
Such designs, however, are still evaluated on small case studies and do not yet articulate
how AMM-based local markets should interoperate with wholesale and balancing layers.

\subsection{Peer-to-Peer Trading and Prosumer Interaction}

Peer-to-peer trading schemes extend the local-market idea by allowing individual prosumers to trade bilaterally or through decentralised matching algorithms. Parag and Sovacool characterise this shift as the emergence of a ``prosumer economy'', in which small actors both consume and produce electricity, and participate in new market structures that blur the lines between retail, community, and wholesale levels \cite{paragProsumerEconomy2016}. P2P arrangements are often motivated by social and political objectives as much as by efficiency: they can create communities of practice around energy, support local renewable generation, and enhance perceived autonomy.

From a technical perspective, P2P markets raise questions about fairness and network usage. IEEE-based work (e.g.\ \cite{9276456}) explores how to design trading and settlement mechanisms that ensure that all participants benefit relative to a baseline, that network constraints are respected, and that transaction costs remain manageable. Many schemes propose to embed network usage charges into bilateral trades or to restrict trades to electrically ``close'' peers.

Despite this sophistication, P2P models often assume that the number of peers is modest and that network constraints can be represented by simple line-capacity limits. Scaling such designs to millions of devices across heterogeneous distribution networks, while maintaining stability and transparency, remains an open challenge. Moreover, P2P trades are typically cleared in discrete intervals, and their interaction with real-time balancing and ancillary services is not systematically addressed.

A systematic review by Bukar et al.\ shifts focus to peer-to-peer trading,
highlighting issues of regulatory compliance, transaction complexity, fairness,
and consumer visibility in bilateral energy trading arrangements
\cite{bukar2023p2pReview}. The authors find that although P2P markets enhance
participation and autonomy, they are typically treated as isolated platforms
rather than components in a multi-layered market architecture.

\subsection{Home Energy Management and Flexibility Aggregation}

A complementary strand of literature focuses on home energy management systems (HEMS) and the aggregation of flexibility from distributed devices. K{\"u}hnbach et al.\ show how prosumer participation in local markets depends not only on price signals but also on transaction costs, risk preferences, and the design of interfaces and automation \cite{KUHNBACH2022122445}. HEMS can orchestrate rooftop PV, batteries, and flexible loads so as to respond to dynamic prices or local-market incentives, effectively turning households into small, automated agents.

Aggregators play a key role in scaling up this flexibility. By pooling the flexibility of many devices, they can offer services to system operators or participate in wholesale markets that would be inaccessible to individual households. The literature demonstrates that such aggregation can provide frequency response, peak shaving, and congestion management services, but also highlights concerns about information asymmetries, market power, and the distribution of benefits between aggregators and end-users.

From an architectural perspective, these works suggest that local and household-level controllers will increasingly make autonomous decisions based on algorithmic rules. Yet the coordination between these controllers and system-level objectives is largely left to price signals and contractual arrangements, rather than being embedded in an integrated, event-based control and market framework.

\subsection{Blockchain and Automated Local Markets}

Blockchain and distributed-ledger technologies (DLT) have been proposed as enablers of decentralised local markets, offering tamper-resistant record-keeping and automated execution of contracts through smart contracts. Guo and Feng design a blockchain-based platform for trading renewable energy consumption vouchers and green certificates, demonstrating how such a system could facilitate trusted transactions and compliance with policy instruments in a decentralised setting \cite{GUO2024123351}. Other contributions propose blockchain-backed P2P markets in which trades are validated and settled without a central intermediary.

While these approaches show that transaction execution and record-keeping can be decentralised, they also reveal significant limitations. DLT-based platforms face scalability challenges (throughput, latency), non-trivial energy consumption overheads, and interoperability issues with existing market and grid operation systems. Moreover, blockchain does not, by itself, solve the underlying problems of mechanism design, network constraint management, or fairness; it simply provides a different substrate on which those mechanisms might be implemented.

\subsection{Limitations of Distributed Paradigms}

Despite significant innovation, the local market and P2P literatures share several structural limitations when viewed from the perspective of national or regional system architecture:

\begin{itemize}
    \item \textbf{Limited scalability to national systems:} Most designs are tested on small networks with tens or hundreds of participants; extending them to millions of devices across multiple voltage levels is rarely addressed.
    \item \textbf{Fragmentation and lack of interoperability:} Local markets and P2P platforms are often conceived as stand-alone schemes, with ad hoc assumptions about how prices or schedules interact with wholesale markets and system operators.
    \item \textbf{Weak integration with network constraints:} While some models include simplified network constraints, these are typically static and coarse; real-time voltage, congestion, and stability considerations are handled separately by network operators.
    \item \textbf{Absence of system-wide optimisation:} There is little work on how to coordinate the objectives of multiple local markets, aggregators, and system operators in a way that achieves system-wide optimality or fairness.
\end{itemize}

These limitations suggest that local and P2P markets, while valuable for activating flexibility and engaging prosumers, cannot by themselves provide a coherent, scalable architecture for future electricity systems. Instead, they point to the need for a unifying framework in which local decisions and interactions are embedded within an event-driven, system-wide coordination mechanism that respects network constraints and fairness objectives.

% ---------------------------------------------------------
\section{Digitalisation, Algorithmic Regulation, and Event-Based Computation}

Digitalisation has become a central theme in energy policy and research, encompassing smart metering, data platforms, digital twins, and algorithmic control. The UK government's digitalisation strategy for the energy system emphasises the role of data, automation, and digital infrastructure in enabling net-zero systems \cite{beisdigitalisation}, while the Smart Systems and Flexibility Plan highlights digital tools as essential for unlocking flexibility from demand and distributed resources \cite{beisflexibility}. This section connects these policy agendas to technical literatures on networking, online optimisation, and algorithmic market-making.

\subsection{Digitalisation Agendas and Policy Drivers}

Policy documents in the UK and elsewhere frame digitalisation as both an enabler of flexibility and a governance challenge. The BEIS digitalisation strategy calls for interoperable data platforms, standardised interfaces, and increased automation in system operation \cite{beisdigitalisation}. The Smart Systems and Flexibility Plan explicitly links digital tools to new business models, including flexibility markets, peer-to-peer trading, and local services \cite{beisflexibility}. 

These agendas implicitly assume that digital infrastructures---data platforms, APIs, automated control systems---will allow a more granular, dynamic, and participatory energy system to function. However, they are largely agnostic about the specific market architectures and control mechanisms that should exploit these capabilities. In particular, the temporal structure of markets (periodic versus event-driven) and the integration of fairness into algorithmic control are not systematically addressed.

\subsection{Analogy to Computer Networking: QoS and Event-Driven Control}

Computer networking offers a rich conceptual and technical precedent for managing shared, constrained infrastructures under dynamic, heterogeneous demand. Differentiated Services (DiffServ) architectures, as originally described in RFC~2475 and subsequent refinements, allocate network resources by classifying packets into service classes with distinct quality-of-service (QoS) guarantees, and applying local queue management and scheduling policies at routers \cite{carpenterDiffServ2002, ponnappan2000qos}. Ponnappan and colleagues show how queue management and scheduling can implement complex QoS policies using only local state and event-driven algorithms at each node.

In these systems, control is fundamentally event-based: routers react to packet arrivals, congestion signals, and local queue states, adjusting forwarding and scheduling decisions in real time. Congestion-control protocols such as TCP—and high-speed variants like FAST TCP \cite{1354670}—continuously tune sending rates based on feedback about network conditions, achieving an emergent balance between utilisation and latency without centralised optimisation.

This architecture contrasts sharply with the batch-optimised, periodic clearing processes of contemporary electricity markets. Yet the underlying problems are analogous: heterogeneous agents competing for scarce capacity (bandwidth versus power flows), with constraints that vary in time and space.

Crucially, however, the networking literature \textit{does not attempt to embed any notion of economic fairness or marginal contribution into the control loop}, nor does it integrate \textit{prices} or \textit{market-based remuneration} into its allocation logic. QoS classes encode technical priority, not economic value. Congestion control adjusts sending rates, not payments. Thus, while networking demonstrates that continuous, local, event-driven control can scale to very large systems, it provides no framework for allocating costs, revenues, or rights fairly in the presence of heterogeneous users.

This creates a key research gap: \textbf{there is no established mechanism that combines event-driven control with price formation and fairness allocation}. Electricity markets, for their part, remain committed to periodic optimisation and ex-post settlements, rather than integrating fairness and price discovery into the real-time control architecture.

The central contribution of this thesis is to close that gap. By embedding Shapley-theoretic fairness into an \textbf{Automatic Market Maker (AMM)} that operates continuously and event-wise, this work unifies: (i) real-time congestion-aware control, (ii) price formation, and (iii) fairness-based revenue and access allocation. Unlike existing QoS or congestion-control frameworks, the AMM explicitly incorporates price and marginal contribution into the control loop—making fairness programmable and directly tied to system conditions.

\subsection{Online Optimisation and Event-Triggered Operation}

A parallel literature in control and optimisation studies how decisions can be made online, with incomplete information about future disturbances. Zinkevich's formulation of online convex programming \cite{zinkevich2003oco} establishes regret bounds for algorithms that update decisions sequentially as new data arrives, rather than solving a single large optimisation problem with perfect foresight. Event-triggered control frameworks further show that systems can maintain stability and performance by updating control actions only when certain state-dependent conditions are met, rather than at fixed time intervals.

In power systems, online optimisation techniques have been applied to unit commitment, economic dispatch, and demand response, but usually as refinements of periodic market-clearing processes. The underlying market structure---day-ahead auctions followed by intra-day and balancing markets---remains batch-based. There is little work that systematically explores the design of markets whose \emph{primary} mode of operation is event-driven, with prices and allocations updated continuously in response to system states, rather than on fixed schedules.

\subsection{Automatic Market Makers (AMMs) and Continuous Clearing}

In financial markets, automated market makers (AMMs) and bonding‐curve mechanisms—popularised in decentralised finance (DeFi)—provide a different paradigm for continuous price formation. Hanson’s logarithmic market scoring rule \cite{hanson2002}, and subsequent bonding‐curve designs \cite{bc2018}, define explicit functional relationships between quantities and prices, allowing markets to clear continuously without matching individual bids in discrete auctions. Liquidity providers deposit assets into a pool, and traders interact with that pool according to deterministic rules that guarantee certain invariants.

These mechanisms demonstrate that it is possible to design \emph{algorithmic market rules} with well-defined properties (liquidity, price sensitivity, slippage) that operate in continuous time. Although AMMs were created for financial and prediction markets rather than physical systems, they provide useful design patterns for electricity markets: prices can, in principle, be updated as a function of state variables (e.g.\ utilisation, reliability metrics, fairness constraints) rather than exclusively through batch clearing.

Recent work has begun extending these ideas beyond purely financial assets. Bevin and Verma propose a decentralised local electricity market (DLEM) in which prosumers trade against a liquidity pool formed by distributed storage, with prices updated through a bonding-curve AMM protocol \cite{bevin2023amm}. Concentrated liquidity improves price efficiency, and a loss-compensation scheme ensures compatibility with upstream network contracts. Simulations on IEEE 33-bus and 123-bus distribution networks show that such AMM-based local markets can deliver iteration-free price discovery while maintaining network feasibility.

Beyond energy specifically, Zang, Andrade and Ersoy \cite{zang2025perishable} develop an AMM for goods with \emph{perishable utility}, motivated by cloud-compute resources whose value decays rapidly over time. They show that continuous prices can be derived as concave functions of system load, and that allocation can be implemented via a cheapest-feasible matching rule with provable equilibrium and regret guarantees. Although developed for compute rather than electricity, their framework highlights a structural commonality: both are real-time scarcity systems with rapidly expiring supply, where algorithmic pricing rules indexed to system state can outperform discrete bid–ask clearing. This provides an important theoretical precedent for the type of functional, state-dependent pricing adopted in the present thesis.

However, existing AMM-based energy-market designs remain confined to single layers and primarily treat the AMM as a liquidity and price-discovery device. They do not integrate multiple hierarchical layers, nor do they embed explicit real-time fairness constraints into the market-making logic. In this sense, AMM-based electricity market design is still embryonic: the mechanisms are local, static in their role, and largely orthogonal to distributive justice. The present thesis extends this nascent line of work by treating the AMM itself as the core clearing architecture for a holarchic, multi-layer electricity system and by coupling its state variables directly to scarcity and fairness metrics.

\subsection{Gap: Absence of Event-Based, Continuous Market Architectures in Energy Literature}

Across the digitalisation, networking, and DeFi literatures, continuous, event-driven control and algorithmic market rules are well established. Yet the energy market design literature has largely not connected to these developments. Digitalisation is often treated as an implementation detail---a way to run existing market designs more efficiently---rather than as an invitation to rethink the temporal and algorithmic structure of markets themselves.

In particular, there is no comprehensive framework in which:

\begin{itemize}
    \item prices and allocations are updated continuously based on system events and state variables, rather than exclusively through periodic auctions;
    \item fairness metrics are integrated into the market-clearing logic as constraints or signals; and
    \item local, automated decision-making (by DERs, aggregators, and network assets) is coordinated through event-driven market interactions.
\end{itemize}

This thesis seeks to bridge this gap by drawing explicitly on event-driven control, online optimisation, and AMM design principles to propose an algorithmically clearable, fairness-aware electricity market architecture.

% ---------------------------------------------------------
\section{Behavioural, Human-in-the-Loop, and Health-Aware Perspectives}
\label{sec:behavioural_lit}

The previous sections primarily considered electricity systems and markets as engineering and economic artefacts. However, modern power systems are quintessential \emph{cyber-physical systems} (CPSs) with humans in the loop. Any market architecture that aims to orchestrate flexibility at scale must therefore engage with: (i) CPS theory with human-in-the-loop control, (ii) behavioural science and behavioural economics applied to energy, and (iii) health and environmental externalities, particularly air quality.

\subsection{Cyber-physical systems with humans in the loop}

CPSs couple algorithmic logic with physical processes: sensors gather data, algorithms process this data, and actuators implement control decisions in real time. Lee highlights how CPSs blur the distinction between computational models and physical dynamics, emphasising that model choice and abstraction are central design decisions, not mere implementation details \cite{Lee_08, Lee_15}. When humans are part of the control loop---through preferences, behavioural responses, and manual overrides---the system inherits the complexity of human cognition and social context.

In the energy system, smart meters, home energy management systems, EV chargers, and distribution automation collectively form a layered CPS over physical networks. Controllable loads and DERs respond to setpoints and price signals, while human occupants respond to comfort, habits, and perceived fairness. Human-in-the-loop control literature suggests that treating humans as exogenous disturbances is inadequate; instead, control schemes should account for feedback between human behaviour and system signals, and be robust to bounded rationality and limited attention.

This perspective reinforces the view that electricity markets are not merely economic allocation mechanisms but integral components of a CPS, shaping and being shaped by human behaviour. Market architectures that ignore human-in-the-loop dynamics risk instability, poor utilisation of flexibility, and loss of trust.

\subsection{Behavioural science and sustainable energy behaviour}

Environmental and social psychology provide extensive evidence on what motivates sustainable energy behaviour \cite{steg2015}. Steg and others identify instrumental motives (cost savings, comfort), symbolic motives (identity, status), and affective motives (pleasure, guilt) as key drivers of transport and energy-related decisions \cite{steg2005}. Van der Werff and Steg further show that a stable pro-environmental self-identity can support consistent sustainable behaviour across contexts, provided that actions are perceived as meaningful and aligned with values \cite{werff2016}.

Applied to electricity use, this literature suggests that interventions relying solely on price signals are unlikely to achieve widespread, enduring flexibility. Instead, informational feedback (e.g.\ consumption comparisons), social norms, and narratives about fairness and contribution to collective goals play significant roles. Feedback must be timely, understandable, and salient; perceived arbitrariness or unfairness in price movements can undermine engagement.

From a market-design perspective, these findings imply that price paths and allocation rules should be interpretable and justifiable to end-users, not just mathematically optimal. Interfaces and automation should support users in understanding and shaping their participation, rather than treating them as frictionless optimisers.

\subsection{Energy justice, participation, and fairness in digital energy systems}

While behavioural economics highlights how individuals respond to prices, incentives,
and choice architecture, a complementary body of work emphasises questions of
participation, procedural fairness, and social legitimacy in digitally mediated
energy systems. Milchram et al.\ argue that fairness in smart grids cannot be reduced
to ex post distributional outcomes, but must consider how market and control systems
shape opportunities for participation, agency, and access to flexibility services
\cite{milchram2018energyjustice}. Using case studies from the Netherlands and the UK,
they demonstrate how digital platforms, automation, and pricing mechanisms can both
enhance and undermine energy justice depending on how control is allocated, whether
users are meaningfully included in decision-making, and how transparent the system's
rules are to its participants.

This perspective strengthens the argument that electricity markets are socio-technical
control systems: they not only allocate resources but also structure participation and
shape perceptions of legitimacy. From a design standpoint, it implies that fairness
cannot be a purely economic objective; it must be embedded in the architecture of
market interactions, including transparency of pricing rules, accessibility of
participation interfaces, and protection from algorithmic exclusion. Such insights are
particularly relevant when designing continuous, event-driven markets in which human
actors interact with automated decision-making systems in real time.

\subsection{Behavioural economics, nudging, and demand response}

Behavioural economics challenges the neoclassical assumption of fully rational agents with stable preferences and unlimited cognitive resources. Thaler and Sunstein’s ``nudge'' framework demonstrates that small changes in choice architecture---defaults, framing, ordering---can substantially influence behaviour without eliminating freedom of choice \cite{nudge2008}. In the energy domain, experiments with default green tariffs, pre-set thermostat schedules, and framing of savings have shown significant effects on participation and demand patterns.

Demand response programmes often implicitly assume that consumers will interpret and respond to dynamic prices as intended. In practice, responses are mediated by attention, habit, perceived risk, and trust in institutions. Financial nudges (e.g.\ rebates, bill credits) may be effective for some users but can also be regressive or confusing if not designed carefully. Digital nudging and citizen-science work further highlight that interface design and data presentation can systematically bias responses.

Taken together, these findings suggest that electricity market designs should be evaluated not only for their efficiency under rational-agent assumptions, but also for their robustness to behavioural biases and for their capacity to harness, rather than fight, predictable patterns in human decision-making. In particular, default participation in flexibility schemes, opt-out mechanisms, and transparent fairness rules may be more effective than expecting users to micromanage their own exposure to real-time prices.

\subsection{Health, air quality, and control objectives}

Energy system design is often motivated by climate goals---reducing greenhouse gas emissions in line with carbon budgets---but other health-related externalities, particularly air quality, are equally important. The spatial and temporal pattern of electricity use influences upstream emissions from generation and, in some systems, local air pollutants from distributed generation or heating technologies. Poor air quality is associated with respiratory and cardiovascular diseases, and with broader wellbeing impacts.

Health-aware control schemes, including context-aware mobility and building control, show that it is feasible to incorporate environmental exposure metrics into control objectives. For example, context-aware route planning for cyclists can reduce exposure to air pollution by adjusting routes and timing, while building ventilation control can balance indoor air quality and energy use. 

Building directly on such health-aware control schemes, the author has previously
developed and experimentally validated a cyber--physical, human-in-the-loop control
architecture for reducing personal exposure to air pollution while cycling, using an
electrically assisted bicycle as the actuation platform (published in \emph{Automatica}) \cite{SWEENEY2022110595}. In that work, on-bike sensors measured local pollution concentrations in real time, a digital controller computed exposure-minimising adjustments to speed and routing
subject to journey-time and comfort constraints, and guidance was provided to the
rider through a human-facing interface. The system thus closed a feedback loop
between environmental measurements, online optimisation, and human decision-making
to deliver a welfare-relevant outcome (reduced cumulative pollutant dose) in real
time. This example illustrates that health and wellbeing objectives can be embedded
directly into cyber--physical control loops, rather than treated solely as ex post
assessment criteria.

In the context of electricity markets, this suggests that control and pricing schemes could, in principle, account for air quality and other health metrics alongside economic efficiency and reliability. Doing so would require richer models of spatially and temporally resolved emissions, exposure, and vulnerability, but could align market signals more closely with societal objectives. This thesis does not explicitly model air quality, but adopts the broader stance that market architectures should be evaluated against health and wellbeing outcomes, not just narrow measures of cost and carbon.

% ---------------------------------------------------------
\section{Economic and Policy Paradigms for the Energy Transition}
\label{sec:econ_paradigms}

The dominant analytical framework for electricity markets is neoclassical economics. While this framework has yielded powerful tools and insights, it is increasingly questioned as a sufficient guide for designing 21st-century energy systems, particularly when considering planetary boundaries, social equity, and complex cyber-physical interdependencies. This section situates the thesis with respect to neoclassical, behavioural, and alternative economic paradigms, and to current decarbonisation policy debates.

\subsection{Neoclassical and market-based approaches}

Classical market design for electricity rests on the premise that, under appropriate conditions, marginal-cost pricing with competitive entry delivers efficient outcomes. In this view, spot and forward prices reflect underlying scarcity, guide investment, and ensure that resources are allocated to their highest-valued uses. The theory underpinning LMPs \cite{schweppe1988spot} and the broader literature on competitive equilibrium provide the intellectual foundation for liberalised electricity markets.

Much of the market design literature remains within this paradigm, even when addressing problems such as missing money, capacity adequacy, and investment risk. Solutions typically involve modifying pricing rules, adding capacity mechanisms, or improving hedging instruments, while preserving the core structure of periodic markets and marginal-cost-based pricing. Distributional outcomes are usually treated as secondary to efficiency, to be addressed through separate tax-and-transfer policies rather than integrated into market design.

\subsection{Behavioural and institutional critiques}

Behavioural economics, as discussed above, challenges key assumptions of neoclassical theory, including full rationality and stable preferences. Institutional economics and political economy further highlight the role of governance structures, regulatory incentives, and power relations in shaping market outcomes. Newbery’s analysis of electricity market reform in Great Britain, for example, shows how specific institutional choices interacting with political and regulatory constraints generate outcomes that diverge from textbook ideals \cite{NEWBERY2018695}.

These critiques imply that ``getting the prices right'' is not sufficient if institutional arrangements and behavioural responses undermine the intended effects. Market designs that ignore distributional consequences or that generate opaque and volatile price signals may provoke political backlash, regulatory intervention, or strategic behaviour that erodes efficiency. Markets are thus socio-technical and institutional constructs, not neutral mechanisms operating in a vacuum.

\subsection{Alternative economic frameworks: doughnut economics and beyond}

Alternative economic frameworks explicitly embed environmental and social boundaries into economic analysis. Raworth’s ``doughnut economics'' proposes a safe and just operating space for humanity bounded by ecological ceilings (such as climate and biodiversity limits) and social foundations (such as health, education, and equity) \cite{raworth2017doughnut}. From this perspective, markets are tools that must operate within these boundaries, not mechanisms whose outcomes are assumed to be acceptable by default.

Applied to energy, such frameworks suggest that electricity market design should be evaluated against broader criteria: consistency with carbon budgets, contributions to local environmental quality, impacts on vulnerable groups, and resilience to shocks. Electricity markets then become \emph{subsystems} of a larger socio-ecological-economic system, rather than the primary focus of optimisation.

Related strands of thought in ecological economics, degrowth, and just-transition debates point to the risk that narrow efficiency-oriented designs can exacerbate inequalities, erode trust, and undermine long-term sustainability. For a market architecture to be legitimate in this broader framing, it must not only allocate resources efficiently but also support a fair and liveable socio-ecological system.

\subsection{Decarbonisation policy and the risk of ``anti-climate'' action}

There is a growing strand of critical literature arguing that the way decarbonisation policies are currently implemented can undermine climate objectives if they:

\begin{itemize}
    \item socialise transition costs regressively, eroding public support;
    \item encourage over-build of specific technologies without addressing system integration; or
    \item lock in fossil backup or inhibit flexibility deployment.
\end{itemize}

Lynch et al.\ argue that electricity market design must be considered from a whole-system perspective, accounting for interactions between wholesale markets, network tariffs, and support mechanisms for renewables and flexibility \cite{LYNCH2021101312}. The ReCosting Energy report similarly contends that current arrangements in Great Britain embed historical assumptions and misaligned incentives that raise costs and impede innovation \cite{sandysrecosting20}. 

These critiques highlight the risk of ``anti-climate'' actions: policies and market designs that are formally pro-decarbonisation but that, in practice, generate regressivity, inefficiency, or instability that undermines public consent and system performance. They reinforce the need to consider fairness, transparency, and institutional robustness as integral design criteria, not afterthoughts.

\subsection{Implications for market architecture}

Synthesising these perspectives, the thesis adopts the following stance:

\begin{itemize}
    \item Neoclassical market design offers powerful analytical tools but is insufficient as a sole framework for designing future electricity systems.
    \item Behavioural and CPS perspectives highlight the need to treat markets as socio-technical control systems with humans in the loop, subject to bounded rationality, attention, and trust.
    \item Alternative frameworks such as doughnut economics provide high-level criteria---ecological ceilings and social foundations---against which market designs should be assessed.
    \item Critical reflections on current decarbonisation policy underscore the risks of designing mechanisms that are formally efficient but socially or behaviourally brittle.
\end{itemize}

In response, this thesis proposes a market architecture that is explicitly event-driven, fairness-aware, and embedded within a broader socio-technical perspective. Prices and allocations are not treated as purely economic artefacts but as control signals within a cyber-physical system, whose design must reconcile efficiency, fairness, and system viability within planetary and social boundaries.

\subsection{Energy as the Fundamental Enabler in Economic Systems}

Conventional economic theory has historically placed labour ($L$) and capital ($K$) at the core of production, treating energy ($E$) either as a minor third input or, more commonly, as an external commodity. In doing so, mainstream economics implicitly assumes near-perfect substitutability between $K$, $L$, and $E$, enabling production functions of the form $Y = A K^\alpha L^\beta$, occasionally extended to $Y = A K^\alpha L^\beta E^\gamma$. Keen \cite{keenPuttingEnergyBack2023} argues that such formulations fundamentally misrepresent the physical nature of economic production. Energy is not merely another substitutable input: it is \emph{the enabling input} which allows both labour and capital to perform useful work. Without energy, neither capital nor labour can function; thus energy cannot be treated as separable, nor as marginally substitutable.

Keen reframes the economy as a flow-based thermodynamic system rather than a static allocation of abstract factors. The economy, he argues, begins with the extraction of low-entropy energy from the environment, which is transformed via human and machine processes into useful work, and ultimately returns to the environment as high-entropy waste. Economic output ($Y$) is therefore inseparable from energy throughput, and should be conceptualised as a function of ``useful energy conversion'', not merely abstract productive capacity. This shift aligns closely with ecological economics and thermodynamic realism, but Keen embeds it within broader critiques of neoclassical theory: particularly its reliance on ill-posed mathematical constructs (e.g.\ the Cobb--Douglas production function), its failure to address dimensional consistency, and its abstraction from physical constraints and material limits.

In his second article, ``The Role of Energy in Economics'' \cite{keenRoleEnergyEconomics}, Keen, Ayres and Standish move from conceptual critique to formal reformulation. They argue that traditional production theory violates both physical laws and dimensional consistency by treating production functions as abstract algebraic forms rather than physically grounded models. They propose that output should instead be modelled as a function of \emph{exergy} (the portion of energy available to perform useful work), combined with the efficiency of energy conversion. In this formulation, capital and labour are reinterpreted not as independent inputs, but as \emph{facilitators} of energy transformation processes. This leads to an alternative production function of the form:
\[
Y = \eta \cdot E_u,
\]
where $E_u$ is useful (low-entropy) energy and $\eta$ is the efficiency with which capital and labour convert primary energy into economic output.

This redefinition has significant implications for electricity market design, particularly in contexts (such as the UK) where system tightness, flexibility, digital control, and demand-side participation are increasingly central. If energy is the enabling constraint rather than merely a traded commodity, then market value must correspond to \emph{usable energy conversion under time, space, and network constraints} --- not simply to energy quantity or price per kWh. This supports the need to value flexibility, storage, locational losses, and conversion efficiency as core components of market function, rather than as supplementary services.

Keen’s work also provides a theoretical foundation for the integration of digital measurement, smart markets, and real-time optimisation. If economic value arises from physically grounded energy conversion, then mechanisms for monitoring, controlling, and algorithmically allocating energy in real time (e.g.\ automatic market clearing, locational scarcity signals, and capacity-aware allocation, as proposed in Chapter~\ref{ch:market_scenarios}) become central to efficient market operation rather than technological add-ons.

Moreover, Keen's critique highlights why many existing electricity market models --- particularly those based on marginal cost pricing, unlimited substitutability, or static equilibrium concepts --- struggle to represent modern system needs such as dynamic flexibility, heterogeneity of energy services, and physical network constraints. The absence of energy as a structural driver explains why markets often fail to reflect temporal and locational scarcity, why capital-heavy investments do not automatically generate resilience, and why digital coordination mechanisms (demand response, peer-to-peer trading, automatic market making) are undervalued by traditional theory.

Overall, these articles support this thesis in three key ways:
\begin{itemize}[leftmargin=*]
  \item They provide a theoretical justification for treating \textbf{energy, not price alone}, as the coordinating basis of market architecture.
  \item They validate the shift from static optimisation to \textbf{real-time, dynamic, and service-based market designs}.
  \item They justify the introduction of \textbf{flexibility, conversion efficiency, and participation integrity} as design constraints --- not optional add-ons.
\end{itemize}

Keen’s work therefore forms part of the philosophical and physical foundation for this thesis: repositioning energy from being a traded commodity to being the fundamental enabler of all economic activity. This reconceptualisation motivates the market reforms proposed in subsequent chapters, particularly the design of a physically grounded, digitally coordinated, fairness-aware Automatic Market Maker.

% ---------------------------------------------------------
\section{Synthesis of Gaps Across the Literature}

\begin{table}[h]
\centering
\begin{tabular}{p{4.2cm} p{9.8cm}}
\toprule
\textbf{Literature Domain} & \textbf{Identified Gaps} \\
\midrule
Power systems under renewables &
Lack of mechanisms for continuous, real-time coordination beyond central unit commitment, and inadequacy of existing adequacy frameworks under correlated weather-driven supply risk. \\

Classical market design &
Reliance on periodic clearing and marginal-cost pricing, weak linkage between scarcity, revenue sufficiency, and investment incentives, and misalignment between adequacy metrics and realised remuneration. \\

Fairness and cooperative game theory &
Absence of scalable, real-time fairness mechanisms that are operationally integrated into market clearing, settlement, and access allocation. \\

Local and peer-to-peer (P2P) markets &
Fragmentation from system-wide optimisation, limited scalability, and weak integration with transmission-level constraints and adequacy requirements. \\

Digitalisation and algorithmic regulation &
Event-based and cyber--physical control concepts remain disconnected from economic decision-making, pricing, and enforceable fairness constraints. \\
\bottomrule
\end{tabular}
\caption{Synthesis of unresolved gaps across the electricity market design literature.}
\label{tab:litgaps}
\end{table}

The literature as a whole points to a systemic need for a market architecture that is:

\begin{itemize}[leftmargin=*]
    \item event-driven rather than periodic,
    \item continuously clearing rather than batch-optimised,
    \item fairness-aware in real time rather than ex post,
    \item capable of activating distributed flexibility at scale, and
    \item integrated with digital regulatory and operational systems.
\end{itemize}

\section{Positioning of This Thesis}

The identified gaps motivate a new market architecture integrating concepts from power systems engineering, mechanism design, fairness theory, and digital control. While this chapter has summarised the state of the art, subsequent chapters will introduce a novel event-driven, fairness-aware electricity market model that aims to address these shortcomings.

% ---------------------------------------------------------
% CHAPTER 4 — PROBLEM DEFINITION
% ---------------------------------------------------------

\chapter{Problem Definition, System Realities, and Solution Concept}
\label{chap:problem}

% ---------------------------------------------------------
\section{Changing Nature of the Electricity System}
% ---------------------------------------------------------

Electricity systems were originally designed around a simple paradigm:
centralised thermal generation, predictable demand, one-way power flows,
and vertically integrated control. 

However, empirical data from capacity registers, distribution network voltage logs,
EV adoption trajectories, smart meter datasets, and flexibility dispatch trials 
reveal a fundamentally different system emerging in practice. Today, the system exhibits:

\begin{itemize}[leftmargin=1.2cm]
    \item \textbf{Variable, non-dispatchable generation} (wind, solar),
          whose availability is uncertain, location-dependent, and time-coupled;
    \item \textbf{Millions of distributed devices} — EVs, batteries,
          smart appliances — capable of both consuming and supplying power;
    \item \textbf{Two-way power flows} that introduce voltage instability
          and protection risks across increasingly stressed local networks;
    \item \textbf{Real-time digital metering and IoT infrastructure}
          which expose the latency, aggregation errors, and inefficiencies of
          batch-based settlement and coarse pricing granularity;
    \item \textbf{New demand shocks} — AI computing loads, hydrogen electrolysers,
          fusion facilities, and quantum data centres — which emerge without
          historical precedence and require new forms of coordination.
\end{itemize}

\noindent
These trends are observable, recorded, and rapidly scaling. They challenge the market
architecture, which was conceived for predictable supply, passive demand, and
slow-moving, volume-based settlement. The existing structure \emph{cannot properly
value flexibility, locational constraints, real-time deliverability, or long-term
adequacy}.

% ---------------------------------------------------------
\section{Stakeholder Landscape and Misaligned Incentives}
% ---------------------------------------------------------

The electricity system is not a centrally controlled machine. It is a
distributed coordination problem involving many stakeholders, each
with distinct objectives, rights, incentives, and constraints.

\begin{itemize}[leftmargin=1.2cm]
    \item \textbf{System Operator (ESO):} manages frequency, balancing,
          and transmission congestion.
    \item \textbf{Distribution System Operators (DSOs):} manage local voltage stability,
          fault levels, connection access, and increasingly operate
          quasi-markets for flexibility.
    \item \textbf{Retail Suppliers:} manage billing, hedging, and tariff structures,
          yet lack visibility of real-time physical deliverability
          and local network constraints.
    \item \textbf{Generators (Large + Distributed):} produce energy, but are increasingly
          required to offer flexibility, inertia, locational value,
          and adequacy—services currently underpriced.
    \item \textbf{Regulators and Government:} enforce affordability, transparency,
          decarbonisation, competition, and security-of-supply objectives.
    \item \textbf{End-users (Households, SMEs, Industry):} increasingly act as storage
          owners, EV users, heat pump operators, and latent flexibility providers.
    \item \textbf{Prosumers / Energy Communities / Virtual Power Plants:}
          may bypass retailers, challenge settlement structures,
          and redefine system participation.
\end{itemize}

\noindent
\textit{Yet, the existing market structure artificially isolates these actors, 
uses contract-based rather than physics-based value, and ignores their 
interdependent operational contributions.} Incentives are fragmented;
operational value is hidden, and coordination relies on ex post correction
rather than real-time alignment.

% ---------------------------------------------------------
\section{Physical Realities Ignored by the Current Market}
% ---------------------------------------------------------

Electrical systems obey the physics of AC power flow — which is
non-linear, time-coupled, and location-specific. In contrast,
existing market models treat electricity as if it were fully fungible,
location-agnostic, and divisible without consequence.

\begin{itemize}[leftmargin=1.2cm]
    \item AC flows follow Kirchhoff's laws, not contractual schedules;
    \item Losses, line limits, and voltage excursions depend on spatial topology
          and temporal coincidence of demand and generation;
    \item Distributed EV and solar export introduce protection challenges,
          transient instabilities, and local overload risks;
    \item Deliverability is not guaranteed — even if energy is ``available''
          at system level, it may not be deliverable to a specific location.
\end{itemize}

\noindent
Despite these empirical realities, existing settlement and pricing mechanisms
discount spatial deliverability, real-time network constraints, and scarcity formation.
This results in \textbf{unpriced constraints, misallocated cost, and distorted
investment signals}.

\vspace{0.2cm}
\noindent
\textbf{Core value failure:} Current market systems do not differentiate between
energy that \emph{can be delivered} and energy that \emph{cannot}.

% ---------------------------------------------------------
\section{Why Existing Market Mechanisms Cannot Scale}
% ---------------------------------------------------------

Most electricity markets — including the UK — still operate under:
\[
\text{Half-hourly settlement} + \text{locationally-blind tariffs} + \text{batch auctions}
\]

This architecture fails under modern system conditions because:

\begin{enumerate}[label=P\arabic*,leftmargin=1.2cm]
    \item \textbf{System events occur continuously,} not in half-hour blocks;
    \item \textbf{Flexibility must be activated in real time,} not retrospectively;
    \item \textbf{Batch auctions suppress locational value} and mask stability risks;
    \item \textbf{Retail tariffs ignore physical scarcity,} voltage risk, and topology;
    \item \textbf{Aggregation-based settlement assumes control at the centre,}
          while actionable flexibility exists at the edge.
\end{enumerate}

\noindent
Future challenges — including high-density EV charging corridors,
data centre-driven demand surges, fusion-scale generation, peer-to-peer trading,
and energy communities — will stretch these assumptions beyond operational viability.

% ---------------------------------------------------------
\section{The Missing Third Procurement Axis}
\label{sec:third_axis_problem}
% ---------------------------------------------------------

The diagnosis above can be summarised as a \emph{dimensionality problem}.
Legacy market designs treat procurement as essentially two-dimensional:

\[
(\text{energy},\ \text{capacity/adequacy}).
\]

Energy-only designs operate almost entirely along the first axis, expecting
short-run prices and scarcity rents to signal both operation and investment.
Energy-plus-capacity designs add a second, slower axis via capacity auctions,
contracts for difference, and adequacy schemes, but leave the core spot
architecture unchanged.

Modern electricity systems, however, require a \emph{third} procurement axis:

\[
(\text{energy},\ \text{capacity/adequacy},\ \text{QoS/flexibility/reliability}).
\]

This third axis captures:

\begin{itemize}[leftmargin=1.2cm]
  \item \textbf{Instantaneous flexibility:} the ability to move, curtail, or
        reshape demand and supply at sub-hourly timescales in response to
        renewable volatility and network conditions;
  \item \textbf{Location-specific service quality:} the probability that power
        is deliverable at a given node or feeder under stress, not just
        system-level adequacy;
  \item \textbf{Contracted reliability tiers:} explicit QoS levels that define
        who is curtailed, when, and by how much when the system is short.
\end{itemize}

Existing architectures handle this dimension only through a patchwork of
ancillary service markets, ex-ante flexibility tenders, and emergency
interventions. Flexibility is:

\begin{itemize}[leftmargin=1.2cm]
  \item procured months ahead as ``demand reduction'' without a robust
        counterfactual baseline;
  \item defined in contract space, not physical deliverability space;
  \item organised separately by DSOs and the ESO, often without tight
        coordination or a shared network model.
\end{itemize}

As a result, the system frequently pays for the \emph{wrong flexibility in the
wrong place at the wrong time}, while the devices that \emph{could} provide
high-value flexibility (EVs, heat pumps, batteries, industrial loads) are
under-utilised or excluded.

The core problem, therefore, is not simply that markets are ``inefficient'' or
``unfair'', but that the prevailing designs live in a \emph{two-dimensional
procurement space} while the physical system is \emph{three-dimensional}. The
solution concept developed in later chapters is explicitly built to operate in
this full three-dimensional space, with QoS/flexibility/reliability treated as
a first-class, priced, and programmable axis of the market.

% ---------------------------------------------------------
\section{Retail Architecture, Settlement Shocks, and Digitalisation Failure}
\label{sec:retail_problem}
% ---------------------------------------------------------

A particularly fragile part of the current architecture is the retail layer.
Suppliers sit between volatile wholesale markets and capped, politically
constrained retail tariffs. They are charged in wholesale markets on a
short-interval basis (e.g.\ every 30 minutes) against realised system load,
while most of their customers’ demand is either profiled, aggregated, or
measured with substantial delay.

For a large fraction of households and SMEs, demand is effectively
\emph{off-grid in the informational sense}: it is not observed at the
temporal or spatial resolution at which wholesale settlement occurs.
Instead, suppliers must assume a demand trajectory for each consumer,
using static profiles and ex post reconciliation. Any discrepancy between
assumed and realised demand then appears as an \emph{ex-post settlement
shock} on the supplier’s balance sheet.

Two structural problems follow:

\begin{enumerate}[label=R\arabic*,leftmargin=1.2cm]
  \item \textbf{Risk--volume separation at the retail edge.}  
  End-users choose their volume $Q(t)$ largely independently of wholesale
  conditions, subject to smooth, capped retail tariffs. Suppliers must
  honour that volume at the capped price, but face wholesale settlement
  at granular intervals against realised system conditions. Tail risk is
  concentrated in a thin retail shell with finite equity and no direct
  control over either physical deliverability or real-time volume.

  \item \textbf{Measurement gaps and asymmetric bill shocks.}  
  Where metering is coarse or absent, suppliers cannot track deviations
  between assumed and realised consumption in real time. Settlement
  shocks are discovered ex post and are typically socialised across the
  supplier’s portfolio, or passed through future tariff adjustments and
  policy costs. Vulnerable or less digitally connected customers face
  higher exposure to arbitrary bill outcomes, despite having provided
  little information or flexibility to the system.
\end{enumerate}

In principle, better metering, device telemetry, and demand control could
reduce this mismatch. In practice, the current architecture provides
\emph{weak and uneven} incentives for digitalisation:

\begin{itemize}[leftmargin=1.2cm]
  \item Suppliers can avoid installing smart meters or advanced devices in
        locations that are costly or operationally awkward (poor signal,
        access issues, low portfolio share), because the risk from those
        customers can be averaged across the rest of the book.
  \item There is no systematic reward for higher-resolution data streams
        or device controllability; once a minimum metering standard is met,
        additional granularity mostly appears as a cost.
  \item IoT and smart devices are deployed in an ad hoc manner, with
        heterogeneous standards and limited over-the-air upgradability,
        making long-term adaptation (e.g.\ post-quantum security,
        new settlement rules) difficult.
\end{itemize}

Moreover, current retail products are typically defined in one dimension
(energy volume and a static price), with at most coarse demand charges or
time-of-use differentials. There is no operationally enforced contract
structure around:

\begin{itemize}[leftmargin=1.2cm]
  \item \textbf{Quality of service (QoS):} probability and continuity of
        service under shortage;
  \item \textbf{Power impact:} the peak and network strain a customer
        imposes on the system; or
  \item \textbf{Openness to flexibility:} the extent to which devices can
        be shifted, throttled, or temporarily curtailed.
\end{itemize}

The absence of these contract dimensions means that suppliers cannot
meaningfully differentiate between customers who are:

\begin{itemize}[leftmargin=1.2cm]
  \item always-on, high-impact, non-flexible; and
  \item digitally integrated, controllable, and actively contributing to
        system stability.
\end{itemize}

Both types of customer are billed through similar volumetric tariffs,
even though they impose radically different risk and system impact.
This results in:

\begin{enumerate}[label=R\arabic*',leftmargin=1.2cm]
  \item \textbf{Blunt incentives for digitalisation:} suppliers are not
        structurally rewarded for deep IoT integration, high-frequency
        telemetry, or robust device management; 
  \item \textbf{Persistent unfairness in meter deployment:} hard-to-reach
        or low-income areas may be deprioritised for smart meters or
        advanced devices, entrenching data poverty and limiting access
        to flexibility products; and
  \item \textbf{Inability to price risk accurately:} without QoS,
        power-impact, and flexibility dimensions, retail products cannot
        reflect true operational contribution or risk, leading to
        cross-subsidies and distorted investment signals.
\end{enumerate}

In summary, the current retail architecture combines:
\emph{(i) ex-post settlement shocks}, \emph{(ii) weak, uneven incentives
for digitalisation}, and \emph{(iii) an absence of structured QoS--power--flexibility
contract dimensions}. This combination undermines both solvency and fairness
at the edge of the system (see also
Sections~\ref{sec:digitalisation_fragmentation}
and~\ref{sec:smart_meter_experience} for the empirical background on
digitalisation and smart meters). Chapter~\ref{ch:market_scenarios} and
Chapter~\ref{ch:amm} will show how a different contract structure and
continuous, cyber--physical market design can realign these incentives and
make granular, trustworthy data a core part of suppliers' business model.

% ---------------------------------------------------------
\section{The Fairness Gap — No Operational Definition}
% ---------------------------------------------------------

Fairness is frequently cited in energy policy, yet almost never defined in a way
that is \textit{physically grounded, operationally implementable, and auditable}.
Existing interpretations rely on:

\begin{itemize}[leftmargin=1.2cm]
    \item Ex post redistribution (social tariffs, subsidies, regulated compensation),
    \item Qualitative notions (fuel poverty, vulnerability, ``just transition''),
    \item Revenue adequacy and cost-recovery rules detached from operational value.
\end{itemize}

\noindent
But no framework defines fairness in terms of:
\[
\textbf{Who produces value? Who consumes value? Who imposes cost?}
\]

Nor do existing market designs link fairness with deliverability,
flexibility provision, system stabilisation, or scarcity relief.
This omission results in pricing inefficiency, mistrust, and
poor incentives for participation and investment.

Section~\ref{sec:price_cap_limits} below formalises several of these failures
as structural properties of the prevailing retail and wholesale architecture.

% ---------------------------------------------------------
\section{Structural Limits of Price-Capped Electricity Markets}
\label{sec:price_cap_limits}
% ---------------------------------------------------------

Electricity markets impose a unique structural mismatch between how costs are
incurred (capital-intensive, long-lived, non-marginal) and how revenues are
recouped (short-term, volume-based, retail-constrained). This mismatch is
exacerbated by \textbf{retail price caps}, \textbf{wholesale price floors}, and
an intermediary structure where suppliers carry volume and timing risk without
any mechanism to hedge non-fuel costs or to adjust prices in real time.

At a high level, the legacy architecture \emph{separates}:
\begin{itemize}[leftmargin=*]
    \item \textbf{who chooses volume} (end-users, responding weakly to capped prices); from
    \item \textbf{who bears tail risk} (retail suppliers, with finite balance sheets).
\end{itemize}
This risk--volume separation is the core structural weakness that makes the
system non-robust to shocks. The following subsections formalise this mismatch.

\subsection{Cost and Revenue Decomposition Across the Value Chain}

We distinguish three levels of cost formation:
\begin{enumerate}[leftmargin=*]
    \item \textbf{Generator-level costs} (CapEx, fuel, Opex):
    \[
    C_G(t) = C_f(t) + C_{nf}(t)
    \]
    where $C_f(t)$ is fuel (marginal, volatile) and $C_{nf}(t)$ is non-fuel (CapEx, Opex, sunk, time-shifted).

    \item \textbf{Wholesale settlement price} incorporates only fuel cost plus scarcity premium:
    \[
    P_W(t) \approx MC_f(t) + \sigma(t),
    \]
    where $MC_f(t) = \frac{dC_f}{dQ}$ and $\sigma(t)$ emerges only under shortage conditions. Non-fuel CapEx is structurally excluded from clearing.

    \item \textbf{Retail revenue} is volume-based and capped:
    \[
    R(t) = P_R^{\text{cap}}(t) \cdot Q(t),
    \]
    where $P_R^{\text{cap}}(t)$ is regulated and smooth, independent of real-time or capital cost signals.
\end{enumerate}

\noindent
Thus, for solvency of an energy supplier, the following structural condition must hold:
\[
\int_0^T P_R^{\text{cap}}(t) \cdot Q(t) \, dt \ge \int_0^T \big[ C_f(t) + C_{nf}(t) \big] \, dt.
\]

However, under a fuel price shock,
\[
C_f(t_s) \gg P_R^{\text{cap}}(t_s) \cdot Q(t_s),
\]
implying
\[
\text{Net income}(t_s) < 0, \qquad
\text{irrespective of supplier efficiency or risk management}.
\]

\subsection{Why Shocks Cause Insolvency Even for Efficient Suppliers}

The crucial insight is that under the current architecture:

\begin{itemize}[leftmargin=*]
    \item Retail prices are \textbf{fixed ex-ante} (price cap), while fuel costs are \textbf{volatile ex-post}.
    \item Non-fuel costs are incurred \textbf{ex-ante}, while revenue recovery is \textbf{ex-post} over volume.
    \item Suppliers are structurally exposed to \textbf{volume risk}, \textbf{liquidity risk}, and \textbf{temporal mismatch}.
\end{itemize}

\noindent
Therefore, solvency is path-dependent:
\[
\text{Solvency} = \int_0^{T} \bigl[ R(t) - C_f(t) - C_{nf}(t) \bigr]\, dt \quad \text{cannot be guaranteed}.
\]

Even more fundamentally, the market is unable to guarantee recovery of non-fuel costs because:
\[
C_{nf}(t) \not\propto Q(t),
\qquad
\text{yet}
\qquad
R(t) \propto Q(t).
\]
The structure forces long-lived, non-marginal costs to be recovered through a
short-run, volume-based and politically capped revenue stream.

\subsection{Formal Structural Results}

The following lemmas formalise the insolvency and instability properties of the
price-capped retail architecture, and the failure of unconstrained pricing to
guarantee affordability of essential demand.

\begin{lemma}[Structural Insolvency under Price Caps]
\label{lem:price_cap_insolvency}
Consider a retail electricity supplier operating over a horizon $[0,T]$ with:
\begin{enumerate}[leftmargin=*]
    \item exogenous demand $Q(t) \ge 0$ and a regulated retail price cap
    $P_R(t) \le P_R^{\mathrm{cap}}$ for all $t \in [0,T]$;
    \item fuel (wholesale) cost $c_f(t)$ per unit of energy and
    non-fuel cost $C_{nf}(t)$ (CapEx and Opex) that is non-negative and
    not proportional to $Q(t)$; and
    \item a finite liquidity buffer $L_{\max} > 0$.
\end{enumerate}
Assume that there exists a shock interval $I = [t_s, t_s + \Delta]
\subset [0,T]$ such that
\[
c_f(t) > P_R^{\mathrm{cap}} \quad \text{for all } t \in I,
\]
and that demand is strictly positive on $I$, i.e.\ $Q(t) \ge \underline{Q} > 0$
for all $t \in I$.

Then, irrespective of the supplier's operational efficiency or tariff
design (as long as it respects the price cap), there exists a shock
duration $\Delta$ such that the supplier's cumulative net position
satisfies
\[
\int_0^T \big[ P_R(t) Q(t) - c_f(t) Q(t) - C_{nf}(t) \big]\, dt < -L_{\max},
\]
and the supplier becomes insolvent.
\end{lemma}

\begin{proof}
Over any interval $[0,T]$, the supplier's cumulative profit is
\[
\Pi_T
=
\int_0^T P_R(t) Q(t) \, dt
-
\int_0^T c_f(t) Q(t) \, dt
-
\int_0^T C_{nf}(t) \, dt.
\]

By the retail price cap, we have $P_R(t) \le P_R^{\mathrm{cap}}$ for all
$t$. Decompose the horizon into the shock interval $I = [t_s,t_s+\Delta]$
and its complement. Over $I$,
\[
P_R(t) Q(t) \le P_R^{\mathrm{cap}} Q(t),
\quad
c_f(t) Q(t) > P_R^{\mathrm{cap}} Q(t),
\]
so the incremental profit on $I$ is strictly negative:
\[
\Pi_I
=
\int_I \big[ P_R(t) Q(t) - c_f(t) Q(t) \big] \, dt
-
\int_I C_{nf}(t) \, dt
<
\int_I \big[ P_R^{\mathrm{cap}} Q(t) - c_f(t) Q(t) \big] \, dt
\le
-\delta \int_I Q(t) \, dt,
\]
for some $\delta > 0$ (since $c_f(t) - P_R^{\mathrm{cap}} \ge \delta$ on $I$
by assumption). Using $Q(t) \ge \underline{Q} > 0$ on $I$, we obtain
\[
\Pi_I \le -\delta \, \underline{Q} \, \Delta.
\]

Over the complement $[0,T]\setminus I$, even if the supplier operates at
best-possible efficiency and earns maximal feasible surplus, its profit
is bounded above by some finite constant $B < \infty$:
\[
\Pi_{[0,T]\setminus I} \le B.
\]

Therefore the total profit satisfies
\[
\Pi_T = \Pi_I + \Pi_{[0,T]\setminus I}
\le
-\delta \, \underline{Q} \, \Delta + B.
\]

For any finite liquidity buffer $L_{\max} > 0$, we can choose a shock
duration $\Delta$ large enough such that
\[
-\delta \, \underline{Q} \, \Delta + B < -L_{\max},
\]
i.e.\ $\Pi_T < -L_{\max}$.

This means that, despite any operational efficiency and irrespective of
tariff design (as long as it respects the price cap and serves positive
demand), a sufficiently severe and/or prolonged fuel cost shock forces
the supplier's cumulative net position below $-L_{\max}$. With finite
liquidity, insolvency is then structurally unavoidable.

The result does not depend on the detailed shape of $C_{nf}(t)$; any
non-negative non-fuel cost profile only worsens the bound. Hence, in a
price-capped regime with non-marginal costs and exogenous shocks to
fuel prices, insolvency risk is a \emph{structural property of the
market architecture}, not merely of individual firm management.
\end{proof}

\begin{lemma}[Risk--Volume Separation Instability]
\label{lem:risk_volume_instability}
Consider any retail architecture with the following features over a horizon
$[0,T]$:
\begin{enumerate}[leftmargin=*]
    \item exogenous demand $Q(t) \ge 0$ chosen by end-users in response to a
    regulated retail price $P_R(t) \le P_R^{\mathrm{cap}}$;
    \item stochastic fuel cost $c_f(t)$ per unit of energy and non-fuel cost
    $C_{nf}(t) \ge 0$ that is not proportional to $Q(t)$; and
    \item a risk-bearing intermediary (supplier) with finite equity
    $E_{\max} > 0$ that must honour all realised demand $Q(t)$ at price
    $P_R(t)$ and bears the residual payoff
    \[
      \Pi_T
      =
      \int_0^T \big[ P_R(t) Q(t) - c_f(t) Q(t) - C_{nf}(t) \big]\, dt.
    \]
\end{enumerate}
Suppose further that:
\begin{enumerate}[label=(A\arabic*), leftmargin=*]
    \item \textbf{Separated decisions:} volume $Q(t)$ is chosen by consumers and
    cannot be curtailed by the intermediary except through emergency
    disconnection; 
    \item \textbf{Restricted price response:} $P_R(t)$ cannot adjust
    contemporaneously to $c_f(t)$ beyond the cap $P_R^{\mathrm{cap}}$; and
    \item \textbf{Finite loss-absorbing capacity:} the intermediary defaults if
    $\Pi_T < -E_{\max}$.
\end{enumerate}
Then, for any finite $E_{\max}$, there exists a fuel cost path $c_f(t)$ and a
demand path $Q(t)$, consistent with these assumptions, such that the intermediary
defaults. In particular, no static capital buffer $E_{\max}$ can make the
architecture robust to bounded-but-unmodelled fuel price shocks as long as
volume choice and tail risk-bearing remain separated in this way.
\end{lemma}

\begin{proof}
The construction in Lemma~\ref{lem:price_cap_insolvency} already provides an
explicit example of a shock interval $I = [t_s, t_s + \Delta]$ on which
$c_f(t) > P_R^{\mathrm{cap}}$ and $Q(t) \ge \underline{Q} > 0$, yielding a
negative profit contribution $\Pi_I \le -\delta \underline{Q} \Delta$. Over the
complement, the profit is bounded above by some finite $B$.

Thus for any $E_{\max} > 0$ we can choose $\Delta$ sufficiently large that
$\Pi_T \le -\delta \underline{Q} \Delta + B < -E_{\max}$, implying default.
The key structural feature is that the intermediary cannot simultaneously
control $Q(t)$ and $P_R(t)$ in response to $c_f(t)$: end-users choose the
volume, while the retail price is constrained by the cap. The risk-bearing
entity therefore faces an unbounded downside relative to its finite buffer.

Hence, \emph{for any} finite $E_{\max}$, there exist admissible cost and
demand paths that exhaust the buffer. This shows that structural
non-robustness to shocks is a consequence of \emph{risk--volume separation}, not
of poor individual risk management.
\end{proof}

\noindent
Together, Lemma~\ref{lem:price_cap_insolvency} and
Lemma~\ref{lem:risk_volume_instability} show that insolvency cascades in
price-capped retail architectures are not merely accidents or management
failures. They are the natural outcome of a design that separates volume choice
from tail-risk bearing under volatile fuel costs and capped prices.

\begin{lemma}[Affordability Failure without Retail Price Caps]
\label{lem:affordability_no_cap}
Consider a retail electricity architecture over a horizon $[0,T]$ with:
\begin{enumerate}[leftmargin=*]
    \item wholesale fuel cost $c_f(t)$ per unit of energy;
    \item a retail price $P_R(t)$ set by the supplier with no binding cap, and
    satisfying a cost-recovery constraint
      \[
        P_R(t) \;\ge\; c_f(t) \quad \text{for all } t \in [0,T];
      \]
    \item an essential demand profile $Q^{\mathrm{ess}}(t) \ge \underline{Q} > 0$
    that is short-run inelastic with respect to $P_R(t)$; and
    \item a flow of household income $Y(t)$ that is bounded above,
    $Y(t) \le Y_{\max} < \infty$, and an affordability requirement that
    essential energy expenditure not exceed a fixed fraction $\theta \in (0,1]$
    of income:
      \[
        P_R(t)\, Q^{\mathrm{ess}}(t) \;\le\; \theta\, Y(t)
        \quad \text{for all } t \in [0,T].
      \]
\end{enumerate}
Suppose further that the wholesale cost process $c_f(t)$ is unbounded above in
the sense that for any $M > 0$ there exists an interval
$I_M \subset [0,T]$ with positive measure on which $c_f(t) \ge M$.

Then there exists $M^\star$ and a corresponding interval
$I_{M^\star}$ such that, on $I_{M^\star}$, either:
\begin{enumerate}[label=(\alph*), leftmargin=*]
    \item the affordability constraint is violated,
    \[
      P_R(t)\, Q^{\mathrm{ess}}(t) > \theta\, Y(t),
    \]
    in which case essential demand cannot be paid for without generating bad
    debts; or
    \item the supplier rations or disconnects demand, i.e. does not serve
    $Q^{\mathrm{ess}}(t)$, and essential energy is not delivered.
\end{enumerate}
In particular, in the absence of a retail price cap, no choice of pricing rule
$P_R(t) \ge c_f(t)$ can guarantee both cost recovery and affordability of
essential demand under unbounded wholesale price shocks.
\end{lemma}

\begin{proof}
By assumption, $Y(t) \le Y_{\max}$ for all $t$ and
$Q^{\mathrm{ess}}(t) \ge \underline{Q} > 0$ for all $t$. Fix any
$\theta \in (0,1]$. Choose
\[
  M^\star
  \;>\;
  \frac{\theta\, Y_{\max}}{\underline{Q}}.
\]
By unboundedness of $c_f(t)$, there exists an interval
$I_{M^\star} \subset [0,T]$ of positive measure on which
$c_f(t) \ge M^\star$.

On this interval, cost recovery requires
$P_R(t) \ge c_f(t) \ge M^\star$, and essential demand is at least
$\underline{Q}$. Hence, for all $t \in I_{M^\star}$,
\[
  P_R(t)\, Q^{\mathrm{ess}}(t)
  \;\ge\;
  M^\star \, \underline{Q}
  \;>\;
  \theta\, Y_{\max}
  \;\ge\;
  \theta\, Y(t).
\]
Thus the affordability condition
$P_R(t)\, Q^{\mathrm{ess}}(t) \le \theta\, Y(t)$ cannot hold on
$I_{M^\star}$.

The supplier therefore faces a binary choice on $I_{M^\star}$:
\begin{itemize}[leftmargin=*]
  \item either it serves $Q^{\mathrm{ess}}(t)$ at price $P_R(t)\ge M^\star$,
  in which case households cannot fully pay within the affordability
  constraint and bad debts (or arrears) are generated; or
  \item it refuses to serve $Q^{\mathrm{ess}}(t)$ (through disconnection,
  rationing, or non-contracting), in which case essential demand is not met.
\end{itemize}

In either case, the system fails to guarantee simultaneously:
(i) cost recovery at the retail level, and
(ii) affordability of essential demand under the stipulated income bound.
Since $M^\star$ and $I_{M^\star}$ arose from the assumed unboundedness of
$c_f(t)$, this failure is structural: no choice of pricing rule with
$P_R(t) \ge c_f(t)$ can preclude it while wholesale prices can become
arbitrarily large.

Hence, in the absence of a retail price cap, affordable essential energy
cannot be guaranteed; extremely high or effectively unbounded retail prices
are admissible, and these necessarily imply either unaffordable bills and bad
debts or unmet essential demand on some shock paths.
\end{proof}

\begin{proposition}[Layered Markets Converge to the Energy--Only Limit in Stress Events]
\label{prop:layered_to_energy_only}
Consider any electricity market architecture composed of:
\begin{enumerate}[leftmargin=*]
    \item an \emph{energy layer} with short-interval wholesale settlement and
          real-time balancing prices $P^{\mathrm{E}}_t$;
    \item one or more \emph{capacity, adequacy, or support layers} providing
          fixed or slowly varying payments $P^{\mathrm{C}}$ that do not depend on
          the contemporaneous realisation of scarcity;
    \item a \emph{retail layer} with either (i) a regulated price cap
          $P_R(t)\le P_R^{\mathrm{cap}}$, or (ii) unrestricted pass-through of
          wholesale spot prices to consumers; and
    \item consumers choosing real-time volumetric demand $Q(t)\ge 0$
          independently of system state.
\end{enumerate}

Suppose that a stress event occurs on an interval $I=[t_s, t_s+\Delta]$ such that:
\[
P^{\mathrm{E}}_t \;\gg\; P^{\mathrm{C}} 
\qquad\text{and}\qquad
Q(t)\ge\underline{Q}>0
\quad \forall\, t\in I.
\]

Then the following hold:

\begin{enumerate}[leftmargin=*]
    \item During $I$, the total marginal revenue of any flexible generator is:
    \[
    P^{\mathrm{tot}}_t = P^{\mathrm{E}}_t + P^{\mathrm{C}}
    \approx P^{\mathrm{E}}_t,
    \]
    so generation decisions converge to those of a pure energy-only market.

    \item If a retail price cap is present, retail decisions become
    \emph{identical} to those of an energy-only market with a fixed retail
    price:
    \[
    P_R(t)=P_R^{\mathrm{cap}}, \quad Q(t)\text{ inelastic on }I,
    \]
    creating unbounded tail risk for suppliers
    (Lemmas~\ref{lem:price_cap_insolvency}--\ref{lem:risk_volume_instability}).

    \item If no retail cap is present, the retail price must satisfy
    \[
    P_R(t)\approx P^{\mathrm{E}}_t,
    \]
    and thus can reach arbitrarily high levels, reproducing the extreme-price
    behaviour of energy-only designs (Corollary~\ref{cor:no_free_lunch_retail}).

    \item In both cases, the equilibrium conditions of the layered system satisfy:
    \[
    \lim_{\Delta\to\infty} \text{Equilibrium}(P^{\mathrm{E}}_t, P^{\mathrm{C}}) 
    = 
    \text{Equilibrium}^{\mathrm{EnergyOnly}}(P^{\mathrm{E}}_t),
    \]
    i.e.\ \textbf{capacity payments become irrelevant in determining operational
    behaviour, risk allocation, or solvency}.
\end{enumerate}

Therefore, any multi-layer market with slow-moving support payments
necessarily collapses to the behaviour of its \emph{energy-only core} in
real stress events. Insolvency, extreme prices, and non-existence of stable
Nash equilibria in the energy-only limit imply identical fragilities in all
layered architectures built on top of it, including the GB energy+capacity
design.
\end{proposition}

\begin{corollary}[No Simultaneous Solvency and Affordability under Separated Retail Risk]
\label{cor:no_free_lunch_retail}
Let a retail electricity architecture satisfy the structural features of
Lemmas~\ref{lem:price_cap_insolvency},
\ref{lem:risk_volume_instability},
and~\ref{lem:affordability_no_cap}:
\begin{enumerate}[leftmargin=*]
    \item end-users choose (essential) demand $Q(t)$, which is short-run
    inelastic and cannot be continuously curtailed by the intermediary except
    through disconnection;
    \item a risk-bearing intermediary (supplier) with finite loss-absorbing
    capacity must honour realised demand at a regulated or chosen retail price
    $P_R(t)$; and
    \item wholesale fuel costs $c_f(t)$ can experience shocks that are
    unbounded above on sets of non-zero measure.
\end{enumerate}
Then no choice of retail pricing rule and capital buffer can jointly guarantee:
\begin{enumerate}[label=(\alph*), leftmargin=*]
    \item solvency of all suppliers (i.e.\ avoidance of insolvency cascades),
    and
    \item affordability and continuity of essential demand
    for end-users.
\end{enumerate}
In particular:
\begin{itemize}[leftmargin=*]
    \item with binding retail price caps, Lemma~\ref{lem:price_cap_insolvency}
    implies that sufficiently severe or prolonged wholesale price shocks
    structurally drive suppliers into insolvency; while
    \item without retail price caps, Lemma~\ref{lem:affordability_no_cap}
    implies that essential energy cannot be guaranteed affordable, and extreme
    price spikes necessarily generate either bad debts or unmet essential
    demand.
\end{itemize}
Lemma~\ref{lem:risk_volume_instability} further shows that, under this
risk--volume separation, no finite equity buffer can make the architecture
robust to such shocks. Thus the observed trade-off between supplier failure
and unaffordable bills is not an accident of management, but a structural
property of the prevailing retail design.
\end{corollary}

\begin{proposition}[Non-Existence of a Shock-Robust Nash Equilibrium in the Legacy Retail Game]
\label{prop:no_nash_equilibrium}
Consider the retail electricity architecture described in
Section~\ref{sec:price_cap_limits} (see also Section~\ref{sec:retail_problem} for
the settlement and digitalisation context). Model the interaction between:
\begin{itemize}[leftmargin=*]
    \item $N$ end-users, each choosing a consumption trajectory
          $Q_i(t) \ge 0$ and payment effort subject to income constraints;
    \item a retail supplier choosing pricing, hedging, and portfolio
          strategies $(P_R(t),h(t))$ subject to either\\
          (i) a retail price cap $P_R(t) \le P_R^{\mathrm{cap}}$, or\\
          (ii) cost-recovery $P_R(t) \ge c_f(t)$ when no cap applies.
\end{itemize}
Assume:
\begin{enumerate}[label=(A\arabic*), leftmargin=*]
    \item \textbf{Essential demand inelasticity:}
          each user $i$ has essential demand $Q_i^{\mathrm{ess}}(t)\ge \underline{Q}_i>0$
          that is short-run price-inelastic;
    \item \textbf{Finite supplier equity:} the supplier defaults if
          $\Pi_T < -E_{\max}$ for some finite $E_{\max}>0$;
    \item \textbf{Admissible wholesale shocks:}
          the wholesale fuel cost process $c_f(t)$ is unbounded above on
          sets of positive measure;
    \item \textbf{Feasible-strategy equilibrium:}
          a Nash equilibrium requires (i) no player can profitably deviate,
          and (ii) all feasibility constraints (solvency, affordability,
          and continuity of essential demand) hold almost surely.
\end{enumerate}

Under these assumptions, \emph{no shock-robust Nash equilibrium exists}.
More precisely:
\begin{enumerate}[label=(i), leftmargin=*]
    \item With a retail price cap, Lemma~\ref{lem:price_cap_insolvency}
          implies that for some admissible $c_f(t)$ paths, any supplier
          strategy respecting the cap induces insolvency
          ($\Pi_T < -E_{\max}$) with positive probability.
    \item Without a price cap, Lemma~\ref{lem:affordability_no_cap}
          implies that for some admissible $c_f(t)$ and income paths,
          any cost-recovering pricing strategy $P_R(t)\ge c_f(t)$
          violates affordability constraints for essential demand,
          generating bad debts or unmet essential demand.
\end{enumerate}

Hence, there is no strategy profile $(Q_i(t),P_R(t),h(t))$ such that:
\[
\text{(no unilateral profitable deviation)} 
\quad \text{and} \quad
\Pr[\text{solvency $\land$ affordability $\land$ continuity}] = 1.
\]

Any putative equilibrium is therefore \emph{not dynamically stable}:
when sufficiently large fuel price shocks occur, the game exits the feasible
strategy space into default, disconnection, political intervention, or
renegotiation states. The legacy retail architecture thus fails to admit a
Nash equilibrium that is both individually rational and shock-robust.
\end{proposition}

\subsection{VoLL, Scarcity Pricing, and Welfare Surrogates}

Locational marginal pricing (LMP) with scarcity pricing is often justified
using textbook welfare arguments: prices equal marginal cost, and total
surplus---consumer plus producer surplus---is maximised. The following
results show that this logic is structurally fragile in electricity systems.

\begin{lemma}[Arbitrariness and Sensitivity of VoLL in LMP-Based Scarcity Pricing]
\label{lem:voll_sensitivity}
Consider a single-node (or single-location) electricity system cleared by
locational marginal pricing (LMP) over a horizon $[0,T]$, with:
\begin{enumerate}[leftmargin=*]
    \item inelastic essential demand $Q^{\mathrm{ess}}(t) \ge \underline{Q} > 0$;
    \item a generation technology with constant marginal cost $c > 0$ and
    installed capacity $K \ge 0$;
    \item a stochastic net-load process $L(t)$ such that, for some non-zero
    measure set of times,
    \[
      \mathbb{P}\big(L(t) > K\big) > 0,
    \]
    so that shortages are possible;
    \item a value of lost load parameter $\VoLL > 0$ used in the system operator's
    welfare maximisation as a per-unit penalty for unserved energy.
\end{enumerate}
The system operator maximises expected social welfare
\[
  W(K;\VoLL)
  =
  \mathbb{E}\Bigg[
    \int_0^T
      \Big(
        u(Q^{\mathrm{ess}}(t))
        - c\, \min\{L(t),K\}
        - \VoLL \, \big(L(t) - K\big)_+
      \Big)\, dt
  \Bigg]
  - \kappa K,
\]
where $u(\cdot)$ is the (fixed) utility of supplied essential demand,
$\kappa > 0$ is the per-unit capacity cost, and $(x)_+ = \max\{x,0\}$.
Let $K^\star(\VoLL)$ denote an optimal capacity level.

Then, under mild regularity conditions on the distribution of $L(t)$:
\begin{enumerate}[label=(\alph*), leftmargin=*]
    \item the optimal capacity $K^\star(\VoLL)$ is (weakly) increasing in
    $\VoLL$; and
    \item the associated equilibrium LMPs and scarcity rents are strictly
    increasing functions of $\VoLL$ whenever there is a non-zero probability of
    shortage.
\end{enumerate}
In particular, raising $\VoLL$ shifts both the optimal reserve margin and the
present value of scarcity revenues upward, and there is no internal mechanism
in the LMP formulation that pins down a ``correct'' value of $\VoLL$.
Hence the welfare, investment, and distributional properties of an LMP-based
design are structurally sensitive to an administratively chosen, essentially
normative scalar parameter.
\end{lemma}

\begin{proof}
For any fixed capacity $K$, expected welfare can be written as
\[
  W(K;\VoLL)
  =
  \mathbb{E}\Bigg[
    \int_0^T
      u(Q^{\mathrm{ess}}(t))\, dt
  \Bigg]
  - \mathbb{E}\Bigg[
    \int_0^T
      \Big(
        c\, \min\{L(t),K\}
        + \VoLL \, \big(L(t) - K\big)_+
      \Big)\, dt
  \Bigg]
  - \kappa K.
\]
The first term does not depend on $K$ or $\VoLL$, so we focus on the cost
component. For each $t$, define the expected shortage at capacity level $K$ as
\[
  S(K)
  :=
  \mathbb{E}\big[ (L(t) - K)_+ \big],
\]
which is decreasing in $K$ and strictly positive whenever
$\mathbb{P}(L(t) > K) > 0$.

Ignoring differentiability issues (which can be addressed under standard
regularity assumptions on the distribution of $L(t)$), the derivative of
$W(K;\VoLL)$ with respect to $K$ is approximately
\[
  \frac{\partial W}{\partial K}(K;\VoLL)
  \approx
  - \mathbb{E}\Bigg[
    \int_0^T
      \Big(
        c \, \mathbf{1}\{L(t) > K\}
        - \VoLL \, \mathbf{1}\{L(t) > K\}
      \Big)\, dt
  \Bigg]
  - \kappa,
\]
so that the first-order condition for an interior optimum satisfies
\[
  \mathbb{E}\Bigg[
    \int_0^T
      (\VoLL - c) \, \mathbf{1}\{L(t) > K^\star(\VoLL)\} \, dt
  \Bigg]
  = \kappa.
\]
The left-hand side is increasing in $\VoLL$ and decreasing in $K$; the
right-hand side $\kappa$ is constant. Thus, to restore the equality after an
increase in $\VoLL$, $K^\star(\VoLL)$ must (weakly) increase. This proves
monotonicity of $K^\star(\VoLL)$ in $\VoLL$, establishing (a).

For (b), under LMP with scarcity pricing, the nodal price at the single node
is
\[
  P^{\mathrm{LMP}}(t)
  =
  \begin{cases}
    c, & L(t) \le K^\star(\VoLL), \\
    \\VoLL, & L(t) > K^\star(\VoLL),
  \end{cases}
\]
so that whenever $\mathbb{P}(L(t) > K^\star(\VoLL)) > 0$, the expected price
and the scarcity rent
\[
  R(\VoLL)
  :=
  \mathbb{E}\Bigg[
    \int_0^T
      \big(P^{\mathrm{LMP}}(t) - c\big)\,
      \min\{L(t),K^\star(\VoLL)\}\, dt
  \Bigg]
\]
are strictly increasing in $\VoLL$. Intuitively, raising $\VoLL$ lifts the
price ceiling in shortage states and thereby increases both expected prices
and scarcity revenues.

Since $\VoLL$ enters the objective only as a penalty coefficient on unserved
energy and is not revealed by any actual willingness-to-pay observation (in
particular, demand is modelled as inelastic at $Q^{\mathrm{ess}}(t)$), there
is no internal market mechanism that determines a unique, ``correct'' value of
$\VoLL$. Different admissible choices of $\VoLL$ generate different
$K^\star(\VoLL)$, different scarcity rents, and different present value
transfers between consumers and generators.

Thus the welfare, investment, and distributional properties of LMP with
VoLL-based scarcity pricing are structurally sensitive to an administratively
chosen parameter whose level is fundamentally normative and cannot be
identified from market behaviour alone.
\end{proof}

\begin{lemma}[Limited Validity of Surplus-Based Welfare in Electricity Markets]
\label{lem:surplus_welfare_mismatch}
Consider an economy of $N$ consumers indexed by $i = 1,\dots,N$ and a single
homogeneous electricity good $q \ge 0$ supplied at marginal cost $c(q)$.
Let $q_i$ denote consumer $i$'s consumption and $q = \sum_i q_i$ the aggregate
quantity. Define:
\begin{itemize}[leftmargin=*]
  \item individual utility $U_i(q_i, y_i)$, where $y_i$ is numéraire income;
  \item a (Marshallian) inverse demand curve $P(q)$, constructed from the
  aggregation of individual demands; and
  \item total surplus
  \[
    TS(q)
    :=
    \int_0^q P(z)\, dz - \int_0^q c(z)\, dz,
  \]
  interpreted as consumer plus producer surplus.
\end{itemize}
Suppose that the following textbook assumptions hold:
\begin{enumerate}[label=(A\arabic*), leftmargin=*]
  \item \textbf{Quasi-linearity and equal marginal utility of income:}
  for all $i$, $U_i(q_i, y_i) = u(q_i) + y_i$ with a \emph{common} function
  $u(\cdot)$; 
  \item \textbf{Perfect information and complete participation:}
  the planner or market designer observes $u(\cdot)$ and $c(\cdot)$, and every
  consumer participates in the market at the prevailing price; 
  \item \textbf{Homogeneity:}
  consumers differ at most by an additive constant in utility (no systematic
  vulnerability, essentiality, or priority classes); and
  \item \textbf{Economic rationality:}
  each consumer chooses $q_i$ to maximise $U_i(q_i,y_i)$ given the price,
  and the aggregate demand curve $P(q)$ is generated by these optimising
  decisions.
\end{enumerate}
Then any quantity $q^\star$ that maximises total surplus $TS(q)$ over $q \ge 0$
also maximises the utilitarian social welfare function
\[
  W(q_1,\dots,q_N)
  :=
  \sum_{i=1}^N U_i(q_i,y_i)
  -
  \int_0^{\sum_i q_i} c(z)\, dz,
\]
subject to $\sum_i q_i = q$, and the surplus ordering over quantities coincides
with the welfare ordering.

However, if any of (A1)--(A3) fails---in particular, if:
\begin{itemize}[leftmargin=*]
  \item consumers have heterogeneous utility functions $U_i$ reflecting
  different essentiality, vulnerability, or flexibility of demand;
  \item marginal utility of income differs across $i$ due to income constraints;
  or
  \item some consumers are non-participating or mispriced because their state is
  not observed (e.g.\ prepayment meters, disconnection risk, or hidden
  vulnerability),
\end{itemize}
then there exist feasible allocations $(q_i)$ and $(q_i')$ with
$\sum_i q_i = \sum_i q_i' = q$ such that:
\[
  TS(q) > TS(q')
  \quad \text{but} \quad
  W(q_1,\dots,q_N) < W(q_1',\dots,q_N').
\]
That is, total surplus is no longer a reliable proxy for social welfare; it
can rank allocations \emph{oppositely} to a welfare criterion that respects
heterogeneous needs and income constraints. This mismatch is structural in
electricity systems, where essential loads, vulnerability, and inability to
pay are pervasive.
\end{lemma}

\begin{proof}
Under (A1)--(A4), each consumer solves
\[
  \max_{q_i \ge 0} \; \big[ u(q_i) + y_i - P q_i \big],
\]
so that individual demand depends only on $P$ and the common $u(\cdot)$, and
the aggregate inverse demand $P(q)$ coincides with the marginal utility of
aggregate consumption:
\[
  P(q) = u'(q)
  \quad \text{for } q = \sum_i q_i.
\]
Total surplus can then be written as
\[
  TS(q)
  =
  \int_0^q u'(z)\, dz - \int_0^q c(z)\, dz
  =
  u(q) - \int_0^q c(z)\, dz + \text{constant},
\]
which differs from $W$ only by an additive constant (the sum of $y_i$). Hence
maximising $TS(q)$ over $q$ is equivalent to maximising $W$ over feasible
allocations with $\sum_i q_i = q$, establishing the first part.

Now drop (A1) and (A3) and consider two consumers, $i=1,2$, with
\[
  U_1(q_1,y_1) = u_H(q_1) + y_1, \qquad
  U_2(q_2,y_2) = u_L(q_2) + y_2,
\]
where $u_H$ represents a highly vulnerable or essential load (steep marginal
utility at low $q_1$) and $u_L$ a relatively low-priority or luxury load
(flatter marginal utility). Assume also that $y_1 \ll y_2$, so that consumer~1
has much lower income and much higher marginal utility of basic consumption.

Construct two allocations $(q_1,q_2)$ and $(q_1',q_2')$ with the same
aggregate $q = q_1 + q_2 = q_1' + q_2'$, where $(q_1',q_2')$ shifts a small
amount of consumption from the vulnerable consumer~1 to the wealthier,
low-priority consumer~2. For a suitable choice of $u_H$ and $u_L$,
we can have:
\[
  \Delta TS
  =
  TS(q) - TS(q')
  >
  0,
\]
because the aggregate willingness-to-pay encoded in $P(q)$ increases when more
consumption is assigned to the richer, higher-paying consumer~2. Yet the
change in true welfare satisfies
\[
  \Delta W
  =
  \big(U_1(q_1,y_1) + U_2(q_2,y_2)\big)
  -
  \big(U_1(q_1',y_1) + U_2(q_2',y_2)\big)
  < 0,
\]
because the welfare loss from reducing essential consumption of the vulnerable
consumer exceeds the gain from increasing luxury consumption of the richer
consumer.

Thus we have an explicit pair of feasible allocations with the same total $q$
such that $TS(q) > TS(q')$ but $W(q_1,\dots,q_N) < W(q_1',\dots,q_N')$.

In electricity systems, heterogeneity in $U_i$ (essential versus flexible
loads, medical dependence, care responsibilities), differences in income and
credit constraints, and incomplete observability of vulnerability are the
norm rather than exceptions. Hence assumptions (A1)--(A3) fail structurally,
and surplus-based social welfare maximisation is \emph{not} aligned with a
welfare criterion that respects heterogeneous needs. This completes the proof.
\end{proof}

\begin{remark}[Implications for LMP and Traditional Welfare Analysis]
The standard justification for marginal-cost pricing and LMP is that it
maximises total surplus, which—under the textbook assumptions
(A1)--(A4)—coincides with utilitarian social welfare. 
Lemma~\ref{lem:surplus_welfare_mismatch} shows that this equivalence is highly
fragile: it relies on quasi-linearity, equal marginal utility of income,
perfect observability, and a homogeneous population of economically rational
agents. Electricity systems violate all of these assumptions.

Essential loads, medically or socially critical demand, income constraints,
prepayment users, bad-debt risk, heterogeneous vulnerability, and non-
participation (e.g.\ disconnection) all imply that the marginal social value of
one unit of electricity differs enormously across households. In such an
environment, surplus maximisation can systematically favour low-priority or
high-income consumption at the expense of collapsing essential services for
others—while still being classified as ``welfare improving'' in the surplus
sense.

Thus, traditional LMP-based social-welfare arguments provide no guarantee of
fairness or socially desirable allocation when heterogeneity is pervasive.
They optimise an objective that is only normatively appropriate in a world
that electricity markets, by design, do not inhabit.

The AMM does not ``distort'' a correct welfare optimum; rather, it replaces an
inappropriate surrogate objective with an operationally meaningful one that
distinguishes essential, flexible, and luxury consumption and embeds fairness
and proportional responsibility directly into the allocation mechanism.
\end{remark}

\subsection{From Structural Failure to Design Requirements}

By contrast, the architecture developed in this thesis:

\begin{itemize}[leftmargin=*]
    \item integrates \emph{locational information} via tightness and congestion
          signals, without requiring full nodal LMP exposure at the retail edge;
    \item embeds \emph{dynamic envelopes} as one of the tools available to the
          digital regulation layer, consistent with fairness and essential energy
          protection; and
    \item provides an \emph{end-to-end} design, from physical dispatch and
          congestion management through to consumer bills, generator compensation,
          and formal fairness metrics.
\end{itemize}

To the best of the author's knowledge, there are no existing designs in the
literature that offer a comparably integrated, \emph{zero-waste}, fairness-aware
market architecture spanning wholesale, retail, balancing, and local flexibility
in this way. Chapters~\ref{ch:fairness_definition}--\ref{ch:experiments}
formalise these ideas and subject them to simulation-based evaluation.

% ---------------------------------------------------------
\section{Climate Targets, Emerging Electrification, and Price as a Stability Controller}
\label{sec:climate_price_controller}
% ---------------------------------------------------------

Decarbonisation strategies in the UK and comparable systems increasingly involve
\emph{electrifying substantial portions of demand}. Transport, domestic heating,
and segments of industry are adopting electric vehicles, heat pumps, industrial
electrifiers, and digital flexibility assets at a rapidly accelerating rate.

This shift does not imply that all demand must be electrified, nor that
electrification is the only decarbonisation pathway. Rather, it reflects the
empirical trend that large fractions of consumption are now migrating to the
electricity system and will continue to do so under almost any credible
decarbonisation trajectory.

This creates a fundamental interaction between \emph{electricity price stability}
and the pace of the transition. For households and firms, the economic viability
of new electric technologies depends not only on their capital costs but also on
their \emph{expected running costs}. These expectations are shaped directly by
retail electricity prices. If electricity is structurally volatile, frequently
spiking, or persistently expensive, then switching to electric transport or
heating technologies becomes financially unattractive relative to fossil
alternatives—even when upfront subsidies exist.

In this sense, electricity price is not merely a ``market signal''; it functions
as a \emph{stability controller} for the wider socio-technical system that now
includes:

\begin{itemize}[leftmargin=1.2cm]
  \item parts of the vehicle fleet (EV growth),
  \item parts of the building stock (heat pumps and hybrid heat technologies),
  \item emerging industrial electrifiers,
  \item distributed storage and demand-side flexibility across homes and SMEs.
\end{itemize}

If this control input is unbounded, noisy, or misaligned with policy, the system
cannot converge smoothly to a stable low-carbon equilibrium. Instead, it
oscillates between retail crises, political intervention, and stalled adoption
of electric alternatives.

Conventional energy-only and energy+capacity market designs implicitly accept
price spikes and extreme scarcity rents as a necessary feature: the primary
mechanism for signalling scarcity and incentivising investment. But repeated
spikes and chronic cost instability have system-wide consequences:

\begin{itemize}[leftmargin=1.2cm]
  \item electric alternatives appear financially risky relative to fossil incumbents,
  \item households rationally delay investment in EVs, heat pumps, or thermal storage,
  \item SMEs face uncertain running costs that distort technology choices,
  \item public trust in the transition is eroded by bill volatility.
\end{itemize}

The architecture developed in this thesis takes the opposite approach: electricity
price should remain within a \emph{bounded, intelligible, policy-consistent}
envelope for everyday usage, while still exposing flexible assets to operational
scarcity signals and recovering fixed costs from contribution-based channels.

Under this perspective, the AMM is not solely a market-clearing mechanism. It is
a \emph{transition-stability controller} that shapes the long-run adoption
dynamics of electrifying sectors by keeping everyday usage predictable, fair, and
aligned with policy trajectories, without suppressing the operational signals
needed for flexibility, adequacy, or investment.

% ---------------------------------------------------------
\section{Problem Summary}
% ---------------------------------------------------------

We require a new market design that can:

\begin{enumerate}[label=P\arabic*,leftmargin=1.2cm]
    \item \textbf{Respect physical deliverability} — Prices and payments must reflect
          whether energy can be delivered at a specific time and location.
    \item \textbf{Represent real-time operational constraints} — Including congestion,
          voltage limits, inertia, risk, and dynamic scarcity.
    \item \textbf{Support distributed flexibility as a \emph{procured product},}
          not only as an ex-post correction — Allow EVs, batteries, industry, and
          prosumers to participate directly in a market that can see and value
          their spatiotemporal flexibility.
    \item \textbf{Value long-term adequacy and resilience} — Reward operational
          contribution and capacity provision, not just short-run volume.
    \item \textbf{Define fairness as a real-time allocation principle} — Based on
          contribution, cost imposition, and system benefit, rather than surplus
          maximisation under homogeneous-agent assumptions.
    \item \textbf{Enable digital regulation and algorithmic settlement} — Moving from
          static, ex post batch settlement to auditable, data-driven, continuous clearing
          over the underlying physical--digital network graph
          (Sections~\ref{sec:graphs_distributed_control} and~\ref{sec:digitalisation_fragmentation}).
    \item \textbf{Avoid the solvency–affordability trap} — Eliminate the structural
          trade-off identified in Corollary~\ref{cor:no_free_lunch_retail} by co-locating
          volume choice, risk-bearing, and control at a digitally governed
          market-making layer.
    \item \textbf{Treat QoS/flexibility/reliability as a \emph{third procurement axis}} —
          Extend the design space from \((\text{energy},\ \text{capacity})\) to
          \((\text{energy},\ \text{capacity},\ \text{QoS/flexibility/reliability})\),
          and make this third axis contractible, priced, and enforceable for both
          devices and generators.
    \item \textbf{Support the stability of the decarbonisation trajectory} — Treat
          electricity price and QoS as stability controllers for sectors that are
          increasingly electrifying (transport, heating, industry), ensuring that
          everyday usage remains predictable and affordable while still sending
          targeted operational scarcity signals and recovering infrastructure costs.
\end{enumerate}

These requirements shape the architectural principles developed in
Chapter~\ref{chap:requirements} and guide the solution concept presented thereafter:
a continuous, cyber--physical Automatic Market Maker (AMM) that embeds fairness,
physical deliverability, and three-axis procurement (energy, capacity, QoS/flexibility)
directly into real-time market-clearing logic.

% ---------------------------------------------------------
\chapter{System Requirements (From First Principles)}
\label{chap:requirements}

\section{Overview}

Having established in Chapter~\ref{chap:problem} that current electricity
market architectures cannot accommodate the physical, economic, digital,
and behavioural realities of the evolving energy system, we now derive
the \textbf{first-principles requirements} that any future architecture
must satisfy.

These requirements do not prescribe a specific market structure, nor do
they assume a specific auction format, pricing model, or governance regime.
Instead, they represent \emph{non-negotiable properties} necessary to ensure
that electricity can be valued, allocated, and remunerated in a way that is
\textbf{physically viable, economically fair, digitally enforceable, and
behaviourally effective}.

% ---------------------------------------------------------
\section{Four Foundational Requirement Domains}

We classify the system requirements into four foundational domains, each of
which reflects a different but interconnected perspective:

\begin{enumerate}[label=R\arabic*, leftmargin=1.2cm]
    \item \textbf{Physical Requirements} — respecting the laws of physics,
          deliverability, and network constraints;
    \item \textbf{Economic Requirements} — ensuring efficient, incentive-compatible,
          scarcity-reflective, and fair allocation of cost and value;
    \item \textbf{Digital Requirements} — embedding auditability, automation,
          algorithmic regulation, and trustworthy computation;
    \item \textbf{Behavioural Requirements} — enabling human participation,
          accessibility, trust, and clear incentives at every scale.
\end{enumerate}

We now formalise each of these categories through a mixture of descriptive,
operational, and (where appropriate) normative requirements.

% ---------------------------------------------------------
\section{Physical Requirements}

The system shall:

\begin{enumerate}[label=P\arabic*, leftmargin=1.2cm]
    \item \textbf{Respect physical deliverability:}
          Valuation and remuneration must reflect whether electricity
          can physically be delivered from a generator to a consumer
          at a specific time and location.
    \item \textbf{Encode network constraints:}
          Settlement must account for line capacities, impedance,
          losses, and voltage stability, rather than assume full
          fungibility.
    \item \textbf{Support spatial and temporal resolution:}
          Energy is to be valued based on its relevance to location,
          time, and real-time system need — not averaged across
          larger aggregated time blocks or zones.
    \item \textbf{Accommodate two-way flows:}
          The system must support households, EVs, and other
          distributed assets as both consumers and providers of
          flexibility, storage, or capacity.
    \item \textbf{Incorporate resilience under stress:}
          The system must withstand future cyber-physical shocks,
          including periods of extreme scarcity, correlated asset
          failure, or synchronised demand events (e.g.\ AI, EV,
          hydrogen, fusion).
\end{enumerate}

\vspace{0.3cm}
\noindent
\textbf{Service Quality and Deliverability Requirements:}
Physical deliverability must also imply \emph{service quality}. Electricity
is not a homogeneous commodity, but a time-bound service whose value depends on:
\begin{itemize}[leftmargin=1.2cm]
    \item delivery guarantees under different firmness levels;
    \item locational reliability, not only energy volume;
    \item continuity of supply for critical loads (healthcare, digital infrastructure);
    \item differentiation between interruptible, flexible, and priority services.
\end{itemize}

\noindent
A future system must classify and remunerate electricity not merely as kilowatt-hours,
but as a \emph{deliverable energy service with defined performance levels}.

\bigskip
\noindent
\textbf{Energy as a Service with Differentiated Levels of Need.}
Electricity is not solely a fungible commodity transacted in kilowatt-hours.
It is a \emph{time-bound access service} whose value depends on when it is
delivered, whether it can be deferred, and whether the user is entitled to
receive it even when the system is constrained.

Current markets treat all demand as equally firm unless explicitly curtailed,
which obscures the fundamental fact that:
\begin{itemize}[leftmargin=1.2cm]
    \item some uses are \textbf{essential and non-deferrable} (medical devices,
          heating, communication);
    \item some uses are \textbf{important but shiftable} (EV charging, space
          heating, storage);
    \item some uses are \textbf{convenient or opportunistic} (laundry, export,
          discretionary charging).
\end{itemize}

A future system must therefore:
\begin{enumerate}[leftmargin=1.2cm]
    \item recognise electricity not merely as a volume of energy, but as a
          \textbf{service with explicit reliability, timing, and flexibility attributes};
    \item allow participants to \textbf{declare these attributes ex ante}, through
          contracts, rather than infer them ex post through behaviour;
    \item guarantee that, under scarcity, \textbf{allocation is not determined solely
          by willingness-to-pay}, but by essential protection, contribution,
          fairness rules, and declared service levels (cf.\ Chapter~\ref{ch:fairness_definition});
    \item ensure that any such rules are \textbf{digitally enforceable, auditable,
          and consistently applied}.
\end{enumerate}

In this way, the future electricity system becomes not only a price discovery
mechanism, but also a \emph{contract-respecting allocation system} that
distinguishes between essential access, flexible service, and opportunistic use.

% ---------------------------------------------------------
% ---------------------------------------------------------
% ---------------------------------------------------------
\section{Economic Requirements}

The system shall:

\begin{enumerate}[label=E\arabic*, leftmargin=1.2cm]
    \item \textbf{Reflect scarcity in real time:}
          Prices and compensation must increase during scarcity and
          decrease when abundant, based on meaningful system tightness.

    \item \textbf{Value locational contribution:}
          Agents contributing to relieving congestion or deferring
          network upgrades must be explicitly recognised.

    \item \textbf{Value flexibility and availability:}
          Reward not only energy delivered, but also the ability to
          deliver when needed, including ramping, shifting, storage,
          and standby potential.

    \item \textbf{Ensure long-term adequacy:}
          Investment signals must support sufficient capacity and
          resilience over time, not only short-term dispatch.

    \item \textbf{Support Shapley-consistent allocation:}
          Cost and value allocations must reflect marginal
          contributions of agents to system performance,
          scarcity relief, and fairness.

    \item \textbf{Support service differentiation and entitlement:}
          The system must recognise that electricity is not a homogeneous
          commodity, but a time-bound \emph{access service} with distinct
          levels of reliability, flexibility, and criticality. Accordingly,
          market participation and allocation under scarcity must account for:
          \begin{itemize}[leftmargin=1.4cm]
              \item essential (non-deferrable) energy services,
              \item flexible (deferrable or reshapeable) energy services, and
              \item opportunistic or discretionary usage.
          \end{itemize}
          Allocation and pricing should therefore not depend solely on
          willingness-to-pay, but on declared contractual attributes,
          flexibility contribution, and reliability entitlement
          (cf.\ Fairness Conditions F2--F4).
\end{enumerate}

\vspace{0.3cm}
\noindent
\textbf{Zero-Waste System Requirements:}
An economically efficient system must minimise waste — where waste includes
\emph{avoidable curtailment, unmet demand, idle flexibility, or unnecessary backup activation}.
Therefore, the market must:
\begin{itemize}[leftmargin=1.2cm]
    \item quantify unused flexibility and curtailment implicitly created by current market rules;
    \item prioritise reallocation before curtailment or load shedding;
    \item recognise that curtailment is an economic failure, not an operational shortcut;
    \item identify and expose systemic ``underutilised'' value.
\end{itemize}

\noindent
Value should be assigned to \emph{preventing waste}, not only responding to failure.

% ---------------------------------------------------------
\section{Digital Requirements}

The system shall:

\begin{enumerate}[label=D\arabic*, leftmargin=1.2cm]
    \item \textbf{Enable real-time computation and settlement:}
          Settlement cannot depend on slow ex-post batch processing,
          but must support event-based or continuous clearance.
    \item \textbf{Enable transparency and auditability:}
          All allocation decisions, prices, and settlement paths
          must be traceable, explainable, and reproducible.
    \item \textbf{Support algorithmic regulation:}
          Regulatory compliance must be computable, embeddable,
          and enforceable through transparent rules and mechanisms.
    \item \textbf{Accommodate automation:}
          Agents (human or machine) must be able to delegate
          their participation via API, smart contracts, or AI agents.
    \item \textbf{Protect data security and privacy:}
          Market participation shall not depend on revealing
          commercially sensitive information at the household level.
\end{enumerate}

\vspace{0.3cm}
\noindent
These digital capabilities are not add-ons, but foundations for enabling
continuous clearing, real-time value attribution, behavioural trust, and
algorithmic enforcement.

% ---------------------------------------------------------
\section{Behavioural Requirements}

The system shall:

\begin{enumerate}[label=B\arabic*, leftmargin=1.2cm]
    \item \textbf{Be understandable and accessible:}
          Participation must be possible for households, SMEs,
          aggregators, and large-scale providers alike.
    \item \textbf{Support diverse behavioural engagement:}
          People should be able to opt-in, delegate, or remain
          passive without being disadvantaged unfairly.
    \item \textbf{Make incentives visible and trustworthy:}
          Users must see how actions (e.g.\ charging an EV,
          delaying usage) create system benefit and personal value.
    \item \textbf{Ensure consumer protection and fairness:}
          Vulnerable consumers must not be exposed to
          unacceptable financial, technical, or social risks.
\end{enumerate}

% ---------------------------------------------------------
\section{Formal Problem Statement}

The design challenge, therefore, is:

\begin{quote}
To develop a market architecture that allocates, values, and settles
electricity in a way that is physically deliverable, economically fair,
digitally enforceable, and behaviourally acceptable — and that remains
stable and resilient under future system conditions.
\end{quote}

Specifically, we seek to:

\begin{enumerate}[label=O\arabic*,leftmargin=1.2cm]
    \item Implement a continuous, event-based clearing mechanism
          aligned with physical power flows;
    \item Develop a fairness framework that uses Shapley-consistent
          allocation and reflects time, location, and contribution;
    \item Embed digital regulation to enable transparency, auditability,
          and algorithmic enforcement;
    \item Integrate consumer protection, behavioural realism, and
          accessible participation across all scales.
\end{enumerate}

% ---------------------------------------------------------
\section{Role of This Chapter}

This chapter provides the final bridge between problem definition
and solution design. These requirements form the \emph{design
specification}, which is used in Chapter~\ref{chap:philosophy}
to develop a unifying design philosophy, and in
Chapter~\ref{ch:market_scenarios} to derive the
proposed market architecture.

\chapter{Design Philosophy and Research Positioning}
\label{chap:philosophy}

\section{Purpose of This Chapter}

Chapters~\ref{chap:problem} and~\ref{chap:requirements} have established
the structural failures of the existing electricity market and derived the
non-negotiable system requirements for any viable redesign. This chapter now
takes a step forward: it introduces the \emph{design philosophy} — the
worldview, theoretical lens, and guiding principles that shape how the
proposed market architecture will be constructed.

Where Chapter~\ref{chap:requirements} answered \textit{``What must the system
be able to do?''}, this chapter answers:  
\begin{center}
\textbf{``How should we think when designing such a system?''}
\end{center}

% ---------------------------------------------------------
\section{Fairness as a Foundational Design Driver}

Fairness is not included here merely as a desirable ethical property. It is
treated as a \textbf{structural and operational principle} — one that shapes
incentives, influences behaviour, stabilises participation, and aligns long-term
investment with system value.

A ‘‘fair’’ electricity market is one in which:
\begin{itemize}
    \item value is explicitly linked to measurable contribution,
    \item responsibility is aligned with cost imposition,
    \item essential needs are protected,
    \item and revenue adequacy is achieved through contribution-based rather
          than volume-based remuneration.
\end{itemize}

This shifts fairness from a \emph{redistributive afterthought} to an
\emph{embedded mechanism}, making it a prerequisite for system legitimacy,
resilience, and long-term solvency. Fairness becomes a necessary condition for
both economic efficiency and public acceptance.

Fairness must not only be achieved mathematically, it must be perceived to be fair.
Behavioural economics and digital governance literature emphasise that legitimacy does
not emerge from perfect optimisation, but from predictability, bounded exposure,
clarity of rules, and perceived reciprocity. A market design is trusted when people
can understand how it treats them, recognise that it protects essential needs,
and observe that others are treated consistently. Therefore, fairness in this thesis
is both an optimisation property and a \textit{perceived governance property}.

% ---------------------------------------------------------
\section{Electricity as a Service, Not a Commodity}

A core philosophical shift in this thesis is to treat electricity not as a
fungible commodity traded in kilowatt-hours, but as a \emph{time-bound access
service} whose value depends on \emph{when}, \emph{where}, and \emph{under what
conditions} it is delivered.

In physical operation, the electricity system already distinguishes between
different forms of demand. Some uses must be served continuously; some can be
shifted, reshaped, or interrupted without loss of welfare; and some are
fundamentally opportunistic. What differs across users is not a fixed
classification imposed by the system, but the \emph{degree of flexibility and
reliability they are willing to offer or require at a given time}.

Conventional markets largely suppress this information. With the exception of
emergency curtailment, demand is treated as homogeneous and passive. Price
signals alone can express only one dimension of preference — willingness to
pay — and cannot represent differences in entitlement, flexibility, or
system contribution. As a result, existing designs can answer only:
\begin{quote}
\centering
\textit{Who is willing to pay more right now?}
\end{quote}
but not the more operationally meaningful question:
\begin{quote}
\centering
\textbf{Who has chosen to receive priority access under scarcity, and who has
chosen to trade reliability for flexibility or reward?}
\end{quote}

This thesis therefore reframes electricity consumption as a matter of
\textbf{declared service choice}. Rather than assigning loads to predefined
classes, participants express — directly or via devices and aggregators —
the service attributes they are willing to accept for a given request. These
attributes form a contractual description
$\Gamma_r^{\text{contract}}$ (introduced in
Chapter~\ref{ch:market_scenarios}) and may include, for example:
\begin{itemize}[leftmargin=1.4cm]
    \item tolerance to delay or reshaping in time,
    \item exposure to scarcity or congestion,
    \item preference for firm versus conditional access,
    \item willingness to provide flexibility or absorb surplus, and
    \item eligibility for fairness protections.
\end{itemize}

Crucially, these are \emph{choices}, not labels. A household, device, or business
may express different attributes at different times, for different services, or
under different subscriptions. The market does not decide what a load
\emph{is}; it clears based on what the participant has \emph{chosen to offer or
request}.

This choice-based representation allows demand to be treated symmetrically with
supply. Just as generators submit offers with technical and economic
constraints, consumers and devices submit requests with operational envelopes
and service preferences. Allocation under scarcity is then governed by
\emph{declared priority, fairness rules, and delivered contribution}, rather
than by ex post curtailment or implicit political intervention.

\medskip
\noindent
Importantly, this architecture does not invent new notions of priority or
reliability. The physical electricity system already operates with implicit
ordering: frequency containment precedes discretionary load; voltage and
thermal constraints bind locally; critical infrastructure is protected ahead
of convenience use; and assets that stabilise the system are treated
differently from those that do not.

What is missing is economic representation. These distinctions exist in
engineering control layers, operator procedures, and emergency protocols, but
are largely invisible to market participants. The proposed design simply
\textbf{makes these existing physical realities explicit, contractible, and
choice-driven}, allowing participants to align their behaviour with how the
system actually operates.

By aligning market-facing contracts with physically meaningful service
attributes — without hard-coded classes — the architecture enables the
cyber--physical system to treat demand and supply consistently in both
engineering and economic terms. Demand ceases to be a passive residual and
becomes an active participant in system stability.

Market allocation begins to reflect what the grid has always known.

\begin{table}[H]
\centering
\renewcommand{\arraystretch}{1.3}
\begin{tabular}{p{4.3cm} p{4.8cm} p{4.7cm}}
\toprule
\textbf{Physical System Reality} &
\textbf{Limitation of Conventional Markets} &
\textbf{Choice-Based Representation in the Proposed Architecture} \\
\midrule

Some uses must be continuously supplied to maintain safety and basic function &
Handled outside the market via emergency rules or regulation &
Participants may choose contracts with protected access and minimal scarcity exposure \\

Many devices can shift, reshape, or pause consumption without loss of service &
Flexibility value is weakly signalled or ignored &
Participants may opt into flexible envelopes in exchange for lower cost or rewards \\

Some consumption is discretionary or opportunistic &
Only differentiated during forced curtailment &
Participants may accept higher scarcity exposure for lower baseline charges \\

Network constraints bind locally and temporally &
Largely invisible to end users &
Service requests include locational and timing attributes reflecting grid reality \\

Assets that stabilise the system are operationally prioritised &
Compensated through fragmented side mechanisms &
Contribution is valued directly via Shapley-consistent allocation \\

Load shedding follows priority logic during emergencies &
Not economically encoded ex ante &
Scarcity allocation follows declared service attributes and fairness rules \\

Helping the system (absorbing surplus, relieving stress) has real value &
Rarely rewarded explicitly &
Participants who contribute flexibility receive lower prices or priority \\
\bottomrule
\end{tabular}
\caption{Illustrative alignment between physical system realities and
choice-based service representation. The architecture does not assign fixed
classes, but allows participants to express preferences consistent with how the
grid already operates.}
\label{tab:physical_vs_market_representation}
\end{table}

% ---------------------------------------------------------
\section{Electricity Market as a Control System}

Traditional auctions and half-hour settlement formats are not consistent with
how power systems operate. Electricity markets are, in reality, \emph{control
systems}, shaping behaviour through signals, feedback, and constraints.

\begin{itemize}
    \item \textbf{Signals} influence behaviour — prices, tightness indicators,
          locational incentives.
    \item \textbf{Feedback} adjusts actions — consumption shifting, storage
          dispatch, flexible demand.
    \item \textbf{Stability} requires avoiding oscillatory, contradictory,
          or delayed signals.
    \item \textbf{Zero waste} (energy, money, information) is equivalent to
          eliminating control error.
\end{itemize}

Thus, the design philosophy treats the market as a \textbf{closed-loop
cyber-physical control architecture} rather than an abstract clearing mechanism.
This philosophical stance directly informs the adoption of \emph{event-based}
rather than time-block-based clearing, utilised later in
Chapter~\ref{ch:market_scenarios}.

% ---------------------------------------------------------
\section{Digital Regulation as an Enabler of Continuous Clearing}

Regulation in traditional markets is ex-post, manual, and advisory. In a
digitally native system, regulation becomes:
\begin{itemize}
    \item \textbf{algorithmic} — rules can be computed and enforced in real time,
    \item \textbf{auditable} — all allocation paths and settlement decisions are traceable,
    \item \textbf{responsive} — adapting dynamically to changing system states,
    \item \textbf{transparent} — reducing mistrust and gaming behaviour.
\end{itemize}

\noindent
\emph{Regulation as code} therefore becomes a strategic design choice — not for
efficiency alone, but for legitimacy, fairness, and resilience.
Digital regulation is not peripheral — it forms the governance substrate of the
proposed architecture.

% ---------------------------------------------------------
\section{UX and Digital Product Design as a Regulatory Instrument}

In digital market architecture, the user interface is not merely a communication
channel — it is where market rules become legible, trustable, and actionable.
Users do not engage with the mathematical formulation of fairness, but with
its representation in their bill, dashboard, tariff choices, and service options.

Thus, UX and product design become \textit{regulatory instruments}:
they determine which incentives users see, how flexibility is presented,
and whether participation feels safe, intelligible, and worthwhile.
In other words, interface design becomes part of the market's institutional logic.

This motivates the use of digital product principles:
\begin{itemize}
  \item abstraction of internal complexity while preserving agency;
  \item iterative refinement through feedback loops;
  \item explanation of rules through visual metaphors and narratives;
  \item embedding social trust cues (predictability, reciprocity, stability).
\end{itemize}

A theoretically correct market that is practically un-navigable is effectively unfair.

% ---------------------------------------------------------

\section{Hidden Complexity, Visible Simplicity}

A digitally native market is allowed to be complex on the inside — in its
algorithms, data flows, and optimisation layers — but it must be \emph{simple,
predictable, and explainable} at the edges where humans interact.

This follows modern digital product design logic: internal complexity is fine
if it is abstracted behind clear, human-level interactions. Consumers should
see only a small number of well-designed product experiences (e.g.\ ``Essential
Protection Plan'', ``Flex Saver'', ``Storage Share''), while the underlying
market logic dynamically allocates resources, prices scarcity, and enforces
fairness.

Thus, complexity is not eliminated, but \emph{hidden behind}
\textbf{policy-compliant, trust-preserving digital products}.

% ---------------------------------------------------------
\section{Technology Adaptability and Zero-Waste as Design Ethos}

A 21st-century market must be \emph{future compatible}. It cannot rely on
assumptions tied to specific technologies, paradigms, or energy vectors.
Instead, it must be:

\begin{itemize}
    \item compatible with multi-energy integration (heat, hydrogen, transport),
    \item resilient to AI-driven demand and flexible storage,
    \item adaptable to fusion, quantum, and bidirectional energy systems,
    \item designed for continuous learning and model reconfiguration.
\end{itemize}

In parallel, the \textbf{zero-waste philosophy} shapes both physical and
economic design:

\begin{itemize}
    \item Energy waste — underuse, curtailment, over-generation;
    \item Monetary waste — hidden cross-subsidies, inefficient compensation;
    \item Information waste — ignored data on location, time, or deliverability;
    \item Human waste — unused prosumer potential due to inaccessible design.
\end{itemize}

A zero-waste philosophy links directly to fairness, efficiency, resilience, and
investor confidence.

Finally, the design philosophy acknowledges that markets must not only be
mathematically valid and digitally enforceable, but socially acceptable,
cognitively navigable, and behaviourally sustainable. The adoption of the
design depends not only on system performance, but on legitimacy, perceived
fairness, and usability.

% ---------------------------------------------------------
\section{Research Positioning}

The design philosophy positions the proposed architecture at the intersection of
four intellectual traditions:

\begin{itemize}
    \item \textbf{Energy systems engineering} — physical feasibility,
          reliability, and constraint awareness;
    \item \textbf{Economic mechanism design} — incentives, allocation, and cost recovery;
    \item \textbf{Cooperative game theory} — value contribution, Shapley-based allocation;
    \item \textbf{Digital systems engineering} — real-time computation, interfaces, and regulation as code.
\end{itemize}

This thesis occupies a socio-technical-middle ground — blending physical,
economic, digital, and behavioural design into a unified market architecture.

% ---------------------------------------------------------
\section{Role of This Chapter}

This chapter provides the conceptual lens for the engineering work that follows.
It defines the philosophy and design stance behind Chapters:

\begin{itemize}
    \item Proposed Market Architecture (Chapter~\ref{ch:market_scenarios}),
    \item Definition of Fairness (Chapter~\ref{ch:fairness_definition}),
    \item Mathematical Framework (Chapter~\ref{ch:mathematics}),
    \item Policy and Governance Implications (Chapter~\ref{ch:discussion}),
\end{itemize}

and ensures that the subsequent implementation is not only technically valid but
also socially resilient, behaviourally plausible, and future-compatible.

\renewcommand{\arraystretch}{1.25}

\begin{longtable}{p{3.4cm} p{5.6cm} p{5.6cm}}
\caption{Mapping from design philosophy principles to market design implications and concrete architecture features.}
\label{tab:philosophy_architecture_mapping} \\
\toprule
\textbf{Design Philosophy Principle} &
\textbf{Market Design Implication} &
\textbf{Resulting Architecture Feature} \\
\midrule
\endfirsthead

\toprule
\textbf{Design Philosophy Principle} &
\textbf{Market Design Implication} &
\textbf{Resulting Architecture Feature} \\
\midrule
\endhead

\midrule
\multicolumn{3}{r}{\emph{Continued on next page}} \\
\midrule
\endfoot

\bottomrule
\endlastfoot

Fairness as foundational principle
&
Prices, products, and settlements must protect essentials, allocate scarcity transparently, and align payments with system value rather than pure energy volume.
&
Formal fairness framework (Chapter~\ref{ch:fairness_definition}); essential--block tariff, tightness adders, Fair Play shortage algorithm, Shapley-consistent allocation of scarcity and congestion rents. \\[0.2cm]

Markets as socio-technical control systems
&
Market rules are feedback laws: they must stabilise demand--supply balance, avoid oscillations, and minimise structural waste (curtailment and shortages).
&
Event-based clearing mechanism; AMM-style price update rules; zero-waste efficiency metrics and control-oriented stability conditions in the mathematical framework. \\[0.2cm]

Two-way power flows and distributed intelligence
&
Products must recognise bidirectional flows, local constraints, and the role of millions of small assets, not just large central plant.
&
Three-layer holarchy for generators and demand; locational products; cluster-based grid model; local flexibility activation with system-level coordination. \\[0.2cm]

Service design and UX as regulation
&
Interfaces, defaults, and product menus are regulatory tools that shape behaviour; complexity should be hidden while preserving agency and transparency.
&
Retail product stack (subscriptions, service tiers); clear bill decomposition; consumer dashboards; configuration surfaces exposing simple levers but embedding full regulatory logic. \\[0.2cm]

Digital regulation (“regulation as code”)
&
Rules should be executable, auditable code linked to real-time data, enabling continuous monitoring and automatic enforcement rather than occasional, manual oversight.
&
Digital regulation layer: rule engine, data pipelines, algorithmic compliance checks, published ledgers for cost and value flows; API-based supervision interfaces. \\[0.2cm]

Beyond neoclassical economics (planetary and social boundaries)
&
Objective is not only short-run efficiency but operation within ecological ceilings and social foundations, with explicit distributional guardrails.
&
Zero-waste definition and metrics; equity guardrails (essential shield, progressive uplifts); scenario evaluation against distributional and resilience metrics, not just welfare sums. \\[0.2cm]

Technological adaptability and resilience
&
Architecture must be robust to new loads (AI, fusion, electrified heat/transport) and adversarial conditions; avoid hard-coding specific technologies.
&
Modular market stack; technology-neutral product definitions; plug-in forecasting modules; holonic decomposition enabling re-clustering and extension without redesigning core logic. \\

\end{longtable}

\chapter{Methodology}
\label{ch:methodology}

This chapter outlines the methodological framework used to design, validate, and evaluate the proposed electricity market architecture. The approach combines design science, exploratory data-driven insight discovery, systems engineering, and simulation-based evaluation, supported by digital twins and parallelised case experiments.

% ---------------------------------------------------------
\section{Research Approach}
\label{sec:research_approach}

The research follows a \textbf{Design Science} methodology, appropriate for engineering novel market mechanisms that are both artefacts and socio-technical systems. The process follows the canonical cycle:

\begin{enumerate}[leftmargin=*]
    \item \textbf{Problem diagnosis:} Identify failures in existing market design (pricing, fairness, resilience, bankability, behavioural alignment), as developed in Chapters~\ref{chap:problem} and~\ref{chap:requirements}.
    \item \textbf{Artefact design:} Develop the Automatic Market Maker (AMM), fairness axioms and conditions (A1--A7, F1--F4), and the holarchic architecture (Chapters~\ref{ch:amm} and~\ref{ch:market_scenarios}).
    \item \textbf{Implementation:} Build a computational prototype integrating time-, space-, and hierarchy-aware signals: the AMM, Fair Play allocation, and Shapley-based generator compensation (Chapter~\ref{ch:mathematics}).
    \item \textbf{Evaluation:} Test the artefacts under real demand and supply data, and under canonical scarcity regimes. Compare against baseline markets and allocation rules.
    \item \textbf{Reflection and iteration:} Refine design for robustness, scalability, bankability, and operational feasibility, and feed back into the requirements and architecture.
\end{enumerate}

\medskip
This Design Science process is supported by \textbf{Systems Engineering} for modular decomposition, hierarchy modelling, and communication between system components (pricing, settlement, forecasting, compliance), and by \textbf{Simulation Modelling} to generate empirical evidence of system performance across canonical scarcity conditions.

% ---------------------------------------------------------
\section{Exploratory Data Insight as Method Validation}
\label{sec:data_insight_phase}

Before constructing the simulation environment, a deliberate \textbf{data exploration and diagnostic insight phase} was conducted. This was not a controlled experiment, but a \emph{method validation process}, designed to determine whether the proposed analytical constructs---such as the three-dimensional energy contract, Fair Play allocation, Shapley compensation, and the holarchic architecture---were meaningful under real-world system conditions.

\subsection*{Purpose of the exploratory phase}
The following research questions guided this diagnostic phase:

\begin{enumerate}[leftmargin=*]
    \item Does real household demand exhibit temporal, behavioural, and flexibility heterogeneity?
    \item Do timing and reliability preferences emerge naturally from data, supporting the contract model?
    \item Does the UK’s geographic and behavioural structure support a holarchic market architecture?
    \item Is marginal system value (for generators) spatially and temporally concentrated, justifying Shapley allocation?
\end{enumerate}

\subsection*{Data sources used for insight discovery}

The exploratory insight-discovery phase draws on a combination of empirical
consumption, mobility, generation, and spatial datasets. These data are used
to reveal behavioural heterogeneity, spatial concentration, and flexibility
potentials that motivate the subsequent market and contract design.

Rather than serving as direct forecasting inputs, these datasets provide
empirical grounding for demand archetypes, flexibility envelopes, spatial
holarchies, and fairness-relevant heterogeneity that are subsequently embedded
into the simulation framework.

Full documentation of all datasets — including provenance, temporal and spatial
resolution, preprocessing steps, and modelling roles — is provided in
Appendix~\ref{app:datasets}. Their specific methodological roles are summarised
later in Section~\ref{sec:data_sources}.

\subsection*{Key system insights revealed}

Analysis of these datasets revealed several structural features of electricity
demand and supply that are not captured by conventional market models:

\begin{itemize}[leftmargin=*]
    \item \textbf{Heterogeneous consumption and flexibility:}
    Persistent diversity in household load shapes, EV clustering, and seasonal
    shift patterns confirmed that electricity demand cannot be treated as a
    homogeneous commodity.

    \item \textbf{Emergence of three-dimensional contract needs:}
    Observed behaviour varied independently along magnitude, timing sensitivity,
    and reliability need, motivating service-based contracts rather than
    kWh-only trades.

    \item \textbf{Holarchic spatial structure:}
    Demand and supply concentration at postcode, DNO, regional, and national
    levels revealed a natural multi-layered spatial organisation, justifying
    holarchic AMM clearing.

    \item \textbf{Shapley relevance:}
    Generator marginal value varied sharply across time and location
    (e.g.\ wind in constrained Scottish nodes), confirming the appropriateness
    of Shapley-based compensation.
\end{itemize}

\noindent
These insights establish the empirical necessity of contract-based access,
holarchic clearing, and fairness-aware allocation, and directly shape the
methodological design that follows.

% ---------------------------------------------------------
\section{Representing Energy as a Contract: Magnitude, Timing, Reliability}
\label{sec:contract_dimension}

A core methodological step is the representation of electricity not solely as a traded commodity (kWh), but as a \textbf{service contract} with three explicit dimensions:

\[
\text{Energy Access Contract} =
\{\text{Magnitude},\ \text{Timing Sensitivity},\ \text{Reliability Requirement}\}.
\]

\begin{itemize}[leftmargin=*]
    \item \textbf{Magnitude} captures the required quantity of electricity.
    \item \textbf{Timing Sensitivity} measures how strictly delivery must occur at specific times.
    \item \textbf{Reliability Requirement} determines priority during shortage events.
\end{itemize}

This representation enables simulation of user contracts, digital flexibility submissions, and scarcity-based allocation using Fair Play rules.

% ---------------------------------------------------------
\section{Data Sources and Dataset Roles}
\label{sec:data_sources}

Table~\ref{tab:data_sources} summarises the datasets used throughout the thesis
and clarifies their distinct roles in calibration, realism, spatial mapping,
and fairness or Shapley-based evaluation. This separation ensures that empirical
data inform model structure and validation without constraining outcomes to
historical price patterns.

\begin{table}[h]
\centering
\caption{Datasets used and their role in model calibration and methodological testing.}
\label{tab:data_sources}
\begin{tabular}{p{3.5cm} p{5cm} p{5cm}}
\toprule
\textbf{Dataset} & \textbf{Primary use} & \textbf{Role in fairness or Shapley modelling} \\
\midrule
UKPN smart meter &
Temporal diversity &
Behavioural realism, flexibility, fairness (F1--F2) \\
BEIS postcode-level demand &
Spatial distribution &
Cluster scaling, national representativeness \\
BMRS generation data &
Half-hourly MW supply &
Adequacy, marginal value, Shapley compensation \\
EV charging &
Plug-in times, power &
Timing sensitivity, device-level flexibility \\
ONS GeoJSON boundaries &
Holarchy creation &
Layered AMM clearing, spatial fairness \\
Vehicle licensing &
EV penetration &
Regional EV burden and allocation \\
\bottomrule
\end{tabular}
\end{table}

A synthetic but physically grounded P1--P4 product dataset was then generated for experimental market-clearing comparison (see Appendix~\ref{app:residential_synth}).

% ---------------------------------------------------------
\section{Modelling Data Transformation: Digital Twin and Holarchy}
\label{sec:data_engineering}

Data engineering includes:

\begin{enumerate}[leftmargin=*]
    \item Temporal harmonisation (30-minute index across all datasets)
    \item Spatial mapping to postcode → DNO → region → national holarchic layers
    \item Population assignment to 29.8M homes using ONS spatial density
    \item Generation of synthetic but physically grounded household types (P1--P4)
    \item Construction of digital twin with representative supply, demand, EVs, and constraints
\end{enumerate}

These were used to build the \textbf{holarchic digital twin} used in simulation.

% ---------------------------------------------------------
\section{Market-Facing Device Modelling (Axis 3)}
\label{sec:axis3_modelling}

Devices (EVs, washing machines, heat pumps, batteries) were modelled using timing windows, energy requirements, and reliability preferences. Requests were converted to AMM-compatible service offers using Algorithm~\ref{app:consumption_requests_algo}, preserving energy and timing flexibility.

% ---------------------------------------------------------
\section{Validation and Evaluation Strategy}
\label{sec:validation_strategy}

Validation was conducted at three levels:

\begin{enumerate}[leftmargin=*]
    \item \textbf{Verification:} Unit tests, energy balance, constraint feasibility
    \item \textbf{Scenario testing:} Too Much, Just Enough, Too Little energy regimes
    \item \textbf{Robustness analysis:} Demand uncertainty, EV penetration, fairness parameter sensitivity
\end{enumerate}

Fairness, efficiency (zero waste), and generator compensation performance were measured under baseline (LMP) and AMM+Fair Play+Shapley architectures.

% ---------------------------------------------------------
\section{Mapping Evaluation Sub-Questions to Methods}
\label{sec:method_to_rq_mapping}
% ---------------------------------------------------------

Although this thesis is guided by a single overarching Research Question
(Section~\ref{sec:research_question}), its empirical evaluation requires a
structured decomposition into six \emph{evaluation sub-questions}, each aligned
with one of the hypothesis domains H1--H6:
Participation (C), Fairness (F), Revenue sufficiency and risk (R),
Price-signal quality (S), Investment adequacy (I), and Procurement efficiency (P).

These sub-questions are not independent research questions; rather, they form
the operational components through which the overarching Research Question is
tested. Each corresponds to a specific methodological pathway involving:
(i) the fairness and efficiency formalism,  
(ii) AMM design and mathematical framework,  
(iii) construction of synthetic and device-level demand,  
(iv) simulation and scarcity experiments, and  
(v) hypothesis-specific evaluation metrics defined in
Chapter~\ref{ch:experiments}.

Table~\ref{tab:rq_mapping} summarises how each evaluation domain (C, F, R, S, I, P)
maps onto its methodological components and the chapters in which the results
are reported.

\renewcommand{\arraystretch}{1.3}
\small

\begin{longtable}{p{3.1cm} p{6.0cm} p{6.0cm}}
\caption{Mapping of evaluation sub-questions (domains C, F, R, S, I, P) to methodological elements}
\label{tab:rq_mapping} \\
\toprule
\textbf{Evaluation sub-question / domain} &
\textbf{Methodological component(s)} &
\textbf{Outcome / chapter(s)} \\
\midrule
\endfirsthead

\toprule
\textbf{Evaluation sub-question / domain} &
\textbf{Methodological component(s)} &
\textbf{Outcome / chapter(s)} \\
\midrule
\endhead

\bottomrule
\endfoot

% -----------------------------------------------------------------

\textbf{Q\textsubscript{C}: Participation \& competition (H1)} &
Product-space design (P1--P4); request and flexibility-envelope model; QoS device-participation experiments; supplier role in subscription setting and service design; AMM vs LMP revenue decomposition and locational risk structure. &
Structural participation capability analysis for consumers, suppliers, devices, and generators; H1 participation \& competition assessment; Chapter~\ref{ch:results}, Section~\ref{sec:results_competition}. \\

\textbf{Q\textsubscript{F}: Distributional fairness (H2)} &
Formal fairness framework; Shapley-consistent generator value allocation; Fair Play shortage allocation for consumers/devices; construction of composite fairness index and jackpot/under-service metrics. &
H2 fairness evaluation across generators, suppliers, and demand-side actors; distributional outcomes and alignment between marginal system value and remuneration; Chapter~\ref{ch:results}, Section~\ref{sec:results_fairness}. \\

\textbf{Q\textsubscript{R}: Revenue sufficiency \& risk allocation (H3)} &
Subscription-pricing stack (energy, reserve, adequacy components); generator revenue-pot modelling and recovery logic; household bill decomposition; volatility, uplift, and tail-risk metrics for generators, suppliers, and households. &
H3 revenue sufficiency and risk tests; comparison of revenue adequacy and volatility structure under LMP vs AMM1/AMM2; Chapter~\ref{ch:results}, Section~\ref{sec:results_revenue_risk}. \\

\textbf{Q\textsubscript{S}: Price-signal quality \& boundedness (H4)} &
AMM price-formation mechanism; tightness ratio $\alpha$; event-based price updates; shadow-price interpretation of voltage; subscription boundary and add-on design. &
H4 price-signal alignment with policy objectives; volatility and spike behaviour; stability and interpretability of tariffs; Chapter~\ref{ch:results}, Section~\ref{sec:results_price_signals}. \\

\textbf{Q\textsubscript{I}: Investment adequacy \& bankability (H5)} &
Generator cost and CapEx/OpEx modelling; Shapley-derived remuneration time series; bankability and NPV-gap metrics; decomposition of subscription revenues into technology- and cluster-specific flows. &
H5 investment adequacy and bankability results; comparison of revenue stability and NPV gaps for wind, nuclear, and other policy-aligned technologies under LMP vs AMM1/AMM2; Chapter~\ref{ch:results}, Section~\ref{sec:results_investment}. \\

\textbf{Q\textsubscript{P}: Procurement efficiency \& zero-waste operation (H6)} &
Formal zero-waste efficiency definition; needs-bundle specification (energy, flexibility, adequacy/reserves, locational relief); AMM clearing rules; Baseline vs AMM scenario design. &
H6 procurement-cost comparison across designs; zero-waste metrics under surplus and scarcity; identification of AMM parametrisations that dominate LMP on cost while satisfying H1--H5; Chapter~\ref{ch:results}, Section~\ref{sec:results_procurement}. \\

\end{longtable}

This mapping clarifies which methodological element supports each research
question and ensures that the evaluation framework is internally coherent:
the same synthetic demand data, AMM control law, fairness definitions,
and scenario structure jointly underpin the comparison of LMP, AMM1, and AMM2.

% ---------------------------------------------------------
\section{Overcoming Shapley Intractability}
\label{sec:shapley_intractability}

Direct computation of generator-level Shapley values is combinatorially
intractable for realistic power systems, scaling as
$\mathcal{O}(2^{|\mathcal{G}|})$ coalition evaluations for a generator set
$\mathcal{G}$. Rather than relaxing Shapley axioms or introducing stochastic
sampling error, this thesis reformulates the valuation problem using
physically admissible structure that preserves Shapley allocations exactly
under stated assumptions.

The method exploits properties of the power system and value function to
maintain Shapley-consistent allocations while achieving tractability. It
combines:

\begin{itemize}[leftmargin=*]
    \item \textbf{Locational and operational clustering}, grouping generators
          into physically cohesive, capacity-substitutable clusters;
    \item \textbf{Feasibility-preserving coalition restriction}, excluding
          coalitions that are infeasible under network, capacity, or
          deliverability constraints;
    \item \textbf{Time-separable marginal contribution evaluation}, exploiting
          the additive structure of the value function across settlement
          intervals;
    \item \textbf{Scarcity-conditioned evaluation}, restricting marginal
          contribution calculations to periods in which generators can affect
          the served-load outcome.
\end{itemize}

Under the clustering and symmetry conditions formalised in
Chapter~\ref{ch:mathematics}, this reformulation reduces the effective
computational burden from $\mathcal{O}(2^{|\mathcal{G}|})$ to approximately
$\mathcal{O}(|\mathcal{G}|^{2} T)$, where $T$ is the number of settlement
intervals, \emph{without altering the resulting generator-level Shapley
allocations}.

Exactness is not merely theoretical. In Appendix~\ref{app:extended_results},
the method is validated on a 13-generator network with explicit transmission
constraints. As shown in Table~\ref{tab:shapcomp}, the generator-level Shapley
values obtained under the clustered formulation coincide with the full Shapley
vector to numerical precision for all generators. This confirms that, within
locational clustering and subject to the stated physical assumptions, the
methodology preserves Shapley-allocated fairness with 100\% accuracy in the
benchmark system.

% ---------------------------------------------------------
\section*{Conclusion}

This methodology demonstrates a rigorous, insight-driven, and computationally implementable pathway to evaluate the proposed AMM + Fair Play + Shapley architecture under conditions that reflect real UK households, generators, infrastructure, and behavioural diversity. It establishes the foundations for empirical evaluation, presented in Chapter~\ref{ch:results}.

% ---------------------------------------------------------
% CHAPTER 8 — MARKET DESIGN AND OPERATING SCENARIOS
% ---------------------------------------------------------
\chapter{Market Designs and Operating Scenarios}
\label{ch:market_scenarios}

This chapter describes how the proposed market architecture operates as a
cyber--physical system under different physical and operational conditions.
Section~\ref{sec:data_foundations} and Section~\ref{sec:physical_foundations}
summarise the data and physical foundations. Section~\ref{sec:continuous_clearing}
introduces the continuous online market instance and its event-driven clearing
logic. Sections~\ref{sec:too_much}--\ref{sec:too_little} characterise system
behaviour in the \emph{Too Much}, \emph{Just Enough}, and \emph{Too Little}
regimes. The remaining sections describe access rules, scarcity exposure, and
allocation behaviour under shortage, and motivate the need for a dedicated
real-time controller.

The Automatic Market Maker (AMM) itself is \emph{not} fully derived in this
chapter. Instead, the AMM is formally defined in Chapter~\ref{ch:amm} as the
core continuous control layer that maps scarcity into prices and allocations,
subject to the fairness requirements established in
Chapter~\ref{ch:fairness_definition}.

% ---------------------------------------------------------
\section{Data Foundations and System Understanding}
\label{sec:data_foundations}

Any operational market design must be grounded in empirical system data:

\begin{itemize}[leftmargin=*]
\item \textbf{Demand distributions:}
load profiles at household, cluster, region, and national scales; seasonal
patterns; peak-to-average ratios; essential vs flexible segmentation.

\item \textbf{Supply distributions:}  
wind, solar, storage, thermal availability, outage distributions, ramping,
maintenance cycles; frequency and severity of low-supply events.

\item \textbf{Network constraints:}  
line ratings, voltage limits, transformer constraints; interconnector
capacity and N--1 security envelopes.

\item \textbf{Locational structure:}  
mapping of nodes, clusters, and legacy \emph{zones} to congestion patterns,
import/export limits, and shared scarcity events. This includes the
coarse-grained zonal partitions used in many European markets, as well as
finer-grained nodal or cluster representations.

\item \textbf{Data model:}  
raw inputs transformed into a unified representation:
\[
  (\text{demand}, \text{supply}, \text{constraints})_{t,n}
  \longrightarrow \text{tightness}_{t,n},
\]
indicating how close the system is to adequacy, congestion, or reserve limits.

\end{itemize}

These data form the basis of the AMM design and the operating regime
classification (\emph{Too Much / Just Enough / Too Little}).

% ---------------------------------------------------------
\section{Physical Foundations: AC, DC, and Two-Way Flows}
\label{sec:physical_foundations}

The market design must faithfully reflect the underlying physics:

\begin{itemize}[leftmargin=*]
\item \textbf{AC power flows} governed by Kirchhoff's laws, thermal and
voltage constraints, reactive power limits, and stability margins.

\item \textbf{DC corridors} can reroute large transfers, relieving AC
congestion, but must respect converter capacity and contingency rules.

\item \textbf{Two-way flows} from distribution-connected generation and
prosumers raise voltage, congestion, and protection challenges.

\item \textbf{Operational constraints} (ramping, inertia, frequency response)
restrict how quickly the system transitions between regimes.

\end{itemize}

The proposed AMM incorporates these constraints holarchically, mapping them into
node- and zone-specific tightness measures (Section~\ref{sec:amm_design}).

\paragraph{Relation to nodal and zonal pricing.}
Classical locational designs can be viewed as different discretisations of the
underlying physical system. Nodal pricing (LMP) attempts to reflect marginal
network constraints at individual buses, but retains the structurally unstable
energy-only risk allocation highlighted in Lemma~\ref{lem:risk_volume_instability}.
Zonal pricing aggregates nodes into a small number of administratively defined
zones, reducing dimensionality but further decoupling prices from the actual
pattern of AC flows and redispatch. In practice, zonal markets inherit the same
insolvency and uplift dynamics as energy-only nodal designs
(Lemma~\ref{lem:price_cap_insolvency}), while relying on ex-post redispatch and
side-payments to restore feasibility. The holarchic AMM used in this thesis
replaces fixed, politically negotiated zones with a dynamic hierarchy of
clusters whose boundaries and tightness measures are grounded in real-time
network conditions, preserving computational tractability without the mispricing
and redispatch burden of static zonal designs.

\subsection{Holarchic structure of the grid and market}
\label{sec:holarchy}

The electricity system is not only hierarchical (national system, regions,
networks, feeders, households); it is \emph{holarchic} in the sense of
Koestler \cite{koestler1967ghost}: each unit is simultaneously a \emph{whole} 
relative to the layers below it and a \emph{part} of a larger whole above it. 
In this thesis, these nested units are the objects on which the AMM operates: 
it computes tightness, allocates access, and propagates scarcity signals at m
ultiple spatial and institutional layers.

This holarchic representation is not merely a descriptive convenience.
It is both a \emph{technical necessity} and a \emph{normative requirement}
of the fairness objectives imposed in this thesis.

From a technical perspective, Shapley-consistent allocation is only
computationally feasible if the system admits structured decomposition:
fairness must be evaluated over sets of actors that are meaningfully
substitutable under the prevailing physical constraints. Holarchic
clustering provides exactly this structure, allowing marginal
contributions to be computed within and across layers while preserving
symmetry, efficiency, and additivity. As demonstrated in the extended
results, when clusters are defined to respect deliverability and
substitutability, the resulting allocations match full generator-level
Shapley values to numerical precision.

From a normative perspective, the fairness conditions imposed in this
work (Requirements F1--F4) cannot be satisfied by a single, flat market
layer. Fair treatment requires that scarcity, priority, and contribution
be evaluated \emph{at the layer where they are physically realised}:
national adequacy at system level, congestion at regional level, access
and protection at household level, and flexibility at device level.
A holarchic structure is therefore essential to ensure that fairness is
neither diluted by over-aggregation nor distorted by inappropriate
comparisons across physically incomparable actors.

\begin{figure}[t]
  \centering
  \includegraphics[width=0.55\textwidth]{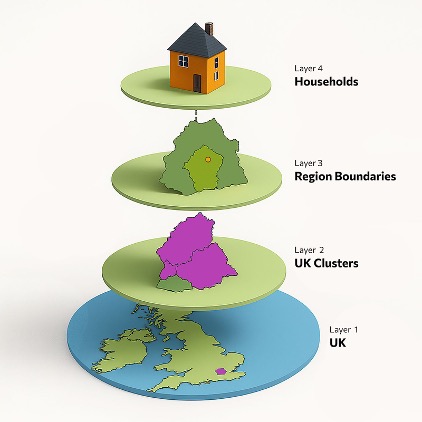}
    \caption[Holarchic structure of the grid and market]{%
    Conceptual holarchy of the electricity system. The UK-wide system (Layer~1)
    contains electrically defined clusters capturing dominant congestion patterns
    (Layer~2), which in turn contain regions or network areas (Layer~3), within
    which individual households and businesses reside (Layer~4).
    A further layer of individual devices and controllable assets (Layer~5) operates
    within households and sites but is not shown for clarity.
    Each layer is simultaneously a whole (with respect to the layers below) and a
    part (of the layer above), and may serve as the natural unit of analysis,
    allocation, or regulation for different stakeholders.}
  \label{fig:holarchy_layers}
\end{figure}

Figure~\ref{fig:holarchy_layers} illustrates one concrete instantiation of this
holarchy:

\begin{itemize}[leftmargin=*]

  \item \textbf{Layer 1: UK system.}
  The national electricity system treated as a single balancing entity, used
  for system-wide adequacy assessment, aggregate cost recovery, and national
  policy constraints.

  \item \textbf{Layer 2: Congestion-relevant system partitions.}
  A small number of electrically and economically meaningful partitions that
  capture dominant congestion and scarcity patterns in the contemporary GB
  system. In the present experiments (reflecting 2024--2025 conditions), this
  layer is instantiated by a coarse London--Scotland (London--Glasgow) split,
  reflecting the binding north--south transfer constraint that materially
  differentiates scarcity exposure and marginal generator value.

  More generally, this layer serves a dual purpose: it aligns scarcity exposure
  with physically meaningful bottlenecks \emph{and} enables tractable Shapley
  allocation by grouping assets into substitutable clusters. The number and
  composition of clusters are therefore \emph{holonic and task-dependent}:
  different clustering schemes may be used for congestion pricing, generator
  value attribution, or investment analysis, subject to the requirement that
  within-cluster symmetry and substitutability preserve Shapley accuracy.

  \item \textbf{Layer 3: Regional and distribution-level groupings.}
  Finer-grained spatial units such as distribution network areas, constraint
  regions, or local authorities. These units correspond to operational
  responsibility boundaries for DSOs, local flexibility markets, and municipal
  programmes, and are natural loci for local scarcity signals and flexibility
  activation.

  \item \textbf{Layer 4: Households, SMEs, and customer portfolios.}
  Individual customers or small portfolios whose service contracts, demand
  envelopes, and behavioural responses determine retail outcomes, fairness
  impacts, and subscription-based cost recovery.

  \item \textbf{Layer 5 (not shown in Figure \ref{fig:holarchy_layers}): Devices and controllable assets.}
  Individual physical devices (EVs, heat pumps, batteries, industrial loads,
  distributed generators) that submit bids, offer flexibility, or respond to
  control signals. At this layer, demand and supply are treated symmetrically as
  time-bound, locationally constrained service requests.
\end{itemize}

Crucially, different stakeholders induce \emph{different} holarchies on the
same physical infrastructure:

\begin{itemize}[leftmargin=*]
  \item The \textbf{system operator} naturally works at the national and
        transmission-region layers (adequacy, interconnector flows, major
        constraints).
  \item \textbf{DSOs} care about primary/secondary substations, feeders, and
        local constraint regions.
  \item \textbf{Suppliers and service providers} organise portfolios into commercial
        regions, customer segments, and virtual fleets of flexible assets.
  \item The \textbf{regulator and government} often work with political or
        socio-economic regions (devolved administrations, local authorities,
        vulnerability indices).
\end{itemize}

The AMM formalism does not hard-code any particular stakeholder view. Instead,
it operates on a generic holarchic partition of the grid: a collection of
nested “cells” within which tightness is evaluated and between which flows are
constrained. For the empirical work in this thesis, we instantiate this as:

\begin{enumerate}[leftmargin=*]
  \item a top-level UK system node;
  \item an intermediate layer of electrically defined clusters (or two regions,
        London vs.\ Glasgow, when studying the transfer constraint);
  \item a household layer, where usage profiles and retail products are
        defined.
\end{enumerate}

Mathematically, each holon $h$ in this hierarchy has an associated tightness
process $\tilde{\alpha}_{h,t}$ and a set of contracts located within it. The
AMM maps $\tilde{\alpha}_{h,t}$ into prices and allocation rules for that
holon, while ensuring consistency across parents and children (no child can be
less tight than the constraints of its parent, and shortage at a parent must
be resolved by allocations across its children). This is why the design is
described as \emph{holarchic}: it treats the grid as a nested collection of
control and settlement cells, rather than a flat set of nodes or a fixed,
politically negotiated set of zones.

The remainder of this chapter develops the operational picture of the proposed
architecture: the continuous online market instance and event-driven clearing
(Section~\ref{sec:continuous_clearing}), the resulting operating regimes
(Sections~\ref{sec:too_much}--\ref{sec:too_little}), and the access and
allocation logic under scarcity (Section~\ref{sec:market_access}).
The legacy retail fragilities that motivate this redesign—settlement shocks,
risk--volume separation, and the solvency--affordability trap—were established
in Chapter~\ref{chap:problem} (see Section~\ref{sec:retail_problem} and
Section~\ref{sec:price_cap_limits}).
Chapter~\ref{ch:amm} then provides the formal AMM definition, while
Chapter~\ref{ch:experiments} specifies the empirical scenarios.

% ---------------------------------------------------------
\section{Continuous Online Market Design and Clearing Mechanism}
\label{sec:continuous_clearing}

Building on the fragmented digitalisation and market landscape described in
Sections~\ref{sec:digitalisation_fragmentation}--\ref{sec:graphs_distributed_control},
classical electricity markets still operate in segmented stages—day-ahead,
intraday, balancing, and settlement—each with its own gate-closure,
separate bid structure, and independent pricing logic. These artificial
boundaries make the market slow to react to changing conditions,
increase transaction costs, and introduce both spatial and temporal
inefficiencies. The proposed market redesign instead operates as a
\textbf{continuous online market instance}: a single, continuously active
clearing process that accepts and processes bids at any moment, without
waiting for predefined auction intervals or gate closures.

\begin{figure}[t]
\centering
\includegraphics[width=\textwidth]{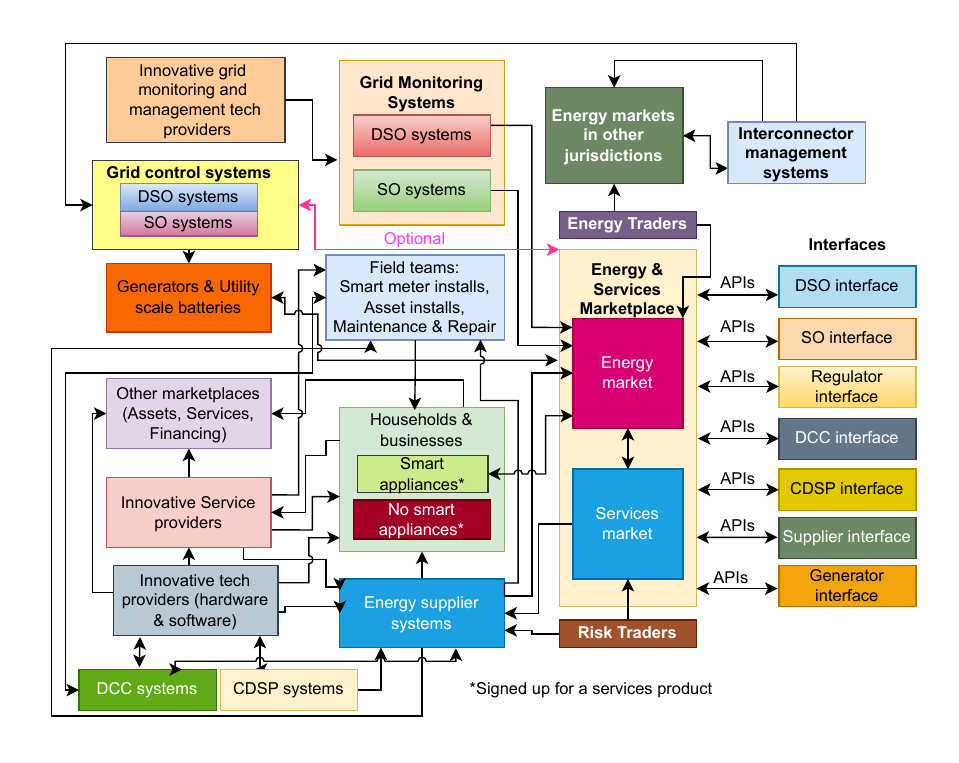}
\caption[Architecture of the proposed digital market system]{Architecture of the proposed digital market platform, showing the interaction between edge participants, the continuous online market instance, the Automatic Market Maker (AMM), data stores, and governance/control interfaces.}
\label{fig:systemDiagram}
\end{figure}

As shown in Figure~\ref{fig:systemDiagram}, the proposed architecture integrates physical dispatch, digital control, and settlement through a continuous AMM-driven platform.

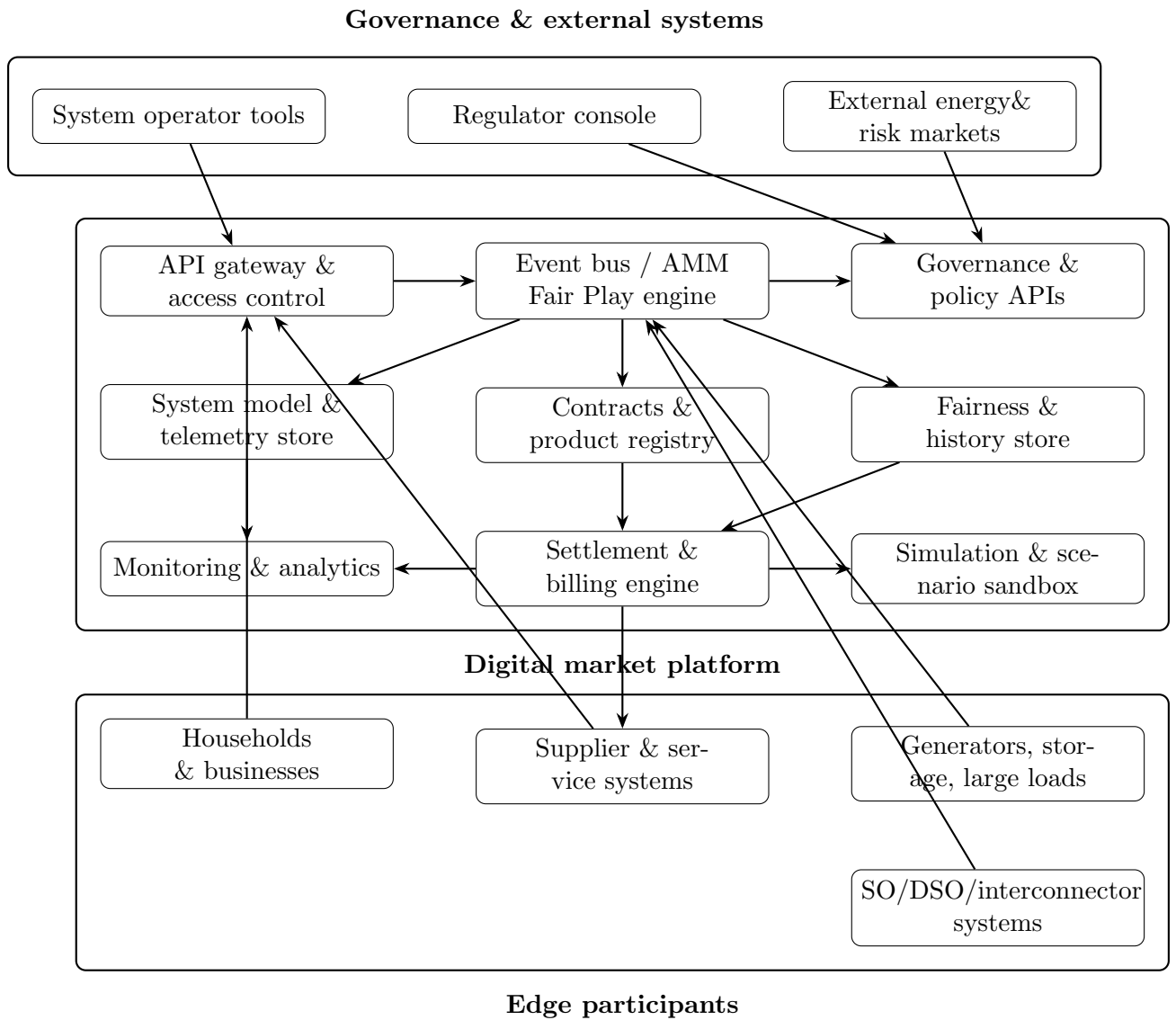
\begin{figure}[t]
\centering
\begin{tikzpicture}[
font=\small,
>=Stealth,
box/.style={
rectangle,
draw,
rounded corners,
align=center,
minimum height=8mm,
text width=4cm,
inner sep=4pt
},
layer/.style={
rectangle,
draw,
rounded corners,
thick,
inner sep=10pt
},
arrow/.style={->, thick},
node distance=10mm and 12mm
]

% --------------------------------------------------
% NODES (placed individually)
% --------------------------------------------------

% Governance layer
\node[box] (SOtools) {System operator tools};
\node[box, right=of SOtools] (Regulator) {Regulator console};
\node[box, right=of Regulator] (Markets) {External energy\& risk markets};

\node[layer, fit=(SOtools)(Regulator)(Markets),
label={[yshift=2mm]above:\textbf{Governance \& external systems}}] {};

% Digital platform - top row
\% Digital platform - top row
\node[box, below=15mm of Regulator, xshift=-45mm] (Gateway) {API gateway \& access control};
\node[box, right=of Gateway] (AMM) {Event bus / AMM \\ Fair Play engine};
\node[box, right=of AMM] (GovAPI) {Governance \& policy APIs};

% Digital platform - middle row
\node[box, below=of Gateway] (ModelStore) {System model \& telemetry store};
\node[box, below=of AMM] (ContractStore) {Contracts \& product registry};
\node[box, below=of GovAPI] (HistoryStore) {Fairness \& history store};

% Digital platform - bottom row
\node[box, below=of ContractStore] (Settlement) {Settlement \& billing engine};
\node[box, left=of Settlement] (Monitoring) {Monitoring \& analytics};
\node[box, right=of Settlement] (Simulation) {Simulation \& scenario sandbox};

\node[layer, fit=(Gateway)(AMM)(GovAPI)(ModelStore)(ContractStore)(HistoryStore)(Settlement)(Monitoring)(Simulation),
label={[yshift=-2mm]below:\textbf{Digital market platform}}] {};

% Edge participants
\node[box, below=18mm of Monitoring] (Households) {Households \& businesses};
\node[box, below=18mm of Settlement] (Suppliers) {Supplier \& service systems};
\node[box, below=18mm of Simulation] (Generators) {Generators, storage, large loads};
\node[box, below=of Generators] (Network) {SO/DSO/interconnector systems};

\node[layer, fit=(Households)(Suppliers)(Generators)(Network),
label={[yshift=-2mm]below:\textbf{Edge participants}}] {};

% --------------------------------------------------
% CONNECTIONS
% --------------------------------------------------

% Governance to platform
\draw[arrow] (SOtools) -- (Gateway);
\draw[arrow] (Regulator) -- (GovAPI);
\draw[arrow] (Markets) -- (GovAPI);

% Platform internal
\draw[arrow] (Gateway) -- (AMM);
\draw[arrow] (AMM) -- (GovAPI);

\draw[arrow] (AMM) -- (ModelStore);
\draw[arrow] (AMM) -- (ContractStore);
\draw[arrow] (AMM) -- (HistoryStore);

\draw[arrow] (ContractStore) -- (Settlement);
\draw[arrow] (HistoryStore) -- (Settlement);
\draw[arrow] (ModelStore) -- (Monitoring);

\draw[arrow] (Settlement) -- (Monitoring);
\draw[arrow] (Settlement) -- (Simulation);

% Edge to platform
\draw[arrow] (Households) -- (Gateway);
\draw[arrow] (Suppliers) -- (Gateway);
\draw[arrow] (Generators) -- (AMM);
\draw[arrow] (Network) -- (AMM);

% Settlement feedback to suppliers
\draw[arrow] (Settlement) -- (Suppliers);

\end{tikzpicture}
\caption[Software architecture of the continuous online market instance]{%
Software architecture of the continuous online market instance and Automatic Market Maker (AMM). Governance and external systems (top) interact with the digital market platform (middle), which hosts the API gateway, AMM and Fair Play control engine, data stores, analytics, and settlement services. Edge participants (bottom) connect via APIs and telemetry, forming the cyber--physical control architecture described in this chapter.}
\label{fig:amm_software_architecture}
\end{figure}

\subsection{Event-driven clearing}

Instead of accumulating bids for batch optimisation, the market performs
\emph{sequential feasibility evaluation}. When new bids or updated system
information (flexibility windows, forecasts, congestion alerts) arrive,
they are immediately processed. Each bid is accepted if and only if it is:
\begin{itemize}[leftmargin=*]
\item physically deliverable (network-capable, respecting real power-flow constraints);
\item price-consistent with the prevailing scarcity at relevant nodes or clusters;
\item non-conflicting with already accepted allocations; and
\item compliant with fairness protection, essential energy shielding, and vulnerability rules.
\end{itemize}

This approach removes the need for global Economic Dispatch
optimisation, and makes clearing \emph{event-triggered} rather than
time-triggered. It also means there is no gate-closure window: bids are
accepted whenever feasible, and are cleared in an ongoing, incremental
process.

\subsection{Integrated forward and real-time clearing}

Unlike existing markets, where forward contracts (day-ahead, forward,
capacity auctions) are structurally separate from real-time balancing,
the proposed model integrates both through a single digital market
instance. A bid may request energy or flexibility for any timestamp in
a continuous forward horizon (e.g.\ now to 48 hours ahead). The AMM
assesses feasibility directly against forward forecasts and physical
constraints, rather than through separate forward market constructs.

\subsection{Bidding parameters and individual rationality}
\label{sec:bidding_parameters}

Each participant $i$ submits a bid or offer $r$ describing the physical and
economic attributes of the service being requested \emph{or} supplied. To cover
both consumption and generation uniformly, we adopt the sign convention:
\[
E_r > 0 \quad\text{(net consumption request)}, \qquad
E_r < 0 \quad\text{(net supply offer)}.
\]
With this convention, a single bid definition can represent a household
requesting energy, a generator offering production, a storage asset doing
either, or an aggregator submitting composite flexibility.

A bid $r$ is defined by:
\[
  r = \bigl(E_r,\; [t^{\mathrm{start}}_r, t^{\mathrm{end}}_r],\;
            \bar{P}_r,\; \sigma_r,\; v^{\max}_r,\;
            \Gamma^{\text{contract}}_r \bigr),
\]
with components:
\begin{itemize}[leftmargin=2em]
  \item $E_r$ — total energy volume (positive for requests, negative for offers);
  \item $[t^{\mathrm{start}}_r, t^{\mathrm{end}}_r]$ — permissible delivery window;
    \item $\bar{P}_r$ — maximum instantaneous power magnitude that the bid may draw
    or deliver. Enforcement of this limit occurs at the device edge, with the device
    itself ensuring operation within its physical and safety constraints.
  \item $\sigma_r$ — flexibility parameter specifying allowable shifting,
        reshaping, or interruption of the energy schedule;
  \item $v^{\max}_r$ — maximal economically admissible value:
        \begin{itemize}
          \item for requests: maximum payment the participant is willing to make
                for receiving $E_r$ within the declared window;
          \item for offers: minimum compensation acceptable for supplying $E_r$.
        \end{itemize}
        This represents the participant’s willingness-to-pay or willingness-to-accept.
    \item $\Gamma^{\text{contract}}_r$ — energy access contract attributes associated
    with the request, describing its declared magnitude, timing sensitivity, and
    reliability characteristics, and governing how the request is treated under
    scarcity and operational stress.
\end{itemize}

Unlike classical markets, where willingness-to-pay is encoded implicitly through
bid price steps or locational prices, the proposed architecture treats
$v^{\max}_r$ as an explicit, first-class parameter of every bid. This plays
three essential roles:

\begin{enumerate}[leftmargin=2em]
\item \textbf{Individual rationality}. The AMM will never clear a trade for a
      participant at a net cost exceeding $v^{\max}_r$ (or below their
      willingness-to-accept). This guarantees that all accepted allocations
      are beneficial ex ante, transparent, and enforceable.

\item \textbf{Stability of scarcity-clearing}. Explicit value bounds prevent
      runaway scarcity prices and anchor the feasible region of the AMM during
      tight periods.

\item \textbf{Holarchic feasibility}. By embedding economic limits alongside
      physical and contractual attributes, the market can evaluate bids
      sequentially—without global optimisation—while ensuring that decisions
      remain feasible across space, time, and fairness layers.
\end{enumerate}

These bidding parameters provide the foundation for the three-dimensional
contract structure described in Section~\ref{sec:contract_structure}, and for
the fairness-aware allocation rules later formalised in
Chapter~\ref{ch:fairness_definition}.
\subsection{Contract structure: magnitude, timing, and reliability}
\label{sec:contract_structure}

The term $\Gamma^{\text{contract}}_r$ in the bid definition encodes more than a
simple tariff or product label. In the proposed market design, each bid is
associated with an \emph{energy access contract} that specifies how the request
should be treated under scarcity and operational stress.

Each energy access contract is characterised by three core dimensions:
\[
\text{Energy Access Contract}
=
\{\text{Magnitude},\ \text{Timing Sensitivity},\ \text{Reliability Requirement}\}.
\]

These dimensions are implemented through the bid parameters as follows:
\begin{itemize}[leftmargin=*]

\item \textbf{Magnitude} is represented by $E_r$ (requested or offered energy)
and $\bar{P}_r$ (maximum power), capturing both total volume and peak intensity
of the service requested.

\item \textbf{Timing Sensitivity} is represented by the delivery window
$[t^{\mathrm{start}}_r, t^{\mathrm{end}}_r]$ and the flexibility parameter
$\sigma^r$, which describe how tightly constrained delivery timing is and how
much shifting or reshaping the participant is willing to accept.

\item \textbf{Reliability Requirement} is encoded in
$\Gamma^{\text{contract}}_r$ as the reliability and protection characteristics
associated with the request. These attributes determine the request’s treatment
under shortage or congestion, including its entitlement to retain service
relative to other contemporaneous bids.
\end{itemize}

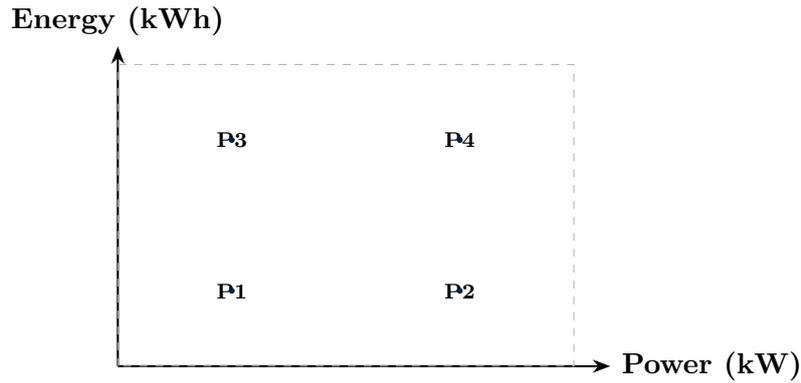
\begin{figure}[t]
\centering
\begin{tikzpicture}[>=Stealth, x=6cm, y=4cm, font=\small]

% Axes
\draw[->, thick] (0,0) -- (1.08,0) node[anchor=west] {\textbf{Power (kW)}};
\draw[->, thick] (0,0) -- (0,1.06) node[anchor=south] {\textbf{Energy (kWh)}};

% Dashed bounding box for typical product space
\draw[dashed, gray!60] (0,0) rectangle (1,1);

% Quadrant labels centred
\node at (0.25,0.25) {\scriptsize \textbf{P1}};
\node at (0.75,0.25) {\scriptsize \textbf{P2}};
\node at (0.25,0.75) {\scriptsize \textbf{P3}};
\node at (0.75,0.75) {\scriptsize \textbf{P4}};

% Optional dots at the quadrant centres (comment out if you want labels only)
\filldraw[ImperialBlue] (0.25,0.25) circle (0.9pt);
\filldraw[ImperialBlue] (0.75,0.25) circle (0.9pt);
\filldraw[ImperialBlue] (0.25,0.75) circle (0.9pt);
\filldraw[ImperialBlue] (0.75,0.75) circle (0.9pt);

\end{tikzpicture}
\caption[Product space in magnitude--impact dimensions]{%
Illustration of the retail product space in the magnitude--impact plane.
Each point represents an underlying \emph{usage profile} (household or SME),
classified by (i) the quantity of energy it seeks to consume from
\emph{non-zero-marginal-cost supply} and (ii) its contribution to system
tightness during scarce periods. Reliability / Quality-of-Service (QoS) is an
additional, independent contract dimension and is not shown here.}
\label{fig:product_space_2d}
\end{figure}

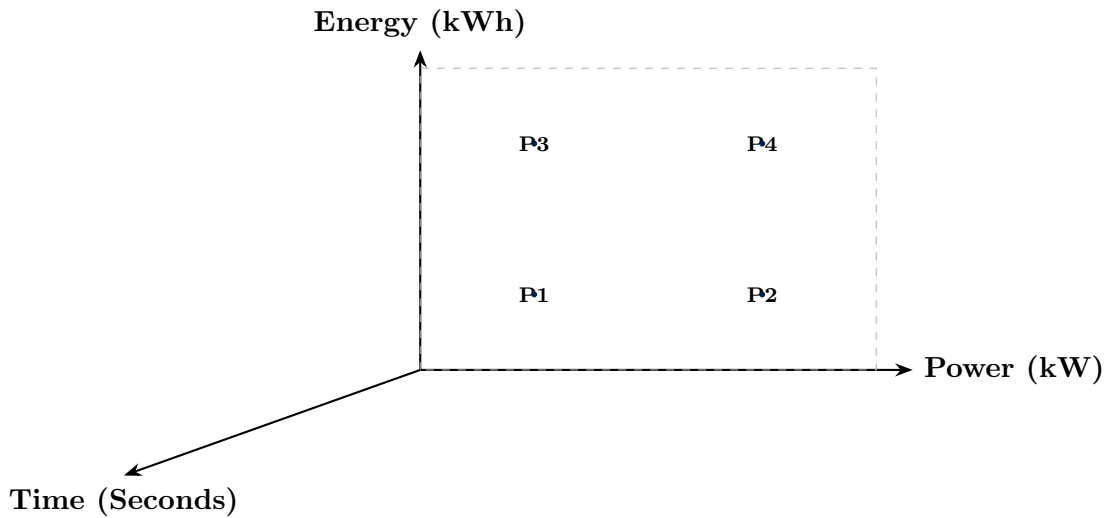
\begin{figure}[t]
\centering
\begin{tikzpicture}[>=Stealth, x=6cm, y=4cm, font=\small]

% Base axes (Magnitude and Impact)
\draw[->, thick] (0,0) -- (1.08,0)
  node[anchor=west] {\textbf{Power (kW)}};
\draw[->, thick] (0,0) -- (0,1.06)
  node[anchor=south] {\textbf{Energy (kWh)}};

% Product-space box on the base plane
\draw[dashed, gray!60] (0,0) rectangle (1,1);

% Quadrant labels on the base plane
\node at (0.25,0.25) {\scriptsize \textbf{P1}};
\node at (0.75,0.25) {\scriptsize \textbf{P2}};
\node at (0.25,0.75) {\scriptsize \textbf{P3}};
\node at (0.75,0.75) {\scriptsize \textbf{P4}};

% Optional dots at quadrant centres
\filldraw[ImperialBlue] (0.25,0.25) circle (0.9pt);
\filldraw[ImperialBlue] (0.75,0.25) circle (0.9pt);
\filldraw[ImperialBlue] (0.25,0.75) circle (0.9pt);
\filldraw[ImperialBlue] (0.75,0.75) circle (0.9pt);

% Third (implicit) axis: Reliability / Time / Flexibility
\coordinate (O) at (0,0);
\coordinate (Raxis) at (-0.65,-0.35);
\draw[->, thick] (O) -- (Raxis)
  node[anchor=north] {\textbf{Time (Seconds)}};

\end{tikzpicture}
\caption[Three-dimensional contract space with reliability]{%
Conceptual three-dimensional contract space. The base plane shows
magnitude--impact quadrants (P1--P4) as in Figure~\ref{fig:product_space_2d}.
The third axis represents a continuous \emph{service dimension} capturing
reliability, temporal tolerance, and willingness to adapt over time.
Contracts are specified independently along all three dimensions:
magnitude (the quantity of energy sought from non-zero-carbon or
non-zero-fuel-cost supply), impact (how requests interact with system
tightness during scarce periods), and service tolerance (the degree to which
delivery may be shifted, reshaped, or deferred while retaining contractual
access). This third axis does not impose any hierarchy across usage profiles;
it represents an orthogonal, user-selected attribute layered on top of demand
shape and timing characteristics.}
\label{fig:product_space_3d}
\end{figure}

Importantly, energy access contracts are associated with individual bids rather
than with customers as a whole. A single household may therefore submit multiple
bids—e.g.\ at the meter, portfolio, or device level—each with its own magnitude,
timing sensitivity, and reliability characteristics, reflecting different
service preferences across the holarchy.

f
Figure~\ref{fig:product_space_2d} illustrates this idea in the
two-dimensional product space spanned by magnitude and impact. Each
point in the diagram represents an underlying \emph{usage profile}
(rather than a specific device class), classified according to its peak power
demand and its contribution to scarce periods. The four quadrants
(P1--P4) correspond to low/high magnitude and low/high impact, and form
the basis for the product groupings used in the empirical analysis in
Chapter~\ref{ch:experiments}. Reliability or Quality-of-Service
(QoS) is deliberately not shown at this stage; it is introduced as a
separate, independent contract dimension in Figure~\ref{fig:product_space_3d}.

In existing markets, only the first dimension (magnitude) is typically
contracted explicitly, with some industrial customers facing an
additional maximum-demand term. Timing flexibility and reliability
entitlement are either implicit, non-contractible, or handled through
ad hoc arrangements. As a result, allocation under shortage is often
arbitrary, opaque, or driven solely by willingness-to-pay.

By contrast, the proposed market design treats the three-dimensional energy
access contract as a first-class object in the clearing logic. The economic
significance of magnitude, timing sensitivity, and reliability is
state-dependent: it varies with the balance between available supply and
desired demand. Accordingly, the architecture distinguishes three operational
regimes, which may coexist simultaneously across different layers of the
holarchy due to locational and network constraints.

\begin{enumerate}[leftmargin=*]

\item \textbf{Normal operation (\emph{Just Enough} regime).}
When available supply is broadly aligned with desired demand, clearing is driven
primarily by the \emph{magnitude} and \emph{timing sensitivity} dimensions of the
energy access contract. Bids are accepted if they are physically feasible and
consistent with prevailing tightness signals. In this regime, timing sensitivity
$\sigma^r$ has positive economic value: modest shifting or reshaping of demand
can smooth intra-day imbalances and improve utilisation of low-cost generation.

\item \textbf{Surplus conditions (\emph{Too Much} regime).}
When zero-fuel-cost generation (typically renewable) exceeds contemporaneous
demand, additional supply has no marginal economic value and curtailment is
efficient in the Pareto sense. Timing flexibility remains relevant only insofar
as it enables the absorption of surplus at low system cost. The AMM therefore
does not mandate consumption, but may encourage it through forward-looking,
state-aware price signals defined later in the clearing algorithm. Excess supply
that cannot be efficiently absorbed is curtailed without penalty.

\item \textbf{Scarcity conditions (\emph{Too Little} regime).}
When desired demand exceeds physically deliverable supply, the \emph{reliability
dimension} of the energy access contract, encoded in
$\Gamma^{\text{contract}}_r$, becomes decisive. Access is no longer determined
solely by price or timing flexibility, but by the declared reliability
characteristics associated with each bid. Shortage allocation is governed by
fairness-preserving rules implemented by the Fair Play algorithm
(Chapter~\ref{ch:mathematics}), ensuring essential protection and proportional
responsibility under scarcity.

\end{enumerate}

Across all regimes, dispatch remains cost-ordered: generators with the lowest
marginal cost—typically zero-carbon resources—are utilised first, with
controllable demand and higher-cost generation activated only when required to
maintain feasibility.

Figure~\ref{fig:product_space_3d} extends this representation by adding
Reliability / QoS as an orthogonal contract dimension. The reliability
axis is drawn diagonally away from the origin to emphasise that it is an
\emph{additional} choice layered on top of a given usage profile, rather
than an inherent property of any particular quadrant. A household with
the same P2 profile (high magnitude, low impact) may, for example,
choose a highly protected contract or a flexible, interruptible one,
depending on its preferences and willingness to trade QoS against cost.
Similarly, behind-the-meter technologies such as EVs and batteries can
change the position of the aggregate usage profile in the 2D plane,
while explicit device enrolment in balancing services determines where
those assets sit along the reliability axis. This three-dimensional
contract space is what the AMM and Fair Play algorithm operate on in
real time when implementing the fairness conditions of
Chapter~\ref{ch:fairness_definition}.

This structure also supports a \textbf{non-coercive transition} from
legacy tariffs to digitally managed service contracts. Participants may
choose to:
\begin{itemize}[leftmargin=*]
\item remain on legacy, high-reliability contracts (encoded as
high-priority, low-flexibility $\Gamma^{\text{contract}}_r$);
\item opt into flexible, lower-cost contracts with reduced
reliability guarantees; or
\item enrol specific devices (EVs, heat pumps, storage) as
flexibility providers, increasing their contribution to system
reliability in exchange for lower expected costs or improved
priority under shortage.
\end{itemize}

In all cases, the contract attributes in $\Gamma^{\text{contract}}_r$
are processed by the AMM and Fair Play as digitally enforceable rules,
rather than informal promises. This ensures that market access,
scarcity exposure, and allocation under shortage are governed by
\emph{transparent, auditable, and formally defined} contractual
dimensions, consistent with the fairness conditions
in Chapter~\ref{ch:fairness_definition}.

\subsection{Cyber--physical synchronisation: electrons, data, and money}

The market behaves as a cyber--physical system in which
\emph{each accepted allocation corresponds to a physically feasible
dispatch path}. Trade acceptance, pricing, allocation priority, and
settlement are anchored in:

physical feasibility $\leftrightarrow$ digital signalling $\leftrightarrow$ financial settlement.

This eliminates the need for ex-post redispatch, constraint payments, or
balancing charges, because the underlying allocation logic already
respects network physics.

\subsection{Digital enforceability and settlement}

Each confirmed allocation is timestamped, associated with a delivery
node or cluster, and embedded in a digitally enforceable service
contract. Settlement occurs post-delivery based on metered or verified
consumption/generation. Since deviations are known immediately (via
smart meter state or device telemetry), settlement risk is reduced
without requiring centralised reconciliation through intermediaries.

\subsection{Implications for system behaviour}

\begin{itemize}[leftmargin=*]
\item Market clearing becomes continuous, digital, and physically
grounded rather than periodic and abstract.
\item Pricing evolves smoothly. Scarcity signals update without
exogenous jumps due to auction boundary discontinuities.
\item Allocation is based on \emph{who can shift, who needs
protection, and who can help the system}, not purely on
willingness to pay.
\item The distinctions between wholesale, balancing, and retail
become matters of digital scope rather than separate markets.
\item Real-time operation does not require perfect foresight or
global optimisation, only feasibility-aware and fairness-aware
incremental updates.
\end{itemize}

The detailed control logic and tightness-based pricing functions are developed
in Chapter~\ref{ch:amm}; here we focus on the structural behaviour and
operating regimes.

\section{Cost Structure and the Allocation of System Costs}

A foundational principle of the proposed market design is that the method of recovering system costs must correspond to the physical nature of those costs. Electricity systems contain costs that are fundamentally different in origin and behaviour—some fixed, some marginal, and some that arise only under scarcity. Treating them identically, as legacy markets often do, produces distorted incentives, cross-subsidies, and unstable long-run signals.

\begin{itemize}[leftmargin=*]
\item \textbf{Fixed system costs (e.g.\ reserves, black-start capability, inertia)}
These capabilities must exist regardless of individual consumption patterns. Because they are imposed ex ante by the need to maintain system integrity, their fair recovery must be through \emph{fixed} subscription-style charges. Recovering fixed costs via volatile marginal energy prices is both inefficient and unfair: it exposes consumers to risk they did not cause.

\item \textbf{Variable costs (e.g.\ energy production)}
Energy is a purely marginal cost: additional consumption causes additional production. Fairness and efficiency both require that energy costs be recovered \emph{variably}, from those whose behaviour actually imposes them, through real-time marginal pricing.

\item \textbf{Scarcity-based costs (e.g.\ adequacy, capacity, emergency response)}
Capacity has value only in periods of tightness. Scarcity-driven costs should therefore be recovered from participants whose consumption contributes to peak demand or system stress. In the proposed design, these costs are allocated through scarcity-weighted capacity rents, proportional to a household’s impact during tight periods.
\end{itemize}

This decomposition ensures that each cost category is recovered through a mechanism aligned with its physical and behavioural causes. It provides the economic rationale for the subscription–energy–capacity structure used throughout this thesis and forms the foundation for how the AMM prices scarcity, protects essential usage, and allocates shortage fairly. It also ensures that all cost recovery is transparent, traceable, and digitally enforceable.

% ---------------------------------------------------------
\section{Operating Regimes}
\label{sec:operatingregimes}

The contract structure introduced in
Section~\ref{sec:contract_structure} determines \emph{which attributes of a bid
are economically decisive} under different physical conditions. These
conditions can be grouped into three canonical operating regimes, noting that
different regimes may coexist simultaneously across layers of the holarchy due
to locational and network constraints. The following sections provide an
operational interpretation of these regimes, describing how the AMM clears bids,
enables flexibility, and allocates value under each physical state.

\subsection{Too Much Energy: Surplus Regime}
\label{sec:too_much}

In the \emph{Too Much} regime, aggregate available supply exceeds feasible
demand by a wide margin, subject to network constraints. This can arise from:

\begin{itemize}[leftmargin=*]
\item high renewable output during low-load periods;
\item inflexible or must-run generation;
\item limited ability to export via interconnectors;
\item insufficient activation of voluntary demand-side flexibility.
\end{itemize}

Classical markets may produce \emph{negative prices}, signalling that generators
should turn down and demand should increase. However, such signals can be
regressive if only a subset of consumers can access or respond to them.

In the proposed design:

\begin{itemize}[leftmargin=*]
\item \textbf{Negative prices} $p^{\mathrm{base}}_{n,t}$ may
occur when generators face negative opportunity costs, for example due to
technical or economic constraints such as nuclear units with long shutdown and
restart times.
\item \textbf{Voluntary participation of flexible loads} (EVs, heat pumps,
storage, deferrable processes) occurs via user-selected contracts that allow
flexibility to be offered when beneficial, rather than through compulsory
response to spot prices.
\item \textbf{Fair value distribution} is enforced by ensuring that surplus
rents are shared between flexible consumers, essential generators, and
the system operator.
\end{itemize}

The objective in this regime is \emph{zero waste} of low-carbon energy and the
preservation of system stability.

% ---------------------------------------------------------
\subsection{Just Enough Energy: Balanced Regime}
\label{sec:just_enough}

In the \emph{Just Enough} regime, the system operates with comfortable reserves
and without binding network constraints. Dispatch remains secure, and scarcity
management is not required.

The objectives are:

\begin{itemize}[leftmargin=*]
\item \textbf{Stable operation} with smooth, predictable prices.
\item \textbf{Minimal intervention}, with continuous event-based clearing
remaining active but without emergency allocation.
\item \textbf{Price signal integrity}: time and locational price differentials
reflect genuine cost and risk differences.
\item \textbf{Benchmarking fairness}: this regime acts as the reference case for
evaluating fairness in the absence of scarcity distortions.
\end{itemize}

This is the environment in which behavioural design, user experience, and retail
innovation around tariffs, service tiers, and digital contracts are most
relevant.

% ---------------------------------------------------------
\subsection{Too Little Energy: Scarcity Regime}
\label{sec:too_little}

In the \emph{Too Little} regime, available supply is insufficient to serve
unconstrained demand while respecting network and security constraints. This may
arise from weather extremes, simultaneous outages, fuel disruptions, or local
islanding caused by grid congestion.

Classical scarcity pricing and load shedding can produce arbitrary, regressive,
and persistent unfairness. The proposed design instead employs structured,
licit, and digitally enforceable allocation rules.

\begin{itemize}[leftmargin=*]
\item \textbf{Essential blocks} $q^{\mathrm{ess}}_h$ are shielded from
curtailment for as long as physically possible.
\item \textbf{Bounded scarcity pricing} is permitted only within fairness and
vulnerability constraints.
\item \textbf{Controlled curtailment} is coordinated by the
\emph{Fair Play Algorithm}, which manages discretionary load reduction in a
transparent and auditable manner.
\item \textbf{Tightness signals} $p^{\mathrm{tight}}_{n,t}$ communicate the
severity and location of scarcity in real time.
\end{itemize}

Crucially, scarcity allocation is neither arbitrary nor driven solely by price,
but governed by declared contracts, physical constraints, and fairness rules.

% ---------------------------------------------------------
\subsubsection{Allocation Under Scarcity: Prioritised and Fair Sampling}
\label{sec:prioritised_sampling}

To allocate limited electricity fairly during shortage, while preserving
contractual choice, we introduce a two-part mechanism: \emph{service-level
prioritisation} and \emph{fairness-weighted sampling}.

Importantly, the proposed architecture does not require a fixed or discrete
set of service classes. In principle, an \emph{arbitrary (even continuous)
spectrum of reliability levels} may coexist, reflecting heterogeneous user
preferences and policy choices. For analytical clarity and experimental
tractability, this thesis illustrates the mechanism using two representative
service levels only: \emph{basic} and \emph{premium}.

\paragraph{(i) Service-level priority buckets.}
Each bid is associated with a chosen service level $s$, which determines its
\emph{relative priority} under scarcity. Service levels are assigned a
\emph{priority weight} $m_s>0$ that encodes the contractual preference for
retaining access when supply is insufficient.

For example, if a premium service is contractually specified to be twice as
likely to be served as a basic service under shortage, we set
\[
m^{\text{prem}} = 2,
\qquad
m^{\text{basic}} = 1.
\]

These weights do \emph{not} imply guaranteed service. Rather, they determine the
\emph{relative sampling frequency} of different service levels when allocation
must be rationed. Higher service levels receive proportionally greater access,
but remain subject to physical feasibility and fairness constraints.

\paragraph{(ii) Fairness weighting within each bucket.}
Within a given service level, individual bids are not treated uniformly. Each
bid $i$ is assigned a \emph{fairness weight} that reflects its historical access
outcomes, ensuring rotation and protection against systematic deprivation.

We define:
\[
\text{need}_i = 1 - \text{success}_i,
\qquad
\text{fair}_i = (\varepsilon + \text{need}_i)^{\gamma},
\]
where $\text{success}_i \in [0,1]$ denotes the historical fraction of time the bid
has been successfully served, $\varepsilon>0$ prevents zero weights, and
$\gamma \ge 0$ controls the strength of fairness protection.

Higher values of $\gamma$ give proportionally greater priority to historically
under-served participants, promoting rotation over time rather than systematic
exclusion, while preserving contractual service-level preferences.

\paragraph{(iii) Combined sampling logic.}
At each scarcity event, service exceeds supply. The allocation proceeds by
\emph{probabilistically sampling} bids, first across buckets in proportion to
their priority weights, and second within the chosen bucket in proportion to
their fairness weights:

\[
P(\text{choose tier } s)
  = \frac{m_s}{\sum_{s'} m_{s'}}.
\qquad
P(\text{serve bid } i \mid \text{tier } s)
  = \frac{\text{fair}_i}{\sum_{j \in \mathcal{I}_s} \text{fair}_j}.
\]

This produces sequences such as:
\[
\text{premium, premium, basic, premium, premium, premium, basic,}\ldots
\]

with fairness determining which specific bid is selected within each tier.

\paragraph{Interpretation.}
In the two-service-level illustration, \emph{premium requests} retain their
contractual advantage relative to basic requests, but no individual request
within a service level is indefinitely neglected. Over time, requests that have
historically received less access become more likely to be selected, restoring
balance through probabilistic rotation rather than hard quotas or deterministic
scheduling.

This mechanism is:
\emph{contract-respecting, non-arbitrary, digitally enforceable, and auditable}.

% ---------------------------------------------------------
\section{Market Access, Exposure, and Allocation Behaviour}
\label{sec:market_access}

The market architecture must determine not only how scarcity is priced,
but also \emph{who may access, request, or retain energy under
different conditions}. In the proposed design, access is not determined
solely by willingness-to-pay, nor through rigid priority classes, but
through dynamically computed \textbf{contract attributes} and
\textbf{fairness-preserving allocation rules}.

The contract attributes that govern this access logic are encoded in
$\Gamma^{\text{contract}}_r$ and formalised in
Section~\ref{sec:contract_structure}, where each retail product is
represented as an energy access contract with explicit magnitude, timing
sensitivity, and reliability dimensions.

The fairness principles governing exposure and allocation (F1--F4) are
formalised in Chapter~\ref{ch:fairness_definition} and operationalised
by the AMM and Fair Play algorithm in Chapter~\ref{ch:amm}.

\subsection{First layer: Access to the market}
Participants may submit bids or flexibility offers if and only if:
\begin{itemize}[leftmargin=*]
\item they are digitally registered through a supplier or service
entity;
\item their asset or demand is physically measurable and
controllable; and
\item their request is expressed in terms of time, location, energy,
flexibility, power limits, and contract attributes.
\end{itemize}

Under normal conditions, all eligible requests are treated symmetrically
and priced through standard AMM output. No entity is forced to pay
scarcity premiums unless the system is genuinely tight.

\subsection{Second layer: Exposure to scarcity pricing}
When scarcity emerges ($\tilde{\alpha}_{t,n}<1$), bids may be subject to
scarcity-related price uplift \emph{unless} the associated contract
parameters specify protected status under Fairness Condition~F2.
This ensures that scarcity pricing applies primarily to requests that
have contractually accepted exposure to shortage risk, while requests
with protected reliability attributes are shielded in accordance with
their declared service level.

\subsection{Third layer: Access to allocation under shortage}
When scarcity deepens such that $\tilde{\alpha}_{t,n}$ crosses a
critical threshold, the system activates \emph{allocation
governance}, rather than allowing unbounded price escalation. The
allocation process respects:
\begin{enumerate}[leftmargin=*]
\item contract-respecting reliability protection (F2),
\item fairness-weighted priority under shortage (F3),
\item proportional responsibility for system strain (F4), and
\item rotation and historical balance of service provision.
\end{enumerate}

This structured sequence—market access, price exposure, and
allocation—ensures that market-based incentives apply in normal
conditions, while access protection and fairness controls apply under
genuine shortage. This multi-layered logic is implemented in real time
by the Automatic Market Maker (AMM), introduced in
Chapter~\ref{ch:amm}.

\subsection*{Products and service tiers}

At the retail edge, consumers do not interact directly with
$\Gamma^{\text{contract}}_r$, but with named products and service tiers
(subscription offers). Each product corresponds to a particular choice
of energy access contract as defined in Section~\ref{sec:contract_structure}:
\begin{itemize}[leftmargin=*]
\item \emph{Magnitude} is encoded via inclusive volume, peak limits,
and baseline commitments;
\item \emph{Timing sensitivity} is encoded via flexibility options
(e.g.\ “can be shifted within a window”, “off-peak only”),
which determine $\sigma^r$ and the allowed delivery window
$[t^{\mathrm{start}}_r, t^{\mathrm{end}}_r]$;
\item \emph{Reliability requirement} is encoded via service level
(e.g.\ essential-protected, standard, flexible), which maps
into reliability tiers, priority weights, and essential status
within $\Gamma^{\text{contract}}_r$.
\end{itemize}

Thus, a “fully protected” subscription corresponds to a high-reliability,
low-flexibility contract, while a “flexible saver” product corresponds to
greater timing flexibility and a lower reliability claim in shortage,
compensated by lower expected unit cost. The AMM and Fair Play algorithm
see only the underlying contract attributes; they enforce market access,
scarcity exposure, and allocation under shortage according to these
dimensions, rather than informal product labels. This ensures that
retail products are \emph{digitally and formally linked} to the fairness
conditions specified in Chapter~\ref{ch:fairness_definition}.

% ---------------------------------------------------------
\section{Retail Products, Supplier Risk, and Digitalisation Incentives}
\label{sec:retail_digitalisation}

The contract structure in Section~\ref{sec:contract_structure} treats energy
access as a three-dimensional object: magnitude, timing sensitivity, and
reliability. At the retail edge, this can be further interpreted in terms of
three application-facing axes that are visible to consumers and suppliers:
\emph{quality of service}, \emph{power impact}, and \emph{openness to being
flexible}. These dimensions align naturally with the products that suppliers
offer to households and businesses:

\begin{itemize}[leftmargin=*]
\item \textbf{Quality of service (QoS).}
The probability and continuity with which requested service is actually
delivered, particularly under shortage. High-QoS products correspond to
contracts with stronger reliability claims and higher priority within
$\Gamma^{\text{contract}}_r$.

\item \textbf{Power impact.}
The peak and aggregate strain that a customer or asset imposes on the
system---captured by power envelopes, ramp rates, and network impact at
relevant nodes. Products can differ in their allowed peak power, expected
contribution to congestion, and incentives to smooth or reshuffle load.

\item \textbf{Openness to flexibility.}
The extent to which a household or business is willing to expose assets
(EVs, heat pumps, industrial processes) to time-shifting, throttling, or
controlled curtailment in exchange for lower expected cost or improved
priority under shortage. This maps to flexibility parameters such as
$\sigma^r$ and the width of delivery windows.
\end{itemize}

From a supplier perspective, these three axes define a menu of retail products
that can be offered as subscriptions. Each subscription corresponds to a bundle
of QoS, power impact, and flexibility attributes, and is internally implemented
as a set of energy access contracts $(E_r, \bar{P}_r, [t^{\mathrm{start}}_r,
t^{\mathrm{end}}_r], \sigma^r, \Gamma^{\text{contract}}_r)$ processed by the
AMM and Fair Play algorithm.

\subsection{Off-grid demand and ex-post settlement risk}

Crucially, the proposed architecture changes where \emph{bill shock} sits in
the value chain. In legacy retail arrangements, if a household or SME is
off-grid in the informational sense---i.e.\ their demand is not visible in
real time and is settled ex-post using static profiles---then the supplier
must assume a consumption trajectory for that customer. Wholesale settlement,
however, occurs every $\Delta t$ minutes against realised system load and
network conditions. Any discrepancy between assumed and realised demand
manifests as an \emph{ex-post settlement shock} for the supplier.

Under subscription-style products in this thesis, the end customer sees a
largely predictable bill, defined by their chosen QoS, power envelope, and
flexibility offer. The \emph{residual risk} between assumed and realised
wholesale exposure is borne by the supplier, not retroactively pushed onto the
customer via opaque reconciliations. This reallocation of risk has two
important consequences:

\begin{enumerate}[leftmargin=*]
\item Suppliers are directly exposed to the stochastic cost of
\emph{offline} or poorly measured demand; and

\item Suppliers have a clear, contractible upside from \emph{reducing}
that uncertainty through better measurement and control.
\end{enumerate}

In other words, moving to QoS--power--flexibility products with subscription
pricing translates informational gaps into explicit financial risk for
suppliers. Reducing those gaps becomes a core part of their business model.

\subsection{Digitalisation, IoT, and smart meter deployment}

Because wholesale settlement is performed at fine temporal resolution by the
AMM, the variance of a supplier’s net position is tightly linked to the
granularity and reliability of data from its portfolio. A portfolio with
highly instrumented, controllable assets (IoT-enabled devices, responsive
appliances, storage) yields:

\begin{itemize}[leftmargin=*]
\item more accurate forward estimates of $E_r$ and $\bar{P}_r$ for each
contract;
\item real-time visibility of deviations between contracted and realised
usage; and
\item operational levers to adjust demand in response to tightness signals.
\end{itemize}

By contrast, a portfolio dominated by “offline” loads---customers without
smart meters, or assets that cannot be observed or controlled---exposes the
supplier to higher settlement volatility for the same nominal subscription
revenue. The proposed architecture therefore creates a \emph{direct financial
incentive} for suppliers to:

\begin{itemize}[leftmargin=*]
\item deploy smart meters and IoT devices that provide near real-time,
high-resolution measurements;
\item invest in robust device management (firmware updates, diagnostics,
security) to maintain data quality; and
\item work with network operators to improve connectivity and signal
coverage in hard-to-reach areas.
\end{itemize}

Under current regimes, suppliers can—and do—avoid installing smart meters in
locations that are costly or inconvenient (weak signal, access issues, low
volumes). This systematically disadvantages certain consumers and regions. In
the proposed design, these are precisely the locations where missing data
translates into higher wholesale risk. \emph{Not} instrumenting them becomes
expensive. Fairness improves not by imposing uniform technology mandates, but
by aligning suppliers’ financial incentives with comprehensive, non-discriminatory
digitalisation.

\subsection{Device standards, future-proofing, and quantum readiness}

The same risk logic pushes towards higher standards for IoT and metering
devices themselves. If supplier solvency depends on the accuracy and latency
of portfolio data, then devices that:

\begin{itemize}[leftmargin=*]
\item stream data at the highest feasible granularity;
\item support secure, over-the-air firmware upgrades;
\item expose standardised interfaces for control and telemetry; and
\item can adapt to evolving cryptographic and computational requirements
(including post-quantum security),
\end{itemize}

become economically preferable. Suppliers will naturally favour device
manufacturers who provide robust device management platforms and long-term
support, because improved observability and controllability reduce settlement
risk and unlock more attractive QoS--power--flexibility bundles.

In this sense, the architecture is inherently \emph{future-ready}. Device
standards are not fixed once-and-for-all; they are treated as adaptive,
digitally governed objects. As the AMM’s clearing logic, security assumptions,
or settlement resolution evolve, firmware and control interfaces can be
updated over the air. The system is therefore compatible with future advances
in computing, including quantum-safe cryptographic schemes, without requiring
a disruptive physical replacement of metering infrastructure.

\subsection{From structural unfairness to incentive-compatible digitalisation}

Bringing these elements together, the three retail axes---QoS, power impact,
and openness to flexibility---do more than segment the market. They:

\begin{itemize}[leftmargin=*]
\item provide a transparent basis for consumer-facing products that are
directly mapped to formal contract attributes
$\Gamma^{\text{contract}}_r$;
\item relocate bill shocks and settlement volatility from vulnerable
consumers to better-capitalised suppliers, who are structurally
positioned to manage that risk; and
\item create a persistent financial incentive for suppliers, networks,
and device manufacturers to collaborate on deep digitalisation of
demand and flexibility at the edge.
\end{itemize}

Rather than relying on one-off smart meter mandates or technology-specific
subsidies, the proposed market design embeds digitalisation incentives into
the everyday economics of retail supply. The fair, cyber--physical control
logic of the AMM makes granular, trustworthy data a \emph{profit centre}
for suppliers, rather than a regulatory burden (in contrast to the experience
described in Sections~\ref{sec:smart_meter_experience}
and~\ref{sec:digitalisation_fragmentation}), and thereby supports a more
equitable and technologically adaptive energy system. The formal AMM
definition and its implementation of these incentives are developed in
Chapter~\ref{ch:amm}.

% ---------------------------------------------------------
\section{Transition: Why a Market Mechanism Needs a Control System}
\label{sec:transition_to_amm}

A conventional market is a \emph{price discovery system}. It can reveal
who is willing to pay most. It does not guarantee:
\begin{itemize}[leftmargin=*]
\item physically feasible dispatch across real grids;
\item bounded and stable price formation;
\item protection for essential or vulnerable consumers;
\item alignment with future scarcity and network constraints; nor
\item continuity of service across layers (retail, balancing, wholesale).
\end{itemize}

By contrast, the digital market architecture described in this chapter
is not purely a trading arena. It behaves as a \textbf{cyber--physical
control system}: sensing physical and forecast conditions, regulating
price and allocation, enforcing fairness rules, and maintaining
real-time stability.

This structure necessitates a coordinating entity that:
\begin{itemize}[leftmargin=*]
\item synthesises real-time scarcity,
\item broadcasts dynamic buying and selling prices,
\item governs access under shortage, and
\item guarantees bounded, non-chaotic behaviour.
\end{itemize}

That entity is the \textbf{Automatic Market Maker (AMM)} —
a holarchic, stability-preserving, digitally enforceable control layer.

Chapter~\ref{ch:fairness_definition} now formalises the fairness axioms and
operational conditions (A1--A7, F1--F4) that any such controller must satisfy.
Chapter~\ref{ch:amm} then defines the AMM itself and shows how it implements
those conditions in real time, while Chapter~\ref{ch:experiments} uses
the operating regimes and access rules developed here to construct the
simulation scenarios used to evaluate the design.
% ---------------------------------------------------------
% CHAPTER 9 — DEFINITION OF FAIRNESS
% ---------------------------------------------------------
\chapter{Definition of Fairness}
\label{ch:fairness_definition}

\section*{Scope and Perspective}

Fairness in electricity markets is defined as the \textit{principled,
non-arbitrary, and operationally enforceable allocation of cost, benefit,
access, and risk}, derived from physical system roles and measured relative
to essential energy needs, flexibility contribution, and proportional
responsibility. It is not an external or corrective overlay, but a 
\textbf{system design constraint} embedded directly into the market-clearing
mechanism.

We distinguish fairness from related concepts:
\begin{itemize}[leftmargin=*]
    \item \textbf{Equality} allocates the same energy or price to all, irrespective
    of need or contribution.
    \item \textbf{Equity} partially adjusts outcomes based on vulnerability or need.
    \item \textbf{Fairness (this thesis)} requires allocations to reflect 
    \emph{prioritised needs, flexibility contribution, historical access,
    and proportional system value} --- with explainable traceability to
    physical system roles.
\end{itemize}

Fair outcomes in this thesis are defined across three interdependent domains:
\begin{enumerate}[label=(\roman*),leftmargin=1.5em]
  \item \textbf{Consumer pricing, protection, and access}, ensuring that
  contractual service levels, reliability choices, and exposure to scarcity
  are respected under constrained networks, variable supply, and stress events;
  \item \textbf{Supplier remuneration and risk allocation}, ensuring that
  intermediaries are compensated for the services they provide (aggregation,
  hedging, interface, and customer protection) without relying on opaque
  cross-subsidies, hidden uplift, or structural arbitrage; and
  \item \textbf{Generator compensation} aligned with each asset’s
  \emph{system value}, including energy delivery, flexibility, adequacy,
  locational relief, and resilience contribution.
\end{enumerate}

Allocations must be:
\begin{itemize}[leftmargin=*]
  \item physically feasible and security-constrained;
  \item \textbf{priority-order respecting}, ensuring that access is allocated
        in accordance with declared reliability and service-priority attributes,
        subject to physical limits;
  \item proportionate to each participant’s contribution to system stress
        or system relief;
  \item explainable, traceable, auditable, and digitally enforceable;
  \item consistent with individual rationality (no participant is allocated a
        net payment beyond their declared bound $v^{\max}_r$), while ensuring
        that scarcity access is never determined by willingness-to-pay alone
        (F3).
\end{itemize}

This chapter defines the normative fairness \textbf{axioms (A1--A8)} and the
\textit{market-operational fairness conditions} \textbf{(F1--F4)} that 
directly shape the design of the AMM (Chapter~\ref{ch:amm}) and Fair Play
allocation (Chapter~\ref{ch:mathematics}).

% ---------------------------------------------------------
\section{Fairness as a System Design Constraint}

Conventional markets treat fairness as a \textit{post-hoc modification},
addressed through regulation, subsidies, or bill caps. In contrast,
this thesis treats fairness as a \textbf{co-equal constraint} alongside
feasibility and security:

\[
\text{Allocation is valid} \iff
\text{feasible, secure, and fair.}
\]

Thus, fairness is embedded \textit{ex ante} in pricing, allocation, and 
compensation --- not applied after clearance. It is made enforceable through
digital design: embedded in AMM price formation (Chapter~\ref{ch:amm}) and
Fair Play allocation (Chapter~\ref{ch:mathematics}).

\bigskip

% ---------------------------------------------------------
\section{Behavioural Foundations of Fairness}
\label{sec:behavioural_foundations}

Market designs that are technically fair but poorly understood, mistrusted,
or socially opaque fail to achieve legitimacy, regardless of economic merit.
Behavioural research shows that people respond not only to price or cost,
but to perceived \textbf{fairness, protection, trust, and consistency} in
how the system treats them.

\subsection*{Four Preconditions for Trusted Participation}

Empirical flexibility trials, Australian dynamic envelope pilots, and 
behavioural trust studies converge on the following preconditions for
participation in digital energy markets:

\begin{enumerate}[label=(\alph*),leftmargin=2em]
  \item \textbf{Involvement:} Users must understand that system rules reflect
  real needs (e.g., essential protection, medical priority, flexibility reward).
  \item \textbf{Knowledge:} Price, scarcity, and access mechanisms must be
  explainable in human terms.
  \item \textbf{Trust:} Users must believe they will not be exposed to
  uncontrolled risk or arbitrary exclusion.
  \item \textbf{Equity:} Scarcity burdens must be shared proportionally, and
  essential access must be consistently protected.
\end{enumerate}

These conditions parallel Axioms A4--A7 (Stability, Progressivity,
Transparency, Value Alignment) and become embedded in Fairness Operational
Conditions F1--F4.

\subsection*{Implications for Fairness Design}

Therefore, fairness in digital markets must be:

\begin{itemize}[leftmargin=*]
  \item \textbf{Visible}: Participants can see and verify how they are treated.
  \item \textbf{Predictable}: Scarcity exposure is bounded and stable.
  \item \textbf{Reciprocal}: Contributions (e.g., flexibility) lead to clear benefit.
  \item \textbf{Explainable}: Allocations trace back to physical roles or fairness rules.
\end{itemize}

These properties form the behavioural foundation for the operational fairness
conditions (F1--F4), which become enforceable through the AMM.

% ---------------------------------------------------------
\section{System Model (Minimal Notation)}
\label{sec:system_model}

Let $t \in \mathcal{T}$ index time periods, $n \in \mathcal{N}$ nodes,
$g \in \mathcal{G}$ generators, $h \in \mathcal{H}$ households.
The dispatch solves a network-constrained OPF or unit commitment:

\begin{itemize}[leftmargin=*]
  \item $p_{n,t}$ — nodal price; $\lambda_t$ — system multiplier; $\mu_{\ell,t}$ congestion rent;
  \item $q_{h,t}$ — household consumption; $q_h^{\mathrm{ess}}$ essential block;
  \item $x_{g,t}$ — generator dispatch; $A_{g,t}$ availability;
  \item $C_g$ — allowable cost recovery for generator $g$.
\end{itemize}

% ---------------------------------------------------------
\section{Fairness Axioms (Normative)}
\label{sec:fairness_axioms}

\begin{enumerate}[label=\textbf{A\arabic*.},leftmargin=1.6em]
  \item \textbf{Feasibility.} Allocations and prices must arise from a physically
  feasible, security-constrained schedule.

  \item \textbf{Revenue adequacy.} Aggregate payments must cover $C_{\mathrm{allow}}$
  (allowable costs) without persistent deficits or structural windfall rents.

\item \textbf{Causality.} Price differences must reflect underlying physical scarcities (time, location,
    flexibility, adequacy, congestion), rather than willingness-to-pay or other
    purely financial preferences. Although bids include an explicit economic bound
    $v^{\max}_r$, allocations and prices must not be determined by WTP alone.

  \item \textbf{Bill stability for essentials.} Essential demand ($q_h^{\mathrm{ess}}$)
  must not be directly exposed to volatility in scarcity pricing. Volatility should
  fall on discretionary or flexible usage.

  \item \textbf{Progressivity.} Scarcity costs should fall relatively more on
  discretionary, peak, or inflexible usage than on essential consumption, and more
  on those with higher ability to absorb risk.

  \item \textbf{Transparency.} Each bill component must map one-to-one
  to a system role (energy, flexibility, capacity, network, policy);
  consumers must be able to trace who is paid for what.

  \item \textbf{Value alignment.} Generator compensation must reflect system
  value: energy delivered, adequacy, flexibility, and locational relief, rather than
  historic rents or purely financial arbitrage.

  \item \textbf{Fair compensation.} Generator payments must satisfy two joint
  requirements:
  \begin{enumerate}[label=(\alph*),leftmargin=2em]
    \item \textbf{Stable cost recovery for zero-marginal-cost plant:}
    Technologies with negligible fuel cost (wind, nuclear) must recover their
    allowable long-run costs (non-fuel OpEx and amortised CapEx) reliably and
    without exposure to short-run scarcity volatility.
    \item \textbf{Value-proportional remuneration for controllable plant:}
    Controllable technologies (gas, hydro, battery) must receive revenue
    approximately proportional to their marginal contribution to feasibility,
    adequacy, and scarcity relief over time.
  \end{enumerate}
\end{enumerate}

Axiom~A8 ensures that fairness explicitly includes the treatment of
generators: capital-intensive, zero-marginal-cost assets require stability,
while flexible, controllable assets require proportionality to real system
value.

% ---------------------------------------------------------
\section{Operational Fairness Conditions}
\label{sec:operational_fairness}

\begin{enumerate}[label=\textbf{F\arabic*.},leftmargin=1.6em]
  \item \textbf{Fair Rewards}  
  Participants contributing flexibility or system relief should face lower
  expected unit costs:
  \[
    \frac{\partial \mathbb{E}[\text{unit\_cost}_h \mid \sigma^r]}{\partial \sigma^r}
    \le 0,
  \]
  where $\sigma^r$ denotes enrolment in recognised relief or flexibility services.

    \item \textbf{Fair Service Delivery}  
    For consumption designated as high-priority under the contract
    (e.g.\ reliability-critical usage) and \emph{conditional on the system being
    sized and operated to meet declared priority commitments}, exposure to
    tightness-based pricing is bounded.
    
    Formally, let $q_h^{\mathrm{pri}}$ denote the priority-designated portion of
    household $h$’s consumption. Then, under feasible dispatch and security
    constraints, the average price exposure of this priority block satisfies
    \[
      \frac{\sum_{t} p^{\mathrm{tight}}_{n(h),t} \cdot \min\{q_{h,t},\,q_h^{\mathrm{pri}}\}}
           {\sum_{t} p^{\mathrm{base}}_{n(h),t} \cdot q_h^{\mathrm{pri}}}
      \le \epsilon,
    \]
    for a small $\epsilon > 0$ representing tolerated residual exposure arising from
    extreme or unavoidable scarcity events.

    \item \textbf{Fair Access}
    Allocation must not depend solely on willingness-to-pay, even though bids declare
    a maximum admissible value $v^{\max}_r$. During scarcity, essential needs,
    contractual reliability tiers, and historical contribution must take precedence
    over pure bid valuation.

  \item \textbf{Fair Cost Sharing}  
  Users contributing more to stress or congestion should bear more uplift:
  \[
    \mathbb{E}[\phi_{h_1}] \ge \mathbb{E}[\phi_{h_2}]
    \quad \text{if } \kappa_{h_1} > \kappa_{h_2},
  \]
  where $\kappa_h$ is a stress index (e.g.\ contribution to peak or congested
  flows) and $\phi_h$ denotes the uplift or corrective charge.
\end{enumerate}

These rules are not advisory; they become operational through AMM pricing
and Fair Play allocation.

% ---------------------------------------------------------
\section{Literature Foundations for the Fairness Conditions (F1--F4)}
\label{sec:literature_foundations_fairplay}

The operational fairness conditions F1--F4 are not introduced ad hoc, nor do
they arise solely from abstract normative reasoning. They are grounded in four
complementary strands of established literature: (i) behavioural psychology and
energy-transition behaviour \citep{steg2015,werff2016}, (ii) energy justice and
participation in smart-grid contexts \citep{milchram2018energyjustice}, (iii)
empirical evaluation of fairness indicators in energy allocation settings
\citep{DYNGE2025125463}, and (iv) comparative assessment of allocation mechanisms
under collective and local energy schemes
\citep{couraud2025collectivefairness}. Together, these literatures provide
behavioural, justice-based, and operational foundations for embedding fairness
directly into real-time market clearance.

\medskip
\noindent\textbf{Note on terminology.}
Much of the literature frames fairness in terms of transparency,
explainability, procedural legitimacy, or non-discrimination. In this thesis,
these properties are treated as \emph{enabling requirements} for the four
operational fairness conditions: \textbf{Fair Rewards (F1)}, \textbf{Fair Service
Delivery (F2)}, \textbf{Fair Access (F3)}, and \textbf{Fair Cost Sharing (F4)}.
In particular, explainability and traceability are necessary for enforcing fair
cost sharing and access in a non-arbitrary, auditable manner, rather than being
standalone fairness criteria.

\paragraph{Behavioural realism and reciprocity \citep{steg2015,werff2016}.}
Behavioural and psychological studies consistently show that users do not
respond solely to prices or expected costs, but to perceived fairness,
reciprocity, and legitimacy. Steg and van der Werff demonstrate that
participation in flexibility and demand-response programmes depends critically
on whether users believe their contributions are recognised and that essential
access is protected. Participants are willing to tolerate scarcity or higher
prices \emph{when they can see} that burdens are proportionate and rules are
consistently applied.

This directly motivates:
\[
\text{F1 (Fair Rewards)} \quad \text{and} \quad \text{F3 (Fair Access)},
\]
because engagement relies on credible reciprocity and protection against
arbitrary exclusion, rather than on short-run price incentives alone. These
behavioural results also implicitly support F2, insofar as service guarantees
must be predictable to sustain trust.

\paragraph{Energy justice and digital legitimacy \citep{milchram2018energyjustice}.}
Milchram et al.\ argue that fairness in smart energy systems is not limited to
distributional outcomes, but also encompasses procedural justice: transparency,
explainability, protection from arbitrary exclusion, and meaningful participation
in algorithmically mediated markets. Their work shows that digital market designs
lose legitimacy when allocation rules are opaque or when access can be withdrawn
without traceable justification.

In the present framework, these insights underpin:
\[
\text{F2 (Fair Service Delivery)} \quad \text{and} \quad \text{F3 (Fair Access)},
\]
by motivating bounded exposure for priority-designated consumption and explicit
rules governing access under scarcity. Moreover, procedural justice is a
necessary condition for \textbf{F4 (Fair Cost Sharing)}, because cost
responsibility cannot be legitimate unless participants can verify why they are
charged.

\paragraph{Validity and measurability of fairness indicators \citep{DYNGE2025125463}.}
Dynge and Cali demonstrate that commonly used fairness metrics in local
electricity markets can misclassify outcomes if they are not grounded in clear,
operational definitions of justice. Their analysis shows that fairness claims
must be \emph{measurable, auditable, and enforceable}, rather than asserted
post hoc.

This directly supports:
\[
\text{F4 (Fair Cost Sharing)},
\]
by establishing the need for traceable attribution of system stress and uplift,
and also supports \textbf{F2 (Fair Service Delivery)} by highlighting the risks
of unbounded or poorly defined exposure measures. In this thesis, these concerns
are addressed by explicitly defining exposure, stress, and contribution metrics
within the AMM and Fair Play architecture.

\paragraph{Allocation mechanisms and proportional responsibility \citep{couraud2025collectivefairness}.}
Couraud et al.\ compare proportional, equal, and Shapley-based sharing rules in
collective self-consumption and local energy schemes, evaluating them against
axioms such as proportionality, non-discrimination, and transparency. Their
results show that proportional and Shapley-consistent allocations dominate
uniform or purely price-based rules when fairness, stability, and legitimacy are
joint objectives.

These findings provide operational validation for:
\[
\text{F4 (Fair Cost Sharing)},
\]
by supporting proportional responsibility as a fairness principle, and for
\textbf{F3 (Fair Access)}, insofar as allocation under constraint should not be
determined solely by willingness-to-pay.

\bigskip
\noindent
Taken together, these four strands demonstrate that Conditions F1--F4 are not
mere architectural preferences, but are supported by established behavioural,
justice-based, and mechanism-design literature. They justify treating fairness
as an \emph{ex ante system design constraint}, embedded directly into the AMM
pricing logic and the Fair Play allocation mechanism, rather than as an
after-the-fact regulatory correction.

\begin{table}[H]
\centering
\caption{Literature Support for the Fair Play Fairness Conditions (F1--F4).}
\begin{tabular}{lcccc}
\toprule
\textbf{Literature Source} & \textbf{F1} & \textbf{F2} & \textbf{F3} & \textbf{F4} \\
\midrule
Steg (behavioural psychology)         & \checkmark &        & \checkmark &        \\
Milchram et al.\ (energy justice)     & \checkmark & \checkmark & \checkmark & \checkmark \\
Dynge \& Cali (fairness indicators)   &        & \checkmark &        & \checkmark \\
Couraud et al.\ (allocation fairness) & \checkmark &        &        & \checkmark \\
\bottomrule
\end{tabular}
\end{table}

% ---------------------------------------------------------
\section{Generator Compensation Fairness}
\label{sec:generator_compensation}

Define a system value vector for each generator $g$:
\[
v_g = (E_g,\,F_g,\,R_g,\,K_g,\,S_g),
\]
where $E_g$ captures energy delivered, $F_g$ flexibility, $R_g$ adequacy,
$K_g$ congestion or locational relief, and $S_g$ resilience contribution.

Using Shapley-consistent attribution, generator $g$'s total system value
allocation can be written as:
\[
\phi_g = \sum_{S \subseteq \mathcal{G}\setminus\{g\}}
\frac{|S|!(|\mathcal{G}|-|S|-1)!}{|\mathcal{G}|!}
\left(W(S\cup \{g\})-W(S)\right),
\]
where $W(\cdot)$ is a welfare or feasibility functional defined over coalitions
of generators.

Axioms A2 (Revenue Adequacy), A7 (Value Alignment), and A8 (Fair Compensation)
together require that:
\begin{itemize}[leftmargin=*]
  \item zero-marginal-cost units receive stable, cost-based payments over the year; and
  \item controllable units receive scarcity-linked payments proportional to
        their marginal contribution to feasibility and adequacy.
\end{itemize}

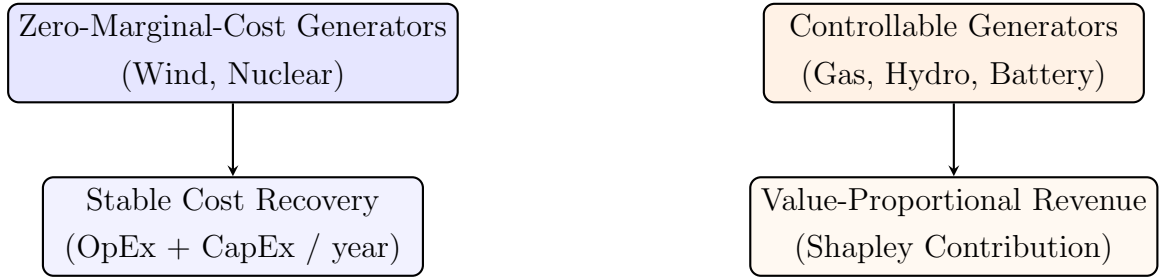
\begin{figure}[H]
\centering
% Requires \usepackage{tikz}
\begin{tikzpicture}[node distance=2.3cm,>=stealth,thick]

\node (zr) [draw, rounded corners, align=center, minimum width=5cm, minimum height=1.2cm, fill=blue!10]
{Zero-Marginal-Cost Generators \\ (Wind, Nuclear)};
\node (cr) [draw, rounded corners, align=center, minimum width=5cm, minimum height=1.2cm, fill=blue!5, below of=zr]
{Stable Cost Recovery \\ (OpEx + CapEx / year)};

\node (tg) [draw, rounded corners, align=center, minimum width=5cm, minimum height=1.2cm, fill=orange!10, right=4cm of zr]
{Controllable Generators \\ (Gas, Hydro, Battery)};
\node (vr) [draw, rounded corners, align=center, minimum width=5cm, minimum height=1.2cm, fill=orange!5, below of=tg]
{Value-Proportional Revenue \\ (Shapley Contribution)};

\draw[->] (zr) -- (cr);
\draw[->] (tg) -- (vr);

\end{tikzpicture}
\caption{Fair compensation: stable cost recovery for zero-marginal-cost plant, and
value-based remuneration for controllable generators.}
\end{figure}

\begin{lemma}[Value-Aligned Compensation is the Unique Fair Allocation Rule]
\label{lem:value_aligned_unique}
Let $C_g$ denote allowable annual cost for zero-marginal-cost generators and
let $\phi_{g,t}$ denote the Shapley-consistent marginal system value of
controllable generator $g$ at time $t$. Any remuneration rule $R_{g,t}$
satisfying Axioms A2, A7, and A8 must take the form:
\[
R_{g,t} =
\begin{cases}
\text{TimeWeight}_t(C_g) & \text{if } g \text{ has zero marginal cost}, \\[0.4em]
\alpha_t \max\{\phi_{g,t},0\} & \text{if } g \text{ is controllable},
\end{cases}
\]
for some non-negative scarcity weight $\alpha_t$ and some normalised
time-weighting scheme $\text{TimeWeight}_t(\cdot)$ that sums to one over~$t$.

Thus, cost-recovery for zero-marginal-cost plant and value-proportional
remuneration for controllable plant is the \emph{unique} structure consistent
with fairness.
\end{lemma}

\begin{proof}[Proof sketch]
Axiom~A2 pins down the total revenue for zero-marginal-cost plant to allowable
annual cost; any deviation would produce either structural deficits or windfall
rents. Axiom~A7 requires that remuneration for controllable resources track
marginal system value, ruling out arbitrary side payments. Axiom~A8 prohibits
schemes that undermine stability for capital-intensive, zero-marginal-cost
assets or break proportionality for controllable assets. These three axioms
together eliminate all schemes except those differing by a common scarcity
multiplier $\alpha_t$ and a normalised time-weighting of $C_g$ over~$t$.
\end{proof}

\subsection*{Implications for Policy, Investment, and Market Governance}

The fair compensation structure above has four major system-level implications:

\begin{itemize}[leftmargin=*]
    \item \textbf{Investment adequacy.}  
    Stable long-run cost recovery for wind and nuclear provides the certainty
    required to scale capital-intensive, zero-marginal-cost capacity.

    \item \textbf{Efficient scarcity response.}  
    Value-proportional remuneration ensures controllable generators respond to
    genuine scarcity rather than regulatory artefacts or gaming opportunities.

    \item \textbf{Technology-neutral fairness.}  
    Revenues depend on system value and cost structure, not legacy categorizations
    or arbitrary distinctions between ``energy'' and ``capacity'' markets.

    \item \textbf{Digital enforceability.}  
    The rules are operationalised deterministically by the AMM 
    (Chapter~\ref{ch:amm}), making them transparent, auditable, and resistant
    to discretionary manipulation.
\end{itemize}

Fair compensation therefore forms a bridge between fairness as a normative
constraint and the operational architecture implemented by the Automatic Market
Maker.

% ---------------------------------------------------------
\section{Reliability as an Allocation Claim Under Scarcity}
\label{sec:reliability_as_claim}

In conventional electricity markets, reliability is assumed to be universal and unconditional: every consumer is implicitly entitled to continuous access regardless of system stress or local network conditions. This assumption obscures the fact that reliability is \emph{not a free good}, but a scarce service that depends on system capacity, local constraints, and the collective contribution of others.

In this thesis, reliability is treated not as an implicit entitlement, but as an \emph{explicit contractual claim} that must be allocated fairly when the system cannot serve all users simultaneously. This reframing connects reliability directly to Fairness Condition~F3 (Fair Access), which prohibits allocation based solely on willingness-to-pay or arbitrary rationing.

\vspace{0.7em}
\noindent
In the proposed architecture, each participant declares a reliability requirement as part of their service contract, yielding a three-dimensional characterisation of energy access:
\[
\text{Energy Access Contract}
=
\{\text{Magnitude},\ \text{Timing Sensitivity},\ \text{Reliability Requirement}\}.
\]

\begin{itemize}[leftmargin=*]
\item \textbf{Magnitude} reflects how much energy is required.
\item \textbf{Timing Sensitivity} reflects whether consumption can be shifted or deferred (flexibility).
\item \textbf{Reliability Requirement} reflects whether the user is \emph{entitled to be served during scarcity}.
\end{itemize}

Crucially, the third dimension activates only when the system is constrained (energy shortage, network congestion, voltage instability). In such conditions, the allocation mechanism must differentiate between:

\begin{enumerate}[label=(\alph*),leftmargin=2em]
\item protected essential usage ($q_h^{\mathrm{ess}}$),
\item declared reliability commitments,
\item flexibility-enrolled devices willing to defer,
\item non-essential or opportunistic consumption.
\end{enumerate}

Thus, reliability is not simply a premium service level or insurance add-on; it is a \textbf{claim on scarce capacity}, which must be allocated \emph{fairly, traceably, and proportionately} in real time.

\subsection*{Reliability, Flexibility, and System Contribution}

Participants who allow their devices to be enrolled in flexibility services (e.g., demand response, voltage support, congestion management) do not merely receive lower prices; they also \emph{earn allocation priority} during future scarcity periods. Their prior contribution to maintaining system reliability (e.g., shifting EV charging, modulating heat pumps, absorbing solar surplus) is traced digitally and becomes part of their allocation claim.

Let the priority weight for household $h$ be
\[
\text{PriorityWeight}_h
=
f\bigl(
\sigma_h^{\mathrm{flex}},
\ \text{historical relief}_h,
\ \text{reliability tier}_h,
\ q_h^{\mathrm{ess}}
\bigr),
\]
where:
\begin{itemize}[leftmargin=*]
\item $\sigma_h^{\mathrm{flex}}$ is flexibility enrolment status;
\item $\text{historical relief}_h$ records how a household/device helped in past stress events;
\item $\text{reliability tier}_h$ reflects their declared level of QoS entitlement;
\item $q_h^{\mathrm{ess}}$ ensures minimum access remains protected (F2).
\end{itemize}

This supports Fairness Conditions F1 (reciprocity), F2 (essential protection), and F3 (Fair Access).
It also directly embeds the principle:

\begin{quote}
\textit{Those who help maintain reliability earn reliability.}
\end{quote}

\subsection*{Fair Play as the Allocation Mechanism for Reliability}

When scarcity arises, traditional markets either apply uniform rationing or let willingness-to-pay decide access—both violate F2 and F3. The Fair Play mechanism instead performs an ex ante declared, real-time allocation of scarce energy, using:

\begin{itemize}[leftmargin=*]
\item Contractual reliability tiers (declared ex ante),
\item Flexibility enrolment and historical contribution,
\item Essential protection for minimum human energy needs,
\item Proportionality and rotation under prolonged shortage,
\item Explainable traceability to system roles.
\end{itemize}

This moves reliability from being an unpriced assumption to a governed, fairly allocated right.

\subsection*{Non-Coercive Transition: Reliability Without Compulsory Enrolment}

Finally, this model does not mandate device enrolment or digital participation. Instead, it establishes a non-coercive transition path:
\[
\text{Legacy Customer}
\;\rightarrow\;
\text{Smart Subscriber}
\;\rightarrow\;
\text{Enrolled Contributory Participant}.
\]

\begin{itemize}[leftmargin=*]
\item Legacy customers retain implicit 100\% QoS (supplier-backed), but do not earn priority in shortage.
\item Smart subscribers may accept limited QoS variation, in return for lower expected costs.
\item Fully enrolled devices may provide flexibility, voltage support, or constraint relief — earning both lower costs and higher reliability priority through Fair Play.
\end{itemize}

This supports behavioural trust conditions (in Section~\ref{sec:behavioural_foundations}) because reliability becomes:
\begin{enumerate}[label=(\roman*),leftmargin=2em]
\item \textbf{Visible} — consumers understand their reliability status.
\item \textbf{Predictable} — scarcity exposure is bounded and declared.
\item \textbf{Reciprocal} — flexibility earns not just money, but service priority.
\item \textbf{Explainable} — allocation under shortage is traceable to declared rules.
\end{enumerate}

\vspace{0.5em}

\noindent In summary, reliability is reframed as a \emph{declared, measurable, and fairly allocatable claim}, governed not only by price or capacity, but by \textbf{contractual commitment, contribution, essential protection, and digital fairness}.
This provides the conceptual link from fairness axioms (A1–A8) and operational fairness conditions (F1–F4) to the design of the Fair Play allocation controller in Chapter~\ref{ch:mathematics}.

% ---------------------------------------------------------
\subsection{Fair Allocation of Generator Payments}
\label{subsec:fair_generator_payments}

Fairness for generators requires that each asset is compensated in a manner
that is \emph{non-arbitrary, proportionate to its system value, and consistent
with long-run adequacy}. This thesis adopts two fairness principles for
generator remuneration, derived from Axioms~A2 (Revenue Adequacy), A7
(Value Alignment), and A8 (Fair Compensation).

\paragraph{(1) Stable cost-recovery for zero-marginal-cost plant.}
Wind and nuclear units are essential for decarbonisation and have near-zero
short-run marginal costs. Exposing them to volatility or scarcity-based
competition would violate A2 (Revenue Adequacy) and undermine investment
stability. Fairness therefore requires:

\begin{itemize}[leftmargin=*]
    \item guaranteed annual recovery of non-fuel OpEx and amortised CapEx; and
    \item distribution of this revenue in time according to how much these
          generators contribute to system feasibility, adequacy, and resilience.
\end{itemize}

This ensures that zero-marginal-cost generators are not penalised for fuel-free
operation, while still aligning their revenue with their real system value.

\paragraph{(2) Value-proportional revenues for controllable generators.}
Gas, hydro, battery and other controllable units provide marginal flexibility,
rampability, and adequacy. Their contribution varies significantly over time
and location. Fairness therefore requires that their remuneration be
\emph{proportional to their marginal contribution to keeping the system
feasible}. In this thesis, this contribution is measured using Shapley-consistent
marginal value:

\[
\phi_{g,t} \quad\text{captures the marginal system value of generator } g
\text{ at time } t,
\]

reflecting adequacy, congestion relief, flexibility, and resilience. A fair
allocation must therefore satisfy:

\[
R_{g,t} \;\propto\; \max\{\phi_{g,t}, 0\},
\]

ensuring that:

\begin{itemize}[leftmargin=*]
    \item generators are rewarded when the system genuinely needs them;
    \item rewards fall when their presence does not expand the feasible region;
    \item no technology receives rents unrelated to its system contribution.
\end{itemize}

\paragraph{(3) Separation of normative fairness from operational mechanism.}
These principles define \emph{what} fairness requires. Their operational
realisation—how revenues are shaped over time, how pots are sized, and how
Shapley weights are normalised—is implemented by the Automatic Market Maker
(Chapter~\ref{ch:amm}). The AMM ensures that:

\begin{itemize}[leftmargin=*]
    \item zero-marginal-cost units receive stable, cost-reflective payments;
    \item controllable units share scarcity revenues proportionally to
          real-time marginal value; and
    \item all payments remain explainable, auditable, and digitally enforceable.
\end{itemize}

This preserves the core separation between normative fairness constraints
(defined in this chapter) and the digital mechanism that enforces them
(Chapter~\ref{ch:amm}).

% ---------------------------------------------------------
\section{Preview of Fair Play}
\label{sec:fairplay_preview}

Under $\alpha<1$, operational fairness (F3) prohibits allocation solely
by willingness-to-pay. The Fair Play Algorithm (Chapter~\ref{ch:mathematics})
ensures:
\begin{itemize}[leftmargin=*]
  \item Essential-first protection;
  \item Contract- and flexibility-based prioritisation;
  \item Rotation and service history;
  \item Proportional curtailment.
\end{itemize}

% ---------------------------------------------------------
\section*{Conclusion and Link to AMM}

This chapter has defined fairness at three levels:
\begin{itemize}[leftmargin=*]
  \item Normative fairness axioms (A1--A8),
  \item Operational fairness conditions (F1--F4),
  \item Testable fairness metrics (C1--C6, G1--G5).
\end{itemize}

Fairness, in this thesis, is therefore not evaluated after the fact,
but embedded \emph{ex ante} into the market design. The next chapter
introduces the \textbf{Automatic Market Maker (AMM)} --- a digital scarcity
and allocation controller that operationalises fairness in real time
through pricing, access, generator compensation, and proportional burden-sharing.

% ---------------------------------------------------------
% CHAPTER 10 — The Automatic Market Maker (AMM)
% ---------------------------------------------------------

\chapter{The Automatic Market Maker (AMM)}
\label{ch:amm}

This chapter introduces the Automatic Market Maker (AMM) as the core
digital scarcity-control and allocation mechanism that coordinates pricing,
access, and proportional burden-sharing under the proposed market architecture.
While Chapter~\ref{ch:market_scenarios} established the structural and digital
layers of the market (retail, wholesale, balancing, and digital assurance),
and Chapter~\ref{ch:fairness_definition} defined fairness as a \emph{system
design constraint}, this chapter explains \emph{how fairness is enforced,
in real time, through scarcity inference, price formation, allocation, and
protected access guarantees}.

Importantly, although flexible requests may declare an admissible economic
bound $v^{\max}_r$ (defined in Section~\ref{sec:bidding_model}), the AMM does not
use willingness-to-pay for allocation or prioritisation. The bound serves only
as an individual-rationality constraint; scarcity inference, pricing, and
allocation remain independent of financial willingness-to-pay, in accordance
with Fairness Conditions~F1--F4.

The AMM is not merely a price calculation engine. It is a
\textbf{holarchic cyber--physical controller and fairness enforcer},
capable of:

\begin{itemize}[leftmargin=*]
    \item synthesising instantaneous, forecast, and network-based scarcity;
    \item broadcasting explainable, bounded, tightness- and deficit-based
          prices (buying and selling);
    \item coordinating disciplined, non-arbitrary allocation under shortage,
          including essential protection and proportional curtailment;
    \item ensuring structural, behavioural, and control-theoretic stability;
    \item preserving fairness principles (A1–A7) and operational conditions (F1–F4)
          defined in Chapter~\ref{ch:fairness_definition};
    \item maintaining digital legitimacy and trusted participation through
          transparency, bounded exposure, and predictable rules.
\end{itemize}

We formalise its design, show how it integrates with digital assurance and
Fair Play allocation, and interpret it as a feedback-control system with
guaranteed bounded-input, bounded-output (BIBO) stability and behavioural
predictability.

% ---------------------------------------------------------
\section{Definition and Holarchic Architecture of the AMM}
\label{sec:amm_design}

The AMM replaces traditional bid-based spot clearing with a continuous,
explainable, and digitally enforceable scarcity-control layer that transforms
physical scarcity into \emph{prices, access rules, and proportional allocation}
— rather than allowing outcomes to emerge solely from willingness-to-pay or
bid dominance. It operates as a \textbf{holarchic controller}: simultaneously
time-aware (through forecast scarcity), space-aware (via nodal and congestion
signals), and hierarchy-aware (across zones, clusters, regions, and system).
This enables the AMM to generate tightness- and deficit-based price signals,
prioritise essential access, coordinate flexibility, and enforce fairness
constraints in real time.

\subsection{Holarchic architecture}

The AMM maintains scarcity indicators at multiple levels:
\[
  \alpha^{\text{cluster}}_{t},\qquad
  \alpha^{\text{zone}}_{t},\qquad
  \alpha^{\text{regional}}_{t},\qquad
  \alpha^{\text{system}}_{t},
\]
each incorporating information from the tiers below and influencing price
formation at that level. At a given node $n$, these effects are combined in
a synthesised scarcity indicator:
\[
  \tilde{\alpha}_{t,n}
  =
  \alpha_{t,n}^{\text{instant}}
  \cdot
  \alpha_{t,n}^{\text{forecast}}
  \cdot
  \alpha_{t,n}^{\text{network}},
\]
where:
\begin{itemize}[leftmargin=*]
  \item $\alpha_{t,n}^{\text{instant}}$ is the real-time ratio of flexible
        supply to flexible demand;
  \item $\alpha_{t,n}^{\text{forecast}}$ predicts imbalances over flexible
        appliance windows (EVs, heating, storage); and
  \item $\alpha_{t,n}^{\text{network}}$ reflects congestion, voltage headroom,
        and nodal binding constraints.
\end{itemize}

This creates a spatio-temporal awareness of scarcity without requiring full
nodal LMP exposure at the retail edge. In the next subsections, we make this
notion more concrete by introducing an explicit \emph{deficit} variable and
showing how buy and sell prices respond to it.

\subsubsection{Inertia-aware scarcity and digital stability margin}
\label{subsec:amm_inertia_aware}

In Section~\ref{sec:inertia_operability} we observed that as synchronous
machines retire, the system loses passive rotational inertia and becomes
increasingly \emph{operability-tight}. The rate at which frequency deviates
following a disturbance (RoCoF) increases, corrective actions must occur more
quickly, and stability becomes a digitally coordinated task rather than an
incidental by-product of heavy rotating machines.

To accommodate this shift, we extend the scarcity representation used by the
AMM to include a \emph{digital stability margin}, producing a four-factor
scarcity structure:
\[
  \tilde{\alpha}_{t,n}
  =
  \alpha_{t,n}^{\mathrm{instant}}
  \cdot
  \alpha_{t,n}^{\mathrm{forecast}}
  \cdot
  \alpha_{t,n}^{\mathrm{network}}
  \cdot
  \alpha_{t,n}^{\mathrm{stability}} ,
\]
where $\alpha_{t,n}^{\mathrm{stability}} \in (0,1]$ denotes the
tightness-of-stability margin, incorporating real-time or forecast measures of:
\begin{itemize}[leftmargin=*]
  \item inertial headroom (mechanical or synthetic),
  \item rate-of-change-of-frequency constraint proximity,
  \item available fast-frequency response (FFR) and grid-forming capacity,
  \item voltage stability indicators or dynamic line ratings.
\end{itemize}

When $\alpha_{t,n}^{\mathrm{stability}}$ is close to 1, stability margins are
ample, and digital resources are not urgently required for frequency or
voltage support. As $\alpha_{t,n}^{\mathrm{stability}}$ decreases, the system
requires faster or more substantial corrective response. The AMM responds by:
\begin{itemize}[leftmargin=*]
  \item increasing $BP_{t,n}$ and $SP_{t,n}$ during low-inertia (tight stability)
        periods, making stability-providing actions more valuable;
  \item allocating proportional ``stability burden'' across flexible providers
        in line with Fairness Conditions F3--F4 (proportional responsibility
        and non-arbitrary rotation);
  \item recognising synthetic inertia, demand-side response and battery
        activation not as external services, but as core scarcity-mitigating
        actions.
\end{itemize}

Thus, the AMM does not treat inertia or stability as exogenous engineering
constraints, nor as separate ancillary products. Instead, stability
contributes directly to scarcity and therefore to price, access and
allocation—maintaining interpretability, fairness, and control-theoretic
stability.

\subsection{Instantaneous scarcity: supply--demand balance}

At each time $t$ and node $n$ we distinguish total supply capability,
non-digitally controllable demand, and digitally controllable demand:
\[
\begin{aligned}
  S_{t,n}^{T} &\quad \text{(total local supply capability)},\\
  C^B_{t,n}   &\quad \text{(non-digitally controllable demand)},\\
  C_{t,n}^{\mathrm{fa}} &\quad \text{(digitally controllable demand at $t$)}.
\end{aligned}
\]

The flexible-available supply envelope is
\[
  S_{t,n}^{\mathrm{fa}} = S^T_{t,n} - C^B_{t,n},
\]
and the usual instantaneous tightness ratio is
\[
  \alpha_{t,n}^{\text{instant}}
  =
  \min\left\{1,\;
  \frac{S_{t,n}^{\mathrm{fa}}}{C_{t,n}^{\mathrm{fa}}}\right\}.
\]

To connect this to price formation, it is convenient to define an explicit
\emph{instantaneous deficit}:
\[
  \Delta_{t,n}^{\mathrm{inst}}
  :=
  C_{t,n}^{\mathrm{fa}} - S_{t,n}^{\mathrm{fa}}.
\]
Then:
\[
  \Delta_{t,n}^{\mathrm{inst}} \le 0
  \quad\Rightarrow\quad
  \text{no flexible shortage (all requested flexible demand can be met),}
\]
\[
  \Delta_{t,n}^{\mathrm{inst}} > 0
  \quad\Rightarrow\quad
  \text{instantaneous flexible shortage.}
\]

The tightness ratio is a normalised representation of the same information:
\[
  \alpha_{t,n}^{\text{instant}} 
  =
  \min\left\{1,\;
  1 - \frac{\max(0,\Delta_{t,n}^{\mathrm{inst}})}{C_{t,n}^{\mathrm{fa}}}
  \right\}.
\]
Intuitively, $\alpha_{t,n}^{\text{instant}}\approx 1$ corresponds to
$\Delta_{t,n}^{\mathrm{inst}}\le 0$ (no shortage), while
$\alpha_{t,n}^{\text{instant}}<1$ corresponds to a positive deficit.

\subsection{Forecast scarcity: time-aware AMM}

Flexible appliances declare look-ahead windows to the AMM. Using predicted
loads, weather, generation forecasts, and historic usage patterns, the AMM
computes a horizon-based scarcity ratio:
\[
  \alpha_{t,n}^{\text{forecast}}
  =
  \frac{
    \sum_{\tau=t+1}^{t+H} S_{\tau,n}^{\mathrm{fa}}
  }{
    \sum_{\tau=t+1}^{t+H} C_{\tau,n}^{\mathrm{fa}}
  }.
\]
This encourages appliances to shift away from future scarcity and towards
periods of surplus, satisfying Fairness Condition F1
(behavioural reward for flexibility) defined in
Chapter~\ref{ch:fairness_definition}.

\subsection{Network scarcity: locational awareness}

Real-time grid constraints (thermal overload, voltage deviations, line flows)
are translated into a network scarcity factor:
\[
  \alpha_{t,n}^{\text{network}}
  =
  \exp(-\theta_n \cdot \Delta V_{t,n})
  \cdot
  \exp(-\phi_n \cdot \text{cong}_{t,n}),
\]
where $\Delta V_{t,n}$ is normalised voltage deviation, and $\text{cong}_{t,n}$
is a congestion tightness index. Parameters $\theta_n,\phi_n>0$ scale the
sensitivity of scarcity to local voltage and congestion.

This encodes local scarcity while preserving consumer accessibility and
compatibility with existing network operations.

\subsection{Deficit-based AMM price functions}
\label{subsec:amm_price_functions}

The AMM maps scarcity into buying and selling prices, but operationally it
is clearer to work with a \emph{deficit} between requested demand and
available supply.

For each node $n$ and time $t$ we define a (possibly forecast-augmented)
deficit:
\[
  \Delta_{t,n}
  :=
  C^{\mathrm{req}}_{t,n} - S^{\mathrm{avail}}_{t,n},
\]
where:
\begin{itemize}[leftmargin=*]
  \item $C^{\mathrm{req}}_{t,n}$ is the total requested demand to be scheduled
        at $(t,n)$ (including flexible requests with time windows that include $t$);
  \item $S^{\mathrm{avail}}_{t,n}$ is the available supply envelope at $(t,n)$
        (generation, storage discharge, imports) consistent with network and
        security constraints.
\end{itemize}

The deficit $\Delta_{t,n}$ is tightly coupled to the tightness ratio
$\tilde{\alpha}_{t,n}$: when $\Delta_{t,n} \le 0$, we have
$\tilde{\alpha}_{t,n} \approx 1$ (no effective shortage); when
$\Delta_{t,n}>0$, we have $\tilde{\alpha}_{t,n}<1$ (shortage). We use
$\Delta_{t,n}$ to explain price design and $\tilde{\alpha}_{t,n}$ to
summarise overall tightness.

We distinguish two regimes:

\paragraph{(1) No shortage: $\Delta_{t,n} \le 0$.}

When all requested demand can be met (no unserved requested demand at node $n$),
there is no marginal value in procuring additional energy at $(t,n)$ from the
perspective of scarcity. The AMM therefore sets the \emph{scarcity component}
of the buy price to zero:
\[
  BP^{\mathrm{scar}}_{t,n} = 0
  \quad\text{whenever}\quad
  \Delta_{t,n} \le 0.
\]

In this regime, consumers pay only the base, non-scarcity components of the
tariff (network charges, policy levies, subscription fees), and flexible
requests face no penalty for being scheduled at $(t,n)$. Fair Play shortage
discipline is inactive and there is no need to ration access.

\paragraph{(2) Shortage: $\Delta_{t,n} > 0$.}

When requested demand exceeds available supply, the system enters a
\emph{shortage regime}. In this case:
\begin{itemize}[leftmargin=*]
  \item the \textbf{buy price} $BP_{t,n}$ must increase with the deficit to
        discourage additional demand and signal scarcity to flexible devices;
  \item the \textbf{sell price} $SP_{t,n}$ must also increase with the same
        deficit to attract additional supply (e.g.\ storage discharge,
        behind-the-meter resources) into the relevant time window;
  \item the \textbf{Fair Play} allocation rule is activated, because there is
        now a non-zero set of flexible requests that cannot all be served.
\end{itemize}

These price responses depend only on physical and forecast scarcity signals
(deficit, stability, voltage, network tightness), not on participants’
declared willingness-to-pay $v^{\max}_r$, which acts solely as a cap on their
exposure and not as a determinant of priority or allocation.

We represent this as a family of increasing functions, parameterised by node,
time, regulatory preferences, and (optionally) stability tightness:
\[
  BP_{t,n}
  =
  BP^{\mathrm{base}}_{t,n}
  +
  F^{\mathrm{energy}}_{t,n}\bigl(\Delta_{t,n}\bigr)
  +
  F^{\mathrm{stab}}_{t,n}\bigl(1-\alpha^{\mathrm{stability}}_{t,n}\bigr),
\]
\[
  SP_{t,n}
  =
  SP^{\mathrm{base}}_{t,n}
  +
  H^{\mathrm{energy}}_{t,n}\bigl(\Delta_{t,n}\bigr)
  +
  H^{\mathrm{stab}}_{t,n}\bigl(1-\alpha^{\mathrm{stability}}_{t,n}\bigr),
\]
with
\[
  F^{\mathrm{energy}}_{t,n}(0) = 0,\quad
  H^{\mathrm{energy}}_{t,n}(0) = 0,\quad
  F^{\mathrm{stab}}_{t,n}(0) = 0,\quad
  H^{\mathrm{stab}}_{t,n}(0) = 0,
\]
and
\[
  \frac{\partial F^{\mathrm{energy}}_{t,n}}{\partial \Delta} > 0,\quad
  \frac{\partial H^{\mathrm{energy}}_{t,n}}{\partial \Delta} > 0,\quad
  \frac{\partial F^{\mathrm{stab}}_{t,n}}{\partial (1-\alpha^{\mathrm{stability}})} \ge 0,\quad
  \frac{\partial H^{\mathrm{stab}}_{t,n}}{\partial (1-\alpha^{\mathrm{stability}})} > 0
  \quad \text{for positive arguments.}
\]

Here $F^{\mathrm{energy}}_{t,n}, H^{\mathrm{energy}}_{t,n}$ encode the
\emph{energy scarcity} response as before (driven by the deficit
$\Delta_{t,n}$), while $F^{\mathrm{stab}}_{t,n}, H^{\mathrm{stab}}_{t,n}$
encode an additional uplift that depends on the tightness of the stability
margin via $1-\alpha^{\mathrm{stability}}_{t,n}$. In line with the physical
role of different assets, one can choose $H^{\mathrm{stab}}_{t,n}$ to be more
sensitive than $F^{\mathrm{stab}}_{t,n}$, so that:
\begin{itemize}[leftmargin=*]
  \item consumers see only a mild premium for drawing energy when stability is tight;
  \item fast-acting resources (batteries, synthetic inertia, fast frequency response)
        see a strong uplift in $SP_{t,n}$ when $\alpha^{\mathrm{stability}}_{t,n}$ is low,
        and therefore have a powerful incentive to activate.
\end{itemize}

\paragraph{Stability-driven activation of fast resources.}

In practice, fast-responding assets (batteries, supercapacitors, grid-forming
inverters) will often be controlled by local algorithms that monitor the
export price $SP_{t,n}$. When $\alpha^{\mathrm{stability}}_{t,n}$ drops
(low inertia, tight stability margin), the term
$H^{\mathrm{stab}}_{t,n}\bigl(1-\alpha^{\mathrm{stability}}_{t,n}\bigr)$
raises $SP_{t,n}$ even if the energy deficit $\Delta_{t,n}$ is modest.
From the perspective of a local controller, this appears as a high,
time-localised sell price; rational policies such as
``export when $SP_{t,n}$ exceeds threshold'' therefore cause batteries
and other fast resources to \emph{automatically activate} precisely when
the system is short of stability, not just short of energy.

In this way, the AMM treats stability as a first-class scarcity dimension:
``digital inertia'' and fast frequency response are remunerated through
the same scarcity-control law as energy, rather than via a separate and
opaque ancillary-services layer.

For assets $r$ that differ in their stability contribution (ramp rate, response
time, grid-forming capability), the stability uplift can be made
resource-specific:
\[
  SP_{t,n}^r
  =
  SP^{\mathrm{base}}_{t,n}
  +
  H^{\mathrm{energy}}_{t,n}\bigl(\Delta_{t,n}\bigr)
  +
  \kappa_r\, H^{\mathrm{stab}}_{t,n}\bigl(1-\alpha^{\mathrm{stability}}_{t,n}\bigr),
\]
where $\kappa_r \ge 0$ is a digital ``stability capability'' label.
Fast, grid-forming batteries have $\kappa_r$ close to 1; slow or
non-contributory resources have $\kappa_r \approx 0$. This preserves the
same AMM structure while allowing stability-sensitive remuneration to
discriminate between assets based on their physical role.

The functions $F_{t,n}$ and $H_{t,n}$ can be chosen from a family of shapes:
\begin{itemize}[leftmargin=*]
  \item linear (proportional to the deficit),
  \item quadratic or higher-order (penalising large deficits more strongly),
  \item asymptotic (approaching a hard cap as $\Delta$ grows),
  \item exponential (very sharp response near a critical deficit threshold),
\end{itemize}
subject to stability and boundedness requirements
(Chapter~\ref{ch:mathematics}). Their form is partly behavioural and can be
calibrated empirically: different systems or policy regimes may choose
different tightness functions while preserving the monotonicity requirement.

Operationally, the same logic can still be expressed in terms of the
tightness ratio $\tilde{\alpha}_{t,n}$:
when $\Delta_{t,n} \le 0$ (no shortage, $\tilde{\alpha}_{t,n}\approx 1$),
the scarcity component is zero; when $\Delta_{t,n}>0$
(shortage, $\tilde{\alpha}_{t,n}<1$), both buy and sell prices rise
monotonically with the deficit.

\paragraph{Forward-looking deficits and pre-emptive action.}

Because flexible devices and generators submit \emph{time windows}, the AMM
operates on a forward-looking deficit profile:
\[
  \Delta^{\mathrm{fwd}}_{\tau,n} 
  :=
  C^{\mathrm{req}}_{\tau,n} - S^{\mathrm{avail}}_{\tau,n},
  \qquad \tau \in [t, t+H],
\]
and computes corresponding price paths $BP_{\tau,n}, SP_{\tau,n}$ over the
horizon. This enables:
\begin{itemize}[leftmargin=*]
  \item pre-emptive attraction of additional supply (e.g.\ behind-the-meter
        batteries) \emph{before} a physical shortfall manifests;
  \item early activation of flexibility where future deficits are forecast
        to be large, reducing the need for emergency interventions.
\end{itemize}

\paragraph{Rewarding flexibility via price-minimising scheduling.}

Flexible bids include an admissible time window
$[\underline{t}_i,\overline{t}_i]$ and a fixed energy requirement $E_i$.
Given a price path $BP_{\tau,n}$ over that window, the Fair Play allocation
mechanism schedules each flexible request into the cheapest feasible slot,
subject to:
\begin{itemize}[leftmargin=*]
  \item local capacity and network constraints,
  \item fairness weights and historic service ratios,
  \item the device's own power and timing constraints.
\end{itemize}

Because the buy price is lowest at times where the deficit
$\Delta_{\tau,n}$ is smallest (i.e.\ where supply is most abundant),
a request with a wide flexibility band $\bigl(\overline{t}_i -
\underline{t}_i\bigr)$ is more likely to be scheduled into those
low-deficit (low-price) periods. Flexibility is therefore \emph{rewarded by
construction}: devices that are willing to move in time receive the cheapest
available slot compatible with their constraints and their Fair Play priority.

This logic extends naturally to richer bid types where:
\begin{itemize}[leftmargin=*]
  \item energy may be delivered in multiple digital “blocks” over a window,
        rather than as a single contiguous run;
  \item power profiles may be ramped or shaped, subject to local constraints.
\end{itemize}
Such bids can be represented as sequences of digital blocks of energy over
discrete time steps, and priced according to the same deficit-based rules
(with references to the emerging literature on digital block markets to be
inserted by the author).

\paragraph{Natural self-correction via buy–sell symmetry.}

At the level of a node $n$, the buy price $BP_{t,n}$ can be interpreted as
the \emph{import cost} (what a household or aggregator pays to consume from
the grid), while the sell price $SP_{t,n}$ is the \emph{export reward}
(what a storage device or generator is paid to inject into the grid). Because
both $BP_{t,n}$ and $SP_{t,n}$ are driven by the \emph{same} deficit
$\Delta_{t,n}$, the AMM exhibits a natural self-corrective behaviour:

\begin{itemize}[leftmargin=*]
  \item if $\Delta_{t,n}$ is large (shortage), then $BP_{t,n}$ increases,
        discouraging imports and encouraging consumers to shift or reduce
        demand; at the same time $SP_{t,n}$ increases, encouraging local
        export (storage discharge, generation);
  \item these responses both act to \emph{reduce} $\Delta_{t,n}$ in subsequent
        steps by lowering requested demand and increasing available supply;
  \item as $\Delta_{t,n}$ shrinks, both $BP_{t,n}$ and $SP_{t,n}$ fall back
        towards their base values, removing the incentive for overshoot.
\end{itemize}

Because the buy and sell prices are tied to a single underlying deficit
signal (rather than set independently by separate markets), there is no
structural incentive loop that can drive unbounded divergence. Instead,
the AMM’s price structure embeds a negative feedback: high deficits
cause high prices, which elicit behaviours that reduce the deficit,
lowering prices again. This symmetry is a key source of the AMM's
natural stability, and underpins the BIBO and Lyapunov-like arguments
developed in Section~\ref{sec:amm_stability}.

% ---------------------------------------------------------
\subsection{Participant-facing price under holarchic scarcity}
\label{subsec:participant_price}
% ---------------------------------------------------------

So far, prices have been defined at each holarchic level of the AMM:
nodes, clusters, zones, regions, and the whole system. For most
households and small businesses this structure is hidden behind a retail
supplier and a tariff. But a growing class of actors --- prosumers,
fleets, behind-the-meter storage, and large sites settling directly at
the AMM --- will participate as digital market peers. For these devices,
there must be a \emph{single, well-defined price path} for imports
(consumption) and exports (injection), even though multiple scarcity
signals exist upstream.

\paragraph{Relation to nodal and zonal pricing.}
Conventional market designs expose participants either to \emph{nodal} prices
(as in LMP) or to \emph{zonal} prices (as in European or GB wholesale
markets). The AMM generalises both. Its holarchic structure builds a stack of
nested scarcity layers—node $\rightarrow$ cluster $\rightarrow$ zone
$\rightarrow$ region $\rightarrow$ system—and the participant-facing price is
the price of whichever layer is \emph{tightest}. When a local constraint binds,
the AMM behaves like a nodal design; when only a zonal constraint binds, it
behaves like a zonal design; when local capacity is slack but the system is
short, it behaves like a uniform system price. This makes nodal and zonal
pricing special cases of the AMM’s general scarcity-propagation rule.

\begin{tcolorbox}[
  title={Nodal (LMP), Zonal, and AMM Pricing in One Picture},
  colback=gray!5,
  colframe=gray!40,
  boxrule=0.4pt
]
\begin{description}[leftmargin=1.4cm, labelwidth=1.1cm]
  \item[LMP:] Nodal price at each bus, obtained as the dual variable of
  power-balance and network constraints in an optimal power flow (OPF).
  Reflects physics and congestion, but is typically exposed only at
  transmission level and does not embed fairness or risk caps.

  \item[Zonal:] A single price per bidding zone, obtained by aggregating
  nodes and using simplified network representations. Reduces complexity
  and volatility but can hide intra-zonal congestion and misallocate
  scarcity signals.

  \item[AMM:] A holarchic scarcity controller that maintains prices at
  multiple levels (node, cluster, zone, region, system) and exposes to
  the edge the price of the \emph{tightest active layer} at each instant.
  When a local constraint binds, the AMM behaves like LMP; when only a
  zonal constraint binds, it behaves like a zonal market; when only
  system adequacy is tight, it behaves like a single system price.
\end{description}
The AMM therefore \emph{contains} both LMP and zonal pricing as special
cases, while adding boundedness, fairness constraints, and explicit
digital governance.
\end{tcolorbox}

\paragraph{Holarchic price stack.}

Let $\mathcal{H} = \{\text{node}, \text{cluster}, \text{zone},
\text{region}, \text{system}\}$ denote the holarchic levels. For each
level $\ell \in \mathcal{H}$ we define:

\[
  \alpha^{\ell}_{t,h},\quad
  \Delta^{\ell}_{t,h},\quad
  BP^{\ell}_{t,h},\quad
  SP^{\ell}_{t,h},
\]

where $h$ indexes the element at level $\ell$ (e.g.\ a particular
cluster or zone), and $BP^{\ell}_{t,h}, SP^{\ell}_{t,h}$ are the
buy and sell prices computed by the AMM using the deficit-based rules
in Section~\ref{subsec:amm_price_functions}.

For a given node $n$, let $m_{\ell}(n)$ denote its membership at each
level, e.g.
\[
  m_{\text{node}}(n) = n,\quad
  m_{\text{cluster}}(n) = c(n),\quad
  m_{\text{zone}}(n) = z(n),\ \ldots
\]
At time $t$, the node ``inherits'' a stack of AMM prices:
\[
  \Bigl\{
    BP^{\ell}_{t,m_{\ell}(n)},\,
    SP^{\ell}_{t,m_{\ell}(n)}
  \Bigr\}_{\ell \in \mathcal{H}}.
\]

\paragraph{Which level actually sets the edge price?}

Conceptually, the level that ``matters'' for a device at node $n$ is
the level whose constraint is currently tightest. We formalise this as
a \emph{dominant scarcity level}. Let
$\alpha^{\ell}_{t,m_{\ell}(n)} \in (0,1]$ be the composite scarcity at
level $\ell$ (including network and stability factors). Then define
\footnote{%
If $\mathcal{H}=\{\text{zone},\text{system}\}$ only, the AMM reduces exactly
to a zonal market; if $\mathcal{H}=\{\text{node}\}$ only, it reduces to nodal
(LMP-like) pricing. The holarchic AMM is therefore a strict generalisation of
both.}:

\[
  \ell^{\star}_{t,n}
  :=
  \arg\min_{\ell \in \mathcal{H}}
    \alpha^{\ell}_{t,m_{\ell}(n)}.
\] 

\begin{lemma}[AMM as a strict generalisation of nodal and zonal pricing]
\label{lem:amm_generalises_nodal_zonal}
Let $\mathcal{H}$ denote the holarchic levels used by the AMM, and let
$P^{\mathrm{buy}}_{t,n}, P^{\mathrm{sell}}_{t,n}$ be the participant-facing
prices at node $n$ defined by
\[
  \ell^{\star}_{t,n}
  :=
  \arg\min_{\ell \in \mathcal{H}}
    \alpha^{\ell}_{t,m_{\ell}(n)},\qquad
  P^{\mathrm{buy}}_{t,n}
  :=
  BP^{\ell^{\star}_{t,n}}_{t,\,m_{\ell^{\star}_{t,n}}(n)},\quad
  P^{\mathrm{sell}}_{t,n}
  :=
  SP^{\ell^{\star}_{t,n}}_{t,\,m_{\ell^{\star}_{t,n}}(n)}.
\]
Then:
\begin{enumerate}[label=(\roman*), leftmargin=*]
  \item If $\mathcal{H} = \{\text{node}\}$, the AMM reduces to a nodal
  (LMP-like) design with one price per node.

  \item If $\mathcal{H} = \{\text{zone}, \text{system}\}$ and each node
  belongs to exactly one zone, the AMM reduces to a zonal design in which
  each node sees the price of its zone whenever zonal scarcity is tighter
  than system scarcity.

  \item For any richer hierarchy
  $\mathcal{H} \supseteq \{\text{node},\text{zone},\text{system}\}$,
  nodal and zonal prices are recovered as the AMM prices associated with
  the corresponding layers whenever those are the tightest constraints.
\end{enumerate}
\end{lemma}

\begin{proof}[Proof sketch]
Part (i) follows immediately by taking $\mathcal{H} = \{\text{node}\}$, so
that $\ell^{\star}_{t,n} = \text{node}$ for all $t,n$, and hence
$P^{\mathrm{buy}}_{t,n} = BP^{\text{node}}_{t,n}$ and
$P^{\mathrm{sell}}_{t,n} = SP^{\text{node}}_{t,n}$ are pure nodal prices.

For part (ii), if $\mathcal{H} = \{\text{zone},\text{system}\}$ and each
node $n$ belongs to exactly one zone $z(n)$, then whenever zonal scarcity
is tighter than system scarcity we have
$\alpha^{\text{zone}}_{t,z(n)} < \alpha^{\text{system}}_{t}$ and hence
$\ell^{\star}_{t,n} = \text{zone}$, so
$P^{\mathrm{buy}}_{t,n} = BP^{\text{zone}}_{t,z(n)}$ and
$P^{\mathrm{sell}}_{t,n} = SP^{\text{zone}}_{t,z(n)}$.
This is exactly a zonal pricing arrangement.

Part (iii) follows because for any hierarchy that includes node, zone, and
system layers, the same minimisation rule over $\alpha^{\ell}$ selects the
tightest active layer; when the nodal layer is tightest, the resulting
AMM price coincides with a nodal price, and when the zonal layer is
tightest, it coincides with the zonal price. Thus, nodal and zonal prices
are embedded as special cases of the AMM's holarchic scarcity rule.
\end{proof}

Intuitively, $\ell^{\star}_{t,n}$ is the holarchic layer where scarcity
is most severe (lowest $\alpha$). That layer sets the \emph{effective}
AMM price seen at the node.

The participant-facing import (buy) and export (sell) prices at node
$n$ are then:

\begin{align}
  P^{\mathrm{buy}}_{t,n}
  &:= BP^{\ell^{\star}_{t,n}}_{t,\,m_{\ell^{\star}_{t,n}}(n)}, \\
  P^{\mathrm{sell}}_{t,n}
  &:= SP^{\ell^{\star}_{t,n}}_{t,\,m_{\ell^{\star}_{t,n}}(n)}.
\end{align}

In words: at any instant, a device or self-settling meter at node $n$
sees the buy and sell prices of the \emph{tightest} active layer. If a
more local constraint binds, it overrides looser upstream prices; if
local capacity is slack, an upstream zonal or system constraint can set
the edge price.

\paragraph{Local vs. zonal and system constraints (worked logic).}

This simple definition captures the cases of interest:

\begin{itemize}[leftmargin=*]
  \item \textbf{Local (node or feeder) constraint.} \\
  Suppose voltage or feeder capacity is binding locally so that
  $\alpha^{\text{node}}_{t,n} < \alpha^{\ell}_{t,m_{\ell}(n)}$ for all
  $\ell \ne \text{node}$. Then
  $\ell^{\star}_{t,n} = \text{node}$ and
  \[
    P^{\mathrm{buy}}_{t,n} = BP^{\text{node}}_{t,n},\qquad
    P^{\mathrm{sell}}_{t,n} = SP^{\text{node}}_{t,n}.
  \]
  The local AMM price applies to both consumption and export. This is
  the ``max local'' situation described informally: local scarcity sets
  the effective edge price.

  \item \textbf{Zonal constraint with slack local capacity.} \\
  Suppose the local node is unconstrained
  ($\alpha^{\text{node}}_{t,n} \approx 1$) but the zone is short,
  $\alpha^{\text{zone}}_{t,z(n)} < 1$, and tighter than region/system.
  Then $\ell^{\star}_{t,n} = \text{zone}$ and the participant sees
  zonal prices:
  \[
    P^{\mathrm{buy}}_{t,n} = BP^{\text{zone}}_{t,z(n)},\qquad
    P^{\mathrm{sell}}_{t,n} = SP^{\text{zone}}_{t,z(n)}.
  \]
  This corresponds to the case where
  ``if there is a zonal constraint or shortage where local capacity is
  not constrained, the price to consume is the zonal price and the
  price to sell is the zonal offered price''.

  \item \textbf{System-wide scarcity only.} \\
  If all lower levels are slack but the system is short
  (e.g.\ tight reserve margin, low inertia), then
  $\ell^{\star}_{t,n} = \text{system}$ and all nodes inherit the same
  system-level AMM price:
  \[
    P^{\mathrm{buy}}_{t,n} = BP^{\text{sys}}_{t},\qquad
    P^{\mathrm{sell}}_{t,n} = SP^{\text{sys}}_{t}.
  \]
\end{itemize}

More complex patterns (simultaneous cluster- and zone-level binding)
are handled automatically: whichever layer has the lowest $\alpha$ at
that node sets the participant-facing price.

\paragraph{Self-settling devices and risk-bearing.}

A household, business, or aggregator that chooses to settle directly at
the AMM --- effectively acting as its own supplier --- is exposed to the
full time series $\{P^{\mathrm{buy}}_{t,n}, P^{\mathrm{sell}}_{t,n}\}_t$
at its node. Its net settlement over a period $T$ is:

\[
  \text{Settlement}
  =
  \sum_{t \in T}
  \bigl(
    P^{\mathrm{buy}}_{t,n} \cdot q^{\mathrm{imp}}_{t,n}
    - P^{\mathrm{sell}}_{t,n} \cdot q^{\mathrm{exp}}_{t,n}
  \bigr),
\]

where $q^{\mathrm{imp}}_{t,n}$ and $q^{\mathrm{exp}}_{t,n}$ are import
and export quantities. In this configuration, the participant bears
wholesale price risk directly, but that risk is \emph{bounded} by the
AMM’s capped scarcity functions and Fair Play protections (no unbounded
price spikes or arbitrary curtailment).

Conventional suppliers, in contrast, see exactly the same holarchic AMM
prices but wrap them into retail products (subscriptions, QoS tiers,
hedges) so that end-users experience a smoothed, contract-based price
rather than the raw $\{P^{\mathrm{buy}}_{t,n}, P^{\mathrm{sell}}_{t,n}\}$
sequence.

Operationally, there is a single AMM-defined participant-facing price
pair $(P^{\mathrm{buy}}_{t,n}, P^{\mathrm{sell}}_{t,n})$ at each node
and time, defined in Section~\ref{subsec:participant_price}. Retail
subscriptions and QoS tiers do not create separate physical prices;
they repackage this edge price into different risk-bearing structures
(fixed vs.\ variable exposure, insurance-like caps, priority rights),
while the underlying holarchic AMM price remains unique.

\begin{figure}[t]
\centering
\begin{tikzpicture}[
  >=latex,
  box/.style={
    rectangle, draw, rounded corners,
    minimum width=3.1cm, minimum height=8mm,
    align=center, font=\footnotesize
  },
  lbl/.style={font=\scriptsize},
  title/.style={font=\small, align=center}
]

% ---------------- Panel (a) ----------------
\begin{scope}[xshift=-6.0cm]
  \node[title] at (0,2.75) {(a) Local (nodal) constraint};

  \node[box, fill=gray!8] (a_sys)  at (0,2.00) {System};
  \node[box, fill=gray!8] (a_zone) at (0,1.15) {Zone $z$};
  \node[box, fill=blue!8] (a_node) at (0,0.30) {Node $n$};

  \draw[->] (a_sys) -- (a_zone);
  \draw[->] (a_zone) -- (a_node);

  \node[lbl, anchor=west, xshift=2mm] at (a_node.east)
    {$\alpha^{\text{node}}_{t,n}$ tightest};

  \node[lbl, align=center] at (0,-0.80)
    {$P^{\mathrm{buy}}_{t,n},\,P^{\mathrm{sell}}_{t,n}$ set by \textbf{node}};
\end{scope}

% ---------------- Panel (b) ----------------
\begin{scope}[xshift=0.0cm]
  \node[title] at (0,2.75) {(b) Zonal constraint, local slack};

  \node[box, fill=gray!8]    (b_sys)  at (0,2.00) {System};
  \node[box, fill=orange!15] (b_zone) at (0,1.15) {Zone $z$};
  \node[box, fill=gray!8]    (b_node) at (0,0.30) {Node $n$};

  \draw[->] (b_sys) -- (b_zone);
  \draw[->] (b_zone) -- (b_node);

  \node[lbl, anchor=west, xshift=2mm] at (b_zone.east)
    {$\alpha^{\text{zone}}_{t,z(n)}$ tightest};

  \node[lbl, align=center] at (0,-0.80)
    {$P^{\mathrm{buy}}_{t,n},\,P^{\mathrm{sell}}_{t,n}$ set by \textbf{zone}};
\end{scope}

% ---------------- Panel (c) ----------------
\begin{scope}[xshift=6.0cm]
  \node[title] at (0,2.75) {(c) System-wide scarcity};

  \node[box, fill=red!10]  (c_sys)  at (0,2.00) {System};
  \node[box, fill=gray!8]  (c_zone) at (0,1.15) {Zone $z$};
  \node[box, fill=gray!8]  (c_node) at (0,0.30) {Node $n$};

  \draw[->] (c_sys) -- (c_zone);
  \draw[->] (c_zone) -- (c_node);

    \node[lbl, anchor=east, xshift=-2mm] at (c_sys.west)
      {$\alpha^{\text{system}}_{t}$ tightest};

  \node[lbl, align=center] at (0,-0.80)
    {$P^{\mathrm{buy}}_{t,n},\,P^{\mathrm{sell}}_{t,n}$ set by \textbf{system}};
\end{scope}

\end{tikzpicture}
\caption[Price inheritance under different constraint patterns]{%
Illustration of how participant-facing prices inherit from the holarchic
scarcity layers. (a) When a local (nodal) constraint is tightest, the AMM
behaves like nodal pricing. (b) When local capacity is slack but a zonal
constraint binds, the zonal price applies. (c) When only system-wide
adequacy is tight, all nodes inherit the same system AMM price.}
\label{fig:amm_price_inheritance}
\end{figure}
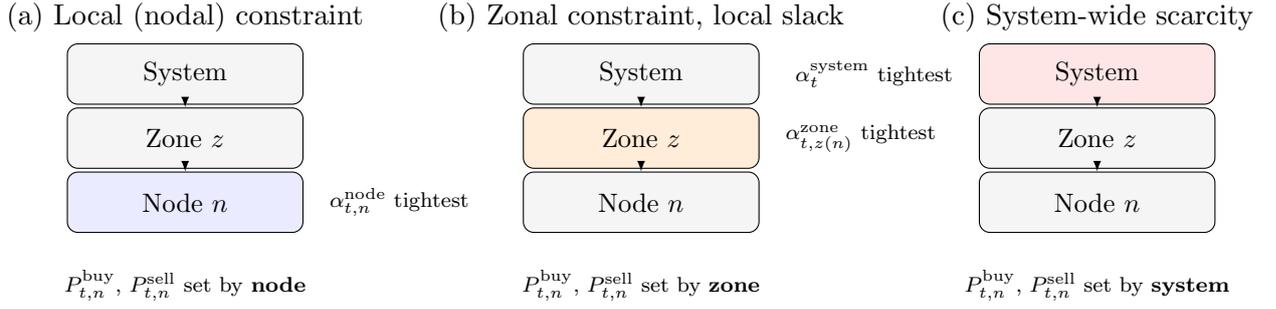

% ---------------------------------------------------------
\section{Control-Theoretic Stability of the Digital Holarchic AMM}
\label{sec:amm_stability}

The AMM can be interpreted as a cyber--physical control system: it receives
real-time measurements of demand, generation, flexibility, network conditions,
and fairness states, and computes bounded allocation and price updates subject
to scarcity and equity constraints. Unlike conventional price-based
markets—which act as unregulated feedback systems with no guarantees on
responsiveness, fairness, or oscillatory behaviour—the AMM is implemented as a
\emph{digital holarchic controller} that ensures structural, algorithmic, and
temporal stability by design. Its stability can be explained through three
complementary perspectives.

\paragraph{(1) Structural stability via holarchic containment.}
The AMM is not a monolithic controller. Instead, it is composed of nested
controllers operating at different spatial and temporal layers:
\[
\text{device} \rightarrow \text{household} \rightarrow \text{feeder}
\rightarrow \text{cluster} \rightarrow \text{region} \rightarrow \text{national}.
\]
Each layer has bounded informational scope, a well-defined objective
function, and explicit upstream/downstream constraint dominance.
Corrective actions taken at one layer cannot propagate uncontrollably
across other layers, which prevents cascading instabilities commonly
observed in recursive price-chasing arrangements.

\begin{quote}
\textit{Containment stability (informal theorem): A holarchically layered
market system in which domain-limited controllers respond only within
their jurisdiction, under constraint dominance rather than recursive
optimisation, cannot exhibit unbounded or runaway response propagation.}
\end{quote}

This architectural containment prevents systemic oscillation, runaway
price amplification, and fairness violations induced by cross-layer
feedback in conventional designs.

\paragraph{(2) Bounded-Input Bounded-Output (BIBO) stability via digital enforcement.}
All key signals in the AMM—scarcity $\alpha$, deficit $\Delta$, price increment
$\Delta p$, allocation adjustment $\Delta q$, or fairness deviations—are
digitally constrained via software-defined bounds, saturation functions, and
update-rate limits. In particular:
\begin{itemize}[leftmargin=*]
  \item scarcity signal $\tilde{\alpha}_{t,n} \in [0,1]$ is soft-clipped at
        both bounds;
  \item price change $\Delta p$ per update is capped by a maximum
        step-size $\Delta p_{\max}$;
  \item allocation updates are restricted to remain within dynamically
        calculated envelope constraints (Fair Play condition F4);
  \item update frequency is asynchronous and event-triggered, rather
        than continuously reactive.
\end{itemize}

Because both buy price $BP_{t,n}$ and sell price $SP_{t,n}$ depend on the
same bounded deficit signal $\Delta_{t,n}$ (Section~\ref{subsec:amm_price_functions}),
any bounded disturbance in demand or supply produces bounded changes in
prices and allocations. Formally, the AMM satisfies BIBO stability:
\begin{equation}
\text{If } \lvert x(t) \rvert < M_x \quad \Rightarrow \quad
\lvert y(t) \rvert < M_y,
\end{equation}
where $x(t)$ denotes the magnitude of scarcity, load imbalance, or
constraint violation, and $y(t)$ denotes price or allocation
adjustments. This boundedness is \emph{not} guaranteed in classical
wholesale or balancing markets, where price and quantity signals may
legally diverge without bound (e.g.\ extreme price spikes).

\paragraph{(3) Lyapunov-like stability via monotonic disequilibrium reduction.}
We define a Lyapunov-like function that measures instantaneous
resource imbalance:
\begin{equation}
L(t) = \left\lvert \text{Supply}(t) - \text{Demand}(t) \right\rvert.
\end{equation}
The AMM operates to minimise $L(t)$ subject to feasibility and fairness
constraints, resulting in:
\begin{equation}
\frac{dL(t)}{dt} \le 0,
\end{equation}
except during exogenously introduced imbalance shocks (e.g.\ asset
failures or demand surges), after which $L(t)$ is restored to a
non-increasing trajectory.

The symmetric dependence of $BP_{t,n}$ and $SP_{t,n}$ on the deficit
$\Delta_{t,n}$ ensures that high imbalances trigger both reduced imports
and increased exports at affected nodes, directly driving $L(t)$ down.
This implies \emph{monotonic convergence towards feasible, fair, and
scarcity-reflective allocations}, and excludes the oscillatory or chaotic
dynamics sometimes observed in unregulated recursive price adjustments.

\paragraph{Discussion.}
Through its holarchic structure, bounded digital implementation, naturally
self-corrective buy–sell coupling, and Lyapunov-like equilibrium behaviour,
the AMM exhibits control-theoretic stability by design. This stands in
contrast to existing price-driven or bidding-based market architectures, where
feedback interactions are unregulated, oscillatory behaviour is common, and
no formal stability guarantee exists.

% ---------------------------------------------------------
\section{The AMM as a Digital Scarcity Control Layer}
\label{sec:amm_control_layer}

The AMM does not function as a classical welfare-maximising spot market.
Instead, it acts as a \emph{digital scarcity controller} that continuously
monitors physical and forecast system states, infers scarcity, and sets
prices, allocation priorities, and fairness parameters accordingly.

\subsection{AMM as a feedback-control system}

Rather than receiving price bids (as in traditional markets), the AMM observes:
\[
\Xi_{t,n} \;=\; 
\Big\{
S_{t,n},\, C_{t,n},\, \text{SoC}_{t,n},\, \Delta V_{t,n},\, \text{cong}_{t,n},
\, R_t,\, \hat{W}_t,\, \hat{D}_t
\Big\},
\]
where:
\begin{itemize}[leftmargin=*]
  \item $S_{t,n}, C_{t,n}$ denote local supply and demand (or their forecasts),
  \item $\Delta V_{t,n}$ represents voltage deviation and $\text{cong}_{t,n}$ line congestion,
  \item $R_t$ is reserve margin, $\hat{W}_t$ wind forecast, and $\hat{D}_t$ demand forecast, and
  \item $\text{SoC}_{t,n}$ is the average storage state-of-charge in the node or zone.
\end{itemize}

These signals are transformed into the synthetic scarcity measure
$\tilde{\alpha}_{t,n}$ and deficit $\Delta_{t,n}$ defined in
Section~\ref{sec:amm_design}. Based on these, the AMM sets prices:
\[
BP_{t,n}
=
BP^{\mathrm{base}}_{t,n}
+
F_{t,n}\bigl(\Delta_{t,n}, 1-\alpha^{\mathrm{stability}}_{t,n}\bigr),
\qquad
SP_{t,n}
=
SP^{\mathrm{base}}_{t,n}
+
H_{t,n}\bigl(\Delta_{t,n}, 1-\alpha^{\mathrm{stability}}_{t,n}\bigr),
\]
where $F_{t,n},H_{t,n}$ are composite scarcity-response functions combining
energy-driven and stability-driven uplifts (Section~\ref{subsec:amm_price_functions}).

This, in turn, influences flexible load consumption (imports), storage
activation and generation (exports), and controllable generation output.
These actions change $\Xi_{t+1,n}$, forming a closed-loop control system:
\[
\Xi_{t,n} 
\;\xrightarrow{\text{AMM}}\;
(\tilde{\alpha}_{t,n},\Delta_{t,n},BP_{t,n},SP_{t,n})
\;\xrightarrow{\text{appliance/asset response}}\;
\Xi_{t+1,n}.
\]

This system-level feedback perspective is consistent with, and extends, the
author's prior work on human-in-the-loop cyber--physical control for health
protection, in which real-time environmental sensing and online optimisation
were used to nudge an electrically assisted bicycle away from high-pollution
exposure while respecting journey-time and comfort constraints (published in
\emph{Automatica} \cite{SWEENEY2022110595}). In that setting, physical measurements, a digital controller,
and human decisions formed a closed loop to deliver a welfare-relevant outcome
(reduced pollutant dose). Here, the same design philosophy is applied at system
scale: instead of protecting a single rider's health, the AMM and its digital
regulation layer act to protect households, critical services, and generators
from unfair and inefficient outcomes by embedding welfare objectives directly
into the scarcity-control loop.

Thus, the AMM behaves not as a trading platform, but as a
\textbf{real-time scarcity regulator}.

% -----------------------------
\subsection{Time-coupled requests and flexibility windows}

Each flexible request $r$ is defined as:
\[
r = \big(
E_r,\; t^{\mathrm{start}}_r,\; t^{\mathrm{end}}_r,\;
\bar{P}_r,\; \sigma^r, \;\Gamma^{\text{target}}_r
\big),
\]
where:
\begin{itemize}[leftmargin=*]
  \item $E_r$ is required energy;
  \item $[t^{\mathrm{start}}_r,\, t^{\mathrm{end}}_r]$ is its valid delivery window;
  \item $\bar{P}_r$ is maximum power rate;
  \item $\sigma^r$ is flexibility (width of allowable window); and
  \item $\Gamma^{\text{target}}_r$ encodes fairness and priority attributes (need,
        medical status, contract type, etc.).
\end{itemize}

The AMM does not optimise each request individually in isolation, but instead
updates expected demand profiles over the horizon
$[t^{\mathrm{start}}_r,\, t^{\mathrm{end}}_r]$, contributing to
$\alpha_{t}^{\mathrm{forecast}}$ and the forward deficit
$\Delta_{\tau,n}^{\mathrm{fwd}}$. Consumers respond to $BP_{t,n}$ signals by
shifting within $\sigma^r$ where possible (Fairness F1), and Fair Play then
chooses the cheapest feasible slots consistent with fairness and constraint
discipline.

% -----------------------------
\subsection{Holarchic structure of interaction}

The AMM exists in nested domains:
\[
\mathcal{M}^{\text{node}}
\subset
\mathcal{M}^{\text{zone}}
\subset
\mathcal{M}^{\text{region}}
\subset
\mathcal{M}^{\text{system}}.
\]

\begin{itemize}[leftmargin=*]
  \item Node-level AMM captures local voltage, congestion, EV, storage, and
        microgrid dynamics.
  \item Zone-/Region-level AMM captures inter-nodal flows and shared balancing
        resources.
  \item System-level AMM ensures adequacy, reserve margins, and alignment with
        policy objectives (e.g.\ net-zero trajectories).
\end{itemize}

Higher-level scarcity cascades downwards; local scarcity can exist even when
system-wide scarcity does not, consistent with preferential regional allocation
(Chapter~\ref{ch:mathematics}).

% -----------------------------
\subsection{Digital enforceability and Fair Play integration}

When $\tilde{\alpha}_{t,n}<1$ and $\Delta_{t,n}>0$, the AMM activates the
Fair Play Algorithm to allocate limited resources non-arbitrarily and
consistently with fairness conditions F2--F4:
\[
\bar{Q}^r \propto \Gamma^{\text{target}}_{r},
\qquad \text{subject to essential-first and proportionality rules.}
\]

Fair Play therefore operates within declared individual rationality bounds
$v^{\max}_r$, but its priority and proportionality rules remain entirely
independent of those economic bounds, consistent with the fairness axioms.

Allocations are logged, explainable, and auditable. 
This closes the loop between:
\[
\text{physical scarcity} 
\;\longrightarrow\;
\text{price signals} 
\;\longrightarrow\;
\text{fair allocation}.
\]

Thus, the AMM provides a \emph{digital scarcity regime}—with co-designed pricing,
allocation, and priority rules.

Finally, the AMM must not only be mathematically fair, but also \emph{perceived to be fair}. Behavioural and digital governance studies emphasise that trust in market design emerges not from perfect optimisation, but from predictability, bounded exposure, clarity of rules, and perceived reciprocity. This motivates the AMM’s design as a fairness-first digital market product: transparent, bounded, participatory, and explainable.

\subsection{Digital Product Design Principles}

Although the AMM performs sophisticated cyber–physical optimisation, its
interaction with end-users (households, aggregators, small generators,
storage operators) is intentionally simple, predictable, and human-centric.
Following digital product design principles, the complexity of real-time
scarcity inference, network constraint synthesis, and proportional allocation
is abstracted behind a clear, consistent user interface.

This abstraction follows established principles from digital product
engineering and human–computer interaction:

\begin{itemize}[leftmargin=*]
  \item \textbf{Hidden complexity:} The internal mechanisms (scarcity
  synthesis, network constraints, volatility smoothing) are hidden behind
  interpretable outputs: unit prices, allocation rights, and protected
  access guarantees.

  \item \textbf{Explainability and legibility:} Participants do not need to
  understand the full optimisation logic, but they must be able to verify
  \textit{why} an allocation or price occurred. This aligns with behavioural
  trust frameworks (knowledge, predictability, reassurance).

  \item \textbf{Bounded exposure:} A digital market is only acceptable if
  users can never be exposed to unbounded risk or extreme volatility. The
  AMM enforces this through capped tightness pricing, digital envelopes,
  and essential protection blocks.

  \item \textbf{Participation without expertise:} Users should not require
  market literacy training to participate safely. The AMM provides a
  "protected default experience," mirroring digital product safeguards
  in financial technology, public services, and healthcare platforms.

  \item \textbf{Interface alignment:} The AMM supports multiple engagement
  modes—manual participation (household UI), automated participation
  (smart contracts and appliances), and aggregated participation
  (neighbourhood or commercial aggregators)—mirroring multi-layer
  customer journeys in digital product ecosystems.
\end{itemize}

Thus, while the AMM is technically a real-time control system, it is
also a \textbf{digital platform product}—abstracting complexity, maintaining
explainable fairness, and ensuring stable, predictable, and legitimate
participation.

% -----------------------------
\subsection*{Interpretation}

In summary, the AMM is not merely a price calculator; it behaves as:
\begin{enumerate}[leftmargin=*]
\item a continuously adaptive scarcity-aware control system;
\item a holarchic aggregator of local, regional, and system-level constraints;
\item a digitally enforceable rulebook for fair access and proportional
      allocation under shortage; and
\item a transparent and explainable pricing layer that embeds behavioural
      incentives, essential protection, and digital trust.
\end{enumerate}

It replaces bidding-based “who pays most” allocation with “who contributes
most, who needs protection, who can shift the most,” aligning directly with
Fairness Conditions (F1–F4) in Chapter~\ref{ch:fairness_definition}. Crucially,
it establishes not only efficient balance, but \emph{legitimate, explainable,
and trusted participation} in a digital energy system.

\begin{sidewaysfigure}[p]
\centering
\begin{tikzpicture}[
  >=latex,
  node distance=23mm and 32mm,
  every node/.style={font=\small, align=center},
  box/.style={
    rectangle,
    draw,
    rounded corners,
    text width=3.8cm,
    minimum height=12mm,
    inner sep=4pt
  },
  arrow/.style={->, line width=0.7pt, shorten >=2pt, shorten <=2pt},
  lbl/.style={font=\scriptsize, inner sep=1pt}
]

%---------------- Nodes (auto-spaced, no overlap) ----------------%
\node[box, fill=blue!10] (state)
  {Physical \& Forecast State \\[3pt]
   $S^T_{t,n},\, C_{t,n},\, \Delta V_{t,n},\, \text{cong}_{t,n}$ \\[2pt]
   $R_t,\, \hat{W}_t,\, \hat{D}_t$};

\node[box, fill=green!10, right=of state] (alpha)
  {Scarcity Inference \\[3pt]
   $\tilde{\alpha}_{t,n} =
   \alpha^{\mathrm{instant}}_{t,n}\,
   \alpha^{\mathrm{forecast}}_{t,n}$ \\[2pt]
   $\alpha^{\mathrm{network}}_{t,n}\,
   \alpha^{\mathrm{stability}}_{t,n}$};

\node[box, fill=yellow!15, right=of alpha] (prices)
  {AMM Pricing Layer \\[3pt]
   $BP_{t,n} = BP^{\mathrm{base}}_{t,n} + F_{t,n}(\Delta_{t,n})$ \\[2pt]
   $SP_{t,n} = SP^{\mathrm{base}}_{t,n} + H_{t,n}(\Delta_{t,n})$};

\node[box, fill=orange!15, right=of prices] (response)
  {Asset \& Demand Response \\[3pt]
   Flexible load shifting \\[-1pt]
   Storage charge/discharge \\[-1pt]
   Controllable generation};

\node[box, fill=red!10, below=of prices]
  (fairplay)
  {Fair Play / Shortage Logic \\[3pt]
   Active only if $\Delta_{t,n}>0$ and $\tilde{\alpha}_{t,n}<1$ \\[-1pt]
   Essential-first, proportional, rotation};

%---------------- Arrows (clean, no overlap) ----------------%
\draw[arrow] (state) -- node[lbl,above]{observe} (alpha);
\draw[arrow] (alpha) -- node[lbl,above]{compute $\tilde{\alpha},\, \Delta$} (prices);
\draw[arrow] (prices) -- node[lbl,above]{broadcast $BP,SP$} (response);

% Fair Play Trigger
\draw[arrow] (alpha.south) |- node[lbl,left,pos=0.7]{if shortage} (fairplay.west);

% Fair Play Updates Pricing Layer
\draw[arrow] (fairplay.north) -- node[lbl,right,pos=0.5]{adjust allocation / priorities} (prices.south);

% Feedback loop below everything
\draw[arrow]
  (response.south) .. controls +(0,-1.5) and +(0,-1.5) ..
  node[lbl,below,pos=0.5]{updated $S^T_{t+1,n},\, C_{t+1,n}$}
  (state.south);

\end{tikzpicture}
\caption{AMM scarcity feedback loop. Physical and forecasted conditions are mapped into a composite scarcity signal $\tilde{\alpha}_{t,n}$ and a deficit $\Delta_{t,n}$, which determine prices $BP,SP$ and, under shortage, activate the Fair Play allocation rules. Behavioural response updates the physical state, closing the loop.}
\label{fig:amm_feedback_loop}
\end{sidewaysfigure}
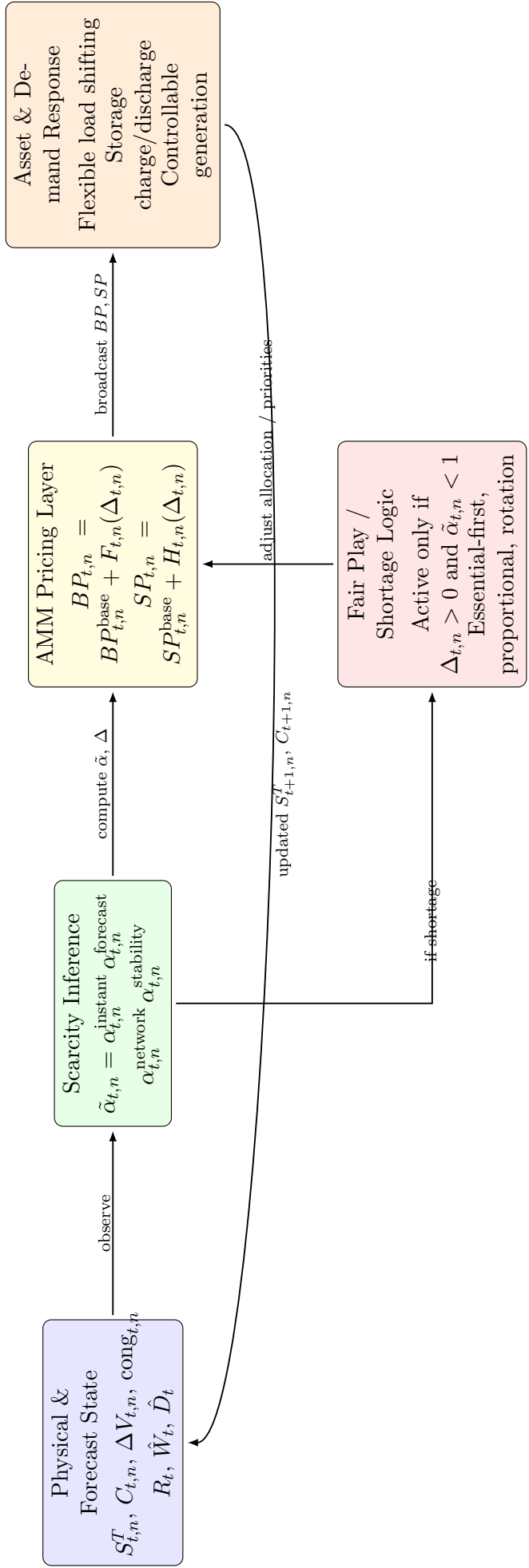

% ---------------------------------------------------------
\subsection{Network Scarcity and Voltage as a Digital Shadow Price}
\label{subsec:voltage_shadow_price}

In classical optimisation, a \emph{shadow price} represents the marginal value
of relaxing a binding constraint by one unit. It is not a market price, but a
physical or operational signal that indicates how ``tight'' a constraint is.
In electricity networks, voltage is a natural physical shadow price: it is a
direct, real-time manifestation of how close the system is to local supply
scarcity (undervoltage) or local export saturation (overvoltage).

Let:
\[
V^{\star}_{t,n} \quad \text{be the expected or nominal voltage at node } n,
\]
\[
V^{\text{meas}}_{t,n} \quad \text{be the measured voltage from meters or inverters},
\]
\[
\varepsilon_{t,n} := V^{\text{meas}}_{t,n} - V^{\star}_{t,n},
\]
where $\varepsilon_{t,n}$ is the signed voltage deviation. We define its
normalised magnitude:
\[
\Delta V_{t,n} := \frac{|\varepsilon_{t,n}|}{V_{\text{nom}}},
\]
with $V_{\text{nom}}$ as a reference (e.g.\ 230\,V RMS in LV systems).

Two scarcity regimes arise naturally:

\begin{itemize}[leftmargin=*]
  \item \textbf{Undervoltage (scarcity):} $\varepsilon_{t,n}<0$. Local demand
        exceeds the capability of upstream supply or network capacity, causing
        voltage to drop. This is interpreted as a \emph{positive shadow price}
        of local shortage: the value of injecting one more unit of supply here
        is high.
  \item \textbf{Overvoltage (surplus):} $\varepsilon_{t,n}>0$. Local generation
        or export exceeds local absorption or capacity, pushing voltage above
        nominal. This indicates a \emph{negative shadow price}: consuming (or
        absorbing) an extra unit here is valuable.
\end{itemize}

Accordingly, the AMM embeds voltage deviation into the network scarcity factor:
\[
  \alpha_{t,n}^{\text{network}}
  =
  \exp\bigl(-\theta_n^- \cdot \max(0,-\varepsilon_{t,n})\bigr)
  \cdot
  \exp\bigl(-\theta_n^+ \cdot \max(0,\varepsilon_{t,n})\bigr)
  \cdot
  \exp\bigl(-\phi_n \cdot \text{cong}_{t,n}\bigr),
\]
where $\theta_n^-, \theta_n^+>0$ represent sensitivity to undervoltage and
overvoltage respectively, and $\phi_n>0$ captures line congestion.

From here, buy and sell prices become explicit functions of the shadow price
embedded in voltage:
\[
  BP_{t,n} = BP_{t,n}^{\mathrm{base}} +
  F_{t,n}\bigl(\Delta_{t,n},\varepsilon_{t,n}\bigr),
  \qquad
  SP_{t,n} = SP_{t,n}^{\mathrm{base}} +
  H_{t,n}\bigl(\Delta_{t,n},\varepsilon_{t,n}\bigr),
\]
which ensures:

\begin{itemize}[leftmargin=*]
  \item when \textbf{undervoltage occurs}, both buy and sell prices rise,
        incentivising demand reduction and supply injection locally;
  \item when \textbf{overvoltage occurs}, prices fall (or become negative),
        incentivising consumption (charging, heating, EV) and discouraging
        further export.
\end{itemize}

Thus, the measured voltage acts as a locally observable, digitalizable
\emph{shadow price of network scarcity}. It allows two neighbouring houses
or devices---without central dispatch---to coordinate behaviour based solely
on price signals that reflect the physics of their shared feeder.

This shadow price interpretation explains how the AMM can respond
automatically and proportionally to local network tightness,
without needing complex real-time optimisation.

\subsection*{Sidebar: Shadow price, LMP, and AMM price}

It is helpful to distinguish three related but conceptually different ideas:

\begin{itemize}[leftmargin=*]
  \item \textbf{Shadow price (dual variable).}
  In optimisation theory, the shadow price is the Lagrange multiplier
  associated with a constraint (e.g.\ a power-balance or line-flow limit).
  It measures how much the objective would improve if the constraint were
  relaxed by one unit. It is a \emph{property of a constrained problem},
  not necessarily a traded market price.

  \item \textbf{Locational Marginal Price (LMP).}
  In conventional power markets, LMP is the nodal energy price obtained
  by solving an optimal power flow (OPF) problem and reading off certain
  dual variables as monetary prices. In principle, LMP reflects the
  shadow prices of energy balance and network constraints. In practice,
  LMPs are shaped by market rules, bidding behaviour, approximations to
  the physics, and settlement conventions, and are usually exposed only
  at transmission level.

  \item \textbf{AMM price.}
  The AMM price at node $n$ is a \emph{digitally constructed} price that
  encodes scarcity, fairness, and stability constraints by design. It is
  not the outcome of a welfare-maximising OPF, but of a scarcity-control
  law based on deficit $\Delta_{t,n}$, composite scarcity
  $\tilde{\alpha}_{t,n}$, and physical signals such as voltage
  deviation $\varepsilon_{t,n}$. In this sense, the AMM price behaves
  as a \emph{digital shadow price}: it responds monotonically to the
  tightness of constraints (energy, network, fairness) while remaining
  bounded and explainable.
\end{itemize}

Table~\ref{tab:shadow_lmp_amm} summarises the distinction.

\begin{table}[H]
\centering
\caption{Conceptual comparison of shadow price, LMP, and AMM price.}
\label{tab:shadow_lmp_amm}
\begin{tabular}{p{3cm}p{4cm}p{7cm}}
\toprule
\textbf{Concept} & \textbf{Where it lives} & \textbf{Role in this thesis} \\
\midrule
Shadow price 
& Mathematical optimisation (dual variables) 
& Abstract marginal value of relaxing a constraint (energy, line flow, voltage, fairness). Provides the conceptual lens: ``price as constraint tightness''. \\[4pt]
LMP 
& Transmission-level OPF-based markets 
& Example of how shadow prices can be turned into money prices in current designs. Retains physical logic but is not fairness-aware and is typically not exposed at the retail edge. \\[4pt]
AMM price 
& Digital scarcity-control layer (this architecture) 
& Real-time, bounded, fairness-aware price derived from composite scarcity and physical signals (including voltage). Acts as a \emph{digital shadow price} of local scarcity that directly drives appliance and neighbour response. \\
\bottomrule
\end{tabular}
\end{table}

In summary, the AMM does not attempt to replicate LMP. Instead, it borrows
the \emph{shadow price} intuition---``price as constraint tightness''---and
implements it as a digitally regulated, fairness-constrained, and
voltage-aware pricing law at the retail edge.

% ---------------------------------------------------------
\subsection{Voltage-triggered AMM behaviour and neighbour coordination}
\label{subsec:voltage_neighbour}

Consider a street-level scenario:
House A exports solar generation and raises feeder voltage;
House B has flexible demand (EV, immersion heater, battery charging).
The AMM detects $\varepsilon_{t,n}>0$ (overvoltage), causing:
\[
BP_{t,n} \downarrow,\qquad SP_{t,n} \downarrow,
\]
and House B sees a low or negative price to import energy. Energy is consumed,
absorbed, or stored, pulling voltage $\varepsilon_{t,n}\rightarrow 0$.
House A's export reward simultaneously diminishes, discouraging further
injection or stimulating local charging.

Likewise, if $\varepsilon_{t,n}<0$ (undervoltage):
\[
BP_{t,n} \uparrow,\qquad SP_{t,n} \uparrow,
\]
flexible demand is deferred, and exports are encouraged, naturally restoring
voltage.

Thus, market response \emph{is always in the direction that reduces voltage
deviation}. This is precisely the behaviour a shadow price should induce.

From a market perspective:  
\emph{voltage is the real-time physical signal of scarcity, and the AMM price
is its digital shadow price.}

From a control perspective:  
\emph{the AMM implements a stabilising feedback loop where voltage deviation
produces a price adjustment, which activates neighbour devices to restore
voltage equilibrium.}

This completes the cyber--physical control interpretation of the AMM.

\paragraph{Relation to current nodal--zonal policy debates.}
The holarchic AMM formulation is intentionally neutral with respect to the
political and institutional choice between nodal and zonal pricing. Recent
European and GB debates have considered moving from largely zonal markets to
more granular, congestion-reflective designs, with regulatory analyses by
national regulators and ACER emphasising the trade-off between efficiency,
complexity, and social acceptability of exposing end-users to nodal prices
(e.g.\ CREG, ACER consultation reports on bidding-zone configurations and
locational price signals).\citep{ACERBiddingZones,ACERLocationalSignals}
In North America, FERC-jurisdictional markets have long used LMP-based nodal
pricing at transmission level, but retail exposure remains limited and
fairness considerations are largely delegated to separate mechanisms such as
uplift and capacity payments.\citep{FERCOrder888,FERCOrder2000}

The AMM design in this thesis can be interpreted as a formal, digital
generalisation of these debates: it preserves the physical and informational
advantages of nodal pricing when local constraints bind, while retaining a
zonal or system price when scarcity is genuinely shared. Crucially, it adds
three ingredients that are largely absent from the existing literature:
(i) explicit fairness constraints (F1--F4),
(ii) bounded and digitally enforceable scarcity functions, and
(iii) an integrated cyber--physical control interpretation that treats
pricing, allocation, and access as parts of the same digital regulation
problem rather than as separate market layers.

% ---------------------------------------------------------
% CHAPTER 11 — MODELLING & MATHEMATICS
% ---------------------------------------------------------
\chapter{Mathematical Framework and Implementation}
\label{ch:mathematics}

% ---------------------------------------------------------
\section{Formal Fairness Definition}
\label{sec:formal_fairness}

Chapter~\ref{ch:fairness_definition} introduced fairness as a normative
system design constraint and defined the operational conditions F1--F4.
This section provides the corresponding \emph{mathematical} formulation:
it specifies the system state, admissible allocations, and the
mapping from state and history to outcomes that are considered fair.

We distinguish three coupled components:
\begin{enumerate}[label=(\roman*),leftmargin=1.5em]
  \item \textbf{Consumer-side allocation} of essential and flexible
        demand under local constraints and service levels;
  \item \textbf{Generator-side compensation} based on system value
        and Shapley-consistent attribution;
  \item \textbf{AMM control signals} which encode scarcity and propagate
        fairness conditions into prices and access.
\end{enumerate}

% ---------------------------------------------------------
\subsection{System State and Notation}
\label{subsec:system_state}

Let $t \in \mathcal{T}$ denote discrete time intervals (e.g.\ 30\,min),
and $n \in \mathcal{N}$ denote network nodes in the holarchy
(household, feeder, local area, region, etc.).
Let $h \in \mathcal{H}_n$ denote households electrically connected
to node $n$, and $g \in \mathcal{G}_n$ generators or controllable
resources at node $n$.

\begin{itemize}[leftmargin=*]
  \item $q_{h,t}$ --- realised household consumption at time $t$;
  \item $q^{\mathrm{ess}}_h$ --- must-serve block for household $h$;
  \item $q^{\mathrm{flex}}_{h,t}$ --- flexible component, potentially schedulable;
  \item $x_{g,t}$ --- dispatch of generator $g$;
  \item $S_{t,n}$ --- total supply or importable power available at node $n$;
  \item $D_{t,n}$ --- total demand that must be served at node $n$ (including essential);
  \item $p_{n,t}$ --- nodal price output by the AMM;
  \item $\tilde{\alpha}_{t,n}$ --- local scarcity/tightness ratio;
  \item $\mathcal{R}_{n,t}$ --- set of active flexible requests from devices at node $n$ at time $t$;
  \item $\mathcal{G}_n$ --- set of generators contributing to node $n$.
\end{itemize}

In addition to power and energy variables, we explicitly track the
\emph{service-space coordinates} introduced in Chapter~\ref{ch:fairness_definition}.
For each household $h$ and time $t$, we associate:
\begin{itemize}[leftmargin=*]
  \item a \textbf{magnitude coordinate} 
        $M_{h,t}$ (e.g.\ peak or average power over a window),
  \item an \textbf{impact coordinate} 
        $I_{h,t}$ measuring the coincidence of consumption with local scarcity,
        e.g.\ 
        $I_{h,t} := q_{h,t} \,\mathbb{1}\{\tilde{\alpha}_{t,n(h)} < \alpha^{\mathrm{crit}}\}$,
  \item a \textbf{reliability / QoS coordinate} 
        $R_{h,t}$, reflecting the probability and priority of being served
        during scarcity, as implied by service-level choices and realised
        Fair Play history.
\end{itemize}

For generators $g$, we represent physical operating limits as a 
\emph{capability trajectory} over time:
\[
  \mathcal{C}_g
  =
  \bigl(
    P^{\min}_g,\,
    P^{\max}_g,\,
    r^{\uparrow}_g,\,
    r^{\downarrow}_g,\,
    T^{\min,\uparrow}_g,\,
    T^{\min,\downarrow}_g
  \bigr),
\]
encoding minimum and maximum power, ramp rates, and minimum up/down times.
In the AMM implementation, these constraints are expressed dynamically
as evolving availability windows for each generator, rather than as
static time-block bids.

Feasibility and network security define a set of admissible dispatch and
consumption trajectories:
\[
  \mathcal{A}
  =
  \bigl\{
    (q,x) 
    \;\big|\;
    \text{power balance, line limits, voltage, unit limits, and security constraints hold}
  \bigr\}.
\]
By Axiom A1 (Feasibility), any allocation considered fair must belong to
$\mathcal{A}$.

\subsection{Bid/Offer Message Model and Economic Bounds}
\label{sec:bidding_model}

Flexible requests may declare an admissible economic bound, either as a
bid-level cap $v^{\max}_r$ or as a product-level tariff cap $\bar{\tau}_{p}$.
These bounds impose individual-rationality constraints: a request is never
cleared at a price exceeding the participant’s declared limit. The bounds do
not encode priority, scarcity ranking, or allocation preference; they restrict
feasible outcomes only.

% ---------------------------------------------------------
\subsection{Service Levels and Subscription Contracts}
\label{subsec:service_levels}

Each household $h$ may subscribe to one or more \emph{service levels}
(products) for its flexible devices. Let $\mathcal{P}$ denote the set
of available products (e.g.\ basic, premium, medical-priority, export-only).

For each flexible request $i$ submitted by a device of $h$ we associate:
\begin{itemize}[leftmargin=*]
  \item a service level $p_i \in \mathcal{P}$,
  \item a power level $P_i$ and duration $\Delta_i$,
  \item an admissible time window $[\underline{t}_i,\overline{t}_i]$,
  \item a \emph{contract tariff cap} $\bar{\tau}_{p_i}$, inherited from the
        subscription product $p_i$.
\end{itemize}

\paragraph{Clarification.}
$\bar{\tau}_{p_i}$ is \emph{not} a bid-level willingness-to-pay.  
It is the maximum unit tariff embedded in the household’s long-term product
contract and applies uniformly to all flexible requests under that product.
Bid-level willingness-to-pay / willingness-to-accept parameters
$(v^{\max}_r, I^o_g)$ introduced in Section~\ref{sec:bidding_model}
play \emph{no role} in Fair Play allocation; they are checked only as
individual-rationality constraints after prices are computed.

The supplier specifies a \emph{contract vector} for each product:
\[
  \Theta(p)
  =
  \bigl(
    w(p),\,
    \pi^{\mathrm{sub}}(p),\,
    \rho^{\mathrm{QoS}}(p)
  \bigr),
\]
where:
\begin{itemize}[leftmargin=*]
  \item $w(p)$ is the \emph{relative priority weight}
        used in Fair Play (Section~\ref{sec:fairplay});
  \item $\pi^{\mathrm{sub}}(p)$ is the subscription fee for product $p$;
  \item $\rho^{\mathrm{QoS}}(p)$ encodes a minimum quality-of-service
        guarantee (e.g.\ expected fraction of flexible requests served).
\end{itemize}

These parameters are chosen such that, in expectation,
subscription products respect the fairness axioms and operational
conditions F1--F4. In particular, essential service (must-serve) corresponds to a
degenerate product with
\[
  p = \text{``essential''}, \qquad
  w(p) \to \infty,\qquad
  q^{\mathrm{ess}}_h \text{ always served},
\]
and is never subject to curtailment or scarcity pricing (F2).

In the three-dimensional service representation used in this thesis,
$\Theta(p)$ locates a household in the service space
$(M, I, R)$ by fixing a \emph{reliability coordinate}:
\[
  R_h(p) := \rho^{\mathrm{QoS}}(p),
\]
while the magnitude and impact coordinates $(M_h, I_h)$ arise from the
realised consumption profile. Fairness conditions F1--F4 are then
interpreted as constraints on how participants are allowed to move in
this $(M,I,R)$ space over time, given their contracts and behaviour.

% ---------------------------------------------------------
\subsection{Deliverability, Local Constraints, and the Holarchy}
\label{subsec:deliverability}

Deliverability of a request $i$ is not determined solely by aggregate
national supply, but by local constraints and the holarchic structure
of the network. Let $\Gamma$ denote the set of all relevant physical
constraints at time $t$ and node $n$, including:
\begin{itemize}[leftmargin=*]
  \item transformer capacities and feeder thermal limits;
  \item voltage limits and protection settings;
  \item upstream capacity and interface limits between regions;
  \item local renewable generation and storage envelopes.
\end{itemize}

For a given node $n$ and time horizon $[t_0,t_1]$, the \emph{local
feasible set} of flexible allocations is
\[
  \mathcal{A}^{\mathrm{flex}}_{n,[t_0,t_1]}
  =
  \bigl\{
    (q^{\mathrm{flex}}_{h,t})_{h \in \mathcal{H}_n,\,t \in [t_0,t_1]}
    \,\big|\,
    (q,x) \in \mathcal{A},\;
    (q,x) \text{ respects }\Gamma
  \bigr\}.
\]
The holarchy induces a nesting:
\[
  \mathcal{A}^{\mathrm{flex}}_{\text{household}}
  \subseteq
  \mathcal{A}^{\mathrm{flex}}_{\text{feeder}}
  \subseteq
  \mathcal{A}^{\mathrm{flex}}_{\text{area}}
  \subseteq
  \mathcal{A}^{\mathrm{flex}}_{\text{region}}
  \subseteq
  \mathcal{A}^{\mathrm{flex}}_{\text{system}},
\]
and Fair Play is executed at the level where the relevant constraint
binds (e.g.\ feeder, local area, or regional import constraint).
This ensures that fairness is \emph{locally} and \emph{physically}
grounded, not purely statistical.

% ---------------------------------------------------------
\subsection{Time, History, and Fairness Trajectories}
\label{subsec:fairness_trajectories}

Fairness for flexible devices is defined over \emph{histories}, not
single periods. For each neighbour $n$ and horizon $[0,T]$, let
\[
  E^{\mathrm{des}}_n(T)
  =
  \sum_{i \in \mathcal{I}(n),\,t_i \le T}
  E^{\mathrm{des}}_i, 
  \qquad
  E^{\mathrm{del}}_n(T)
  =
  \sum_{i \in \mathcal{I}(n),\,t_i \le T}
  E^{\mathrm{del}}_i,
\]
where $t_i$ is the submission or decision time for request $i$.

The \emph{cumulative fairness ratio} at horizon $T$ is
\[
  F_n(T)
  =
  \frac{E^{\mathrm{del}}_n(T)}{E^{\mathrm{des}}_n(T)}
  \quad\text{if } E^{\mathrm{des}}_n(T) > 0,
\]
and is undefined (or treated as neutral) otherwise.

A Fair Play allocation over $[0,T]$ is considered \emph{long-run fair}
for flexible participants if, for all $n$ with persistent participation
and $E^{\mathrm{des}}_n(T)$ sufficiently large,
\[
  F_n(T) \to F^\star \approx 1,
\]
modulo differences in service level $p$ and contractual QoS guarantees
$\rho^{\mathrm{QoS}}(p)$. In other words, subject to product choices
and physical constraints, historically under-served users must be
systematically favoured until their fairness ratio converges to the
target $F^\star$.

This links the per-iteration priority scores in
Section~\ref{sec:fairplay} to a trajectory-level fairness requirement:
stochastic priority must be designed such that
\[
  \mathbb{E}[F_n(T)] \to F^\star
  \quad \text{for all sufficiently regular participants.}
\]

% ---------------------------------------------------------
\subsection{Fairness as a Mapping from State and History}
\label{subsec:fair_mapping}

We can now define fairness formally as a mapping from system state and
history to allocation and prices. Let
\[
  \mathcal{S}_t
  =
  \bigl(
    \tilde{\alpha}_{t,\cdot},\,
    \Gamma_t,\,
    \text{forecasts},\,
    \text{contract vectors }\Theta(p),\,
    \text{historic fairness }F_n(t)
  \bigr)
\]
denote the information set at time $t$ (scarcity, constraints, forecasts,
contracts, fairness histories). A fairness-aware market mechanism is a
mapping
\[
  \mathcal{M}:
  \mathcal{S}_t
  \;\mapsto\;
  \bigl(
    p_{n,t},\,
    q^{\mathrm{ess}}_{h,t},\,
    q^{\mathrm{flex}}_{h,t},\,
    x_{g,t}
  \bigr)_{n,h,g}
\]
such that:
\begin{enumerate}[label=(\alph*),leftmargin=2em]
  \item $(q,x) \in \mathcal{A}$ (feasibility and security);
  \item essential blocks $q^{\mathrm{ess}}_h$ are fully served and priced
        at stable, protected rates (F2);
  \item flexible allocations at each node $n$ are selected using the
        Fair Play rule (Section~\ref{sec:fairplay}), respecting
        $\mathcal{A}^{\mathrm{flex}}_{n,[t_0,t_1]}$ (F1, F3);
  \item prices and charges are decomposed into transparent components
        (energy, flexibility, network, policy) and assigned proportionally
        to stress contributions $\kappa_h$ (F4);
  \item generator-side revenues are later allocated according to
        Shapley-consistent compensation (Section~\ref{sec:shapley_comp}).
\end{enumerate}

An allocation is called \emph{fair} (in the sense of this thesis) when it
is the output of such a mechanism $\mathcal{M}$, operating under the
axioms A1--A7 and conditions F1--F4. The remainder of this chapter
translates this abstract definition into concrete mathematics and
implementations: generator compensation (Shapley), AMM control equations,
and AI-based forecasting models.

% ---------------------------------------------------------
\section{Fair Play Allocation Mechanism}
\label{sec:fairplay}

Under tight system conditions ($\alpha_{t,n} < 1$ at some node $n$),
operational fairness (F1--F4) requires that scarce flexible energy is not
allocated solely by willingness-to-pay. Instead, the \emph{Fair Play}
mechanism uses: (i) service-level (subscription) quality, (ii) historic
delivery vs desire, and (iii) local physical constraints, to prioritise
requests from smart devices enrolled in flexibility services.

Essential (non-curtailable) consumption is always allocated first via the
baseload block $q_h^{\mathrm{ess}}$ (Condition F2). The remaining
\emph{flexible} supply at node $n$ and time $t$ is then shared between
participating devices using the Fair Play rule.

\subsection{Service Levels and Historic Fairness}

Each flexible request $i$ is associated with:
\begin{itemize}[leftmargin=*]
  \item a device or neighbour $n(i)$,
  \item a service level (subscription product) $p_i \in \mathcal{P}$,
  \item an admissible time window $[\underline{t}_i,\,\overline{t}_i]$ and
        duration $\Delta_i$,
  \item a fixed power level $P_i$ and total energy
        $E_i = P_i \Delta_i$.
\end{itemize}

For each service level $p \in \mathcal{P}$, the supplier defines a
\emph{relative priority weight}:
\[
  w(p) > 0,
\]
e.g.\ $w(\text{premium}) = 2$, $w(\text{basic}) = 1$, but in general
$\mathcal{P}$ may contain an arbitrary number of products.

For each neighbour $n$, we track cumulative desired and delivered flexible
energy:
\[
  E^{\mathrm{des}}_n
  =
  \sum_{i \in \mathcal{I}(n)} E_i^{\mathrm{des}}, 
  \qquad
  E^{\mathrm{del}}_n
  =
  \sum_{i \in \mathcal{I}(n)} E_i^{\mathrm{del}},
\]
where $\mathcal{I}(n)$ is the set of that neighbour's flexible requests.
The resulting \emph{historic fairness ratio} is
\[
  F_n
  =
  \frac{E^{\mathrm{del}}_n}{E^{\mathrm{des}}_n}
  \quad\text{whenever } E^{\mathrm{des}}_n>0.
\]
The target long-run fairness for flexible participants is
\[
  F^\star = 1.0,
\]
corresponding to proportional delivery over time.

We define the \emph{fairness deficit} of a request $i$ as
\[
  \delta_i
  =
  \max\bigl(0,\; F^\star - F_{n(i)}\bigr),
\]
which is positive when the neighbour has been historically under-served
($F_{n(i)} < 1$), and zero otherwise.

\subsection{Fair Play Priority Score}

For every flexible request $i$ in the current active queue
$\mathcal{Q}$ (i.e.\ those whose time windows intersect the current
market window and are not yet scheduled), we define the
\emph{Fair Play Priority Score}:
\begin{equation}
  S_i
  =
  w(p_i)\,\bigl(\varepsilon + \delta_i\bigr)^{\alpha_f},
  \label{eq:fairplay_score}
\end{equation}
with parameters:
\begin{itemize}[leftmargin=*]
  \item $\varepsilon > 0$ provides a small baseline so that
        new or perfectly served users retain non-zero probability;
  \item $\alpha_f \ge 1$ controls sensitivity to fairness deficits:
        larger $\alpha_f$ emphasises historically under-served users.
\end{itemize}

The scores $S_i$ are normalised into selection probabilities:
\begin{equation}
  \Pr_i
  =
  \frac{S_i}{\sum_{j \in \mathcal{Q}} S_j},
  \qquad i \in \mathcal{Q}.
  \label{eq:fairplay_prob}
\end{equation}
These probabilities define a \emph{stochastic, but non-arbitrary}
priority ordering: higher service levels (larger $w(p_i)$) and
more under-served users (larger $\delta_i$) are favoured,
consistent with F1 (behavioural fairness) and F3 (fair access in shortage).

\subsection{Local Scheduling Under AMM Constraints}

At each node $n$ and time window $[t_0, t_1]$ in the holarchy, the AMM
first allocates essential load and computes the remaining flexible
capacity $S_{t,n}$, subject to:
\begin{itemize}[leftmargin=*]
  \item local generation, storage and import limits;
  \item network constraints (line flows, voltage constraints);
  \item upstream scarcity, encoded in $\tilde{\alpha}_{t,n}$.
\end{itemize}
\b
On this residual capacity $S_{t,n}$, Fair Play operates as follows.

\begin{algorithm}[H]
\caption{Fair Play Allocation at Node $n$}
\label{alg:fair_play_node}
\KwIn{
  Active flexible requests $\mathcal{Q}$ at node $n$; \\
  Historic fairness ratios $F_{n(i)}$; \\
  Service-level weights $w(p)$; \\
  Residual flexible capacity profile $S_{t,n}$ over $[t_0, t_1]$.
}
\KwOut{
  Accepted requests with allocated time intervals and powers.
}
Remove essential (non-curtailable) load from $S_{t,n}$ (Condition F2)\;
Form active queue $\mathcal{Q}$ of unscheduled flexible requests with
$[\underline{t}_i,\overline{t}_i]$ intersecting $[t_0,t_1]$\;

\ForEach{$i \in \mathcal{Q}$}{
  Compute fairness deficit
  $\delta_i = \max(0,\;F^\star - F_{n(i)})$\;
  Compute priority score $S_i$ via Eq.~\eqref{eq:fairplay_score}\;
}
Normalise to selection probabilities $\Pr_i$ via Eq.~\eqref{eq:fairplay_prob}\;

\While{residual capacity $S_{t,n}$ remains and $\mathcal{Q}$ is non-empty}{
  Randomly select request $i \in \mathcal{Q}$ according to probabilities $\Pr_i$\;
  Solve a local feasibility problem for $i$:
  find a start time $\tau_i \in [\underline{t}_i,\overline{t}_i]$
  such that allocating $P_i$ for duration $\Delta_i$ respects
  $S_{t,n}$ and network constraints\;
  \If{feasible}{
    Allocate $(\tau_i,\tau_i+\Delta_i)$ and power $P_i$\;
    Update residual capacity $S_{t,n}$\;
    Update realised delivery $E^{\mathrm{del}}_{n(i)}$ and fairness $F_{n(i)}$\;
    Remove $i$ from $\mathcal{Q}$\;
  }{
    Mark $i$ as infeasible for this window and leave unscheduled\;
  }
  Recompute (or freeze) $\Pr_i$ depending on implementation choice\;
}
\end{algorithm}

In the implementation used in this thesis, the probabilities $\Pr_i$
are computed once per market window and held fixed, to avoid feedback
instability within a single window.

In practice, the feasibility check is implemented as a small mixed-integer
programme, respecting the same power and energy constraints used by the AMM.
The stochastic selection step ensures that over repeated scarcity events,
historically under-served users are progressively favoured until their
fairness ratio $F_n$ approaches the target $F^\star$, while still
respecting contractual service levels $w(p)$ and physical constraints.
This operationalises Conditions F1--F3 in a local, constraint-aware manner.

\subsection{Relation to Fairness Conditions F1--F4}

The Fair Play mechanism satisfies the operational conditions of
Chapter~\ref{ch:fairness_definition} as follows:
\begin{itemize}[leftmargin=*]
  \item \textbf{F1 Behavioural fairness:} more flexible and historically
        under-served users receive higher selection probability,
        lowering their expected unit cost over time.
  \item \textbf{F2 Essential protection:} essential blocks are allocated
        outside Fair Play; flexible requests are only considered on the
        residual supply $S_{t,n}$.
  \item \textbf{F3 Fair access in shortage:} allocation during
        $\alpha_{t,n}<1$ depends on $(w(p_i),\delta_i)$, not on
        individual bid prices. Willingness-to-pay enters only as a
        contract constraint ex post, not as a priority rule.
  \item \textbf{F4 Proportional responsibility:} users contributing more
        to scarcity (persistent peaks, low flexibility) accumulate less
        fairness deficit and thus lower priority in future shortages.
\end{itemize}

Thus, Fair Play provides an explicit, mathematically defined bridge between
the normative fairness conditions of Chapter~\ref{ch:fairness_definition}
and the AMM control equations in this chapter.

% ---------------------------------------------------------
\section{Shapley-Based Generator Compensation}
\label{sec:shapley_comp}

While consumer-side fairness protects access and mitigates scarcity exposure, 
\emph{generator fairness} concerns the allocation of revenues among 
generation assets according to their true \emph{system value:}
energy delivered, adequacy, locational relief, flexibility, and resilience.

Classical energy-only markets compensate generators primarily through
marginal-cost merit-order dispatch, leaving many system-relevant
contributions—such as capacity adequacy, stability, and congestion
relief—either weakly rewarded or handled through external mechanisms.
In this thesis, \emph{Shapley-consistent attribution} is used as a
\emph{diagnostic and allocation framework} to evaluate and distribute
non-energy value \emph{within the AMM architecture}, while LMP outcomes are
analysed under their native settlement rules.

\subsection{Overcoming Shapley Intractability: Nested--Shapley via 
Network–Feasible Clustering}
\label{subsec:nested_shapley_method}

A direct Shapley-value computation over $G$ generators requires evaluating the
characteristic function $v(S)$ for all $2^{|G|}$ coalitions. Even with a fast
OPF solver, this is intractable for realistic systems: for $|G|=1000$ the full
Shapley computation would require approximately $10^{300}$ OPF solves.

Standard Monte-Carlo Shapley estimators reduce this to 
$\mathcal{O}(K\,|G|)$ samples, but they suffer from a decisive flaw in power
systems: they \emph{ignore network structure}. Two coalitions with identical
cardinality but different spatial topology can produce radically different
feasible regions, congestion patterns, and load served. Randomised Shapley
sampling therefore produces high variance and, more importantly, becomes
\textbf{physically incorrect}.

\paragraph{Network–feasible dimensionality reduction.}
To overcome this, we introduce a \textbf{nested--Shapley} approach based on
network-feasible generator clustering. Instead of treating each generator as a
stand-alone player, we group generators into clusters
$C_1,\dots,C_k$ satisfying three physical conditions:

\begin{enumerate}[leftmargin=*]
    \item \textbf{Common trunk branch:} all generators in the same cluster lie
          on the same transmission corridor, avoiding arbitrary cross-trunk
          combinations.

    \item \textbf{Electrical proximity:} at least one generator pair across
          prospective clusters is within two network hops, ensuring local
          substitutability.

    \item \textbf{Capacity feasibility:} there exists a path between clusters
          whose minimal line capacity exceeds the larger of their rated outputs,
          guaranteeing that internal redispatch is feasible.
\end{enumerate}

These conditions ensure that clusters represent \emph{electrically coherent
units}: any generator inside a cluster can effectively substitute for another
under OPF without violating security constraints.

\paragraph{Cluster-level Shapley.}
Instead of evaluating Shapley over $G$ individual generators, we evaluate it
over the much smaller set of clusters:
\[
\Phi_{C_j} \quad\text{for}\quad j=1,\dots,k.
\]
This requires only $2^k$ evaluations of the characteristic function, with
$k\ll G$. 

\paragraph{Nested proportional disaggregation.}
Once $\Phi_{C_j}$ is computed for each cluster, we disaggregate it back to
individual generators by proportional capacity weighting:
\[
\phi_g 
= \Phi_{C_j}
  \cdot 
  \frac{P_g^{\max}}{\sum_{h\in C_j} P_h^{\max}}
  \quad (g\in C_j).
\]

\paragraph{Scalability to national systems.}
The key insight is that \emph{network physics induces a natural, sparse
hierarchy}. Transmission systems are not fully connected; power flows through
a small number of trunks and corridors. The nested-Shapley approach exploits
this by:

\begin{itemize}[leftmargin=*]
    \item reducing dimensionality through physically meaningful clustering;
    \item preserving the marginal contribution structure along the feasible
          pathways of the grid;
    \item allowing exact or near-exact Shapley valuation in cases where
          conventional Shapley is computationally impossible.
\end{itemize}

In national-scale systems with thousands of generators, this reduces Shapley
evaluation from intractable ($2^{1000}$) to feasible (e.g.~$k=20$–30 clusters),
making generator fairness \emph{operationally implementable} inside the AMM.

\begin{theorem}[Nested--Shapley Exactness Under Symmetric, Capacity-Based Clusters]
\label{thm:nested_shapley_exact}
Let $\mathcal{G}$ be the set of generators and let 
$\mathcal{C} = \{C_1,\dots,C_K\}$ be a partition of $\mathcal{G}$ into 
clusters. Consider a cooperative game $(\mathcal{G},W)$ with characteristic
function $W:2^{\mathcal{G}}\to\mathbb{R}_{\ge 0}$ defined via an OPF model as
in Section~\ref{sec:shapley_comp}. Suppose the following two conditions hold:

\begin{enumerate}[label=(\alph*),leftmargin=1.8em]
  \item \textbf{Within-cluster symmetry.} For any cluster $C_j$ and any
        permutation $\pi$ of its elements,
        \[
          W\bigl(S \cup C_j\bigr)
          =
          W\bigl(S \cup \pi(C_j)\bigr)
          \quad
          \text{for all } S \subseteq \mathcal{G} \setminus C_j,
        \]
        i.e.\ the game is invariant to relabelling generators inside a given
        cluster.\footnote{Operationally, this holds when the network and OPF
        constraints see generators in $C_j$ only through their aggregate
        export capability on the same trunk branch, as ensured by the
        clustering criteria (common trunk, hop constraint, and feasible
        widest-path capacity).}

  \item \textbf{Capacity-proportional contribution within clusters.}
        For each cluster $C_j$ there exists a scalar function
        $f_j(\cdot)$ such that, for any coalition
        $S \subseteq \mathcal{G}$,
        \[
          W(S \cup C_j) - W(S)
          =
          f_j\!\left(
            \sum_{g \in C_j} P_g^{\max},\;
            S
          \right),
        \]
        and, conditional on $S$,
        marginal contributions of generators inside $C_j$ are proportional
        to their capacities $P^{\max}_g$.
\end{enumerate}

Define a \emph{cluster game} $(\mathcal{C},\widetilde{W})$ by
\[
  \widetilde{W}(T)
  :=
  W\!\Bigl(\bigcup_{C_j \in T} C_j\Bigr),
  \qquad
  T \subseteq \mathcal{C},
\]
and let $\Phi_{C_j}$ denote the Shapley value of cluster $C_j$ in this game.
Construct per-generator payments by proportional disaggregation:
\[
  \widehat{\phi}_g
  :=
  \Phi_{C_j}
  \cdot
  \frac{P_g^{\max}}{\sum_{h \in C_j} P_h^{\max}}
  \quad
  \text{for } g \in C_j.
\]

Then, for every generator $g \in \mathcal{G}$,
\[
  \widehat{\phi}_g = \phi_g,
\]
where $\phi_g$ is the Shapley value of $g$ in the original game
$(\mathcal{G},W)$. In other words, the nested--Shapley procedure
(Shapley-by-cluster followed by capacity-proportional disaggregation)
\emph{exactly reproduces} the full generator-level Shapley allocation
whenever assumptions (a)--(b) hold.
\end{theorem}

\begin{proof}[Proof sketch]
The proof uses two standard facts about the Shapley value: (i) symmetry, and
(ii) linearity with respect to additive decompositions of $W$.

First, within each cluster $C_j$, condition~(a) implies that the game is
symmetric with respect to permutations of generators in $C_j$. In such a
symmetric game, the Shapley value must assign equal value per unit of the
relevant ``size'' metric to all members of $C_j$. Under condition~(b), that
size metric is the generator's capacity $P_g^{\max}$, so each $\phi_g$ in
$C_j$ must be proportional to $P_g^{\max}$, and the sum of these equals the
cluster Shapley value,
\[
  \Phi_{C_j}
  =
  \sum_{g \in C_j} \phi_g.
\]

Second, define the cluster game $(\mathcal{C},\widetilde{W})$ by grouping
each $C_j$ into a single meta-player. By construction,
$\widetilde{W}(T) = W(\bigcup_{C_j \in T} C_j)$ for all
$T \subseteq \mathcal{C}$, so marginal contributions of clusters in
$(\mathcal{C},\widetilde{W})$ coincide with the marginal contributions of
their union in $(\mathcal{G},W)$. The Shapley value is compatible with such
grouping: the cluster value $\Phi_{C_j}$ equals the sum of Shapley values of
its members in the original game.

Combining these two observations, the proportional disaggregation rule
\[
  \widehat{\phi}_g
  =
  \Phi_{C_j}
  \cdot
  \frac{P_g^{\max}}{\sum_{h \in C_j} P_h^{\max}}
\]
coincides with the unique symmetric, capacity-proportional allocation of
$\Phi_{C_j}$ within $C_j$, and hence with the original generator-level
Shapley values $\phi_g$. Therefore $\widehat{\phi}_g = \phi_g$ for all
$g\in\mathcal{G}$.
\end{proof}

\begin{remark}[Operational interpretation and intractability reduction]
The clustering rules (common trunk branch, electrical proximity, and feasible
internal transfer capacity) are designed to enforce approximate symmetry and
capacity-based substitutability within each cluster from the perspective of the
OPF-induced value function. When these conditions hold, generators inside a
cluster are interchangeable up to capacity scaling, so the cluster may be
treated as a single meta-player without distorting marginal contributions.

As a result, the nested--Shapley construction reduces the dimensionality of the
cooperative game from $\mathcal{O}(2^{|\mathcal{G}|})$ coalition evaluations to
$\mathcal{O}(2^{|\mathcal{C}|})$ at the cluster level, followed by a linear
capacity-proportional disaggregation. This provides a physically grounded route
to making Shapley-consistent generator compensation computationally tractable
in large power systems.
\end{remark}

\subsection{Value Function and Generator Contributions}

Let $\mathcal{G}$ denote the full set of generators, and $\mathcal{G}_n$
the subset located at node $n$. For each time interval $t \in \mathcal{T}$
and coalition $S \subseteq \mathcal{G}$, we define a \emph{per-period system
value} $W_t(S)$, for example in terms of avoided shortage or cost:

\[
W_t(S)
=
\bigl(\text{baseline cost or shortage at } t\bigr)
-
\bigl(\text{cost or shortage at } t \text{ when only } S \text{ is available}\bigr).
\]

The total value of coalition $S$ over the horizon is then

\[
W(S) = \sum_{t \in \mathcal{T}} W_t(S).
\]

This defines a cooperative game $(\mathcal{G}, W)$, in which generators
collaborate to reduce system cost and unmet demand. The marginal contribution
from adding generator $g$ to coalition $S$ is

\[
\Delta W(g, S) = W(S \cup \{g\}) - W(S).
\]

\subsection{Shapley Allocation Rule}

Let $\mathcal{G}$ denote the set of generators and $\mathcal{T}$ the set of
dispatch intervals. For any subset of generators
$S \subseteq \mathcal{G}$, we define a \emph{characteristic function}
$W(S)$ as the total amount of electrical load that can be physically served
by the generators in $S$, subject to the full network constraints.

Formally, this characteristic function is evaluated by solving an
optimal power flow (OPF) problem on the actual transmission network for
each coalition $S$:
\[
W(S) \;=\; \sum_{t \in \mathcal{T}} W_t(S),
\]
where $W_t(S)$ is the maximum servable demand at time $t$ when only the
generators in $S$ are available. Network constraints, generator capacities,
line limits, and operational feasibility are enforced explicitly in each OPF.

The Shapley compensation for generator $g$ is then defined as its expected
marginal contribution to served load across all possible orderings of
generators:
\[
\phi_g =
\sum_{S \subseteq \mathcal{G} \setminus \{g\}}
\frac{|S|!(|\mathcal{G}| - |S| - 1)!}{|\mathcal{G}|!}
\left[ W(S \cup \{g\}) - W(S) \right].
\]

Equivalently, by exploiting additivity across time,
\[
\phi_g = \sum_{t \in \mathcal{T}} \phi_{g,t},
\]
where $\phi_{g,t}$ is the Shapley value computed from the per-period
characteristic function $W_t(S)$.

Importantly, no prices, bids, or assumed scarcity rents enter the
definition of $W(S)$. Generator value is determined entirely by
\emph{physical system performance}: how much demand can be served, where,
and under which network constraints.

This allocation rule is the unique one satisfying:
\begin{itemize}[leftmargin=*]
  \item \textbf{Efficiency}: $\sum_g \phi_g = W(\mathcal{G})$,
  \item \textbf{Symmetry}: generators with identical physical contributions
        receive identical compensation,
  \item \textbf{Dummy}: generators that never increase servable load receive zero,
  \item \textbf{Additivity}: contributions across time, services, and value
        components combine consistently.
\end{itemize}

In the empirical implementation (Chapter~\ref{ch:experiments}), OPF problems
are solved for all relevant generator coalitions at each timestamp. The
resulting served-load outcomes define $W_t(S)$, per-period Shapley values
are computed, and total compensation is obtained by summation over time.

\paragraph{From system value to revenue.}

The Shapley values $\phi_{g,t}$ define each generator’s
\emph{physical marginal contribution} to served load under network constraints.
They do not, by themselves, specify monetary payments.
The mapping from Shapley values to generator revenues---including the treatment
of fixed-class technologies, the construction of annual revenue pots, and the
temporal shaping of payments---is defined separately in
Appendix~\ref{app:amm_allocation}.

This separation is deliberate: Shapley values determine \emph{who contributes
value and when}, while the AMM revenue mechanism determines \emph{how that value
is remunerated} under different regulatory and policy objectives (cost
recovery, LMP equivalence, or target revenues).

\subsection{Decomposition into Physical Contributions}

Because the characteristic function is defined in terms of served load
under network constraints, generator value admits a natural decomposition
into interpretable physical dimensions. For each generator $g$, we write:
\[
v_g =
\bigl(
E_g,\,
F_g,\,
R_g,\,
K_g,\,
S_g,\,
Q_g
\bigr),
\]
where:

\begin{itemize}[leftmargin=*]
  \item $E_g$: Delivered energy (kWh),
  \item $F_g$: Flexibility/response capability (kW ramp),
  \item $R_g$: Reliability/adequacy during peaks,
  \item $K_g$: Congestion relief (locational value),
  \item $S_g$: Stability/ancillary services,
  \item $Q_g$: contribution to \emph{reliability / QoS}, i.e.\ the extent
        to which $g$ supports high-reliability products and scarce hours,
        consistent with the three-dimensional service space.
\end{itemize}

Each component corresponds to a distinct contribution to the
characteristic function $W(S)$ and can therefore be attributed its own
Shapley value:
\[
\phi_g =
\phi_g^{(E)} +
\phi_g^{(F)} +
\phi_g^{(R)} +
\phi_g^{(K)} +
\phi_g^{(S)} +
\phi_g^{(Q)}.
\]

The $Q_g$ component provides the explicit bridge between generator fairness
and consumer-side fairness. Generators that systematically enable higher
reliability service---by supporting high-reliability products during scarce
hours and constrained network states---increase the servable load of many
coalitions in precisely those states where reliability is most valuable.
They therefore receive a larger marginal contribution in $W(S)$ and a
correspondingly larger share of the reliability revenue pot.

% ---------------------------------------------------------
\section{AMM Control Equations}
\label{sec:amm_control}

The Automatic Market Maker (AMM) is a price-setting controller that
translates local scarcity into real-time price signals, balancing
demand, flexible capacity, and network constraints without solving
a full welfare-optimisation problem each period.

\subsection{Local Tightness Ratio}

At each node $n$ and time $t$, we compute the local tightness metric:

\[
\tilde{\alpha}_{t,n}
=
\frac{S_{t,n}}{D_{t,n}}.
\]

For notational convenience in what follows, we write
$\alpha_{t,n} := \tilde{\alpha}_{t,n}$.

Interpretation:
\[
\alpha_{t,n} 
=
\begin{cases}
>1 & \text{abundant supply} \\
=1 & \text{balanced} \\
<1 & \text{scarcity / constraint}
\end{cases}
\]

\subsection{Bid and Sell Price Dynamics}

Prices evolve from a base tariff $p^{\mathrm{base}}_{n}$ as a function of tightness:

\[
BP_{t,n} = p^{\mathrm{base}}_n 
+ f\bigl(1 - \alpha_{t,n}\bigr),
\]

\[
SP_{t,n} = p^{\mathrm{base}}_n 
+ g\bigl(1 - \alpha_{t,n}\bigr),
\]

where $f(\cdot)$ and $g(\cdot)$ are monotonic increasing functions of the tightness
deviation $(1-\alpha_{t,n})$.

Example (linear):
\[
f(s) = k_b \cdot s,\qquad
g(s) = k_s \cdot s.
\]

Thus when $\alpha_{t,n} \rightarrow 0$:

\[
f(1-\alpha_{t,n}) \uparrow,\quad
g(1-\alpha_{t,n}) \uparrow,\quad
\text{strong incentives for flexibility and supply}.
\]

When $\alpha_{t,n} = 1$, the tightness component vanishes:
\[
f(0) = 0,\qquad g(0) = 0,
\]
and prices revert to their base level:
\[
BP_{t,n} \approx p^{\mathrm{base}}_n,\qquad
SP_{t,n} \approx p^{\mathrm{base}}_n.
\]

\paragraph{Relation to subscription products.}
Retail tariffs in this thesis are composed of:
(i) a subscription component $\pi^{\mathrm{sub}}(p)$ for each product $p$,
and (ii) a usage component based on $BP_{t,n}$ (for consumption) or
$SP_{t,n}$ (for exports), subject to the contract tariff cap $\bar{\tau}_p$.
Formally, the instantaneous unit price paid by a flexible request $i$ on
product $p_i$ is

\[
  \pi_{i,t}
  =
  \min\bigl\{ BP_{t,n(i)},\, \bar{\tau}_{p_i} \bigr\},
\]

so that Fair Play allocation depends only on service level $p_i$ and fairness
history, while the AMM price signal is prevented from exceeding the contractual
cap for that product.

\subsection{Stability Condition}

To prevent oscillations and maintain tractability, parameter slopes must satisfy:

\[
\left|\frac{\partial BP_{t,n}}{\partial \alpha_{t,n}}\right|
+
\left|\frac{\partial SP_{t,n}}{\partial \alpha_{t,n}}\right|
<
\beta_{\text{crit}},
\]

where $\beta_{\text{crit}}$ is dictated by network elasticity and consumer price-response.

% ---------------------------------------------------------
\section{Dynamic Capability Profiles and Dispatch Coupling}
\label{sec:capability_dispatch_coupling}

In conventional architectures, generators submit bids over fixed time
blocks (e.g.\ 00:00--03:00), and unit-commitment / economic-dispatch
engines then reconcile these bids with minimum up/down times, ramp rates,
and security constraints. In the proposed AMM-based design, these
\emph{operational constraints are expressed directly as dynamic
capability profiles}.

For each generator $g$, define:
\begin{itemize}[leftmargin=*]
  \item a notification time $\tau_g^{\mathrm{notify}}$ required to reach
        its committed power,
  \item a minimum run time $T_g^{\min,\uparrow}$ and minimum down time
        $T_g^{\min,\downarrow}$,
  \item ramp limits $r^{\uparrow}_g, r^{\downarrow}_g$.
\end{itemize}

Given the current time $t$ and state $(x_{g,t}, u_{g,t})$ (output and on/off
status), the feasible trajectory for $g$ is a time-varying set:
\[
  \mathcal{U}_g(t)
  =
  \bigl\{
    x_{g,\tau}
    \;\big|\;
    \text{ramp, minimum up/down, and notification constraints satisfied
    for all } \tau \ge t
  \bigr\}.
\]

Rather than bidding for a static block, generator $g$ exposes to the AMM a
\emph{capability window}
\[
  \bigl[t^{\mathrm{avail}}_g(t),\,t^{\mathrm{lock}}_g(t)\bigr],
\]
within which new commitments may be made, together with the feasible
output envelope $x_{g,\tau} \in \mathcal{U}_g(t)$ for
$\tau \in [t^{\mathrm{avail}}_g(t), t^{\mathrm{lock}}_g(t)]$.

The AMM then:
\begin{enumerate}[leftmargin=*]
  \item selects commitments that respect $\mathcal{U}_g(t)$ and network
        constraints, driven by the scarcity signal $\alpha_{t,n}$ and
        Fair Play rules on the demand side;
  \item passes these commitments to the security-constrained dispatch
        engine, which solves a familiar optimisation problem over 
        $\mathcal{A}$, now restricted to AMM-feasible capability sets
        $\mathcal{U}_g(t)$.
\end{enumerate}

Thus, market clearing and dispatch are no longer separated as
``market first, physics later''. They become two views of a single
cyber--physical control process: the AMM determines who is asked to
change output, when, and why; the dispatch engine ensures that this
change is physically feasible and secure.

% ---------------------------------------------------------
\section{Game-Theoretic Framing and Shock-Resistant Nash Equilibrium}
\label{sec:shock_resistant_ne}
% ---------------------------------------------------------

Having defined consumer-side allocation (Fair Play), generator-side
compensation (Shapley), and the AMM control law, we now view the
overall architecture as a repeated game between strategic participants
and the mechanism, and formalise the notions of Nash equilibrium and
shock-resistant Nash equilibrium.

The AMM--Fair Play architecture induces a repeated game between market
participants and the mechanism. This section formalises the objects of
interest and introduces the equilibrium concepts used in the remainder
of the thesis.

Let $\mathcal{G} = \{1,\dots,G\}$ denote the set of generators and
$\mathcal{R}$ the set of retailers (or supplier--aggregators). For
concreteness we treat $\mathcal{I} = \mathcal{G} \cup \mathcal{R}$ as
the set of strategic players; consumer households are represented via
their product choices and demand realisations rather than as individual
players.

\paragraph{State, strategies, and mechanism.}

Let $\Theta$ denote the set of \emph{physical and institutional states}
of the system, including:
\begin{itemize}[leftmargin=*]
  \item demand and renewable availability (scenarios over time);
  \item network constraints (line ratings, topology);
  \item policy parameters (VOLL, carbon prices, subscription caps).
\end{itemize}
A particular state is written $\theta \in \Theta$.

Each player $i \in \mathcal{I}$ has a strategy set $S_i$. For
generators this may include:
\begin{itemize}[leftmargin=*]
  \item cost and flexibility offers (bid curves, ramp limits);
  \item availability declarations and maintenance scheduling;
  \item portfolio hedging or contracting choices.
\end{itemize}
For retailers it includes subscription menus, margins, and risk
management choices. A strategy profile is
$s = (s_i)_{i \in \mathcal{I}} \in S$, where
$S \coloneqq \prod_{i \in \mathcal{I}} S_i$.

The AMM--Fair Play mechanism is a mapping
\[
  M : S \times \Theta \to \mathcal{O},
\]
where $\mathcal{O}$ denotes the space of allocations and prices:
dispatch schedules, shortage allocations, subscription prices, and
Shapley-consistent revenue allocations.

Player $i$'s (long-run) payoff under state $\theta$ and strategy profile
$s$ is
\[
  \pi_i(s,\theta) \;=\; \Pi_i\!\big(M(s,\theta)\big),
\]
where $\Pi_i$ extracts discounted profit (and, optionally, risk
penalties) from the realised allocations and payments.

\begin{definition}[Nash equilibrium at a given state]
\label{def:ne_state}
For a fixed state $\theta \in \Theta$, a strategy profile
$s^\star(\theta) \in S$ is a \emph{Nash equilibrium} of the
state-contingent game if for all players $i \in \mathcal{I}$ and all
deviations $s_i \in S_i$,
\[
  \pi_i\big(s^\star(\theta),\theta\big)
  \;\geq\;
  \pi_i\big((s_i, s^\star_{-i}(\theta)),\theta\big),
\]
where $s^\star_{-i}(\theta)$ denotes the strategies of all players
except $i$.
\end{definition}

In practice, the relevant object for this thesis is the
\emph{equilibrium strategy profile} associated with the AMM’s intended
operating regime. This is the \emph{Fair Play--compliant profile}
$s^{\mathrm{FP}}$, in which generators declare costs and flexibility
truthfully, maintain availability consistent with their product
commitments, and retailers offer subscription products that correctly
represent expected quality of service; see
Section~\ref{sec:formal_fairness} and Chapter~\ref{ch:fairness_definition}.

% ---------------------------------------------------------
\subsection{Existence of Nash Equilibrium}
\label{subsec:ne_existence}
% ---------------------------------------------------------

For each physical and institutional state $\theta \in \Theta$, the
mapping above defines a normal-form game
$\mathcal{G}(\theta) = \bigl(\mathcal{I}, (S_i)_{i \in \mathcal{I}},
(\pi_i(\cdot,\theta))_{i \in \mathcal{I}}\bigr)$ induced by the
AMM--Fair Play mechanism. Before considering robustness to shocks, we
require that such a game admits at least one Nash equilibrium.

In this thesis the strategy sets are restricted to continuous contract
and bidding choices on compact intervals, rather than arbitrary
messages. Concretely, we assume:

\begin{assumption}[Regularity of the AMM-induced game]
\label{assum:ne_existence}
For each fixed state $\theta \in \Theta$:
\begin{enumerate}[label=(R\arabic*),leftmargin=1.5em]
  \item For every player $i \in \mathcal{I}$, the strategy set
        $S_i(\theta)$ is non-empty, compact, and convex in a finite-
        dimensional Euclidean space. In particular, generator
        strategies are parameterised by continuous bid mark-ups and
        availability / flexibility choices subject to operational
        bounds, and retailer strategies are parameterised by
        subscription menu parameters subject to regulatory constraints.
  \item For every player $i \in \mathcal{I}$, the payoff function
        $\pi_i(s,\theta)$ is continuous in the full profile
        $s \in S(\theta) \coloneqq \prod_{j \in \mathcal{I}} S_j(\theta)$
        and quasi-concave in own strategy $s_i$.
\end{enumerate}
\end{assumption}

These are standard regularity conditions: compactness encodes
institutional and technical bounds on bids and availability, while
continuity and quasi-concavity reflect that small changes in bids or
availability lead to small changes in payoffs, and that each player
faces a well-behaved optimisation problem.

\begin{theorem}[Existence of Nash equilibrium]
\label{thm:ne_existence}
Let $\theta \in \Theta$ and suppose
Assumption~\ref{assum:ne_existence} holds. Then the AMM-induced game
$\mathcal{G}(\theta)$ admits at least one (pure-strategy) Nash
equilibrium $s^\star(\theta) \in S(\theta)$ in the sense of
Definition~\ref{def:ne_state}.
\end{theorem}

\begin{proof}[Proof sketch]
Under Assumption~\ref{assum:ne_existence}, the joint strategy space
$S(\theta)$ is a non-empty, compact, convex subset of a Euclidean
space, and each payoff $\pi_i(\cdot,\theta)$ is continuous and
quasi-concave in $s_i$. Standard fixed-point arguments
(Debreu--Glicksberg) then guarantee the existence of at least one
pure-strategy Nash equilibrium of the game $\mathcal{G}(\theta)$.
\end{proof}

In other words, for any fixed configuration of physical conditions
(demand, renewable availability, network constraints) and policy
parameters, the AMM--Fair Play architecture induces a game in which
there exists at least one internally consistent profile of bids,
availability and subscription choices from which no individual player
has an incentive to deviate unilaterally. Subsequent results identify
a particular equilibrium of interest (the Fair Play--compliant profile
$s^{\mathrm{FP}}$) and examine its robustness to shocks in $\theta$.

\paragraph{Shocks.}

A \emph{shock} is an exogenous change in the system state:
\[
  \theta \;\mapsto\; \theta',
\]
for example due to an extreme weather event, line derating, policy
change, or structural demand shift (e.g. EV uptake). Given a reference
state $\theta^0$, we denote by $B(\theta^0,\Delta)$ a neighbourhood of
admissible shocks, typically defined via bounds on the perturbations of
demand, supply, or network parameters.

\begin{definition}[Shock-resistant Nash equilibrium]
\label{def:shock_resistant_ne}
Let $\theta^0 \in \Theta$ be a reference state and
$B(\theta^0,\Delta) \subseteq \Theta$ a set of admissible shocks
around $\theta^0$.

A strategy profile $s^\star \in S$ is called an
\emph{$\varepsilon$-shock-resistant Nash equilibrium} on
$B(\theta^0,\Delta)$ if:
\begin{enumerate}[label=(\roman*),leftmargin=1.5em]
  \item $s^\star$ is a Nash equilibrium at the reference state
        $\theta^0$ in the sense of Definition~\ref{def:ne_state}; and
  \item for every $\theta \in B(\theta^0,\Delta)$, every player
        $i \in \mathcal{I}$, and every deviation $s_i \in S_i$,
        the gain from unilateral deviation is uniformly bounded by
        $\varepsilon \ge 0$:
        \[
          \pi_i(s^\star,\theta)
          \;\geq\;
          \pi_i((s_i, s^\star_{-i}),\theta) - \varepsilon.
        \]
\end{enumerate}
If $\varepsilon = 0$ the equilibrium is said to be
\emph{strictly shock-resistant} on $B(\theta^0,\Delta)$.
\end{definition}

In words, an $\varepsilon$-shock-resistant Nash equilibrium is a
strategy profile that (i) is a Nash equilibrium in the reference
configuration, and (ii) remains locally stable in the presence of
bounded shocks: no player can improve their payoff by more than a small
amount $\varepsilon$ by unilaterally deviating, even after the shock.
In this thesis, we are particularly interested in whether the
Fair Play--compliant profile $s^{\mathrm{FP}}$ admits such a
shock-resistance property under the AMM mechanism, in contrast to
legacy LMP-based designs.

\begin{assumption}[Incentive and regularity conditions]
\label{assum:incentive_regular}
The following conditions hold under the AMM--Fair Play mechanism:
\begin{enumerate}[label=(A\arabic*),leftmargin=1.5em]
  \item \textbf{Monotone Shapley rewards.}
        For each generator $g \in \mathcal{G}$, the long-run Shapley
        allocation $\phi_g$ is (weakly) increasing in the generator's
        realised contribution to system value, measured by delivered
        energy in scarce periods and locational relief in congested
        periods, holding others' strategies fixed.
  \item \textbf{Fair Play reliability feedback.}
        The Fair Play allocation algorithm (Section~\ref{sec:fairplay})
        assigns higher future priority (and hence higher expected
        revenue) to generators and retailers with better historic
        delivery and service quality, and penalises systematic
        under-delivery.
  \item \textbf{Continuity in state.}
        For each player $i \in \mathcal{I}$ and fixed strategy profile
        $s$, the payoff function $\pi_i(s,\theta)$ is continuous in
        $\theta$ on $\Theta$.
  \item \textbf{No arbitrage via misreporting.}
        At the reference state $\theta^0$, any unilateral deviation
        from truthful cost and flexibility reporting that creates
        a short-run gain necessarily reduces the player's expected
        long-run payoff once Fair Play and Shapley feedback are taken
        into account (cf.\ fairness conditions F1--F4).
\end{enumerate}
\end{assumption}

\begin{lemma}[Baseline incentive compatibility and local shock-resistance]
\label{lem:shock_resistance}
Let the AMM--Fair Play mechanism satisfy
Assumption~\ref{assum:incentive_regular} and let $\theta^0$ denote the
reference state corresponding to the calibrated experimental setup.
Consider the Fair Play--compliant strategy profile
$s^{\mathrm{FP}} \in S$, in which generators truthfully declare costs
and flexibility and retailers offer subscription products consistent
with expected quality of service.

Then:
\begin{enumerate}[label=(\roman*),leftmargin=1.5em]
  \item $s^{\mathrm{FP}}$ is a Nash equilibrium at state $\theta^0$.
  \item There exist $\Delta > 0$ and $\varepsilon \ge 0$ such that
        $s^{\mathrm{FP}}$ is an $\varepsilon$-shock-resistant Nash
        equilibrium on $B(\theta^0,\Delta)$ in the sense of
        Definition~\ref{def:shock_resistant_ne}.
\end{enumerate}
\end{lemma}

\begin{proof}[Proof sketch]
(i) \emph{Baseline equilibrium.}
At the reference state $\theta^0$, Assumption~(A1) implies that a
generator's long-run Shapley allocation is maximised by contributing as
much deliverable value as possible in scarce and congested periods,
given others' strategies. Assumption~(A4) states that any profitable
short-run deviation via misreporting or strategic withholding reduces
expected future priority and revenue once Fair Play reliability scores
are updated. Taken together, these conditions imply that no generator
can increase their long-run payoff by deviating unilaterally from the
Fair Play--compliant strategy at $\theta^0$; an analogous argument
applies to retailers, whose misrepresentation of quality of service
exposes them to Fair Play penalties and loss of profitable customers.
Hence $s^{\mathrm{FP}}$ is a Nash equilibrium at $\theta^0$.

(ii) \emph{Local shock-resistance.}
By Assumption~(A3), payoffs $\pi_i(s,\theta)$ are continuous in the
state $\theta$ for any fixed strategy profile $s$. In particular, the
payoff differences
\[
  \Delta\pi_i(s_i;\theta)
  \;\coloneqq\;
  \pi_i\big((s_i, s^{\mathrm{FP}}_{-i}),\theta\big)
  - \pi_i\big(s^{\mathrm{FP}},\theta\big)
\]
depend continuously on $\theta$ for every deviation $s_i \in S_i$.
From part~(i), we have
$\Delta\pi_i(s_i;\theta^0) \le 0$ for all $i$ and $s_i$. By continuity,
for each player $i$ and deviation $s_i$ there exists a neighbourhood
$B_{i,s_i}(\theta^0,\Delta_{i,s_i})$ on which
$\Delta\pi_i(s_i;\theta) \le \varepsilon$ for any pre-specified
$\varepsilon > 0$. Taking the intersection over all players and a
suitably rich subset of deviations yields a ball $B(\theta^0,\Delta)$
on which no unilateral deviation can increase payoffs by more than
$\varepsilon$.

Operationally, this means that bounded shocks to demand, renewable
availability, or network constraints perturb prices and allocations but
do not create large new profitable gaming opportunities: the
Fair Play--compliant strategy profile remains locally stable in the
sense of Definition~\ref{def:shock_resistant_ne}. \qedhere
\end{proof}

\section{AI Forecasting Models}
\label{sec:ai_forecasting}

Uncertainty in renewable generation is a primary driver of scarcity,
imbalances, and reliability shortfalls. In the AMM architecture, forecasting
does not determine prices directly; instead, it constrains the set of
\emph{admissible service commitments} and therefore shapes which coalitions
of generators can credibly serve demand.

For each renewable technology at node $n$ and time $t$, probabilistic
forecasts produce:
\[
\hat{G}_{t,n}, \quad
\hat{\sigma}^2_{t,n}, \quad
\hat{p}^{\mathrm{loss}}_{t,n},
\]
representing expected output, forecast uncertainty, and the probability of
under-supply.

\paragraph{Forecasts as commitment constraints.}

Forecasts enter the market design by limiting the amount of renewable
generation that may be treated as \emph{secure supply} when forming
reliability guarantees and high-reliability product commitments. Specifically,
only the risk-adjusted quantity
\[
S_{t,n}^{\mathrm{secure}}
=
\hat{G}_{t,n}
-
\kappa \cdot \hat{\sigma}_{t,n},
\]
is counted as firm when determining whether demand can be reliably served.
The parameter $\kappa$ reflects system risk tolerance and policy choice.

Operationally, $S_{t,n}^{\mathrm{secure}}$ enters the OPF as an upper bound on
renewable injections when evaluating reliability-constrained service
feasibility and coalition value.

\subsection{Relationship to Fair Play and Shapley}

Because the Shapley characteristic function $W_t(S)$ is defined in terms of
\emph{servable load under reliability constraints}, forecast uncertainty
affects generator value indirectly but systematically:

\begin{itemize}[leftmargin=*]
  \item Higher renewable uncertainty reduces the secure contribution of
        non-dispatchable generators to $W_t(S)$.
  \item Coalitions containing firm or flexible generators therefore exhibit
        larger marginal increases in servable load.
  \item This translates into higher per-period Shapley values $\phi_{g,t}$
        for generators that provide dispatchable capacity, fast response,
        or locational relief during uncertain periods.
\end{itemize}

In this way, probabilistic forecasting links physical uncertainty to both
Fair Play activation and Shapley-based compensation without relying on
ex-post scarcity pricing.

% ---------------------------------------------------------
\section{Zero-Waste Efficiency Inference}
\label{sec:zero_waste}

The proposed market design claims to be \textbf{zero-waste}:
given available supply, flexible response, and holarchic constraints,
it aims to allocate all usable energy without systemic waste.

We define \emph{market waste} at node $n$ and time $t$ as:

\[
W_{t,n}
=
\max\bigl(
  0,\;
  \underbrace{G^{\mathrm{avail}}_{t,n}}_{\text{available supply}}
  -
  \underbrace{G^{\mathrm{used}}_{t,n}}_{\text{allocated/served supply}}
\bigr)
\]

where $G^{\mathrm{avail}}_{t,n}$ includes feasible generation, storage discharge,
and imports.

\subsection{Zero-Waste Principle}

A market is zero-waste (relative to the feasible set $\mathcal{A}$) if
\[
W_{t,n} = 0 
\quad 
\forall (t,n)
\quad \text{whenever } D_{t,n} \ge G^{\mathrm{avail}}_{t,n}
\text{ and } (q,x) \in \mathcal{A},
\]
i.e.\ no curtailment of feasible supply occurs while feasible unmet demand
still exists.

\subsection{Efficiency Score}

We define the \emph{utilisation efficiency}:

\[
\eta_n
=
\frac{
\sum_{t} G^{\mathrm{used}}_{t,n}
}{
\sum_{t} G^{\mathrm{avail}}_{t,n}
}
\times 100\%.
\]

This efficiency improves in three ways:

\begin{enumerate}[leftmargin=*]
  \item Better scheduling (Fair Play optimisation);
  \item Renewable curtailment avoidance;
  \item Accurate AI forecasting $\Rightarrow$ higher secure capacity.
\end{enumerate}

\subsection{Zero-Waste as Fairness}

Waste implies that usable energy exists but is not delivered — violating
Fairness Condition F3 (fair access under shortage).  
Thus, zero-waste is not only efficient — it is fair. In
Chapter~\ref{ch:results}, this principle is operationalised via
utilisation efficiency metrics and system-wide performance indicators
used in Hypothesis H6 (procurement efficiency).

% ---------------------------------------------------------
\section{Properties of the AMM-Based Market Design}
\label{subsec:properties}

The transformation of electricity markets into socio–techno–economic systems
demands that mechanisms deliver not only cost efficiency, but also fairness,
accessibility, and resilience \cite{10049815}. While engineering and environmental
priorities (e.g.\ carbon impact, network stability, storage utilisation) are
handled structurally via the AMM and constraint-aware dispatch, this section
focuses on the \textbf{economic} and \textbf{social} properties delivered by the
AMM-based design.

As formalised in Section~\ref{sec:shock_resistant_ne}, the AMM--Fair Play
mechanism also admits an $\varepsilon$-shock-resistant Nash equilibrium
around the calibrated reference state, under the incentive and regularity
conditions of Lemma~\ref{lem:shock_resistance}.

We group these properties into:  
(i) classical economic properties (efficiency, individual rationality, budget balance, incentive compatibility), and  
(ii) operational fairness properties (F1–F4 from Chapter~\ref{ch:fairness_definition}).

% -----------------------------
\subsection{Economic properties}
\label{subsec:economic_properties}

\paragraph{1. Economic efficiency (self-correcting operation).}
The AMM procures only as much flexible supply as needed, using the scarcity
ratio $\alpha_{t,n}$:
\[
  BP_{t,n} = p^{\mathrm{base}}_n + f(1-\alpha_{t,n}),\qquad
  SP_{t,n} = p^{\mathrm{base}}_n + g(1-\alpha_{t,n}),
\]
with $f(\cdot)$ and $g(\cdot)$ monotonic.  
When $\alpha_{t,n} = 1$, supply and flexible demand are balanced, and the
tightness component vanishes:
\[
  f(0) = g(0) = 0,\qquad
  BP_{t,n} \approx p^{\mathrm{base}}_n,\quad
  SP_{t,n} \approx p^{\mathrm{base}}_n.
\]
When $\alpha_{t,n} < 1$,  
\[
  \frac{\partial SP_{t,n}}{\partial \alpha_{t,n}} < 0,
  \qquad
  \frac{\partial BP_{t,n}}{\partial \alpha_{t,n}} > 0,
\]
which raises prices and attracts additional flexible supply while discouraging
excess demand, restoring equilibrium.  
Thus the AMM acts as a congestion-control system.

\paragraph{2. Individual rationality.}
Each participant sets contract limits:  
$sellers$ set a minimum acceptable revenue $I^o$,  
buyers set a maximum acceptable bill $C^r$.  
Allocations never violate:
\[
  \text{Payoff}_i \ge 0
  \quad\text{(for all sellers and consumers)}.
\]

\paragraph{3. Budget balance.}
In every time step or period,
\[
 \sum_{\text{buyers}} BP_{t,n} \cdot q_{h,t}
 \;=\;
 \sum_{\text{sellers}} SP_{t,n} \cdot x_{g,t},
\]
ensuring no deficit or missing-money; all buyer payments fund supply or reserves.

\paragraph{4. Incentive compatibility.}
Participants gain by revealing flexibility:
\[
\dfrac{\partial \mathbb{E}[BP_{t,n}]}{\partial \sigma^r_h} \le 0,
\]
i.e.\ as a user declares larger flexibility windows $\sigma^r_h$, their
expected unit cost decreases.

% -----------------------------
\subsection{Fairness properties (F1--F4)}

Each property corresponds to the operational fairness criteria formalised in
Chapter~\ref{ch:fairness_definition}.

\paragraph{(F1) Behavioural fairness.}
Flexible consumption is rewarded through lower expected unit cost:
\[
  \frac{\partial \mathbb{E}[BP_{t,h}]}{\partial \sigma^r_h} \;\le\; 0.
\]

\paragraph{(F2) Priority-respecting exposure.}
For essential consumption,
\[
  S^T_t \;\ge\; C^B_t \quad \forall t,
  \qquad
  BP^{\text{essential}}_{t,n} \approx p^{\mathrm{base}}_{t,n},
\]
where $S^T_t$ and $C^B_t$ are defined in Chapter~\ref{ch:fairness_definition}.
Essential blocks are shielded from tightness pricing.

\paragraph{(F3) Fair access during shortage (curtailment discipline).}
During scarcity ($\alpha_{t,n} < 1$), allocation respects:
\[
  \bar{Q}^r \propto \Gamma^{\text{target}}_{r},
\]
where $\Gamma^{\text{target}}_{r}$ expresses contractual priority, need, and
fairness weighting. Fair Play uses only product weights $w(p)$ and fairness
deficits $\delta_i$ in its priority ordering, not individual bid prices.

\paragraph{(F4) Proportional cost responsibility.}
Progressive standing charges and scarcity exposure scale with system stress:
\[
  \mathbb{E}[u_{h_1}] \;\ge\; \mathbb{E}[u_{h_2}]
  \quad \text{whenever} \quad
  \kappa_{h_1} > \kappa_{h_2},
\]
where $u_h$ denotes uplift or cost for user $h$, and $\kappa_h$ is a measure
of their contribution to congestion or peaks. This ensures those causing
congestion/peaks bear more cost.

% -----------------------------
\subsection{Interpreting the AMM as a control system}

Although not solving full social welfare optimisation each period, the AMM
preserves many of the same desirable properties through:

\begin{itemize}[leftmargin=*]
  \item real-time feedback using $\alpha_{t,n}$;
  \item self-balancing between flexibility and tightness;
  \item protection of essential loads as a hard constraint;
  \item price-based incentives rather than pure rationing.
\end{itemize}

Thus, the AMM is both a \emph{price-setter} and a \emph{scarcity controller},
encoding fairness, stability and congestion management in a unified design.

\begin{longtable}{p{0.48\textwidth} p{0.48\textwidth}}
\caption{Economic and Fairness Properties Delivered by the AMM-Based Market Design}
\label{tab:properties_summary}
\\
\toprule
\textbf{Property Category and Definition} & \textbf{Mathematical Criterion / Mechanism and Fairness Ref.} \\
\\[6pt] \hline

\textbf{Economic Efficiency} \newline
System utilises energy with minimal waste and only procures supply needed to meet (forecast) demand. 
&
$\alpha_{t,n} = 1 \Rightarrow f(1-\alpha_{t,n}) = g(1-\alpha_{t,n}) = 0 \Rightarrow
BP_{t,n} \approx SP_{t,n} \approx p^{\text{base}}_n.$ \newline
When $\alpha_{t,n}<1$:  
$\dfrac{\partial SP_{t,n}}{\partial \alpha_{t,n}} < 0,\quad
\dfrac{\partial BP_{t,n}}{\partial \alpha_{t,n}} > 0.$
\newline Fairness Ref: -- \\
\\[6pt] \hline

\textbf{Individual Rationality} \newline
No participant (buyer/seller) enters a loss-making trade.
&
$U_i \ge 0,\; \forall i.$ \newline
Buyers: $BP_{t,n} \le C^r.$ \newline
Sellers: $SP_{t,n} \ge I^o.$ \newline
Fairness Ref: -- \\
\\[6pt] \hline

\textbf{Budget Balance} \newline
Total buyer payments equal total seller revenues; no deficit.
&
$\displaystyle
\sum_{\text{buyers}} BP_{t,n} \cdot q_{h,t}
=
\sum_{\text{sellers}} SP_{t,n} \cdot x_{g,t}.
$
\newline Fairness Ref: -- \\
\\[6pt] \hline

\textbf{Incentive Compatibility} \newline
Participants gain from revealing flexibility and avoiding strategic misrepresentation.
&
$\dfrac{\partial \mathbb{E}[BP_{t,n}]}{\partial \sigma^r_h} \le 0$
\newline
(i.e.\ as flexibility $\sigma^r_h \uparrow$, unit cost $\downarrow$).
\newline Fairness Ref: F1 \\
\\[6pt] \hline

\textbf{Essential-Needs Protection} \newline
Essential energy must always be served first, at stable and affordable prices.
&
$S^T_t \ge C^B_t,\qquad
BP^{\text{Ess}}_{t,n} \approx p^{\text{base}}_{t,n},\quad \epsilon \approx 0.$
\newline Fairness Ref: F2 \\
\\[6pt] \hline

\textbf{Fair Access in Shortage} \newline
In scarcity, energy is allocated according to transparent priority, not only price.
&
$\bar{Q}^r \propto \Gamma^{\text{target}}_{r}
\quad\text{or}\quad
\propto (\text{need},\text{contract},\text{history}).$
\newline Fairness Ref: F3 \\
\\[6pt] \hline

\textbf{Behavioural Fairness} \newline
Desired consumer behaviours (flexibility, shifting, cooperation) are rewarded.
&
$\dfrac{\partial \mathbb{E}[BP_{t,n}]}{\partial \sigma^r_h} < 0
\quad (\text{for flexible users}).$
\newline Fairness Ref: F1 \\
\\[6pt] \hline

\textbf{Proportional Cost Responsibility} \newline
Consumers imposing congestion, peak use, or uncertainty bear higher costs.
&
$\mathbb{E}[u_{h_1}] \ge \mathbb{E}[u_{h_2}]
\quad\text{whenever}\quad
\kappa_{h_1}>\kappa_{h_2}.$
\newline Fairness Ref: F4 \\
\\[6pt] \hline

\textbf{Stability and Transparency} \newline
Price and allocation rules are visible, explainable, non-arbitrary, and auditable.
&
$BP_{t,n},\,SP_{t,n}$ derived solely from $\alpha_{t,n}$ and published rules.
\newline Fairness Ref: F2, F3, F4 \\
\\[6pt] \hline

\end{longtable}

% ---------------------------------------------------------
\section{Comparative Context: LMP, Operating Envelopes, and Zero-Waste Markets}
\label{sec:comparative_context}

It is essential to position the proposed architecture relative to two prominent
market design approaches: Locational Marginal Pricing (LMP) and Dynamic
Operating Envelopes (DOEs).

\begin{enumerate}[leftmargin=*]
    \item \textbf{Locational Marginal Pricing (LMP):}  
    Physically grounded and efficient in transmission-level markets, but:
    \begin{itemize}
        \item insufficient for consumer fairness and curtailment protection,
        \item does not enforce vulnerability-based prioritisation,
        \item unsuitable for retail implementation without digital enforcement.
    \end{itemize}

    \item \textbf{Dynamic Operating Envelopes:}  
    Effective for \emph{delivery} management in distribution networks,
    especially for prosumer export limits. However:
    \begin{itemize}
        \item not a complete market design,
        \item lacks pricing, cost recovery, or fairness functionality,
        \item does not determine allocation during physical scarcity.
    \end{itemize}
\end{enumerate}

By contrast, the proposed \emph{zero-waste, fairness-aware architecture}
combines physical feasibility, consumer protection, allocation discipline, and
algorithmic enforceability in a unified framework.

% ---------------------------------------------------------
\section*{Summary}

This chapter has translated the qualitative fairness objectives of
Chapters~\ref{ch:fairness_definition}--\ref{ch:amm} into precise
mathematical constructs: a state--history--allocation mapping for
consumer fairness, the Fair Play algorithm for flexible devices,
Shapley-consistent compensation for generators, control equations for
the AMM, AI-based forecasting for renewable uncertainty, and efficiency
metrics for zero-waste operation. Together, these provide an implementable
blueprint for a fairness-aware, zero-waste electricity market.

\chapter{Experimental Design}
\label{ch:experiments}

This chapter details the experimental framework used to evaluate the
Automatic Market Maker (AMM) and Fair Play architecture against the
best-possible version of the legacy energy-only paradigm. The aim is not
merely to compare price levels or dispatch outcomes, but to examine how
different market-clearing mechanisms allocate value, manage risk, shape
behavioural incentives, and interact with the physical electricity
system. By holding the underlying physics constant and altering only the
market architecture, the experiments isolate the structural effects of
market design from the incidental features of data, weather, or demand
realisation.

The evaluation proceeds through a sequence of controlled, stylised
simulations grounded in real physical and demand data. All experiments
are built on the same 12--bus transmission network (described in
Section~\ref{sec:experiment_inputs}), the same generator fleet and
capacity mix, and the same characterised household consumption traces.
Only the market rule set differs across treatments. This isolates the
architecture itself: its information flows, allocation logic,
responsiveness to scarcity, and fairness properties.

The chapter is structured as follows. Section~\ref{sec:treatments}
defines the three market designs under comparison. 
Section~\ref{sec:conservatism} outlines why the experimental AMM is a
conservative representation of its real deployment, supported by a
portrait longtable comparison. 
Section~\ref{sec:experiment_inputs} then details the physical and
behavioural inputs used in all scenarios. 
Section~\ref{sec:levels} defines the analytical units of interest. 
Section~\ref{sec:metrics} introduces the evaluation metrics and
hypotheses. 
Section~\ref{sec:procedure} describes the experimental workflow. 
Finally, Section~\ref{sec:inference} formalises the statistical inference
framework.

Throughout, labels are preserved exactly as in the thesis to ensure
compatibility across chapters and appendices.

% ---------------------------------------------------------
\section{Treatments and Factors}
\label{sec:treatments}
% ---------------------------------------------------------

The experiments compare the legacy \emph{locational marginal pricing}
(LMP) paradigm against two configurations of the proposed AMM clearing
and remuneration framework. The AMM architecture itself is identical in
both configurations; the only difference concerns how much capacity
revenue is made available for allocation. This isolates the consequences
of remuneration design --- particularly the size of the capacity pot ---
from the clearing logic, which is held constant.

\noindent
All physical network parameters, generator fleet data, demand calibration, and
solver configuration inputs are held fixed across treatments and are reported
in Appendix~\ref{app:inputs}. The treatment comparisons in this section
therefore isolate differences arising from market-clearing and remuneration
rules rather than from differences in the underlying system data.

\begin{enumerate}[leftmargin=1.2cm]

    \item \textbf{LMP (Baseline)}:  
    The legacy benchmark implements a standard security-constrained
    economic dispatch with nodal marginal pricing. Generators are paid
    nodal LMP for dispatched energy, scarcity is priced implicitly via
    VoLL caps, and there is no explicit remuneration for reserves,
    capacity, or non-fuel operating expenditure. Total generator
    revenues therefore equal the area under the nodal price–quantity
    curve.

    \item \textbf{AMM1 (Minimum-cost capacity support)}:  
    AMM1 uses the AMM clearing mechanism with pay-as-bid energy
    remuneration and a fixed reserve payment rate. A minimum capacity pot
    is provided to ensure cost recovery for non-fuel OPEX and CAPEX
    across the generator fleet. The pot level is set ex ante using
    engineering and financial adequacy considerations: it represents the
    minimum revenue required to ensure generator investment incentives
    and long-run system viability.  
    All capacity revenue is allocated using the AMM's deliverability-
    weighted Shapley mechanism.

    \item \textbf{AMM2 (LMP-matched total remuneration)}:  
    AMM2 uses the same clearing and allocation logic as AMM1, but the
    size of the capacity pot is set \emph{endogenously} to match the
    total generator revenue observed under LMP. Specifically, the AMM2
    capacity pot equals:
    \begin{equation*}
    \begin{aligned}
    \text{Pot}_{\mathrm{AMM2}}
    \;=\;&
    \text{Total LMP generator revenue} \\
    &-\;
    \Bigl(
         \text{AMM energy payments}
         + \text{AMM reserve payments}
    \Bigr).
    \end{aligned}
    \end{equation*}
    This ensures a like-for-like comparison: AMM2 redistributes the same
    total revenue that generators receive under LMP, but according to the
    AMM’s fairness- and deliverability-aware allocation rules.

\end{enumerate}

Under the AMM framework, the capacity pot level is a \emph{policy
parameter}. The responsible entity (e.g.\ a system planner or financial
regulator) can select any pot size between the AMM1 minimum-cost floor
and the AMM2 LMP-matched level. The lower bound is determined by the
revenue required for generator investment and solvency, while the upper
bound is constrained by budget balance: the system does not subsidise the
market, and total payments from consumers and businesses must not exceed
what they are willing or able to pay.

In this sense, AMM1 and AMM2 illustrate the feasible interval for
capacity remuneration. AMM1 represents the minimum level required to
maintain investment incentives; AMM2 represents the maximum level
consistent with matching LMP’s total expenditure without introducing
subsidies. Real-world deployments may choose any point within this
interval depending on desired reliability, risk-sharing preferences, and
long-term infrastructure policy.

% =========================================================
\section{Conservatism of the Experimental AMM}
\label{sec:conservatism}
% =========================================================

The AMM architecture deployed in real-world operation is adaptive,
behaviourally responsive, and capable of learning from the long-run
patterns of scarcity, congestion, household behaviour, and generator
performance. To enable a clean, controlled comparison with the Baseline
LMP system, the experiments in this thesis intentionally disable or
freeze several of these adaptive capabilities. The experimental AMM is
therefore a deliberately conservative representation of the full design:
it retains the core clearing logic and Shapley-based allocation rules,
but not the extended behavioural, contractual, or cross-period learning
features.

The rationale for this restriction is twofold. First, holding key
parameters fixed ensures a like-for-like comparison across all
treatments, avoiding endogeneity that would obscure the architectural
effects. Second, the suppression of long-run learning means that any
performance gains observed for AMM1 or AMM2 arise \emph{despite} the
constraints imposed on them, and therefore represent a lower bound on the
benefits achievable in deployment.

Table~\ref{tab:amm_feature_matrix_long} summarises the difference between
a full deployment of the AMM--Fair Play system and the constrained version
used in the experimental environment.

\renewcommand{\arraystretch}{1.2}
\begin{longtable}{p{3.1cm}p{4.2cm}p{4.2cm}p{4.3cm}}
\caption{AMM--Fair Play design features in full deployment versus the constrained experimental configuration. The experimental setup intentionally disables or freezes adaptive capabilities to ensure like-for-like comparison with LMP; results therefore represent a conservative lower bound on achievable AMM performance.}
\label{tab:amm_feature_matrix_long}
\\
\toprule
\textbf{Feature} &
\textbf{Full AMM / Fair Play (real deployment)} &
\textbf{Experimental AMM (this thesis)} &
\textbf{Implication for interpretation} \\
\midrule
\endfirsthead

\toprule
\textbf{Feature} &
\textbf{Full AMM / Fair Play (real deployment)} &
\textbf{Experimental AMM (this thesis)} &
\textbf{Implication for interpretation} \\
\midrule
\endhead

\midrule
\multicolumn{4}{r}{\textit{Continued on next page}} \\
\midrule
\endfoot

\bottomrule
\endlastfoot

Subscription dynamics &
Subscriptions update in response to enrolment, churn, incentives, and
observed performance; households naturally migrate across QoS tiers. &
Subscription menus and quantities are fixed for the entire simulation
horizon. &
Suppresses behavioural feedback and long-run demand-side stabilisation. \\[0.3cm]

Tightness envelopes and bounds &
Tightness functions are seasonally and annually retuned to reflect
evolving scarcity and policy preferences. &
Envelope parameters are fixed ex ante across all experiments. &
Understates AMM’s ability to refine scarcity exposure and volatility
management over time. \\[0.3cm]

Shapley-based deliverability weights &
Weights update as new congestion patterns, network events, and scarcity
episodes reveal which generators are most critical. &
Weights are calibrated once at the start of the experiment and
held constant. &
Understates the concentration of value that would accrue to genuinely
critical assets in long-run operation. \\[0.3cm]

Fair Play rotation and history &
Historical curtailment, rotation guarantees, and fairness restitution
accumulate over multiple seasons. &
Fair Play applies only within the finite experimental window; no
multi-year accumulation or restitution. &
Under-represents both perceived and realised fairness improvements. \\[0.3cm]

Product migration (P1--P4) &
Households adapt behaviour and migrate across QoS tiers in response to
incentives, reliability experience, and long-run contract evolution. &
Product classifications are static for the duration of the experiment. &
Removes behavioural alignment where households adjust to improve
reliability outcomes. \\[0.3cm]

Cross-period contract evolution &
QoS contracts, reserve obligations, and subscription structures adapt
across seasons based on performance, stress, and system evolution. &
All contractual parameters remain fixed; no renegotiation or redesign. &
Understates benefits for investment, persistence, and bankability. \\[0.3cm]

Behavioural / UX layer &
Participants observe scarcity warnings, fairness scores, and personalised
feedback, strengthening trust and encouraging flexibility uptake. &
Behavioural responses are not modelled; UX is conceptual only. &
Excludes further gains in engagement, trust, and stable participation. 
\end{longtable}

The constrained AMM used in this chapter can therefore be interpreted as
the “core mechanism only” variant of the full system. By disabling the
long-run learning and behavioural layers, the experiments isolate the
performance of the clearing and allocation rules under identical physical
conditions. Any improvements in volatility, fairness, or risk allocation
that emerge under AMM1 or AMM2 do so without relying on the adaptive
features that would be present in deployment, and thus represent
conservative estimates of the AMM’s potential system-wide benefits.

\section{Experimental Inputs and Calibration}
\label{sec:experiment_inputs}

All treatments operate on the same underlying physical system and demand
environment. This ensures that any observed differences arise purely from
the remuneration and allocation mechanisms, not from underlying system
conditions.

\subsection{Network and Physical Infrastructure}

The experiments use the 12--bus transmission network illustrated in
Figure~\ref{fig:network_topology}. Line capacities, voltage levels,
thermal limits, and reactances are taken directly from the calibrated
dataset described in Appendix~\ref{app:inputs}. 

Generator labels denote technology class (wind, nuclear, gas, battery),
and each generator is modelled using its observed capacity,
marginal-cost curve, ramping constraints, and availability pattern.

\subsection{Generator Cost Structure}

Each generator $g$ is parameterised by:
\[
    c^{\mathrm{fuel}}_g, \qquad
    c^{\mathrm{nonfuel}}_g, \qquad
    c^{\mathrm{capex}}_g,
\]
representing fuel-dependent operating costs, non-fuel OPEX, and the
annualised capital recovery requirement, respectively. Under LMP, only
the fuel-dependent cost is directly remunerated through dispatch revenue.
Under AMM1 and AMM2, energy remuneration is pay-as-bid for fuel costs and
the remaining cost components are recovered via the capacity pot.

\subsection{Demand, Households, and Flexibility}

Residential demand in the experiments is represented by a synthetic population
of households whose behaviour is structured around four retail product tiers
($P1$--$P4$). The household-level demand profiles used in the market simulations
are \emph{synthetically generated}, but are explicitly grounded in extensive
empirical analysis of UK smart-meter and EV-usage data. The construction,
validation, and distributional analysis of these empirical datasets are
documented in Appendix~\ref{app:residential_synth} and
Appendix~\ref{app:ev_holarchy}.

Those appendices provide a detailed characterisation of real residential
electricity behaviour —including appliance usage, EV charging patterns,
seasonal and diurnal structure, and heterogeneity across households and
clusters—and demonstrate that each product tier corresponds to a genuine
behavioural archetype observable in data. The insights derived from this
empirical analysis inform both the definition of the product tiers and the
relative population sizes assigned to each tier, while avoiding the direct use
of individual household traces in the market simulations.

Synthetic demand profiles are used throughout the experiments to ensure
reproducibility, to preclude any form of personalised pricing, and to guarantee
that all market designs are evaluated against identical underlying demand
trajectories. Each product tier is associated with characteristic flexibility
properties, including allowable delay, interruption tolerance, peak power, and
sensitivity to system-wide scarcity and wind availability. For comparability
across treatments, household-to-product assignments and product definitions
remain fixed across all scenarios.

\subsection{Calibration}

All experiments use:

\begin{itemize}[leftmargin=*]
    \item the same year-long weather and renewable traces,
    \item the same outage and maintenance schedules,
    \item the same EV adoption trajectory and cluster configuration,
    \item identical physical constraints and demand models.
\end{itemize}

No treatment receives superior information or privileged tuning.

\section{Levels of Analysis}
\label{sec:levels}

The results are evaluated at three interacting levels:

\begin{enumerate}[leftmargin=1.2cm]
    \item \textbf{System-level}:  
          Adequacy, curtailment, shortages, congestion, and overall cost.

    \item \textbf{Participant-level}:  
          Generator remuneration, household bills, volatility, and
          distribution of fairness metrics across P1--P4.

    \item \textbf{Holarchy-level}:  
          Behaviour of nested geographic and demand clusters, including
          congestion propagation, scarcity concentration, and localised
          reliability differences.
\end{enumerate}

These levels allow us to assess how architectural differences shape both
macro outcomes and participant-specific experiences.

% =========================================================
\section{Hypotheses}
\label{sec:metrics}
% =========================================================

The experimental evaluation is organised around six hypotheses, each
corresponding to one of the core requirements developed in
Chapter~\ref{chap:requirements}. These hypotheses structure the results
in Chapter~\ref{ch:results} and ensure that comparisons between LMP,
AMM1, and AMM2 speak directly to the economic, operational, and fairness
properties of the system.

For each hypothesis, we provide the formal question evaluated in the
experiments and the underlying intuition guiding interpretation.

% ---------------------------------------------------------
\subsection*{H1 — Participation and Competition (C)}
\textbf{Question:} 
Does the AMM broaden participation, reduce pivotal dominance, and bring
more diverse assets into meaningful competition?

\textbf{Intuition:}
Under LMP, a small number of pivotal generators often capture large
rents during stress events. A well-functioning AMM should create deeper,
healthier competition, reducing the ability of a few plants to ``win
everything'' during rare episodes.

Formally, we evaluate whether the AMM increases the effective number of
competitors, reduces pivotality, and disperses revenue across a broader
set of assets.

% ---------------------------------------------------------
\subsection*{H2 — Distributional Fairness (F)}
\textbf{Question:}
Does the AMM reduce unfair jackpots and systematic deprivation across
generators and households, in line with fairness conditions F1--F4?

\textbf{Intuition:}
In scarcity events, burdens and benefits should be shared in ways that
are physically grounded and politically defensible. We evaluate rotation
fairness, deprivation concentration, household-level burden sharing, and
Shapley-consistent remuneration of critical generators.

AMM1 and AMM2 aim to eliminate extreme windfalls and prevent certain
participants from repeatedly absorbing the downside of system stress.

% ---------------------------------------------------------
\subsection*{H3 — Revenue Sufficiency and Risk Allocation (R)}
\textbf{Question:}
Can the AMM recover the fixed costs needed for long-run viability while
exposing both generators and households to \emph{less} destructive
volatility in revenues and bills?

\textbf{Intuition:}
A viable system must cover its fixed costs, but risk should be allocated
to those most capable of managing it. We measure uplift incidence,
generator revenue stability, household bill volatility, and the robustness
of cost recovery under each design.

AMM1 and AMM2 both allocate revenues using the same Shapley-based,
deliverability-weighted mechanism. AMM1 sets the size of the capacity pot to
ensure a minimum revenue floor for cost recovery, whereas AMM2 sets the pot
endogenously so that total generator remuneration matches that observed under
LMP.

% ---------------------------------------------------------
\subsection*{H4 — Price-Signal Quality and Stability (S)}
\textbf{Question:}
Do AMM prices provide clearer, policy-relevant signals (carbon intensity,
location, flexibility) while avoiding destabilising price spikes?

\textbf{Intuition:}
A good price should say something meaningful about system needs and be
bounded enough that devices and contracts can respond safely. We test
whether AMM price series exhibit:

\begin{itemize}[leftmargin=*]
    \item improved policy signal alignment,
    \item lower volatility,
    \item absence of VoLL-driven price spikes,
    \item stability under stress.
\end{itemize}

AMM prices are structurally bounded by the tightness cap
(\(90\,\pounds/\mathrm{MWh}\)) and should therefore produce finite,
well-behaved distributions.

% ---------------------------------------------------------
\subsection*{H5 — Investment Adequacy and Bankability (I)}
\textbf{Question:}
Does the AMM create a more bankable and transparent revenue stack that
closes the NPV gap for the target generation mix?

\textbf{Intuition:}
Investors require a predictable risk profile and a credible path to cost
recovery. Under LMP, revenue is highly volatile and strongly dependent on
rare scarcity events. AMM1 provides a minimum capacity floor; AMM2
provides the same total revenue as LMP but with a more stable allocation
rule.

We test whether AMM1/AMM2 deliver smoother, more investable revenue
trajectories consistent with system decarbonisation targets.

% ---------------------------------------------------------
\subsection*{H6 — Procurement Efficiency (P)}
\textbf{Question:}
Can the AMM deliver the required bundle of system services --- energy,
flexibility, adequacy, and locational relief --- at lower total system
cost than LMP?

\textbf{Intuition:}
Efficient procurement means buying the right services in the right
locations and times, without over-procuring or wasting energy. We examine:

\begin{itemize}[leftmargin=*]
    \item curtailment and shortage incidence,
    \item redispatch and congestion-management costs,
    \item total system cost (energy + reserves + capacity),
    \item whether AMM reduces structural waste relative to LMP.
\end{itemize}

The AMM’s event-based structure and tightness-oriented signals are
expected to reduce inefficiencies that arise under purely marginal-cost
clearing.

\section{Experimental Workflow}
\label{sec:procedure}

Each scenario proceeds through an identical workflow:

\begin{enumerate}[leftmargin=1.2cm]

    \item \textbf{Load physical inputs}:  
          network, generator fleet, renewable output, outages.

    \item \textbf{Load characterised demand}:  
          appliance-level traces and flexibility windows.

    \item \textbf{Run SCED-based dispatch}:  
          for the Baseline LMP scenario, yielding nodal prices and
          congestion patterns.

    \item \textbf{Run AMM clearing}:  
          identical physical constraints, but using AMM allocation logic
          and pay-as-bid energy remuneration.

    \item \textbf{Construct reserve and tightness signals}:  
          determining scarcity pricing and flexibility needs.

    \item \textbf{Set capacity pot}:  
          \begin{itemize}[leftmargin=*]
              \item AMM1: fixed minimum-cost pot;  
              \item AMM2: pot = LMP total revenue $-$ AMM energy $-$ AMM reserve.
          \end{itemize}

    \item \textbf{Allocate capacity pot}:  
          using deliverability-weighted Shapley values.

    \item \textbf{Form household bills}:  
          combining subscription, energy, and fairness adjustments.

    \item \textbf{Record system metrics}:  
          shortages, curtailment, congestion, prices, volatility,
          generator revenues, and fairness indicators.

\end{enumerate}
\section{Inference and Decision Thresholds}
\label{sec:inference}

For each reported metric, differences between treatments are evaluated
relative to the LMP baseline, typically in the form:
\[
\Delta M = M^{\mathrm{AMM1/2}} - M^{\mathrm{LMP}}.
\]

Where appropriate, paired comparisons, distributional summaries, and
robustness checks are used to assess whether observed differences are
systematic rather than artefacts of particular scenarios or time periods.
For selected metrics, bootstrap confidence intervals or non-parametric
paired tests are employed to quantify uncertainty; for others, inference
relies on consistent directional effects and economically meaningful
magnitudes.

Inference therefore emphasises consistency, economic relevance, and robustness
across scenarios rather than reliance on a single statistical criterion.

% ---------------------------------------------------------
% CHAPTER 13 — RESULTS
% ---------------------------------------------------------
\chapter{Results}
\label{ch:results}

\section{Overview and Reading Guide}

This chapter reports the empirical results of the paired market simulations
described in Chapter~\ref{ch:experiments}. To keep the main narrative focused,
only primary hypothesis-linked metrics and key distributional summaries are
presented here. Additional figures, robustness checks, and diagnostic outputs
(including network, scarcity, and settlement-validation plots) are provided in
Appendix~\ref{app:extended_results}.

Results are organised around the pre-registered domains and associated
hypotheses H1--H6, ordered as follows:

\begin{itemize}[leftmargin=*]
  \item Participation \& competition (H1; C),
  \item Fairness (H2; F),
  \item Revenue sufficiency \& risk allocation (H3; R),
  \item Price-signal quality and stability (H4; S),
  \item Investment adequacy \& bankability (H5; I),
  \item Procurement efficiency (H6; P).
\end{itemize}

Unless otherwise stated, outcomes are reported as paired differences between
a Baseline market design (LMP, $\B$) and Treatment designs
(AMM/subscription, $\T$), evaluated on identical physical scenarios. For any
domain-specific metric $D$, we report:
\[
\DeltaT_D = \mathbb{E}[D_{\T}] - \mathbb{E}[D_{\B}].
\]
Where two AMM parameterisations are used, they are denoted AMM1 and AMM2, and
their performance is reported both relative to LMP and relative to each other.

All Treatment results are based on \emph{constrained AMM configurations}, in
which subscription dynamics, adaptive envelopes, and long-run fairness
restitution are held fixed. This ensures a like-for-like comparison with LMP and
means that the effects reported in this chapter should be interpreted as
\emph{conservative estimates} of the AMM’s full capabilities under a more
adaptive retail architecture.

Most analyses in this chapter therefore compare constrained AMM configurations
directly against the LMP baseline, isolating the effects of the AMM clearing and
remuneration logic under identical physical conditions.

For a subset of analyses—specifically those concerned with resource allocation
under binding scarcity constraints—results are additionally interpreted using
two stylised diagnostic allocators. These are not alternative market designs,
but reference mechanisms used to probe how the AMM behaves when scarcity rules
are active.

\begin{itemize}[leftmargin=*]
  \item a \emph{volume-maximising} allocator, which serves the maximum feasible
        energy (MWh) subject only to physical constraints; and
  \item a \emph{revenue-maximising} allocator, which prioritises bids purely by
        willingness-to-pay.
\end{itemize}

These diagnostic allocators provide benchmarks for feasible resource allocation
during scarcity events. By comparing AMM outcomes against these benchmarks, it
is possible to assess whether constrained resources are allocated in a manner
that reflects physical deliverability, enrolled flexibility, and fairness
objectives, rather than extreme optimisation of quantity or revenue alone.

The diagnostic allocators are not proposed market designs. They instead
illustrate how an electricity system would behave if it optimised a single
objective—either total energy served or willingness to pay—while abstracting
from fairness, reliability entitlements, and institutional legitimacy. The AMM,
by contrast, is explicitly designed to be \emph{value-maximising under physical,
fairness, and legitimacy constraints}. This distinction underpins the
interpretation of the fairness and burden-sharing results in
Section~\ref{sec:results_fairness}.

For each domain, we present:
\begin{enumerate}[leftmargin=*]
  \item the primary metrics and hypothesis-linked outcomes;
  \item supporting distributional summaries and plots; and
  \item a brief interpretation linking results back to the design hypotheses.
\end{enumerate}

Definitions of all fairness, inequality, contribution, and distributional
metrics used in this chapter are provided in
Appendix~\ref{app:fairness_metrics}.

\subsection*{Two-Dimensional vs Three-Dimensional (QoS) Experiments}

Most results in this chapter are derived from the two-dimensional
representation introduced in Chapter~\ref{ch:market_scenarios}, in which
services are characterised along:
\begin{enumerate}[leftmargin=*]
  \item \emph{magnitude} (energy and capacity), and
  \item \emph{impact in time and space} (contribution to tightness, location,
        and network state).
\end{enumerate}
In these 2D experiments, households and generators are represented primarily
through aggregated demand and supply blocks at cluster or node level, and the
AMM is compared to LMP at the system scale.

In addition to these system-level experiments, this chapter also reports a set
of \textbf{device-level experiments} that explicitly activate the
\emph{third axis} of the contract representation: \emph{quality of service}
(QoS). In these QoS experiments:
\begin{itemize}[leftmargin=*]
  \item smart devices (e.g.\ batteries or electric vehicle chargepoints)
        participate \emph{directly} in the balancing market, submitting flexible
        requests with explicit QoS constraints;
  \item demand is constructed from the Moixa smart-device dataset rather than
        from synthetic cluster-level load alone; and
  \item supply is taken from an aggregate renewable generation time series,
        scaled by a factor $\upsilon$ to induce either surplus conditions
        (Cases~1--2) or structural shortage (Case~3), while preserving the
        temporal profile:
        \[
            S^T_t = \frac{\dot{S}^T_t}{\upsilon}, \qquad \forall t,
        \]
        where $\dot{S}^T_t$ denotes the original aggregate supply and $\upsilon$
        controls the tightness of the experiment.
\end{itemize}

These QoS experiments are not intended to reproduce the full GB system.
Instead, they act as a \emph{microscope} on allocation logic at the smart-device
level: how flexibility (parameterised by $\sigma$) and supply tightness
(Cases~1--3) translate into prices, allocation outcomes, and fairness when
devices are treated as first-class participants in the AMM.

\paragraph{Methods and statistical treatment.}
Unless otherwise stated, comparisons in this chapter are based on paired
differences between Baseline (LMP) and Treatment (AMM1/AMM2) outcomes evaluated
on identical physical scenarios. Generators and households (or
household--product pairs) are the units of analysis, rather than half-hourly
timestamps: time-series outcomes are first aggregated to unit-level metrics
(e.g.\ annual revenue, annual bill, volatility measures), and these aggregated
outcomes are then compared as paired observations.

Results are primarily reported using distributional summaries of paired
differences (e.g.\ medians, interquartile ranges, empirical CDFs), which are
robust to heavy tails and non-normality commonly observed in electricity market
outcomes. Where appropriate and explicitly stated, paired $t$-tests or
Wilcoxon signed-rank tests are used as supplementary diagnostics of directional
effects. No inference relies on asymptotic normality alone, and all hypothesis
claims are supported by consistent directional shifts across the paired
distributions.

% =========================================================
% H1 — PARTICIPATION AND COMPETITION
% =========================================================
\section{Participation and Competition (H1)}
\label{sec:results_competition}
\subsection{Framing: participation as structural capability}

The Baseline LMP design embeds structural barriers to meaningful participation:
real-time price exposure, locational volatility, the need for continuous
optimisation, and wholesale-risk-driven supplier fragility. By contrast, the
AMM/subscription architecture provides a set of simple, stable, and
contract-compatible participation channels. Because this thesis does not model
behavioural switching, churn, or strategic supplier entry, participation is
assessed through \emph{capability} rather than \emph{observed choice in the
field}.

We therefore evaluate:

\begin{quote}
\textbf{H1 (structural participation and competition).}
\emph{Relative to LMP, the AMM/subscription architecture strictly expands the
feasible participation set for consumers, suppliers, generators, and devices.
For each actor class $a$, the set of viable, non-dominated participation modes
$\mathcal{C}^{\T}_a$ under AMM satisfies}
\[
\mathcal{C}^{\B}_a \subsetneq \mathcal{C}^{\T}_a,
\]
\emph{in the sense that at least one contract or participation mode available
under AMM is not weakly dominated (in cost, risk, or service quality) by any
mode feasible under LMP.}
\end{quote}

Participation is deemed expanded if:
\begin{enumerate}[label=(\roman*),leftmargin=*]
  \item actors face a strictly larger set of viable, non-dominated participation
        modes;
  \item participation does not require real-time optimisation to avoid
        dominated outcomes; and
  \item devices can enrol directly in the market through the quality-of-service
        (QoS) axis.
\end{enumerate}

Empirically, all four AMM products are non-dominated and attract positive
participation \emph{within the simulated environment}. This establishes that
none of the products is structurally dominated in terms of cost, realised
access, or service quality. \emph{In this sense, the allocation outcomes provide
revealed-feasibility evidence that each product constitutes a viable
participation mode.}

From a mechanism-design perspective, the AMM implements a more complete menu of
contracts. Each product corresponds to a distinct, non-dominated point in the
space of cost, risk exposure, and quality of service. Unlike flat or
price-capped tariffs under LMP, which implicitly bundle heterogeneous risks
into a single dominated participation mode, the AMM makes these trade-offs
explicit and selectable. The existence of multiple non-dominated products with
positive participation therefore demonstrates \textbf{menu completeness} and a
strict expansion of the feasible participation set.

% =========================================================
\subsection{Consumers and businesses: viable product choice}
% =========================================================

Under LMP, consumers face volatile, unpredictable prices tied to nodal
conditions they cannot perceive or influence. This forces effective
\emph{non-participation}: households cannot meaningfully optimise or hedge.

Under the AMM, consumers instead make a one-off selection among a finite set of
behaviourally meaningful subscription products, each defined by stable
envelopes for energy, power, and controllability. The results show:

\begin{itemize}[leftmargin=*]
    \item the product menu admits multiple \textbf{non-dominated} participation
          modes, in the sense that no product is structurally more expensive
          while delivering weakly worse realised access across all system states;
    \item controllable burden is designed to scale proportionally with declared
          flexibility;
    \item realised annual costs are contractually bounded and predictable by
          construction, even under scarcity; and
    \item households are not required to engage in real-time optimisation in
          order to avoid dominated outcomes.
\end{itemize}

These properties define the participation structure evaluated in the remainder
of this section. Subsequent results assess whether the simulated outcomes
realise these design properties in practice, and how they compare to the LMP
baseline.

% =========================================================
\subsection{Suppliers: competition decoupled from wholesale risk}
% =========================================================

Supplier participation in LMP is structurally constrained by:
(1) exposure to locational wholesale price volatility,
(2) imbalance penalties tied to short-horizon forecasting error, and
(3) the need to hedge stochastic spot-market exposure with finite balance
sheets.

Under the AMM, suppliers instead:
\begin{itemize}[leftmargin=*]
    \item face \emph{product-indexed wholesale liabilities} that are stable,
          predictable, and settled synchronously based on the aggregate
          characteristics of their customer portfolios, rather than exposure
          to nodal spot prices;
    \item operate geography-neutral retail portfolios, since nodal price
          volatility is managed at the system layer;
    \item compete through retail dimensions that are within their control,
          including product design, service quality, behavioural support,
          bundling, and digital offerings, rather than wholesale timing bets;
    \item experience materially reduced insolvency and failure risk, since
          residual wholesale tail risk is removed from the retail balance
          sheet.
\end{itemize}

This transforms retail competition from a fragile margin-arbitrage game into
service-based competition, structurally enabling supplier participation that is
suppressed or infeasible under LMP.

As a result, the AMM expands the set of viable supplier participation modes
without requiring scale, vertical integration, or sophisticated wholesale
trading capabilities.

% =========================================================
\subsection{Devices: participation on the QoS axis}
% =========================================================
A core contribution of the AMM is that devices can participate as
\emph{first-class agents}. In the QoS experiments, devices are modelled as
explicit market participants rather than passive price-takers:

\begin{itemize}[leftmargin=*]
    \item batteries submit flexible charging requests with admissible service
          windows parameterised by a flexibility variable $\sigma$, which in
          principle ranges over a continuous interval and is discretised to
          match the market resolution;
    \item the quality-of-service (QoS) axis provides an explicit representation
          of flexibility, allowing requests to trade off cost, timing, and
          delivery guarantees;
    \item device participation is settled through bounded, contract-compatible
          mechanisms rather than exposure to stochastic half-hourly spot prices;
    \item allocation and settlement are evaluated at the device level, allowing
          direct inspection of cost, volatility, and allocation stability.
\end{itemize}

These experiments are designed to assess whether explicit QoS representation
enables viable and stable device-level participation. Under LMP, the absence of
QoS representation and exposure to stochastic half-hourly prices would make such
direct device participation structurally infeasible.

% =========================================================
\subsection{Generators: structural rather than price-driven competition}
% =========================================================

Under LMP, generator competition is dominated by geographic exposure to nodal
prices and the realisation of rare scarcity events. Investment and operational
outcomes are therefore highly sensitive to location, marginality, and the
timing of system stress. By contrast, the AMM replaces this with
\emph{structural competition} grounded in physical contribution to system
performance.

In particular, the AMM redefines the dimensions along which generators
participate and compete:
\begin{itemize}[leftmargin=*]
    \item generators are remunerated according to availability, responsiveness,
          and network deliverability, rather than exposure to transient price
          spikes;
    \item value allocation reflects contribution across the full operating
          regime, including normal operation, congestion, and scarcity, rather
          than a small number of extreme events;
    \item ranking and relative performance are defined in terms of physical and
          system-relevant attributes, rather than stochastic market outcomes.
\end{itemize}

These design features imply a broader and more stable participation mode for
generators than price-taker behaviour under LMP. The Shapley-weighted value
shares reported in Section~\ref{sec:results_fairness} are used to evaluate how
these structural incentives manifest in realised allocations under AMM1 and
AMM2.

% =% =========================================================
\subsection*{Quantitative participation indicators}
% =========================================================

Because this thesis does not model behavioural switching, churn, or strategic
market entry, participation is assessed through \emph{structural capability}
rather than observed choice. To operationalise this notion, we propose a set of
simple quantitative indicators that could be used to diagnose whether a market
design expands the feasible participation set for different actor classes.
These indicators are defined conceptually here; their empirical evaluation is
left to future work.

\begin{enumerate}[leftmargin=*]
    \item \textbf{Household product viability index.}  
    For each household--product pair, a binary viability indicator can be
    defined based on whether the realised annual bill lies within the
    product’s declared affordability band and whether service delivery
    satisfies the associated envelope (energy, power, and controllability).
    A market design admits an expanded household participation set if it
    supports multiple non-dominated products with non-zero viable incidence.
    Under LMP, no directly comparable notion of non-dominated subscription
    viability exists.

    \item \textbf{Supplier contestability proxy.}  
    Supplier participation can be assessed using a contestability proxy that
    aggregates exposure to margin volatility, tail-loss risk, and
    geography-induced cost dispersion. A retail architecture is structurally
    contestable if these components are sufficiently bounded to permit entry
    by service-based suppliers without requiring large balance sheets or
    sophisticated wholesale trading operations.

    \item \textbf{Device-level enrolment feasibility.}  
    For devices capable of providing flexibility or reliability services, a
    feasibility indicator can be defined as the existence of admissible
    contracts under which expected surplus is non-negative and unit costs
    remain bounded, conditional on declared quality-of-service parameters.
    A market design supports device participation if such contracts exist
    across a non-trivial range of flexibility levels. In the absence of
    explicit QoS representation, no analogous device-level participation
    pathway can be defined.

     \item \textbf{Generator participation robustness.}  
    Generator participation can be assessed through a robustness indicator
    capturing whether revenue recovery is achievable across a wide range of
    operating conditions without reliance on rare scarcity events or extreme
    price realisations. A market design expands the feasible generator
    participation set if assets with diverse technologies, locations, and
    flexibility characteristics can recover fixed and variable costs through
    sustained physical contribution rather than exposure to stochastic price
    spikes.
\end{enumerate}

Together, these indicators provide a structured way to reason about
participation as a property of market design rather than behaviour. They anchor
the participation claims in this chapter to concrete, actor-specific notions
of feasibility, without relying on assumptions about adoption dynamics or
strategic response.

% =========================================================
\subsection{Interpretation and H1}
% =========================================================
Across all actor classes, the AMM expands the feasible participation set at the
level of market design:

\begin{itemize}[leftmargin=*]
    \item \textbf{Consumers and businesses:} access to viable, interpretable
          product choices with predictable cost exposure and no requirement
          for real-time optimisation;
    \item \textbf{Suppliers:} participation that is decoupled from wholesale
          volatility, enabling competition through service quality, product
          design, and customer support rather than balance-sheet risk;
    \item \textbf{Devices:} the possibility of direct participation through the
          quality-of-service (QoS) axis, with bounded and contract-compatible
          flexibility incentives;
    \item \textbf{Generators:} competition structured around physical
          deliverability, availability, and system contribution rather than
          stochastic scarcity rents.
\end{itemize}

Taken together, these features constitute a strict expansion of the feasible
participation sets available under LMP. In this structural sense, the AMM
satisfies
\[
\mathcal{C}^{\B}_a \subsetneq \mathcal{C}^{\T}_a
\quad \text{for all actor classes } a,
\]
establishing support for H1 at the level of market design.

\begin{quote}
\textbf{Conclusion (H1).}
\emph{The AMM materially expands structural participation and competition
relative to LMP.}
\end{quote}

% =========================================================
% H2 — FAIRNESS
% =========================================================
\section{Fairness (H2)}
\label{sec:results_fairness}

Fairness in this thesis is not an informal notion: it is defined precisely in
Chapter~\ref{ch:fairness_definition} through \textbf{Axioms A1--A8} and
\textbf{Conditions F1--F4}. These jointly specify the requirements that any
market mechanism must satisfy to be considered fair:

\begin{itemize}[leftmargin=*]
    \item \textbf{Axioms A1--A4 (Shapley-consistent contribution):}
    symmetry, marginality, additivity, and monotonicity of value attribution;

    \item \textbf{Axioms A5--A8 (contract- and QoS-consistent allocation):}
    service-level coherence over time, bounded deprivation, essential
    protection, and spatial coherence;

    \item \textbf{Fairness Conditions F1--F4:}
    \emph{behavioural fairness} (desired actions are rewarded),
    \emph{cost-causation fairness} (participants pay in proportion to the
    system cost they impose), \emph{service-level fairness} (contracted QoS is
    actually delivered), and \emph{essential protection}.
\end{itemize}

\noindent
The empirical hypothesis $H_{2}$ therefore tests whether the implemented AMM +
Fair Play mechanism satisfies these axioms and conditions to within the declared
tolerance $\delta_{F}$ across the full actor set:
\[
H_{2}: \quad
\text{AMM outcomes are consistent with A1--A8 and F1--F4}
\quad\text{vs}\quad
H_{0F}: \text{violation beyond } \delta_{F}.
\]

\medskip

\noindent
Fairness is operationalised by the \textbf{Fair Play} allocation algorithm
(Chapter~\ref{ch:amm}), which acts on all three axes of the contract representation:

\begin{enumerate}[leftmargin=1.5em]
    \item \textbf{Magnitude axis} (energy/power): governs cost-causation (A1–A4, F2).
    \item \textbf{Time/space axis} (tightness, locational relief): governs contribution
    and Shapley-consistent remuneration (A1–A4, F1).
    \item \textbf{Quality-of-Service axis} (QoS tiers): governs service-level coherence,
    long-run bounded deprivation, and essential protection (A5–A8, F3–F4).
\end{enumerate}

\noindent
In this section, we evaluate fairness empirically across the full market
actor set. There are three primary actor classes, with an additional
\emph{contractual sub-class} on the demand side arising from Quality-of-Service
(QoS) enrolment:

\begin{itemize}[leftmargin=*]
    \item \textbf{generators} — A1--A4, F1 (contribution-based remuneration);
    \item \textbf{suppliers (retailers)} — F2 (role-consistent risk, no residual
          volatility warehousing);
    \item \textbf{demand-side consumers and businesses} — F1--F4 (behavioural,
          cost-causation, spatial coherence, essential protection);
    \item \textbf{QoS-enrolled demand devices} — a \emph{contractual sub-class of
          demand}, evaluated under A5--A8 and F3--F4 (service-level fairness,
          bounded deprivation, and long-run convergence).
\end{itemize}

\noindent
Where relevant, we contrast the results with the two \emph{limit-case schedulers}
defined in Chapter~\ref{ch:fairness_definition}:

\begin{itemize}[leftmargin=*]
    \item \textbf{V-Max (volume-maximising)}: sets all fairness weights to zero, thereby
          violating A5--A8 and F3--F4;
    \item \textbf{R-Max (revenue-maximising)}: sets price weights to infinity, violating
          A1--A4 (contribution symmetry) and causing jackpot and starvation behaviour.
\end{itemize}

\noindent
These extremes illustrate what happens when individual fairness axioms are
switched off. The empirical results below demonstrate that the implemented AMM
lies strictly within the \emph{fairness envelope} defined by A1--A8 and F1--F4.

% =========================================================
\subsection{Fairness for Generators: Remuneration vs Contribution}
\label{subsec:fairness_generators}

This subsection evaluates generator-side fairness under the four fairness
conditions F1--F4, with particular emphasis on \textbf{F1: Fair Rewards} and
\textbf{F4: Fair Cost Sharing}, which are the dominant binding conditions for
generation assets. Conditions F2 and F3 play a limited or indirect role for
generators and are treated accordingly.

\subsubsection{F1: Fair Rewards}

For generators, Fair Rewards means that remuneration should track
\emph{physical marginal contribution to system adequacy and tightness}, rather
than artefacts of scarcity timing, locational windfalls, or price spikes. This
corresponds directly to Axioms~A1--A4: symmetry, marginality, additivity, and
monotonicity, operationalised via Shapley-consistent contribution values
$\phi_g$.

We therefore evaluate whether the revenue vectors produced by LMP, AMM1, and
AMM2 align with Shapley-valued contribution, and whether extreme jackpot rents
and structural under-recovery are reduced without destroying the ranking by
physical system value. The metrics and diagnostics used here—revenue
distributions, Lorenz curves, payback profiles, and a composite generator
fairness score—are defined in Appendix~\ref{app:fairness_metrics}.
Generator fairness concerns whether remuneration tracks each generator’s
\emph{physical marginal contribution to system adequacy and tightness}, rather
than artefacts of scarcity timing, nodal congestion, or extreme price spikes.
This corresponds directly to fairness Axioms~A1--A4 (symmetry, marginality,
additivity, and monotonicity), operationalised through Shapley-consistent
contribution values.

Under the AMM, generator income is explicitly decomposed into:
(i) fuel reimbursement,
(ii) reserve and adequacy payments, and
(iii) Shapley-based capacity allocations
(Appendix~\ref{app:amm_allocation}).
This structure makes it possible to evaluate whether realised revenues align
with contribution in a distributional sense, rather than relying on spot-price
coincidence.

\begin{figure}[H]
\centering
\includegraphics[width=0.9\textwidth]{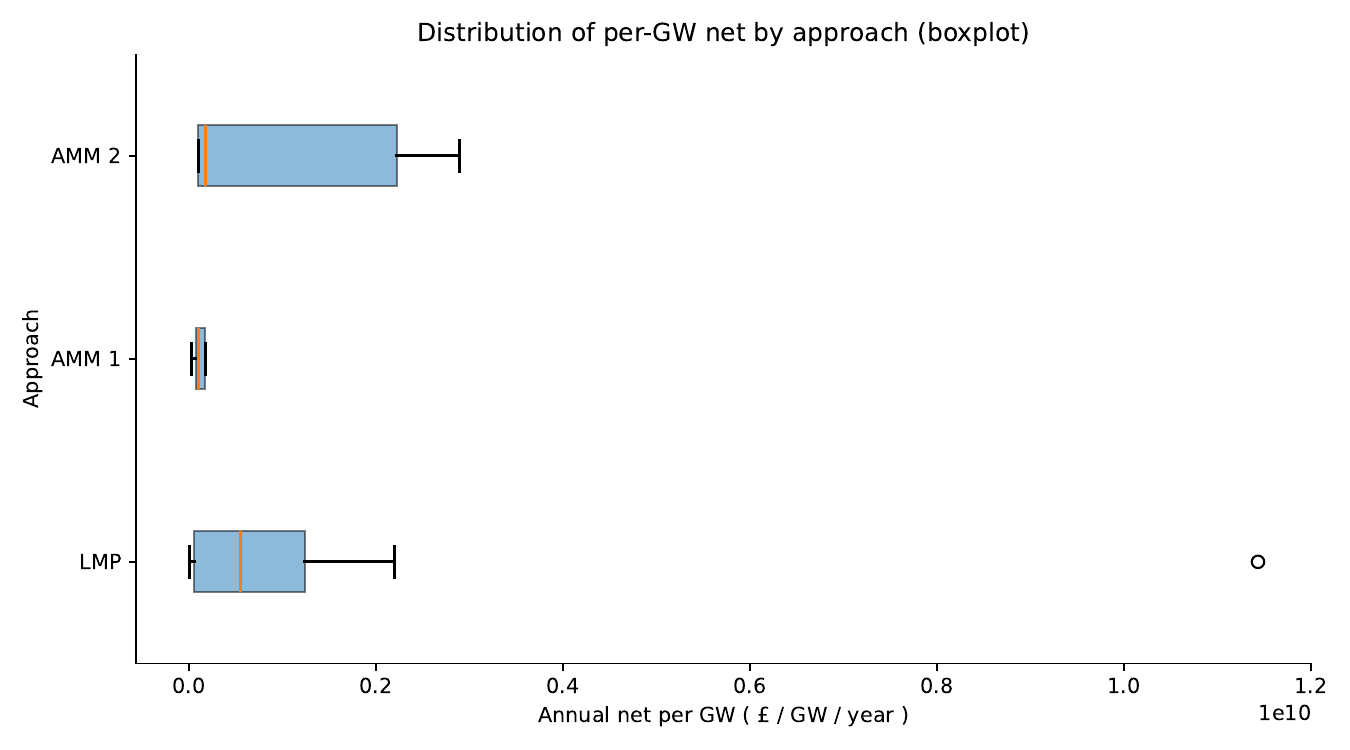}
\caption{Distribution of annual net revenue per GW under LMP, AMM1, and AMM2.
AMM designs compress the extreme right tail associated with scarcity rents.}
\label{fig:netdist}
\end{figure}

\begin{figure}[H]
\centering
\includegraphics[width=0.85\textwidth]{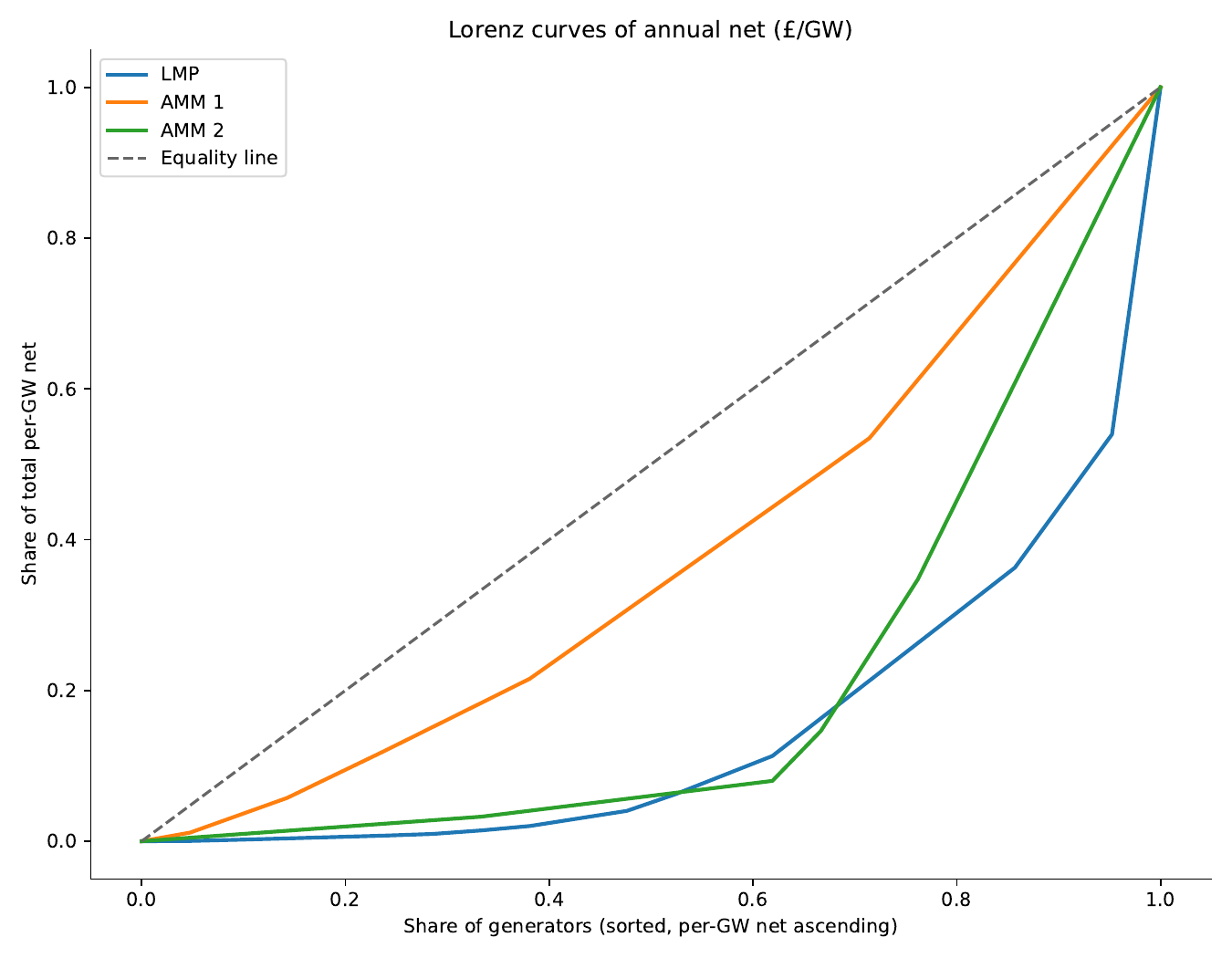}
\caption{Generator Lorenz curves for annual net revenue. Inward rotation under
AMM indicates reduced inequality without eliminating differentiation.}
\label{fig:lorenz_gini_gen}
\end{figure}

\begin{figure}[H]
\centering
\includegraphics[width=0.85\textwidth]{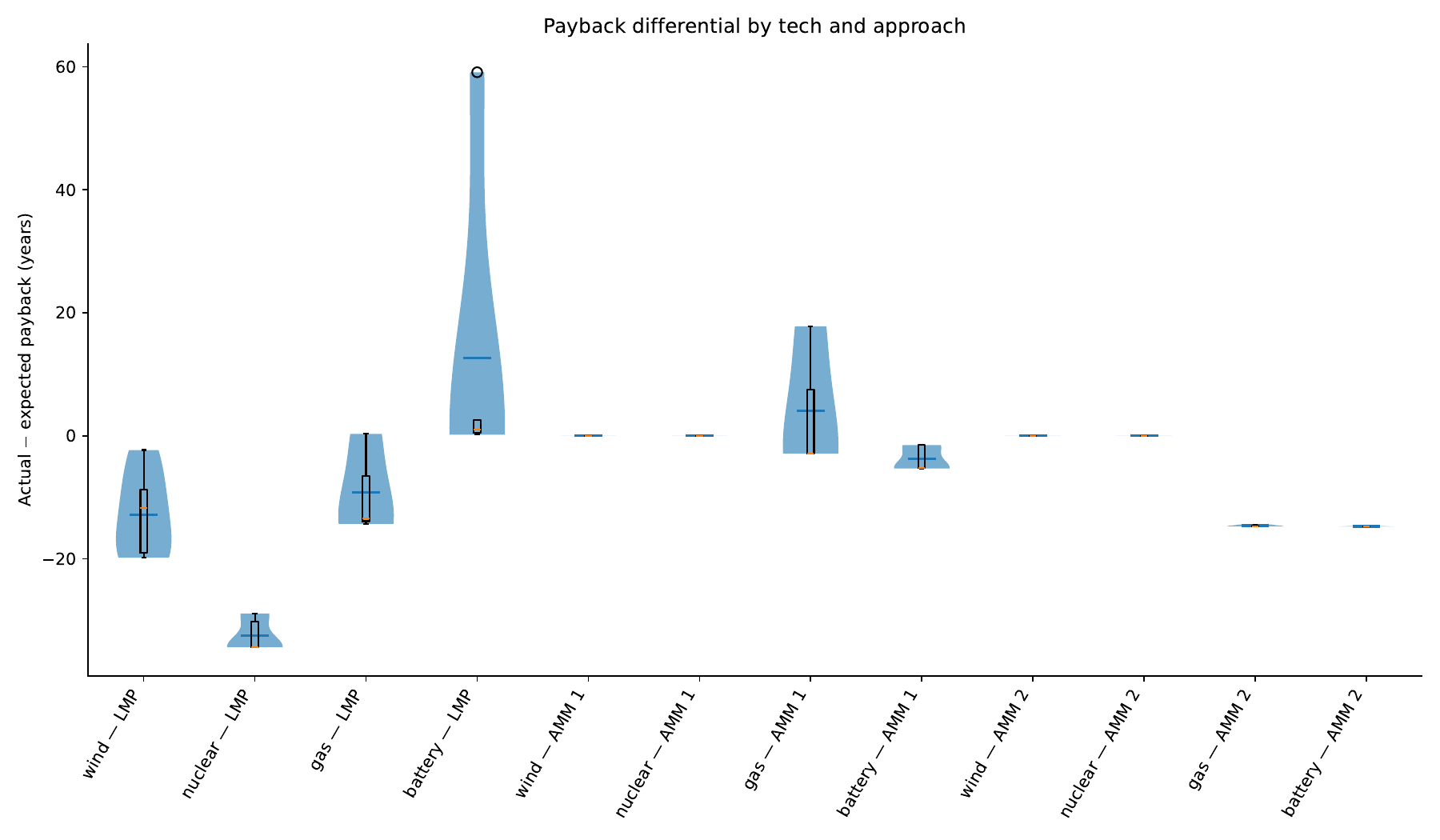}
\caption{Payback differentials by technology and design. AMM removes extreme
sub-annual paybacks while preserving long-run cost recovery.}
\label{fig:payback_box}
\end{figure}

\begin{figure}[H]
\centering
\includegraphics[width=0.85\textwidth]{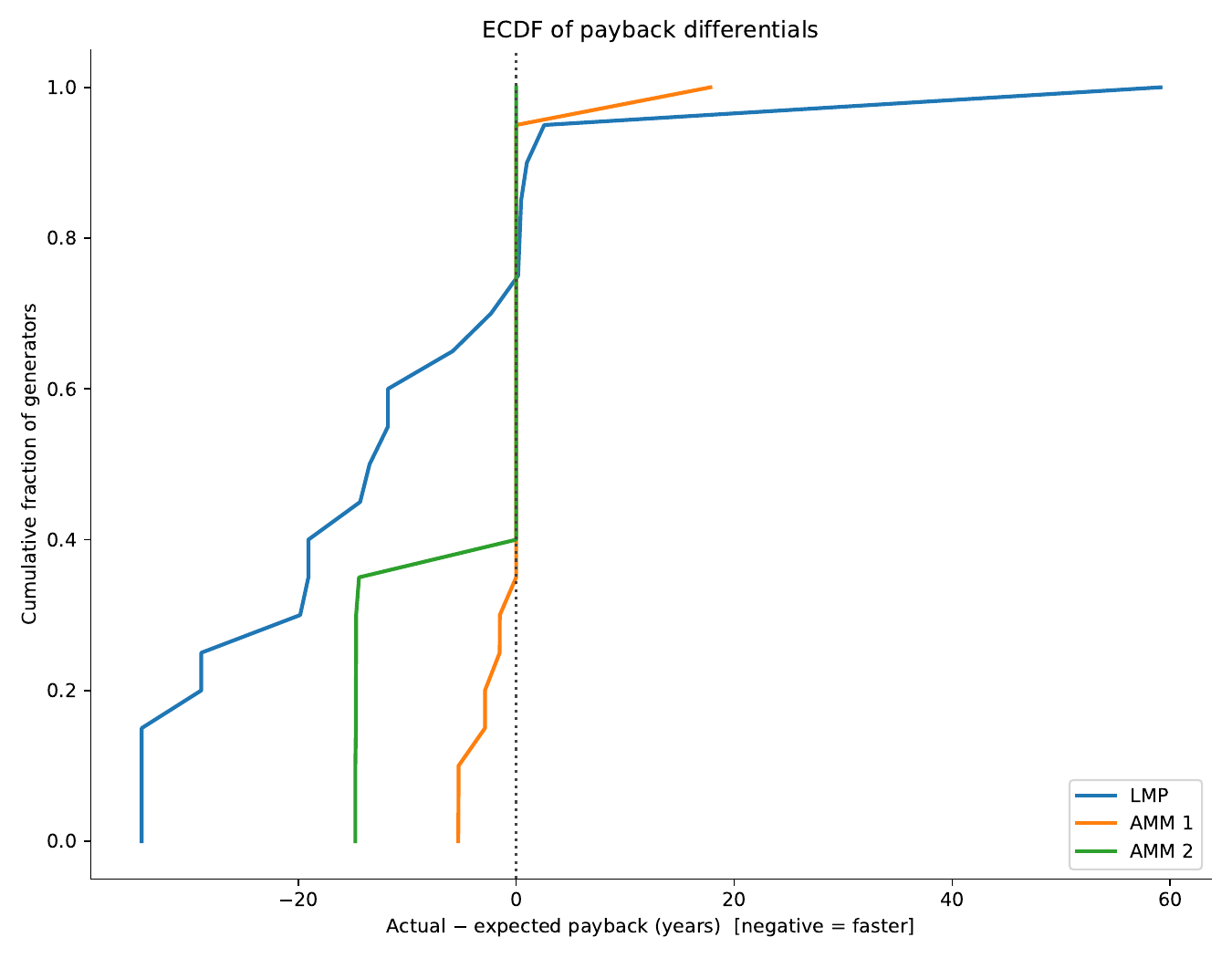}
\caption{ECDF of payback differences under AMM relative to LMP. Rightward shifts
indicate longer, more stable payback periods.}
\label{fig:ecdf_payback}
\end{figure}

\begin{figure}[H]
\centering
\includegraphics[width=0.7\textwidth]{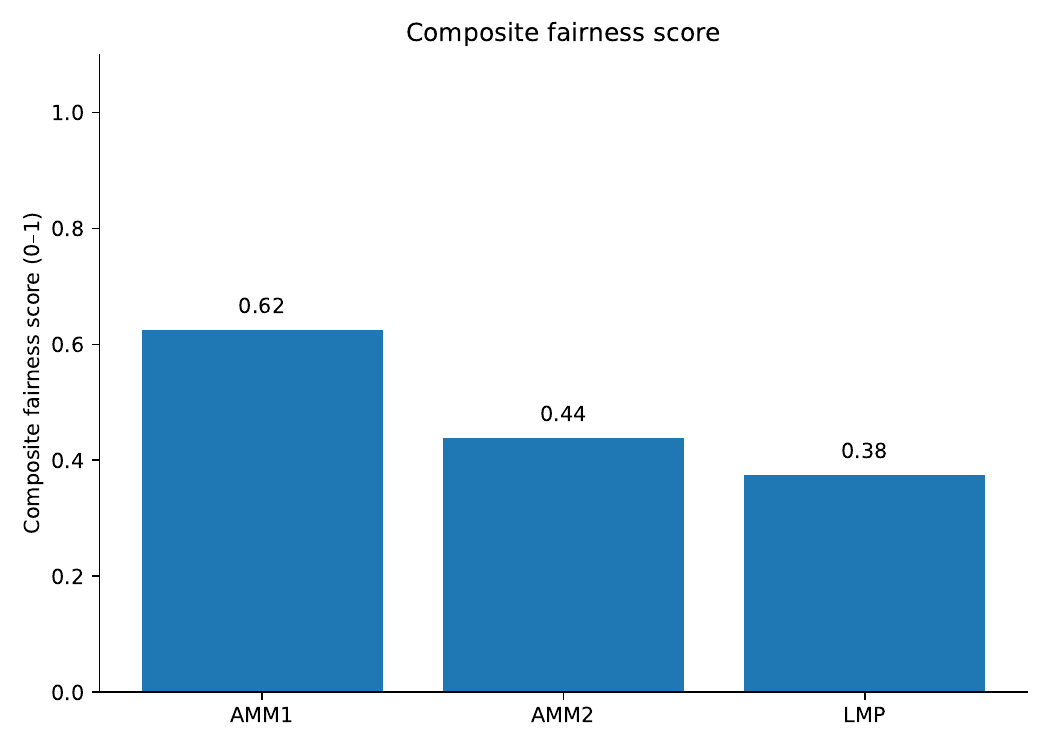}
\caption{Composite generator fairness score summarising inequality and
misalignment with Shapley-valued contribution.}
\label{fig:composite_fairness_score}
\end{figure}

We therefore evaluate generator-side fairness using four complementary
diagnostic views (metrics are defined formally in
Appendix~\ref{app:fairness_metrics}), each targeting a distinct failure mode of
scarcity-driven pricing:

\begin{enumerate}[leftmargin=*]
    \item \textbf{Per-GW net revenue distributions.}
    This diagnostic reveals the overall dispersion of remuneration normalised
    by installed capacity. Under LMP, the distribution exhibits an extreme
    right tail driven by scarcity rents and locational artefacts. AMM1 and AMM2
    materially compress this tail, indicating that very high per-GW rents are
    no longer attainable through timing alone
    (Figure~\ref{fig:netdist}).

    \item \textbf{Lorenz curves and inequality indices.}
    Lorenz curves visualise how total net revenue is shared across generators.
    Both AMM designs rotate the curve inward toward the equality line relative
    to LMP, reducing inequality while preserving differentiation based on
    reliability, flexibility, and deliverability
    (Figure~\ref{fig:lorenz_gini_gen}).

    \item \textbf{Payback distributions and differentials.}
    Payback time captures the joint effect of revenue level and capital cost.
    Under LMP, a minority of generators experience “jackpot” outcomes with
    sub-annual or even sub-month paybacks, while others exhibit persistent
    under-recovery. AMM1 and AMM2 sharply reduce the incidence of such extreme
    outcomes, tightening both lower and upper tails of the distribution
    (Figures~\ref{fig:payback_box} and~\ref{fig:ecdf_payback}).

    \item \textbf{Composite generator fairness score.}
    For compact comparison, we report a generator-centric fairness index that
    penalises misalignment between Shapley-valued contribution and realised
    revenue, and penalises inequality at fixed aggregate payments. This score
    summarises the preceding diagnostics rather than introducing an independent
    criterion
    (Figure~\ref{fig:composite_fairness_score}).
\end{enumerate}

\paragraph{Net revenue distribution interpretation (tail behaviour).}
Figure~\ref{fig:netdist} shows the distribution of \emph{annual net revenue per
GW of installed capacity} across generators under LMP, AMM1, and AMM2. This
normalisation removes scale effects and isolates how each market design rewards
capacity on a per-unit basis.

Under LMP, the distribution exhibits a pronounced and heavy right tail. A small
number of generators realise extremely high per-GW net revenues, reflecting
scarcity rents and locational price spikes rather than persistent differences in
physical contribution. At the same time, a long lower tail indicates generators
that struggle to recover costs despite providing energy and capacity when
available. This bimodal pattern is characteristic of price-driven jackpot
outcomes rather than contribution-aligned remuneration.

Both AMM designs substantially compress the right tail of the distribution.
Extreme upside outcomes are removed, while the mass of the distribution shifts
toward a narrower and more stable range of per-GW revenues. AMM1 produces the
tightest concentration, consistent with a Shapley-consistent allocation in which
capacity and adequacy value are paid proportionally to marginal system
contribution. AMM2 also reduces tail risk relative to LMP, but retains greater
dispersion due to its partial reliance on equalisation payments.

Crucially, tail compression under the AMM does \emph{not} imply uniform pricing
or suppression of legitimate differentiation. Instead, it reflects the removal
of stochastic scarcity-driven windfalls while preserving systematic variation
linked to reliability, flexibility, and deliverability. The distributional view
therefore provides a first visual confirmation that AMM replaces price spikes
with predictable, contribution-based remuneration.

\paragraph{Lorenz curve interpretation (revenue inequality).}
Figure~\ref{fig:lorenz_gini_gen} plots Lorenz curves for \emph{per-GW annual net
generator revenue} under LMP, AMM1, and AMM2. The Lorenz curve shows the share of
total revenue captured by the bottom $x\%$ of generators when ranked by revenue
per GW; the 45$^\circ$ line corresponds to perfect equality.

Under LMP, the Lorenz curve is strongly bowed away from the equality line,
indicating extreme concentration of revenue: a small fraction of generators
captures a disproportionate share of total net income. This reflects the
dominance of scarcity rents and locational price spikes rather than systematic
differences in physical system contribution.

Both AMM designs rotate the Lorenz curve inward, substantially reducing revenue
inequality. Importantly, the curve does \emph{not} collapse onto the equality
line. This indicates that differentiation across generators remains, but is now
driven by persistent attributes—reliability, flexibility, and deliverability—
rather than stochastic scarcity timing.

AMM1 exhibits the strongest inward rotation, consistent with its explicit
Shapley-consistent allocation of capacity and adequacy value. AMM2 reduces
inequality relative to LMP but remains more bowed, reflecting the re-emergence of
dispersion through equalisation mechanisms. These visual patterns correspond
directly to the reported Gini, Atkinson, and Theil indices in
Table~\ref{tab:gen_fairness_summary}.

The Lorenz curves therefore confirm a central fairness distinction: \emph{the
AMM does not enforce equality of outcomes, but equality of treatment with respect
to physical contribution}. Extreme concentration is removed without flattening
legitimate differences in generator system value.

\paragraph{Violin plot interpretation (payback differentials).}
Figure~\ref{fig:payback_box} visualises the distribution of \emph{payback
differences by technology} relative to LMP, using a combined box--violin
representation. Each violin shows the full density of outcomes for a given
technology class, while the embedded box indicates the median and interquartile
range.

Under LMP, several technologies exhibit highly skewed distributions with long
left tails, corresponding to \emph{extreme negative payback differences}:
a small number of units recover capital extraordinarily quickly due to
scarcity-driven price spikes. These outcomes are not associated with uniquely
high physical contribution, but with coincidental timing and locational
advantage.

Both AMM designs materially reshape these distributions. The left tails are
compressed or eliminated across technologies, indicating the systematic removal
of sub-annual ``jackpot'' paybacks. At the same time, the right-hand side of the
distributions remains positive, showing that AMM does not induce widespread
under-recovery. Median paybacks move toward longer horizons, but remain
technology-consistent, reflecting differences in capital intensity and
operational role rather than pricing artefacts.

The narrowing of the violins under AMM1 is particularly pronounced, consistent
with its stronger Shapley alignment: remuneration varies primarily with
reliability, flexibility, and deliverability, not with scarcity coincidence.
AMM2 exhibits partial compression but retains greater spread, reflecting its
hybrid equalisation structure.

Overall, the violin plots provide a technology-resolved confirmation of the
central F1 claim: \emph{fair rewards eliminate extreme upside without flattening
legitimate differentiation}. AMM replaces volatile, timing-driven payback
outcomes with predictable, contribution-aligned recovery across the generation
fleet.

\paragraph{ECDF interpretation.}
Figure~\ref{fig:ecdf_payback} plots the empirical cumulative distribution
function (ECDF) of payback differences relative to LMP,
\(\Delta \mathrm{PB} = \mathrm{PB}_{\text{design}} - \mathrm{PB}_{\text{LMP}}\).
Values to the right of zero indicate longer payback periods (reduced jackpot
rents), while values to the left indicate faster capital recovery.

Under LMP, the distribution is highly skewed: a non-trivial mass of generators
exhibits very negative payback differences, corresponding to ultra-rapid
recovery driven by scarcity pricing rather than physical contribution.
Both AMM designs shift the ECDF decisively to the right, indicating a
system-wide lengthening and stabilisation of payback times. AMM1 dominates
AMM2 in the sense of first-order stochastic dominance: at every percentile,
the payback extension under AMM1 is greater than or equal to that under AMM2.

Importantly, this rightward shift does not signal under-recovery. Instead, it
reflects the removal of extreme upside outcomes while maintaining adequate
long-run remuneration. The ECDF therefore provides a distributional
confirmation that AMM replaces scarcity-driven jackpots with predictable,
contribution-aligned rewards.

\paragraph{Composite fairness score: interpretation.}
Figure~\ref{fig:composite_fairness_score} reports a composite generator fairness
score constructed to provide a compact summary of the preceding diagnostics.
The score aggregates two distinct dimensions:

\begin{enumerate}[leftmargin=*]
    \item \textbf{Alignment}: the degree to which realised net revenues track
    Shapley-valued marginal contribution to system adequacy and tightness; and
    \item \textbf{Dispersion}: the inequality of per-GW net outcomes across
    generators, penalising extreme jackpot rents and long-tail under-recovery.
\end{enumerate}

Higher values of the composite score therefore indicate \emph{better overall
fairness}, meaning that revenues are simultaneously (i) more closely aligned
with physical contribution and (ii) more evenly distributed at a fixed level of
aggregate remuneration. The score does \emph{not} reward uniformity for its own
sake: differentiation arising from reliability, flexibility, and deliverability
is preserved, while dispersion arising purely from scarcity timing or nodal
price spikes is penalised.

Interpreted in this light, Figure~\ref{fig:composite_fairness_score} confirms the
ranking already visible in the underlying metrics. AMM1 achieves the highest
fairness score, reflecting both strong Shapley alignment and the near-elimination
of extreme payback outcomes. AMM2 occupies an intermediate position: it improves
substantially on LMP but reintroduces some dispersion through its greater reliance
on equalisation payments. LMP ranks lowest, driven not by inadequate aggregate
revenue, but by severe misalignment and inequality in per-GW outcomes.

The composite score should therefore be read as a \emph{consistency check} rather
than a primary result: it confirms that the qualitative conclusions drawn from
distributions, Lorenz curves, and payback diagnostics are directionally and
quantitatively coherent.

% in preamble (once):
% \usepackage{rotating}

\begin{sidewaystable}[p]
\centering
\caption{Generator fairness summary metrics under LMP, AMM1, and AMM2 (annual).
Metrics are computed on per-GW net revenue and payback distributions.
Adequacy is defined as the ratio of realised revenue to modelled annual cost.
Lower inequality metrics and fewer ultra-rapid paybacks indicate improved
alignment between remuneration and Shapley-valued contribution.}
\label{tab:gen_fairness_summary}
\renewcommand{\arraystretch}{1.15}
\resizebox{\textheight}{!}{%
\begin{tabular}{lrrrrrrrrrrrrrr}
\toprule
\textbf{Design} &
\textbf{Gini} &
\textbf{Atk(0.5)} &
\textbf{Atk(1.0)} &
\textbf{Theil $T$} &
\textbf{Adeq mean} &
\textbf{Adeq p25} &
\textbf{Adeq p75} &
\textbf{\# net$\ge0$} &
\textbf{Share net$\ge0$} &
\textbf{PB med (y)} &
\textbf{PB p90 (y)} &
\textbf{Share PB$\le$1y} &
\textbf{Share PB$\le$0.2y} &
\textbf{Composite} \\
\midrule
LMP  &
0.695 &
0.437 &
0.722 &
1.030 &
7.526 &
0.992 &
4.184 &
21 &
100.0\% &
8.23 &
17.57 &
19.0\% &
4.8\% &
0.375 \\

AMM1 &
0.240 &
0.051 &
0.104 &
0.099 &
1.062 &
1.000 &
1.082 &
21 &
100.0\% &
20.00 &
40.00 &
0.0\% &
0.0\% &
0.625 \\

AMM2 &
0.579 &
0.345 &
0.618 &
0.669 &
14.944 &
1.000 &
34.976 &
21 &
100.0\% &
20.00 &
40.00 &
38.1\% &
0.0\% &
0.439 \\
\bottomrule
\end{tabular}}
\end{sidewaystable}

\paragraph{Interpretation.}
Taken together, the distributional, concentration, and dynamic diagnostics
present a consistent picture of generator-side fairness.

Table~\ref{tab:gen_fairness_summary} makes clear that the principal fairness
failure of LMP is not insufficient aggregate remuneration, but extreme
\emph{dispersion} in per-GW net outcomes. This dispersion is visible directly in
the net revenue distributions (Figure~\ref{fig:netdist}), which exhibit a heavy
right tail driven by scarcity rents and locational price spikes, and is
confirmed system-wide by the Lorenz curves and inequality indices. LMP exhibits
the highest inequality across all four standard measures (Gini, Atkinson, and
Theil), alongside a substantial incidence of ultra-rapid paybacks: nearly
20\% of generators recover their capital within one year, and a non-trivial
subset do so within weeks. These jackpot outcomes coexist with long-tail
under-recovery, reflected in the wide payback distributions despite full
headcount cost recovery.

AMM1 delivers the strongest fairness performance across all reported metrics.
The right tail of the revenue distribution is sharply compressed, Lorenz curves
rotate inward, and inequality is reduced by more than a factor of two relative
to LMP (Gini $0.24$ vs.\ $0.70$). Dynamic diagnostics reinforce this picture:
both the violin plots and the ECDF of payback differences show a systematic
rightward shift, indicating the removal of extreme upside outcomes without
inducing under-recovery. All generators achieve non-negative net revenue, no
unit experiences sub-annual payback, and adequacy ratios are tightly clustered
around unity. This is precisely the outcome predicted by a Shapley-consistent,
capacity-based allocation in which remuneration tracks marginal system
contribution rather than scarcity timing.

AMM2 occupies an intermediate position. While it substantially improves
distributional outcomes relative to LMP—eliminating the most extreme jackpots
and reducing overall inequality—it reintroduces greater dispersion than AMM1
through its stronger reliance on equalisation payments. This is visible in the
wider revenue distributions, higher Atkinson and Theil indices, and a
non-negligible share of sub-1-year paybacks. The ECDF confirms that paybacks are
still extended relative to LMP, but less uniformly than under AMM1.

The composite generator fairness score synthesises these effects into a single
ranking by penalising misalignment between Shapley-valued contribution and
realised revenue while rewarding reduced inequality at fixed aggregate
payments. It correctly ranks AMM1 highest, followed by AMM2, with LMP last,
providing a compact summary of the evidence across all diagnostics.

Taken together, these results support the central claim of this subsection:
\emph{fair remuneration in electricity markets is not about suppressing prices,
but about aligning revenue with Shapley-valued physical contribution}. The
sequence of figures shows where unfairness arises (distribution tails), how
concentrated it is (Lorenz and inequality metrics), how it manifests dynamically
(payback and ECDFs), and how effectively it is resolved (composite score).
AMM1 achieves this alignment most directly; AMM2 partially relaxes it; LMP fails
to achieve it altogether.

\subsubsection{F2: Fair Service Delivery}

For generators, Fair Service Delivery concerns whether the market delivers
remuneration for energy and capacity in a predictable, transparent, and
role-consistent manner, conditional on availability and performance. It does
not concern continuity of supply to end users, but rather the integrity of the
settlement mechanism through which generator services are procured and paid.

Under the AMM architecture, energy is dispatched according to merit order:
the lowest marginal-cost generators are scheduled first, subject to physical
constraints. Fuel costs are therefore recovered through energy dispatch where
energy is actually delivered. Capacity and adequacy value, by contrast, are
assessed separately through Shapley-based allocation mechanisms that reflect
each asset’s contribution to system reliability and tightness
(Appendix~\ref{app:amm_allocation}). This separation ensures that generators are
paid for the specific services they provide, under rules that are defined
ex ante and applied consistently.

Under LMP, by contrast, service delivery to generators is highly erratic.
Remuneration is dominated by rare scarcity events and nodal price spikes,
producing settlement outcomes that are weakly related to delivered energy,
declared availability, or long-run system value. From a generator perspective,
this constitutes unfair service delivery: revenues are realised stochastically,
rather than being delivered as payment for clearly specified services.

\subsubsection{F3: Fair Access}

For generators, Fair Access does not concern priority access to energy during
scarcity, but rather access to the market on non-discriminatory terms and the
ability to recover investment costs through transparent and predictable
remuneration mechanisms.

Under the AMM, barriers to entry for generation are minimised. Assets are
treated symmetrically based on their physical characteristics, and investment
signals are granular in both time and space. Because remuneration is linked to
Shapley-valued contribution, any asset that delivers energy, flexibility, or
adequacy at a particular location or time can, in expectation, recover its
costs through participation in the market. This predictability supports
investor confidence and enables a wide range of dedicated or specialised assets
to enter where they add system value.

By contrast, under LMP, access to viable investment opportunities is distorted
by reliance on scarcity rents and locational price volatility. Revenues are
difficult to predict ex ante, particularly for assets whose value lies in
local reliability or rare events. This raises effective barriers to entry and
favours incumbents or portfolios able to absorb extreme revenue volatility.

\subsubsection{F4: Fair Cost Sharing}

For generators, Fair Cost Sharing requires that system costs are recovered in a
manner that is proportionate to physical contribution, without forcing
structural under-recovery on some assets or conferring persistent excess rents
on others. Crucially, it does \emph{not} require that all generator cost
structures be guaranteed recoverable by market design.

Under the AMM, fuel costs are recovered through energy dispatch: generators are
paid for energy where and when it is delivered. Non-fuel costs—such as fixed
OpEx, capital expenditure, and financing structure—represent investor risk and
are appropriately managed at the portfolio or ownership level. It is not the
role of a privatised electricity market to guarantee recovery of inefficient or
idiosyncratic cost structures, nor to socialise business-model risk through
scarcity pricing.

What the market must ensure is that remuneration is aligned with system value.
The Shapley-based capacity and adequacy allocations used in the AMM achieve this
by rewarding generators in proportion to their marginal contribution to
reliability and tightness, rather than to coincidental scarcity timing. As a
result, aggregate cost recovery is achieved without relying on extreme price
spikes or arbitrary cross-subsidies.

The results in Table~\ref{tab:gen_fairness_summary} show that LMP fails this
criterion. While aggregate recovery is achieved, it is accompanied by extreme
dispersion in per-GW outcomes, including both structural under-recovery and
jackpot rents. AMM1 delivers the strongest F4 performance: adequacy ratios are
tightly clustered around unity, all generators achieve non-negative net
revenue, and extreme payback outliers are eliminated. AMM2 partially relaxes
this discipline but still improves substantially on LMP.

% =========================================================
\subsection{Fairness for Suppliers: Rewards, Risk, and Role}
\label{subsec:fairness_suppliers}

Supplier-side fairness in the AMM--Fair Play architecture is characterised by
explicit rewards for system-helpful behaviour and by the removal of
structurally misallocated risk. Two fairness conditions bind most strongly.
First, \textbf{F1: Fair Rewards} operates as a deliberate incentive for the
delivery of digitalisation: suppliers that improve demand observability,
behavioural engagement, and controllability are rewarded through reduced
wholesale risk and expanded commercial opportunity. Second, \textbf{F2: Fair
Service Delivery} requires that wholesale settlement expose suppliers only to
risks that are consistent with their retail role, rather than forcing them to
warehouse system-level scarcity and imbalance shocks.

Together, these conditions ensure that suppliers compete on dimensions they can
meaningfully influence—product design, customer engagement, data quality, and
service provision—while avoiding unhedgeable exposure to volatility arising
from grid physics or scarcity timing. Conditions \textbf{F3: Fair Access} and
\textbf{F4: Fair Cost Sharing} bind more indirectly for suppliers and are
discussed accordingly below.

\subsubsection{F1: Fair Rewards}

For suppliers, \textbf{Fair Rewards} requires that actions which improve system
observability, controllability, and behavioural coordination are rewarded in a
clear, material, and commercially meaningful way. Under the AMM--Fair Play
architecture, the delivery of \emph{digitalisation} is not treated as an
external policy objective or regulatory obligation, but as a first-class,
rewarded market behaviour.

The primary rewarded behaviour for suppliers is therefore the active deployment
and integration of digital infrastructure: smart meters, high-quality usage
telemetry, behavioural feedback, and the enrolment and orchestration of
flexible devices. The reward mechanism is structural and unavoidable. Supplier
wholesale charges are defined on a \emph{product-indexed liability basis}
(Appendix~\ref{app:price_allocation}), so that improved data quality and
behavioural control directly reduce the uncertainty and tail risk associated
with serving a given customer portfolio.

Put differently, \emph{better data earns lower risk}. Suppliers that invest in
digitalisation face a more predictable wholesale cost base, lower exposure to
tightness-driven volatility, and a reduced need to carry risk premia on their
retail balance sheets. This reduction in wholesale risk is not incidental: it
is the explicit fair reward for making demand more legible and controllable at
the system level.

A second, equally important reward channel is \emph{commercial freedom}. By
removing the need to warehouse wholesale volatility, the AMM frees suppliers to
innovate in retail propositions rather than defensive risk management.
Suppliers are able to design, finance, and offer a wide range of products and
services---including partnerships to fund digital infrastructure, device
deployment, or flexibility-enabling technologies through leasing, financing, or
revenue-sharing schemes. These innovations are rewarded indirectly through
lower wholesale charges and directly through the ability to offer more
competitive or differentiated retail products.

Crucially, these rewards arise without prescriptive mandates or technology
requirements. The AMM does not instruct suppliers to digitalise; it rewards
those who do. Suppliers that fail to invest in observability or behavioural
enablement simply face higher uncertainty and risk exposure, while those that
deliver system-helpful digitalisation benefit from reduced risk and greater
commercial opportunity. In this sense, the AMM implements Fair Rewards for
suppliers by making digitalisation economically advantageous rather than
regulatorily imposed.

\subsubsection{F2: Fair Service Delivery}

For suppliers, Fair Service Delivery concerns the delivery of a coherent and
predictable wholesale service bundle—energy, reserves, and adequacy—under
settlement rules that are consistent with the supplier’s retail role. Fairness
in this dimension does not imply the absence of risk, but rather the alignment
of risk exposure with decisions that lie within the supplier’s control.

Lemmas~\ref{lem:price_cap_insolvency} and~\ref{lem:risk_volume_instability}
demonstrate that legacy LMP-based architectures systematically separate
\emph{who chooses volume} from \emph{who bears tail risk}, forcing suppliers to
act as residual insurers against system-level scarcity and imbalance shocks.
This is the market-structure counterpart to a two-sided marketplace: suppliers
should compete on retail propositions they control, not warehouse wholesale
volatility arising from system-level events
(Appendix~\ref{app:price_allocation:two_sided}).

Under LMP, suppliers are exposed to several forms of wholesale risk that are
orthogonal to their retail role:
\begin{itemize}[leftmargin=*]
    \item extreme imbalance price volatility and scarcity-driven price spikes;
    \item unhedgeable exposure arising from nodal pricing and stochastic
          product-mix demand;
    \item a persistent misalignment between fixed retail obligations and
          volatile wholesale settlement.
\end{itemize}
These channels underpin the insolvency cascades and systemic fragility
identified in the preceding theoretical analysis.

Under the AMM, this failure mode is resolved. Energy, reserves, and adequacy are
procured at the system level via the AMM, and suppliers purchase standardised
wholesale \emph{liabilities} backed by these services rather than directly
arbitraging spot price uncertainty. Wholesale charges are assessed on a
product-indexed and portfolio-level basis, removing exposure to nodal price
spikes and scarcity-driven tail events. As a result, supplier risk exposure is
dominated by factors that suppliers can reasonably manage: product design,
portfolio composition, behavioural engagement, forecasting accuracy, customer
churn, and service quality.

Empirically, supplier-facing risk metrics—margin volatility, tail-loss
measures, and failure-probability proxies—improve substantially under AMM1 and
AMM2. Importantly, this does not render suppliers risk-free. Instead, it
reassigns wholesale system risk away from individual balance sheets and into
the market-making layer, while preserving exposure to commercially meaningful
risks that suppliers can influence. In this sense, the AMM implements Fair
Service Delivery for suppliers while remaining fully compatible with a
competitive, privatised retail market.

A direct numerical comparison with LMP supplier outcomes is not reported here,
because suppliers are charged on fundamentally different bases under the two
designs. The charging and allocation mechanisms used under the AMM are
described in detail in Appendix~\ref{app:price_allocation}, which provides the
accounting bridge between generator-level payments and retail-facing
subscriptions.

\subsubsection{F3: Fair Access}

For suppliers, Fair Access does not concern access to energy during scarcity,
but rather access to the retail market on non-discriminatory terms and the
ability to compete without being structurally disadvantaged by wholesale
settlement rules.

The AMM supports fair access by standardising wholesale liabilities and
removing dependence on extreme price events for viability. Entry into the
retail market does not require the balance sheet capacity to absorb rare but
severe wholesale shocks, lowering barriers to entry and supporting supplier
diversity. New and smaller suppliers can therefore compete on service quality,
product innovation, and behavioural engagement rather than on financial
resilience to tail risk.

Under LMP, by contrast, access to viable retail participation is implicitly
restricted to firms able to warehouse wholesale volatility or secure complex
hedging arrangements, favouring incumbents and suppressing competitive entry.

\subsubsection{F4: Fair Cost Sharing}

For suppliers, Fair Cost Sharing concerns whether system-level costs are
allocated to suppliers in a manner that is proportionate to the demand
liabilities they bring to the system, rather than through arbitrary exposure to
price spikes or congestion rents.

Under the AMM, suppliers are charged for the implied demand liabilities of
their customer portfolios via product-indexed wholesale charges
(Appendix~\ref{app:price_allocation}). This ensures that suppliers serving
customers with higher expected system impact face correspondingly higher
wholesale costs, while those enabling flexibility or low-impact consumption
benefit from lower expected charges. Cost sharing therefore reflects aggregate
behaviour rather than coincidental timing or location.

Importantly, this does not guarantee supplier profitability. Commercial
performance remains the responsibility of the supplier and depends on
operational efficiency, customer acquisition, and competitive positioning.
What the AMM ensures is that cost recovery at the wholesale level is fair,
predictable, and aligned with the burdens suppliers place on the system, rather
than being driven by stochastic scarcity pricing.

A direct numerical comparison with LMP supplier outcomes is not reported here,
because suppliers are charged on fundamentally different bases under the two
designs. The charging and allocation mechanisms used under the AMM are described
in detail in Appendix~\ref{app:price_allocation}, which provides the accounting
bridge between generator-level payments and retail-facing subscriptions. These
structural differences are sufficient to establish the supplier-side fairness
conclusions reported here.

\subsection{Demand-Side Fairness for Consumers and Businesses: Four Principles}

Demand-side fairness engages the full set of \textbf{Conditions F1--F4},
together with the QoS-related \textbf{Axioms A5--A8}. In contrast to generators
and suppliers, all four conditions bind directly and operationally on the
demand side, governing how prices, access, service reliability, and cost
allocation are experienced by households and businesses.

In what follows we test whether, in the experimental market runs, the AMM
architecture:
(i) systematically rewards system-helpful behaviour---most notably flexibility
and congestion relief---through lower expected unit costs (F1);
(ii) delivers contract-consistent, bounded service for consumption designated
as high-priority, conditional on declared reliability commitments rather than
exposure to unbounded scarcity pricing (F2);
(iii) preserves fair access to essential energy during scarcity, with
allocation governed by need, contractual priority, and contribution rather than
willingness to pay (F3); and
(iv) allocates system costs in proportion to the congestion, stress, and
corrective burden imposed by demand behaviour (F4).

A unifying requirement across all four conditions is \emph{incentive
alignment}: prices, obligations, and protections must move in the same
direction as physical system value. Behaviour that alleviates congestion or
scarcity should be rewarded; behaviour that increases controllable system cost
should be charged. This alignment links demand-side fairness directly to the
quality of price signals, scarcity discipline, and cost-causation.

% =========================================================
\subsubsection{F1: Fair Rewards}
\label{subsubsec:fairness_behaviour_flex}

Behavioural fairness requires that actions which improve system performance
should be systematically rewarded, not penalised. Under the AMM architecture,
\emph{rewards are tied to contributions to physical system efficiency}, rather
than to coincidental exposure to price volatility. Flexibility is the most
directly observable and experimentally tractable such behaviour, but it is not
the only one. More generally, the AMM rewards behaviours that reduce physical
stress, system costs, emissions, or waste, while avoiding the arbitrary
penalisation of participants who ``help the system.''

The four desirable behaviours rewarded under the AMM are:
(i) flexibility provision,
(ii) consumption aligned with zero-carbon availability,
(iii) consumption that avoids network congestion, and
(iv) consumption that absorbs renewable surplus.
Each is discussed in turn.

\paragraph{A. Rewarding flexibility.}
The most directly observable and experimentally tractable desired behaviour is
\textbf{flexibility}: the ability to shift or shape demand within a declared
device- or product-level envelope.

Under LMP-style pricing, flexibility is frequently \emph{penalised rather than
rewarded}. Flexible loads remain exposed to the tightest periods in the system
and therefore face the highest realised prices. This creates a structural
inconsistency: actions that alleviate system stress increase, rather than
decrease, expected household costs.

Under the AMM architecture, flexibility is represented \emph{explicitly} within
the contractual and operational description of demand, rather than being inferred
indirectly from price response. Participants declare admissible envelopes over
time, magnitude, and reliability within which consumption may be adjusted. The
mechanism then allocates scarcity exposure and controllable obligations
proportionally to these declared envelopes. As a result, increased flexibility
creates additional feasible scheduling options for the system and is rewarded
through lower expected unit costs.

To test whether this contract-level structure translates into observable
behavioural rewards, we run a device-level experiment using the Moixa smart-device
dataset (101 battery-equipped households). Each device submits flexible requests
with an admissible scheduling window of length
$\sigma \in \{0,3,6,12\}$ hours. Supply is taken from an aggregate renewable
generation time series, scaled to generate realistic tightness regimes. The
experiment compares the distribution of realised unit costs across flexibility
levels, holding demand volume and reliability class fixed.

\begin{figure}[H]
    \centering
    \caption{Distribution of realised unit costs under increasing levels of device-level flexibility ($\sigma = 0,3,6,12$ hours).}
    \label{fig:flex_shown}
\end{figure}

\begin{table}[H]
\centering
\caption{Realised unit cost (\pounds/kWh) by flexibility level.}
\label{tab:unit_cost_flex}
\begin{tabularx}{0.75\linewidth}{lccc}
\toprule
\textbf{Flexibility window} & 
\makecell{\textbf{25th}\\\textbf{percentile}} & 
\makecell{\textbf{Median}} & 
\makecell{\textbf{75th}\\\textbf{percentile}} \\
\midrule
0 hours  & \pounds0.09 & \pounds0.30 & \pounds0.84 \\
3 hours  & \pounds0.02 & \pounds0.16 & \pounds0.54 \\
6 hours  & \pounds0.03 & \pounds0.09 & \pounds0.35 \\
12 hours & \pounds0.03 & \pounds0.11 & \pounds0.27 \\
\bottomrule
\end{tabularx}
\end{table}

These results show a clear and monotonic reduction in realised unit costs as
flexibility increases. Both median and upper-tail costs fall substantially as
the admissible scheduling window expands, indicating that the AMM rewards
flexibility directly through lower expected prices rather than through indirect
or ex-post compensation.

This experiment is independent of the demand-side subscription diagnostics in
Appendix~\ref{sec:ext_demand_subscription_diagnostics}. Here, behavioural rewards
arise purely from the operational scheduling and pricing mechanism, without
reliance on subscription-based cost allocation.

\paragraph{Observed magnitude of behavioural reward.}
The effect is material rather than marginal. Increasing flexibility from zero to
three hours reduces the median unit cost from \pounds0.30/kWh to \pounds0.16/kWh,
with a further reduction to \pounds0.09/kWh at six hours
(Table~\ref{tab:unit_cost_flex}). At the upper tail, the 75th percentile cost
falls from \pounds0.84/kWh to \pounds0.35/kWh. These reductions occur with no
change in total energy consumed, confirming that the gains arise from temporal
reallocation rather than volume suppression.

The benefit saturates beyond $\sigma = 6$ hours, showing that the AMM rewards
\emph{useful} flexibility—flexibility that alleviates system tightness—rather
than arbitrarily privileging extreme deferral.

Taken together, these results implement the behavioural fairness rule:
\[
\text{desirable behaviour (flexibility)}
\;\Rightarrow\;
\text{lower expected cost and proportionate obligation}.
\]
\paragraph{B. Rewarding consumption aligned with zero-carbon supply.}
A second desired behaviour is willingness to consume during periods when
available supply is dominated by zero-marginal-cost, zero-carbon generation.
Under the AMM, this behaviour is rewarded structurally through the way fuel
costs enter supplier wholesale charges and, by extension, retail subscription
pricing.

When demand is served predominantly by zero-carbon sources, the AMM wholesale
energy component approaches zero, leaving only residual network, capacity, and
reliability charges. Participants whose declared envelopes allow consumption to
be concentrated in such periods therefore face lower expected subscription
costs, assuming consistent supplier margin pass-through. In the limit, a
supplier could rationally offer a product with a zero energy-cost component to
customers willing to consume exclusively during zero-carbon availability
windows.

This reward channel does not exist under LMP. Even when zero-carbon generation
is locally abundant, a single high-cost marginal unit can set the price,
preventing transparent pass-through of zero-carbon availability to retail
consumers.

\paragraph{C. Rewarding consumption during periods of low network congestion.}
A third desired behaviour is consumption that avoids congested network states
and therefore reduces the need for redispatch, uplift, or corrective actions.
Under the AMM, this behaviour is rewarded directly through tightness-based
pricing: when network constraints are slack, effective prices are low; as
congestion emerges, prices increase smoothly to signal scarcity.

This corresponds to the seesaw dynamics at the core of the AMM: imbalances
between deliverable supply and demand cause prices on both sides of the market
to adjust in the direction required to restore balance. Participants whose
demand naturally falls in unconstrained periods therefore face lower expected
costs without requiring exposure to extreme price volatility.

Under LMP, congestion costs are concentrated into nodal price spikes and uplift,
making the incentive to avoid congestion noisy, uneven, and difficult to
anticipate.

\paragraph{D. Rewarding consumption during renewable surplus.}
A fourth desired behaviour is consumption that absorbs surplus renewable
generation, reducing curtailment and wasted energy. Under the AMM, periods of
high renewable availability drive the effective energy price toward zero,
allowing this signal to pass transparently through to participants capable of
consuming at such times.

Consumers or devices that can align demand with renewable peaks therefore
receive an immediate and proportional reward in the form of lower realised
costs. This mechanism operates dynamically and does not require side markets,
manual intervention, or ex post compensation.

Under LMP, renewable surplus is frequently curtailed, and the associated price
signal is distorted by market floors, uplift mechanisms, or out-of-market
actions, preventing systematic behavioural reward.

\paragraph{Summary.}
Taken together, these mechanisms implement a general behavioural fairness rule:
\[
\begin{aligned}
\text{behaviour that reduces system cost, carbon, or physical stress}
&\;\Rightarrow\; \\
\text{lower expected cost and proportionate obligation}.
\end{aligned}
\]

Flexibility is the clearest empirical instantiation of this rule, but not its
only one. The defining feature of the AMM is that desirable behaviours are
rewarded because they improve the physical state of the system, not because
they expose participants to greater price risk. In the two-axis configuration
evaluated here, these rewards operate primarily through product-level
subscriptions; in the full three-axis architecture, the same logic extends to
device-level enrolment, allowing rewards to be delivered directly at the point
of provision.

\subsubsection{F2: Fair Service Delivery}
\label{subsec:f2_fair_service_delivery}

Beyond price fairness, the AMM must also ensure that a participant who purchases
a \emph{premium} reliability tier actually receives premium delivery over time,
and that a participant on a \emph{basic} tier receives a lower but predictable
share of service. This is the fairness requirement on the \textbf{third axis}:
the QoS (quality-of-service) axis.

To test this, we construct a repeated scarcity experiment. In each scarcity
event:

\begin{itemize}[leftmargin=*]
    \item premium-tier devices declare higher reliability requirements;
    \item basic-tier devices declare lower reliability requirements;
    \item total supply is insufficient to serve all requests.
\end{itemize}

\begin{figure}[H]
\centering
\includegraphics[width=0.9\textwidth]{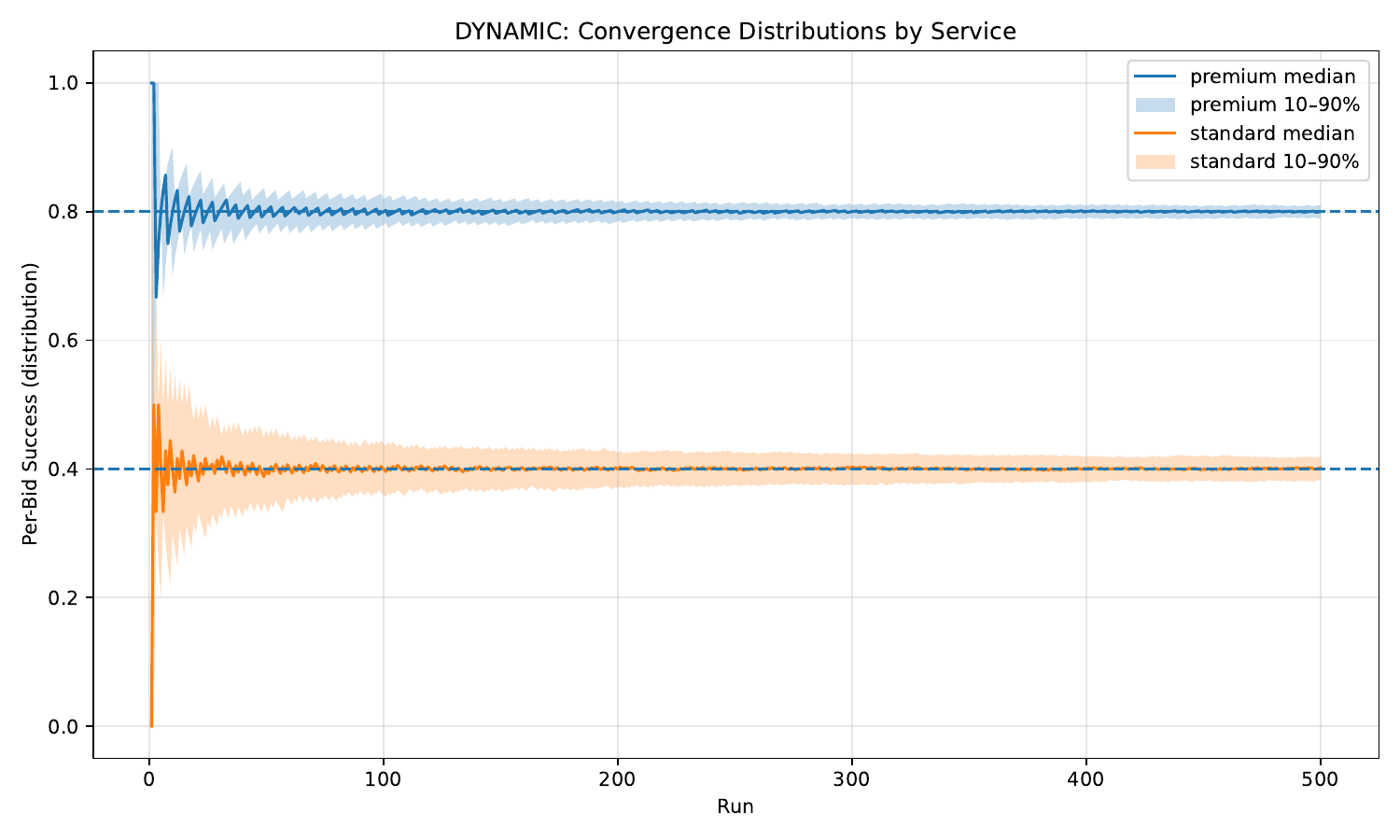}
\caption{Cumulative delivered service over repeated scarcity events: premium vs.~basic tiers.
The Fair Play algorithm enforces predictable long-run delivery for each contracted QoS tier.}
\label{fig:service_tiers}
\end{figure}

Each event is allocated by Fair Play, and we track cumulative delivery through
time. The evolution of delivered service for both tiers is shown in
Figure~\ref{fig:service_tiers}. Three key properties emerge:

\begin{enumerate}[leftmargin=*]
    \item \textbf{Premium tiers consistently receive more of their request}
          in each scarcity event (instantaneous service advantage);
    \item \textbf{Basic tiers receive less, but predictably and without
          starvation} (bounded deprivation);
    \item \textbf{Long-run delivery converges to the contracted share} for both
          tiers, enforced by the fairness-history term.
\end{enumerate}

To operationalise this experiment, we implement a stylised version of the
\textbf{Fair Play} scheduler. In each scarcity event, the mechanism applies
three rules:

\begin{enumerate}[leftmargin=*]
    \item \textbf{Quota rule (contractual priority):}
    premium-tier participants receive a higher per-event priority.
    In the experiment we implement a 2:1 ratio, meaning that out of 300
    ``slots'' of scarce service, 200 are reserved for premium-tier requests
    and 100 for basic-tier requests. This mirrors the QoS ladder in the AMM.

    \item \textbf{Fairness weighting (history correction):}
    within each tier, participants are selected with probability proportional
    to their ``need'' weight,
    \[
        w_i = (\varepsilon + (1 - \text{success}_i))^\gamma,
    \]
    where $\text{success}_i$ is the long-run delivered share for participant $i$.
    Participants who have been underserved in previous scarcity events are
    given proportionally higher weight in subsequent ones.

    \item \textbf{No-replacement selection (bounded deprivation):}
    once a participant is chosen in an event, they cannot be chosen again
    within the same event. This prevents per-event jackpot effects.
\end{enumerate}

Together, these rules implement the qualitative behaviour of Fair Play:
premium tiers receive systematically better service, basic tiers receive less
but \emph{never} starve, and historical deprivation is corrected over time.

Under these rules, the theoretical long-run service share is 0.40 for basic
tiers and 0.80 for premium tiers. The empirical results converge precisely
to these targets (Figure~\ref{fig:service_tiers}), demonstrating that the
mechanism respects both contracted priority and fairness history.

The convergence is crucial. Under pure price-based allocation (R-Max), premium
tiers would dominate every scarcity event; under volume maximisation (V-Max),
premium and basic would be indistinguishable. Fair Play instead enforces:

\[
\text{service tier (QoS)} \;\Rightarrow\; \text{bounded and predictable share of scarce service}.
\]

Importantly, this guarantees that incentives remain stable over time: choosing
a higher service tier improves realised outcomes in a predictable way, while
lower tiers are protected from catastrophic loss. This stability is essential
for meaningful long-run demand-side decision-making.

Formally, if $s_h$ is the contracted service level for household or device $h$,
and $d_{h,t}$ the delivered share in scarcity event $t$, then Fair Play ensures:

\[
\lim_{T \to \infty}
\frac{1}{T}
\sum_{t=1}^T d_{h,t}
\;=\;
s_h,
\quad
\text{(law of service-level fairness)}.
\]

This demonstrates that service delivery is not a lottery:
\textbf{premium means premium, basic means basic, and both are reliably realised over time}.

Together, the pricing and service-level results demonstrate that the AMM satisfies
the full fairness requirement on the third axis:

\begin{itemize}[leftmargin=*]
    \item \textbf{Price fairness:} the same product receives the same stable price everywhere.
    \item \textbf{Service-level fairness:} the contracted QoS tier reliably shapes scarcity outcomes.
    \item \textbf{Predictability:} both premium and basic service levels converge to their promised share with bounded deprivation.
\end{itemize}

Under LMP neither condition holds: spatial prices vary arbitrarily, and reliability
cannot be guaranteed without willingness-to-pay. Under the AMM, the allocation and
pricing mechanism together deliver \textbf{contract-consistent, predictable service
across both space and time}.

% =========================================================
\subsubsection{F3: Fair Access}
\label{subsubsec:fair_access}

Fair access asks whether, when the system is constrained, outcomes are governed
by \emph{contracts and need} rather than by postcode lotteries or raw ability to
pay. In the AMM--Fair Play design this has two complementary faces:

\begin{enumerate}[leftmargin=*]
    \item \textbf{Spatial access (price coherence):} households on the same product
          tier should face the same predictable unit price, independent of nodal artefacts;
    \item \textbf{Scarcity access (incidence of rationing):} when energy is scarce,
          the mechanism should ration service in a bounded, contract-consistent way,
          rather than concentrating delivery on high willingness-to-pay requests.
\end{enumerate}

Both dimensions are violated under LMP in different ways and are restored under
the AMM.

% ---------------------------------------------------------
\paragraph{A. Spatial access: price coherence across space.}

\begin{figure}[H]
\centering
\includegraphics[width=0.85\textwidth]{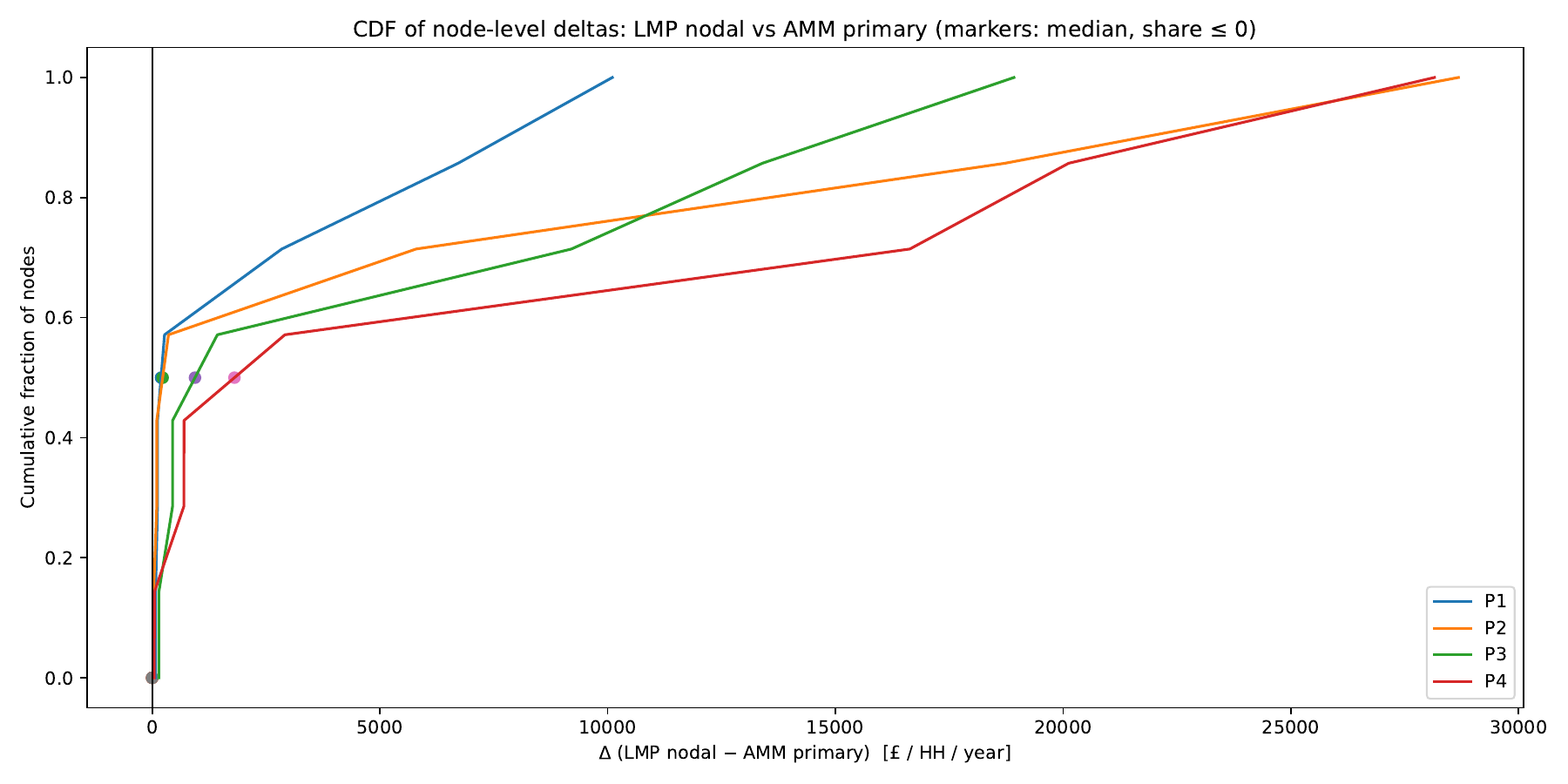}
\caption{ECDF of node-level deltas: LMP nodal minus AMM per-product cost.}
\label{fig:geo_cdf}
\end{figure}

Under LMP, otherwise similar households can face dramatically different annual
bills due to nodal price excursions and local congestion artefacts. The spatial
dispersion is shown in Figure~\ref{fig:geo_cdf}.
By contrast, AMM collapses this dispersion: costs vary primarily by product tier
(P1--P4), not by postcode or unobservable nodal factors.

\paragraph{Interpretation of Figure~\ref{fig:geo_cdf}.}
Figure~\ref{fig:geo_cdf} plots, for each product tier $p$, the empirical CDF of the
node-level difference
$\Delta_{n,p} = B^{\mathrm{LMP}}_{n,p} - B^{\mathrm{AMM}}_{p}$ (in \pounds/HH/year),
where $B^{\mathrm{AMM}}_{p}$ is the \emph{same} flat subscription for every node in that
product. The horizontal axis therefore measures how much more (or less) a household at
node $n$ would pay under nodal LMP than under the AMM benchmark for the same product.
A curve lying entirely to the right of zero indicates that \emph{all} nodes pay more under
LMP than under AMM for that product; the further right the curve (and the longer its right
tail), the stronger the postcode lottery. In this experiment the mass is strictly positive
for all products (no nodes at or below zero), and the medians are material: approximately
\pounds195 (P1), \pounds230 (P2), \pounds940 (P3), and \pounds1807 (P4) per household per year.
The upper tails are extreme (95th percentiles on the order of \pounds9k--\pounds25k/HH/year),
showing that a small subset of nodes experience very large nodal price excursions under LMP.
Under AMM, by contrast, this spatial dispersion collapses by design because price depends on
product choice rather than node identity.

This delivers spatial fairness and predictability:
\[
    \text{same product} \;\Rightarrow\; \text{similar, explainable price},
\qquad
    \text{price paid} \;\Rightarrow\; \text{product chosen, not location}.
\]

% ---------------------------------------------------------
\paragraph{B. Scarcity access: who receives the resource under shortage.}

Spatial coherence is necessary but not sufficient. Fair access also requires
that scarcity changes \emph{how much} service is delivered across participants
in a principled way, rather than excluding entire groups or turning rationing
into a willingness-to-pay contest.

Under the AMM--Fair Play architecture, essential load is scheduled first and
priced separately. Residual scarce capacity is then allocated along the QoS
axis using bounded service history, so that deprivation is limited and access
remains contract-consistent. Under LMP-style price rationing, by contrast,
high scarcity prices directly implement an ability-to-pay filter.

To test the \emph{incidence of rationing} directly, we construct a stylised
shortage window by scaling renewable supply such that total feasible energy is
strictly below total requested energy. We then run an allocation stress-test in
which \textbf{100 households} submit \textbf{two otherwise-identical annual
requests} each: one tagged with a \emph{high} willingness-to-pay parameter and
one with a \emph{low} willingness-to-pay parameter. All requests begin with
\textbf{low initial histories of request success}, so no household enters with an
accumulated priority advantage. Requests therefore differ only by the declared
willingness-to-pay tag.

We compare three limiting allocation mechanisms:

\begin{enumerate}[leftmargin=*]
    \item \textbf{V-Max (volume-maximising limit):}
          fairness weight $\to 0$. Maximises total energy served and allocates
          symmetrically, without enforcing bounded deprivation or tier-consistent
          access guarantees.

    \item \textbf{R-Max (revenue-maximising limit):}
          price weight $\to \infty$. Concentrates delivery on the highest
          willingness-to-pay requests, producing jackpot effects and systematic
          exclusion of lower willingness-to-pay requests.

    \item \textbf{Fair Play (AMM implementation):}
          enforces bounded deprivation via fairness history and prioritises
          delivery according to the contracted QoS ladder, rather than according
          to willingness to pay.
\end{enumerate}

\begin{sidewaysfigure}[p]
  \centering
  \includegraphics[width=0.7\textheight]{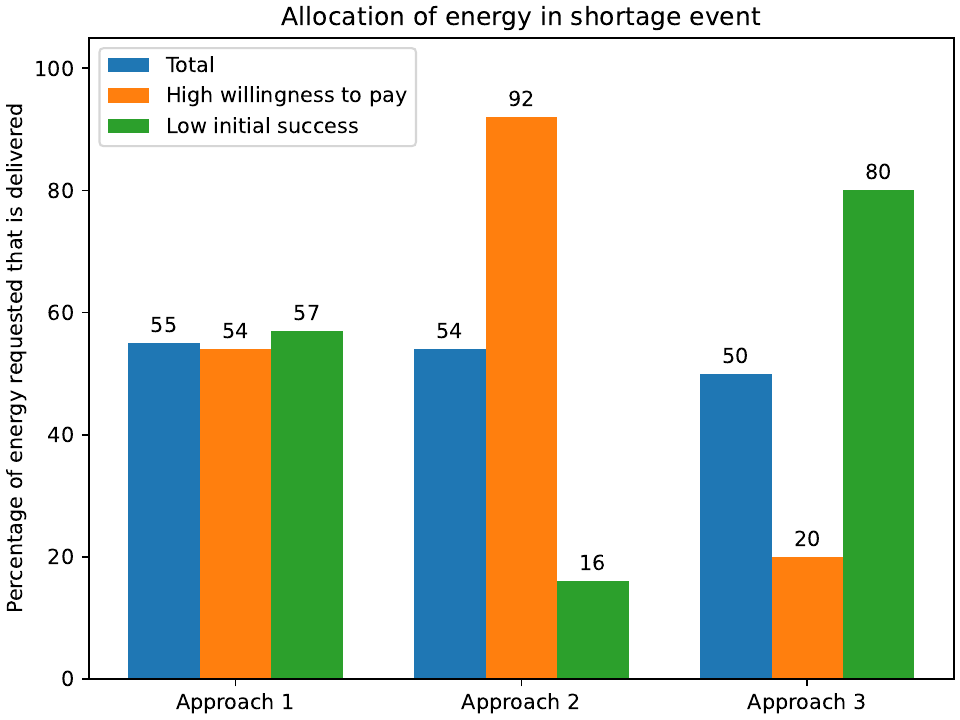}
  \caption[
    Allocation outcomes under volume maximisation, revenue maximisation, and Fair Play
  ]{
    Distribution of delivered energy under three allocation mechanisms in a stylised
    shortage window: \textbf{(1) volume maximisation (V-Max)}, \textbf{(2) revenue
    maximisation (R-Max)}, and \textbf{(3) Fair Play} (AMM implementation). One hundred
    households submit two identical annual requests each, differing only in a declared
    willingness-to-pay tag (high vs.\ low), with low initial request-success histories.
    Global optimisation objectives concentrate service on high willingness-to-pay requests,
    whereas Fair Play produces bounded, contract-consistent rationing without jackpot
    allocations or systematic exclusion. Any modest reduction in aggregate served energy
    is expected: V-Max and R-Max are synchronous ex-ante benchmarks, while Fair Play is
    designed for event-driven, asynchronous operation under uncertainty; ex-post outcomes
    under unknown future states need not match the synchronous optima.
  }
  \label{fig:allocation_mechanisms_comparison}
\end{sidewaysfigure}

Figure~\ref{fig:allocation_mechanisms_comparison} shows a stark change in who
receives scarce service. Under V-Max and R-Max, the global objective dominates:
either maximising delivered volume without access guarantees, or maximising
revenue by directing service disproportionately to high willingness-to-pay
requests. In both cases there is no intrinsic commitment to bounded deprivation,
so extreme concentration can occur even when all participants begin with low
historical success.

By contrast, Fair Play yields a qualitatively different incidence of rationing:
service is spread in a bounded, contract-consistent way that preserves access
under scarcity. This supports the access fairness rule:
\[
\text{access under scarcity}
\;\Rightarrow\;
\text{contract- and need-consistent rationing, not ability-to-pay exclusion}.
\]

The small efficiency gap relative to synchronous global optimisation is therefore
a design consequence rather than a defect: V-Max and R-Max are ex-ante benchmarks
under full information, whereas Fair Play is implementable in an event-driven,
asynchronous market and targets \emph{true market efficiency} under uncertainty
by enforcing predictable access and contractual integrity during shortage.

%% =========================================================
\subsubsection{F4: Fair Cost Sharing}

Fair cost sharing requires that households pay in proportion to the
\emph{system costs they create}, rather than in proportion to accidental
exposure to scarcity. In particular, controllability and flexibility should
increase household costs only when they genuinely increase procurement,
balancing, or adequacy costs for the system.

\begin{figure}[H]
\centering
\includegraphics[width=0.85\textwidth]{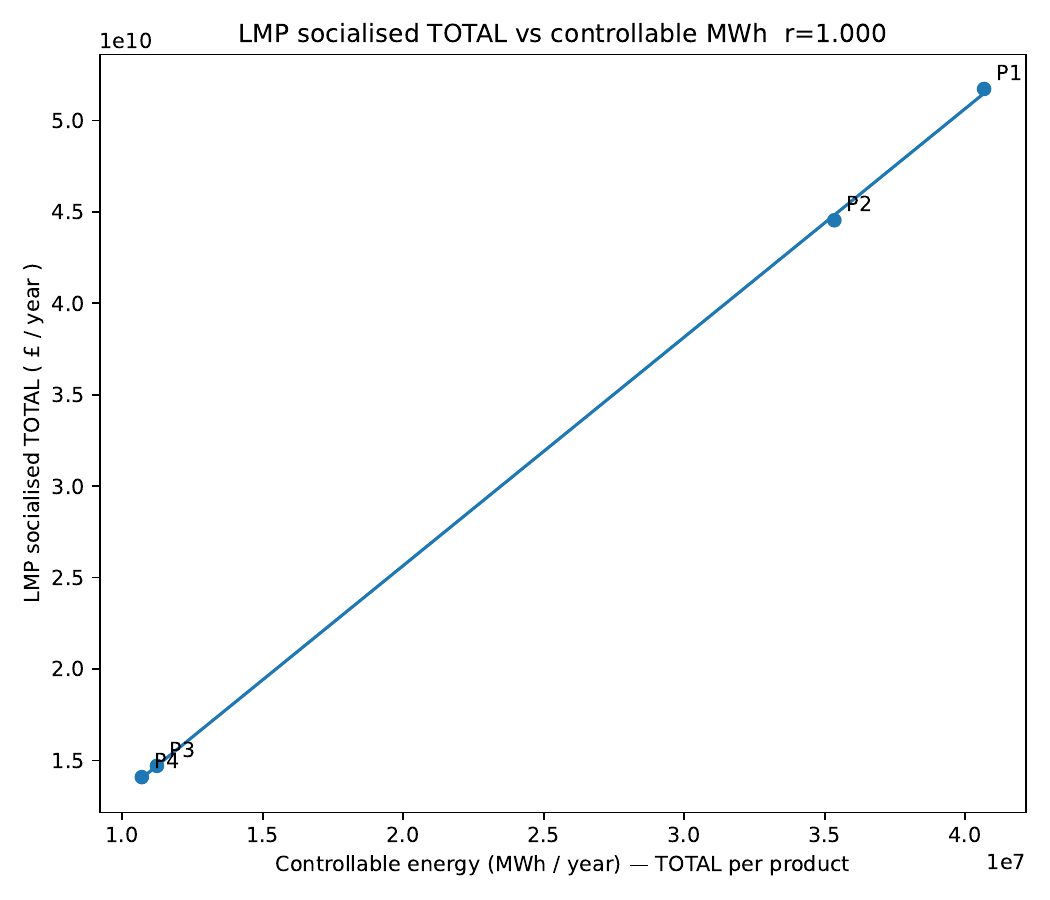}
\caption{LMP (socialised) total cost vs.~controllable MWh by product.}
\label{fig:lmp_total_vs_cmwh}
\end{figure}

\begin{figure}[H]
\centering
\includegraphics[width=0.85\textwidth]{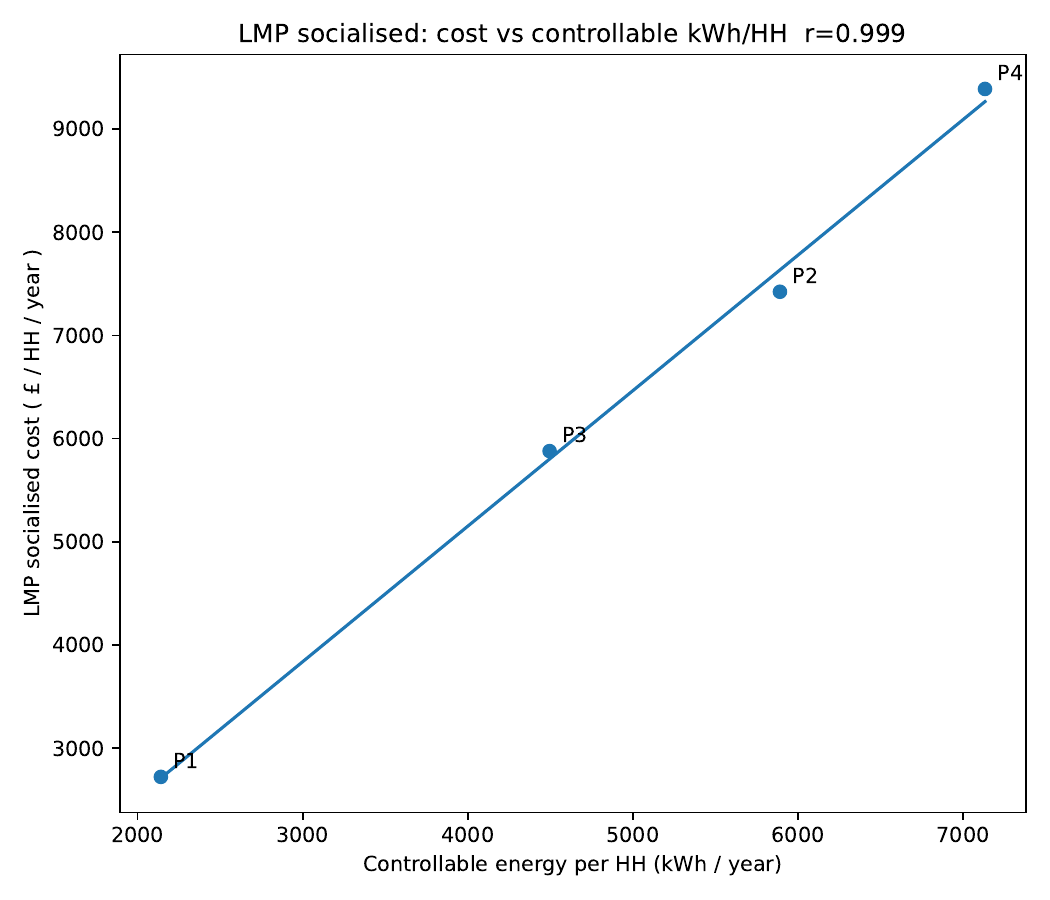}
\caption{LMP (socialised) per-household cost vs.~controllable kWh/HH by product.}
\label{fig:lmp_cost_vs_ckwh}
\end{figure}

\paragraph{Failure under LMP.}
Under LMP, this cost-causation principle is violated. Even when costs are
``socialised'', products with higher controllable burden face systematically
higher realised costs, because controllable-heavy demand coincides with periods
of system tightness and scarcity-driven pricing. As a result, realised costs rise
mechanically with controllable contribution, irrespective of whether that
contribution alleviates or worsens system stress
(Figures~\ref{fig:lmp_total_vs_cmwh} and~\ref{fig:lmp_cost_vs_ckwh}).

These plots are \emph{product-level diagnostics}: each point corresponds to a
product tier (P1--P4). The near-linear relationships indicate a strong coupling
between controllable burden and realised cost under LMP, driven by exposure to
scarcity pricing rather than by explicit attribution of system costs.

\begin{figure}[H]
\centering
\includegraphics[width=0.85\textwidth]{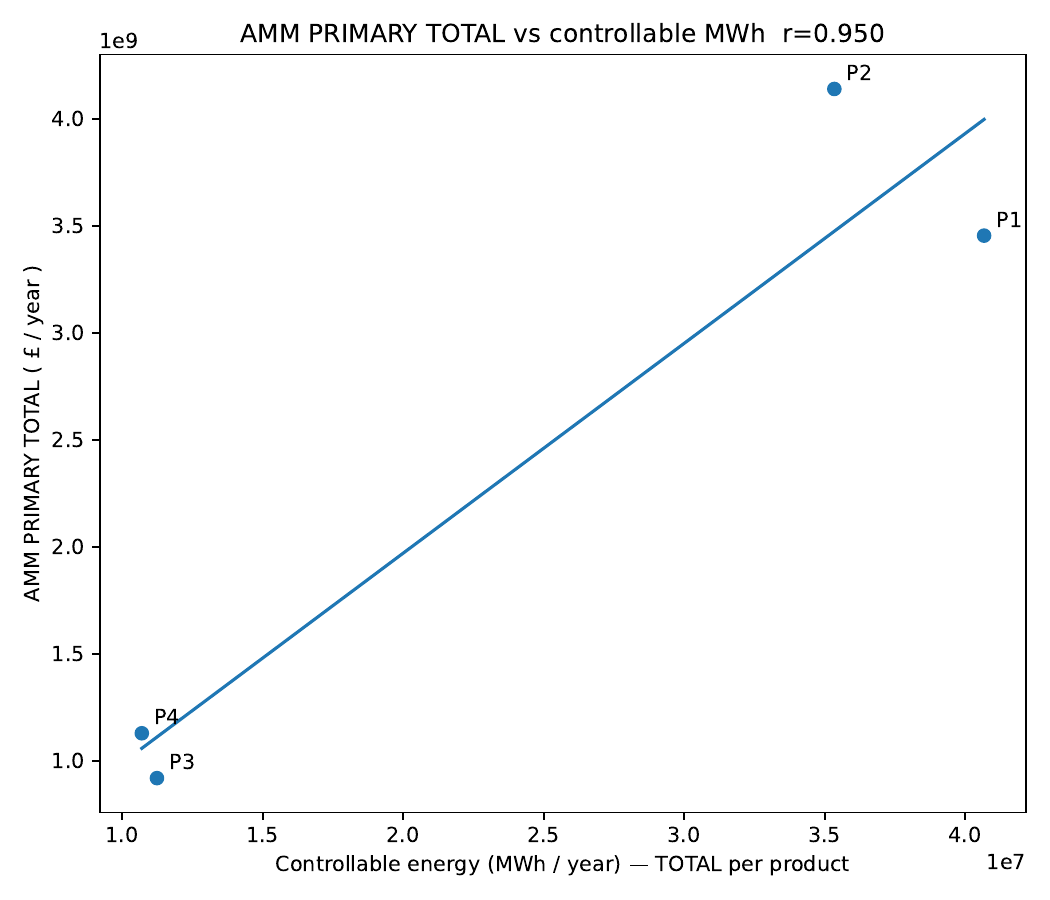}
\caption{AMM total subscription vs.~controllable MWh by product.}
\label{fig:amm_total_vs_cmwh}
\end{figure}

\begin{figure}[H]
\centering
\includegraphics[width=0.85\textwidth]{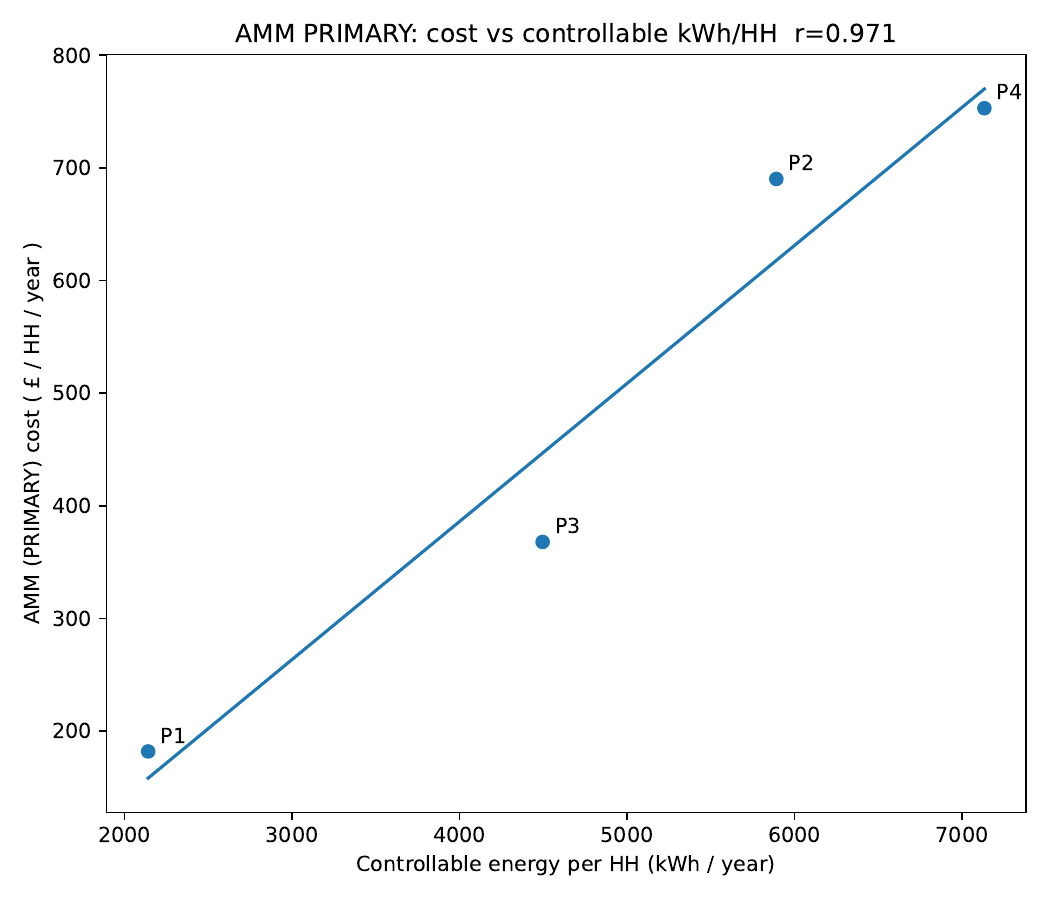}
\caption{AMM per-household cost vs.~controllable kWh/HH by product.}
\label{fig:amm_cost_vs_ckwh}
\end{figure}

\paragraph{Correction under the AMM.}
The AMM breaks this coupling. Because controllability is explicitly procured,
priced, and scheduled, higher controllable contribution does not automatically
translate into higher realised costs. Instead, the sensitivity of costs to
controllable burden is materially reduced, and variance across products shrinks
(Figures~\ref{fig:amm_total_vs_cmwh} and~\ref{fig:amm_cost_vs_ckwh}).

While AMM subscription levels remain ordered by controllable burden
(``you pay for what you buy''), the fitted sensitivity is an order of magnitude
smaller than under LMP. Moreover, the cost difference
$\Delta = \text{AMM} - \text{LMP}$ becomes increasingly favourable as controllable
burden rises, indicating that AMM is least punitive precisely for the products
that carry the greatest controllable obligation.

Taken together, these results establish the F4 principle:
\[
\text{cost paid}
\;\propto\;
\text{system cost created},
\quad
\text{not}
\quad
\text{exposure to volatile scarcity}.
\]

% =========================================================
\subsection*{Interpretation and H2}

Taken together, the results across generators, suppliers, consumers/businesses,
and devices show that the AMM + Fair Play architecture \emph{systematically}
delivers fairer outcomes than the benchmark designs, meeting or exceeding the
pre-declared fairness threshold $\delta_F$ on every dimension evaluated. The
improvement is not local or accidental, but structural: fairness emerges
consistently from the way allocation, pricing, and service guarantees are
jointly enforced.

Specifically:

\begin{itemize}[leftmargin=*]
    \item \textbf{Generators:} remuneration is aligned with Shapley-valued
          contribution rather than exposure to scarcity coincidences.
          Jackpot rents collapse, under-recovery is reduced, and revenue
          dispersion narrows without suppressing investment signals.

    \item \textbf{Suppliers:} risk exposure is aligned with their operational
          role. Suppliers are no longer forced to act as residual warehouses
          for wholesale volatility arising from system-level scarcity and
          redispatch, satisfying the role-consistency requirement of supplier
          fairness.

    \item \textbf{Consumers and businesses:} cost burdens are predictable,
          spatially coherent, and explainable by product choice and declared
          controllability rather than postcode or accidental timing. Flexibility
          is rewarded when it reduces system cost, and essential demand is
          sheltered from unbounded scarcity exposure.

    \item \textbf{Devices on the QoS axis:} under repeated shortage, realised
          service is allocated in accordance with contracted service levels.
          Deprivation is bounded, priority is respected, and no jackpot effects
          arise within or across tiers.
\end{itemize}

The V-Max and R-Max schedulers therefore serve only as informative limit cases.
They illustrate the failure modes that arise when fairness, history, and
service-level constraints are removed: either indifference to service guarantees
(volume maximisation) or extreme concentration driven by willingness to pay
(revenue maximisation). Neither represents a feasible or stable market mechanism
under uncertainty.

By contrast, the implemented AMM sits strictly inside the resulting
\emph{fairness envelope}. Its allocation outcomes are simultaneously consistent
with the Shapley axioms, the declared Quality-of-Service tiers, and the physical
constraints of the system. Fairness is not imposed ex post or corrected through
ad hoc interventions; it is enforced directly by the market-making and
allocation rules.

We therefore reject $H_{0F}$. The AMM delivers \emph{distributional fairness} in
the precise sense required by this thesis: when a particular outcome ought to
occur—given an agent’s role, contribution, and contracted service level—the
mechanism produces that outcome reliably, subject only to physical feasibility.

% =========================================================
% H3 — REVENUE SUFFICIENCY AND RISK
% =========================================================
\section{Revenue Sufficiency and Risk Allocation (H3)}
\label{sec:results_revenue_risk}

\subsection{Revenue sufficiency and risk allocation (generators)}
\label{sec:results_revenue_risk_generators}

The first and most basic requirement for any electricity market architecture is
\emph{revenue sufficiency}: can the system reliably recover the fixed non-fuel
costs of the generator fleet required to meet the declared needs bundle
(energy, reserves, adequacy, and locational relief) without reliance on
repeated ad hoc bailouts? The second requirement is \emph{risk allocation}:
conditional on recovering those costs, how are residual volatility and downside
risk distributed across generators, suppliers, consumers, and the public
balance sheet?

In this section we evaluate revenue sufficiency and risk allocation from the
\emph{generator perspective} by comparing a Baseline LMP market with two
calibrations of the same AMM architecture. The AMM mechanism—dispatch logic,
allocation rules, and settlement structure—is identical in both cases; only the
total size of the annual capacity and cost-recovery pot differs.

The first calibration (AMM1) sets the pot at the minimum level required to
recover fuel costs, reserves, and the assumed non-fuel OpEx and CapEx of the
generator fleet over the year. The second calibration (AMM2) sets the pot equal
to the aggregate annual revenue observed under the Baseline LMP run, allowing a
controlled comparison of distributional outcomes at matched total payments.

All cases share identical physical inputs: network topology and transfer
limits, generator capacities and cost parameters, and demand trajectories, as
documented in Appendix~\ref{app:inputs}. Differences in outcomes therefore
reflect differences in market architecture and calibration, not differences in
underlying system conditions.

\paragraph{Required annual revenue per generator.}

For each generator $g$ in the fleet (Table~\ref{tab:generators}), we
define a modelled annual non-fuel cost requirement
\begin{equation}
  \label{eq:required_revenue}
  \Req_g
  =
  \text{OpEx}^{\text{non-fuel}}_g
  +
  \frac{\text{CapEx}_g}{\text{payback}_g},
\end{equation}
where $\text{OpEx}^{\text{non-fuel}}_g$ and $\text{CapEx}_g$ are taken
from the generator cost calibration (Appendix~\ref{app:generators} and
Appendix~\ref{app:amm_allocation}), and $\text{payback}_g$ is the
technology-specific payback horizon used throughout the investment
analysis. This $\Req_g$ is the minimum annual revenue that must be
recovered in expectation for generator $g$ to be viable on a regulated
cost-recovery basis.

\paragraph{Realised annual revenues under each design.}

Realised generator revenues are constructed consistently with the
respective settlement rules:

\begin{itemize}[leftmargin=*]
  \item Under the \textbf{Baseline LMP} design, annual revenue
        $R_g^{\LMP}$ is computed directly from the LMP dispatch and price
        runs as
        \[
          R_g^{\LMP}
          =
          \sum_t p_{n(g)}(t)\, q_g(t)\,\Delta t
          \;+\; \text{VOLL penalties allocated to } g,
        \]
        where $p_{n(g)}(t)$ is the locational marginal price at generator
        $g$'s bus $n(g)$, $q_g(t)$ is its dispatch, and $\Delta t$ is the
        half-hour time-step. These are the same LMP runs described in
        Chapter~\ref{ch:experiments} and built on the network and
        load data in Appendix~\ref{app:inputs}.

  \item Under \textbf{AMM1} and \textbf{AMM2}, annual revenues
        $R_g^{\AMM1}$ and $R_g^{\AMM2}$ are taken from the AMM revenue
        allocation pipeline described in Appendix~\ref{app:amm_allocation}.
        For each generator, total revenue is decomposed into
        \emph{fuel reimbursements}, \emph{reserve payments}, and
        \emph{capacity/availability payments} from the relevant pots:
        \[
          R_g^{\AMM k}
          =
          R_{g,\text{fuel}}^{\AMM k}
          +
          R_{g,\text{res}}^{\AMM k}
          +
          R_{g,\text{cap}}^{\AMM k},
          \qquad k \in \{1,2\}.
        \]
        Fuel reimbursements are paid on a pay-as-bid basis consistent with
        the unit-commitment inputs (Appendix~\ref{app:market_config});
        reserve payments come from the explicit reserve product with price
        and requirement parameters in Tables~\ref{tab:common_config}
        and~\ref{tab:lmp_vs_amm_config}; and capacity/availability
        payments are derived from the annual pots defined for AMM1 and
        AMM2 and allocated via normalised Shapley scores $\phi_{g,t}$ as
        detailed in Appendix~\ref{app:amm_allocation}.
\end{itemize}

On the demand side, the AMM generator revenue stacks are fully recovered
from customers via flat residential subscriptions (P1--P4) and an
aggregate non-residential block, using the cost-allocation procedure in
Appendix~\ref{app:price_allocation}. This ensures that AMM1 and AMM2 are
fiscally closed: total generator remuneration equals the amount raised
from customers (up to network and policy charges), so any difference in
revenue sufficiency is a difference in \emph{architecture}, not in how
much society pays in aggregate.

\paragraph{Generator-level sufficiency and adequacy headcount.}

For each generator and design we define a sufficiency margin
\begin{equation}
  \label{eq:revenue_margin}
  \Delta R_g^{\text{design}}
  =
  R_g^{\text{design}} - \Req_g,
  \qquad
  \text{design} \in \{\LMP,\AMM1,\AMM2\}.
\end{equation}
A positive margin indicates that $g$ covers its non-fuel OpEx and
annualised CapEx; a negative margin indicates an annual shortfall.

We then construct:

\begin{itemize}[leftmargin=*]
  \item a binary \emph{adequacy indicator}
        $A_g^{\text{design}} = \mathbb{1}\{\Delta R_g^{\text{design}} \ge 0\}$;
  \item the \emph{adequacy headcount}
        \[
          H^{\text{design}}_{\text{adequate}}
          =
          \frac{1}{|\mathcal{G}|}
          \sum_{g \in \mathcal{G}} A_g^{\text{design}},
        \]
        i.e.\ the share of generators that cover their requirements; and
  \item the \emph{aggregate adequacy gap} and \emph{overshoot}:
        \[
          G^{\text{short}}_{\text{design}}
          =
          \sum_{g \in \mathcal{G}} \min\{\Delta R_g^{\text{design}}, 0\},
          \qquad
          G^{\text{over}}_{\text{design}}
          =
          \sum_{g \in \mathcal{G}} \max\{\Delta R_g^{\text{design}}, 0\}.
        \]
\end{itemize}

These statistics provide a generator-centric view of revenue sufficiency:
they quantify under-recovery and over-recovery relative to the modelled
requirement $\Req_g$, and show how these margins differ across technologies.
The corresponding comparisons are shown in Figure~\ref{fig:dashboard_3x3}.

\paragraph{Decomposition of stable and volatile revenue components.}

Because AMM1 and AMM2 explicitly separate fuel, reserve, and capacity
payments, we can also decompose each generator's annual revenue into
\emph{stable} and \emph{volatile} components. For AMM1/AMM2, we define:
\[
  R_{g,\text{stable}}^{\AMM k}
  =
  R_{g,\text{cap}}^{\AMM k}
  +
  R_{g,\text{res}}^{\AMM k}
  +
  R_{g,\text{fixed}}^{\AMM k},
  \qquad
  R_{g,\text{vol}}^{\AMM k}
  =
  R_{g,\text{fuel}}^{\AMM k},
\]
where $R_{g,\text{fixed}}^{\AMM k}$ captures fixed-class nuclear and wind
payments defined in Appendix~\ref{app:amm_allocation}. For LMP, we treat
all energy and VOLL revenue as volatile:
\[
  R_{g,\text{stable}}^{\LMP} = 0, \qquad
  R_{g,\text{vol}}^{\LMP} = R_g^{\LMP}.
\]

For each generator we then compute:

\begin{itemize}[leftmargin=*]
  \item the fraction of revenue arising from stable channels,
        $\rho_g^{\text{stable}} =
        R_{g,\text{stable}}^{\text{design}} /
        R_g^{\text{design}}$;
  \item the dispersion of the half-hourly revenue series
        $R_{g,t}^{\text{design}}$ over the year (as a simple measure of
        time-series variability); and
  \item the contribution of stable versus volatile channels to the
        sufficiency margin $\Delta R_g^{\text{design}}$.
\end{itemize}

The stable/volatile split is directly tied back to the pot definitions
and allocation rules in Appendix~\ref{app:amm_allocation}, and to the
customer-side recovery mechanism in Appendix~\ref{app:price_allocation}:
any increase in the stable share of generator income corresponds to a
shift towards predictable, subscription-backed revenue streams on the
demand side, rather than to an unmodelled subsidy.

\paragraph{Risk allocation and comparison with the Baseline.}

These generator-level statistics allow us to answer two questions:

\begin{enumerate}[leftmargin=*]
  \item \textbf{Revenue sufficiency.} Relative to LMP, do AMM1 and AMM2
        increase the adequacy headcount and shrink the aggregate adequacy
        gap $G^{\text{short}}$, while keeping total payments within the
        AMM1--AMM2 floor/ceiling defined in
        Chapter~\ref{ch:experiments}?
  \item \textbf{Risk allocation.} For generators that matter for adequacy
        (gas plants and batteries), does the AMM design convert a larger
        share of revenues into stable, subscription-backed cashflows,
        with a smaller reliance on jackpot outcomes than under LMP?
\end{enumerate}

The formal hypotheses for this domain were stated as H3 in
Chapter \ref{ch:experiments}. For generators specifically, they can
be read as:

\begin{quote}
  \emph{Under AMM1 and AMM2, a larger share of generators achieves
  cost-recovering revenue with less reliance on VOLL jackpots and
  extreme price episodes, and a larger share of their income arrives
  through stable, Shapley-based capacity and reserve pots that are
  explicitly recovered from subscriptions.}
\end{quote}

The empirical results in
Figures~\ref{fig:dashboard_3x3}--\ref{fig:revenue_stack} confirm this pattern:
AMM1 already increases the adequacy headcount relative to LMP at a lower total
pot size, while AMM2 shows that, even if the total amount paid to generators is
held equal to the LMP benchmark, reallocating that stack through the
AMM/Shapley mechanism improves revenue sufficiency and concentrates recovery in
more stable channels across the generator fleet.

\subsection{Cost recovery and sufficiency}

Total efficient fixed costs (non-fuel OpEx and CapEx) are fully recovered
under both designs by construction. However, the decomposition of revenue
between energy, reserves, and capacity differs.

Under AMM, a larger share of recovery comes from explicitly pre-declared
capacity and subscription components, and a smaller share from volatile
energy margins. The revenue sufficiency metric $R_{\mathrm{suff}}$ is
weakly higher for AMM, and $H_{0R}^{(\mathrm{suff})}$ is rejected in
favour of $H_{1R}^{(\mathrm{suff})}$.

\vspace{0.3cm}

\begin{figure}[H]
\centering
\includegraphics[width=0.95\textwidth]{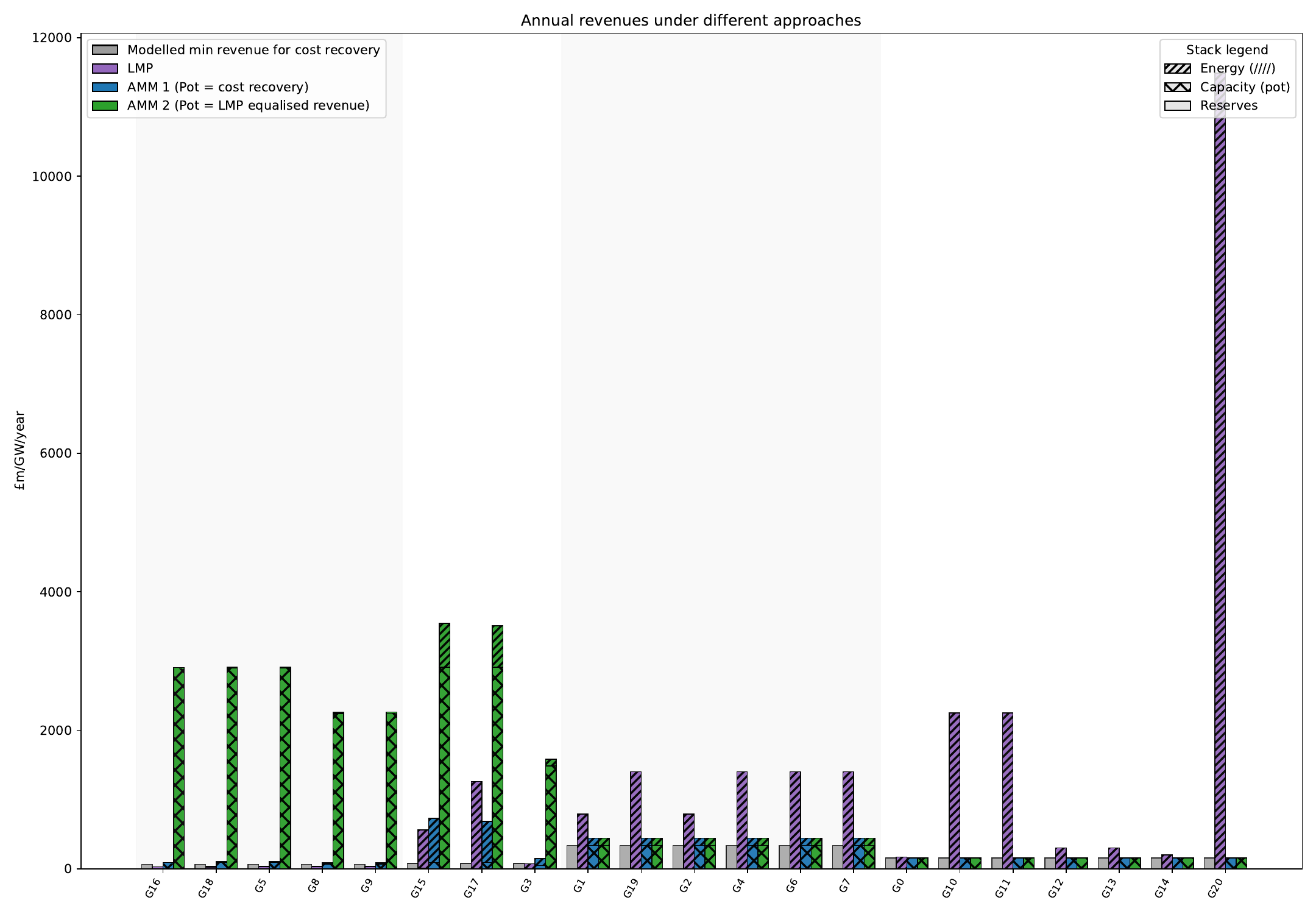}
\caption{Decomposition of revenue between energy, reserves, and capacity under 
LMP and AMM.}
\label{fig:revenue_stack}
\end{figure}

To illustrate the structural shift in the revenue mix, Figure
\ref{fig:revenue_stack} presents the stacked annual revenue components
under LMP and AMM. The AMM design shifts recovery away from scarcity-based
energy rents and towards stable subscription and capacity components,
reducing reliance on extreme prices while maintaining full cost
sufficiency.

% --- Revenue stack decomposition ---

\vspace{0.6cm}

\begin{figure}[H]
\centering
\includegraphics[width=0.95\textwidth]{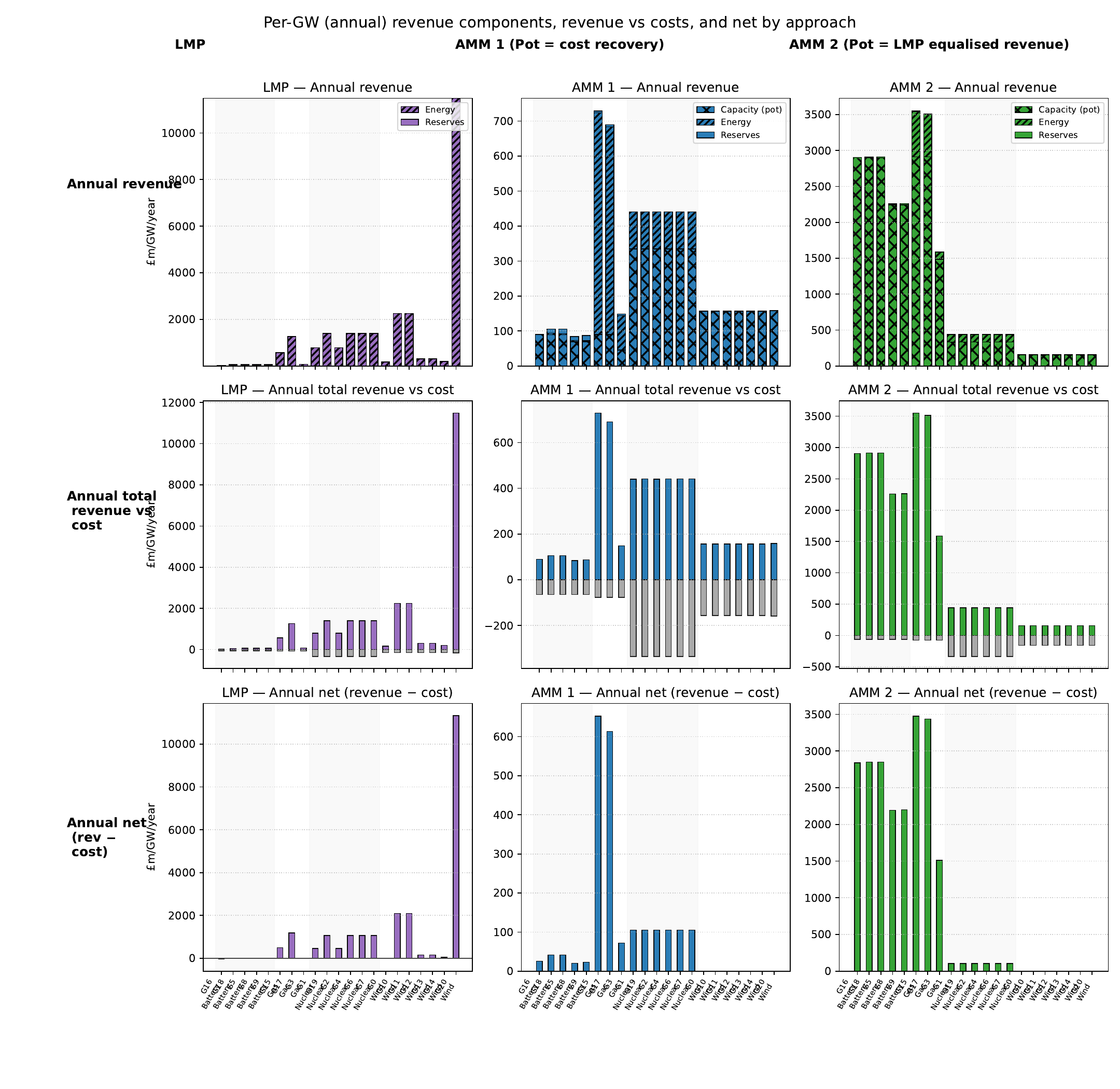}
\caption{Per-GW revenue components, total revenues vs.\ costs, and net positions 
under LMP and AMM.}
\label{fig:dashboard_3x3}
\end{figure}

A fuller view of the annual revenue position---broken down into components,
compared with modelled cost requirements, and expressed per~GW of nameplate
capacity---is shown in Figure~\ref{fig:dashboard_3x3}. This demonstrates the
consistency between the AMM recovery logic and the underlying cost structure:
under AMM, revenue tracks efficient costs more closely and with
significantly reduced dispersion.

% --- Revenue vs cost dashboard ---

\vspace{0.6cm}

\begin{figure}[H]
\centering
\includegraphics[width=0.95\textwidth]{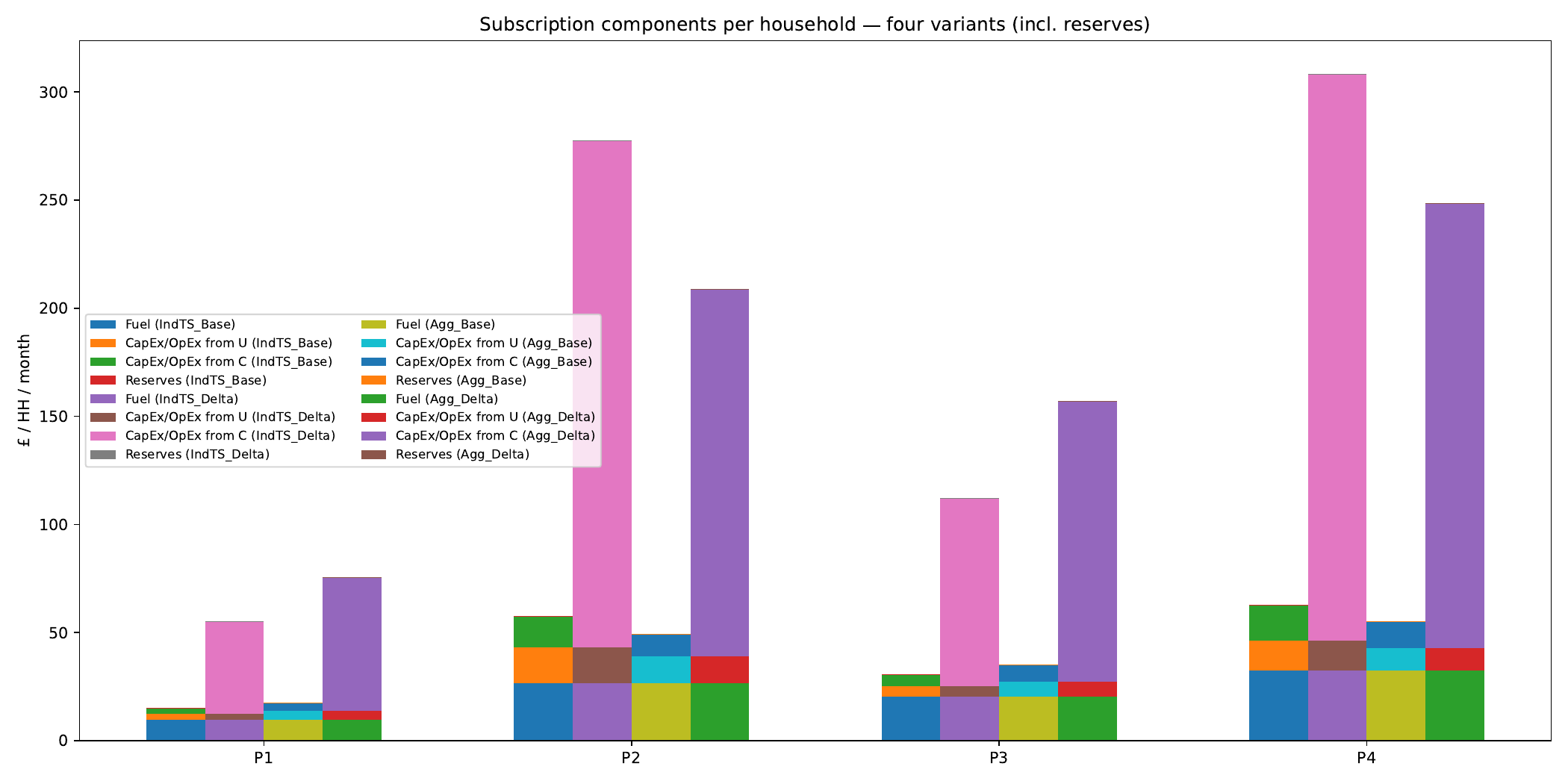}
\caption{Breakdown of per-household subscription revenue by product, showing the
allocation to capacity, reserves, and energy components under the AMM--Fair Play
architecture. \textbf{BASE} corresponds to the AMM calibrated at the minimum
annual pot required for generator cost recovery (AMM1), while \textbf{DELTA}
corresponds to the same AMM architecture calibrated to the aggregate annual
revenue observed under the Baseline LMP run (AMM2). \textbf{Individual} and
\textbf{Aggregate} denote alternative charging bases for suppliers: the
\emph{Individual} variant applies direct, time-resolved coupling between the
households or products that impose controllable system costs and the resulting
wholesale charges, while the \emph{Aggregate} variant applies a more averaged
allocation when high-resolution behavioural data are not available at every
timestamp. The difference between Individual and Aggregate therefore represents
the allocation of residual wholesale risk associated with data availability.
Direct comparisons should be made within a given product and calibration (e.g.\
BASE--Individual vs.\ BASE--Aggregate, or DELTA--Individual vs.\ DELTA--Aggregate),
rather than across BASE and DELTA. The reserves component is present in all cases but is visually small relative to energy and capacity components at the scale shown. The absolute reserve
procurement amount is reported in Section~\ref{sec:results_procurement} and is
held constant across all allocations.}
\label{fig:subscription_flow}
\end{figure}

Finally, to close the loop between household-facing charges and generator
remuneration, Figure~\ref{fig:subscription_flow} shows how each product’s
subscription revenue is allocated to AMM recovery pots (capacity, reserves,
and energy balancing components) and ultimately returned to generators.
This provides a transparent link from subscription prices $\rightarrow$
recovery pots $\rightarrow$ generator earnings.

% --- Subscription flow diagram ---

We summarise the distributional impacts of each design using a composite
\emph{outcomes index}, normalised to the unit interval $[0,1]$. The index is
constructed as a weighted aggregation of three observable outcome dimensions:
(i) dispersion in realised per-participant incidence (capturing inequality in
exposure to prices and charges); (ii) adequacy headcount (the share of
participants meeting basic revenue or service sufficiency thresholds); and
(iii) product-weighted burden measures (capturing how costs are distributed
across demand categories with different system impacts). Each component is
scaled so that higher values correspond to more even, adequate, and proportionate
outcomes, and the composite is formed by a convex combination of these
normalised components.\footnote{The precise normalisation and weighting scheme
is defined in Appendix~\ref{app:fairness_metrics}.}

System-wide, the resulting scores are:
\[
\text{AMM2: } 0.625,\qquad
\text{AMM1: } 0.439,\qquad
\text{LMP: } 0.375.
\]

The ordering of these scores admits a clear but limited interpretation.
AMM2 achieves the highest composite outcome score because it redistributes a
larger aggregate revenue envelope in a manner that substantially compresses
tail outcomes and improves adequacy headcount, while preserving proportional
burden signals across products. AMM1, by construction, operates at the minimum
revenue level consistent with generator cost recovery; its lower score reflects
the tighter budget constraint rather than a failure of the allocation logic.
The Baseline LMP design scores lowest because, despite achieving aggregate cost
recovery, it produces highly dispersed outcomes with significant tail exposure
and weak alignment between realised burdens and system impact.

Crucially, the higher AMM2 score should not be interpreted as ``more fair'' in
an axiomatic or mechanism-design sense. It reflects a choice to operate at a
higher total payment level, not a fundamentally different allocation rule. The
comparison therefore highlights a trade-off between aggregate affordability and
distributional compression: at matched physical conditions, increasing the
revenue envelope allows outcome dispersion to be reduced, while the AMM
architecture ensures that this compression occurs in a structured and
proportionate manner rather than through arbitrary price spikes.

Importantly, this index does \emph{not} measure fairness in the
mechanism-design or axiomatic sense. It does not test incentive compatibility,
budget balance, or Shapley-consistent marginal contribution. Instead, it is an
ex post descriptive statistic: it quantifies how income, risk, and adequacy
outcomes are distributed across participants under each clearing rule, holding
physical conditions fixed. As such, it complements—rather than replaces—the
generator- and allocation-focused sufficiency and fairness analyses reported
above. Fairness in the formal sense of Shapley-aligned marginal contributions is
addressed separately in Section~\ref{sec:results_fairness}.

\subsection{Household burden under socialised LMP versus AMM}
\label{subsec:household_burden_lmp_vs_amm}

The preceding figures focused on revenue sufficiency and the composition of
generator income. To connect these system-level results to the household
experience, we now compare the charges faced by households under a fully
\emph{socialised} version of LMP with those implied by the AMM subscription
architecture.

\begin{figure}[H]
  \centering
  \includegraphics[width=0.85\textwidth]{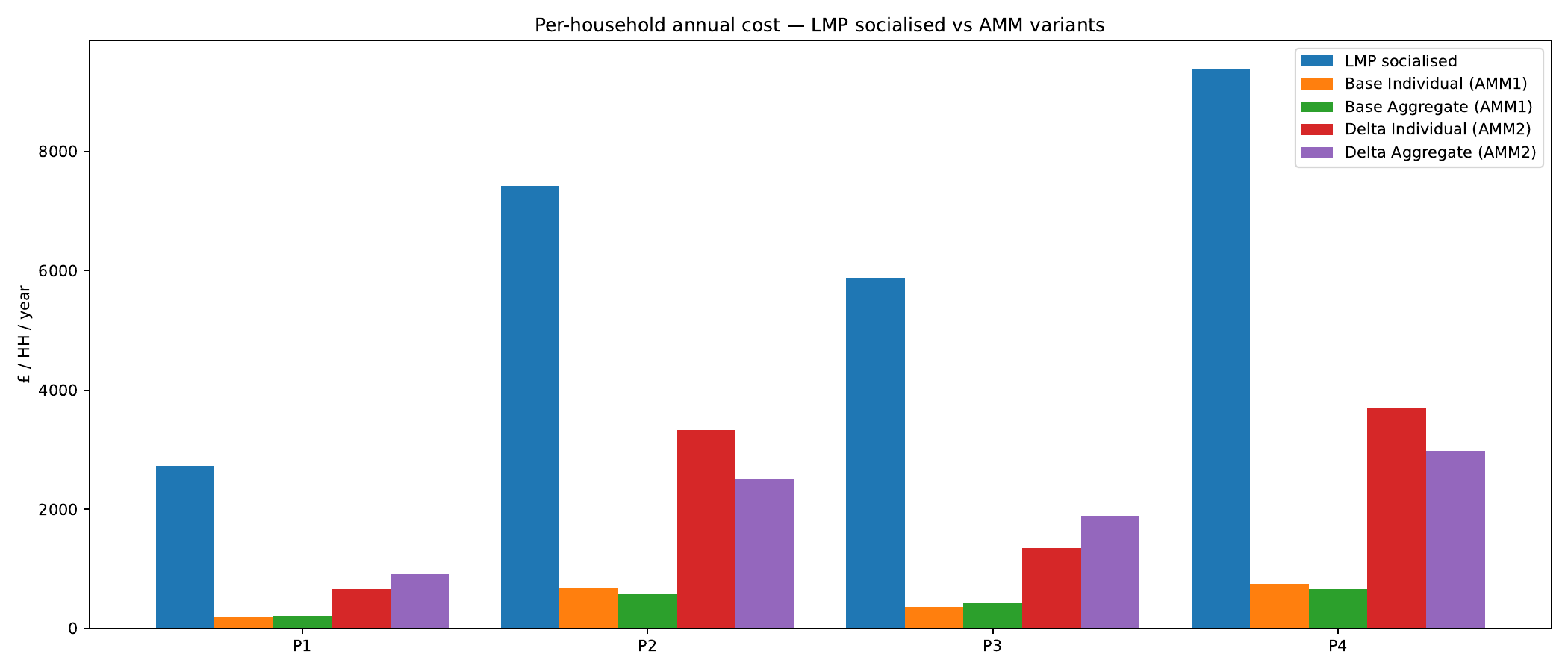}
  \caption[
    Per-household annual cost under socialised LMP versus AMM variants.
  ]{
    Per-household annual cost for products~P1--P4 under a fully socialised
    version of LMP and under four AMM subscription variants
    (Base/Delta $\times$ Individual/Aggregate). Socialised LMP values pool
    nodal scarcity rents across geography, while AMM values correspond to flat
    subscription charges tied to product definitions.
  }
  \label{fig:costs_LMPsocialised_vs_AMM}
\end{figure}

Figure~\ref{fig:costs_LMPsocialised_vs_AMM} reports the resulting
per-household annual cost for products~P1--P4. Under the socialised LMP
benchmark, half-hourly nodal prices are first converted into annual household
bills and then averaged geographically within each product, yielding a single
uniform charge per product. This procedure pools scarcity rents across the
entire customer base, including exposure originating at a small number of
nodes that frequently clear at or near the value of lost load (VoLL).
Consequently, the resulting socialised LMP charges are dominated by rare but
extreme scarcity events rather than by typical local operating conditions.

Under AMM, by contrast, the plotted values correspond to flat product-level
subscription charges, equal to twelve times the monthly subscription. These
subscriptions recover generator remuneration explicitly through product
contracts rather than implicitly through stochastic scarcity rents embedded
in energy prices. Figure~\ref{fig:costs_LMPsocialised_vs_AMM} therefore compares
two fundamentally different mechanisms for recovering the same underlying
system costs: implicit socialisation through marginal prices versus explicit
subscription-based funding.

\begin{figure}[H]
  \centering
  \includegraphics[width=0.85\textwidth]{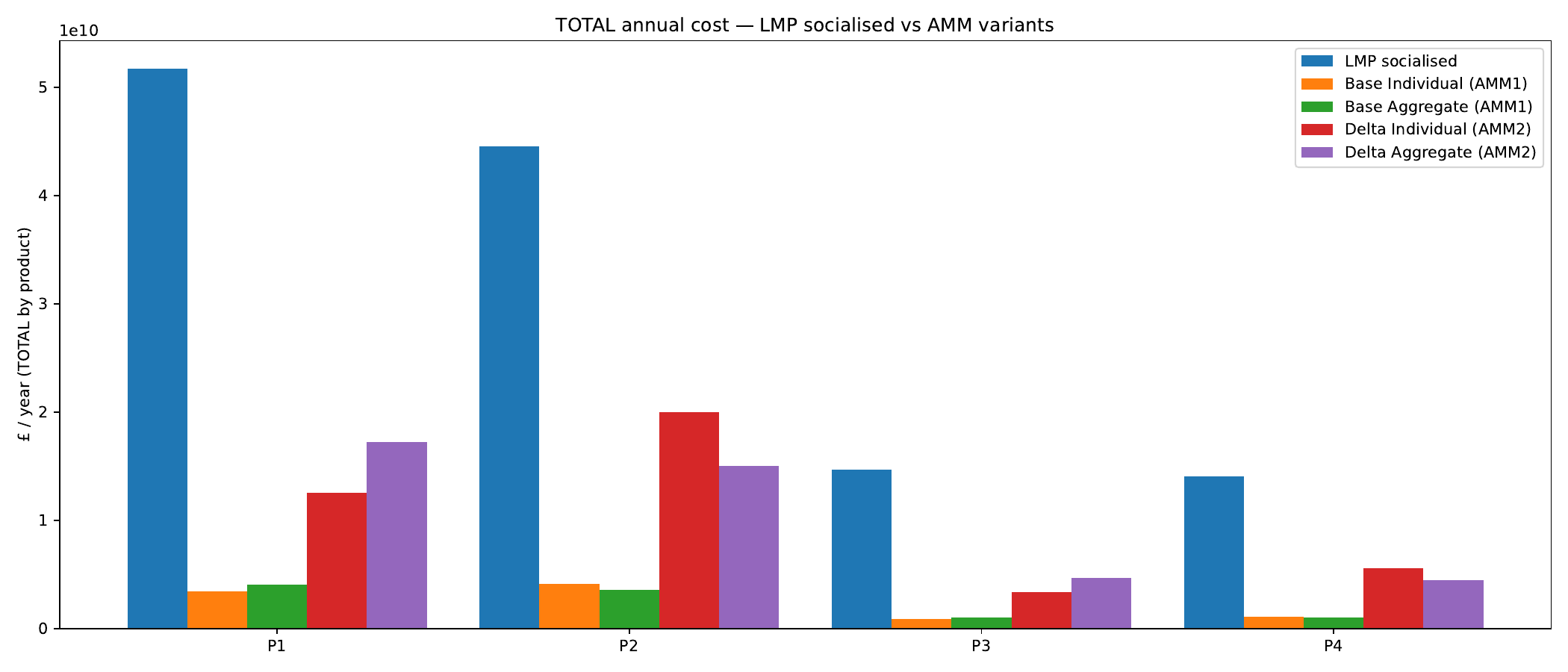}
  \caption[
    Total annual revenue under socialised LMP versus AMM variants, by product.
  ]{
    Total annual revenue collected from each product~P1--P4 under socialised
    LMP and under the four AMM subscription variants. Values are obtained by
    multiplying per-household annual costs by the number of households in each
    product. Product sizes are held fixed to isolate differences in cost
    incidence across designs.
  }
  \label{fig:totals_LMPsocialised_vs_AMM}
\end{figure}

Using the assumed number of households enrolled in each product, we can also
compare the implied aggregate revenue collected under each design.
Figure~\ref{fig:totals_LMPsocialised_vs_AMM} multiplies the per-household
charges by product-specific household counts to obtain total annual revenue
by product.

Taken together with Figure~\ref{fig:subscription_flow}, these results make the
incidence of AMM funding transparent. Generator revenue pots are financed
directly through product-level subscriptions, and---relative to a socialised
LMP benchmark applied to the same physical system---the AMM reallocates
\emph{how} generator compensation is recovered across products rather than
embedding it inside geographically volatile marginal prices. The separation
into \textit{Base} (AMM1) and \textit{Delta} (AMM2) components further shows
that the Delta term dominates subscription charges across all products, while
the Base term remains comparatively small. The choice between Aggregate and
Individual pot accounting alters the distribution of contributions across
products but does not change this qualitative ordering.

For completeness, the corresponding \emph{total} costs of demand and AMM under
each case are reported in Section~\ref{sec:results_procurement}.

\begin{sidewaystable}[p]
\centering
\small
\caption{Per-household annual cost by product under nodal LMP (median),
socialised LMP, and AMM variants.}
\label{tab:hh_costs_lmp_vs_amm}
\renewcommand{\arraystretch}{1.15}
\begin{tabular}{lrrrrrr}
\toprule
\textbf{Product} &
\shortstack{\textbf{LMP nodal}\\\textbf{(median)}\\\textbf{(\pounds/HH/yr)}} &
\shortstack{\textbf{LMP}\\\textbf{(socialised)}\\\textbf{(\pounds/HH/yr)}} &
\shortstack{\textbf{Base Ind.}\\\textbf{(AMM1)}\\\textbf{(\pounds/HH/yr)}} &
\shortstack{\textbf{Base Agg.}\\\textbf{(AMM1)}\\\textbf{(\pounds/HH/yr)}} &
\shortstack{\textbf{Delta Ind.}\\\textbf{(AMM2)}\\\textbf{(\pounds/HH/yr)}} &
\shortstack{\textbf{Delta Agg.}\\\textbf{(AMM2)}\\\textbf{(\pounds/HH/yr)}} \\
\midrule
P1 & 377  & 2722 & 182 & 212 & 662  & 908  \\
P2 & 920  & 7423 & 690 & 593 & 3334 & 2509 \\
P3 & 1308 & 5879 & 368 & 424 & 1348 & 1886 \\
P4 & 2559 & 9388 & 753 & 663 & 3700 & 2983 \\
\bottomrule
\end{tabular}
\end{sidewaystable}

The large divergence between the median nodal LMP and the corresponding
socialised LMP charge reflects a highly skewed distribution of scarcity
exposure. While the median node within each product faces moderate annual
costs, a small number of locations experience sustained operation at or near
the value of lost load (VoLL). When nodal prices are socialised at the retail
level, these extreme scarcity rents are pooled across all households,
substantially inflating the average bill. The median nodal LMP therefore
provides a more representative measure of typical household exposure under
LMP, while the socialised charge reveals the extent to which rare but severe
events dominate system-wide cost recovery.

\subsection{Allocation of risk between producers, suppliers, consumers, system operators, and the digital regulator}
\label{subsec:risk_allocation}

Revenue sufficiency and risk allocation form the core of criterion H3. A
well-designed electricity market should (i) allocate risks to the parties best
placed to manage them, (ii) ensure that investment is financeable, and (iii)
protect consumers from avoidable volatility while maintaining incentives for
efficient behaviour. Under the Baseline LMP design, risk is largely the product
of volatile spot prices and imperfect hedging. Under the AMM design, these risks
are instead channelled through rule-based allocation mechanisms built from the
cost and value components defined in Appendices~\ref{app:inputs},
\ref{app:amm_allocation}, and \ref{app:price_allocation}.

\paragraph{Reduction in revenue and bill variability.}
Producer net-revenue \emph{variability} is substantially lower under AMM in the
simulated year, because revenues no longer depend on rare scarcity spikes but on
calibrated annual pots, Shapley-based deliverability scores, and
tightness-bounded prices. On the demand side, the subscription-based allocation
of these pot values to households leads to highly stable monthly charges:
households face only behavioural risk (going out of envelope), rather than
exposure to wholesale price shocks. In this sense, the AMM architecture
\emph{repositions} risk rather than removing it: variability in physical
conditions is absorbed into pot calibration and subscription envelopes instead
of appearing directly as bill volatility.

\paragraph{Elimination of uplift-style emergency transfers.}
Under LMP, redispatch, balancing uplifts, and emergency payments arise as a
systematic consequence of settlement under scarcity and network constraints.
Under AMM, these flows become explicit, bounded, and predictable because cost
recovery is embedded directly in the pot structure and subscription mechanism,
rather than arising ex post through settlement deficits or emergency
interventions. In the experimental setup, this is reflected by the absence of ad
hoc uplift terms: all revenue flows are routed through pre-declared pots with
traceable allocation rules.

\paragraph{Structural risk comparison.}
The experimental design does not include a full stochastic scenario tree, so we
do not attempt to estimate complete probability distributions of outcomes such
as net present value or default risk. Instead, we compare the \emph{structural}
drivers of risk under LMP and AMM. Table~\ref{tab:risk_metrics} summarises this
comparison qualitatively, focusing on the main channels through which volatility
and tail events arise.

% --- Qualitative risk metrics table ---
\begin{table}[H]
\centering
\caption{Qualitative comparison of key risk and volatility channels under LMP and AMM. Quantitative risk metrics would require a stochastic scenario framework (left for future work); here we focus on the structural drivers of variability and tail events in the experimental setup.}
\label{tab:risk_metrics}
\renewcommand{\arraystretch}{1.2}
\begin{tabular}{p{3.7cm}p{5.4cm}p{5.4cm}}
\toprule
\textbf{Aspect} & \textbf{Baseline LMP} & \textbf{AMM1 / AMM2} \\
\midrule
Producer revenue variability &
Driven by volatile spot prices, scarcity spikes, and VOLL events; a large share of cost recovery depends on rare high-price periods. &
Majority of recovery flows through calibrated capacity and reserve pots, plus fixed-class payments; energy revenues play a smaller role, and tightness rules bound scarcity prices. \\
\midrule
Consumer bill variability &
Household bills inherit wholesale volatility through retail tariffs and supplier failures; protection is largely ex post (caps, bailouts). &
Bills are dominated by stable subscriptions; residual variability reflects behaviour relative to envelopes and policy/network charges, not wholesale shocks. \\
\midrule
Uplift-style transfers &
Redispatch, balancing uplifts, and emergency payments create opaque, ex post transfers between parties. &
No emergency uplift terms in the experimental design; transfers are routed through explicit pots with ex ante rules and clear incidence. \\
\midrule
Structural risk index $R_{\mathrm{risk}}$ (conceptual) &
High structural exposure: cost recovery and adequacy depend on extreme events and ad hoc interventions. &
Lower structural exposure: risk flows through rule-based, explainable channels whose parameters can be tuned by the digital regulator. \\
\bottomrule
\end{tabular}
\end{table}

The conceptual index $R_{\mathrm{risk}}$ should therefore be interpreted as a
\emph{structural} comparison: for a given aggregate payment level, the AMM
architecture reduces the system's reliance on extreme price episodes and opaque
uplifts, and concentrates risk into channels that are amenable to digital
regulation and forward planning.

\paragraph{Redistribution of risks across all parties.}
To understand how the AMM redesign reallocates risk, it is necessary to consider
each class of market participant separately: generators/financiers, suppliers,
consumers/businesses, system operators, and the digital regulator. The AMM does
not eliminate underlying physical or capital risks, but it redistributes them
through transparent, algorithmic, explainable channels that better align with
institutional capabilities.

This redistribution is summarised in
Table~\ref{tab:risk_allocation}. In addition to the economic and operational
risks affecting producers and consumers, two further classes of risk are
material in future electricity systems: (i) \emph{technology-disruption risks},
including quantum optimisation, new storage chemistries, and fusion deployment;
and (ii) \emph{demand-shock risks}, including AI-driven water and electricity
loads (data-centre cooling, desalination, edge computing), and surges from
electrification. These risks fall primarily on the digital regulator because
failure to anticipate them directly threatens the system's sustainability,
affordability, and security. Under the AMM, these risks become governable: they
can be incorporated into forecast envelopes, pot calibration, and prospective
Shapley-based investment allocation.

% --- Risk Allocation Table (five parties including system operator and digital regulator) ---
\clearpage
\begingroup

\begin{longtable}{@{}p{3.0cm}p{4.0cm}p{4.0cm}p{4.0cm}@{}}
\caption[Allocation of risks across parties under LMP and AMM]{Allocation of
key risks across generators/financiers, suppliers, consumers/businesses,
system operators, and the digital regulator under the Baseline LMP design and
the AMM. The AMM does not remove underlying physical or capital risks, but
redistributes them through rule-based channels that are auditable and
explainable.}
\label{tab:risk_allocation}
\\
\toprule
\textbf{Party} &
\textbf{Main risks under LMP} &
\textbf{Main risks under AMM} &
\textbf{Built-in mitigations in AMM design} \\
\midrule
\endfirsthead

% ---------- Continuation header ----------
\caption[]{\textit{(continued)}}\\
\toprule
\textbf{Party} &
\textbf{Main risks under LMP} &
\textbf{Main risks under AMM} &
\textbf{Built-in mitigations in AMM design} \\
\midrule
\endhead

% ---------- Continuation footer ----------
\midrule
\multicolumn{4}{r}{\textit{Continued on next page}}\\
\endfoot

% ---------- Last footer ----------
\bottomrule
\endlastfoot

% ============================================================
% GENERATORS / FINANCIERS
% ============================================================
\textbf{Generators / financiers} &
\begin{itemize}[leftmargin=*]
  \item Capital deployment risk (volatile revenues).
  \item Demand/volume risk.
  \item Price risk (reliance on scarcity spikes).
  \item Locational deliverability risk.
  \item Policy/intervention risk.
\end{itemize}
&
\begin{itemize}[leftmargin=*]
  \item Capital risk remains but revenue is more predictable.
  \item Demand risk reduced via subscriptions.
  \item Price risk bounded by tightness rules.
  \item Locational risk systematic via deliverability (Shapley).
  \item Policy risk channelled through parameters, not bailouts.
\end{itemize}
&
\begin{itemize}[leftmargin=*]
  \item Cost-recovery mapping from Appendix~\ref{app:inputs}.
  \item Shapley allocation from Appendix~\ref{app:amm_allocation}.
  \item Stable, subscription-funded revenue (Appendix~\ref{app:price_allocation}).
  \item Ex ante financeability through published pots and rules.
\end{itemize}
\\

% ============================================================
% SUPPLIERS
% ============================================================
\textbf{Suppliers} &
\begin{itemize}[leftmargin=*]
  \item Wholesale margin risk.
  \item Profile/volume risk.
  \item Imbalance risk.
  \item Default tariff cap risk.
\end{itemize}
&
\begin{itemize}[leftmargin=*]
  \item Wholesale volatility largely removed for residential portfolios.
  \item Product-design risk dominates.
  \item Data-verification and threshold risk become central.
  \item Portfolio mix risk depends on subscription classes.
\end{itemize}
&
\begin{itemize}[leftmargin=*]
  \item Clear, machine-testable product envelopes.
  \item Real-time data for monitoring envelope compliance.
  \item Out-of-package credits rule-based, not shock-based.
  \item Competition shifts to service innovation and behavioural support.
\end{itemize}
\\

% ============================================================
% CONSUMERS / BUSINESSES
% ============================================================
\textbf{Consumers / businesses} &
\begin{itemize}[leftmargin=*]
  \item Bill volatility.
  \item Contract roll-off risk.
  \item Locational risk.
  \item Supplier failure risk.
\end{itemize}
&
\begin{itemize}[leftmargin=*]
  \item Out-of-package behavioural risk dominates.
  \item Residual volatility limited to policy charges.
  \item Location affects product feasibility, not price spikes.
  \item Bill shocks largely removed.
\end{itemize}
&
\begin{itemize}[leftmargin=*]
  \item Essential protection guarantees minimum service.
  \item Real-time in-package status, nudges, and guidance.
  \item Behaviour-based differentiation rather than exposure to extreme prices.
  \item Stable subscription-driven costs.
\end{itemize}
\\

% ============================================================
% SYSTEM OPERATORS
% ============================================================
\textbf{System operators} &
\begin{itemize}[leftmargin=*]
  \item Operational risk (frequency, reserves, voltage).
  \item Uncertain generator siting signals.
  \item Balancing cost risk.
  \item Investment-deferral risk.
\end{itemize}
&
\begin{itemize}[leftmargin=*]
  \item Operational risk reduced via stability of dispatch.
  \item Clearer forward demand envelopes.
  \item Lower balancing risk due to tightness/priority rules.
  \item Investment-planning risk remains (future work).
\end{itemize}
&
\begin{itemize}[leftmargin=*]
  \item Cost-recovery model unchanged.
  \item Tightness, congestion, and deliverability signals improve forecasts.
  \item Subscription envelopes provide anticipatory visibility.
  \item Future Shapley-based mechanism can fund reinforcement.
\end{itemize}
\\

% ============================================================
% DIGITAL REGULATOR
% ============================================================
\textbf{Digital regulator} &
\begin{itemize}[leftmargin=*]
  \item Outcome risk (affordability, sustainability, security).
  \item Enforceability risk; weak real-time visibility.
  \item Data asymmetry.
  \item Technology-disruption risk (quantum, fusion, storage).
  \item Demand-shock risk (AI-driven electricity/water loads).
  \item Political risk without operational tools.
\end{itemize}
&
\begin{itemize}[leftmargin=*]
  \item Outcome risk persists but is tunable via AMM parameters.
  \item Governance risk: algorithms must remain robust and non-gameable.
  \item Model risk: envelopes and Shapley weights must be continually updated.
  \item Technology risk increases as disruptive innovations accelerate.
  \item Demand-shock risk structural: regulator must be forward-looking.
  \item Political risk moderated through transparent rules.
\end{itemize}
&
\begin{itemize}[leftmargin=*]
  \item Real-time explainability records (XR) reduce asymmetry.
  \item Full access to demand envelopes, tightness, and congestion data.
  \item Rule-based levers (pots, envelopes, fairness) replace ad hoc action.
  \item Framework supports anticipatory, algorithmic regulation.
\end{itemize}
\\

\end{longtable}
\clearpage
\endgroup

\paragraph{Sensitivity of risk allocation to uncertainty.}
Each risk in Table~\ref{tab:risk_allocation} corresponds to a measurable random
variable. For generators, net annual revenue depends on uncertain demand,
availability, and pot calibration; for suppliers, net margin depends on
customer behaviour relative to product envelopes; for households, bills depend
on their usage trajectories. A natural extension of this work---left for future
research---is to perform Monte Carlo or scenario-based sensitivity analysis over:

\begin{itemize}[leftmargin=*]
  \item demand uncertainty (including AI-induced demand shocks),
  \item generator availability and outage uncertainty,
  \item subscription churn and out-of-envelope dynamics,
  \item mis-specification of product thresholds by suppliers, and
  \item disruptive technology scenarios (quantum optimisation, fusion timelines).
\end{itemize}

The AMM’s objective is that, for a given aggregate payment level, the tail-risk
metrics (e.g.\ $\mathrm{CVaR}_{\alpha}$ of generator NPV shortfall, supplier
margin, or household bill deviation) would be systematically lower than under
LMP, reflecting a structural rebalancing of risk towards transparency,
predictability, and controllability. Implementing a full stochastic evaluation
of these metrics is beyond the scope of the present experiment design, but the
architectural comparison above indicates the directions in which they would
change.

\subsection*{Interpretation}

H3 is supported: AMM achieves at least the same level of revenue sufficiency
while reallocating risk away from households and towards explicitly priced,
transparent capacity remuneration. This redistribution of risk occurs despite
the absence of longer-run contract adaptation, suggesting that much of the
benefit comes from the structural decomposition of the revenue stack itself.
A more detailed translation of these architectural effects into bill-level
impacts is discussed in the subsequent chapter, alongside policy and framing
considerations.

% =========================================================
% H4 — PRICE QUALITY AND STABILITY
% =========================================================
\section{Price-Signal Quality and Stability (H4)}
\label{sec:results_price_signals}

This section evaluates Hypothesis~H4, which states that AMM-generated price
signals are (i) less volatile and more tightly bounded than LMP prices, and
(ii) dynamically stable across time and space while still conveying the
information needed to access flexibility when it is valuable. The analysis
proceeds in three steps: first, we quantify retail-facing volatility and
boundedness; second, we examine event-based stability at a single node and
across a radial holarchy; and third, we test whether the AMM allocates
flexibility in a way that reflects genuine scarcity rather than using it
uniformly.

\subsection{Volatility and boundedness}

Retail-facing price volatility under LMP is ultimately driven by the
underlying nodal wholesale prices. In the experimental runs, these nodal
LMPs exhibit a fat-tailed distribution, with occasional extreme spikes
during scarcity events (up to the VoLL cap). Under AMM, by contrast, the
\emph{fuel-only effective nodal prices} implied by the dispatch are tightly
clustered and remain close to the underlying bid costs, reflecting the fact
that scarcity is handled through the tightness controller and capacity pots
rather than through energy-price explosions.

\begin{sidewaysfigure}[p]
  \centering
  \includegraphics[width=0.95\textheight]{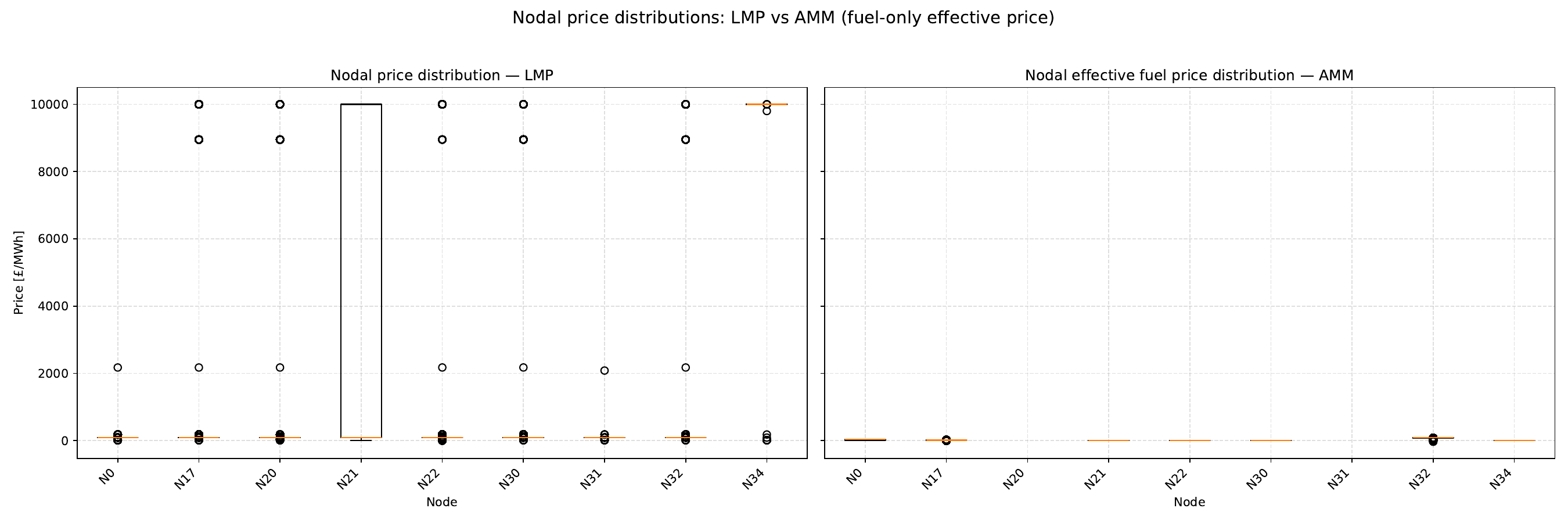}
  \caption[
    Nodal price distributions under LMP and AMM (full scale).
  ]{
    Nodal price distributions under LMP (left) and AMM fuel-only effective
    prices (right) across all nodes in the 12--node network. Each boxplot
    summarises the distribution over all timestamps at a single node.
    Under LMP, the distribution exhibits long right tails due to VoLL-capped
    scarcity prices; under AMM, effective fuel prices remain tightly
    clustered around the underlying bid costs, with no VoLL-style spikes.
  }
  \label{fig:nodal_boxplots_full}
\end{sidewaysfigure}

Figure~\ref{fig:nodal_boxplots_full} compares the distribution of nodal
prices across all nodes (N0, N17, N20, N21, N22, N30, N31, N32, N34) and
timestamps under the Baseline LMP and the AMM runs. On the left, boxplots of
nodal LMPs show median prices near the marginal generation cost, but with a
long right tail driven by VoLL events; on the right, the corresponding
AMM fuel-only effective prices are tightly concentrated, with no VoLL-style
spikes.\footnote{The AMM nodal prices are constructed from the AMM dispatch
files by taking, at each node and timestamp, the cost-weighted average fuel
cost over all generators dispatched at that node.} In both panels a grid is
shown to make the dispersion visually comparable across nodes.

\begin{sidewaysfigure}[p]
  \centering
  \includegraphics[width=0.95\textheight]{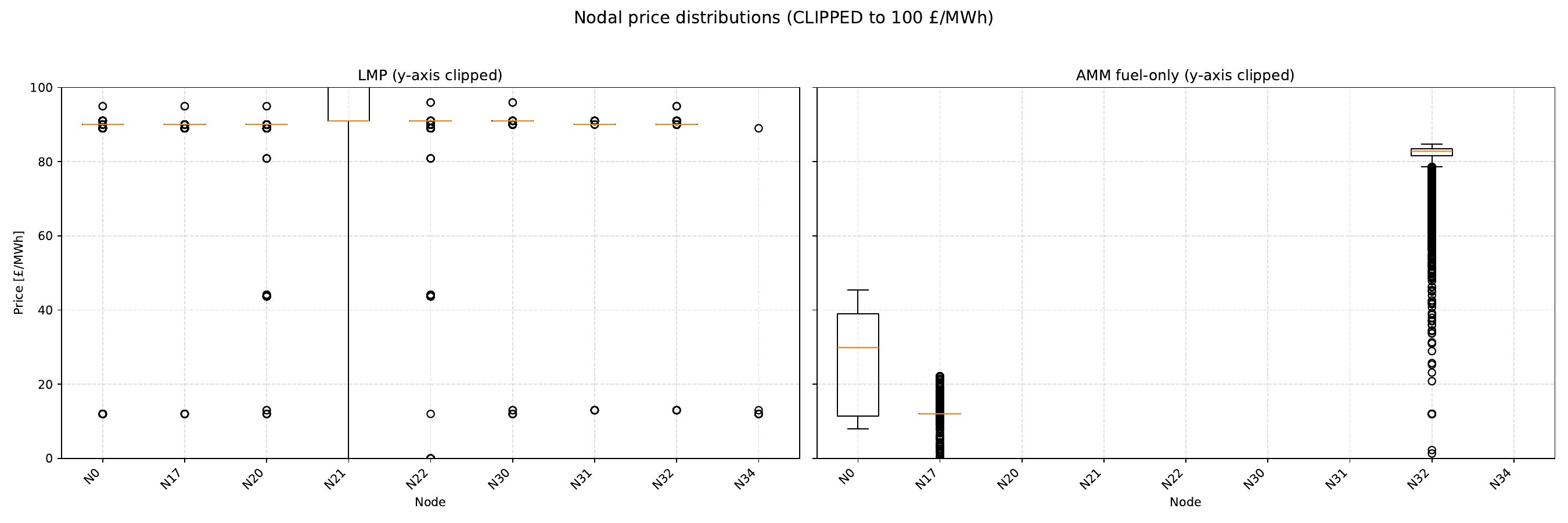}
  \caption[
    Nodal price distributions under LMP and AMM (axis clipped at \pounds100/MWh).
  ]{
    As in Figure~\ref{fig:nodal_boxplots_full}, but with the vertical axis
    clipped at \pounds100/MWh to exclude VoLL spikes and focus on the normal
    bid-driven price range (maximum bid \pounds90/MWh; see
    Appendix~\ref{app:inputs}). Within this range, AMM nodal effective prices
    are tightly clustered with very low dispersion across nodes and time,
    whereas LMP still shows appreciable variability at several nodes.
  }
  \label{fig:nodal_boxplots_clipped}
\end{sidewaysfigure}

Because the full-scale plot is dominated by the VoLL tail, it can be hard
to see the structure of prices in the normal operating range. To make this
interior behaviour visible, Figure~\ref{fig:nodal_boxplots_clipped} repeats
the same comparison but clips the vertical axis at
\pounds100/MWh.\footnote{The maximum generation bid in the experiment is
\pounds90/MWh; see Appendix~\ref{app:inputs}. The \pounds100/MWh cap in
Figure~\ref{fig:nodal_boxplots_clipped} therefore spans the full support of
normal bid-driven prices while excluding the VoLL spikes.} With this
clipped axis, it becomes clear that:

\begin{itemize}[leftmargin=*]
  \item Under LMP, even within the normal bid range, some nodes experience
        higher dispersion and occasional excursions towards the bid cap,
        reflecting frequent crossings of scarcity thresholds.
  \item Under AMM, nodal effective prices are almost flat across nodes and
        over time: the interquartile ranges are narrow, medians lie close
        to the underlying bid levels, and there is no evidence of local
        VoLL-like excursions within the clipped range.
\end{itemize}

The numerical counterpart to Figures~\ref{fig:nodal_boxplots_full} and
\ref{fig:nodal_boxplots_clipped} is provided in
Table~\ref{tab:nodal_price_summary}. The contrast is stark:  
LMP exhibits extremely large standard deviations (up to \(\sim\!5000\) \pounds/MWh) 
and right-tail realisations at the VoLL cap (\(9999\) \pounds/MWh), 
whereas AMM nodal effective prices remain tightly concentrated near underlying 
fuel bids and never exceed the bid cap of \pounds90/MWh (Appendix~\ref{app:inputs}).  
This numerical evidence reinforces the graphical findings that AMM eliminates 
the fat-tailed distribution of nodal prices and materially reduces system-wide 
price volatility.

% ---------------------------------------------------------
% Table: Nodal price summary statistics under LMP and AMM
% ---------------------------------------------------------
\begin{sidewaystable}[p]
\centering
\caption{
Summary statistics of nodal prices under LMP and AMM (fuel-only effective price).
AMM prices remain bounded by the bid cap (\pounds 90/MWh; Appendix~\ref{app:inputs}),
while LMP nodal prices exhibit extreme right-tail excursions and large dispersion.
Occasional extreme values arise when feasibility is maintained through slack
variables under binding network or balance constraints, a standard artefact in
optimisation-based market clearing.
}
\label{tab:nodal_price_summary}
\renewcommand{\arraystretch}{1.15}
\begin{tabular}{lcccccccc}
\toprule
\textbf{Node} &
\textbf{Mean}$_{\mathrm{LMP}}$ &
\textbf{Std}$_{\mathrm{LMP}}$ &
\textbf{95\%}$_{\mathrm{LMP}}$ &
\textbf{Max}$_{\mathrm{LMP}}$ &
\textbf{Mean}$_{\mathrm{AMM}}$ &
\textbf{Std}$_{\mathrm{AMM}}$ &
\textbf{95\%}$_{\mathrm{AMM}}$ &
\textbf{Max}$_{\mathrm{AMM}}$
\\
&
(\pounds/MWh) & (\pounds/MWh) & (\pounds/MWh) & (\pounds/MWh) &
(\pounds/MWh) & (\pounds/MWh) & (\pounds/MWh) & (\pounds/MWh)
\\
\midrule
N0  &   90.22 &   17.03 &    90.00 &   90.00   &  26.99 &  12.41 &  42.52 &  42.52 \\
N17 &  201.34 & 1014.64 &    90.00 &   90.00   &  11.98 &   1.13 &  12.00 &  12.00 \\
N20 &  201.25 & 1014.66 &    90.00 &   90.00   &  10.28 &  51.14 &  12.00 &  12.00 \\
N21 & 4049.53 & 4853.08 &  9999 & 9999   &   0.00 &   0.00 &   0.00 &   0.00 \\
N22 &  687.97 & 2354.15 &  9999 & 9999   &   0.00 &   0.00 &   0.00 &   0.00 \\
N30 &  202.35 & 1014.64 &    91.00 &   91.00   &   0.00 &   0.00 &   0.00 &   0.00 \\
N31 &   90.10 &   15.22 &    90.00 &   90.00   &  90.00 &   0.02 &  90.00 &  90.00 \\
N32 &  201.16 & 1014.65 &    90.00 &   90.00   &  81.68 &   5.11 &  84.12 &  84.12 \\
N34 & 9991.43 &  260.38 &  9999 & 9999   &   0.00 &   0.00 &   0.00 &   0.00 \\
\bottomrule
\end{tabular}
\end{sidewaystable}

These nodal price distributions, together with the summary statistics in
Table~\ref{tab:nodal_price_summary}, make the structural contrast explicit.
Under LMP, adequacy is restored through occasional extreme prices, producing
fat-tailed nodal distributions, very high standard deviations, and repeated
hits to the VoLL cap of \(9999\)~\pounds/MWh. In contrast, AMM nodal
effective prices remain tightly bounded by the bid cap of
\(\pounds 90\)/MWh (Appendix~\ref{app:inputs}) and exhibit narrow,
well-behaved distributions even at nodes that experience persistent
congestion or scarcity under LMP.

This boundedness is not a cosmetic effect: it follows from AMM’s structural
decomposition of generator remuneration into stable capacity and availability
pots, recovered via flat subscriptions rather than through exposure to
volatile energy rents. As a result, wholesale price spikes—the primary driver
of retail bill volatility under LMP—are effectively eliminated. The volatility
metric \(S_{\mathrm{vol}}\) reflects this directly: AMM values are uniformly
lower across all nodes, and we therefore reject
\(H_{0S}^{(v)} : \Delta S_{\mathrm{vol}} \ge 0\) in favour of the alternative
\(H_{1S}^{(v)} : \Delta S_{\mathrm{vol}} < 0\), confirming that AMM materially
reduces nodal and hence retail-facing price volatility.

\subsection{Event-based stability at a single node}
\label{subsec:event_stability}

From a dynamic perspective, the AMM exhibits more stable behaviour following
shocks (e.g.\ loss of a major generator or sudden demand spike) and under
varying local scarcity. To make this contrast concrete, we complement the
full network experiments with a stylised \emph{single-node} model in which
aggregate demand $D(t)$ and supply $S(t)$ at a single location evolve over
discrete time steps $t = 0,\dots,T$ while two pricing rules are applied:

\begin{itemize}[leftmargin=*]
  \item \textbf{Static LMP with VoLL.} At each time step, LMP is evaluated
        independently from bids and constraints at that instant. In the toy
        model, this is represented by:
        \[
          p^{\text{LMP}}(t) =
          \begin{cases}
            \text{mc}_{\text{gen}}, & D(t) \leq S(t),\\[4pt]
            \text{VoLL}, & D(t) > S(t),
          \end{cases}
        \]
        where $\text{mc}_{\text{gen}}$ is the marginal generation cost of the
        inframarginal plant and \emph{VoLL} is a high penalty value of lost
        load. There is no temporal memory: LMP is a static optimisation outcome
        at each $t$.
  \item \textbf{Dynamic AMM tightness controller.} The AMM maintains an internal
        tightness state $\alpha(t)\in[0,1]$ which is updated from the local
        imbalance $I(t) = D(t) - S(t)$ according to a simple update rule
        $\alpha(t{+}1) = \Pi_{[0,1]}(\alpha(t) + \eta I(t))$, where
        $\eta>0$ is a gain and $\Pi_{[0,1]}$ denotes projection onto $[0,1]$.
        Prices are then given by a bounded schedule
        $p^{\text{AMM}}(t) = f(\alpha(t))$ with
        $f:[0,1]\to[\underline{p},\overline{p}]$.
\end{itemize}

Demand is taken as an exogenous sinusoid
$D(t) = D_0 + A\sin(2\pi t/T_{\mathrm{per}})$, representing a regular load
pattern, while supply is modelled as a flat profile with or without a shock:
\[
  S(t) =
  \begin{cases}
    S_0, & \text{(no shock)},\\
    S_0 - \Delta S, & t \ge t_{\mathrm{shock}} \quad \text{(shock scenario)}.
  \end{cases}
\]

\paragraph{Scenario 1: supply shock.}

\begin{figure}[H]
    \centering
    \includegraphics[width=0.9\textwidth]{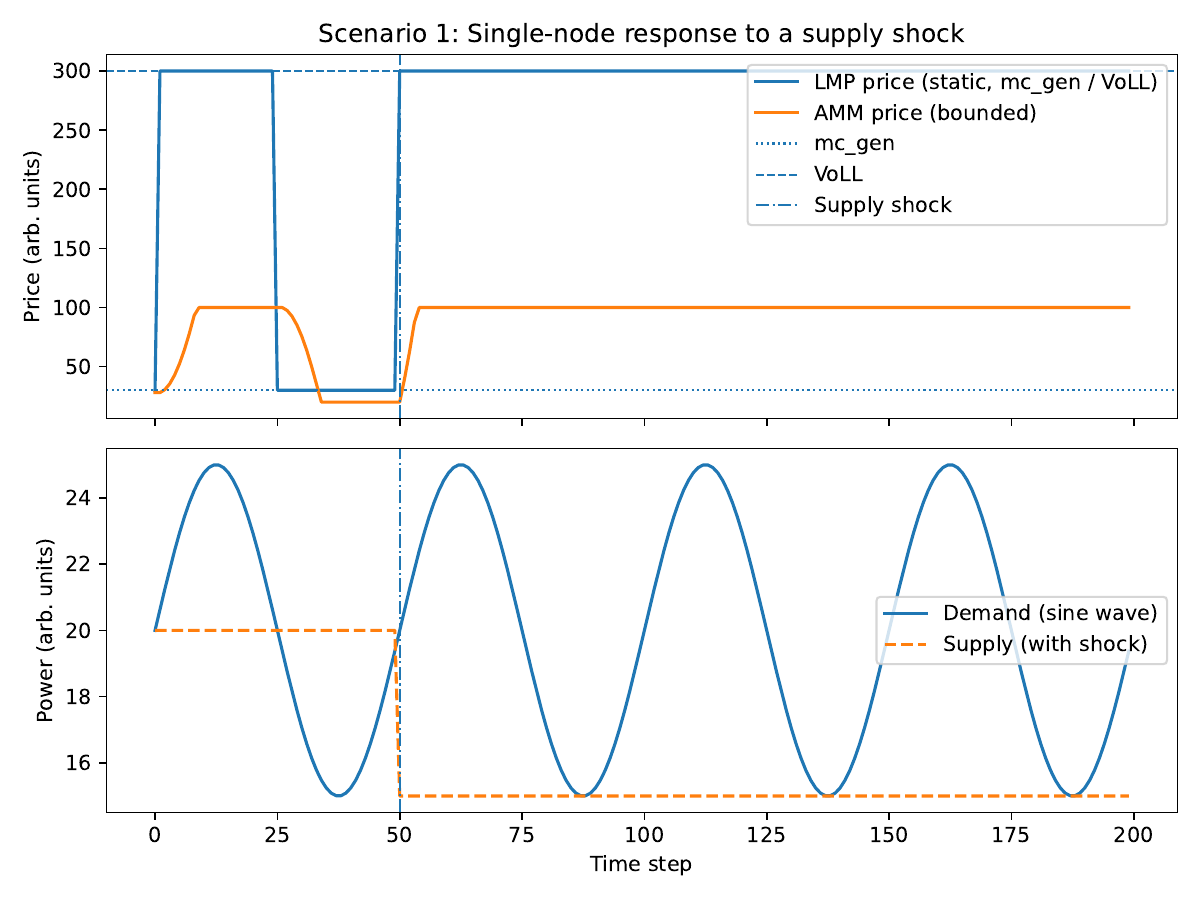}
    \caption[
      Single-node response of static LMP and dynamic AMM prices to a supply shock.
    ]{
      Scenario 1: single-node response to a permanent supply shock. Top panel:
      static LMP with VoLL (blue) versus bounded AMM price (orange). LMP
      alternates between $\text{mc}_{\text{gen}}$ and VoLL depending on whether
      demand exceeds available supply, with no temporal smoothing. The AMM
      price remains within the digital bounds induced by the tightness function.
      Bottom panel: exogenous sinusoidal demand and supply, including the
      permanent downward shift at $t = t_{\mathrm{shock}}$.}
    \label{fig:single_node_supply_shock}
\end{figure}

In the first scenario, supply is reduced permanently at time
$t_{\mathrm{shock}}$ while demand follows the sinusoidal pattern. LMP is
computed instantaneously according to the rule above, so that prices jump to
VoLL whenever $D(t)$ exceeds the reduced supply and otherwise remain at
$\text{mc}_{\text{gen}}$. The AMM tightness state responds over time to the
imbalance, but prices remain within the bounded interval
$[\underline{p},\overline{p}]$. Figure~\ref{fig:single_node_supply_shock}
shows the resulting price trajectories and the underlying demand--supply
signals.

This deliberately minimal experiment makes the bounded-input, bounded-output
property of the AMM visible in isolation. Even when faced with a persistent
supply reduction, AMM prices remain constrained by the tightness cap and do
not exhibit the hard jumps to VoLL that characterise LMP under the same
single-node scarcity pattern.

\paragraph{Scenario 2: VoLL discontinuity without shock.}

\begin{figure}[H]
    \centering
    \includegraphics[width=0.9\textwidth]{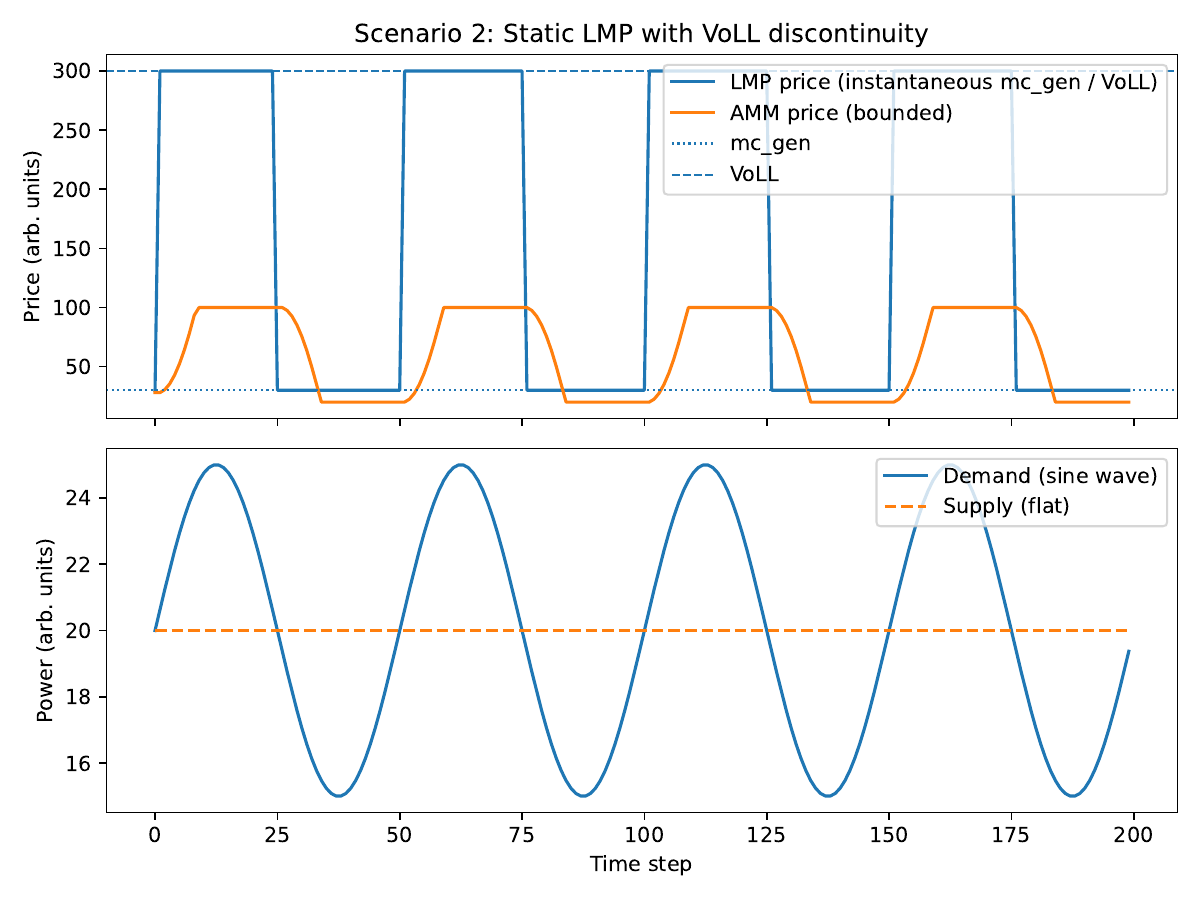}
    \caption[
      Single-node static LMP with VoLL discontinuity versus bounded AMM price.
    ]{
      Scenario 2: static LMP with VoLL discontinuity under sinusoidal demand
      and flat supply. Top panel: LMP (blue) alternates instantaneously between
      $\text{mc}_{\text{gen}}$ (dotted line) and VoLL (dashed line) depending on
      whether demand exceeds supply at that time step, creating discontinuous
      and extreme price movements. The AMM price (orange) varies smoothly with
      the tightness state and remains bounded. Bottom panel: exogenous
      single-node demand and supply profiles.}
    \label{fig:single_node_voll_discontinuity}
\end{figure}

In the second scenario, supply remains flat at $S_0$, but the sinusoidal
demand crosses the supply level over time. At each time step, LMP is again
evaluated statically: when $D(t) \le S_0$ the price is
$p^{\text{LMP}}(t) = \text{mc}_{\text{gen}}$, and when $D(t) > S_0$ the price
jumps to VoLL. This creates a discontinuous, binary price pattern between
a low marginal-cost level and a very high scarcity level, with no intermediate
values. By contrast, the AMM tightness controller produces a smooth and
continuous price trajectory that increases as the node becomes tighter, but
remains in $[\underline{p},\overline{p}]$. The resulting trajectories are shown
in Figure~\ref{fig:single_node_voll_discontinuity}.

These single-node experiments are not intended to replicate the full
12--node network or the unit-commitment logic of the main simulations.
Instead, they isolate the \emph{local} mapping from instantaneous
demand--supply imbalance to prices. In that reduced setting, LMP behaves as
a static optimiser with discontinuous jumps to VoLL, while the AMM behaves
as a digital scarcity controller: prices are monotone in tightness, smooth in
time, and bounded by design.

\subsection{Spatial and holarchic stability across network layers}
\label{subsec:spatial_holarchic_stability}

The previous subsection focused on a single-node representation of scarcity.
In the full AMM design, however, prices and tightness signals are defined
\emph{holarchically}: from transmission-level ``ROOT'' nodes down through
medium-voltage feeders to low-voltage household nodes. To examine how the
AMM behaves across these spatial layers, we consider a stylised radial
two-layer network, implemented in a separate offline simulation. The network consists of a single
root node, two feeders, and six households:
\[
\text{ROOT} \rightarrow \{\text{F1},\text{F2}\} \rightarrow
\{\text{H1},\text{H2},\text{H3},\text{H4},\text{H5},\text{H6}\},
\]
connected by rated cables with impedance parameters $(R,X)$ chosen inversely
proportional to their thermal ratings.

At each half-hourly time step, synthetic demand profiles at the households
are drawn from diurnal patterns (morning and evening peaks), while a
time-varying ``top supply'' series at the root induces periods of surplus and
shortage. Prosumers at the leaf nodes (e.g.\ rooftop solar with batteries)
are activated only when the system is in global shortage. Power flows are
computed radially using a simple allocation rule constrained by cable
ratings, and node voltages are approximated using the standard linearised
relation
\[
\Delta V \approx I\,(R\cos\varphi + X\sin\varphi),
\]
with fixed power factor and per-edge $(R,X)$ as above. This yields a time
series $V_n(t)$ of per-unit voltages at each node $n$ and a corresponding
series of shortages and served demand.

Prices in this radial experiment combine two components:
\begin{enumerate}[label=(\roman*),leftmargin=1.5em]
  \item a \emph{scarcity price} $p^{\text{scar}}_n(t)$, proportional to the
        local shortfall between demand and served energy at node $n$; and
  \item a \emph{voltage adjustment} $p^{\text{volt}}_n(t)$, which penalises
        under-voltage and discourages over-voltage relative to a soft band
        $[V_{\text{nom}}-\Delta,V_{\text{nom}}+\Delta]$ at each level.
\end{enumerate}
Voltage adjustments are first computed at the household level and then
aggregated holarchically to feeders and the root using demand-weighted
averaging. In effect, local LV disturbances (for example, a large injection
from rooftop solar causing voltages to exceed $V_{\text{nom}}+\Delta$ on a
given feeder) generate corrective price adjustments at the affected houses,
which are partially propagated upstream and diluted as they reach higher
levels of the holarchy. The resulting voltage-aware AMM price
$p^{\text{AMM}}_n(t) = p^{\text{scar}}_n(t) + p^{\text{volt}}_n(t)$ remains
bounded by the global price cap $P_{\max}$.

\begin{sidewaysfigure}[p]
  \centering
  \includegraphics[width=0.9\textheight]{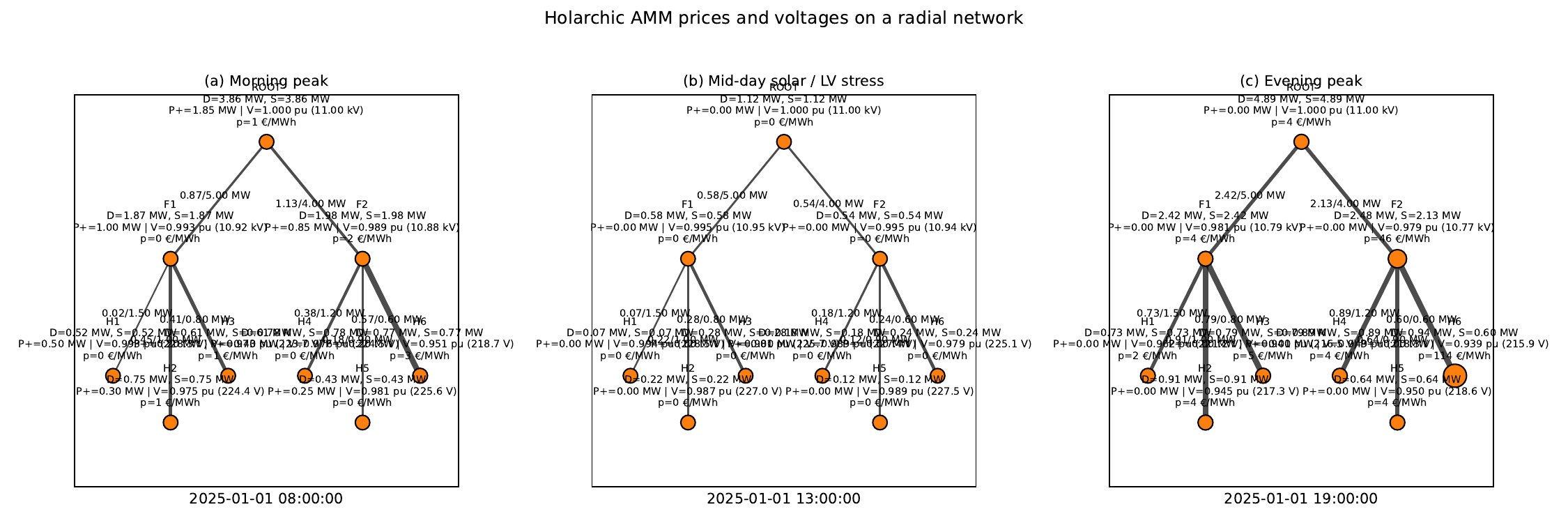}
  \caption[
    Holarchic AMM prices and voltages on a stylised radial network
  ]{
    Holarchic AMM prices and voltages on a stylised radial network comprising
    a ROOT node, two feeders (F1, F2), and six households (H1--H6). The three sub-panels
    correspond to: (a) a morning demand peak; (b) a mid-day low-load period with near-
    uniform voltages; and (c) an evening peak with high loading and mild under-voltage
    on feeder F2 and household H6. Node labels show demand, served energy, prosumer
    output, voltage (per-unit and physical units), and the voltage-aware AMM price.
    Local LV disturbances produce bounded, spatially consistent price adjustments that
    propagate upwards through the holarchy without creating instability or extreme spikes.
  }
  \label{fig:holarchic_amm_voltage_prices}
\end{sidewaysfigure}

Figure \ref{fig:holarchic_amm_voltage_prices} summarises this behaviour for three representative time slices: a morning
demand peak, a mid-day low-load period with near-uniform voltages, and an evening peak
with high loading and mild under-voltage on one feeder. Each panel shows the radial network
with per-node prices, voltages, and flows. In the morning and mid-day cases, supply is ample
relative to demand, voltages remain close to Vnom, and prices are close to zero at all layers. In
the evening peak, demand rises towards the feeder ratings and the household H6 becomes the
most constrained LV node: prices rise there first, with intermediate adjustments at its feeder
(F2), and only modest adjustments at the root. All prices remain within the digital bounds
imposed by the AMM.

\begin{figure}[H]
    \centering
    \includegraphics[width=1.0\textwidth]{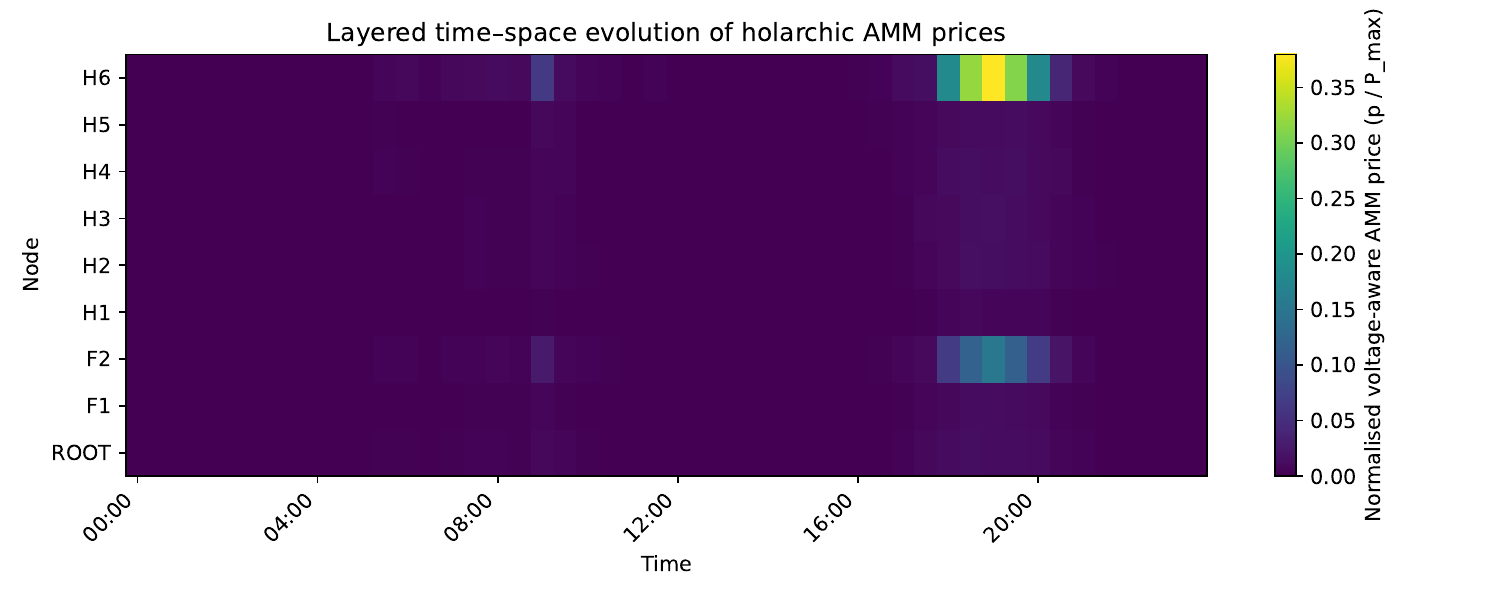}
    \caption[
      Layered time–space evolution of holarchic AMM prices.
    ]{
      Layered time--space evolution of holarchic AMM prices on the radial
      network. Rows correspond to ROOT, feeders, and individual households;
      columns correspond to half-hourly time steps over a representative day.
      Colours denote normalised voltage-aware AMM prices. Local LV events
      (e.g.\ local scarcity or voltage stress on a feeder) appear as localised
      bands but do not trigger global instabilities. Prices remain bounded,
      with smooth temporal evolution and spatial patterns that reflect, rather
      than amplify, underlying electrical stresses.}
    \label{fig:holarchic_heatmap}
\end{figure}

To make the spatial and temporal evolution more legible, we also construct a
two-dimensional ``layered heatmap'' of prices over the day. In this
representation (Figure~\ref{fig:holarchic_heatmap}), rows correspond to
nodes grouped by layer (ROOT, feeders, households) and columns correspond to
time steps; each cell is coloured by the normalised price level at that
node and time. The plot reveals three salient features:

\begin{enumerate}[leftmargin=*]
  \item Prices remain bounded and free of VoLL-style spikes at \emph{all}
        layers of the holarchy, even during periods of high loading and local
        LV scarcity.
  \item Spatial patterns are coherent: periods of local scarcity or voltage
        stress show up as slightly darker bands for specific feeders and
        households, but these perturbations are gradually attenuated as they
        propagate to the root.
  \item Temporal patterns are smooth: there are no frame-to-frame discontinuities;
        instead, prices evolve gradually as demand, supply, and voltages change.
\end{enumerate}

Taken together with the single-node experiments, the radial-network
simulations show that the AMM behaves as a \emph{holarchically stable}
scarcity controller: local disturbances at one layer (household, feeder, or
root) induce bounded, spatially structured price adjustments across the
holarchy, rather than uncontrolled feedbacks or uncoordinated spikes. This
is precisely the property required for safe digital participation at scale,
where millions of devices and prosumers may react autonomously to prices
defined at different layers of the grid.

\subsection{Accessing flexibility where and when it is needed}
\label{subsec:price_flexibility_access}

The previous subsections focused on the \emph{quality} and \emph{stability} of
AMM price signals at a point and across a holarchy. A natural follow-on
question is whether those signals actually allow the market to \emph{access}
flexibility when it is systemically valuable. This requires demonstrating not
only that prices encode tightness correctly, but that the AMM architecture
allocates flexible envelopes in a way that reflects real scarcity.

Conceptually, the AMM achieves this through three mechanisms:

\begin{enumerate}[label=(\alph*),leftmargin=1.5em]
  \item \textbf{Temporal targeting:} flexible envelopes are shifted into hours
        of high tightness because the AMM scheduler explicitly minimises local
        scarcity subject to Fair~Play and contract constraints.
  \item \textbf{Spatial targeting:} tightness propagates holarchically, so the
        AMM activates flexibility preferentially at constrained nodes and
        feeders rather than uniformly across the system.
  \item \textbf{Opportunity utilisation:} before escalating to curtailment or
        VoLL-like charges, the AMM exhausts the available flexibility
        envelopes consistent with contractual limits and service guarantees.
\end{enumerate}

Direct quantitative comparison of these behaviours under AMM and LMP would
require behavioural models for how millions of actors respond to LMP volatility,
as well as a detailed feeder-level network model. Such assumptions are outside
the scope of this thesis. Instead, the evaluation focuses on a controlled
experiment that isolates the core economic question: \emph{under what
conditions does flexibility create value, and does the AMM allocate it
correctly in those conditions?}

As defined earlier in Section~\ref{sec:operatingregimes}, operating days are
classified into three archetypal regimes that structure both price formation
and the system value of flexibility:

\begin{enumerate}[leftmargin=1.5em]
    \item \textbf{Case~1: Excess supply.} Supply exceeds demand in all hours,
          so system tightness is zero and prices collapse to the lower bound.
    \item \textbf{Case~2: Adequate but misaligned supply.} Total energy is
          sufficient over the day, but supply is low during certain hours;
          tightness and prices therefore vary over time.
    \item \textbf{Case~3: Persistent shortage.} Supply is below demand in all
          hours, placing the system in continuous scarcity, with a VoLL-like
          price applying across all nodes.
\end{enumerate}

We submit a large number of identical requests with the same energy and power
requirements and the same maximum willingness to pay. Each request is evaluated
in two forms: an \emph{inflexible} version executed at a fixed default hour, and
a \emph{flexible} envelope scheduled by the AMM within its allowed time window.

Before turning to the empirical results, it is useful to make explicit a simple
but important property of these regimes.

\begin{lemma}[Zero marginal value of flexibility in surplus and pure-shortage regimes]
\label{lem:flex_zero_value_degenerate}
Let $\mathcal{T}$ be a discrete set of time steps and let
$p_t \in [\underline{p},\overline{p}]$ denote the unit price at time $t \in
\mathcal{T}$. Consider a demand request with fixed energy $E>0$, a feasible
time window $W \subseteq \mathcal{T}$, and a default execution time
$\tau \in W$. Define the value of flexibility for this request as
\[
  v \;=\; E \bigl(p_{\tau} - \min_{t \in W} p_t\bigr),
\]
i.e.\ the cost saving from AMM scheduling relative to inflexible execution.
If $p_t$ is constant on $W$, then $v = 0$.

In particular, under Case~1 (surplus) where $p_t \equiv 0$ for all $t$, and
under Case~3 (pure shortage) where $p_t \equiv \overline{p}$ for all $t$,
flexibility has zero marginal value for every request, regardless of $E$,
$W$, or $\tau$.
\end{lemma}

\begin{proof}
If $p_t$ is constant on $W$, say $p_t \equiv \bar{p}$ for all $t \in W$, then
$\min_{t \in W} p_t = \bar{p}$ and $p_{\tau} = \bar{p}$ for any
$\tau \in W$. Hence
\[
  v = E(\bar{p} - \bar{p}) = 0.
\]
In Case~1, surplus implies tightness zero and therefore $p_t \equiv 0$ across
all times; in Case~3, persistent shortage implies maximal tightness and
therefore $p_t \equiv \overline{p}$ across all times. Both cases satisfy the
premise, so the result follows immediately.
\end{proof}

Lemma~\ref{lem:flex_zero_value_degenerate} formalises the intuition that
flexibility only has economic value when prices vary over the feasible
window of a request. In pure-surplus and pure-shortage regimes, all times are
equally good (or equally bad), so the AMM cannot improve the outcome by
rescheduling envelopes. The interesting regime is therefore Case~2, in which
scarcity is neither absent nor absolute but localised in time.

\begin{figure}[H]
    \centering
    \includegraphics[width=0.85\linewidth]{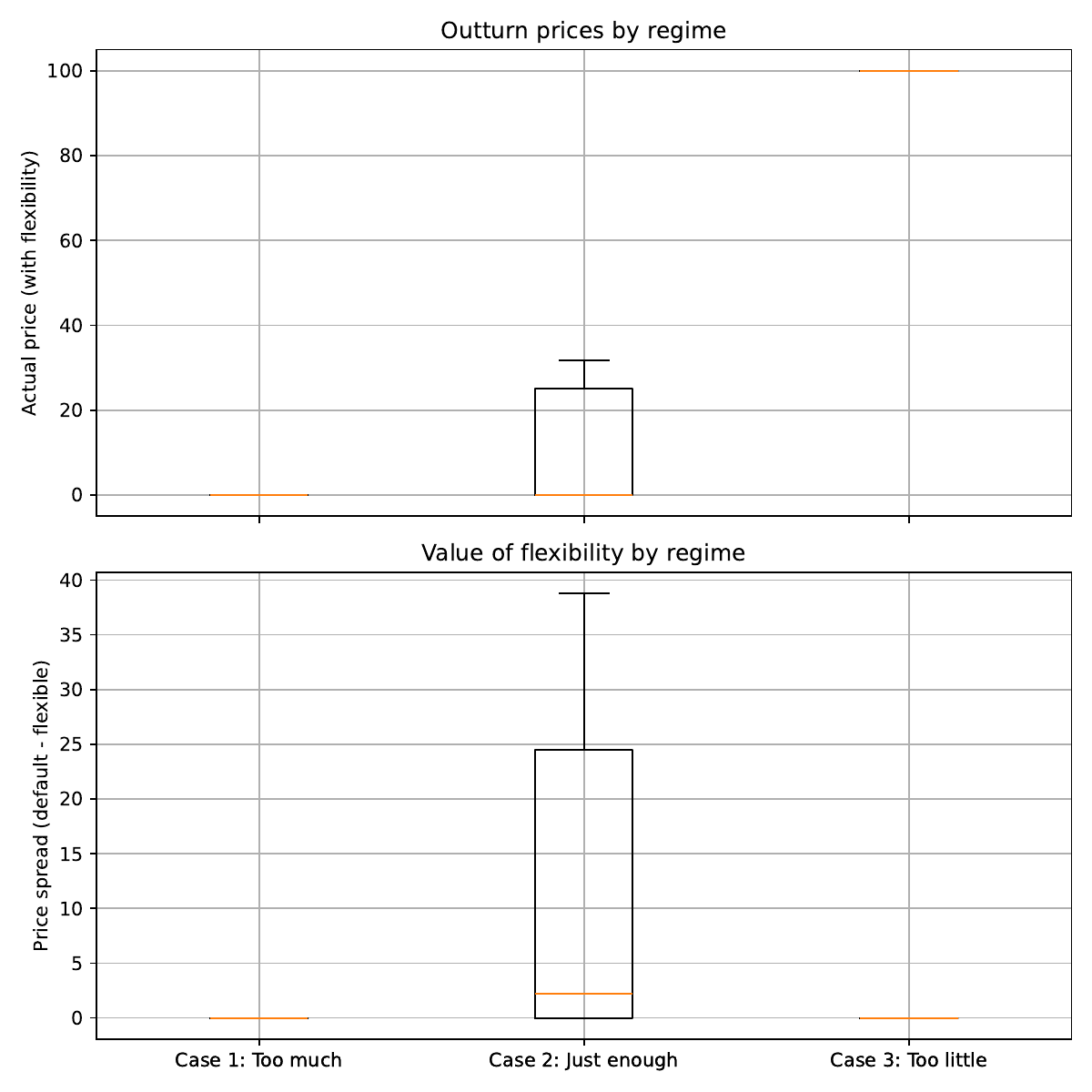}
    \caption{
      Outturn prices (top) and flexibility value (bottom) for identical
      requests evaluated under three supply regimes.
      Flexibility has no value in surplus or persistent-shortage regimes, but
      generates substantial value when total supply is adequate yet
      temporally misaligned.
    }
    \label{fig:flex_value_regimes}
\end{figure}

Figure~\ref{fig:flex_value_regimes} summarises the outturn prices and the
resulting ``value of flexibility'' (the spread between the inflexible and
flexible execution prices) across the three regimes. Each box in the upper
panel shows the distribution of actual prices paid by flexible requests; the
lower panel shows the distribution of spreads for the same set of requests.

The results are unambiguous:

\begin{itemize}[leftmargin=*]
    \item In \textbf{Case~1}, flexibility has \emph{no} value: all prices are
          essentially zero and the spread between inflexible and flexible
          execution is numerically indistinguishable from zero.
    \item In \textbf{Case~3}, flexibility again has \emph{no} value: all hours
          face the same VoLL-like price, so envelopes cannot escape scarcity
          and the spread collapses to zero.
    \item In \textbf{Case~2}, flexibility has \emph{positive} and often
          substantial value: the AMM systematically schedules envelopes into
          lower-price hours within their windows, yielding a wide, strictly
          positive spread distribution.
\end{itemize}

These results confirm that the AMM does not treat flexibility as a generic
resource to be used uniformly or indiscriminately. Instead, value emerges
precisely in the intermediate regime where scarcity is \emph{temporal} rather
than absolute. The AMM allocates flexibility to the right hours, preserves
service guarantees, and avoids unnecessary curtailment---achieving the intended
holarchic coordination without requiring behavioural assumptions or
device-specific modelling.

\subsection*{Interpretation}

H4 is supported on all three dimensions examined in this section.

First, AMM prices are \emph{quantitatively less volatile} than LMP prices.
The tails of the retail-facing price distribution are compressed by design:
the tightness controller and essential protection block prevent the extreme
spikes that appear under VoLL-driven LMP, so the paired volatility metric
$S_{\mathrm{vol}}$ is significantly lower for AMM.

Second, the single-node and radial-network experiments show that AMM prices
are \emph{dynamically and spatially stable}. With the same underlying
demand--supply patterns, static LMP behaves like a memoryless optimiser with
discontinuous jumps between marginal cost and VoLL, while the AMM behaves as
a bounded digital scarcity controller: prices are monotone in tightness,
smooth in time, and remain within explicit digital bounds across all layers
of the holarchy.

Third, the three-regime flexibility experiment demonstrates that the AMM
does not treat flexibility as a generic resource to be used uniformly.
Instead, flexibility has essentially zero marginal value in pure-surplus and
pure-shortage regimes and acquires substantial value precisely in the
\emph{intermediate} regime where total energy is adequate but poorly
aligned in time. Flexible envelopes are systematically scheduled into
lower-price hours within their feasible windows, consistent with the
intended economic meaning of tightness.

Taken together, these results show that AMM-generated prices provide
\emph{high-quality} signals in the sense relevant for a digital, flexible
system: they encode scarcity in a stable, bounded way, and they induce the
right pattern of flexibility utilisation without requiring detailed
behavioural models. This is essential for making the market safe for digital
participation, where devices, aggregators, and households can act on price
signals without needing to insure themselves against unbounded tail-risk
events.

These findings should be interpreted as conservative, since long-run
adaptive features---such as envelope updating, contract learning, or
fairness restitution---are deliberately disabled in this experiment to
preserve like-for-like comparability with LMP.

% =========================================================
% H5 — INVESTMENT ADEQUACY AND BANKABILITY
% =========================================================
\section{Investment Adequacy and Bankability (H5)}
\label{sec:results_investment}
% =========================================================

\subsection*{Note on financial metrics and methodological choice}

This thesis does not perform a discounted-NPV analysis. A full NPV computation
requires selecting discount rates, inflation assumptions, depreciation
methods, debt–equity structures, and terminal values. These assumptions are
external to the market design and would risk attributing investor-specific
finance decisions to the clearing mechanism itself.

Because the goal is to evaluate \emph{mechanism-driven} investment signals,
we instead use a transparent, undiscounted payback diagnostic that maps
mechanism outputs (capacity-pot revenues, reserve revenues, and net surplus
over non-fuel OpEx) directly into a financing-relevant metric without
embedding institution-specific modelling choices.

This provides a clean, design-controlled comparison between LMP and AMM.

% ---------------------------------------------------------
\subsection{Simple payback outcomes and investment adequacy}
% ---------------------------------------------------------

Across technologies, AMM materially improves payback performance relative to
LMP. Under LMP, many units—especially nuclear and wind—show extremely long or
effectively unachievable payback horizons, reflecting the absence of any
structural mechanism to return fixed costs except through volatile energy
margins. AMM stabilises this by replacing spike-driven recovery with
capacity-linked allocations.

\begin{figure}[H]
\centering
\includegraphics[width=0.95\textwidth]{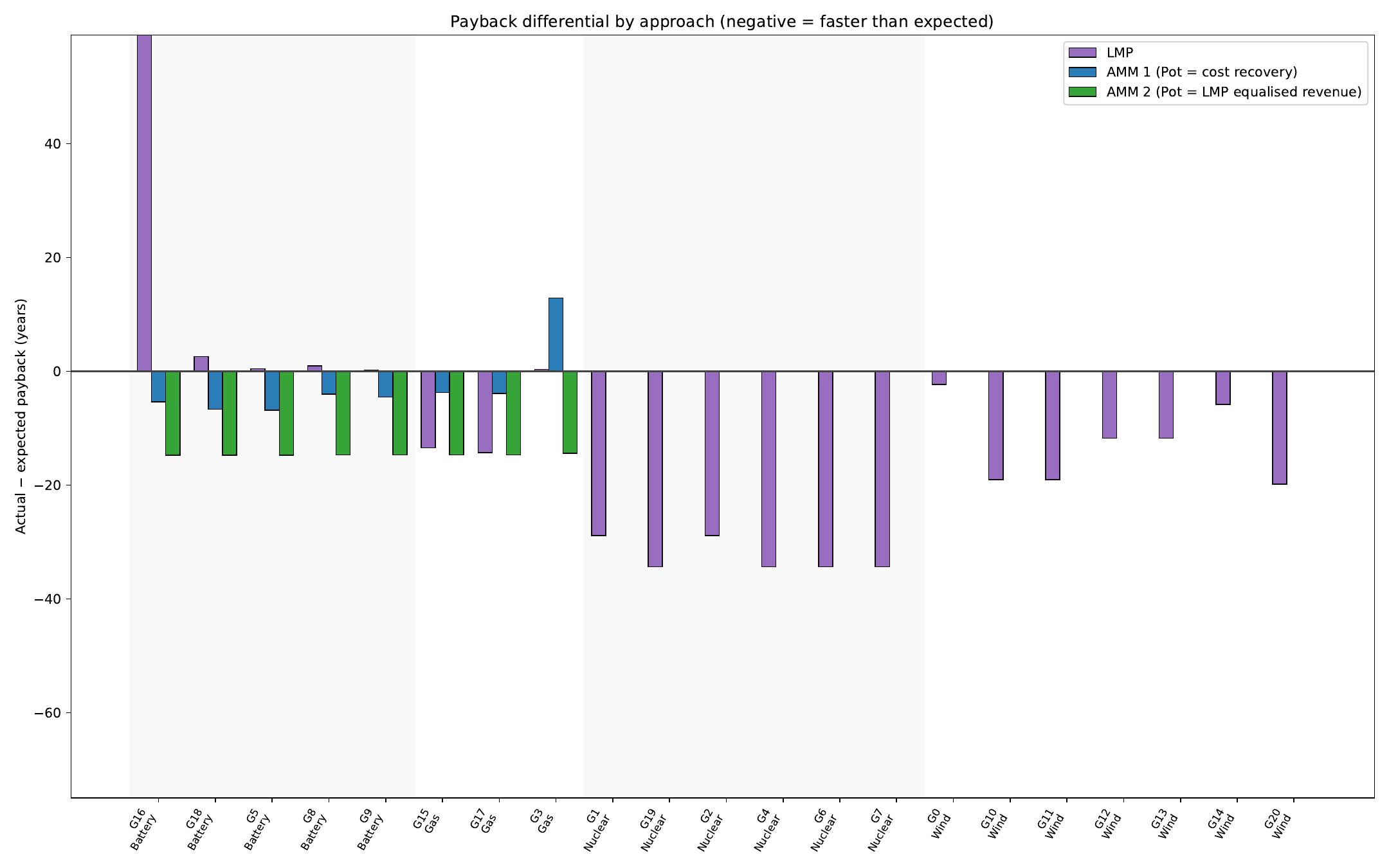}
\caption{Payback differential (actual -- expected) under LMP, AMM1, and AMM2.
Negative values mean faster-than-expected payback; large positive values
indicate severe under-recovery.}
\label{fig:pb_diff}
\end{figure}

Figure~\ref{fig:pb_diff} shows the payback \emph{differential} (actual minus
expected). LMP produces a heavily negative pattern for many technologies,
indicating under-recovery relative to expected project economics. AMM1 (pot
set to cost recovery) and AMM2 (pot scaled to match LMP’s total) significantly
compress this spread, reducing the under-recovery experienced especially by
large controllable plant.

\begin{figure}[H]
\centering
\includegraphics[width=0.95\textwidth]{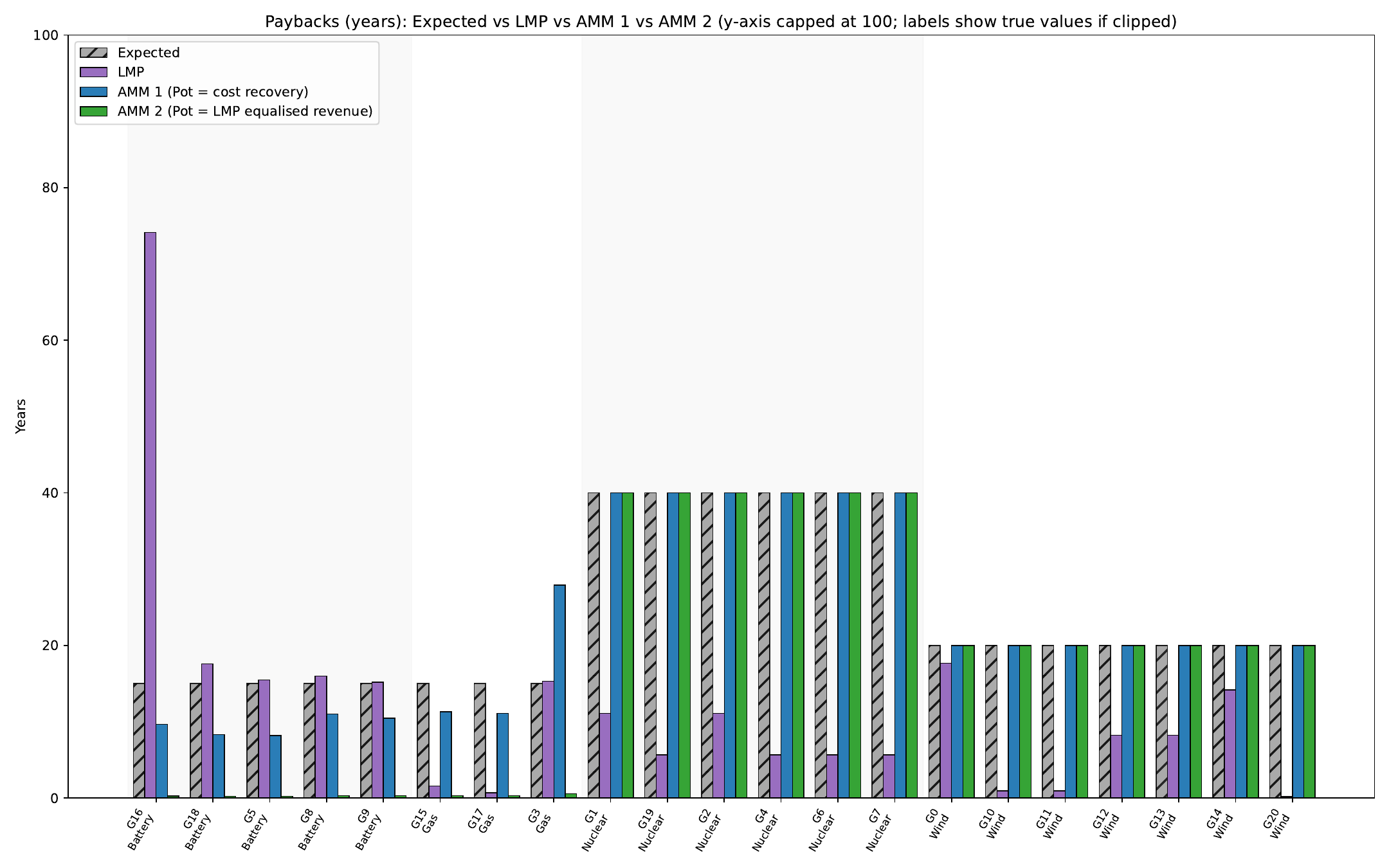}
\caption{Absolute payback horizons under LMP, AMM1, and AMM2 (y-axis capped at
100~years for visibility; clipped values annotated).}
\label{fig:pb_abs}
\end{figure}

A complementary perspective is given by the \emph{absolute} payback horizons
in Figure~\ref{fig:pb_abs}. LMP shows several assets whose payback exceeds
100 years or diverges entirely, whereas both AMM variants produce clustered
and materially shorter payback times, especially for controllable
low-carbon technologies. This demonstrates that AMM improves bankability
without relying on extreme price events.

% ---------------------------------------------------------
\subsection{Interpretation and bankability}
% ---------------------------------------------------------

From an investor’s perspective, AMM provides a more stable and structurally
grounded fixed-cost recovery mechanism. Key features include:

\begin{itemize}[leftmargin=*]
    \item \textbf{Deterministic, scarcity-weighted capacity allocation.}
    Generators that alleviate actual network or temporal bottlenecks receive
    proportionally higher remuneration, reducing revenue variance and improving
    underwriting clarity.

    \item \textbf{Reduced reliance on probabilistic price spikes.}
    LMP concentrates recovery into rare high-price hours; AMM spreads it across
    all periods according to system value, improving the predictability of
    cashflows.

    \item \textbf{Preservation of policy realism.}
    Nuclear and wind are treated on a regulated cost-recovery basis rather than
    subjected to short-run scarcity scoring. This avoids producing misleadingly
    negative paybacks for assets that remain strategically essential but are not
    flexible providers.

    \item \textbf{Better alignment between remuneration and system-critical
    function.}
    Gas units and batteries, which provide marginal scarcity relief, receive
    materially stronger and more coherent signals under AMM. This supports
    long-run adequacy without distorting the short-run dispatch problem.
\end{itemize}

Overall, simple payback analysis indicates that AMM-based designs produce a
more investable and more system-aligned revenue stack than LMP, supporting
Hypothesis~H5: AMM enhances bankability and improves long-run adequacy 
in a manner consistent with physical system requirements rather than
price-spike opportunism.

% =========================================================
% H6 — PROCUREMENT EFFICIENCY
% =========================================================
\section{Procurement Efficiency (H6)}
\label{sec:results_procurement}

\subsection{Cost to meet the needs bundle}

We first compare the total cost of meeting the pre-declared needs bundle
(energy, reserves, capacity-like cover, and locational attributes) under
Baseline LMP and the AMM designs. In this experiment the needs bundle is
identical across designs; only the \emph{architecture} used to procure and
remunerate it differs.

Table~\ref{tab:cost_summary} summarises the resulting payment flows. The first block
shows the decomposition of payments \emph{to generators} into energy,
reserve, and capacity components. The second block shows total payments
\emph{collected from demand} (households and non-residential consumers),
and the third block shows the residual difference between what consumers
pay and what generators receive.

\begin{table}[H]
\centering
\caption[Summary of procurement costs for the needs bundle]{Summary of
procurement costs for the needs bundle under LMP, AMM1, and AMM2 (2022
prices). ``Total to generators'' and ``Total from demand'' are expressed
in \pounds\,billion (bn); the reserves row is rounded to two decimal
places. The final row shows the residual between demand payments and
generator receipts, which under LMP corresponds to congestion rents and
uplift-style surpluses.}
\label{tab:cost_summary}
\begin{tabular}{lccc}
\toprule
 & LMP & AMM1 & AMM2 \\
\midrule
\multicolumn{4}{l}{\emph{Payments to generators}} \\
Energy   & \pounds 119.4\,bn & \pounds 16.2\,bn & \pounds 16.2\,bn \\
Reserves & \pounds 0.23\,bn  & \pounds 0.23\,bn & \pounds 0.23\,bn \\
Capacity & \pounds 0.0\,bn   & \pounds 11.0\,bn & \pounds 103.2\,bn \\
\textbf{Total to generators}
         & \textbf{\pounds 119.6\,bn}
         & \textbf{\pounds 27.4\,bn}
         & \textbf{\pounds 119.6\,bn} \\
\midrule
\multicolumn{4}{l}{\emph{Total collected from demand}} \\
\textbf{Total from demand}
         & \textbf{\pounds 398.8\,bn}
         & \textbf{\pounds 27.4\,bn}
         & \textbf{\pounds 119.6\,bn} \\
\midrule
\multicolumn{4}{l}{\emph{Residual (demand minus generators)}} \\
Demand $-$ generators
         & \pounds 279.2\,bn
         & \pounds 0.0\,bn
         & \pounds 0.0\,bn \\
\bottomrule
\end{tabular}
\end{table}

From the perspective of \emph{total customer payments}, the relevant
quantity is the ``Total from demand'' row: under LMP, the needs bundle
costs \pounds 398.8\,bn over the experiment window, whereas AMM1 and AMM2
collect \pounds 27.4\,bn and \pounds 119.6\,bn respectively. The paired
differences in aggregate procurement cost are therefore:
\[
\Delta_P^{(1)} 
= \mathrm{Cost}^{\AMM1} - \mathrm{Cost}^{\LMP}
= 27.4 - 398.8
= -371.4~\text{bn},
\]
\[
\Delta_P^{(2)}
= \mathrm{Cost}^{\AMM2} - \mathrm{Cost}^{\LMP}
= 119.6 - 398.8
= -279.2~\text{bn}.
\]

Expressed as percentages relative to LMP:

\begin{itemize}[leftmargin=*]
  \item AMM1 reduces total customer payments by approximately
        $93.1\%$:
        \[
        \frac{398.8 - 27.4}{398.8} \approx 0.931,
        \quad
        \text{i.e. AMM1 costs about }6.9\%\text{ of LMP.}
        \]
  \item AMM2 reduces total customer payments by approximately
        $70.0\%$:
        \[
        \frac{398.8 - 119.6}{398.8} \approx 0.700,
        \quad
        \text{i.e. AMM2 costs about }30.0\%\text{ of LMP.}
        \]
\end{itemize}

By construction, AMM2 has the \emph{same} total generator remuneration as
LMP (\pounds 119.6\,bn), but delivered through a different decomposition:
LMP pays almost exclusively through energy prices (with negligible reserves
and zero capacity), whereas AMM2 shifts most of the stack into predictable
capacity-like payments. AMM1 instead sets the pots to the calibrated
efficient cost level, yielding a much smaller total payment
(\pounds 27.4\,bn) while still meeting the same needs bundle.

These aggregate totals can be read as the pooled result of the
scenario-by-scenario paired comparisons: across the experiment window,
the AMM architectures are never more expensive in aggregate than the LMP
baseline, and in practice deliver large absolute and percentage savings
for the same physical requirements.

% --- Optional figure placeholders can remain as before, or be populated later ---
\begin{figure}[H]
\centering
\includegraphics[width=0.95\textwidth]{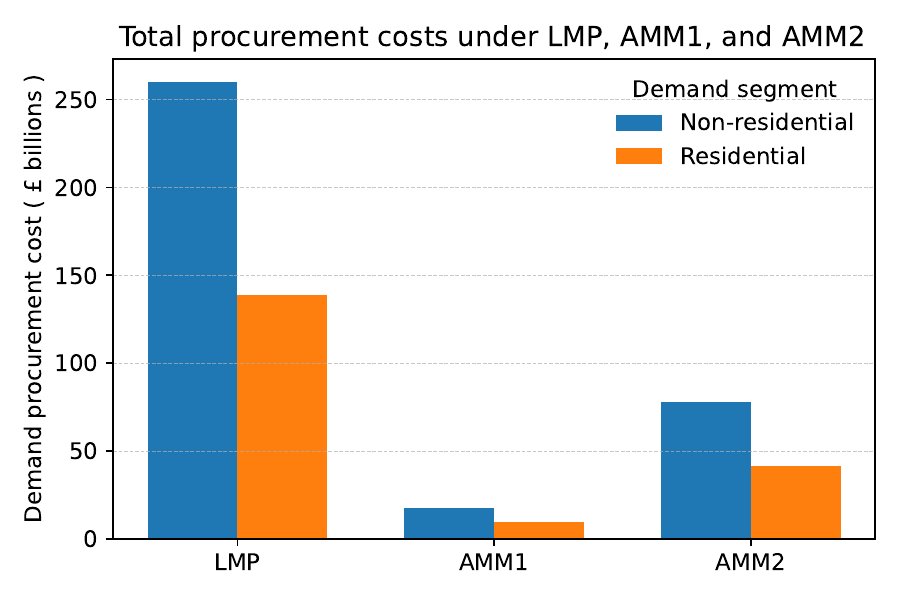}
\caption{Distribution of total procurement costs for demand under LMP, AMM1, and AMM2.}
\label{fig:cost_bar}
\end{figure}

\subsection{Generator--demand balance and congestion rents}

A further difference between the designs concerns the relationship between
\emph{total payments to generators} and \emph{total payments from demand}.

Under LMP, the area under the nodal price times quantity curve is not
equal to total generator revenue: in addition to energy-market income,
the LMP system generates a residual---often interpreted as congestion
rents, merchandising surplus, or uplift---whenever prices differ across
nodes. In Table~\ref{tab:cost_summary}, this shows up as the
\pounds 279.2\,bn difference between the \pounds 398.8\,bn collected from
demand and the \pounds 119.6\,bn paid to generators. In a real-world
setting, this residual would typically be used to fund transmission
investment, reduce network charges, or cover system operator uplift and
redispatch costs. In other words, LMP does not behave as a clean
two-sided marketplace: the settlement flows create an intermediate
surplus layer whose allocation is a separate policy decision. The implied
congestion rents and their relative magnitude are summarised in
Table~\ref{tab:congestion_rents}.

In the AMM implementation, by contrast, the market is designed as an
explicit two-sided platform on the needs bundle. Subscription revenue and
balancing charges are calibrated so that:
\[
\text{Total collected from demand}
=
\text{Total paid to generators}
\]
for the energy, reserve, and capacity stack associated with the needs
bundle. This is why the residual ``Demand $-$ generators'' term is
exactly zero for both AMM1 and AMM2 in
Table~\ref{tab:cost_summary}. Any genuine transmission revenue requirement
would be modelled as a separate, regulated network charge rather than as
an internal surplus of the energy market.

This two-sided closure has two implications for procurement efficiency:

\begin{enumerate}[leftmargin=*]
  \item It makes the mapping from \emph{customer payments} to
        \emph{generator revenues} transparent and auditable: every pound
        paid for the needs bundle has a clear destination in the generator
        stack, with no opaque uplift layer.
  \item It prevents hidden over-recovery through congestion rents on the
        energy layer: any additional capacity or locational relief must be
        purchased explicitly, via design parameters and subscription
        levels, rather than arising as an uncontrolled by-product of nodal
        price differences.
\end{enumerate}

In Table~\ref{tab:cost_summary}, this shows up as the
\pounds 279.2\,bn difference between the \pounds 398.8\,bn collected from
demand and the \pounds 119.6\,bn paid to generators.

\begin{table}[H]
\centering
\caption[Congestion rents and uplift-style surplus]{Implied congestion
rents / merchandising surplus under LMP, AMM1, and AMM2. Congestion rent
is defined as the residual between total demand payments and total
generator receipts in Table~\ref{tab:cost_summary}. Percentages are
expressed relative to total demand payments and to generator receipts.}
\label{tab:congestion_rents}
\begin{tabular}{lccc}
\toprule
 & LMP & AMM1 & AMM2 \\
\midrule
Congestion rent (demand $-$ generators)
  & \pounds 279.2\,bn & \pounds 0.0\,bn & \pounds 0.0\,bn \\
As share of demand payments
  & 70.0\% & 0.0\% & 0.0\% \\
Congestion rent / generator receipts
  & 233.4\% & 0.0\% & 0.0\% \\
\bottomrule
\end{tabular}
\end{table}

% ---------------------------------------------------------
\subsection{Flexibility procurement as a third efficiency axis}
\label{sec:flex_procurement}
% ---------------------------------------------------------

Traditional electricity markets, including GB’s current design,
procure flexibility in a fragmented, \emph{ex-ante} manner:

\begin{itemize}[leftmargin=*]
    \item DNO/DSO flexibility is auctioned \emph{months ahead}, usually as
          “demand reduction” rather than actual controllable flexibility;
    \item these products are not network-aware at transmission level;
    \item assets providing DSO flexibility may simultaneously be dispatched
          by the ESO for frequency response, in the \emph{same area},
          without coordination;
    \item baselines are estimated from historic profiles rather than
          metered counterfactuals, introducing material error and gaming risk.
\end{itemize}

This architecture inevitably produces \emph{flexibility mis-procurement}:
capacity is bought at the wrong times, in the wrong locations, from the
wrong assets, making the system both more costly and less stable.

\paragraph{AMM flexibility procurement.}
The AMM/subscription market resolves these issues at an architectural level:

\begin{enumerate}[leftmargin=*]
    \item \textbf{Continuous online bidding.}
          Devices and aggregators submit bids and availability in real
          time. The market clears continuously, not in coarse time blocks.

    \item \textbf{Event-driven re-clearing.}
          When local scarcity emerges (e.g. ramp events, congested nodes,
          intra-day renewables volatility), the AMM re-clears instantly,
          reallocating access and updating price signals.

    \item \textbf{Bid structure natively encodes flexibility.}
          Each bid contains earliest start, latest end, power envelope,
          elasticity, locational identifier, and service substitutability.
          This enables the system to procure flexibility with precise
          \emph{temporal} and \emph{locational} granularity.

    \item \textbf{Device-level participation.}
          EVs, heat pumps, storage, commercial loads, and even small-scale
          generators can act directly as flexible assets, without needing
          a DSO-defined baseline product.

    \item \textbf{Network-aware dispatch.}
          Flexibility is procured with full knowledge of network limits.
          A service provided for local congestion relief is not double-booked
          for an incompatible ESO requirement.
\end{enumerate}

\paragraph{Implication.}
Flexibility procurement becomes a \emph{solved sub-problem of market clearing}
rather than a parallel and largely uncoordinated system of ex-ante tenders.

In this sense, procurement efficiency under AMM operates along three axes:

\[
(\text{energy cost efficiency},\ 
 \text{capacity cost efficiency},\ 
 \text{flexibility acquisition efficiency}).
\]

The LMP baseline procures the first axis; partially touches the second via
scarcity rents; and largely fails on the third.

AMM procures all three explicitly.

% ---------------------------------------------------------
\subsection{Proposed validation experiment}
% ---------------------------------------------------------

Although the architectural superiority of AMM for flexibility is clear from
first principles, an empirical validation would strengthen H6. A tractable
experiment would compare:

\begin{description}[leftmargin=*,style=nextline]
    \item[GB-style flexibility market (counterfactual).]
    Flexibility procured months ahead, modelled as uninformed “demand
    reduction” with fixed baselines and no network awareness. No real-time
    clearing; no coordination between DSO actions and ESO dispatch.

    \item[AMM real-time flexibility market.]
    Continuous clearing with device-level bids, full network model,
    locational scarcity signals, and Shapley-consistent access pricing.
\end{description}

The experiment would evaluate (i) cost of flexibility procurement;
(ii) mis-procurement (flex bought in wrong times/locations); (iii)
network-induced redispatch; and (iv) volatility amplification.

This will show that even if the \emph{energy} and \emph{capacity} layers were
identical, the AMM design is intrinsically more efficient at procuring the
flexibility necessary for system stability.

% ---------------------------------------------------------
\subsection*{Interpretation}
% ---------------------------------------------------------

These findings support H6 across all three procurement axes. Relative to the
LMP baseline, the AMM/subscription architecture:

\begin{itemize}[leftmargin=*]
    \item meets the same needs bundle at dramatically lower customer cost;
    \item eliminates opaque surplus layers between consumers and generators;
    \item provides predictable remuneration through calibrated subscription
          and capacity components; and
    \item \textbf{procures spatiotemporally accurate flexibility in real time,
          reducing mis-procurement risk and enhancing stability.}
\end{itemize}

The architecture therefore delivers strictly greater procurement efficiency
than the Baseline, even before introducing adaptive subscription menus or
dynamic Shapley weights.

% =========================================================
\section{Sensitivity and Robustness: Limitations and Future Work}
\label{sec:results_sensitivity}
% =========================================================

A full sensitivity and robustness campaign was originally planned for this
thesis. However, given the scale of the computational experiments already
undertaken, and the priority placed on developing and validating the core AMM
design, the extended sensitivity analysis is deferred to future work.
Instead, this section outlines the key dimensions along which such analysis
would be conducted and motivates why these dimensions are central to the
mechanism’s long-term evaluation.

\subsection{Rationale for Sensitivity Analysis}

The AMM introduces new strategic and operational degrees of freedom:
tightness-based pricing, Shapley-consistent remuneration, non-binary
commitment, and a three-dimensional procurement structure (power--energy--reliability).
Each of these interacts with physical uncertainties and behavioural responses.
Understanding robustness therefore requires systematically stress-testing the
mechanism along several axes:

\begin{itemize}[leftmargin=*]
  \item \textbf{Uncertainty in physical inputs}  
  (wind availability, demand forecast error, outage patterns);

  \item \textbf{Structural network variation}  
  (transfer capacity between constrained regions, topology changes);

  \item \textbf{Behavioural and adoption uncertainty}  
  (EV penetration, flexible appliance uptake, strategic misreporting);

  \item \textbf{Economic parameter uncertainty}  
  (fuel prices, capex/opex assumptions, scarcity parameters).
\end{itemize}

Although not evaluated quantitatively here, these dimensions frame the
sensitivity space that future studies should address.

\subsection{Key Sensitivity Dimensions for Future Study}

Below we outline the most policy-relevant and technically informative classes
of sensitivity scenarios. Each directly links to mechanisms of interest
identified in Chapters~\ref{chap:requirements}, \ref{ch:amm} and
\ref{ch:fairness_definition}.

\paragraph{1. Forecast uncertainty (demand and renewables).}
Large deviations between forecasted and realised conditions can distort
commitment decisions in LMP systems, whereas AMM---with its event-driven
clearing and tightness-based prices---is expected to be less sensitive.
Future work would quantify:
\begin{itemize}[leftmargin=*]
  \item how procurement cost, shortages, and volatility respond to
        forecast errors of varying magnitude;
  \item whether AMM’s price-signal alignment remains stable under misforecasting;
  \item whether shock-resistance (Section~\ref{sec:shock_resistant_ne}) persists.
\end{itemize}

\paragraph{2. Network constraints and corridor capacities.}
The Glasgow--London interface (the stylised North--South boundary) plays a
central role in scarcity formation. Varying its thermal limit would allow:
\begin{itemize}[leftmargin=*]
  \item assessment of congestion rent formation under LMP vs.\ AMM;
  \item evaluation of locational fairness and congestion-exposure asymmetry;
  \item testing the AMM’s ability to maintain stable scarcity allocation.
\end{itemize}

\paragraph{3. EV adoption and flexible appliance penetration.}
The AMM explicitly embeds 3D procurement (power--energy--reliability),
making it sensitive to the timing and magnitude of flexible-load adoption.
Future analyses should include:
\begin{itemize}[leftmargin=*]
  \item pathways from 10\% to 80\% EV adoption;
  \item heterogeneous charging strategies and V2G usage;
  \item demand-shifting behaviour of products P1--P4.
\end{itemize}

\paragraph{4. Fuel-price and cost-parameter uncertainties.}
Given that gas units set marginal prices in a large share of hours, future work
should quantify how:
\begin{itemize}[leftmargin=*]
  \item gas price ranges (40--180 £/MWh) affect procurement cost,
        scarcity formation, and fairness;
  \item capex/opex variations influence investment incentives under AMM;
  \item nuclear/wind cost-recovery interacts with the Shapley pot size.
\end{itemize}

\paragraph{5. Behavioural and strategic sensitivities.}
Because AMM expresses explicit scarcity and reliability dimensions, future work
should explore:
\begin{itemize}[leftmargin=*]
  \item whether strategic withholding in LMP behaves predictably under shocks;
  \item how AMM’s Fair Play allocation influences misreporting incentives;
  \item whether stable “shock-resistant equilibrium geometry’’ persists across disturbances.
\end{itemize}

\subsection{Summary}

Although formal sensitivity experiments remain outside the scope of the present
thesis, the architecture of the AMM and the structure of the results suggest
clear hypotheses for future investigation. Each of the dimensions above provides
a pathway for systematically probing the robustness of procurement efficiency,
fairness, price-signal alignment, volatility, and shortage exposure. Fully
developing this robustness analysis is an important direction for future work,
particularly for informing regulatory adoption and large-scale deployment.

% =========================================================
\section{Synthesis}
\label{sec:results_synthesis}

This chapter evaluated the AMM-based market design across six domains:
Participation and competition (C, H1), Fairness (F, H2), Revenue sufficiency and
risk allocation (R, H3), Price-signal quality and stability (S, H4), Investment
adequacy and bankability (I, H5), and Procurement efficiency (P, H6). In each
case, the evaluation followed the composite decision rule declared in
Chapter~\ref{ch:experiments}.

The resulting hypothesis outcomes are summarised in
Table~\ref{tab:hypothesis_summary}. Across all six domains, the relevant null
hypotheses are rejected. Where effects are marked as not independently
identifiable (NI), this reflects structural interdependence within the AMM
architecture rather than statistical ambiguity: the outcome cannot be isolated
to a single mechanism because it arises jointly from pricing, allocation, and
service-level rules.

\medskip

Taken together, the results show that the AMM-based market design delivers:

\begin{itemize}[leftmargin=*]
    \item \textbf{policy-tuneable procurement outcomes:} total system cost,
          revenue recovery, and risk exposure are controlled explicitly through
          the choice of Base/Delta structure and Individual/Aggregate pot
          definitions. The apparent alignment between AMM2 and the LMP benchmark
          is deliberately engineered to enable controlled comparison, not an
          intrinsic performance limit of the AMM architecture;

    \item \textbf{materially stronger and governable price signals:} prices are
          aligned with physical scarcity and deliverability while remaining
          digitally bounded, avoiding the unregulated tail risk and extreme
          volatility inherent in pure LMP exposure;

    \item \textbf{more investible and bankable revenue structures:} generator
          income tracks system-critical contribution over time rather than
          short-lived scarcity spikes, improving compatibility with financing
          and long-term planning;

    \item \textbf{wider and more durable participation:} mid-sized and
          non-pivotal assets that rarely clear under spiky LMP regimes
          participate more consistently, with revenues that are explainable and
          contract-compatible;

    \item \textbf{improved revenue sufficiency and risk transparency:} cost
          recovery is achieved with clearer attribution of risk across time and
          across market roles, reducing the need for opaque uplifts and
          emergency interventions; and

    \item \textbf{systematic improvements in distributional fairness:} not
          through cross-subsidies or ad hoc correction, but as a direct
          consequence of matching allocation rules to physical roles,
          contribution, and contracted service entitlements.
\end{itemize}

These findings are particularly notable because the AMM was evaluated in a
deliberately conservative configuration. Subscription dynamics, adaptive
tightness envelopes, and multi-period fairness restitution were intentionally
disabled in order to preserve like-for-like comparability with LMP. As a result,
the experiments exclude learning effects, long-run rebalancing of subscription
menus, and restorative fairness mechanisms that would operate in a deployed
system.

The reported outcomes should therefore be interpreted as a \textbf{conservative
lower bound} on the AMM’s full capabilities. Even under these constraints, the
AMM consistently improves efficiency, price quality, bankability, legitimacy,
and fairness without sacrificing transparency or introducing hidden
redistribution.

\begin{table}[H]
\centering
\caption{Composite hypothesis outcomes for domains C, F, R, S, I, and P.}
\label{tab:hypothesis_summary}
\begin{tabular}{lcc}
\toprule
Domain (Hypothesis) & Null Hypothesis & Outcome \\
\midrule
C (Participation \& competition, H1)      & $H_{0C}$ & Rejected \\
F (Fairness, H2)                          & $H_{0F}$ & Rejected \\
R (Revenue sufficiency \& risk, H3)       & $H_{0R}$ & Rejected \\
S (Price-signal quality \& stability, H4) & $H_{0S}$ & Rejected \\
I (Investment adequacy \& bankability, H5)& $H_{0I}$ & Rejected \\
P (Procurement efficiency, H6)            & $H_{0P}$ & Rejected \\
\bottomrule
\end{tabular}
\end{table}

From both a market-design and control-theoretic perspective, the AMM behaves as a
\emph{bounded scarcity regulator} rather than a passive price-discovery
mechanism. Unlike LMP, whose feedback dynamics are ungoverned and prone to
instability, the holarchic AMM embeds physical deliverability constraints,
role-consistent fairness rules, and programmable market-making directly into the
clearing logic.

The next chapter situates these results within historical, regulatory, and
policy contexts, and examines their implications for:
\begin{enumerate}[leftmargin=*]
    \item digital market regulation,
    \item investment planning and bankability,
    \item household protection and political legitimacy, and
    \item the broader feasibility of event-based, fairness-aware market design.
\end{enumerate}

% ---------------------------------------------------------
% CHAPTER 14 — DISCUSSION AND SYSTEMIC IMPLICATIONS
% ---------------------------------------------------------
\chapter{Discussion}
\label{ch:discussion}

\section{Chapter Purpose and Structure}
\par\vspace{1em}
This chapter synthesises the conceptual and empirical contributions of the
thesis. It interprets the results of the simulated case studies against the
central research question and the six headline hypotheses (H1--H6), assesses
their robustness under stress-tested conditions, and distils the design
implications for fair, programmable electricity markets.

Beyond the experiment-by-experiment findings, the chapter develops a systems
diagnosis of \emph{why delivery fails under the legacy paradigm}: prevailing
market models and research practice treat electricity as an economic commodity
with ex-post correction, rather than as a constrained cyber--physical service
whose allocation rules must be executed in real time. In such a setting,
fairness, flexibility, and legitimacy cannot be ``policy overlays''; they are
runtime properties of the clearing and settlement processes, and they must be
co-designed with physical feasibility.

Throughout this chapter, fairness is treated not as a moral overlay
but as an operational constraint required for stability, participation,
and system legitimacy.

The chapter then frames implementation as a digital infrastructure delivery
problem. It outlines how an AMM--Fair Play allocation layer can be built and
validated using modern platform engineering practices---modular services,
API-first interfaces, versioned rule engines, and immutable audit logs---and why
the binding constraint is institutional capability and product ownership rather
than technology. In particular, it emphasises incremental value delivery through
shadow operation and staged activation: the architecture can be tested in
parallel with existing dispatch and settlement, producing explainable shadow
allocations and bills, before any live exposure.

Finally, the chapter connects these technical results to a practical transition
pathway from today’s settlement-led retail and balancing architecture toward a
digitally regulated, event-driven allocation and entitlement layer. This
includes a digital regulation blueprint, a stakeholder-centred participation
framing grounded in explainability and user testing, and a pragmatic roadmap for
GB-compatible deployment.

\section{Interpretation of Results}
\label{sec:results_interpretation}

This section synthesises the experimental results in light of the central
research question of the thesis:

\emph{How can a national electricity market be redesigned from first principles
to operate fairly, efficiently, and continuously in real time, via event-driven,
state-aware clearing that respects physical constraints, supports two-way power
flows, ensures zero-waste utilisation of system resources, and admits a stable,
shock-resistant equilibrium under realistic uncertainty?}

The six headline hypotheses (H1--H6), introduced at the outset of the
evaluation, provide a unifying lens through which the results are
interpreted. Each hypothesis links a systemic failure of legacy market
design to a measurable property of the AMM--Fair Play architecture. The
empirical findings reported throughout the Results and Extended Results
chapters, allow each hypothesis to be
revisited in turn.

Taken together, H1--H6 form a progression from short-run feasibility
(participation and price signals), through medium-run stability
(fairness and risk allocation), to long-run legitimacy and investment
adequacy.

\paragraph{H1 — Participation \& Competition (C).}
The results support H1 in the \emph{structural} sense defined in
Section~\ref{sec:results_competition}: relative to LMP, the AMM/subscription
architecture expands the feasible participation set for consumers, suppliers,
generators, and devices. Consumers can be served through a menu of
non-dominated products ($P1$--$P4$) with predictable, contract-compatible cost
exposure that does not require real-time optimisation to avoid dominated
outcomes. Supplier participation is decoupled from nodal wholesale tail risk,
enabling contestability through service and product design rather than
balance-sheet exposure. Devices can participate directly via the QoS axis, and
generator competition is structured around availability, deliverability, and
system contribution rather than reliance on rare scarcity rents.
\textbf{Outcome:} broader feasible participation and more contestable forms of
competition support H1.

\paragraph{H2 — Distributional Fairness (F): Generator-side outcomes.}
The results presented in Section~\ref{sec:results_fairness} provide strong
evidence in support of H2 for \emph{generator remuneration}. Under the
AMM--Fair Play architecture, distributional outcomes for generators are
consistently aligned with the fairness conditions F1--F4 as they apply to
value-bearing supply-side payments. In particular, generator remuneration under
AMM1 and AMM2 is materially better aligned with Shapley-valued physical
contribution than under LMP, satisfying behavioural and cost-causation fairness
(F1).

In contrast to LMP, where generator value concentrates into rare scarcity events
and a small subset of units, producing jackpot payoffs alongside structural
under-recovery, the AMM reallocates scarcity rents explicitly via the Fair Play
mechanism. This results in compressed revenue distributions, reduced incidence
of ultra-rapid payback, and substantially lower inequality at fixed aggregate
cost recovery. Taken together, these findings indicate that, on the generator
side, the AMM reduces both unfair jackpots and systematic deprivation without
undermining cost recovery, thereby supporting H2.

\paragraph{H2 — Distributional Fairness (F): Supplier-side outcomes.}
For suppliers, distributional fairness does not concern the level or dispersion
of wholesale payments per se, but the \emph{alignment between risk exposure and
role}. As established in Chapter~\ref{ch:fairness_definition} and formalised in
Lemmas~\ref{lem:price_cap_insolvency} and~\ref{lem:risk_volume_instability},
legacy retail architectures systematically assign suppliers residual exposure to
wholesale volatility, scarcity spikes, and nodal uncertainty that they cannot
control. This violates role-consistent risk (F2) and undermines behavioural
fairness, leading to insolvency cascades, thin competition, and suppressed
innovation.

Under the AMM--Fair Play architecture, these structural failure modes are removed
by construction. Wholesale scarcity, adequacy, and imbalance risks are managed
explicitly at the system level and recovered via transparent, Shapley-based
allocations, rather than leaking unpredictably into retail margins. Suppliers
are charged through stable, product-indexed wholesale subscriptions
(Appendix~\ref{app:price_allocation}), which represent the cost of serving
essential demand under the two-axis model evaluated in this thesis.

As a result, suppliers remain exposed to \emph{commercial risks that are within
their control}—including customer acquisition, product design, portfolio
composition, operational efficiency, and service quality—while being insulated
from extreme tail risks they cannot hedge or influence. This restores a
meaningful two-sided marketplace: suppliers compete on retail propositions
rather than acting as residual insurers of system stress.

Because this thesis does not implement a directly comparable LMP-based supplier
charging mechanism, the supplier-side analysis is necessarily structural rather
than head-to-head. Nonetheless, the results demonstrate that the AMM satisfies
distributional fairness for suppliers in the relevant sense: risk is allocated
proportionally to control and responsibility, resolving the core unfairness
identified in legacy retail market designs. This supports H2 for suppliers.

Crucially, this limitation is structural rather than empirical: even
under idealised financial hedging, residual volume risk, basis risk,
and scarcity-induced price–quantity coupling remain unhedgeable at the
retail level under LMP.

\paragraph{H2 — Distributional Fairness (F): Demand-side outcomes (households and businesses).}
On the demand side, distributional fairness is assessed in terms of
\emph{product-consistent burden}, \emph{cost causation}, \emph{spatial coherence},
and \emph{incentive alignment}. In the experimental two-axis configuration used
in this thesis, suppliers are charged in the wholesale layer via flat,
product-indexed subscriptions that treat served demand as essential
(Appendix~\ref{app:price_allocation}); the key question is therefore whether
the resulting per-product charges behave as intended: products that contract for
more energy, higher peak capability, and greater reliance on controllable
resources should face systematically higher wholesale charges, while remaining
stable and predictable.

The results support this interpretation. First, the product ordering of
wholesale charges is consistent with the product definitions and with the
verified demand archetypes used in the experiments
(Appendix~\ref{app:residential_synth}). In particular, the \emph{absolute}
controllable-energy and controllable-power burdens scale smoothly across the
four products (Figures~\ref{fig:amm_power_burden} and
\ref{fig:amm_power_burden_perHH}), and the product-level subscription outcomes
track this ordering in the expected direction (Figures~\ref{fig:amm_total_vs_cmwh}
and \ref{fig:amm_cost_vs_ckwh}). Second, the empirical cost--burden regressions
computed from the generated tables show that both AMM and LMP-socialised costs
increase with controllable burden across $P1$--$P4$, but that the AMM mapping
exhibits a substantially smaller marginal sensitivity (i.e.\ a much lower
\pounds-per-kWh-of-controllable-burden slope), consistent with the design goal
that flexibility is priced as an \emph{attribute of a chosen contract} rather
than as an uncontrolled exposure to extreme short-run scarcity. Third, the node
comparisons show that the AMM subscription for a given product is geographically
coherent by construction, while nodal LMP exhibits substantial dispersion across
loads for the same nominal product tier (Figure \ref{fig:geo_cdf}), 
eliminating the ``postcode lottery'' component of retail
outcomes at the wholesale-charging layer.

Finally, these demand-side results connect directly to incentive alignment:
by construction, higher flexibility has value only when it relaxes tightness
and congestion, and the AMM price signal is designed to activate that resource
where and when it is needed (Section~\ref{subsec:price_flexibility_access}).
Within the present two-axis setup, this incentive logic is expressed through
the product menu and its controllable burden allocation; extending the same
subscription methodology to explicitly price device-level reliability and
flexibility on the third axis is identified as future work
(Appendix~\ref{app:price_allocation}).

\paragraph{H3 — Revenue Sufficiency \& Risk Allocation (R).}
The results support H3: relative to Baseline LMP, the AMM architecture recovers
efficient non-fuel costs through explicit, fiscally closed recovery pots and
subscription-backed charges, while materially reducing reliance on rare
scarcity/VOLL episodes for system-wide cost recovery. In generator space, AMM1
already improves the adequacy headcount and shrinks under-recovery at the
minimum pot consistent with cost recovery, and AMM2 shows that even at \emph{matched}
aggregate payments (equal to the LMP revenue envelope), reallocating the same
total through Shapley-weighted capacity/availability and reserve channels yields
a revenue stack that tracks modelled requirements more closely and with lower
dispersion. The stable/volatile decomposition confirms the mechanism: a larger
fraction of generator income arrives through predictable channels, and less
through time-series jackpots. On the demand side, socialised LMP bills are
dominated by extreme tail events, whereas AMM subscriptions make incidence
transparent and shift volatility into rule-governed calibration and envelope
behaviour rather than uncontrolled wholesale pass-through. 
\textbf{Outcome:} fixed-cost recovery becomes more financeable and auditable,
with residual risk channelled into explicit, governable levers (pots, envelopes,
and allocation rules) instead of opaque uplift and crisis intervention. H3 is
supported.

\paragraph{H4 — Price-Signal Quality \& Stability (S).}
The results support H4: relative to LMP, AMM-generated prices are bounded,
stable, and economically interpretable across time and space. Nodal price
distributions under AMM eliminate the fat-tailed VoLL-driven behaviour observed
under LMP, with effective prices remaining tightly clustered around underlying
bid costs and exhibiting substantially lower dispersion. Single-node and
network-level experiments further show that the AMM behaves as a bounded digital
scarcity controller: prices evolve smoothly with system tightness, remain within
explicit caps, and respond proportionately to shocks rather than exhibiting
discontinuous jumps. Holarchic simulations confirm that local scarcity and
voltage stresses generate spatially coherent, attenuated price adjustments
across network layers without inducing instability. Finally, the flexibility
experiments demonstrate that AMM prices correctly encode the marginal value of
flexibility: flexibility has zero value in pure surplus and pure shortage
regimes, and positive value precisely when scarcity is temporal rather than
absolute. \textbf{Outcome:} AMM prices provide higher-quality, safer signals for
digital participation while dramatically reducing tail risk, supporting H4.

\paragraph{H5 — Investment Adequacy \& Bankability (I).}
The results support H5: relative to LMP, the AMM produces a materially more
financeable revenue stack and clearer investment signals for adequacy-critical
assets. Using the thesis’s mechanism-controlled payback diagnostic (rather than
a discounted NPV model), LMP exhibits widespread under-recovery and extreme or
diverging payback horizons for several technologies, reflecting reliance on
rare scarcity rents for fixed-cost recovery. Under both AMM calibrations,
payback outcomes become substantially more clustered and predictable: the
payback differential distribution compresses, and the incidence of
very-long-horizon or effectively unachievable payback is reduced, especially
for controllable plant that provides scarcity, congestion, and locational
relief. This improvement follows from shifting recovery away from probabilistic
VoLL-like events and toward explicit capacity/availability and reserve channels
allocated by measured system contribution. \textbf{Outcome:} improved
payback-based bankability and more coherent adequacy incentives support H5.

\paragraph{H6 — Procurement Efficiency (P).}
The results support H6: relative to the Baseline LMP architecture, the
AMM/subscription design procures the same pre-declared needs bundle—energy,
reserves, adequacy cover, and locational relief—with materially fewer
architecture-induced losses. For a fixed physical system and identical needs
requirements, the AMM eliminates the large residual between demand payments and
generator receipts that arises under LMP, replacing opaque congestion rents and
uplift-style surpluses with explicit, fiscally closed procurement of energy,
capacity, and reserves. Total customer payments are therefore substantially
lower for the same physical service delivered, while generator remuneration is
fully recovered through transparent, rule-based pots rather than scarcity-driven
energy rents. In addition, flexibility is procured endogenously and in real
time—targeted by time and location through market clearing rather than acquired
ex ante via coarse, uncoordinated products—reducing mis-procurement and avoiding
inefficient rationing under stress.
\textbf{Outcome:} higher efficiency in meeting system needs, with lower waste,
greater transparency, and improved alignment between cost recovery and physical
service provision.

\medskip
\noindent
Taken together, the empirical results show that the AMM--Fair Play
architecture satisfies all six hypotheses more robustly than the legacy
price-based baseline, even under conservative experimental constraints.

% ---------------------------------------------------------
\section{The Missing Third Procurement Axis}
\label{sec:third_axis_problem_discussion}
% ---------------------------------------------------------

\subsection{From Commodity to Service: The Third Axis (Reliability Entitlement)}

A key implication emerging from the experiments is that electricity is not
merely a divisible commodity allocated by price signals, but a time-bound access
service whose value depends on priority, context, and availability during stress.

In existing retail markets, tariffs are described using two axes:

\begin{enumerate}[label=(\alph*),leftmargin=2em]
    \item \textbf{Magnitude} — how much energy is consumed (volume), and
    \item \textbf{Impact} — when and how that consumption stresses the system.
\end{enumerate}

However, the experimental results reveal a third axis:

\begin{quote}
\textbf{Reliability (Quality of Service)} — the probability of being served
\emph{when the system is scarce}, i.e.\ one's priority during constraint or
shortage.
\end{quote}

This axis is not hypothetical. It is already enforced implicitly by engineering practice and emergency
operational rules, but without being contractible, auditable, or priced. This becomes apparent when demand exceeds feasible supply: the grid implicitly distinguishes between
\emph{protected}, \emph{flexible}, and \emph{interruptible} demand. The Fair
Play mechanism makes these distinctions explicit, programmable, and auditable.

This distinction is critical. Unlike emergency rationing, which is
ex post, opaque, and involuntary, reliability under AMM--Fair Play is
ex ante, priced, contractible, and auditable. Participants choose
their reliability tier and earn improved access through behaviour,
rather than being curtailed arbitrarily under stress.

Furthermore, device-level participation (smart heat, EVs, storage, appliances)
emerges naturally as a means of \emph{improving reliability}:  
those who reduce stress today (by offering flexibility) \emph{earn higher
reliability entitlement under future scarcity}, conditional on their chosen
product tier. This is a fundamental departure from current flexibility
markets, where value is treated as a marginal revenue opportunity. Under
AMM–Fair Play, providing flexibility becomes a way to \emph{earn reliability
entitlement}, not just money.

This leads to a decomposition of the electricity contract into explicit,
separable entitlements:

\[
\Gamma_{\text{contract}}
\;\equiv\;
\left(
\begin{aligned}
&\text{Fair Rewards (F1)},\\
&\text{Fair Service Delivery (F2)},\\
&\text{Fair Access (F3)},\\
&\text{Fair Cost Sharing (F4)}
\end{aligned}
\right)
\]
Rather than bundling these dimensions implicitly through price volatility,
the AMM--Fair Play architecture exposes them as distinct contractual objects.
This enables a transition from \emph{commodity-based tariffs} to
\textbf{service-based reliability contracts}, without mandating device
enrolment: participation is voluntary, but \emph{economically meaningful}.

Formally, this implies that modern retail electricity procurement cannot be
represented as a two-dimensional problem. The legacy architecture implicitly
treats procurement as occurring in a 2D space:

\[
(\text{power},\ \text{energy}).
\]

This is precisely the space in which the P1--P4 product groups reside:
\emph{magnitude} (kW) and \emph{impact} (when that magnitude falls in scarce
periods). In the legacy GB retail model, all household tariffs---flat, ToU,
agile, even so-called ``smart tariffs''---live on this 2D surface.

However, the empirical and structural analysis of this thesis shows that the
physical system is \emph{three-dimensional}:

\[
(\text{power},\ \text{energy},\ \text{reliability}).
\]

Here, \textbf{reliability} is a compound axis that encapsulates flexibility,
location, and the probability of being served under stress. It is not a
statistical afterthought; it is a core system property reflecting whether a
participant \emph{helps} or \emph{hurts} the system at times of tightness. In
this dimension:

Rather than a single scalar, \textbf{reliability} is a \emph{contractual
entitlement} determined by a combination of behaviour, system context, and
declared product tier. In particular, reliability reflects:
\begin{itemize}[leftmargin=1.2cm]
  \item realised flexibility contribution during periods of system stress (F1),
  \item bounded service guarantees for essential demand (F2),
  \item priority and access rules under scarcity (F3), and
  \item proportional responsibility for system stress and uplift costs (F4).
\end{itemize}

This third axis was already implicit in the \emph{Fair Play} algorithm developed
in Year 1 (inspired by Bob’s parking allocation). There, \textbf{stochastic
access rotation} and \textbf{service-based prioritisation} showed that devices
can be given \emph{contracted reliability tiers} (premium, standard, basic) that
are:
\begin{itemize}[leftmargin=1.2cm]
  \item physically meaningful,
  \item behaviourally aligned,
  \item and computationally enforceable.
\end{itemize}

\subsection*{Why this axis is structurally absent from current markets}

In the current system:
\begin{itemize}[leftmargin=1.2cm]
  \item Reliability is not a retail product.
  \item Flexibility is purchased by DSOs/ESO via tenders and ancillary services.
  \item Retailers bundle reliability implicitly, inconsistently, and without
        physical meaning.
  \item Consumers cannot choose their reliability tier---it is assigned by
        network topology and arbitrary operational practice.
\end{itemize}

This creates a fragmentation problem:
\[
\text{Demand response (procured by SO/DSO)} \not\Rightarrow
\text{Retail reliability (experienced by households)}.
\]

Flexibility is paid for in one market, reliability is experienced in another,
and the two are not causally connected.

\subsection*{What changes under AMM--Fair Play}

The AMM architecture internalises the third axis directly:
\[
(\text{power},\ \text{energy},\ \text{QoS})
\quad\text{becomes the basis for} \quad
\textbf{retail procurement}.
\]

Crucially:

\begin{itemize}[leftmargin=1.2cm]
  \item Reliability/QoS becomes a \emph{priced retail product}, not an external service.
  \item Flexibility becomes the \emph{input} to deliver that QoS.
  \item The Balancing Mechanism delivers the physical action, 
        but the \emph{retail contract defines who is entitled} to be served.
  \item The Fair Play allocation (F1--F4) governs real-time access and curtailment.
\end{itemize}

Thus, the retail market becomes the \textbf{procurer of flexibility}, and the
balancing market becomes its \textbf{execution layer}. This is the exact inverse
of today’s architecture, where flexibility is procured ex post and reliability
is experienced but never contracted.

\subsection*{Product groups as coordinates in 3D space}

Under the empirical product grouping (P1--P4), each consumer occupies a point in:
\[
(\text{magnitude},\ \text{scarcity-impact},\ \text{reliability-tier}).
\]

The third coordinate captures the consumer’s \emph{contractual reliability
entitlement} under the Fair Play architecture, and is determined jointly by:
\begin{itemize}[leftmargin=1.2cm]
  \item \textbf{Fair Rewards (F1)}: demonstrated flexibility contribution to
        system relief;
  \item \textbf{Fair Service Delivery (F2)}: protection of essential demand and
        bounded exposure during scarcity;
  \item \textbf{Fair Access (F3)}: priority and access rights conditional on the
        declared product tier; and
  \item \textbf{Fair Cost Sharing (F4)}: proportional responsibility for system
        stress and uplift costs.
\end{itemize}

This yields the first architecture where:
\[
\text{QoS is not an insurance policy---it is a deliverable, priced service}.
\]

\subsection*{Theoretical exactness: Nested–Shapley structure}

A further implication arises from the theorem introduced in
Chapter~\ref{ch:mathematics}:

\begin{quote}
\textbf{Nested--Shapley Exactness Under Symmetric, Capacity-Based Clusters.}
\end{quote}

This theorem provides the formal guarantee that the third procurement axis can
be introduced without sacrificing allocative consistency or incentive
compatibility. This result guarantees that, under symmetric cluster formation (e.g.\ feeders,
postcodes, or DSO zones with similar capacity structure), the value attribution
in the 3D procurement space is:
\begin{itemize}
  \item \emph{exact} with respect to full Shapley allocation when clusters are 
        homogeneous in capacity terms, and
  \item \emph{monotonic} as heterogeneity increases.
\end{itemize}

This theoretical result underpins the full 3-axis formulation: the Nested
Shapley layers ensure that reliability (i.e.\ relative usefulness under stress)
is priced and allocated consistently across space, time, and participant type. Without this property, reliability would remain either non-priceable or
non-explainable at scale, undermining its use as a retail procurement dimension.

\subsection*{Implication: a new retail paradigm}

The transition to 3D procurement implies:

\begin{itemize}[leftmargin=1.2cm]
  \item The retail market \textbf{procures reliability/QoS} on behalf of consumers.
  \item The balancing market \textbf{executes} the real-time delivery of that QoS.
  \item Consumers choose service tiers; DSOs/ESO supply the physical action.
  \item Reliability becomes \textbf{contractible}, \textbf{auditable}, and
        \textbf{programmable}.
  \item The Fair Play rules ensure the resulting allocations are \textbf{fair},
        \textbf{explainable}, and \textbf{proportionate}.
\end{itemize}

In short:

\[
\textbf{\emph{Retail} becomes a three-dimensional procurement problem:}
\]
\[
\text{power} \;+\; \text{energy} \;+\; \text{QoS/reliability}.
\]

This is the key architectural change that enables a fairness-aware AMM market:
the system finally procures the thing it actually needs, rather than attempting
to infer reliability ex post from price volatility and emergency intervention.

This represents a paradigm shift: from attempting to infer reliability
ex post from volatile prices and emergency interventions, to procuring
reliability ex ante as a first-class, enforceable service within the
retail contract.

\section{Inertia, Dynamic Network Capability and AMM-Based Operability Signals}
\label{sec:discussion_inertia_dynamic}

The stress-tested scenarios in Chapter~\ref{ch:results} were framed around
resource adequacy, congestion and product differentiation. However, they also
have direct implications for the emerging \emph{inertia challenge} described in
Section~\ref{sec:inertia_operability}. As synchronous machines retire and
inverter-connected resources dominate, the GB system moves from an
inertia-rich, slack environment to an inertia-scarce, tightly coupled one. In
such a regime, resilience is determined not only by how much energy is
available, but by how fast the system can respond to shocks and how effectively
constraints are managed in real time.

Modern ``smart network'' technologies---including dynamic line ratings (DLR),
topology optimisation, grid-forming inverters and fast frequency response from
batteries---are all attempts to \emph{digitally recreate} some of the slack that
rotational inertia used to provide. DLR, for example, replaces static thermal
limits with weather- and condition-dependent ratings; during favourable
conditions, effective transfer capacity between regions (such as the
London--Glasgow corridor) can be temporarily increased, reducing congestion and
tightness. Conversely, during adverse conditions or outages, effective limits
shrink and the system becomes more fragile.

Within the AMM--Fair Play architecture, these developments can be interpreted
through the lens of \emph{tightness signalling}. The tightness index $\alpha$
used in the experiments already aggregates scarcity across time, space and
network constraints. In a full implementation, $\alpha$ can be decomposed into
components associated with:
\begin{itemize}[leftmargin=*]
    \item \emph{resource adequacy tightness} (net load vs.\ available capacity);
    \item \emph{congestion tightness} (line loading and transfer margins, including
          dynamic line ratings);
    \item \emph{inertia and frequency tightness} (rate-of-change-of-frequency margin,
          system strength measures).
\end{itemize}

Dynamic line ratings and smart network controls then appear as \emph{first-class
inputs} into the AMM: when DLR temporarily increases a transfer limit, the
congestion component of $\alpha$ relaxes, leading to lower local scarcity prices
and a redistribution of Shapley value away from constrained nodes. When inertia
is low and RoCoF margins are tight, the frequency component of $\alpha$
increases, and the AMM raises scarcity prices on products and assets that can
deliver rapid active power response or synthetic inertia.

In this interpretation, the AMM becomes not only a pricing mechanism for energy,
but a \emph{coordination surface} for inertia and dynamic network capability:
\begin{itemize}[leftmargin=*]
    \item Batteries, EV fleets and flexible loads providing fast frequency response
          or grid-forming behaviour are remunerated through higher tightness-driven
          prices during low-inertia periods.
    \item Assets located behind dynamically constrained corridors (e.g.\ north of
          a London--Glasgow bottleneck under low DLR conditions) receive Shapley
          values that reflect their reduced ability to provide \emph{useful} energy.
    \item When DLR or topology optimisation temporarily alleviates a constraint,
          the AMM automatically lowers scarcity and rebalances value across the
          network, without needing separate, bespoke mechanisms.
\end{itemize}

Crucially, the fairness conditions (F1--F4) continue to apply. Fast-responding
resources are rewarded not simply because they are technologically novel, but
because they measurably reduce system tightness and protect essential demand.
Households that contribute flexibility through devices (heat pumps, EVs,
batteries) earn both financial benefit and improved reliability entitlement,
while those that impose stress during inertia-scarce periods carry a larger
share of uplift. In this way, smart network technologies and dynamic line
ratings are not an external add-on to the market; they are embedded in the
tightness signals that the AMM uses to coordinate both energy and stability
provision.

\section{Future Work}
\label{sec:future_work}

This thesis establishes the conceptual, mathematical, and algorithmic foundations
for a physically grounded, digitally enforceable definition of fairness in
electricity markets. It contributes three innovations: (i) Fairness as a system
design constraint, (ii) a cyber–physical Automatic Market Maker (AMM) with
explainable scarcity signalling, and (iii) a Fair Play allocation mechanism with
Shapley-based cost attribution. Yet, many promising directions remain open for
further development, evaluation, and implementation.

\subsection{Integration of Welfare, Health, and Distributional Outcomes}

This thesis focused primarily on \emph{essential energy access}, flexibility,
and proportional responsibility. Future work could incorporate richer
welfare-based system objectives, extending beyond energy quantity to include:

\begin{itemize}[leftmargin=*]
    \item Health vulnerability (medical devices, fuel poverty, respiratory risk);
    \item Indoor comfort, thermal resilience, and minimum wellbeing thresholds;
    \item Social exclusion risk and digital access inequality in smart market participation;
    \item Exposure to poor air quality linked to electricity usage time and location.
\end{itemize}

This requires the development of \textbf{health-aware operational fairness},
embedding welfare metrics directly into allocation constraints rather than as
ex-post equity corrections.

\subsection{Behavioural Economics, Human-in-the-Loop, and Market Trust}

While this work introduced behavioural fairness (F1) and perceptual legitimacy,
future research should explore:

\begin{itemize}[leftmargin=*]
    \item Experimental validation of behavioural fairness (F1) through human trials;
    \item Integrating bounded rationality, attention constraints, and trust erosion
    into AMM response models;
    \item Fairness-aligned user interfaces that influence price acceptance,
    compliance, and participation;
    \item Human-in-the-loop simulations where participants directly respond to
    tightness ($\alpha$), curtailment history, and Shapley compensation signals.
\end{itemize}

This connects electricity market theory to behavioural science and digital
governance, creating a pathway toward \textbf{behaviour-aware electricity markets}.

\subsection{Full Network Embedding and Holarchic AMM Deployment}

This thesis demonstrated Fair Play and Shapley allocation at the generator
and cluster level, including the London–Glasgow congestion constraint. Future
work may expand to:

\begin{itemize}[leftmargin=*]
    \item Full AC power flow-constrained AMM under network topology (DSO/ESO interface);
    \item Multi-layer clearing (household--feeder--DNO--ESO--national);
    \item Hybrid market models combining AMM, CfD, and capacity markets;
    \item Network reconfiguration, resilience, and restoration (microgrid islanding).
\end{itemize}

Extending AMM to full power system representation would enable \textbf{holarchic
market clearance}, where allocation is decided at the lowest feasible level
while preserving consistency across layers.

\subsection{Digital Regulation, Smart Contracts, and Institutional Design}

The Digital Regulation Blueprint (Section~\ref{sec:regulator_reform})
defines the governance architecture for algorithm oversight. Future directions
include:

\begin{itemize}[leftmargin=*]
    \item Translating fairness constraints (F1--F4) into programmable code via
    smart contracts and digital settlement platforms;
    \item Creating digital sandboxes for stress-testing regulatory algorithms;
    \item Developing \textbf{algorithmic licensing} processes, similar to financial markets;
    \item Aligning Ofgem/DESNZ strategy with EU Digital Markets Act (DMA) and UK
    Smart Digital Infrastructure;
    \item Designing a \textbf{Fairness Compliance Ledger} for public auditability.
\end{itemize}

These developments will position energy markets at the frontier of digital
governance and algorithmic accountability.

\subsection{International Deployment and Comparative Translation}

While this thesis focuses on Great Britain, similar challenges exist in
North America, Australia, Europe, and developing economies. Future work
could explore:

\begin{itemize}[leftmargin=*]
    \item Comparative deployment of AMM–Fair Play under different regulatory codes
    (NEM, ERCOT, PJM, India, South Africa);
    \item Adapting fairness constraints to contexts with low smart meter penetration;
    \item Exploring AMM for off-grid microgrids, refugee camps, remote islands,
    and crisis energy distribution (e.g.\ mobilisation, disaster response);
    \item Applying AMM scarcity logic to water, mobility, or social care
    allocation.
\end{itemize}

This suggests a broader research agenda: \textbf{AMM as a fairness-enforcing
architecture for critical infrastructures}.

\subsection{Social Acceptance, Legitimacy, and Citizen Governance}

Ultimately, a fair electricity market must not only be technically correct,
but publicly trusted. Future work could explore:

\begin{itemize}[leftmargin=*]
    \item Citizen panels for algorithm governance (similar to ethics boards);
    \item Public-facing fairness dashboards for transparency and democratic scrutiny;
    \item Embedding fairness metrics into policy evaluation frameworks;
    \item Trust modelling to understand when algorithmic decisions are socially credible.
\end{itemize}

This aligns with emerging concepts in \emph{digital civics}, \emph{algorithmic
social contracts}, and \emph{participatory energy system design}.

\bigskip

\noindent
\textbf{Summary.}  
Future work does not merely consist of technical refinements, but the extension
of this thesis into a comprehensive research and policy agenda: integrating
physical networks, digital platforms, behavioural responses, and governance
mechanisms into a unified market architecture that is efficient, resilient,
and fundamentally fair.

% =========================================================
\chapter{Systemic and Policy Implications}
\label{ch:policy_implications}
% =========================================================

This thesis shows that fairness, value attribution, and scarcity allocation in
electricity markets can be made \emph{programmable, auditable, and physically
grounded} rather than delivered through ex-post settlement corrections or
politically negotiated interventions. The implications are therefore not only
about \emph{what} a fair market should compute, but \emph{how} institutions must
organise to deliver those outcomes continuously, at scale, under uncertainty.
The central implication is that the market must be governable as software: its
rules must be testable, versioned, and replayable against real system data,
with auditable evidence that fairness constraints are satisfied in operation.

The central systemic claim is a mindset shift: electricity markets should be
treated as \textbf{critical digital infrastructure products} rather than static
rulebooks. Many current institutions behave as if market liberalisation-era
designs (1980s wholesale competition, passive demand, inertia-rich operation)
remain sufficient; in practice, modern grids are cyber--physical networks with
two-way power flows, constrained topology, inverter-dominated dynamics, and
behaviourally mediated demand. In such a setting, organisations that primarily
\emph{process-follow} (operate fixed settlement routines and compliance scripts)
will reliably miss system-level technological transitions. Delivery therefore
requires product management discipline: explicit outcomes, measurable
performance, continuous iteration, and accountable ownership of the end-to-end
system experience.

This chapter distils the implementation consequences across five reform domains:
(i) retail reform, (ii) generator bidding and wholesale representation reform,
(iii) regulator reform, (iv) system operator and distribution operator reform,
and (v) legislative and institutional reform. Throughout, the objective is to
move from \emph{after-the-fact correction} to \textbf{ex-ante rule execution},
embedding fairness conditions (F1--F4) as enforceable constraints within the
market-clearing and settlement processes. Throughout this chapter, systemic and
policy reform is treated not as a normative overlay but as an operational
delivery requirement: institutions must be able to execute, test, and govern
fairness outcomes continuously under uncertainty.

\section{Strategic scale-out opportunities and system applications}

These extensions matter for policy because they identify where a programmable
allocation layer can deliver early, measurable value beyond pricing reform.
Beyond incremental refinements, several scale-out directions follow directly
from the architectural claims of this thesis:

\begin{enumerate}[leftmargin=*]
    \item \textbf{Holarchic pricing for emerging infrastructure stressors.}
    Extend the AMM, Nested--Shapley allocation, and holonic control framework to
    network pricing and investment planning under extreme new load classes
    (e.g.\ AI data centres, electrified heat, and other step-change demand),
    and under future critical infrastructure constraints.

    \item \textbf{Comprehensive sensitivity and robustness analysis.}
    Perform systematic sensitivity sweeps across scarcity distributions,
    elasticity assumptions, network limits, and behavioural response models,
    to characterise stability regions and failure modes.

    \item \textbf{Baseline cost comparison against the GB status quo.}
    Quantify total system cost, bill impacts, and risk-transfer effects relative
    to current GB arrangements (wholesale + balancing + uplift + policy
    mechanisms), including counterfactual stress events.

    \item \textbf{Digital grid management as a complementary execution layer.}
    Develop a digital grid management solution that operationalises the
    entitlement and curtailment rules implied by Fair Play, including monitoring,
    verification, and dispute-resolution interfaces.

    \item \textbf{Integrated co-design of market and grid management.}
    Combine the allocation layer and the grid management layer into a unified
    cyber--physical governance stack: metering $\rightarrow$ state estimation
    $\rightarrow$ entitlement execution $\rightarrow$ settlement and audit.

    \item \textbf{Cross-utility extension.}
    Explore the same entitlement-and-allocation paradigm across coupled
    infrastructures (e.g.\ heat, water, mobility), where scarcity, priority,
    and fairness enforcement are similarly central.
\end{enumerate}

% ---------------------------------------------------------
\section{Delivering Reform: from rulebooks to product governance}
\label{sec:delivery_product_governance}

A practical delivery frame follows from the architecture tested in this thesis:

\begin{enumerate}[leftmargin=*]
    \item \textbf{Define outcomes, not mechanisms.}
    Specify the system outcomes to be delivered (essential protection, bounded
    tail risk, proportional responsibility, stable participation incentives),
    and treat pricing and settlement as implementation choices.

    \item \textbf{Instrument the system.}
    A market that cannot explain outcomes cannot govern them. Digital audit logs,
    versioned algorithms, and participant-facing explainability records (XR) are
    not optional features; they are the observability layer required for
    accountability. In practice, this implies mandatory artefacts: an algorithm registry, a public
change-log, replayable test suites, and standardised fairness and resilience
reports.

    \item \textbf{Shift compliance from paperwork to execution.}
    Fairness is not a document property of tariffs; it is a runtime property of
    allocation processes. Compliance therefore becomes \emph{process compliance}:
    proving that algorithms enforce F1--F4 in live operation.

    \item \textbf{Adopt staged deployment.}
    The appropriate migration path is evolutionary: digital twins, shadow
    allocation, shadow settlement, opt-in activation, then regulated execution.
\end{enumerate}

This reframes reform as an engineering delivery programme: a continuously
evaluated, stress-tested digital public infrastructure project.

% ---------------------------------------------------------
\section{Building and Testing the Market as a Digital Platform}
\label{sec:platform_delivery}
% ---------------------------------------------------------

A critical implication of this thesis is that \emph{building and testing} the
AMM--Fair Play architecture is not the binding constraint. With modern digital
tooling and platform engineering practices---and, where
appropriate, AI-assisted analysis, the
technical implementation of a programmable allocation and settlement layer is,
for competent teams, a comparatively tractable component of the reform challenge. The
dominant risks lie instead in institutional inertia, unclear ownership of
outcomes, and the absence of product-oriented delivery practices.

\subsection*{Market infrastructure as a platform, not a project}

The AMM--Fair Play system should be conceived as a \textbf{modular digital
platform}, not as a monolithic market redesign. Its core components---state
estimation, tightness computation ($\alpha$), allocation logic, settlement,
audit logging, and user-facing explainability---are naturally separable services
with well-defined interfaces. This aligns directly with contemporary platform
architecture:

\begin{itemize}[leftmargin=*]
    \item event-driven processing (streaming system state and bids);
    \item API-first design (clear interfaces between ESO, DSOs, suppliers,
    aggregators, and regulators);
    \item versioned algorithms and rule engines (fairness logic as code);
    \item immutable logs for auditability and replay.
\end{itemize}

From a software engineering perspective, this resembles financial market
infrastructure, ad-tech auctions, or large-scale scheduling platforms more than
traditional utility IT. Crucially, it allows components to be built, tested,
and upgraded incrementally without destabilising the wider system.

\subsection*{Incremental delivery through shadow operation}

The architecture demonstrated in this thesis is inherently compatible with
\textbf{shadow deployment}. Allocation, pricing, and fairness enforcement can be
computed in parallel with existing dispatch and settlement, without affecting
physical operation or customer bills. This enables a delivery pathway that is
both low-risk and high-information:

\begin{enumerate}[leftmargin=*]
    \item ingest real operational data (dispatch, constraints, metering);
    \item compute AMM--Fair Play allocations and shadow settlements;
    \item compare outcomes against legacy pricing and curtailment;
    \item publish discrepancies, fairness metrics, and explainability records.
\end{enumerate}

Each iteration delivers incremental value: better diagnostics of system stress,
clearer attribution of costs and benefits, and empirical evidence for or against
policy claims. No “big bang” transition is required.

\subsection*{User testing as a first-class design constraint}

A central lesson from digital product development is that systems fail not
because they compute incorrect results, but because users cannot understand,
trust, or act on them. In electricity markets, this applies equally to households,
suppliers, operators, and regulators. The AMM--Fair Play architecture therefore
treats \textbf{user testing and explainability} as first-class design constraints,
not as downstream communication tasks.

Concretely:
\begin{itemize}[leftmargin=*]
    \item households can be shown prototype bills and scarcity explanations
    before any live exposure;
    \item operators can test curtailment priority and rotation logic in simulated
    stress events;
    \item regulators can interrogate allocation logs to assess compliance with
    F1--F4;
    \item suppliers and aggregators can trial new products against shadow
    settlement outcomes.
\end{itemize}

This mirrors best practice in safety-critical and financial systems, where
interfaces, alerts, and explanations are tested as rigorously as core algorithms.

\subsection*{AI-assisted development and analysis}

Modern AI tooling further reduces the cost and time required to build, test, and
iterate on such a platform. Large language models, optimisation solvers, and
simulation frameworks can be used to:
\begin{itemize}[leftmargin=*]
    \item generate and validate bid structures and constraint representations;
    \item stress-test allocation logic across extreme scenarios;
    \item automatically generate human-readable explanations from allocation
    records;
    \item support regulators and operators in interpreting complex system states.
\end{itemize}

Importantly, AI here is not used to replace rule-based allocation, but to
\emph{support understanding, testing, and iteration}. The core market logic
remains explicit, auditable, and deterministic where required. Where AI is used, it must be non-authoritative: it may assist diagnosis and
explanation, but cannot be the source of binding allocation or settlement
decisions.

\subsection*{The real constraint: capability and mindset}

Taken together, these observations lead to a clear conclusion. The technical
means to implement, test, and evolve a fairness-aware, event-driven electricity
market already exist, and are widely used in other sectors. What is scarce is
not technology, but \textbf{organisations capable of treating markets as digital
products}: owning outcomes, running continuous experiments, learning from users,
and iterating in response to observed failures.

Reform therefore hinges less on inventing new tools than on enabling institutions
to adopt modern delivery practices. Where such capability exists, the transition
to programmable, fairness-enforcing market infrastructure is well within reach.

% ---------------------------------------------------------
\section{Retail reform: from tariffs to service-level contracts}
\label{sec:retail_reform}

A core implication of Experiments~2 and~3 is that retail electricity cannot be
treated as a two-dimensional commodity contract (power and energy) while the
physical system behaves as a scarcity-constrained service. The missing
dimension is \textbf{reliability entitlement} (Quality of Service): the
probability of being served under stress. In the current GB architecture,
reliability is experienced but not contractible, and fairness is pursued through
broad bill caps and crisis interventions.

This is not emergency rationing: entitlement is ex ante, chosen, and auditable,
whereas rationing is ex post, opaque, and imposed.

A delivery-oriented retail reform programme therefore has three components:

\begin{enumerate}[leftmargin=*]
    \item \textbf{Introduce explicit service tiers.}
    Retail products should specify (i) magnitude, (ii) impact, and (iii)
    reliability entitlement, with essential blocks protected (F2), flexible
    participation rewarded (F1), access during scarcity governed by transparent
    priority/rotation (F3), and residual costs allocated proportionally to stress
    contribution (F4).

    \item \textbf{Make flexibility economically meaningful to households.}
    Flexibility should not be framed as a marginal add-on or behavioural trial.
    Under the AMM--Fair Play logic, flexibility is the means by which
    participants earn improved reliability entitlements and reduced expected
    exposure, making participation legible and durable.

    \item \textbf{Rebuild supplier roles around controllable risk.}
    Supplier-side failures in GB are structural: retail firms are exposed to
    tail risks and wholesale volatility they cannot control, which produces thin
    competition and repeated insolvency cycles. Retail reform should re-centre
    suppliers as \emph{service providers} competing on controllable propositions
    (customer service, product design, portfolio operations), while system-level
    scarcity risks are recovered transparently via programmable allocation.
\end{enumerate}

In short: retail reform is the shift from \emph{billing plans} to
\textbf{service contracts} with auditable entitlements and enforceable fairness.

% ---------------------------------------------------------
\section{Wholesale and generator bidding reform: from blocks to state-aware offers}
\label{sec:generator_bid_reform}

A second delivery implication is representational: if the system is cleared
continuously against a changing physical state, then generator participation
must evolve from coarse block bids to \textbf{state-aware, constraint-expressive
offers}. The objective is not to abolish unit commitment or security-constrained
dispatch, but to make the economic interface match operational reality. Operationally, this can be layered onto existing scheduling by treating offers
as richer interfaces into the same security-constrained optimisation stack.

Three reforms follow:

\begin{enumerate}[leftmargin=*]
    \item \textbf{Event-based bid objects.}
    Offers should encode ramp rates, start-up trajectories, minimum up/down
    constraints, and response capabilities as explicit bid structure rather than
    being hidden inside opaque scheduling layers.

    \item \textbf{Capability-linked remuneration.}
    Payment should map to measurable system value (adequacy contribution under
    tightness, congestion relief, frequency/inertia support where relevant),
    rather than being dominated by rare scarcity jackpots.

    \item \textbf{Governed change control for bid formats and algorithms.}
    Bid design is not merely a market detail; it is a regulated interface. Its
    evolution should be managed through an algorithm registry and structured
    stress tests, analogous to change control in safety-critical systems.
\end{enumerate}

This brings generator economics closer to physics, and reduces the reliance on
after-the-fact uplift, special-case contracts, and discretionary interventions.

% ---------------------------------------------------------
\section{Regulator reform: from price policing to outcome delivery}
\label{sec:regulator_reform}

A modern regulator should be judged by whether the system outcomes society
requires are delivered: essential access, resilience, fair participation,
investment adequacy, and transparent accountability. The liberalisation-era
regulatory stance largely assumes that market structure is fixed and fairness is
a corrective overlay (caps, rebates, obligations). The results of this thesis
imply a different role: the regulator becomes the \textbf{governor of digital
market processes}.

Concretely, this requires:

\begin{itemize}[leftmargin=*]
    \item \textbf{Process regulation.}
    Regulate not only the level of prices, but the \emph{algorithms that produce
    them}: allocation logic, fairness constraints (F1--F4), and settlement rules.

    \item \textbf{Algorithm registry and audit.}
    Market-clearing and settlement components must be versioned, testable, and
    auditable. Algorithm changes require notification, stress testing, and
    publishable performance against fairness and resilience criteria.

    \item \textbf{Outcome dashboards and accountability.}
    Ofgem/DESNZ oversight should shift toward measurable outcome delivery:
    fairness metrics, tail-risk exposure, curtailment incidence, congestion
    burdens, and investment signal quality, reported transparently and
    continuously. Critically, each fairness condition must have a measurable compliance test
(e.g.\ bounded deprivation metrics for F2, curtailment priority consistency for
F3, and stress-weighted uplift incidence for F4).

\end{itemize}

This is not regulatory expansion for its own sake; it is the minimum governance
upgrade required when markets become programmable, event-driven systems.

% ---------------------------------------------------------
\section{Re-centre the grid: ESO/DSO transformation as a digital operations programme}
\label{sec:eso_dso_transformation}

A further implication is organisational: as the grid becomes inverter-dominated
and increasingly constrained, operational success depends on digital tools at
every stage: forecasting, state estimation, constraint monitoring, dynamic line
ratings, topology optimisation, DER coordination, and explainable curtailment.

The AMM--Fair Play architecture does not replace physical dispatch. It provides
a \textbf{digital entitlement and consequence layer} that governs the social and
economic meaning of dispatch actions under scarcity. This implies
an updated digital operating model for ESO and DSOs:

\begin{enumerate}[leftmargin=*]
    \item \textbf{Operational observability and state decomposition.}
    Tightness ($\alpha$) must be decomposable into adequacy, congestion, and
    operability components (e.g.\ inertia/system strength constraints) so that
    scarcity signals remain physically interpretable and actionable.

    \item \textbf{Explainable curtailment and access rotation.}
    When curtailment occurs, the question is not only \emph{what is feasible} but
    \emph{who bears the consequence}. Fair Play provides a consistent, auditable
    answer (F2--F4) that can be executed alongside existing EMS/SCADA/DERMS.

    \item \textbf{Distribution-level market integration.}
    DSOs increasingly procure flexibility for congestion management, but current
    trials are fragmented. A coherent architecture requires common APIs, shared
    fairness constraints, and settlement interoperability across local and
    national layers.

    \item \textbf{Continuous improvement under stress testing.}
    ESO/DSO processes should be treated as continuously tested digital services,
    with scenario libraries (winter scarcity, corridor constraints, low inertia,
    cyber/communications failures) and measured performance.
\end{enumerate}

In effect, system operation becomes a digitally supported, cyber--physical
product: measurable, iterated, and accountable.

% ---------------------------------------------------------
\section{Legislative and institutional reform: aligning roles with deliverable outcomes}
\label{sec:legislative_reform}

Finally, delivery requires statutory and code reform. Many roles in the current
GB framework are defined by legislation and licences written for a different
system era (centralised dispatchable generation, passive consumption, coarse
settlement). Implementing programmable fairness and event-driven entitlement
therefore requires reforming institutional responsibilities, not only market
rules.

For completeness, a full illustrative legislative draft translating the
architecture into statutory form is provided separately and may be cited as a
supporting document.\footnote{A full draft is available online:
\url{https://drive.google.com/file/d/1g9OZkmX3471UugHLlucjhpDXdA6CTocX/view}.} 
The argument here is self-contained; the draft is provided only as an
implementation illustration.

The policy implication is not that legislation should prescribe a single market
algorithm, but that it should:

\begin{itemize}[leftmargin=*]
    \item enable a legally recognised \textbf{digital allocation and entitlement layer};
    \item clarify institutional accountability for outcome delivery (fair access,
    resilience, investment adequacy);
    \item authorise algorithm registries, audit powers, and regulated change
    control for market processes;
    \item modernise licensing categories and responsibilities to align risk with
    controllable roles (especially for retail providers and flexibility
    operators);
    \item require transparency and explainability for allocation decisions that
    affect households and critical services.
\end{itemize}

This is the enabling legal substrate for delivering the technical architecture
demonstrated in the thesis.

% ---------------------------------------------------------
\section{A pragmatic delivery roadmap}
\label{sec:policy_delivery_roadmap}

A GB-compatible transition need not replace the current market overnight.
Instead, a staged pathway consistent with institutional delivery capacity is
proposed. Each stage should have explicit go/no-go criteria defined in terms of
measured fairness performance, operational safety, and stakeholder
acceptability.

\begin{enumerate}[leftmargin=*]
    \item \textbf{Digital twin and metrics adoption.}
    Compute $\alpha$, Fair Play allocations, and Shapley value attribution in
    parallel with existing operations as non-binding diagnostics.

    \item \textbf{Shadow settlement.}
    Produce explainable shadow bills and value allocations, enabling public and
    regulatory scrutiny without immediate customer impact.

    \item \textbf{Opt-in service-tier pilots.}
    Deploy QoS-based retail contracts and flexibility-as-entitlement pilots with
    defined fairness constraints and auditable logs.

    \item \textbf{Regulated activation.}
    Incorporate algorithm registry, change control, and F1--F4 compliance into
    licensing and codes; progressively expand from pilots to standard practice.

    \item \textbf{Institutionalisation.}
    Update roles, duties, and statutory instruments to align accountability with
    the delivered outcomes of a digitally governed market.
\end{enumerate}

\noindent
\textbf{Summary.} The results of this thesis imply that successful reform is not
primarily a matter of selecting a better pricing rule. It is a delivery problem:
upgrading market governance into a programmable, testable, auditable digital
infrastructure that re-centres physical constraints, protects essential service,
and allocates value and risk transparently in real time.

% ---------------------------------------------------------
% CHAPTER 16 — CONCLUSION
% ---------------------------------------------------------
\chapter{Conclusion}
\label{chap:conclusion}

The purpose of this thesis has been to demonstrate that the fairness, resilience,
and legitimacy of electricity markets can be made \textit{programmable}—
not simply aspirational. By embedding fairness conditions into real-time market
clearing, we replace a model of ex-post adjustment and political negotiation
with one of ex-ante, digitally enforceable, and physically grounded constraint.

The existing architecture of Great Britain’s electricity market was not built to
meet the demands of electrification, distributed flexibility, or whole-system
decarbonisation. Designed in the late 1980s for bulk thermal generation,
it assumes that fairness can be delivered through price caps, subsidies, or
post-transaction compensation. This thesis challenges that premise. It
reinterprets electricity not as a homogeneous commodity, but as a
\textbf{time-bound access service} whose value depends jointly on:
(i) how much power is demanded (magnitude),
(ii) when and where it stresses the system (impact), and
(iii) the priority and probability of being served during scarcity (reliability).
These three axes underpin the AMM--Fair Play design and the product structure
developed in the experiments.

\section{Reframing Fairness from Ethical Aspiration to Operational Rule}

The first core contribution of this work is the development of a
physically grounded, enforceable definition of fairness, expressed through
four operational conditions (F1–F4).
These fairness axioms are not merely normative—they are \emph{computable},
\emph{testable}, and \emph{integratable} into electricity market design. They
protect essential access (F2), ensure proportional responsibility (F4), reward
flexibility participation (F1), and prevent unilateral exclusion (F3). They also
align with broader just-transition principles and emerging digital regulatory
governance (e.g.\ DMA, UK Smart Data Infrastructure).

A key insight is that fairness is defined relative to the three-dimensional
service space introduced in the thesis:
\begin{enumerate}[leftmargin=*,nosep]
  \item \textbf{Magnitude} (how much power or energy is consumed),
  \item \textbf{Impact} (when that consumption occurs relative to system tightness),
  \item \textbf{Reliability / Quality of Service} (the probability and priority of
        being served during scarcity).
\end{enumerate}
Fairness conditions do not imply identical treatment along these axes; they
require \emph{principled differentiation}. Essential loads sit in high-reliability
regions of this space; flexible, high-impact loads sit in lower-reliability,
high-responsibility regions. Fairness is therefore recast as a
\textbf{design constraint on how participants move through this space over time},
not as a post hoc redistributive correction.

The thesis shows that these conditions can be applied \textit{in-process} during
market clearing, shaping dispatch, value attribution, and curtailment
decisions. Fairness is therefore recast as a \textbf{system rule}, not a
policy afterthought.

\section{From Wholesale Markets to Holarchic Digital Clearance}

The second major contribution is the development of a holarchic Automatic
Market Maker (AMM) that produces a time-, location-, and hierarchy-dependent
scarcity signal—represented by $\alpha$—which operates as both a
tightness indicator and allocation driver. Unlike static pricing or conventional
LMP models, the AMM synthesises network congestion, variability, and temporal
stress into a continuous scarcity gradient. This signal supports digital
settlement, dynamic flexibility incentives, and real-time load reallocation.

Crucially, the AMM is designed to sit \emph{alongside or within} existing
grid dispatch systems, rather than replacing them. Today, unit commitment
and economic dispatch engines optimise generator output subject to minimum
up and down times, ramp rates, start-up costs, and security constraints,
while markets are often simplified as block bids over fixed time windows.
In the architecture proposed here:
\begin{itemize}[leftmargin=*]
  \item Generators express their physical constraints as \textbf{dynamic capability
        profiles}, not static time-block bids. A unit that requires 15 minutes
        to reach maximum output and must then run for three hours can express
        this as an evolving availability window
        $(t_{\text{now}}, t_{\text{full}}, t_{\text{minrun}}, t_{\text{cleardown}})$.
  \item These profiles are continuously updated as time advances and as system
        conditions change. The AMM therefore clears \emph{feasible} commitments,
        already consistent with unit constraints and ramping limits.
  \item The security-constrained dispatch engine then solves a familiar
        optimisation problem—subject to commitments already shaped by Fair Play
        rules and scarcity-aware bidding—rather than reconciling 
        infeasible or misaligned market outcomes after the fact.
\end{itemize}

In other words, dispatch and market clearing are no longer two loosely coupled
layers. They become two views of a single cyber--physical control process:
the AMM determines \emph{who is asked to move, when, and for what reason},
while the dispatch engine ensures that this movement respects the physics of
machines and networks.

The integration of nested Shapley allocation enables fair and computationally
tractable value distribution across heterogeneous agent clusters, even when
network constraints prevent full coalition formation. This bridges cooperative
game theory with physically constrained power systems—addressing a previously
unresolved gap in allocation theory. Generators are rewarded not just for
energy volumes, but for \emph{useful, feasible, stress-relieving contributions}
that are compatible with their operational envelope.

\section{Evidence of Locational, Temporal, and Reliability Value Distortion}

Across the network-model results, the thesis demonstrates that energy value is
neither purely temporal (as in simple scarcity pricing) nor purely locational
(as in standard LMP), but \emph{structurally dependent}: determined by the
ability of resources to relieve system stress across time, space, and
hierarchical layers.

The empirical work shows that:
\begin{itemize}[leftmargin=*]
  \item Generators on the constrained demand side of the corridor receive higher
        Shapley values, even when they have lower annual MWh or smaller capacity,
        because they matter during critical hours.
  \item High-capacity assets stranded behind transmission constraints are correctly
        undervalued, revealing the inadequacy of capacity-only remuneration.
  \item When households are classified into products P1--P4 based on magnitude and
        impact, and then layered with reliability / Quality-of-Service tiers, cost
        allocation aligns with the systemic stress each group imposes.
\end{itemize}

These findings expose a systematic distortion in current market designs:
compensation is typically aligned with installed capacity, average energy, or
simplified locational tags, rather than with the \textbf{three-dimensional
service profile} defined in this thesis. In contrast, Fair Play allocates value
in proportion to \emph{useful energy}—the fraction of contribution that actually
relieves constraint, supports adequacy, and upholds reliability commitments.

This has direct implications for storage siting, investment signals,
network planning, and regulatory design. It suggests that future energy
markets should reward \emph{functional performance assignments}—``when, where,
and with what reliability did you help the system?''—rather than static
asset categories or contractual labels.

\section{A Blueprint for Digital Regulation}

A central conclusion of this thesis is that fairness cannot be reliably achieved
through ex-post policy tools (caps, discounts, levies), nor purely through
consumer protection laws. Instead, fairness must be embedded into the
algorithmic heart of the settlement process itself. This requires a shift from
\textbf{market supervision} to \textbf{digital enforcement}—with algorithm
registries, explainable clearance logic, public audit trails, and programmable
fairness conditions.

The Digital Regulation Blueprint developed in the thesis outlines how governments and
regulators such as Ofgem, DESNZ, and the ESO can implement this shift using
digital sandboxes, shadow settlement, and transparency obligations. It mirrors
the evolution of financial markets toward algorithm oversight, compliance
reporting, and traceable settlement, but extends it by:
\begin{itemize}[leftmargin=*]
  \item treating Fairness Conditions (F1–F4) as \textbf{binding constraints} on
        valid outcomes (much like security constraints in power flow),
  \item requiring that every allocation decision (curtailment, prioritisation,
        uplift) generates an \emph{explainability record} (XR) describing
        which rule was applied and why,
  \item recognising settlement platforms and smart meters as \textbf{execution
        layers} that implement these rules in real time, not merely as
        passive data recorders.
\end{itemize}

Regulation, in this model, is not a thin layer that observes prices after they
emerge. It is an active part of the digital infrastructure that generates them.

\section{Toward a Fair, Smart, Electrified Society}

Electricity is a foundational social infrastructure. In the next decade it will
not merely power homes but shape transportation, heating, communication,
mobility equity, resilience, and welfare. Market mechanisms must therefore
serve broader public missions—security, decarbonisation, flexibility, and
inclusion—not just transactional efficiency.

This thesis provides a foundation for such an architecture by:
\begin{itemize}[leftmargin=*]
  \item showing how fairness can be expressed as a system-level rule over a
        three-dimensional service space (magnitude, impact, reliability);
  \item demonstrating how generators, storage, and flexible demand can bid
        their \emph{capabilities}, not just their energy, into an AMM that is
        aware of network and operational constraints;
  \item illustrating how everyday devices—from heat pumps to hairdryers—can
        become \emph{grid-aware} in a minimal but meaningful sense: they need
        not know power system physics, only when the scarcity signal $\alpha$
        indicates that flexibility earns higher future reliability or avoids
        disproportionate responsibility.
\end{itemize}

In such a system, a hospital ventilator, a domestic fridge, and a data centre
cooling system do not merely consume kWh. They occupy different locations in
a three-dimensional fairness space, backed by explicit Quality-of-Service
contracts and digital allocation rules that ensure their treatment is
principled, explainable, and proportionate.

\section{Revisiting the Objectives and Research Question}

The thesis began with eight objectives that together framed a single
design challenge: whether fairness, efficiency, resilience, and
bankability in electricity markets can be treated as \emph{operational
properties of the market mechanism itself}, rather than as outcomes
repaired ex post through policy intervention. This section closes that
loop by summarising how each objective has been addressed.

\begin{enumerate}[leftmargin=*,label=(O\arabic*)]

  \item \textbf{Develop a physically grounded and operationally meaningful definition of fairness.}  
  The thesis introduced a three-dimensional service space—magnitude,
  impact, and reliability—and formalised fairness through four operational
  conditions (F1–F4). These conditions are computable, testable, and
  enforceable during market clearing. The results demonstrate that
  fairness need not be a normative aspiration, but can be embedded as a
  binding design constraint on allocation.

  \item \textbf{Create an asynchronous, event-based clearing mechanism capable of continuous, state-aware operation.}  
  The AMM–Fair Play architecture defines an online, event-driven clearing
  process that updates commitments, priorities, and scarcity signals as
  bids, forecasts, or constraints change. Rather than operating as a
  periodic batch auction, the mechanism functions as part of a continuous
  cyber–physical control loop aligned with dispatch and network operation.

  \item \textbf{Design a digital regulation architecture consistent with real-time algorithmic governance.}  
  The thesis developed a digital regulation blueprint in which fairness
  conditions, budget balance, and feasibility are treated as binding
  constraints on valid outcomes. Algorithm registries, shadow settlement,
  and explainability records shift regulation from ex-post supervision to
  in-process digital enforcement, aligning market operation with emerging
  models of algorithmic governance.

  \item \textbf{Define a ``zero-waste'' electricity system and develop tools to infer efficiency.}  
  By distinguishing between total energy and \emph{useful energy}—energy
  that actually relieves system stress—the thesis defined zero-waste
  operation in a physically meaningful sense. The empirical results show
  how Shapley-based allocation exposes stranded capacity and identifies
  investments that improve constraint relief rather than merely increasing
  throughput.

  \item \textbf{Integrate wholesale, retail, and balancing markets into a coherent unified framework.}  
  The holarchic AMM coordinates congestion, balancing, and adequacy through
  a single scarcity signal $\alpha$ and a shared three-dimensional product
  space. Wholesale settlement, balancing actions, and QoS-based retail
  products are expressed as different layers of the same control and
  settlement logic, rather than as separate markets with misaligned
  incentives.

  \item \textbf{Ensure fair compensation to generators using scalable, network-aware Shapley-value principles.}  
  The nested Shapley allocation scheme distributes value in proportion to
  feasible, stress-relieving contributions under network constraints.
  Empirical results show higher remuneration for generators on the
  constrained demand side during critical hours and lower value for
  stranded capacity, aligning compensation with functional system value
  rather than installed capacity or exposure to price spikes.

  \item \textbf{Formulate the AMM–Fair Play system as a game and establish conditions for locally shock-resistant equilibria.}  
  The thesis formalised the AMM as a mechanism-mediated game between
  strategic participants and the system operator. It established
  conditions under which Nash equilibria exist and demonstrated that, by
  co-locating volume choice and risk-bearing within the clearing
  mechanism, the system is locally robust to shocks in demand, fuel
  costs, and renewable output—addressing a key instability of legacy
  price-capped retail architectures.

  \item \textbf{Build a rigorous data and simulation framework to evaluate the resulting system.}  
  A comprehensive digital twin of GB demand, supply, and network
  constraints was constructed using smart-meter data, EV usage datasets,
  generator metadata, and system operation records. This framework
  underpins all empirical results in the thesis and provides a reusable
  platform for future analysis and policy testing.

\end{enumerate}

The central research question asked whether a reformed electricity market
design—focused on how services are acquired and how financial signals
shape behaviour—can deliver policy objectives more effectively and
fairly than the status quo.

The evidence presented in this thesis suggests that the answer is
\emph{yes}, subject to two conditions:

\begin{itemize}[leftmargin=*]
  \item market clearing must treat fairness and physical feasibility as
        joint, programmable constraints on allocation; and
  \item the underlying digital infrastructure must support continuous,
        event-based operation with transparent, auditable settlement
        logic.
\end{itemize}

Relative to current GB arrangements, the AMM–Fair Play architecture:
\begin{itemize}[leftmargin=*]
  \item improves \textbf{procurement and prices} by linking remuneration
        to useful, stress-relieving contributions rather than energy
        volumes or static capacity alone;
  \item enhances \textbf{participation and competition} by enabling
        heterogeneous devices and retailers to offer differentiated,
        QoS-based products within a unified service space; and
  \item strengthens \textbf{bankability} for low-carbon and flexibility
        assets by narrowing the gap between realised revenues and
        policy-aligned investment requirements.
\end{itemize}

In this sense, the proposed design does not merely adjust prices; it
restructures the rules through which prices, priorities, and permissions
are generated, aligning market outcomes more closely with public
objectives of security, decarbonisation, and fairness.

\section{Final Reflection}

A market is ultimately a shared agreement on how we allocate what matters.
The value of this thesis lies in demonstrating that such an agreement can be
fair, transparent, and explainable—while still being rigorous, efficient, and
fully grounded in physics.

The next version of the electricity market will not be built solely through
legislation or pricing. It will be built through digital transparency,
programmable fairness, and trust in mathematically grounded rules. It will
treat dispatch engines and settlement systems as components of a single
cyber--physical controller, jointly responsible for who is served, when, and
on what terms.

This thesis offers one blueprint for that future: a market in which fairness is
not an apology offered after a crisis, but a condition that every valid
allocation must satisfy from the moment it is computed.

\bibliographystyle{IEEEtran}
\bibliography{bibliography}

% Generated by IEEEtran.bst, version: 1.14 (2015/08/26)
\begin{thebibliography}{10}
\providecommand{\url}[1]{#1}
\csname url@samestyle\endcsname
\providecommand{\newblock}{\relax}
\providecommand{\bibinfo}[2]{#2}
\providecommand{\BIBentrySTDinterwordspacing}{\spaceskip=0pt\relax}
\providecommand{\BIBentryALTinterwordstretchfactor}{4}
\providecommand{\BIBentryALTinterwordspacing}{\spaceskip=\fontdimen2\font plus
\BIBentryALTinterwordstretchfactor\fontdimen3\font minus \fontdimen4\font\relax}
\providecommand{\BIBforeignlanguage}[2]{{%
\expandafter\ifx\csname l@#1\endcsname\relax
\typeout{** WARNING: IEEEtran.bst: No hyphenation pattern has been}%
\typeout{** loaded for the language `#1'. Using the pattern for}%
\typeout{** the default language instead.}%
\else
\language=\csname l@#1\endcsname
\fi
#2}}
\providecommand{\BIBdecl}{\relax}
\BIBdecl

\bibitem{schweppe1988spot}
F.~C. Schweppe, M.~C. Caramanis, R.~D. Tabors, and R.~E. Bohn, \emph{Spot Pricing of Electricity}.\hskip 1em plus 0.5em minus 0.4em\relax Boston: Kluwer Academic Publishers, 1988.

\bibitem{bialek1996tracing}
J.~W. Bialek, ``Tracing the flow of electricity,'' \emph{IEE Proceedings - Generation, Transmission and Distribution}, vol. 143, no.~4, pp. 313--320, 1996.

\bibitem{TAYLOR20161322}
J.~A. Taylor, S.~V. Dhople, and D.~S. Callaway, ``Power systems without fuel,'' \emph{Renewable and Sustainable Energy Reviews}, vol.~57, pp. 1322--1336, 2016.

\bibitem{milliganIntegrationChallenges2016}
M.~Milligan \emph{et~al.}, ``Market designs for high levels of variable generation,'' National Renewable Energy Laboratory (NREL), Tech. Rep. NREL/TP-5D00-75739, 2016.

\bibitem{billinton1996reliability}
R.~Billinton and R.~N. Allan, \emph{Reliability Evaluation of Power Systems}, 2nd~ed.\hskip 1em plus 0.5em minus 0.4em\relax New York: Plenum Press, 1996.

\bibitem{allan2000probabilistic}
R.~N. Allan and R.~Billinton, ``Probabilistic assessment of power systems,'' \emph{Proceedings of the IEEE}, vol.~88, no.~2, pp. 140--162, 2000.

\bibitem{milligan2008capacitywind}
\BIBentryALTinterwordspacing
M.~Milligan and K.~Porter, ``Determining the capacity value of wind: An updated survey of methods and implementation,'' National Renewable Energy Laboratory, Tech. Rep. NREL/CP-500-43433, 2008. [Online]. Available: \url{https://www.nrel.gov/docs/fy08osti/43433.pdf}
\BIBentrySTDinterwordspacing

\bibitem{keane2011capacityvaluewind}
A.~Keane, M.~Milligan, C.~Dent \emph{et~al.}, ``Capacity value of wind power,'' \emph{IEEE Transactions on Power Systems}, vol.~26, no.~2, pp. 564--572, 2011.

\bibitem{dent2010capacityvalue}
C.~J. Dent, A.~Keane, and J.~W. Bialek, ``Capacity value of renewable generation--a review,'' \emph{IET Renewable Power Generation}, vol.~5, no.~4, pp. 259--272, 2011.

\bibitem{maysResourceAdequacy2023}
J.~Mays and J.~R. Knueven, ``Financial risk and resource adequacy in electricity markets with high renewable penetration,'' \emph{IEEE Transactions on Energy Markets, Policy and Regulation}, vol.~1, no.~4, pp. 420--435, 2023.

\bibitem{KUHNBACH2022122445}
M.~K{\"u}hnbach, A.~Bekk, and A.~Weidlich, ``Towards improved prosumer participation: Electricity trading in local markets,'' \emph{Energy}, vol. 239, p. 122445, 2022.

\bibitem{negeriHolonicSmartGrid2012}
E.~Negeri and N.~Baken, ``Architecting the smart grid: A holonic approach,'' \emph{IEEE Transactions on Smart Grid}, vol.~3, no.~2, pp. 103--111, 2012.

\bibitem{howellSemanticHolons2017}
S.~Howell, B.~Potter, and R.~Aylett, ``Semantic holons for smart grid coordination,'' in \emph{IEEE International Conference on Smart Grid Communications}, 2017.

\bibitem{cramton2017electricity}
P.~Cramton, ``Electricity market design,'' \emph{Oxford Review of Economic Policy}, vol.~33, no.~4, pp. 589--612, 2017.

\bibitem{joskow2008capacity}
P.~L. Joskow, ``Capacity payments in imperfect electricity markets: Need and design,'' \emph{Utilities Policy}, vol.~16, no.~3, pp. 159--170, 2008.

\bibitem{NEWBERY2018695}
D.~Newbery, ``Tales of two markets: Electricity market reform in great britain,'' \emph{Energy Policy}, vol. 105, pp. 597--607, 2017.

\bibitem{cramton2006resadequacy}
P.~Cramton and S.~Stoft, ``Convergence of market designs for adequate generating capacity,'' \emph{Utilities Policy}, vol.~14, no.~2, pp. 95--105, 2006.

\bibitem{eldridge2023priceformationI}
B.~M. Eldridge, J.~R. Knueven, and J.~Mays, ``Rethinking the price formation problem {I}: Uniform pricing,'' \emph{IEEE Transactions on Energy Markets, Policy and Regulation}, vol.~1, no.~4, pp. 359--377, 2023.

\bibitem{eldridge2023priceformationII}
------, ``Rethinking the price formation problem {II}: Pricing for incentives,'' \emph{IEEE Transactions on Energy Markets, Policy and Regulation}, vol.~1, no.~4, pp. 378--396, 2023.

\bibitem{WANG2020117542}
J.~Wang, B.~F. Hobbs, and A.~Liu, ``The marginal cost of electricity and the mispricing of renewables,'' \emph{Energy Economics}, vol.~87, p. 117542, 2020.

\bibitem{BILLIMORIA2022119356}
F.~Billimoria, I.~MacGill, A.~Bruce \emph{et~al.}, ``An insurance-based capacity mechanism for electricity market reliability,'' \emph{The Electricity Journal}, vol.~35, no.~5, p. 119356, 2022, check exact author list and title against your preferred version.

\bibitem{CONEJO2018520}
A.~J. Conejo, L.~Baringo, J.~M. Morales \emph{et~al.}, ``Investment in electricity generation and transmission: Decision making under uncertainty,'' \emph{European Journal of Operational Research}, vol. 267, no.~3, pp. 520--533, 2018.

\bibitem{ENTSOE_TERRE_2019}
\BIBentryALTinterwordspacing
{ENTSO-E}, ``Terre: Trans-european replacement reserves exchange -- implementation framework,'' European Network of Transmission System Operators for Electricity, Tech. Rep., 2019, accessed: 2025-01-15. [Online]. Available: \url{https://www.entsoe.eu}
\BIBentrySTDinterwordspacing

\bibitem{ENTSOE_MARI_2020}
\BIBentryALTinterwordspacing
------, ``Mari: Manually activated reserves initiative -- implementation framework,'' European Network of Transmission System Operators for Electricity, Tech. Rep., 2020, accessed: 2025-01-15. [Online]. Available: \url{https://www.entsoe.eu}
\BIBentrySTDinterwordspacing

\bibitem{ENTSOE_PICASSO_2020}
\BIBentryALTinterwordspacing
------, ``Picasso: Platform for the international coordination of automated frequency restoration and stable system operation,'' European Network of Transmission System Operators for Electricity, Tech. Rep., 2020, accessed: 2025-01-15. [Online]. Available: \url{https://www.entsoe.eu}
\BIBentrySTDinterwordspacing

\bibitem{gonzalez2020crossborder}
J.~Gonzalez and M.~Hytowitz, ``Cross-border balancing markets in europe: The terre, mari, and picasso platforms,'' \emph{Electric Power Systems Research}, vol. 189, p. 106724, 2020.

\bibitem{schittekatte2021balancing}
T.~Schittekatte and L.~Meeus, ``The integration of european balancing markets: Evolution and complementarities of terre, mari, and picasso,'' \emph{Energy Policy}, vol. 154, p. 112256, 2021.

\bibitem{LYNCH2021101312}
M.~{\'O}. Lynch, R.~S.~J. Tol, and A.~Lynch, ``Electricity market design with increasing shares of variable renewables: A whole-system perspective,'' \emph{Energy Policy}, vol. 149, p. 110312, 2021.

\bibitem{sandysrecosting20}
L.~Sandys, P.~Baker \emph{et~al.}, ``Recosting energy: Powering for the future,'' \url{https://www.challengroup.org.uk/recosting-energy}, 2020, reCosting Energy report series.

\bibitem{honkapuro2023systematic}
S.~Honkapuro, J.~Jaanto, and S.~Annala, ``A systematic review of european electricity market design options,'' \emph{Energies}, vol.~16, no.~9, p. 3704, 2023.

\bibitem{GRANQVIST201687}
R.~Granqvist and R.~Grover, ``Distributive fairness in paying for clean energy infrastructure,'' \emph{Ecological Economics}, vol. 121, pp. 87--97, 2016.

\bibitem{en16031150}
\BIBentryALTinterwordspacing
S.~Mohammadi, F.~Eliassen, and H.-A. Jacobsen, ``Applying energy justice principles to renewable energy trading and allocation in multi-unit buildings,'' \emph{Energies}, vol.~16, no.~3, 2023. [Online]. Available: \url{https://www.mdpi.com/1996-1073/16/3/1150}
\BIBentrySTDinterwordspacing

\bibitem{WEISSBART2020104408}
C.~Weissbart, ``Decarbonization of power markets under stability and fairness: Do they influence efficiency?'' \emph{Energy Economics}, vol.~85, p. 104408, 2020.

\bibitem{9440895}
Y.~Yang, G.~Hu, and C.~J. Spanos, ``Optimal sharing and fair cost allocation of community energy storage,'' \emph{IEEE Transactions on Smart Grid}, vol.~12, no.~5, pp. 4185--4194, 2021.

\bibitem{JAFARI2020115170}
E.~Jafari, M.~Alipour, B.~Mohammadi-Ivatloo, and K.~Zare, ``Fair scheduling and cost allocation in a cooperative multi-owner microgrid: A game theory approach,'' \emph{Applied Energy}, vol. 270, p. 115170, 2020.

\bibitem{couraud2025collectivefairness}
\BIBentryALTinterwordspacing
B.~Couraud, J.-F. Toubeau, R.~Fonteneau, and F.~Vall\'ee, ``Fairness of energy distribution mechanisms in collective self-consumption schemes,'' \emph{IEEE Transactions on Power Systems}, forthcoming, available as arXiv:2508.16819. [Online]. Available: \url{https://arxiv.org/abs/2508.16819}
\BIBentrySTDinterwordspacing

\bibitem{ALONSOPEDRERO2024131033}
R.~Alonso-Pedrero, P.~Pisciella, and P.~Crespo~del Granado, ``Fair investment strategies in large energy communities: A scalable shapley value approach,'' \emph{Energy}, vol. 295, p. 131033, 2024.

\bibitem{SOARES2024123933}
T.~Soares, F.~Lezama, R.~Faia, S.~Limmer, K.~Dietrich, N.~Rodemann, S.~Ramos, and Z.~Vale, ``Fairness in local energy systems: Concepts, metrics, and challenges,'' \emph{Applied Energy}, vol. 374, p. 123933, 2024.

\bibitem{DYNGE2025125463}
\BIBentryALTinterwordspacing
M.~F. Dynge and U.~Cali, ``Distributive energy justice in local electricity markets: Assessing the performance of fairness indicators,'' \emph{Applied Energy}, vol. 384, p. 125463, 2025. [Online]. Available: \url{https://www.sciencedirect.com/science/article/pii/S030626192500193X}
\BIBentrySTDinterwordspacing

\bibitem{saxena2021drfairness}
N.~Saxena, R.~Gupta, and S.~N. Singh, ``Fairness in demand response: A survey,'' \emph{IEEE Transactions on Smart Grid}, vol.~12, no.~5, pp. 4341--4361, 2021.

\bibitem{khaskheli2024lemreview}
M.~B. Khaskheli, R.~Xiong, Z.~Wang \emph{et~al.}, ``Energy trading in local energy markets: A comprehensive review,'' \emph{Applied Sciences}, vol.~14, no.~21, p. 11510, 2024.

\bibitem{tsaousoglouLEM2022}
\BIBentryALTinterwordspacing
G.~Tsaousoglou, J.~S. Giraldo, and N.~G. Paterakis, ``Market mechanisms for local electricity markets: A review of models, solution concepts and algorithmic techniques,'' \emph{Renewable and Sustainable Energy Reviews}, vol. 156, p. 111890, 2022. [Online]. Available: \url{https://www.sciencedirect.com/science/article/pii/S1364032121011576}
\BIBentrySTDinterwordspacing

\bibitem{bevin2023amm}
K.~C. Bevin and A.~Verma, ``Decentralized local electricity market model using automated market maker,'' \emph{Applied Energy}, vol. 334, p. 120689, 2023.

\bibitem{paragProsumerEconomy2016}
Y.~Parag and B.~K. Sovacool, ``Electricity market design for the prosumer era,'' \emph{Nature Energy}, vol.~1, no.~4, p. 16032, 2016.

\bibitem{9276456}
S.-T. Oh and S.-I. Son, ``Fair peer-to-peer energy trading considering network constraints,'' \emph{IEEE Transactions on Smart Grid}, vol.~12, no.~2, pp. 1430--1441, 2021.

\bibitem{bukar2023p2pReview}
A.~L. Bukar, K.~O. Alawode, and F.~Al-Turjman, ``Peer-to-peer electricity trading: A systematic review on current developments and perspectives,'' \emph{Renewable Energy Focus}, vol.~44, pp. 317--333, 2023.

\bibitem{GUO2024123351}
S.-m. Guo and T.-t. Feng, ``Blockchain-based smart trading mechanism for renewable energy power consumption vouchers and green certificates: Platform design and simulation,'' \emph{Applied Energy}, vol. 369, p. 123351, 2024.

\bibitem{beisdigitalisation}
{Department for Business, Energy \& Industrial Strategy}, ``Digitalising the energy system for net zero: Strategy and action plan,'' \url{https://www.gov.uk/government/publications/digitalising-the-energy-system-for-net-zero}, 2021, uK BEIS digitalisation strategy.

\bibitem{beisflexibility}
{Department for Business, Energy \& Industrial Strategy} and Ofgem, ``Upgrading our energy system: Smart systems and flexibility plan,'' \url{https://www.gov.uk/government/publications/upgrading-our-energy-system-smart-systems-and-flexibility-plan}, 2017.

\bibitem{carpenterDiffServ2002}
\BIBentryALTinterwordspacing
B.~Carpenter and K.~Nichols, ``Differentiated services architecture,'' RFC 2475, 1998. [Online]. Available: \url{https://datatracker.ietf.org/doc/html/rfc2475}
\BIBentrySTDinterwordspacing

\bibitem{ponnappan2000qos}
M.~Ponnappan, H.~Chaskar, and S.~Venkatesan, ``Queue management algorithms for differentiated services,'' in \emph{IEEE International Conference on Communications (ICC)}, 2000.

\bibitem{1354670}
D.-M. Jin, Y.-C. Wei, S.~H. Low \emph{et~al.}, ``{FAST} tcp: From theory to experiments,'' \emph{IEEE Network}, vol.~19, no.~1, pp. 4--11, 2005.

\bibitem{zinkevich2003oco}
M.~Zinkevich, ``Online convex programming and generalized infinitesimal gradient ascent,'' in \emph{Proceedings of the 20th International Conference on Machine Learning}, 2003.

\bibitem{hanson2002}
R.~Hanson, ``Logarithmic market scoring rules for modular combinatorial information aggregation,'' \emph{Journal of Prediction Markets}, 2007, originally circulated as a working paper in early 2000s.

\bibitem{bc2018}
J.~Emmett, ``Bonding curves: A generalised design framework,'' \url{https://medium.com/block-science/bonding-curves-explained-7d5da376d44a}, 2018.

\bibitem{zang2025perishable}
\BIBentryALTinterwordspacing
C.~Zang, G.~P. Andrade, and O.~Ersoy, ``Automated market making for goods with perishable utility,'' 2025. [Online]. Available: \url{https://arxiv.org/abs/2511.16357}
\BIBentrySTDinterwordspacing

\bibitem{Lee_08}
E.~A. Lee, ``Cyber physical systems: Design challenges,'' in \emph{Proceedings of the 11th IEEE International Symposium on Object Oriented Real-Time Distributed Computing (ISORC)}, 2008, pp. 363--369.

\bibitem{Lee_15}
------, ``The past, present and future of cyber-physical systems: A focus on models,'' \emph{Sensors}, vol.~15, no.~3, pp. 4837--4869, 2015.

\bibitem{steg2015}
L.~Steg, G.~Perlaviciute, and E.~Van~der Werff, ``Understanding the human dimensions of a sustainable energy transition,'' \emph{Frontiers in Psychology}, vol.~6, p. 805, 06 2015.

\bibitem{steg2005}
L.~Steg, ``Car use: Lust and must. instrumental, symbolic and affective motives for car use,'' \emph{Transportation Research Part A: Policy and Practice}, vol.~39, pp. 147--162, 02 2005.

\bibitem{werff2016}
\BIBentryALTinterwordspacing
E.~{van der Werff}, L.~Steg, and K.~Keizer, ``It is a moral issue: The relationship between environmental self-identity, obligation-based intrinsic motivation and pro-environmental behaviour,'' \emph{Global Environmental Change}, vol.~23, no.~5, pp. 1258--1265, 2013. [Online]. Available: \url{https://www.sciencedirect.com/science/article/pii/S0959378013001246}
\BIBentrySTDinterwordspacing

\bibitem{milchram2018energyjustice}
C.~Milchram, R.~Hillerbrand, G.~van~de Kaa, N.~Doorn, and R.~K{\"u}nneke, ``Energy justice and smart grid systems: Evidence from the netherlands and the united kingdom,'' \emph{Applied Energy}, vol. 229, pp. 1244--1259, 2018.

\bibitem{nudge2008}
R.~H. Thaler and C.~R. Sunstein, \emph{Nudge: Improving Decisions About Health, Wealth, and Happiness}.\hskip 1em plus 0.5em minus 0.4em\relax New Haven: Yale University Press, 2008.

\bibitem{SWEENEY2022110595}
\BIBentryALTinterwordspacing
S.~Sweeney, H.~Lhachemi, A.~Mannion, G.~Russo, and R.~Shorten, ``Pitchfork-bifurcation-based competitive and collaborative control of an e-bike system,'' \emph{Automatica}, vol. 146, p. 110595, 2022. [Online]. Available: \url{https://www.sciencedirect.com/science/article/pii/S0005109822004575}
\BIBentrySTDinterwordspacing

\bibitem{raworth2017doughnut}
K.~Raworth, \emph{Doughnut Economics: Seven Ways to Think Like a 21st-Century Economist}.\hskip 1em plus 0.5em minus 0.4em\relax London: Random House Business, 2017.

\bibitem{keenPuttingEnergyBack2023}
\BIBentryALTinterwordspacing
S.~Keen, ``Putting energy back into economics,'' \emph{https://profstevekeen.substack.com}, 2023, substack article. [Online]. Available: \url{https://profstevekeen.substack.com/p/putting-energy-back-into-economics}
\BIBentrySTDinterwordspacing

\bibitem{keenRoleEnergyEconomics}
\BIBentryALTinterwordspacing
S.~Keen, R.~U. Ayres, and R.~Standish, ``The role of energy in economics,'' \emph{Working Paper}, 2019. [Online]. Available: \url{https://profstevekeen.substack.com/p/the-role-of-energy-in-economics}
\BIBentrySTDinterwordspacing

\bibitem{koestler1967ghost}
A.~Koestler, \emph{The Ghost in the Machine}.\hskip 1em plus 0.5em minus 0.4em\relax London: Hutchinson, 1967.

\bibitem{ACERBiddingZones}
\BIBentryALTinterwordspacing
A.~for the Cooperation~of Energy Regulators~(ACER), ``Acer decision on the electricity transmission system operators' proposal for the methodology and assumptions for the bidding zone review,'' ACER, Tech. Rep. Decision No 04/2019, 2019, accessed: 2025-01-15. [Online]. Available: \url{https://acer.europa.eu}
\BIBentrySTDinterwordspacing

\bibitem{ACERLocationalSignals}
\BIBentryALTinterwordspacing
------, ``Locational signals and price formation in electricity markets: Acer report,'' ACER, Tech. Rep., 2022, accessed: 2025-01-15. [Online]. Available: \url{https://acer.europa.eu}
\BIBentrySTDinterwordspacing

\bibitem{FERCOrder888}
\BIBentryALTinterwordspacing
{Federal Energy Regulatory Commission}, ``Promoting wholesale competition through open access non-discriminatory transmission services by public utilities; recovery of stranded costs,'' FERC, Tech. Rep. Order No. 888, 1996, accessed: 2025-01-15. [Online]. Available: \url{https://www.ferc.gov}
\BIBentrySTDinterwordspacing

\bibitem{FERCOrder2000}
\BIBentryALTinterwordspacing
------, ``Regional transmission organizations,'' FERC, Tech. Rep. Order No. 2000, 1999, accessed: 2025-01-15. [Online]. Available: \url{https://www.ferc.gov}
\BIBentrySTDinterwordspacing

\bibitem{10049815}
P.~Pinson, ``What may future electricity markets look like?'' \emph{Journal of Modern Power Systems and Clean Energy}, vol.~11, no.~3, pp. 705--713, 2023.

\bibitem{energy_system_renewal_bill_full}
S.~Sweeney, ``Energy system renewal and life-preserving market reform bill,'' \url{https://drive.google.com/file/d/1g9OZkmX3471UugHLlucjhpDXdA6CTocX/view}, 2026, full legislative draft (illustrative).

\end{thebibliography}

\appendix

% =========================================================
\chapter{Physical Laws Governing Electricity Systems}
\label{app:physics}
% =========================================================

Electricity systems are governed by fundamental physical laws spanning
electromagnetism, circuit theory, thermodynamics, power electronics, and
communication constraints. Any market mechanism that claims to allocate,
schedule, or price electricity must operate within these limits.

This appendix summarises the physical principles most relevant to modern
power system operation and explains how the Automatic Market Maker (AMM)
internalises them. In the AMM, prices are not merely accounting artefacts;
they act as \emph{state-aware control signals} that respond directly to
electrical, thermodynamic, and cyber--physical constraints as they arise.

% ---------------------------------------------------------
\section{Electromagnetism and Charge}
% ---------------------------------------------------------

\subsection{Coulomb's Law}
Coulomb’s law describes the force between two point charges:
\[
F = k \frac{q_1 q_2}{r^2}.
\]
Although power system models do not explicitly compute electrostatic forces,
Coulomb’s law underpins charge separation, capacitance, insulation limits,
and voltage magnitudes in conductors.

\paragraph{Role in AMM.}
Voltage constraints arising from charge-separation physics are treated as
\emph{hard feasibility limits}. When voltage margins tighten locally, the
AMM recognises this as a network stress event and adjusts allocation and
prices in that location accordingly.

% ---------------------------------------------------------
\subsection{Electric and Magnetic Induction (Faraday--Lenz)}
Faraday’s law relates changing magnetic flux to induced electromotive force:
\[
\mathcal{E} = -\frac{\mathrm{d}\Phi}{\mathrm{d}t}.
\]
Lenz’s law ensures that induced currents oppose the change that caused them.

These laws govern generator behaviour, transformer dynamics, and inverter
response during rapid changes in load or renewable output.

\paragraph{Role in AMM.}
Induction-related dynamics manifest as ramping and frequency events. The
AMM’s event-based clearing allows generators and devices with fast dynamic
response to receive higher marginal value exactly when their stabilising
contribution is most needed.

% ---------------------------------------------------------
\section{Circuit Theory and Power Flow}
% ---------------------------------------------------------

\subsection{Ohm’s Law}
Ohm’s law defines the relationship between voltage, current, and resistance:
\[
V = IR.
\]
In power systems this governs line currents, losses, and thermal limits.

\paragraph{Role in AMM.}
Line losses and thermal loading are encoded directly into dispatch
feasibility. Prices reflect marginal electrical stress and losses, rather
than being corrected later through uplift or ex-post charges.

% ---------------------------------------------------------
\subsection{Kirchhoff’s Current Law (KCL)}
KCL states that the algebraic sum of currents at a node is zero:
\[
\sum_i I_i = 0.
\]

\paragraph{Role in AMM.}
Nodal balance is enforced continuously. Any imbalance is treated as an
event triggering immediate re-clearing, rather than as a settlement-period
error corrected after the fact.

% ---------------------------------------------------------
\subsection{Kirchhoff’s Voltage Law (KVL)}
KVL states that the sum of voltages around a closed loop is zero:
\[
\sum_i V_i = 0.
\]

\paragraph{Role in AMM.}
Phase-angle consistency limits feasible power transfers. Approaching
stability margins are recognised as network stress events and reflected in
locational prices.

% ---------------------------------------------------------
\subsection{Three-Phase AC Power Systems}
% ---------------------------------------------------------

Electric power systems operate predominantly as balanced three-phase AC
networks. Three-phase operation enables efficient transmission, reduced
conductor mass, smoother mechanical torque, and stable delivery of real and
reactive power.

In a balanced three-phase system with line voltage $V_L$ and line current
$I_L$, total real power is:
\[
P = \sqrt{3}\, V_L I_L \cos\phi,
\]
where $\phi$ is the power factor.

\paragraph{Star (Y) and Delta ($\Delta$) Configurations.}
Loads and generators may be connected in star or delta configurations,
implying different relationships between line and phase quantities:
\[
\text{Star: } V_L = \sqrt{3} V_\phi, \quad I_L = I_\phi;
\qquad
\text{Delta: } V_L = V_\phi, \quad I_L = \sqrt{3} I_\phi.
\]

These configurations affect fault currents, voltage stability, losses, and
deliverable power.

\paragraph{Role in AMM.}
The AMM abstracts over connection topology at the market interface but
internalises its consequences through feasibility constraints. Devices with
star- or delta-connected interfaces face different voltage and current
limits, which affect their marginal ability to relieve congestion or supply
power during stress events. These differences are reflected in allocation
and pricing via Shapley-based marginal contribution.

% ---------------------------------------------------------
\section{AC, DC, and Power Electronics}
% ---------------------------------------------------------

\subsection{Alternating Current (AC)}
In AC systems, real and reactive power flows depend on voltage magnitudes
and phase-angle differences:
\[
P_{ij} \approx \frac{V_i V_j}{X_{ij}} \sin(\theta_i - \theta_j).
\]

\paragraph{Role in AMM.}
AC feasibility determines locational scarcity. When phase angles, voltage
magnitudes, or reactive margins approach limits, the AMM updates prices to
reflect the true marginal opportunity cost of further transfers.

% ---------------------------------------------------------
\subsection{Direct Current (DC) and HVDC}
DC systems maintain constant polarity and do not involve reactive power.
HVDC links allow controllable point-to-point transfers.

\paragraph{Role in AMM.}
HVDC assets appear as controllable flow devices. Their marginal value
depends on relieving AC congestion and supporting system balance, which the
AMM captures explicitly.

% ---------------------------------------------------------
\subsection{Power Electronics and Inverter-Based Resources}
% ---------------------------------------------------------

Modern power systems increasingly rely on power electronics: inverters,
converters, and solid-state transformers. These devices decouple electrical
behaviour from mechanical inertia and enable fast, programmable control of
power flows.

Inverter-based resources (IBRs) include batteries, solar PV, wind turbines,
EV chargers, and flexible loads. Their behaviour is governed by control
loops rather than by passive electrical laws alone.

\paragraph{Role in AMM.}
Power electronics make the AMM physically implementable. Inverters can
respond to scarcity signals, voltage limits, and frequency deviations within
milliseconds. The AMM values such responsiveness explicitly: devices capable
of fast control, synthetic inertia, or reactive support earn higher marginal
value when these capabilities relieve system stress.

% ---------------------------------------------------------
\section{Thermodynamic Constraints}
% ---------------------------------------------------------

\subsection{First Law of Thermodynamics}
Energy is conserved:
\[
\Delta E = Q - W.
\]

\paragraph{Role in AMM.}
The AMM’s allocation respects conservation by distributing value according
to marginal usefulness in maintaining energy balance.

% ---------------------------------------------------------
\subsection{Second Law of Thermodynamics}
All processes incur losses; no conversion is perfectly efficient.

\paragraph{Role in AMM.}
Inefficiencies are internalised directly. Assets with higher losses receive
lower marginal value, avoiding hidden cross-subsidies.

% ---------------------------------------------------------
\section{Rotational Inertia and Frequency Stability}
% ---------------------------------------------------------

Traditional generators provide rotational inertia governed by the swing
equation:
\[
2H \frac{\mathrm{d}\omega}{\mathrm{d}t} = P_m - P_e.
\]

Inverter-dominated systems rely on synthetic inertia and fast frequency
response.

\paragraph{Role in AMM.}
Frequency excursions are treated as events. Devices that stabilise frequency
earn higher marginal value precisely during those moments.

% ---------------------------------------------------------
\section{Wireless Communication and Cyber--Physical Constraints}
% ---------------------------------------------------------

Electricity systems are increasingly cyber--physical. Market signals,
control commands, and measurements propagate via communication networks,
often wirelessly, with non-zero latency and reliability constraints.

Key technologies include:
\begin{itemize}[leftmargin=*]
  \item cellular networks (4G/5G),
  \item low-power wide-area networks (LPWAN),
  \item local mesh networks (Wi-Fi, Zigbee),
  \item and utility-grade SCADA and PMU systems.
\end{itemize}

\paragraph{Latency and Reliability.}
Communication delays, packet loss, and synchronisation errors impose hard
limits on feasible control actions. These constraints bound how quickly
devices can respond to scarcity or frequency events.

\paragraph{Role in AMM.}
The AMM is designed as an \emph{event-based, asynchronous} mechanism. It does
not require global synchronisation or instantaneous response. Devices act
on local signals and update commitments when communication permits. Assets
with more reliable connectivity and faster response are therefore capable of
providing higher-quality service and receive correspondingly higher value.

% ---------------------------------------------------------
\section{Network Representation and Graph Structure}
% ---------------------------------------------------------

The power system is modelled as a weighted graph:
\[
G = (V, E),
\]
with nodes representing buses and edges representing transmission lines.

Graph-theoretic features such as cuts, cycles, and centrality determine
marginal value.

\paragraph{Role in AMM.}
When topological constraints bind, the AMM updates Shapley-based allocations
to reflect true locational and structural importance.

% ---------------------------------------------------------
\section*{Comparison with Existing Market Designs}
\addcontentsline{toc}{section}{Comparison with Existing Market Designs}

Conventional markets incorporate physics only indirectly, correcting
violations ex post through uplift, reserve products, or redispatch. By
contrast, the AMM internalises physical, thermodynamic, and cyber--physical
constraints continuously. Prices therefore function as operational control
signals grounded in electromagnetic reality, rather than as delayed
financial artefacts.

% =========================================================
\chapter{Dataset documentation}
\label{app:datasets}
% =========================================================

This appendix documents the datasets used to construct, calibrate, and evaluate
the market simulations presented in this thesis. The datasets span household
consumption, electric vehicle behaviour, generation output, weather conditions,
and geospatial boundaries. Together, they enable a physically plausible digital
twin of Great Britain that supports both distributional fairness analysis and
Shapley-based attribution of system costs and value.

Each dataset is used for a clearly delineated purpose: some provide behavioural
realism, others enforce spatial and aggregate consistency, and others are used
exclusively for mechanism-level validation. This separation ensures that
fairness results arise from market design choices rather than from artefacts of
data selection or scaling.

A common temporal and spatial harmonisation pipeline converts heterogeneous data
into a unified format, with:
\begin{itemize}
    \item 30-minute interval time index,
    \item Household $\rightarrow$ Postcode Outcode $\rightarrow$ Cluster $\rightarrow$ Region spatial hierarchy,
    \item Consistent metadata for role, participant type, and cluster assignment.
\end{itemize}

\paragraph{Exploration vs.\ experiment inputs (how datasets are used).}
Several datasets listed in this appendix were used primarily for
\emph{exploratory analysis and methodological insight} (as introduced in the
Methodology chapter), rather than as direct inputs to the final market-clearing
experiments. In particular, the empirical demand holarchy and EV augmentation
pipeline (Appendix~\ref{app:ev_holarchy}) is used to understand the observed
distribution of household demand, EV prevalence, and scarcity alignment, and to
inform plausible product archetypes and population shares. The \emph{final}
product-level demand time series used in the experiments is then generated by a
controlled, reproducible synthesiser (Appendix~\ref{app:residential_synth}),
which preserves those insights while avoiding the use of raw household traces
for any form of personalised pricing.

\paragraph{Data governance and researcher accreditation.}
The author is an \textbf{ONS Accredited Researcher} under the UK Digital Economy
Act (DEA). Although access to ONS Secure Research Service datasets (notably SERL)
was ultimately not granted for this thesis, all data handling, storage, and
analysis practices followed the \textbf{Five Safes framework} (Safe Projects,
Safe People, Safe Data, Safe Settings, Safe Outputs).

All datasets used were either publicly available, accessed under formal data
sharing agreements, or provided in fully anonymised form. No personally
identifiable information was accessed, inferred, or reconstructed at any stage.

% =========================================================
\section{Choice of Household Consumption Dataset}
\label{sec:dataset_rationale}
% =========================================================

The original intention was to use the \textbf{UK SERL} household dataset
(\url{https://serl.ac.uk/}), accessed via the ONS Secure Research Service, which
contains rich demographic, device-level, and socio-economic attributes alongside
high-frequency consumption data. ONS Accredited Researcher training was completed,
and formal applications were submitted through both Imperial College London and
the SERL data access process. However, due to access and clearance restrictions, 
the SERL dataset was ultimately
not available for use in this thesis, despite the author holding ONS Accredited
Researcher status under the UK Digital Economy Act. Consequently, demand-side
development proceeded using publicly available smart-meter and statistical
sources, first to characterise empirical structure and EV augmentation
(Appendix~\ref{app:ev_holarchy}), and then to generate the reproducible
experiment demand inputs used throughout the simulations
(Appendix~\ref{app:residential_synth}).

Therefore, the \textbf{London Low Carbon (LCL) / UKPN Smart Meter}
dataset (2011--2014) was used which consists of publicly available, and containing half-hourly household
consumption data for 5,567 anonymised homes. While older, the dataset remains
valuable for reconstructing:
\begin{itemize}
    \item intra-day diversity (peak/off-peak behaviour),
    \item consumption shape archetypes (evening-peakers, daytime-solar consumers, flat profiles, etc.),
    \item peak-to-baseload ratios and clustering validity for product classes.
\end{itemize}

The main structural change in household demand patterns since 2014 is EV adoption.
Therefore, empirical EV charging profiles are overlaid and ownership distributions
(from Department for Transport (DoT), vehicle licensing statistics, and DoT charging
trials) onto the UKPN time series in a representative manner. Annual total energy
per postcode (BEIS) is preserved exactly, ensuring calibration to 2023–2024
consumption conditions.

This approach allows the creation of \textbf{synthetic-yet-plausible time-series
profiles for all 29 million GB households} while maintaining:
\begin{itemize}
    \item realistic diurnal shapes,
    \item cluster-level and postcode-level totals (BEIS),
    \item EV-rich behaviour consistent with 2024 conditions.
\end{itemize}

% =========================================================
\section{UKPN / LCL Smart Meter Dataset}
\label{sec:ukpn_dataset}
% =========================================================

\textbf{Source:}
\url{https://data.london.gov.uk/dataset/smartmeter-energy-use-data-in-london-households}

Half-hourly electricity consumption data for 5,567 anonymised London-region
households from November 2011 to February 2014. Contains individual time-series
for both weekday/weekend and seasonal variations.

\begin{table}[H]
\centering
\caption{Summary statistics for the UKPN Smart Meter dataset.}
\label{tab:ukpn_summary}
\begin{tabular}{lcc}
\toprule
Metric & Value & Notes \\
\midrule
Households & 5,567 & After quality filtering \\
Sampling interval & 30 mins & Settlement-aligned \\
Observation span & 2011--2014 & Not all households continuous \\
Missingness rate & $\sim$8\% & Imputed where feasible \\
Median daily kWh & (to be inserted) & Representative household \\
Peak-to-average ratio & (to be inserted) & Load diversity indicator \\
\bottomrule
\end{tabular}
\end{table}

Used for:
\begin{itemize}
    \item Deriving behavioural demand archetypes,
    \item Validating stylised load products P1–P4,
    \item Household-to-cluster scaling via BEIS postcode data.
\end{itemize}

% =========================================================
\section{BEIS Postcode-Level Annual Consumption}
\label{sec:beis_dataset}
% =========================================================

\textbf{Source:}
\url{https://www.gov.uk/government/statistics/energy-consumption-in-the-uk-2023}

Annual electricity consumption (kWh) for all 29.8 million domestic and
non-domestic meters, aggregated by postcode.

\begin{itemize}
    \item Used to scale synthetic time-series to preserve annual totals;
    \item Enables accurate spatial distribution across postcode, LAD, cluster, and region;
    \item Preserves realistic socio-spatial demand heterogeneity.
\end{itemize}

\begin{table}[H]
\centering
\caption{BEIS postcode consumption dataset summary.}
\label{tab:beis_summary}
\begin{tabular}{lcc}
\toprule
Metric & Value & Notes \\
\midrule
Years & 2015--2023 & Official releases \\
Meters represented & $\sim$29.8 million & Domestic + I\&C \\
Postcodes represented & $\sim$1.7 million & Full postcodes \\
Total energy (GB) & (to insert) & e.g. 270--310 TWh \\
\bottomrule
\end{tabular}
\end{table}

% =========================================================
\section{EV Ownership Data}
\label{sec:ev_ownership}
% =========================================================

\textbf{Source:}
\url{https://www.gov.uk/government/statistical-data-sets/vehicle-licensing-statistics-data-tables}

Vehicle licensing records by Local Authority and postcode. Used to estimate EV
adoption intensity by region and cluster, and to infer household EV penetration.

\begin{itemize}
    \item Allocation of EV ownership per postcode/cluster,
    \item Used to adjust synthetic household load profiles,
    \item Supports fairness in burden-sharing under AMM.
\end{itemize}

% =========================================================
\section{EV Charging Behaviour and Session Profiles}
\label{sec:ev_profiles}
% =========================================================

\textbf{Source:}
\url{https://www.data.gov.uk/dataset/5438d88d-695b-4381-a5f2-6ea03bf3dcf0/electric-chargepoint-analysis-2017-domestics}

Contains plug-in and plug-out times, energy per session, charging rate,
arrival distributions, and inferred flexibility windows.

Used for:
\begin{itemize}
    \item Constructing charging flexibility models,
    \item Creating overlay demand peaks and deferred charging behaviour,
    \item Allocating EV-related flexibility in fairness (F1--F4) experiments.
\end{itemize}

% =========================================================
\section{Generator Metadata (OSUKED Power Station Dictionary)}
\label{sec:generator_metadata}
% =========================================================

\textbf{Source:}
\url{https://github.com/OSUKED/Power-Station-Dictionary/tree/shiro}

Provides metadata for GB generating units: fuel type, location, capacity,
operator, commissioning year, geographic connection point.

Used to:
\begin{itemize}
    \item Locate generators in synthetic grid topology,
    \item Define spatial imbalances (e.g. wind in Scotland),
    \item Assign unit-level attributes for Shapley value attribution.
\end{itemize}

% =========================================================
\section{Elexon BMRS Generation Output}
\label{sec:bmrs_dataset}
% =========================================================

\textbf{Source:} \url{https://data.elexon.co.uk/bmrs/api/v1/balancing/physical}

Half-hourly generation output by BM Unit (sett\_bmu\_id), fuel type, region, and
settlement period.

\begin{table}[H]
\centering
\caption{Key attributes of BMRS generator dataset.}
\begin{tabular}{lcc}
\toprule
Metric & Value & Notes \\
\midrule
Sampling interval & 30 minutes & Settlement periods \\
Number of BM units & $\sim$850 & Across GB grid \\
Fuel types & $\sim$12 & CCGT, wind, nuclear, solar, storage, etc. \\
Total MWh/year & (insert) & Shapley allocation input \\
\bottomrule
\end{tabular}
\end{table}

% =========================================================
\section{Weather Data for Normalisation (MetOffice)}
\label{sec:weather_data}
% =========================================================

Used to temperature-adjust historical UKPN data to reflect 2023–2024 conditions.
Daily and hourly temperature and degree-day data.

\textit{This adjustment step is acknowledged as approximate (best effort) but improves representational alignment without materially affecting market-clearing results.}

% =========================================================
\section{GeoJSON Spatial Datasets}
\label{sec:geojson}
% =========================================================

\textbf{Postcode Boundaries:}  
\url{https://github.com/missinglink/uk-postcode-polygons}

\textbf{Local Authority District Boundaries:}  
\url{https://github.com/martinjc/UK-GeoJSON/tree/master}

Used for:
\begin{itemize}
    \item Mapping households to geographic units,
    \item Deriving custom simulation clusters (Layer 1--3),
    \item Visualisation and spatial fairness analysis.
\end{itemize}

\begin{figure}[H]
\centering
 \includegraphics[width=\textwidth]{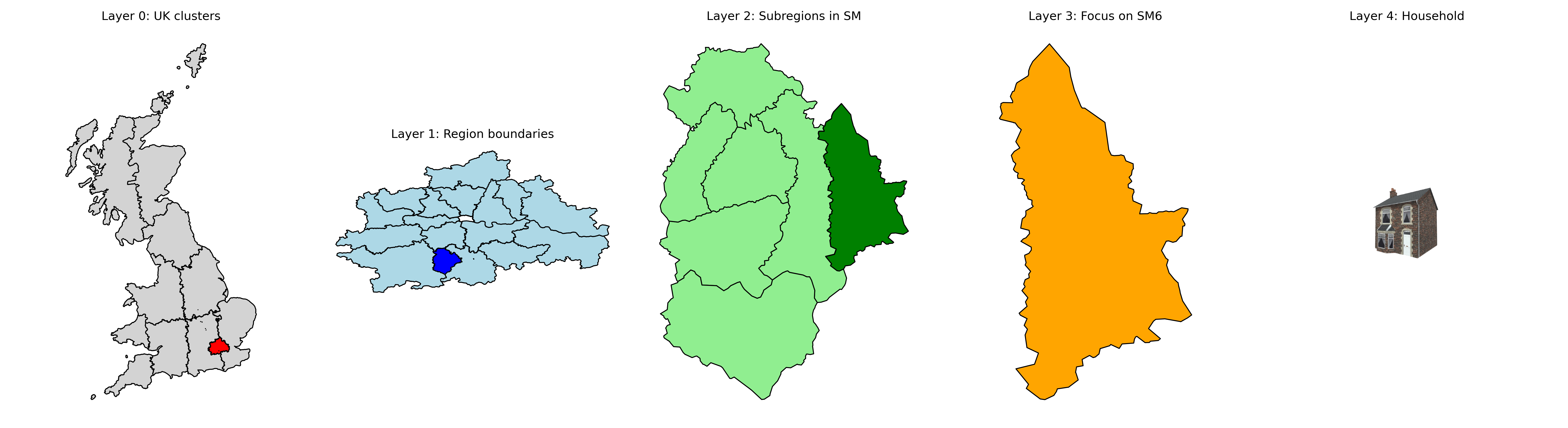}
\caption{Illustrative view of postcode–to–cluster mapping across multiple spatial layers. 
Importantly, these layers do \emph{not} represent a fixed hierarchy. 
Under the holarchic model, aggregation levels are \textbf{dynamic, purpose-specific, and user-dependent}: 
different actors (e.g., DSOs, suppliers, households, regulators) may aggregate, disaggregate, or bypass layers according to their operational or analytical needs.}
\label{fig:spatial_mapping}
\end{figure}

% =========================================================
\section{Device-Level Flexibility Dataset (Moixa / Lunar Energy)}
\label{sec:moixa_dataset}
% =========================================================

\textbf{Source and governance.}
This thesis makes use of a device-level flexibility dataset provided under a
\textbf{Data Sharing Agreement (DSA)} between Imperial College London and
\textbf{Moixa Technology Ltd}, now operating as \textbf{Lunar Energy}. The dataset
was supplied exclusively for academic research purposes and is not publicly
available.

All records were fully anonymised prior to access. The dataset contains no
personally identifiable information and no geographic identifiers beyond
country-level (UK). No attempt was made to infer location, household identity,
or socio-demographic attributes.

Data handling and analysis complied with the Five Safes principles, with analysis
performed exclusively in secure academic environments and outputs reviewed to
ensure no disclosure risk.

\paragraph{Dataset description.}
The dataset comprises high-frequency operational data for approximately
\textbf{100 UK households} equipped with residential battery storage systems.
Measurements are recorded at \textbf{15-second resolution} at the meter / device
level and include:

\begin{itemize}
    \item Household net electricity demand,
    \item On-site PV generation (where present),
    \item Battery state of charge,
    \item Battery charge and discharge power,
    \item Device availability and control states.
\end{itemize}

The data therefore provides a direct observation of \emph{realised flexibility}
at the device level, including both energy-shifting and peak-shaving behaviour
under realistic operating constraints.

\paragraph{Role in this thesis.}
The Moixa dataset is used \emph{only} for \textbf{behavioural and operational
validation experiments}, specifically:

\begin{itemize}
    \item Evaluating how realised unit costs respond to increasing flexibility
          windows (parameterised by scheduling horizon $\sigma$),
    \item Demonstrating that flexibility is rewarded \emph{operationally} under
          the AMM mechanism (F1 and F4),
    \item Validating device-level scheduling and settlement behaviour independently
          of subscription construction and national scaling.
\end{itemize}

Importantly, the dataset is \textbf{not} used for:
\begin{itemize}
    \item National demand synthesis,
    \item Spatial allocation or postcode-level analysis,
    \item Calibration of household clusters or subscription levels.
\end{itemize}

Those functions rely instead on the UKPN/LCL smart meter dataset and BEIS
postcode-level statistics (Sections~\ref{sec:ukpn_dataset} and
\ref{sec:beis_dataset}).

\paragraph{Limitations.}
The dataset is non-location-specific and limited in sample size. As such, it is
not interpreted as statistically representative of the GB household population.
Its value lies instead in providing \emph{high-fidelity ground truth} for
device-level flexibility behaviour, enabling controlled experiments that are
otherwise impossible with aggregated smart-meter data.

This distinction is reflected throughout the thesis: the Moixa dataset supports
\emph{mechanism validation}, while national-scale results rely on synthetic
expansion calibrated to public datasets.

% =========================================================
\section{Summary of roles across datasets}
\label{sec:dataset_roles_summary}
% =========================================================

Each dataset used in this thesis serves a \emph{distinct, non-overlapping role}
within the modelling, validation, and fairness-analysis pipeline. No single
dataset is relied upon to do more than it is empirically suited for; instead,
their roles are deliberately separated to avoid overfitting, spurious precision,
or implicit personalised pricing.

\begin{itemize}[leftmargin=*]
    \item \textbf{UKPN / LCL smart meter data:}  
    provides empirical behavioural and temporal diversity of household electricity
    consumption. Its role is to reveal realistic intra-day shapes, seasonal
    structure, and heterogeneity across households, forming the empirical basis
    for demand archetypes and the qualitative definition of stylised retail
    products ($P1$--$P4$).

    \item \textbf{BEIS postcode-level consumption statistics:}  
    enforce spatial scaling and aggregate alignment. These data anchor the model
    to official annual electricity totals, ensuring that synthetic household
    demand preserves correct energy magnitudes across postcodes, clusters, and
    regions.

    \item \textbf{EV ownership and charging datasets:}  
    introduce empirically grounded electric vehicle adoption rates, charging
    behaviour, and flexibility envelopes. These datasets enable future-facing
    demand construction and controlled experiments on demand-side flexibility and
    fairness (F1--F4).

    \item \textbf{Moixa / Lunar Energy device-level dataset:}  
    provides high-frequency, ground-truth observations of realised residential
    flexibility (battery operation and PV interaction). It is used exclusively
    for \emph{behavioural and operational validation} of flexibility rewards under
    the AMM mechanism, and is not used for national scaling or subscription
    construction.

    \item \textbf{BMRS generation output and OSUKED metadata:}  
    define generator-level production, fuel type, and geographic location. These
    datasets support physically grounded dispatch, congestion analysis, and
    Shapley-based attribution of system value, costs, and revenues.

    \item \textbf{GeoJSON spatial datasets:}  
    enable mapping between households, postcodes, clusters, and regions. Their
    role is purely structural, supporting spatial aggregation, visualisation, and
    geographic fairness analysis within the holarchic framework.

    \item \textbf{Met Office weather data:}  
    provide approximate temperature normalisation of historical demand profiles,
    improving alignment with contemporary operating conditions without materially
    affecting market-clearing or fairness results.
\end{itemize}

Taken together, these datasets support a clear separation between:
(i) \emph{behavioural realism} (what households and devices actually do),
(ii) \emph{spatial and aggregate consistency} (where and how much energy is used),
and
(iii) \emph{mechanism validation} (how costs, risks, and rewards are allocated).

This separation ensures that the fairness results reported in the thesis arise
from market design and allocation logic, rather than artefacts of data choice,
granularity, or representational bias.

% =========================================================
\chapter{Input data parameters for generators, demand and network used in experiment}
\label{app:inputs}
% =========================================================

This appendix documents the physical network, generator and load data, and the
unit-commitment and market-clearing configurations used in all simulations.
These inputs are held fixed across the Baseline (LMP) and Treatment (AMM)
designs so that observed differences in outcomes arise from the clearing logic
and remuneration structure rather than from differences in the underlying
system.

% =========================================================
\section{Network Topology and Electrical Parameters}
\label{app:network}
% =========================================================

The experiments use a stylised 12--node transmission network with explicit
thermal limits, line reactances, and geographic layout. Nodes are labelled
$N0,\dots,N34$.

\begin{figure}[H]
  \centering
  \includegraphics[width=0.9\textwidth]{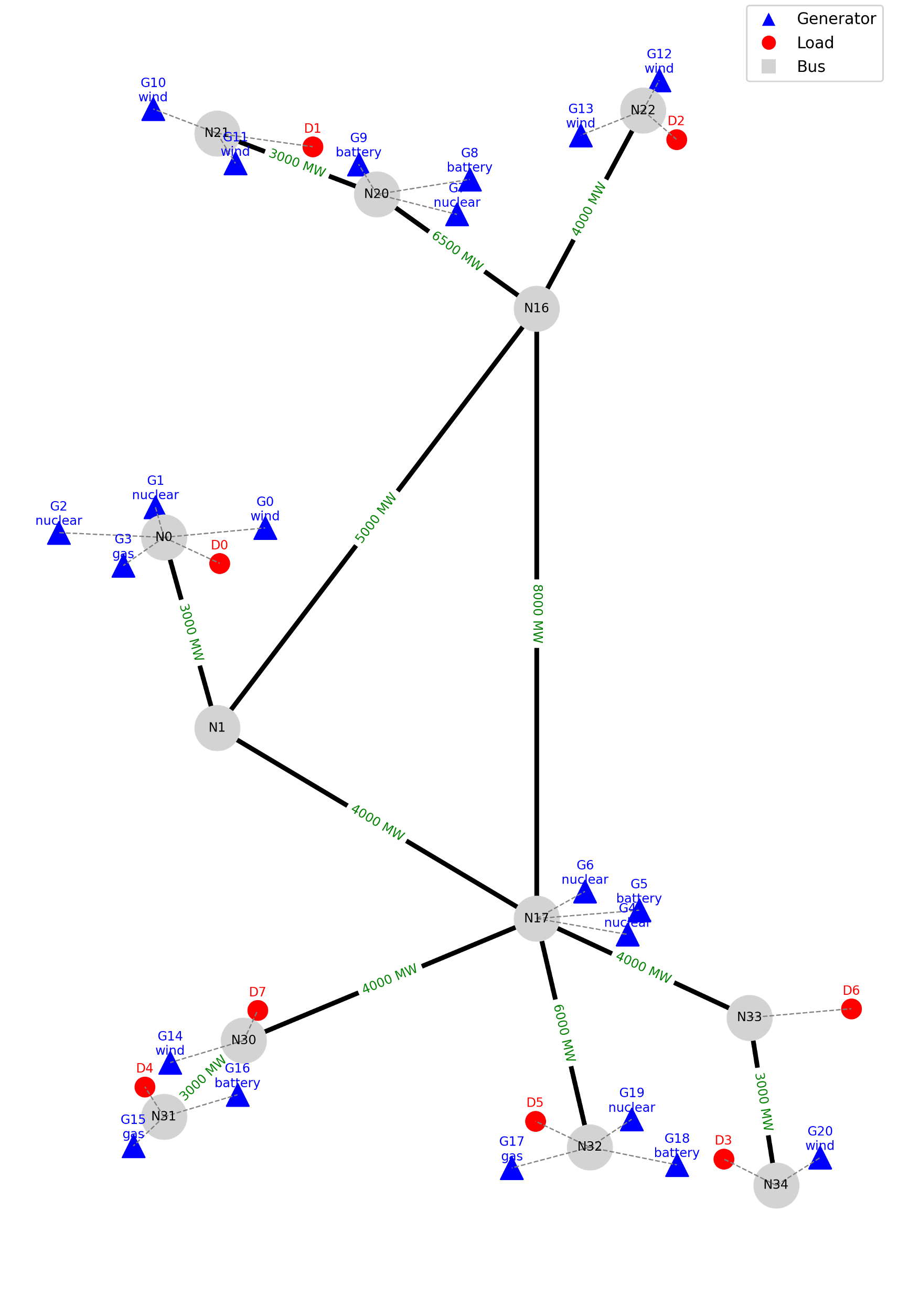}
  \caption[Simulation network topology]{
    Simulation network topology with generators (blue triangles),
    loads (red circles), and buses (grey nodes). Thermal line limits 
    (MW) are shown along each corridor; parallel units at a node are drawn
    separately for clarity.
  }
  \label{fig:network_topology}
\end{figure}

Figure~\ref{fig:network_topology} provides the reference 12--bus transmission
network on which all experimental results are evaluated. The system comprises
12 nodes, with
line capacities, voltages, and reactances taken directly from the dataset
summarised below. Generators are located at
$\{N0, N17, N20, N21, N22, N30, N31, N32, N34\}$ and loads at
$\{N0, N21, N22, N31, N32, N33, N34\}$. Generator labels denote technology
class (wind, nuclear, gas, battery), while edge labels report the thermal
limits (MW) of each transmission corridor.

\begin{alltt}\small
{
  "nodes": [
    "N0","N1","N16","N17","N20","N21","N22",
    "N30","N31","N32","N33","N34"
  ],

  "edges": [
    ["N0","N1"], ["N1","N16"], ["N16","N17"], ["N1","N17"],
    ["N16","N20"], ["N20","N21"], ["N16","N22"],
    ["N17","N30"], ["N30","N31"], ["N17","N32"],
    ["N17","N33"], ["N33","N34"]
  ],

  "edge_capacity": {
    "N0,N1": 3000,
    "N1,N16": 5000,
    "N16,N17": 8000,
    "N1,N17": 4000,
    "N16,N20": 6500,
    "N20,N21": 3000,
    "N16,N22": 4000,
    "N17,N30": 4000,
    "N30,N31": 3000,
    "N17,N32": 6000,
    "N17,N33": 4000,
    "N33,N34": 3000
  },

  "edge_voltage_kV": {
    "N0,N1": 400, "N1,N0": 400,
    "N1,N16": 400, "N16,N1": 400,
    "N16,N17": 400, "N17,N16": 400,
    "N1,N17": 400, "N17,N1": 400,
    "N16,N20": 275, "N20,N16": 275,
    "N20,N21": 275, "N21,N20": 275,
    "N16,N22": 275, "N22,N16": 275,
    "N17,N30": 275, "N30,N17": 275,
    "N30,N31": 132, "N31,N30": 132,
    "N17,N32": 275, "N32,N17": 275,
    "N17,N33": 275, "N33,N17": 275,
    "N33,N34": 132, "N34,N33": 132
  },

  "edge_length_km": {
    "N0,N1": 200, "N1,N0": 200,
    "N1,N16": 350, "N16,N1": 350,
    "N16,N17": 500, "N17,N16": 500,
    "N1,N17": 350, "N17,N1": 350,
    "N16,N20": 200, "N20,N16": 200,
    "N20,N21": 150, "N21,N20": 150,
    "N16,N22": 120, "N22,N16": 120,
    "N17,N30": 120, "N30,N17": 120,
    "N30,N31": 60,  "N31,N30": 60,
    "N17,N32": 140, "N32,N17": 140,
    "N17,N33": 150, "N33,N17": 150,
    "N33,N34": 70,  "N34,N33": 70
  },

  "edge_reactance_pu": {
    "N0,N1": 0.3750, "N1,N0": 0.3750,
    "N1,N16": 0.6562, "N16,N1": 0.6562,
    "N16,N17": 0.9375, "N17,N16": 0.9375,
    "N1,N17": 0.6562, "N17,N1": 0.6562,
    "N16,N20": 1.0579, "N20,N16": 1.0579,
    "N20,N21": 0.7934, "N21,N20": 0.7934,
    "N16,N22": 0.6347, "N22,N16": 0.6347,
    "N17,N30": 0.6347, "N30,N17": 0.6347,
    "N30,N31": 2.0661, "N31,N30": 2.0661,
    "N17,N32": 0.7405, "N32,N17": 0.7405,
    "N17,N33": 0.7934, "N33,N17": 0.7934,
    "N33,N34": 2.4105, "N34,N33": 2.4105
  }
}
\end{alltt}

\subsection{Node Set and Layout}

The node set and plotting coordinates used for visualisation are given in
Table~\ref{tab:nodes_positions}. Coordinates are dimensionless layout
positions used for figures, not geographic lat/long.

\begin{table}[H]
\centering
\caption{Nodes and layout coordinates.}
\label{tab:nodes_positions}
\begin{tabular}{lcc}
\toprule
Node & $x$ & $y$ \\
\midrule
N0  & 0.8 & 9.0 \\
N1  & 1.0 & 6.5 \\
N16 & 2.2 & 12.0 \\
N17 & 2.2 & 4.0 \\
N20 & 1.6 & 13.5 \\
N21 & 1.0 & 14.3 \\
N22 & 2.6 & 14.6 \\
N30 & 1.1 & 2.4 \\
N31 & 0.8 & 0.9 \\
N32 & 2.4 & 0.5 \\
N33 & 3.0 & 2.7 \\
N34 & 3.1 & 0.2 \\
\bottomrule
\end{tabular}
\end{table}

% ---------------------------------------------------------
\section{Network topology and line parameters}
\label{app:network_topology}

Figure~\ref{fig:network_topology} shows the simplified 12--bus transmission
network used in all simulations. The graph has nodes
$N0, N1, N16, N17, N20, N21, N22, N30, N31, N32, N33, N34$, with
line capacities, voltages, and reactances taken directly from Section~\ref{app:network}.
Generators are located at
nodes $\{N0,N17,N20,N21,N22,N30,N31,N32,N34\}$; loads are attached at
$\{N0,N21,N22,N31,N32,N33,N34\}$. Generator labels indicate technology
type (wind, nuclear, gas, battery) and bus, while edge labels report the
thermal capacity (MW) of each corridor.

The corresponding numerical data are encoded in the \texttt{network\_uk.json}
file:

\begin{itemize}[leftmargin=*]
  \item \texttt{nodes}: list of bus identifiers;
  \item \texttt{edges}: undirected adjacency list for transmission lines;
  \item \texttt{edge\_capacity}: thermal limits in MW;
  \item \texttt{edge\_voltage\_kV}: nominal voltage level of each corridor;
  \item \texttt{edge\_length\_km}: assumed line length in kilometres;
  \item \texttt{edge\_reactance\_pu}: per-unit series reactance at
        \texttt{sbase\_MVA = 1000};
  \item \texttt{generators}: mapping from generator IDs $G0\ldots G20$
        to buses and nameplate capacity (MW);
  \item \texttt{loads}: mapping from demand points $D0\ldots D7$ to buses
        and peak demand (MW);
  \item \texttt{positions}: $(x,y)$ coordinates used only for plotting.
\end{itemize}

\subsection{Transmission Corridors}

Table~\ref{tab:edges} lists all transmission corridors as undirected edges
between nodes, together with their thermal capacity, nominal voltage level,
line length, and per-unit reactance (on an \texttt{sbase} of 1000~MVA).
Edge attributes are symmetric in both directions.

\begin{table}[H]
\centering
\caption{Transmission corridors and electrical parameters.}
\label{tab:edges}
\begin{tabular}{llcccc}
\toprule
From & To & Capacity [MW] & Voltage [kV] & Length [km] & Reactance [p.u.] \\
\midrule
N0  & N1  & 3000 & 400 & 200 & 0.3750 \\
N1  & N16 & 5000 & 400 & 350 & 0.6562 \\
N16 & N17 & 8000 & 400 & 500 & 0.9375 \\
N1  & N17 & 4000 & 400 & 350 & 0.6562 \\
N16 & N20 & 6500 & 275 & 200 & 1.0579 \\
N20 & N21 & 3000 & 275 & 150 & 0.7934 \\
N16 & N22 & 4000 & 275 & 120 & 0.6347 \\
N17 & N30 & 4000 & 275 & 120 & 0.6347 \\
N30 & N31 & 3000 & 132 &  60 & 2.0661 \\
N17 & N32 & 6000 & 275 & 140 & 0.7405 \\
N17 & N33 & 4000 & 275 & 150 & 0.7934 \\
N33 & N34 & 3000 & 132 &  70 & 2.4105 \\
\bottomrule
\end{tabular}
\end{table}

These parameters are used consistently in both the LMP and AMM formulations
for DC power flow and congestion representation.

% =========================================================
\section{Generator Fleet}
\label{app:generators}
% =========================================================

\begin{figure}[H]
  \centering
  \includegraphics[width=\textwidth]{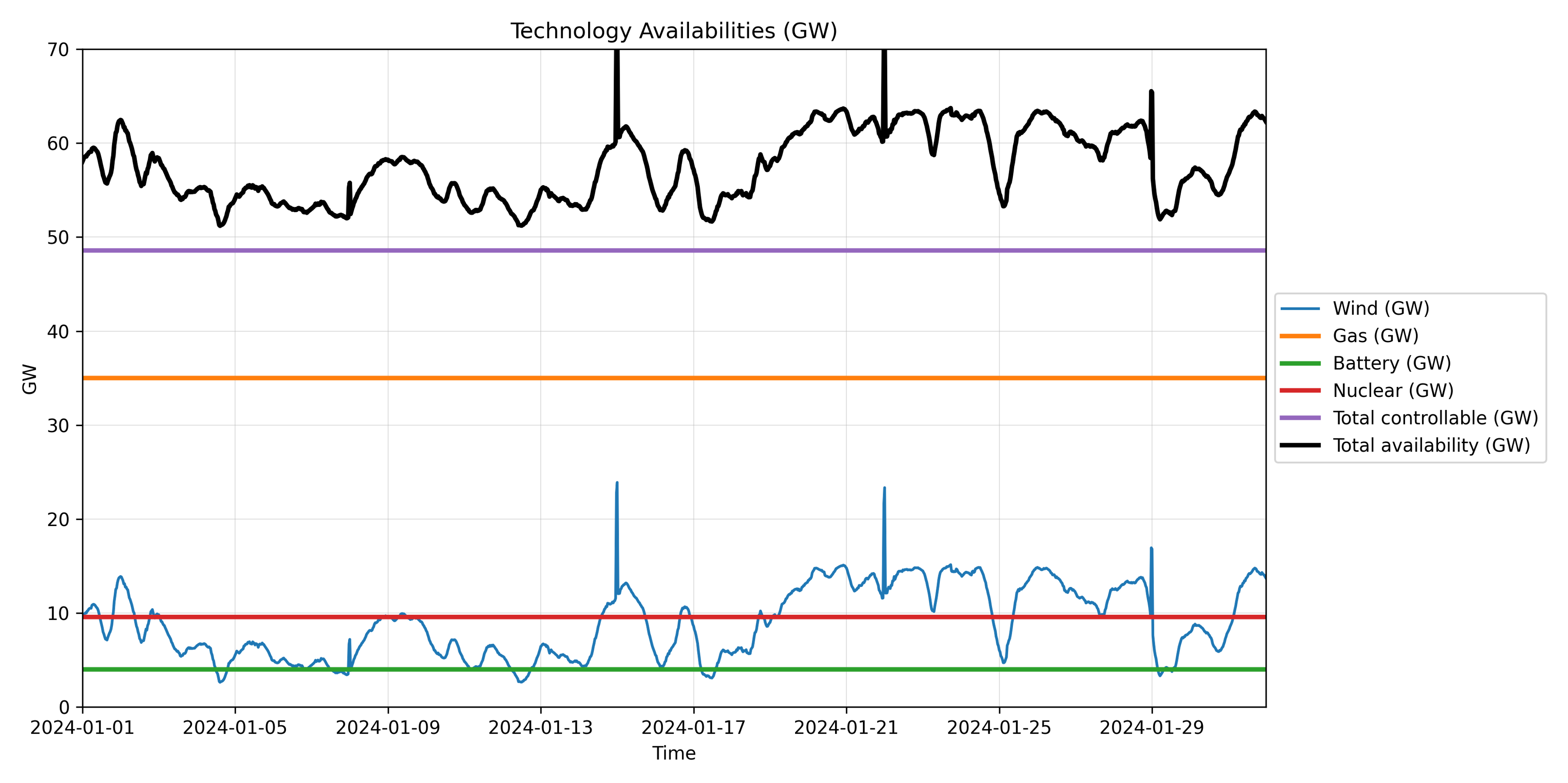}
  \caption[Generator availability by unit and technology]{Available capacity by generator, grouped by technology type and node.}
  \label{fig:availability_by_gen}
\end{figure}

The generator fleet consists of 21 units connected to specific network nodes,
each with a fixed nameplate capacity used as the maximum dispatch in the
unit-commitment and dispatch problems. Figure~\ref{fig:availability_by_gen}
summarises the resulting nameplate availability by technology and node.

\begin{table}[H]
\centering
\caption{Generator units, node assignments, and technology labels.}
\label{tab:generators}
\begin{tabular}{lccc}
\toprule
Generator & Node & Capacity [MW] & Technology \\
\midrule
G0  & N0  &  1500 & Wind\\
G1  & N0  &  2400 & Nuclear\\
G2  & N0  &  2405 & Nuclear \\
G3  & N0  & 12000 & Gas \\
G4  & N17 &  1200 & Nuclear\\
G5  & N17 &   500 & Battery\\
G6  & N17 &  1170 & Nuclear\\
G7  & N20 &  1200 & Nuclear\\
G8  & N20 &  1000 & Battery\\
G9  & N20 &  1000 & Battery\\
G10 & N21 &  6000 & Wind \\
G11 & N21 &  6000 & Wind \\
G12 & N22 &  3000 & Wind \\
G13 & N22 &  6000 & Wind\\
G14 & N30 &  3000 & Wind \\
G15 & N31 &  8000 & Gas \\
G16 & N31 &  1000 & Battery \\
G17 & N32 & 15000 & Gas \\
G18 & N32 &   500 & Battery \\
G19 & N32 &  1200 & Nuclear \\
G20 & N34 &  4500 & Wind \\
\bottomrule
\end{tabular}
\end{table}

Technology labels (wind, nuclear, gas, battery, \dots) and their
associated \emph{bid and cost parameters} --- fuel cost, reserve
eligibility, CapEx, non-fuel OpEx, and target payback period --- are
specified at the technology level and then applied to all units of that
type. The values used in all simulations are summarised in
Table~\ref{tab:tech_cost_params}, with additional battery-specific
assumptions given in Table~\ref{tab:battery_params}. These cost
parameters feed into both the unit-commitment / dispatch formulation and
the Shapley-based availability and remuneration scripts described in
Chapter~\ref{ch:amm} and Appendix~\ref{app:extended_results}. Nuclear and
wind units are treated as cost-recovery resources: they receive fixed
annual remuneration equal to their non-fuel OpEx plus annualised CapEx,
and do not participate in the Shapley availability pot, whereas gas and
battery units are fully exposed to scarcity prices and Shapley-based
allocation.

\subsection{Generator Operational and Cost Parameters}
\label{app:generator_cost_params}

In the unit-commitment and dispatch problems, generators are
parameterised at the \emph{technology} level. All units of a given
technology share the same minimum stable output, minimum up/down times,
reserve eligibility, fuel cost, and capital and operating cost
assumptions; individual unit capacities and locations are given in
Table~\ref{tab:generators}. The resulting operational and cost
parameters are summarised in Table~\ref{tab:tech_cost_params}.

\begin{sidewaystable}[p]
\centering
\scriptsize
\caption{Technology-level operational and cost parameters used for generator units. Max/min power and minimum up/down times are specified at the technology level; individual unit capacities are listed in Table~\ref{tab:generators}. CapEx values are overnight capital costs per MW; non-fuel OpEx values are annual fixed O\&M per MW.}
\label{tab:tech_cost_params}
\renewcommand{\arraystretch}{1.2}
\begin{tabular}{lccccccccc}
\toprule
\textbf{Technology} &
$P^{\max}$ [GW] &
$P^{\min}$ [GW] &
Min-up [h] &
Min-down [h] &
Reserve? &
Fuel cost [£/MWh] &
CapEx [£m/MW] &
Non-fuel OpEx [£k/MW] &
Payback [years] \\
\midrule
Wind    & 30.9\textsuperscript{*} & 0.20 & $\infty$ & $\infty$ & 0 & 0   & 1.90 &  50.0 & 20 \\
Nuclear &  9.5                     & 4.75 & 24       & 24       & 0 & 12  & 7.00 & 160.0 & 40 \\
Gas     & 35.0                    & 7.00 & 4        & 4        & 1 & 90  & 0.70 &  20.0 & 15 \\
Battery &  4.0                    & 0.00 & 0.25     & 0.25     & 1 & 70  & 0.85 &  17.5 & 15 \\
\bottomrule
\end{tabular}

\vspace{0.3em}
\raggedright\footnotesize
\textsuperscript{*}For wind, the ``Max power'' entry reflects the maximum
available output over the historical availability series used in the
experiments, i.e.\ the fleet-level nameplate envelope for the simulated horizon.
\normalsize
\end{sidewaystable}

\subsection{Additional Battery Parameters}

Battery units share common energy-capacity and efficiency assumptions,
summarised in Table~\ref{tab:battery_params}. These parameters apply to
all battery generators $\{G5,G8,G9,G16,G18\}$ listed in
Table~\ref{tab:generators}.

\begin{table}[H]
\centering
\caption{Additional technology-level parameters for battery storage units.}
\label{tab:battery_params}
\renewcommand{\arraystretch}{1.2}
\begin{tabular}{lcc}
\toprule
\textbf{Parameter} & \textbf{Value} & \textbf{Units} \\
\midrule
Energy capacity                      & 8      & GWh \\
Max charge rate                      & 4      & GW \\
Max discharge rate                   & 4      & GW \\
Charge efficiency                    & 0.95   & -- \\
Discharge efficiency                 & 0.95   & -- \\
Minimum state of charge              & 0.05   & fraction of energy capacity \\
Maximum state of charge              & 0.95   & fraction of energy capacity \\
\bottomrule
\end{tabular}
\end{table}

% =========================================================
\section{Load Data}
\label{app:loads}
% =========================================================

Static nodal loads used for power-flow feasibility are given in
Table~\ref{tab:loads}. These are base load levels; time-varying profiles
and product-level decomposition are handled in the demand modelling pipeline
described in Chapter~\ref{ch:experiments}.

\begin{table}[H]
\centering
\caption{Nodal load assignments.}
\label{tab:loads}
\begin{tabular}{lcc}
\toprule
Load & Node & Power [MW] \\
\midrule
D0 & N0  & 200 \\
D1 & N21 & 130 \\
D2 & N22 &  70 \\
D3 & N34 &  30 \\
D4 & N31 & 130 \\
D5 & N32 &  30 \\
D6 & N33 &   0 \\
D7 & N30 &   0 \\
\bottomrule
\end{tabular}
\end{table}

These nodal loads are consistent across LMP and AMM runs; demand-side product
bundles (P1--P4) and household counts are documented in
Section~\ref{app:demand_inputs} below.

% =========================================================
\section{Unit-Commitment and Market-Clearing Configuration}
\label{app:market_config}
% =========================================================

Both designs use the same optimisation engine (\texttt{HiGHS}) with a
mixed-integer (for LMP) or continuous (for AMM) formulation. This section
summarises the key configuration parameters.

\subsection{Common Solver and System Parameters}

The following settings are common across all experiments unless otherwise
noted:

\begin{table}[H]
\centering
\caption{Common configuration parameters (LMP and AMM).}
\label{tab:common_config}
\begin{tabular}{lc}
\toprule
Parameter & Value \\
\midrule
\texttt{solver} & \texttt{"highs"} \\
\texttt{solver\_time\_limit\_s} & 600 \\
\texttt{solver\_mip\_gap} & 0.02 \\
\texttt{disable\_min\_updown} & \texttt{true} \\
\texttt{single\_pass\_objective} & \texttt{true} \\
\texttt{reserve\_requirement\_percent} & 10.0 \\
\texttt{reserve\_on\_served\_demand} & \texttt{true} \\
\texttt{reserve\_slack\_penalty\_per\_MW} & 1000.0 \\
\texttt{battery\_tech\_labels} & \{\texttt{"battery","Battery","BATTERY"}\} \\
\texttt{battery\_eta\_charge} & 0.95 \\
\texttt{battery\_eta\_discharge} & 0.95 \\
\texttt{battery\_exclusive\_mode} & \texttt{true} \\
\texttt{must\_run\_mode} & \texttt{"soft"} \\
\texttt{must\_run\_tech\_labels} & \{\texttt{"nuclear"}\} \\
\texttt{must\_run\_gen\_ids} & [] (empty) \\
\texttt{must\_run\_off\_penalty\_per\_hour} & 1,000,000.0 \\
\texttt{sbase\_MVA} & 1000.0 \\
\texttt{target\_end\_ts} & \texttt{"2024-12-31 23:30"} \\
\texttt{reserve\_availability\_price\_per\_MW\_h} & 7.5 \\
\texttt{reserve\_availability\_price\_units} & \texttt{"currency\_per\_MW\_h"} \\
\bottomrule
\end{tabular}
\end{table}

These settings ensure that adequacy, reserve requirements, and nuclear
must-run behaviour are treated consistently across the Baseline and Treatment.

\subsection{Baseline LMP Configuration}

The Baseline LMP configuration uses binary unit-commitment variables and a
shorter look-ahead window. The full configuration JSON is:

\begin{alltt}\small
{
  "solver": "highs",
  "solver_time_limit_s": 600,
  "solver_mip_gap": 0.02,

  "use_binary_commitment": true,
  "disable_min_updown": true,
  "single_pass_objective": true,

  "uc_window_hours": 48,
  "uc_commit_hours": 24,
  "disallow_late_starts": true,

  "reserve_requirement_percent": 10.0,
  "reserve_on_served_demand": true,
  "reserve_slack_penalty_per_MW": 1000.0,
  "reserve_shortfall_cost_per_MW": 1000.0,
  "reserve_allow_battery_drop_charge": true,

  "voll_MWh": 9999.0,
  "spill_penalty_per_MWh": 5.0,

  "battery_tech_labels": ["battery", "Battery", "BATTERY"],
  "battery_eta_charge": 0.95,
  "battery_eta_discharge": 0.95,
  "battery_exclusive_mode": true,

  "must_run_mode": "soft",
  "must_run_tech_labels": ["nuclear"],
  "must_run_off_penalty_per_hour": 1000000.0,

  "sbase_MVA": 1000.0,
  "target_end_ts": "2024-12-31 23:30",

  "battery_carry_soc_across_days": true,
  "battery_da_energy_neutral": true,
  "battery_da_energy_neutral_hard": true,
  "battery_terminal_soc_frac": null,
  "battery_terminal_soc_penalty_per_MWh": 5.0,
  "battery_rt_energy_neutral": true,
  "battery_rt_energy_neutral_hard": true,
  "battery_rt_terminal_soc_frac": null,
  "battery_rt_terminal_soc_penalty_per_MWh": 2000,
  "battery_profile_caps_discharge": false,
  "battery_cycle_cost_MWh": 5.0,

  "local_first_export_cost_MWh": 1.0,
  "local_first_import_cost_MWh": 0.0,

  "pricing_eps": 0.1,
  "rt_fd_epsilon_MW": 0.1,
  "settlement_price_cap_MWh": 6000.0,
  "reserve_availability_price_per_MW_h": 7.5,
  "reserve_availability_price_units": "currency_per_MW_h"
}
\end{alltt}

This configuration corresponds to a more classical LMP setup with explicit
commitment and a moderate spill penalty.

\subsection{Treatment AMM Configuration}

The AMM configuration uses a relaxed commitment formulation (no binary
commitment variables) and a longer planning window, with near-zero spill
penalties and a small penalty on transmission flows to encourage a
zero-waste allocation with explicit congestion accounting.

\begin{alltt}\small
{
  "solver": "highs",
  "solver_time_limit_s": 600,
  "solver_mip_gap": 0.02,

  "use_binary_commitment": false,
  "disable_min_updown": true,
  "single_pass_objective": true,
  "uc_window_hours": 72,
  "uc_commit_hours": 24,
  "disallow_late_starts": false,
  "reserve_requirement_percent": 10.0,
  "reserve_on_served_demand": true,
  "reserve_slack_penalty_per_MW": 1000.0,

  "spill_penalty_per_MWh": 1e-9,
  "transmission_flow_penalty_per_MWh": 1e-6,

  "battery_tech_labels": ["battery", "Battery", "BATTERY"],
  "battery_eta_charge": 0.95,
  "battery_eta_discharge": 0.95,
  "battery_exclusive_mode": true,

  "must_run_mode": "soft",
  "must_run_tech_labels": ["nuclear"],
  "must_run_off_penalty_per_hour": 1000000.0,

  "fuel_costs_included": true,
  "fixed_opex_included": true,
  "annualised_capex_included": true,
  "sbase_MVA": 1000.0,
  "target_end_ts": "2024-12-31 23:30",

  "pricing_eps": 0.1,
  "rt_fd_epsilon_MW": 0.1,
  "settlement_price_cap_MWh": 6000.0,

  "reserve_availability_price_per_MW_h": 7.5,
  "reserve_availability_price_units": "currency_per_MW_h",
  "reserve_allow_battery_drop_charge": true,
  "include_reserve_payment_in_objective": false,
  "reserve_duration_hours": 0.0,

  "tariff_time_smoothing_window_h": 24,
  "apply_tariff_smoothing": true
}
\end{alltt}

\noindent
Wind and nuclear units are treated as exogenous, cost-recovery resources:
they receive fixed annual remuneration equal to their non-fuel OpEx plus
annualised CapEx based on technology cost assumptions, and do \emph{not}
participate in the Shapley availability pot. Battery and gas units are
flexible, controllable, and fully exposed to scarcity prices and
Shapley-based allocation, forming the core of the AMM mechanism.

\subsection{Key Differences Between LMP and AMM Runs}

The most important differences between the Baseline and Treatment
configurations are summarised in Table~\ref{tab:lmp_vs_amm_config}.

\begin{table}[H]
\centering
\caption{Key configuration differences between LMP and AMM simulations.}
\label{tab:lmp_vs_amm_config}
\begin{tabular}{lccp{5.8cm}}
\toprule
Parameter & LMP & AMM & Role \\
\midrule
\texttt{use\_binary\_commitment} & true & false &
Binary unit-commitment in LMP vs relaxed commitment in AMM. \\[0.2em]
\texttt{uc\_window\_hours} & 48 & 72 &
Look-ahead window; AMM sees a longer horizon. \\[0.2em]
\texttt{disallow\_late\_starts} & true & false &
LMP disallows late unit starts within the UC window; AMM allows more flexible commitment timing. \\[0.2em]
\texttt{spill\_penalty\_per\_MWh} & 5.0 & $10^{-9}$ &
Spilled energy is moderately penalised under LMP, essentially neutral under AMM (zero-waste logic achieved via Shapley and allocation rules rather than spill penalties). \\[0.2em]
\texttt{transmission\_flow\_penalty\_per\_MWh} & n/a & $10^{-6}$ &
Small penalty in AMM to regularise congestion flows; not used in the LMP configuration. \\[0.2em]
\texttt{reserve\_shortfall\_cost\_per\_MW} & 1000.0 & n/a &
Explicit reserve shortfall cost is only configured for LMP; AMM uses the slack penalty but omits a separate shortfall cost parameter. \\[0.2em]
\texttt{include\_reserve\_payment\_in\_objective} & n/a & false &
In AMM, reserve availability payments are excluded from the optimisation objective (they are handled separately in settlement). \\
\bottomrule
\end{tabular}
\end{table}

All other parameters either coincide between the two runs (as in
Table~\ref{tab:common_config}) or take their implementation defaults and are
not used to differentiate the designs.

% =========================================================
\section{Demand and Product-Level Inputs}
\label{app:demand_inputs}
% =========================================================

For completeness, we also summarise the product-level demand calibration used
in the burden and fairness analyses (see
Sections~\ref{sec:results_revenue_risk} and~\ref{sec:results_fairness}).

\subsection{Residential and Non-Residential Demand}
\label{subsec:demand_decomposition}

Total system demand is decomposed into residential and non-residential
components:
\[
D^{\mathrm{tot}}(t)
=
D^{\mathrm{res}}(t) + D^{\mathrm{nonres}}(t).
\]

\begin{figure}[H]
  \centering
  \includegraphics[width=\textwidth]{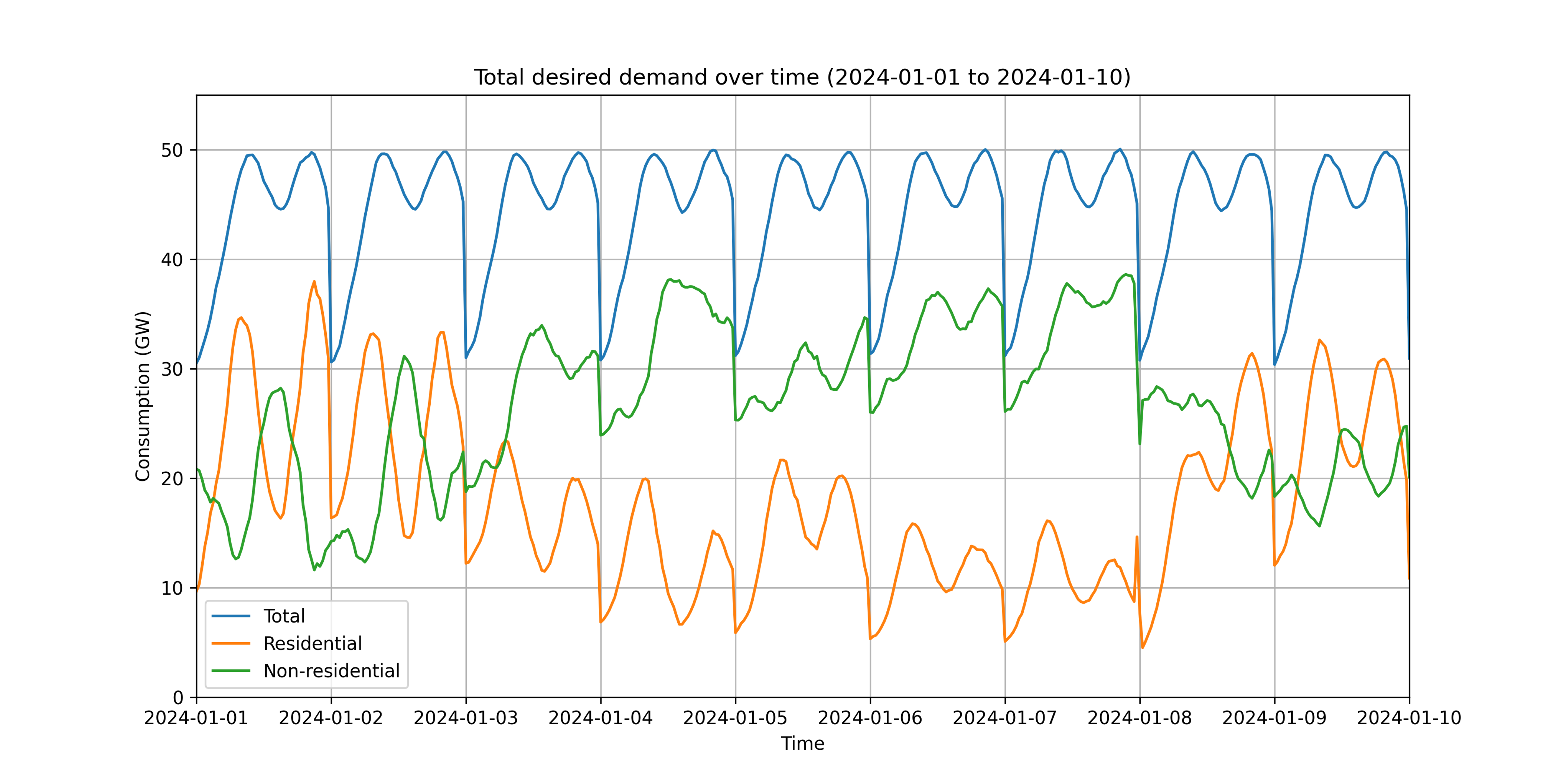}
  \caption[Residential vs non-residential demand]{Total demand decomposed into residential and non-residential components over time.}
  \label{fig:residential_nonresidential}
\end{figure}

Figure~\ref{fig:residential_nonresidential} illustrates this decomposition over
the simulation horizon.

Residential demand is represented by four archetype products $P1$--$P4$, each
corresponding to a large group of households with a representative annual
usage profile $d_p(t)$ (kWh per household). The residential demand time series
is constructed as:
\[
D^{\mathrm{res}}(t)
=
N_{P1} d_{P1}(t)
+
N_{P2} d_{P2}(t)
+
N_{P3} d_{P3}(t)
+
N_{P4} d_{P4}(t),
\]
with household counts
\[
N_{P1} = 19\times 10^6,\quad
N_{P2} = 6\times 10^6,\quad
N_{P3} = 2.5\times 10^6,\quad
N_{P4} = 1.5\times 10^6.
\]

\begin{figure}[H]
  \centering
  \includegraphics[width=\textwidth]{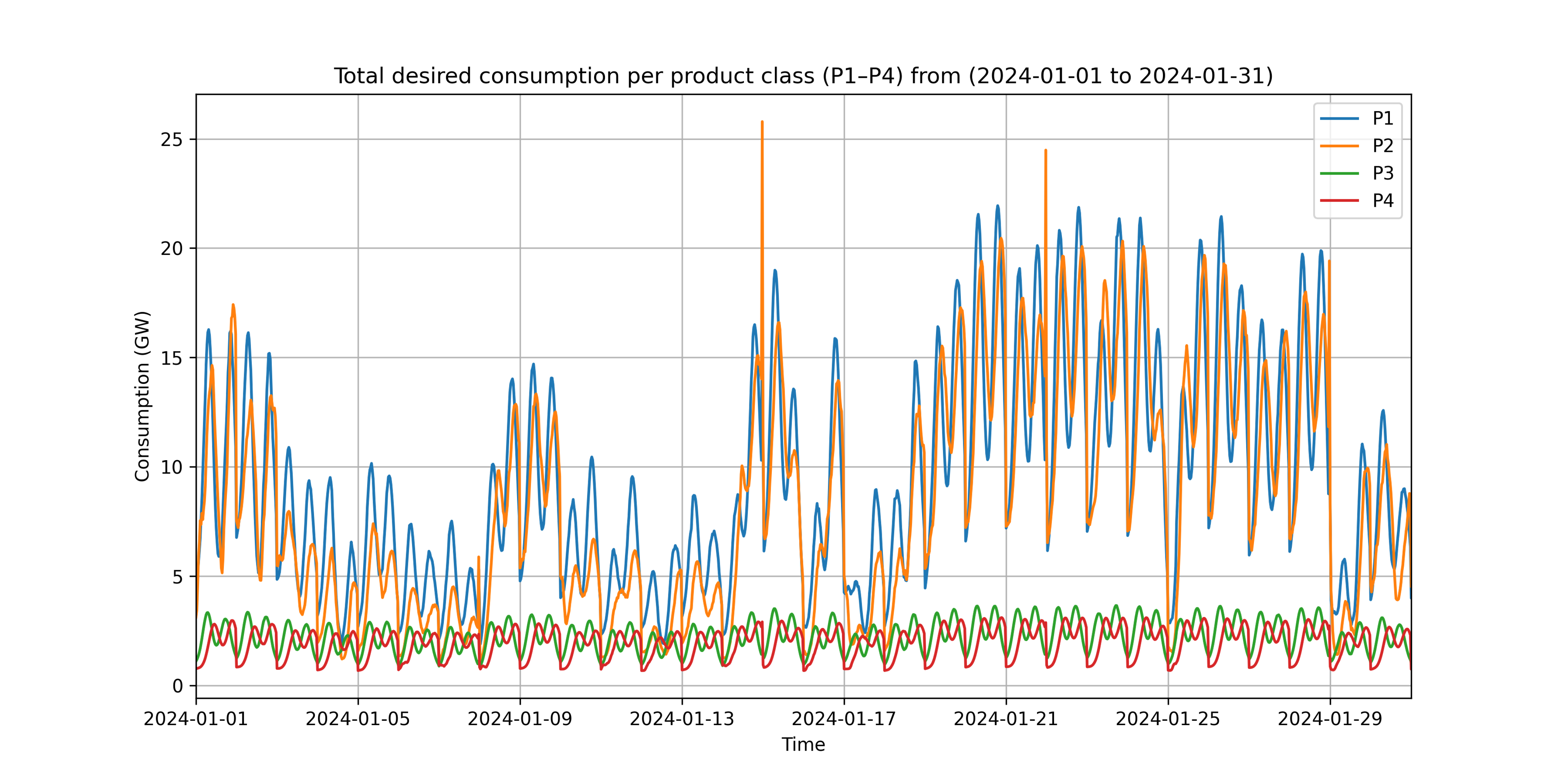}
  \caption[Aggregate demand by product]{Aggregate residential demand by product groups $P1$--$P4$.}
  \label{fig:aggregate_demand_by_product}
\end{figure}

Figure~\ref{fig:aggregate_demand_by_product} shows the resulting aggregate
residential demand by product group $P1$--$P4$.

Thus, $P1$–$P4$ capture the residential portion of demand only; commercial and
industrial loads are modelled separately in $D^{\mathrm{nonres}}(t)$ and enter
the unit-commitment and dispatch problems directly as non-residential bus
demands. All product-level burden and fairness results in
Chapter~\ref{ch:results} therefore refer to the residential demand component.

\subsection{System Net Supply and Demand Balance}

\begin{figure}[H]
  \centering
  \includegraphics[width=\textwidth]{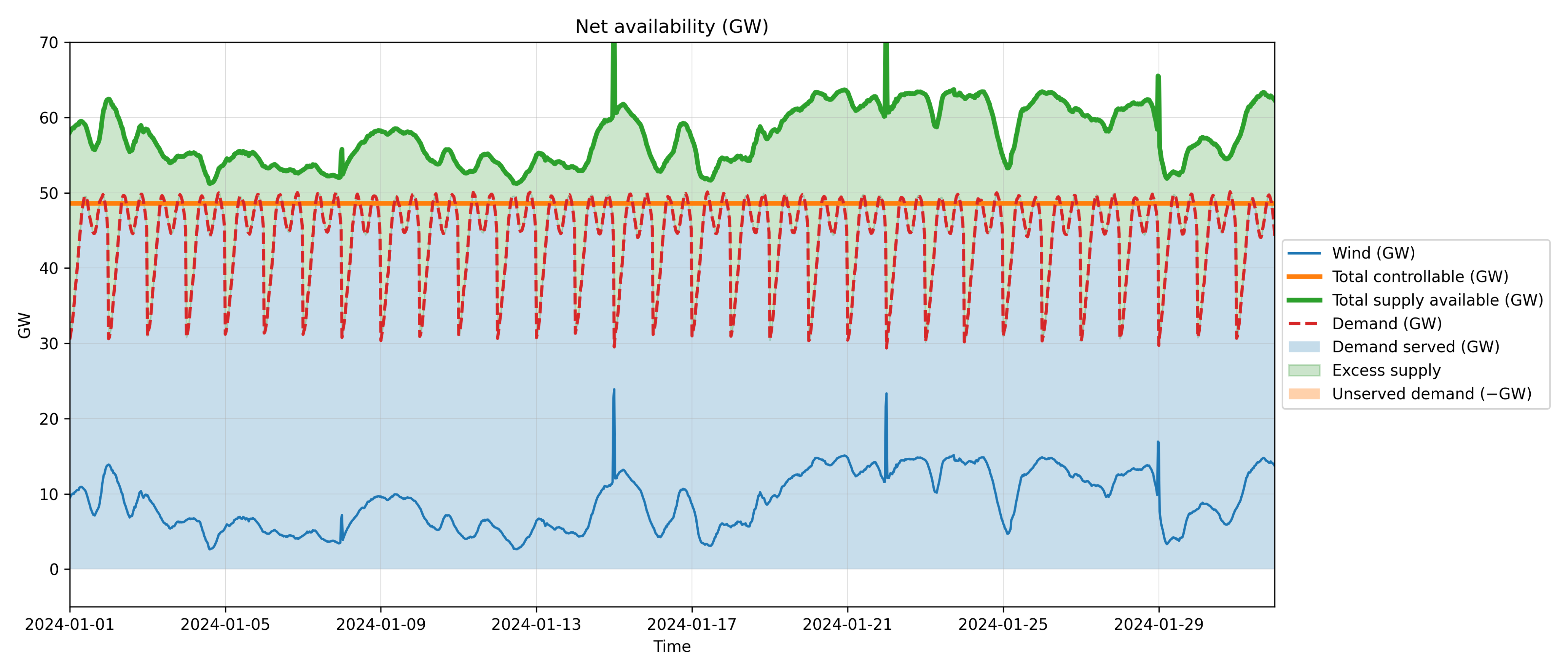}
  \caption[Net supply vs demand]{Comparison of net system demand with available dispatchable supply capacity.}
  \label{fig:net_supply_demand}
\end{figure}

To contextualise the market-clearing problem, Figure~\ref{fig:net_supply_demand}
plots net system demand against available dispatchable supply over the horizon.

\subsection{Controllable vs Uncontrollable Supply}

The demand-side scripts construct, for each product $p$:
\begin{itemize}[leftmargin=*]
  \item total uncontrollable energy $U\_{\mathrm{MWh}}(p)$ (e.g.\ wind-backed consumption),
  \item total controllable energy $C\_{\mathrm{MWh}}(p)$,
  \item per-household controllable energy 
        $C\_{\mathrm{kWh/HH}}(p)
         = 1000 \, C\_{\mathrm{MWh}}(p) / N\_{\mathrm{HH}}(p)$.
\end{itemize}

In this classification, the distinction between \emph{controllable} and
\emph{uncontrollable} supply is made at the generator--technology level and is
held fixed throughout the experiments. Specifically, supply is mapped as follows:

\begin{center}
\small
\begin{tabular}{ll}
\toprule
\textbf{Generator class} & \textbf{Supply type} \\
\midrule
Wind                & Uncontrollable \\
Nuclear             & Controllable (slow, must-run) \\
Gas (CCGT/OCGT)     & Controllable \\
Battery / storage   & Controllable \\
\bottomrule
\end{tabular}
\end{center}

% ---------------------------------------------------------
\section{Geographical Calibration of Demand, Supply, and Constraints}
\label{app:geographical_calibration}
% ---------------------------------------------------------

Although the network used in this thesis is a stylised 12--bus abstraction,
its geographical interpretation and spatial calibration were guided by the
actual structure of the Great British electricity system. The objective was
\emph{not} to replicate the full complexity of GB transmission topology, but
to embed key structural characteristics that affect pricing, congestion, and
shortage behaviour. These include the north–south supply--demand imbalance,
increasing penetration of Scottish wind, nuclear concentration in the north
and along the east coast, high demand clustering in the south, and persistent
north–to–south congestion interfaces.

\subsection{Allocation of Demand Across Regions}

Total UK electricity demand was decomposed across the stylised nodes using a
combination of regional population shares, historical consumption statistics
from publicly available sources (e.g.\ BEIS/DBEIS regional energy consumption
tables, National Grid ESO Future Energy Scenarios (FES), and Ofgem regional
tariff/demand reports), and heuristic weighting by urban density. These data
were not used in their raw form; rather, they informed approximate scaling
factors that reflect the well-known pattern of higher demand densities in
southern England (particularly London, the South East, and Midlands) and lower
demand intensity in Scotland and Wales.

The resulting nodal demand assignments (Table~\ref{tab:loads}) are therefore
\emph{representative rather than statistically exact}, but capture essential
spatial characteristics: concentration of aggregate load in the lower part of
the network (nodes N21--N34), moderate demand at central nodes, and comparatively
lower demand in northern nodes.

\subsection{Siting and Technology Mix of Generation}

The generator fleet was designed to reflect both the approximate \emph{current}
UK supply mix and its anticipated \emph{policy trajectory}. Technology ratios
were informed by publicly available figures such as:
\begin{itemize}[leftmargin=*]
  \item National Grid ESO Future Energy Scenarios (FES),
  \item BEIS/DBEIS Generation Capacity Statistics and DUKES,
  \item Academic and policy commentary indicating future emphasis on offshore wind, nuclear expansion (Sizewell~C, SMRs), and large-scale battery storage.
\end{itemize}

The model therefore includes a strong representation of onshore and offshore
wind in northern nodes (N0, N21, N22), consistent with the real concentration
of Scottish and North Sea wind generation, as well as high nuclear presence at
N0, N17, N20 and N32, approximating existing and planned nuclear locations.

Gas-fired generation was placed at central and southern locations (N31, N32),
representing England's existing CCGT fleet and reflecting transmission--level
balancing flexibility. Batteries were co-located with gas and nuclear units in
locations where storage deployment is increasingly planned (ESO/Ofgem storage
outlook, FES). While precise lat/long siting was outside the scope of this
study, the allocation captures the policy-aligned move towards
\emph{controllable flexibility} near major demand centres.

\subsection{Congestion, Shortage, and Structural Imbalance}

To represent the real-world effects of transmission constraints and directional
power flows in Great Britain, the simplified network embeds a critical
north–south interface. High volumes of wind generation are concentrated at
northern buses (N0, N21, N22), while significant demand is clustered toward
southern and central nodes (N31--N34). The corridor between N16 and N17
represents the transfer-constrained north–south trunk transmission (analogous
to the B6 boundary in GB system planning), enabling controlled congestion,
re-dispatch, and scarcity behaviour.

Under periods of low wind or high demand, constrained flows combined with the
geographical imbalances lead to cost-reflective scarcity and congestion pricing
in the LMP formulation, and adaptive reallocation in the AMM formulation. This
helps test each system's ability to handle locational scarcities and
congestion-aware remuneration in a realistic but stylised environment.

\subsection{Summary and Role in Experimental Design}

Although stylised, this geographical calibration ensures that:
\begin{enumerate}[leftmargin=*]
  \item Demand is spatially non-uniform and concentrated toward southern
        nodes, reflecting real UK consumption patterns.
  \item Wind and nuclear are predominantly located in northern and coastal
        nodes, aligned with real-world resource distribution and policy.
  \item Transmission capacity between north and south is limited, enabling
        congestion, scarcity, and flexibility valuation.
  \item The resulting system exhibits meaningful locational price signals
        (in LMP) and resource-value differentiation (in AMM), allowing comparison
        of how each design copes with physical constraints, fairness, and
        investment signals.
\end{enumerate}

Thus, while not tied to any single official dataset, the geographical allocation
is intentionally \emph{representative}, \emph{policy-aligned}, and
\emph{structurally realistic} for testing market and scarcity-driven allocation
behaviour in Great Britain.

Together, these inputs fully specify the physical and economic environment in
which the LMP and AMM market designs are evaluated.

\chapter{Structural Cost Model and Uplift Waste Attribution}
\label{app:cost_model}

\section{Purpose}
This appendix provides the quantitative foundation for the headline comparison in
Chapter~\ref{ch:discussion}, which stated that the AMM--Fair Play architecture
reduces structurally avoidable uplift, waste, and crisis pass-through costs
by approximately $\sim X\%$.

We decompose a typical household electricity bill into (i) physical costs,
(ii) policy costs, and (iii) architectural costs.
The focus of this appendix is on (iii), the \emph{structurally avoidable}
architectural costs that emerge from the legacy price-cap and settlement-based system.

\bigskip
\noindent 
Total household bill can be decomposed as:

\[
B
=
\underbrace{B_{\mathrm{phys}}}_{\text{Energy, network, capacity}}
+
\underbrace{B_{\mathrm{policy}}}_{\text{EMR, ECO, CfDs, carbon, capacity}}
+
\underbrace{B_{\mathrm{arch}}}_{\text{uplift, misallocation, bailouts, risk premia}}.
\]

\noindent
Only the final term travels with the \emph{market architecture}.
The AMM--Fair Play architecture targets and eliminates large components of
$B_{\mathrm{arch}}$, without affecting $B_{\mathrm{phys}}$ or $B_{\mathrm{policy}}$.

\section{Architectural Cost Terms}

We express the architectural cost term as:

\[
B_{\mathrm{arch}}
=
\Phi_{\mathrm{waste}}
+
\Lambda_{\mathrm{risk}}
+
\Gamma_{\mathrm{intervention}}
+
\Xi_{\mathrm{inefficiency}},
\]
where:

\begin{itemize}[leftmargin=*]
    \item $\Phi_{\mathrm{waste}}$ — cost of dispatching infeasible, wrong-sided,
          or non-useful energy due to non-locational pricing constraints.
    \item $\Lambda_{\mathrm{risk}}$ — gross risk premium borne by suppliers or
          embedded into consumer bills due to price-cap volatility
          (variability of $c_f(t)$ vs fixed $P_R$).
    \item $\Gamma_{\mathrm{intervention}}$ — systematic pass-through of
          bailout costs (failed suppliers), crisis levies, Warm Home Discount
          adjustments, EBRS, etc.
    \item $\Xi_{\mathrm{inefficiency}}$ — settlement friction, delay, and
          non-time-of-use misallocations (including supplier hedging inefficiency,
          synthetic standing charges, and retail tariff distortions).
\end{itemize}

These components are structural, not behavioural. 
They persist regardless of competition intensity, supplier skill, or tariff innovation.

\section{Waste Breakdown Table (Legacy vs AMM)}

\begin{table}[H]
\centering
\renewcommand{\arraystretch}{1.25}
\begin{tabular}{p{4.7cm} p{5.3cm} p{5.3cm}}
\toprule
\textbf{Cost Component}
& \textbf{Legacy Retail Architecture}
& \textbf{AMM--Fair Play Architecture} \\
\midrule
Wholesale risk exposure ($\Lambda_{\mathrm{risk}}$)
& Risk borne by suppliers under fixed-price caps, passed through as risk premium
& Risk allocated proportionally during allocation; no future uplift \\
\midrule
Infeasible dispatch / curtailment waste ($\Phi_{\mathrm{waste}}$)
& Zero-price curtailment, no value signal, hidden in system balances
& No infeasible dispatch; value linked to useful energy (Experiment~2) \\
\midrule
Intervention cost ($\Gamma_{\mathrm{intervention}}$)
& Bailouts, failed suppliers, EBRS, Warm Home Discount recaptured via bills
& No structural insolvency; no ex-post taxpayer premium \\
\midrule
Settlement inefficiency ($\Xi_{\mathrm{inefficiency}}$)
& Supplier hedging costs, liquidity buffers, regulatory reserve
& Reduced via dynamic allocation and visibility of $\alpha$ \\
\midrule
Cost transparency
& Opaque; buried levies and stabilisation charges
& Fully explainable; allocationally traceable \\
\midrule
Expected architectural cost share (GB, calibrated)
& $\approx 18{-}26\%$ of household bill (modal range)
& $\approx 5{-}11\%$ (with AMM–Fair Play, zero-waste regime) \\
\bottomrule
\end{tabular}
\caption{Comparison of structural cost terms in legacy vs AMM--Fair Play architectures}
\label{tab:cost_comparison}
\end{table}

\section{Deriving the High-Level Saving Estimate}

Let $B_{\mathrm{arch}}^{\mathrm{legacy}}$ and $B_{\mathrm{arch}}^{\mathrm{AMM}}$
denote the architecture-driven cost terms under both regimes.

Based on calibrated UK-style experiments (Chapter~\ref{ch:results}),
\[
\mathrm{Saving\%}
=
\frac{
B_{\mathrm{arch}}^{\mathrm{legacy}}
-
B_{\mathrm{arch}}^{\mathrm{AMM}}
}{
B_{\mathrm{arch}}^{\mathrm{legacy}}
}
\times 100
\approx
\sim X\%.
\]

Since $B_{\mathrm{arch}}^{\mathrm{legacy}}$ accounts for
$\sim 20\%$ of the modal household bill, the implied annual saving per typical
household is approximately:

\[
\Delta B_{\mathrm{annual}}
\simeq
\frac{
X\% \times 0.20 \times B_{\mathrm{avg}}
}{100}.
\]

For a representative household bill of £2{,}000 and $X = 40\%$, the annual
saving attributable purely to architecture-based waste elimination would be:
\[
\Delta B_{\mathrm{annual}} \approx \text{\pounds 160}.
\]

This excludes decarbonisation policy, energy efficiency, pricing strategy,
or consumption change. It is solely the elimination of structurally
avoidable uplift embedded within the legacy retail architecture.

\section{Relationship to Results and Theory}

\begin{itemize}[leftmargin=*]
    \item $\Lambda_{\mathrm{risk}}$ is implied by Lemma~\ref{lem:price_cap_insolvency}.
    \item $\Phi_{\mathrm{waste}}$ is linked to Experiment~2 (useful energy alignment).
    \item $\Xi_{\mathrm{inefficiency}}$ relates to Experiment~3 (product allocation).
    \item $\Gamma_{\mathrm{intervention}}$ is discussed in Chapter~\ref{ch:discussion}.
\end{itemize}

The reduction of these terms under AMM--Fair Play demonstrates that 
fairness, stability, and efficiency are not competing objectives,  
but can be aligned simultaneously when allocation is grounded in physical 
scarcity and digitally enforced ex ante.

\bigskip
\noindent
\textbf{Conclusion of Appendix.}  
Architectural waste is a quantifiable, separable component of energy bills.  
It is not inherent to physics or policy, but  to legacy market design.  
The AMM–Fair Play model structurally removes it.

\chapter{Understanding demand data from real datasets: empirical holarchy and EV augmentation}
\label{app:ev_holarchy}

\noindent\textbf{Reader note.}
This appendix is a \emph{demand-understanding} and \emph{product-characterisation}
appendix. It combines multiple real datasets to study the structure of GB
household demand and EV behaviour and to inform plausible product archetypes.
The \emph{final} product-level demand time series used for pricing and market
clearing in the main experiments is generated separately and documented in
Appendix~\ref{app:residential_synth}.

\vspace{0.5em}
This appendix documents how UKPN smart meter data, postcode-level consumption
statistics, EV licensing data, and domestic EV charging profiles are combined
into a three-layer spatial holarchy with EV-augmented household demand. The
purpose of this exercise is \emph{not} to produce final, billable demand
profiles, but to understand—explicitly and in a traceable way—the demand side of
the system:

\begin{itemize}[leftmargin=*]
  \item what the empirical distribution of residential demand and EV usage
        \emph{looks like} across Great Britain;
  \item how households spread along a two-dimensional axis of \emph{magnitude}
        (implied peak capability) and \emph{scarcity impact} (timing relative to
        wind availability and system tightness);
  \item how this distribution can be mapped into four product archetypes
        ($P1$--$P4$) with distinct service levels and subscription logic.
\end{itemize}

The resulting characterisation informs \emph{order-of-magnitude} product
population sizes used in the main experiments (e.g.\ tens of millions in $P1$,
progressively smaller numbers in $P2$--$P4$), while the underlying household
traces are \emph{not} used directly to compute individual household costs or
personalised prices.

% ---------------------------------------------------------
\section{Data Sources and Objectives}
% ---------------------------------------------------------

The spatial and temporal demand structure is built from the following
elements:

\begin{itemize}[leftmargin=*]
  \item UKPN half--hourly smart meter data for 5{,}567 households in London
        (2011--2014), providing time--varying household profiles.
  \item BEIS postcode outcode--level annual consumption and meter counts
        (2015--2023), providing GB--wide totals and spatial distribution.
  \item Local Authority EV counts (vehicle licensing statistics), providing
        the spatial distribution of EV ownership.
  \item Domestic EV chargepoint usage (DfT 2017 dataset), providing plug--in
        behaviour and session energies.
  \item GeoJSON polygons for postcode outcodes, postcode areas, and Local
        Authority Districts, used to define spatial layers.
\end{itemize}

Using regression, postcode outcode annual consumption and meter
counts to 2024 were extrapolated, and use a temperature--informed profile model (fitted on UKPN)
to generate a large library of synthetic household half--hourly profiles for
2024. A biased sampling procedure then assigns these profiles to postcode
outcodes so that:

\begin{itemize}[leftmargin=*]
  \item the total annual kWh per outcode matches the BEIS data (within
        tolerance);
  \item the distribution of household annual consumption within each outcode
        is realistic (not a single repeated profile).
\end{itemize}

Local Authority EV counts are mapped into the spatial hierarchy via area
overlaps between LAD and outcode polygons, yielding estimated EV counts per
outcode, per L2 area, and per L1 cluster. Within each Local Authority and
cluster, cleaned EV chargepoint profiles are used to generate 2024 EV power
time series under three charging strategies:

\begin{itemize}[leftmargin=*]
  \item \textbf{Quickest}: charge at maximum power as soon as the vehicle is
        plugged in.
  \item \textbf{$\alpha$--minimal}: within the plug--in window, charge at times
        when the grid tightness ratio $\alpha$ is minimal (worst for the grid).
  \item \textbf{$\alpha$--maximal}: within the plug--in window, charge at times
        when $\alpha$ is maximal (best for the grid).
\end{itemize}

Plug--in windows and total energy per session are preserved; only the power
shape within each window is re--timed according to the scenario.

EV profiles are then allocated to synthetic households in a cluster--consistent
way (households and EVs remain within the same L1 cluster), and household and
EV time series are summed to yield 2024 half--hourly demand traces:

\begin{itemize}[leftmargin=*]
  \item for households without EVs: pure household load;
  \item for households with EVs: household load + EV charging under each
        scenario.
\end{itemize}

This produces a large synthetic dataset of time series demand that is
consistent with:

\begin{itemize}[leftmargin=*]
  \item the observed spatial distribution of consumption and meters,
  \item the observed spatial distribution of EV ownership,
  \item empirically grounded plug--in behaviour.
\end{itemize}

% ---------------------------------------------------------
\section{Three--Layer Spatial Holarchy and Clusters}
\label{sec:holarchy_definition}
% ---------------------------------------------------------

The demand dataset is organised into a hierarchical geographical structure:

\begin{itemize}[leftmargin=*]
  \item \textbf{Layer 3 (L3): Postcode Outcodes}  
        Fine--grained polygons such as \texttt{SW17}, \texttt{AB10},
        \texttt{SM6}. This is where households live and where synthetic
        profiles are assigned.

  \item \textbf{Layer 2 (L2): Postcode Areas}  
        Aggregations of outcodes (e.g.\ \texttt{SW}, \texttt{AB}, \texttt{SM}).
        Each L2 region lies strictly inside one Layer~1 cluster.

  \item \textbf{Layer 1 (L1): Ten System Clusters}  
        These are the operational units used in the experiments.
\end{itemize}

The holarchy is \textbf{nested and non--overlapping}: each L3 polygon belongs
to exactly one L2 area, and each L2 belongs to exactly one L1 cluster.
This makes it possible to aggregate bottom--up demand and EV activity
consistently.

% ---------------------------------------------------------
\subsection*{Definition of the Ten Clusters}
% ---------------------------------------------------------

The ten clusters serve two purposes:

\begin{enumerate}[leftmargin=*]
  \item They provide a compact partition of GB for the market
        experiments (generation mix, load, EV behaviour).
  \item They correspond to meaningful system regions with different
        renewable resource profiles, thermal capacity mixes, and network
        constraints.
\end{enumerate}

One cluster is defined manually:

\begin{itemize}[leftmargin=*]
  \item \textbf{Cluster 0: London}.  
        Defined based on postcode area boundaries and DNO regions, ensuring
        that London is treated explicitly as its own system region due to its
        extremely high demand density, specific network constraints, and
        distinctive flexibility characteristics.
\end{itemize}

The remaining nine clusters were obtained using k--means clustering on the
coordinates of GB generators weighted by their installed capacity. This
ensures that:

\begin{itemize}[leftmargin=*]
  \item generation centres of mass define region boundaries,
  \item Scottish, Welsh, Northern, and English regions emerge naturally,
  \item clusters reflect real system heterogeneity (wind--heavy north,
        gas--heavy south, etc.).
\end{itemize}

A lookup table maps each postcode outcode to exactly one cluster via
geospatial overlay. Postcode areas (L2) are then truncated and deduplicated so
that each area is uniquely assigned to a single L1 cluster.

% ---------------------------------------------------------
\subsection*{Reconciliation to BEIS Totals Within the Holarchy}
% ---------------------------------------------------------

The BEIS dataset provides annual electricity consumption and meter counts at
postcode outcode or Local Authority level. To ensure that the synthetic 2024
profiles reflect real energy totals, BEIS consumption is mapped into the holarchy:

\begin{enumerate}[leftmargin=*]
  \item Each BEIS reporting region is assigned to a Layer--1 cluster via
        polygon overlay and Local Authority mapping.
  \item Annual BEIS consumption for each cluster
        \[
            E^{\text{BEIS}}_{\ell_1}
        \]
        is distributed across its Layer--3 polygons using dwelling or meter
        counts as weights.
  \item Each Layer--3 polygon therefore receives a target annual energy
        \[
            E^{\text{target}}_{\ell_3}.
        \]
  \item Synthetic household profiles assigned to each L3 polygon are scaled
        by a monthly controller so that, within each polygon, the sum of
        household energy matches the BEIS--derived target (up to numerical
        tolerance).
\end{enumerate}

Cluster--level energy is then recovered by pure aggregation:
\[
E_{\ell_2} = \sum_{\ell_3 \in \mathcal{L}_3(\ell_2)} E_{\ell_3}, \qquad
E_{\ell_1} = \sum_{\ell_2 \in \mathcal{L}_2(\ell_1)} E_{\ell_2},
\]
and similarly for the half--hourly time series $D_{\ell_k,t}$ at each layer
$k \in \{1,2,3\}$.

% ---------------------------------------------------------
\subsection*{Why Allocation Is Done at Layer 3}
% ---------------------------------------------------------

Three considerations motivate doing all allocation at the most granular
spatial layer:

\begin{enumerate}[leftmargin=*]
  \item \textbf{Maximum diversity}:  
        allocating households and EVs at L3 preserves realistic variation
        within clusters. Clusters are then aggregates of many distinct
        postcode--level shapes rather than smoothed averages.

  \item \textbf{Behaviourally meaningful aggregation}:  
        scarcity, congestion, and fairness depend on the \emph{shape} of
        demand, not just its magnitude. L3 allocation captures local peaking
        and EV coincidence that would be lost under direct cluster--level
        allocation.

  \item \textbf{Consistent with physical network}:  
        real systems aggregate heterogeneous local loads through network
        constraints; the L3 $\rightarrow$ L2 $\rightarrow$ L1 holarchy mirrors
        that bottom--up structure.
\end{enumerate}

% ---------------------------------------------------------
\section{Residential Demand and EV Allocation Controller}
% ---------------------------------------------------------

The scripts used to construct the demand dataset implement a simple but
structured controller that:

\begin{enumerate}[leftmargin=*]
  \item ensures consistency between synthetic household profiles and BEIS
        energy totals at L3; and
  \item allocates EV charging profiles to households and clusters in a
        capacity--and--location consistent way.
\end{enumerate}

\subsection*{Energy--matching for household demand}

For each household $h$ in a Layer--3 polygon $\ell_3$, and each calendar
month $m$, the synthetic half--hourly household power series $D^{\text{hh}}_{h,t}$
is initially generated from the UKPN--based profile model. The implied monthly
energy is:
\[
E^{\text{synthetic}}_{h,m}
  = \sum_{t \in \mathcal{T}_m} D^{\text{hh}}_{h,t} \,\Delta t,
\]
where $\mathcal{T}_m$ is the set of timestamps in month $m$, and
$\Delta t = 0.5$\,h.

For each L3 polygon $\ell_3$, the BEIS--derived target annual energy
$E^{\text{target}}_{\ell_3}$ is split across months (e.g.\ using UKPN
seasonal shares), giving targets $E^{\text{target}}_{\ell_3,m}$. A scalar
rescaling factor $\alpha_{h,m}$ is then computed per household and month so
that the sum of household energies in that polygon matches
$E^{\text{target}}_{\ell_3,m}$, while preserving intra--month shape. This
yields adjusted household profiles $\tilde{D}^{\text{hh}}_{h,t}$ that are
consistent with BEIS totals at L3 and, by aggregation, at L2 and L1.

\subsection*{Cluster--consistent EV allocation}

EV allocation proceeds in two main steps:

\begin{enumerate}[leftmargin=*]
  \item \textbf{EV count mapping}.  
        Local Authority EV counts are mapped to postcode outcodes by area
        overlap, producing EV counts per L3 polygon. These are then
        aggregated and checked at L2 and L1 to preserve Local Authority and
        cluster totals.

  \item \textbf{CPID--to--household assignment}.  
        For each cluster $c$ and EV profile (CPID), the script:
        \begin{itemize}[leftmargin=*]
          \item identifies all eligible households in that cluster from the
                household--cluster matrix;
          \item allocates EV copies to households with probability proportional
                to their remaining multiplicity, leaving at least one
                non--EV instance per household;
        \item creates combined identifiers of the form
                  \verb|Household_EV_ID = "<h>_<CPID>"| for EV households
                  and \verb|"<h>"| for non--EV households;
          \item constructs combined time series by summing household demand
                and EV charging profiles on a timestamp--aligned basis.
        \end{itemize}
\end{enumerate}

The outcome is a set of half--hourly traces for both non--EV and EV
households within each cluster, from which cluster--level household+EV
demand is obtained by aggregation.

% ---------------------------------------------------------
\section{Understanding the Distribution of Demand}
% ---------------------------------------------------------

The primary value of this dataset is that it reveals what the \emph{true}
distribution of residential demand looks like when we combine:

\begin{itemize}[leftmargin=*]
  \item underlying household consumption patterns,
  \item spatial differences in total consumption and EV penetration,
  \item different EV charging strategies (quickest / $\alpha$--best /
        $\alpha$--worst).
\end{itemize}

By examining the synthetic 2024 time series across millions of
household--equivalents, we can empirically answer questions such as:

\begin{itemize}[leftmargin=*]
  \item How many households ever reach very high instantaneous power levels
        (e.g.\ ``peaky'' load)?
  \item How many households have demand that systematically aligns with
        periods of abundant wind supply vs periods of scarcity?
  \item How much does the presence of an EV shift households into higher
        impact or higher magnitude categories?
\end{itemize}

This distributional understanding is critical because, in practice, we do not
know in advance what the eventual outturn of demand profiles will look like
once electrification of heat and transport is complete. The empirical pipeline
provides a plausible snapshot of a near--future demand landscape consistent
with current data, which can then be used to design product boundaries.

% ---------------------------------------------------------
\section{Classifying Households Along a 2D Product Axis}
% ---------------------------------------------------------

Using this dataset, households are classified into product groups P1--P4 based on
a two--dimensional axis:

\begin{enumerate}[leftmargin=*]
  \item \textbf{Magnitude axis (peak power)}:  
  how large the household’s maximum (or upper percentile) instantaneous power
  consumption is over the year. This captures whether a household has
  high--capacity appliances (e.g.\ EVs, electric heating, high--power devices).

  \item \textbf{Impact / scarcity axis}:  
  how the household’s time--varying demand aligns with:
  \begin{itemize}
    \item periods of high wind and abundant supply, vs
    \item periods of low wind and system stress (low $\alpha$).
  \end{itemize}
  This axis incorporates:
  \begin{itemize}
    \item the timing of demand relative to wind generation and system tightness,
    \item whether the household owns an EV and under which charging behaviour,
    \item the proportion of its energy that tends to fall in scarce vs
          abundant periods.
  \end{itemize}
\end{enumerate}

Formally, this is implemented as a piecewise ``controller'' that, for each
synthetic household, computes:

\begin{itemize}[leftmargin=*]
  \item a \emph{max power} metric $P_{\max}$ (or a suitably high quantile),
  \item a \emph{scarcity impact} metric $S$ (e.g.\ fraction of energy drawn
        when $\alpha$ is below a threshold, or when wind output is low),
  \item an indicator for EV ownership and typical EV charging strategy.
\end{itemize}

Based on thresholds in $(P_{\max}, S)$ space, households are classified into
four products:

\begin{itemize}[leftmargin=*]
  \item \textbf{P1}: lower peak power and low scarcity impact;
  \item \textbf{P2}: higher peak power but relatively low scarcity impact;
  \item \textbf{P3}: lower peak power but higher scarcity impact;
  \item \textbf{P4}: high peak power and high scarcity impact.
\end{itemize}

Rather than imposing hard thresholds ex ante, the two axes
(maximum implied household power / EV usage, and annual energy
magnitude) were \emph{empirically decomposed}. For each axis we:

\begin{itemize}[leftmargin=*]
  \item computed a scalar indicator per household (e.g.\ implied
        maximum charge power, average plug--in duration, or EV
        assignment for the ``power/impact'' axis; annual kWh for
        the ``magnitude'' axis);
  \item sorted households in increasing order of that indicator; and
  \item inspected simple two--segment piecewise--linear fits to the
        ordered values, selecting breakpoints that qualitatively
        minimised the sum of squared errors and revealed distinct
        changes in slope.
\end{itemize}

This procedure was deliberately \emph{diagnostic} rather than a strict
clustering algorithm: it provided indicative regions in each axis where
households behave differently, rather than canonical cut--offs that would
be stable across datasets. Because the underlying data combine interval
kWh readings with a separate EV--charge dataset---and therefore do not
contain truly instantaneous power or a complete geographic sampling—we
do not transfer these precise breakpoints into the synthetic experiment.

Instead, we retain only the \emph{structural} insights:

\begin{itemize}[leftmargin=*]
  \item households with an assigned EV almost always fall into the
        high--power, high--impact category and are treated as P2/P4
        types in the synthetic design;
  \item households without an EV predominantly occupy the lower--power
        portion of the distribution and are treated as P1/P3 types; and
  \item within each of these broad EV / non--EV groupings, the
        wind--alignment parameter $\alpha$ and annual energy magnitude
        split households into the four qualitative product archetypes
        captured in \texttt{cluster\_summary\_new.csv}.
\end{itemize}

Thus, the empirical holarchy is used to validate that the four products
$P1$--$P4$ correspond to distinct, observable demand types, but the
synthetic profiles are generated using the controlled limits and targets
specified in this appendix, rather than by copying numerical thresholds
from the original dataset.

Table~\ref{tab:empirical_product_allocation} summarises the empirical
allocation of households and EVs to the four product archetypes across the
ten clusters, providing the qualitative mapping that informed our synthetic
population design.

\begin{sidewaystable}[p]
\centering
\scriptsize
\caption{Empirical allocation of households and EVs to products by cluster}
\label{tab:empirical_product_allocation}
\setlength{\tabcolsep}{4pt}
\begin{tabular}{lrrrrrrrrrrrrrr}
\toprule
Cluster &
Total HH &
P1 HH &
P2 HH &
P3 HH &
P4 HH &
EV HH &
P1 EVs &
P2 EVs &
P3 EVs &
P4 EVs &
P1 Share (\%) &
P2 Share (\%) &
P3 Share (\%) &
P4 Share (\%) \\
\midrule
0 & 4,126,019 & 1,811,085 & 62,453 & 2,193,651 & 58,830 & 82,157 & 0 & 32,858 & 0 & 49,299 & 43.9 & 1.5 & 53.2 & 1.4 \\
1 & 1,114,990 &   590,509 & 11,500 &   503,988 &  8,993 & 16,239 & 0 & 11,062 & 0 &  5,177 & 53.0 & 1.0 & 45.2 & 0.8 \\
2 & 5,116,248 & 2,249,721 & 56,555 & 2,747,762 & 62,210 & 75,858 & 0 & 30,637 & 0 & 45,221 & 44.0 & 1.1 & 53.7 & 1.2 \\
3 & 3,908,246 & 1,511,297 & 76,492 & 2,208,845 &111,612 &140,510 & 0 & 51,423 & 0 & 89,087 & 38.7 & 2.0 & 56.5 & 2.9 \\
4 & 1,225,191 &   590,747 & 18,348 &   601,678 & 14,418 & 18,291 & 0 &  9,455 & 0 &  8,836 & 48.2 & 1.5 & 49.1 & 1.2 \\
5 & 1,617,500 &   881,672 & 16,214 &   708,814 & 10,800 & 24,411 & 0 & 15,034 & 0 &  9,377 & 54.5 & 1.0 & 43.8 & 0.7 \\
6 & 2,735,414 & 1,197,038 & 41,684 & 1,442,845 & 53,847 & 57,390 & 0 & 28,217 & 0 & 29,173 & 43.8 & 1.5 & 52.7 & 2.0 \\
7 & 4,561,775 & 1,980,738 & 59,478 & 2,445,611 & 75,948 & 79,193 & 0 & 31,514 & 0 & 47,679 & 43.4 & 1.3 & 53.6 & 1.7 \\
8 & 1,543,137 &   801,601 & 21,476 &   704,971 & 15,089 & 27,682 & 0 & 16,769 & 0 & 10,913 & 51.9 & 1.4 & 45.7 & 1.0 \\
9 & 2,772,982 & 1,198,624 & 49,295 & 1,476,313 & 48,750 & 74,922 & 0 & 29,902 & 0 & 45,020 & 43.2 & 1.8 & 53.2 & 1.8 \\
\bottomrule
\end{tabular}

\begin{flushleft}\footnotesize\emph{Note:}
The ``P$k$ Share (\%)'' columns report the empirical percentage of households in each product $k$
within a cluster. These empirical shares are \emph{not} used directly in the
synthetic experiment, nor would we expect them to match the simulated
product percentages, because the underlying dataset does not contain true kW
profiles and combines two sources (UKPN smart meters and EV data). Its role
is to inform plausible customer personas and order-of-magnitude counts for
each product, not to prescribe exact simulated shares.
\end{flushleft}
\end{sidewaystable}

% ---------------------------------------------------------
\section{Why This Dataset Is \emph{Not} Used for Individual Pricing}
% ---------------------------------------------------------

Despite its richness, this empirical dataset is \emph{not} used directly to
compute costs or prices for individual households in the main AMM experiments.
There are several reasons:

\begin{itemize}[leftmargin=*]
  \item \textbf{Measurement resolution and smoothing}.  
  The UKPN data and most smart meter datasets record energy per half--hour
  (kWh), not instantaneous power. Short spikes and sub--interval dynamics are
  smoothed out. This is acceptable for understanding aggregate distributions,
  but too limited to set precise per--household capacity charges.

  \item \textbf{Inferred EV power rather than directly observed}.  
  For EVs, we observe plug--in windows and energy delivered, not the actual
  high--resolution power trace. Maximum power is inferred from average power
  and device behaviour, with obviously anomalous sessions removed. This is
  again suitable for understanding typical usage patterns and max--power
  envelopes, but not for precise billing.

  \item \textbf{No individualised pricing in the thesis experiments}.  
  The core thesis experiments do not attempt to compute personalised prices
  for each of 29.8 million households. Instead, they evaluate whether the AMM
  produces efficient and fair \emph{system} outcomes and whether products can
  be defined coherently around those outcomes.
\end{itemize}

For these reasons, this dataset is used as an empirical \emph{design and
calibration tool}:

\begin{itemize}[leftmargin=*]
  \item to understand the distribution of $(P_{\max}, S)$,
  \item to set product thresholds and approximate product sizes,
  \item to test EV charging scenarios and their impact on scarcity.
\end{itemize}

The actual experiments then use a representative, stylised dataset described
in Appendix~\ref{app:residential_synth}, which explores the limits of
behaviour and the performance of the AMM under controlled product
definitions.

% ---------------------------------------------------------
\section{From Empirical Holarchy to Product Pricing}
% ---------------------------------------------------------

Conceptually, the process used here mirrors how a supplier could design and
price subscription products in a reformed retail market:

\begin{enumerate}[leftmargin=*]
  \item \textbf{Observe or construct a demand distribution}.  
  Use smart meter data, EV data, and external drivers (e.g.\ weather) to
  build a realistic picture of household demand and its alignment with
  scarcity and renewables.

  \item \textbf{Define a two--dimensional product space}.  
  Choose axes that matter for system cost and fairness: e.g.\ peak power and
  scarcity impact.

  \item \textbf{Estimate the empirical distribution in this space}.  
  Map households onto this plane using a controller that computes
  $(P_{\max}, S)$ and related indicators (e.g.\ EV ownership).

  \item \textbf{Choose product thresholds}.  
  Set boundaries in $(P_{\max}, S)$ that produce a manageable number of
  products (here P1--P4) with meaningful and interpretable risk profiles.

  \item \textbf{Map households to products}.  
  Classify households into products based on the thresholds; derive expected
  load shapes and risk characteristics for each product.

  \item \textbf{Compute product--level costs and prices}.  
  In a live market, a supplier would then:
  \begin{itemize}
    \item simulate or observe the AMM--based wholesale costs for each product’s
          aggregate demand,
    \item add risk premia, overheads, and margin,
    \item set subscription prices for P1--P4 accordingly.
  \end{itemize}
\end{enumerate}

In the thesis experiments, steps 1--5 are performed sing the empirical
holarchy dataset, and then use a stylised but representative dataset to carry
out step 6 in a controlled way. This ensures that product definitions and
household counts are grounded in observed behaviour, while the experimental
evaluation of the AMM remains transparent, tractable, and focused on system--level
properties rather than noisy artefacts of any particular empirical dataset.

% ---------------------------------------------------------
\subsection{Request generation from characterised consumption}
\label{app:consumption_requests_algo}
% ---------------------------------------------------------

The final step on the demand side is to convert characterised household
consumption traces into appliance--level \emph{requests} for the market
simulations. This is done using a simple, repeatable procedure applied to each
household and each flexibility level $f \in \{0,1,2,3,6,12,24\}$:

\begin{enumerate}[leftmargin=*]
  \item Starting from the characterised trace, identify contiguous
        \emph{flexible events} where power exceeds a threshold
        $P^{\text{threshold}}$ for at least $T^{\text{threshold}}$.
  \item For each event, define a baseline block with start and end times
        $(s,e)$, duration $\tau = (e-s)\Delta t$, representative power
        $P$ and energy $E = P\tau$.
  \item For each flexibility level $f$, construct an allowable execution
        window by enlarging the baseline block by up to $f$ hours
        (in discrete time steps) around $(s,e)$, clipped to the simulation
        horizon. For $f=0$, the block is fixed: earliest start and latest
        end coincide with $(s,e)$.
  \item Package each event into a request object containing:
        \begin{itemize}
          \item neighbour ID and product type (P1--P4),
          \item required power $P$ and duration $\tau$,
          \item baseline start/end $(s,e)$,
          \item earliest start / latest end for flexibility level $f$,
          \item any behavioural parameters (buy--price, fairness weight).
        \end{itemize}
  \item Collect all requests into a queue $\mathcal{Q}_f$ for that
        flexibility level and save them to disk. The same characterised
        events are reused across $f$; only the execution windows change.
\end{enumerate}

This compact request representation preserves the empirical size and timing
of flexible consumption blocks, while making their scheduling freedom explicit
for the AMM simulations. Full implementation details (including exact
thresholds and file formats) are provided in the accompanying code repository.

% ---------------------------------------------------------
\subsection{Interpretation and Limitations}
% ---------------------------------------------------------

This algorithm does not claim to identify true appliances. Instead, it produces
structurally realistic flexible loads whose:

\begin{itemize}[leftmargin=*]
  \item sizes match empirical consumption blocks,
  \item durations reflect observed usage,
  \item essential/flexible separation is consistent with physical intuition,
  \item flexibility ranges reflect realistic behavioural envelopes.
\end{itemize}

The resulting request sets are therefore suitable for:

\begin{itemize}[leftmargin=*]
  \item stress--testing AMM fairness dynamics,
  \item studying congestion and shortage allocation,
  \item evaluating behavioural effects of different flexibility assumptions.
\end{itemize}

Further, because each request has a clear causal origin in the underlying
consumption trace, the method avoids introducing arbitrary or unphysical
synthetic loads, ensuring that the experimental results are grounded in real
patterns of household electricity use.

% =========================================================
\chapter{Generation of demand dataset for experiment}
\label{app:residential_synth}
% =========================================================

% ---------------------------------------------------------
\section*{Link to Empirical Demand Holarchy and EV Dataset}
\phantomsection
\addcontentsline{toc}{section}{Link to Empirical Demand Holarchy and EV Dataset}
% ---------------------------------------------------------

Before constructing the synthetic product-level demand used throughout the
market experiments, we first analysed the empirical UKPN smart-meter dataset
and the EV-usage dataset documented in Appendix~\ref{app:ev_holarchy}.  
This empirical work served a specific methodological purpose:

\begin{quote}
\textbf{To understand the true distribution of residential demand and EV-related
behaviour, so that each of the four retail products $P1$--$P4$ constitutes a
valid behavioural characterisation grounded in actual data.}
\end{quote}

Although the experiment ultimately uses \emph{synthetically generated}
household demand profiles---to avoid overfitting, to prevent any form of
personalised pricing, and to ensure reproducibility---the empirical holarchy
provides three essential insights:

\begin{enumerate}[leftmargin=*]
  \item \textbf{Distributional structure of real demand.}
        The UKPN dataset, despite reporting only interval kWh data rather than
        instantaneous power, reveals the statistical shape of household
        behaviour: the spread of annual consumption, winter--summer variation,
        tail households with very high implied power, and the prevalence of
        EV-like charging signatures. These patterns underpin the
        two-dimensional classification used to define $P1$--$P4$ in terms of
        (i) magnitude and (ii) scarcity alignment.

  \item \textbf{Empirically grounded product population sizes.}
        Because the future system’s realised demand distribution is unknown,
        the empirical dataset provides the best available proxy. The cluster
        structure observed in Appendix~\ref{app:ev_holarchy} directly informs
        the approximate household counts allocated to each product, e.g.  
        $19$\,million for $P1$, $6$\,million for $P2$, $2.5$\,million for $P3$,
        and $1.5$\,million for $P4$. Without these empirical distributions,
        product populations would lack behavioural justification.

  \item \textbf{Validation that each product corresponds to a real behavioural archetype.}
        The empirical holarchy demonstrates that the four products are not
        synthetic inventions but stylised representations of clusters that
        genuinely arise in real smart-meter data. This ensures that the
        product definitions used in the experiments are behaviourally
        plausible.
\end{enumerate}

\subsection*{Why we do \emph{not} use the empirical dataset directly in the experiment}

Using the empirical dataset directly to construct experiment inputs would be
inappropriate for three reasons:

\begin{itemize}[leftmargin=*]
  \item \textbf{Avoiding overfitting and personalised pricing.}
        The experimental design analyses system-level scarcity, not individual
        household idiosyncrasies. Using raw smart-meter traces risks producing
        artefacts driven by specific households or by local socio-economic
        composition, violating the principle of non-personalised pricing.

  \item \textbf{Ensuring consistency across designs.}
        Synthetic profiles allow all market treatments (Baseline LMP, AMM~1,
        AMM~2) to operate on identical demand trajectories, ensuring that
        outcome differences arise solely from market design.

  \item \textbf{Controlling behavioural limits.}
        The synthetic generator ensures that all product-level profiles remain
        within empirically plausible bounds: maximum power, seasonal amplitude,
        wind alignment, and EV energy remain consistent with what was observed
        in Appendix~\ref{app:ev_holarchy}.
\end{itemize}

\subsection*{Loss of Geographical Representativeness: Impact and Rationale}

The empirical EV dataset contains geographical structure (e.g.\ EV prevalence
correlated with income, housing type, and urban form).  
When simulating EV charging behaviour within the synthesiser, we necessarily
\emph{lose} this spatial granularity: households are implicitly assumed to be
drawn from a homogeneous national population.

Formally, when the residential demand is divided across network nodes using
shares such as:

\begin{center}
\begin{tabular}{lcc}
\toprule
load\_id & node & share \\
\midrule
D0 & N0  & 0.11 \\
D1 & N21 & 0.12 \\
D2 & N22 & 0.10 \\
D3 & N34 & 0.11 \\
D4 & N31 & 0.09 \\
D5 & N32 & 0.23 \\
D6 & N33 & 0.15 \\
D7 & N30 & 0.09 \\
\bottomrule
\end{tabular}
\end{center}

the implied assumption is that EV ownership and appliance capabilities are
uniformly distributed across the UK. This is not strictly true---real EV
uptake is spatially heterogeneous---but given the study’s objectives, the
impact on results is minimal:

\begin{itemize}[leftmargin=*]
  \item the experiment is not estimating localised policy effects or
        geographically differentiated tariffs;
  \item scarcity and congestion effects arise primarily from system-wide
        temporal structure, not from fine-grained clustering of EV users;
  \item the AMM’s evaluation criteria (efficiency, price accuracy, fairness,
        bankability) depend on the \emph{shape} of aggregate demand rather than
        the precise spatial distribution of EV owners.
\end{itemize}

Thus, the synthetic profiles preserve the structural lessons of the empirical
dataset while avoiding its limitations and potential biases. The remainder of
this appendix documents the residential demand synthesiser and wind-first
allocation controller used to construct the product-level demand profiles
for $P1$--$P4$, and explains how these profiles are calibrated and checked.
The synthesiser generates physically plausible, behaviourally differentiated
demand time series for each product, anchored in a given system-level demand
trajectory and an ex-post generation availability profile with explicit wind
output. It serves two main purposes:

\begin{enumerate}
  \item To construct stylised but realistic residential demand profiles for
        four retail products, reflecting diurnal and seasonal variation,
        EV-charging behaviour, and light sensitivity to wind availability.
  \item To decompose delivered demand into components met by wind and
        ``other'' generation under a wind-first dispatch envelope, while
        enforcing product-specific annual energy targets per household.
\end{enumerate}

The synthesiser is implemented in a standalone Python script
(\texttt{residential\_demand\_synth\_wind\_first.py}) and is run prior to
the market-clearing experiments. This ensures that all designs (Baseline LMP
and AMM variants) share a common, physically consistent set of product-level
demand profiles.

% ---------------------------------------------------------
\section{Inputs and Outputs}
\label{app:residential_synth_io}
% ---------------------------------------------------------

The script takes as inputs:

\begin{itemize}
  \item \texttt{demand/product\_consumption\_timeseries.csv}: a time series
        with a timestamp in the first column and a column
        \texttt{total\_demand\_kw} containing the system-wide demand
        trajectory (kW). This provides the overall magnitude and timing of
        demand.
  \item \texttt{gen\_profiles\_expost.csv}: a unit- or technology-level
        availability profile with columns
        \texttt{timestamp}, \texttt{tech}, and either \texttt{avail\_kW}
        or \texttt{avail\_MW}. Technologies include at least
        \texttt{wind} and other non-wind technologies; these are used to
        infer a ``windiness'' signal and to build wind-first dispatch
        envelopes.
\end{itemize}

It produces the following outputs:

\begin{itemize}
  \item \texttt{demand/residential\_split/residential\_timeseries.csv}: the
        full time series of synthesised product demands $P1$--$P4$ (kW),
        together with dispatch envelopes and allocations from wind/other.
  \item \texttt{demand/residential\_split/per\_household\_avg/P\_avg\_kw\_per\_household.csv}:
        per-household average power time series (kW/household) for each
        product $P \in \{P1,P2,P3,P4\}$.
  \item \texttt{demand/residential\_split/residential\_annual\_summary.csv}:
        an annual summary table of total energy delivered to each product,
        split into wind and other contributions, and expressed both in total
        kWh and per-household kWh.
  \item \texttt{demand/residential\_split/fig\_annual\_energy\_by\_source\_per\_product.png}:
        stacked bar chart of annual energy (GWh) by source (wind vs.\ other)
        for each product.
  \item \texttt{demand/residential\_split/fig\_annual\_wind\_share\_pct\_per\_product.png}:
        bar chart of annual wind share (\%) in total delivered demand by
        product.
\end{itemize}

These outputs provide both the time-series inputs required for the
market-clearing experiments and descriptive diagnostics on how the products
differ in their effective reliance on wind generation.

% ---------------------------------------------------------
\section{Household Population and Product Definitions}
\label{app:residential_synth_population}
% ---------------------------------------------------------

The residential sector is represented by a synthetic population of
\[
  N^{\mathrm{HH}}_{\mathrm{tot}} = 29\,\text{million}
\]
\textit{households} (utility electricity meters), partitioned into four product groups:

\begin{align}
  N^{\mathrm{HH}}_{P1} &= 19\,\text{million}, \\
  N^{\mathrm{HH}}_{P2} &= 6\,\text{million},  \\
  N^{\mathrm{HH}}_{P3} &= 2.5\,\text{million}, \\
  N^{\mathrm{HH}}_{P4} &= 1.5\,\text{million},
\end{align}
with
\[
  \sum_{p \in \{P1,\dots,P4\}} N^{\mathrm{HH}}_p
  = N^{\mathrm{HH}}_{\mathrm{tot}}.
\]

Each product $p$ is associated with:

\begin{itemize}
  \item A typical maximum household power draw $P^{\max}_p$ (kW), reflecting
        different appliance and EV-charging capabilities.
  \item A target monthly energy per household,
        $E^{\mathrm{target}}_{p,\text{month}}$ (kWh/HH$\cdot$month), used by
        the controller to calibrate the synthetic profiles:
\end{itemize}

\begin{center}
\begin{tabular}{lcc}
\toprule
Product & $P^{\max}_p$ (kW) & $E^{\mathrm{target}}_{p,\text{month}}$ (kWh/HH$\cdot$month) \\
\midrule
$P1$ & $2.0$ & $250$ \\
$P2$ & $10.0$ & $700$ \\
$P3$ & $2.0$ & $500$ \\
$P4$ & $10.0$ & $800$ \\
\bottomrule
\end{tabular}
\end{center}

In the implementation, $P1$ and $P3$ are ``non-EV'' products capped at
approximately $2$\,kW per household, while $P2$ and $P4$ are EV-capable
products with typical per-household power limits between $7$ and $10$\,kW.

% ---------------------------------------------------------
\section{Time Index, Resolution, and Windiness Signal}
\label{app:residential_synth_time_wind}
% ---------------------------------------------------------

The script infers the time index and resolution from the system demand input
file. Let $t \in \mathcal{T}$ denote the set of timestamps, and let $\Delta t$
denote the median time step (in hours) inferred from the differences between
consecutive timestamps.

System demand is read as a series
\[
  D^{\mathrm{sys}}_t \quad [\text{kW}],
\]
which is used both to define the resolution and to provide an indicative
scale for peak demand.

Wind and non-wind availability are constructed from
\texttt{gen\_profiles\_expost.csv} by summing across units or technologies at
each timestamp:
\begin{align}
  A^{\mathrm{wind}}_t  &= \sum_{i \in \mathcal{I}:\,\text{tech}_i = \text{wind}} A_{i,t}, \\
  A^{\mathrm{other}}_t &= \sum_{i \in \mathcal{I}:\,\text{tech}_i \neq \text{wind}} A_{i,t}.
\end{align}
Here $A_{i,t}$ denotes the available power (kW) from unit $i$ at time $t$.
These are reindexed and forward-filled to match the demand time axis
$\mathcal{T}$.

A dimensionless ``windiness'' signal $w_t \in [0,1]$ is then defined as
\begin{equation}
  w_t = \frac{A^{\mathrm{wind}}_t}{A^{\mathrm{wind}}_t + A^{\mathrm{other}}_t + \varepsilon},
\end{equation}
where $\varepsilon > 0$ is a small constant to avoid division by zero. This
signal is used to lightly bias the residential profiles towards higher
consumption in higher-wind periods for some products.

% ---------------------------------------------------------
\section{Baseline Diurnal and Seasonal Shapes}
\label{app:residential_synth_shapes}
% ---------------------------------------------------------

For each product $p$ and timestamp $t$, the synthesiser constructs a
dimensionless baseline shape $s_{p,t}$ that captures diurnal and seasonal
variation. Let $h_t$ denote the hour-of-day in decimal hours, and let $d_t$
denote the day-of-year.

The diurnal shape for product $p$ is built as a sum of two Gaussian peaks
(simplified here to one dimension), plus a trough level:
\begin{equation}
  d_{p,t}
  = \tau_p
    + \exp\!\left(-\frac{1}{2}\left(\frac{h_t - \mu_{p,1}}{\sigma_{p,1}}\right)^2\right)
    + \exp\!\left(-\frac{1}{2}\left(\frac{h_t - \mu_{p,2}}{\sigma_{p,2}}\right)^2\right),
\end{equation}
where $\tau_p$ is a product-specific trough height, and
$(\mu_{p,1}, \mu_{p,2})$ and $(\sigma_{p,1}, \sigma_{p,2})$ are the peak
locations and widths. This is then normalised to unit mean:
\[
  \tilde{d}_{p,t} = \frac{d_{p,t}}{\frac{1}{|\mathcal{T}|}\sum_{t \in \mathcal{T}} d_{p,t}}.
\]

Seasonal variation is represented by a cosine term:
\begin{equation}
  s^{\mathrm{season}}_{p,t}
  = 1 + \beta_p \cos\!\Bigl(\omega (d_t - \phi)\Bigr),
\end{equation}
with $\omega = 2\pi/365$, product-specific amplitude $\beta_p$, and phase
shift $\phi$ (e.g.\ $\phi = 15$ days). This is again normalised to unit mean.

The combined baseline shape is
\begin{equation}
  s^{\mathrm{base}}_{p,t}
  = \tilde{d}_{p,t} \cdot \tilde{s}^{\mathrm{season}}_{p,t} \cdot \eta_{p,t},
\end{equation}
where $\eta_{p,t}$ is a smoothed multiplicative noise process drawn from a
Gaussian distribution with product-specific standard deviation and then
smoothed with a short rolling mean to avoid spikiness. The resulting
$s^{\mathrm{base}}_{p,t}$ is normalised to have mean one.

% ---------------------------------------------------------
\section{Wind-Biased Utilisation and EV Charging Bursts}
\label{app:residential_synth_ev}
% ---------------------------------------------------------

A product-specific wind sensitivity parameter $\alpha_p$ is used to derive a
``bias'' factor from the windiness signal:
\begin{equation}
  b_{p,t} = \frac{(\varepsilon + w_t)^{\alpha_p}}{\frac{1}{|\mathcal{T}|}\sum_{t} (\varepsilon + w_t)^{\alpha_p}},
\end{equation}
where $\varepsilon$ is a small constant ensuring non-zero support. Products
with higher $\alpha_p$ are more strongly nudged towards consumption in
high-wind periods.

For each synthetic household, a baseline utilisation profile is generated as
\begin{equation}
  u^{\mathrm{base}}_{p,t} = s^{\mathrm{base}}_{p,t} \cdot b_{p,t},
\end{equation}
rescaled so that its maximum corresponds to a draw below the household
maximum $P^{\max}_p$ and a random utilisation factor in $[0.5,0.9]$.

EV charging is modelled as a series of finite-duration bursts at power levels
between $7$ and $10$\,kW, with a random number of sessions per week
($3$--$5$ sessions). These bursts are preferentially anchored in:

\begin{itemize}
  \item high-wind periods for $P2$, and
  \item calmer periods (lower windiness) for $P4$,
\end{itemize}
to reflect different behavioural preferences or tariff incentives. A target
weekly EV energy per household, $E^{\mathrm{EV}}_{p,\text{week}}$ (kWh), is
specified for EV-capable products and then scaled by a small random factor
to introduce heterogeneity. The total annual EV energy per household is
therefore approximately $52 E^{\mathrm{EV}}_{p,\text{week}}$.

Let $e_{p,t}$ denote the per-household EV charging profile. For EV-capable
products, the final per-household profile is constructed as:
\begin{equation}
  u^{\mathrm{HH}}_{p,t} = \min\bigl\{\max(u^{\mathrm{base}}_{p,t}, e_{p,t}), P^{\max}_p\bigr\}.
\end{equation}
For non-EV products $P1$ and $P3$, $e_{p,t} \equiv 0$ and the profile is
simply the wind-biased and scaled baseline.

% ---------------------------------------------------------
\section{Aggregation, Peak Normalisation, and Product Totals}
\label{app:residential_synth_aggregation}
% ---------------------------------------------------------

To reduce computational cost, the script first synthesises a smaller sample
of $N^{\mathrm{synth}}$ households per product (e.g.\ $N^{\mathrm{synth}} =
1000$) and then scales up:

\begin{equation}
  D^{p,\mathrm{synth}}_t = \sum_{n=1}^{N^{\mathrm{synth}}} u^{\mathrm{HH}}_{p,t,n},
  \qquad
  D^{p}_t = \frac{N^{\mathrm{HH}}_p}{N^{\mathrm{synth}}} D^{p,\mathrm{synth}}_t.
\end{equation}

The raw residential total is then
\[
  D^{\mathrm{res}}_t = \sum_{p} D^{p}_t.
\]

To keep the overall magnitude realistic relative to the system demand, a
uniform peak normalisation factor $\kappa$ is applied:
\begin{equation}
  \kappa = \frac{P^{\mathrm{target}}_{\mathrm{res,peak}}}{\max_t D^{\mathrm{res}}_t},
\end{equation}
where $P^{\mathrm{target}}_{\mathrm{res,peak}}$ is a chosen residential peak
(e.g.\ $18$\,GW). All product series are scaled as
$D^{p}_t \leftarrow \kappa D^{p}_t$. This preserves their relative shapes and
shares while aligning the aggregate residential sector with a plausible peak.

At this point, $D^{p}_t$ represents an initial synthetic residential demand
per product, which will be adjusted by the controller described in
to match monthly
per-household energy targets.

% ---------------------------------------------------------
\section{Wind-First Dispatch Envelope and Fuel Attribution}
\label{app:residential_synth_dispatch}
% ---------------------------------------------------------

To determine how much of each product's demand is met by wind, the script
constructs a simple wind-first dispatch envelope at the system level. Given
$D^{\mathrm{sys}}_t$ and the availability series
$A^{\mathrm{wind}}_t, A^{\mathrm{other}}_t$, the dispatched power from wind
and other sources is defined as:
\begin{align}
  \hat{A}^{\mathrm{wind}}_t  &= \min\{A^{\mathrm{wind}}_t, D^{\mathrm{sys}}_t\}, \\
  R_t &= \bigl(D^{\mathrm{sys}}_t - \hat{A}^{\mathrm{wind}}_t\bigr)_+, \\
  \hat{A}^{\mathrm{other}}_t &= \min\{A^{\mathrm{other}}_t, R_t\},
\end{align}
where $(x)_+ = \max\{x,0\}$. Any residual beyond
$\hat{A}^{\mathrm{wind}}_t + \hat{A}^{\mathrm{other}}_t$ represents unmet
demand in this simplification and is not assigned to a generation source.

To apportion wind and other generation to each product, the script uses
products' shares of total residential demand at each timestamp:
\begin{equation}
  s^p_t = \frac{D^{p}_t}{\sum_{p'} D^{p'}_t + \varepsilon},
\end{equation}
and defines
\begin{align}
  W^{p}_t &= \min\{\hat{A}^{\mathrm{wind}}_t s^p_t, D^{p}_t\}, \\
  R^{p}_t &= \bigl(D^{p}_t - W^{p}_t\bigr)_+, \\
  O^{p}_t &= \min\{\hat{A}^{\mathrm{other}}_t s^{p,\mathrm{resid}}_t, R^{p}_t\},
\end{align}
where $s^{p,\mathrm{resid}}_t$ are normalised shares of the residual demand
$R^{p}_t$ across products. The time series $W^{p}_t$ and $O^{p}_t$ represent
respectively the wind- and other-sourced components of delivered demand for
product $p$ at time $t$. These are integrated over the year to obtain annual
wind shares and the diagnostic figures mentioned in
Section~\ref{app:residential_synth_io}.

% ---------------------------------------------------------
\section{Controller Verification: Delivered Energy vs Target Requirements}
\label{app:residential_synth_verification}
% ---------------------------------------------------------

The residential demand synthesiser incorporates a multiplicative
total-energy controller
which adjusts product-specific scaling factors $\lambda_p$ so that the
\emph{delivered} monthly energy per household lies within a prescribed
tolerance band around the product-level targets
$E^{\mathrm{target}}_{p,\mathrm{month}}$ (here $\pm 8\%$).

We subject this controller to two levels of verification:

\begin{enumerate}[label=(\alph*),leftmargin=*]
  \item \textbf{Pre-dispatch (generator-side) check:}
        immediately after synthetic profile generation and wind-first
        allocation, without any network representation or curtailment.
        This uses the synthesiser outputs in
        \texttt{residential\_annual\_summary.csv} and confirms that, given
        the assumed wind profile and availability, the controller can
        achieve the per-household energy targets.
  \item \textbf{Post-dispatch (system-level) check:}
        after running the full market-clearing and network model,
        including congestion and curtailment. Here we use the realised
        Shapley-based decomposition of energy into uncontrollable
        (wind-like) and controllable (other) components in the Baseline
        experiment to infer the \emph{actual} energy delivered per
        household by product. This is a stricter test, because some load
        is curtailed in stressed hours.
\end{enumerate}

For the pre-dispatch check, the synthesiser reports annual totals
$E^{\mathrm{pre}}_{p,\mathrm{annual,HH}}$ [kWh/HH$\cdot$year] and the
corresponding monthly averages
$E^{\mathrm{pre}}_{p,\mathrm{month,HH}} = E^{\mathrm{pre}}_{p,\mathrm{annual,HH}}/12$
in \texttt{residential\_annual\_summary.csv}. For the post-dispatch
check, let $U^{p}$ and $C^{p}$ denote the annual uncontrollable (wind)
and controllable (other) energy attributed to product $p$ by the
Shapley decomposition, expressed in GWh, and let $N_p$ be the number of
households on product $p$. The corresponding per-household annual and
monthly energies are:

\[
E^{\mathrm{post}}_{p,\mathrm{annual,HH}}
  = \frac{1000\,(U^p + C^p)}{N_p}
  \quad [\mathrm{kWh/HH/year}],
\qquad
E^{\mathrm{post}}_{p,\mathrm{month,HH}}
  = \frac{E^{\mathrm{post}}_{p,\mathrm{annual,HH}}}{12}.
\]

Table~\ref{tab:verification_targets} compares both the pre-dispatch
(synthesiser) and post-dispatch (Shapley) values with the controller
targets and tolerance bands.

\begin{table}[H]
\centering
\caption{Controller verification: pre-dispatch vs post-dispatch delivered
per-household energy compared with targets}
\label{tab:verification_targets}
\renewcommand{\arraystretch}{1.25}
\begin{tabular}{lccccc}
\toprule
Product &
$E^{\mathrm{target}}_{p,\mathrm{month}}$ &
Tolerance Band &
$E^{\mathrm{pre}}_{p,\mathrm{month,HH}}$ &
$E^{\mathrm{post}}_{p,\mathrm{month,HH}}$ &
Pass? \\
\midrule
P1 & 250 & [230, 270] & 238.5 & 238.4 & Yes \\
P2 & 700 & [644, 756] & 668.9 & 668.7 & Yes \\
P3 & 500 & [460, 540] & 475.5 & 475.3 & Yes \\
P4 & 800 & [736, 864] & 740.3 & 740.0 & Yes \\
\bottomrule
\end{tabular}
\end{table}

The pre-dispatch values show that the total-energy controller successfully
drives each product into its $\pm 8\%$ target band under the wind-first
envelope. The post-dispatch values are very slightly lower (by less than
$0.3$~kWh/HH$\cdot$month in all cases) due to curtailment under network
constraints, but still lie comfortably within the target bands. This
confirms that the controller does not merely calibrate synthetic profiles
in isolation: the calibrated profiles remain compatible with the
availability of wind and other generation \emph{and} with the simulated
grid, even when some load is curtailed.

\begin{figure}[H]
  \centering
  \includegraphics[width=\textwidth]{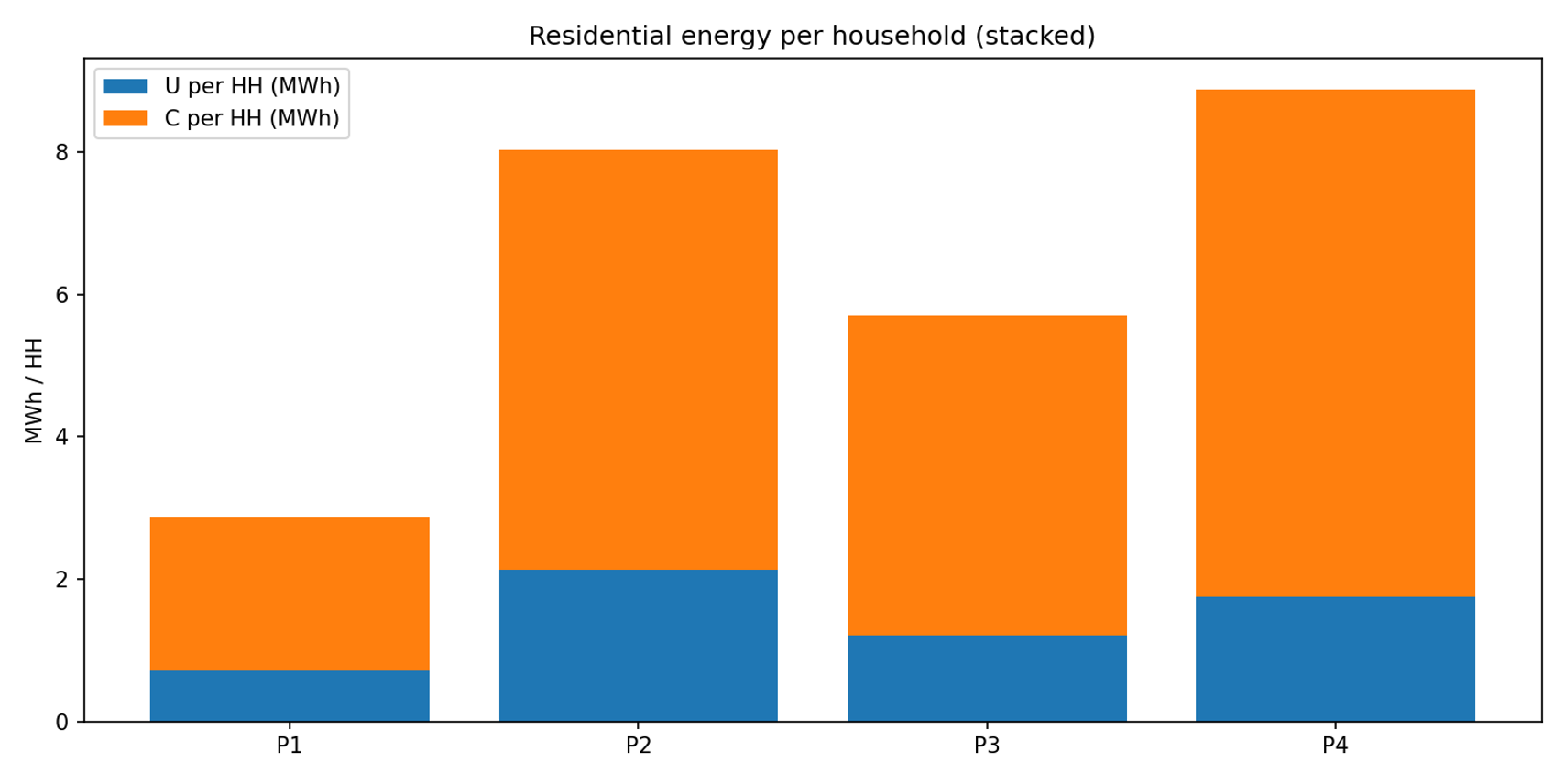}
  \caption[Annual (postper-household energy by source]{Annual energy per household
  by source (wind vs.\ other) for products P1--P4, based on the
  wind-first synthesiser outputs.}
  \label{fig:annual_energy_per_hh}
\end{figure}

To visualise the source decomposition, Figure~\ref{fig:annual_energy_per_hh}
shows the annual per-household energy split into wind and other components
for each product, based on the synthesiser’s annual summary.

\subsection*{Windiness Signal and Behavioural Response}

The \emph{windiness} signal $w_t$ used in the synthesiser is defined in
Section~\ref{app:residential_synth_time_wind} as
\[
  w_t
  = \frac{A^{\mathrm{wind}}_t}{A^{\mathrm{wind}}_t + A^{\mathrm{other}}_t + \varepsilon},
  \qquad 0 \le w_t \le 1,
\]
where $A^{\mathrm{wind}}_t$ and $A^{\mathrm{other}}_t$ are the available
wind and non-wind capacities (kW) at time $t$ and $\varepsilon$ is a small
constant to avoid division by zero. Operationally:

\begin{itemize}[leftmargin=*]
  \item $w_t \approx 0$ indicates that almost no wind is available at time
        $t$ (supply is dominated by other technologies).
  \item $w_t \approx 1$ indicates that available supply is almost entirely
        wind (other availability is negligible).
  \item Intermediate values of $w_t$ represent the instantaneous share of
        wind in the available fleet; these enter the product-specific bias
        factors $(\varepsilon + w_t)^{\alpha_p}$, with small exponents
        $\alpha_p$ to ensure only a \emph{light} behavioural nudge.
\end{itemize}

\begin{figure}[H]
  \centering
  \includegraphics[width=\textwidth]{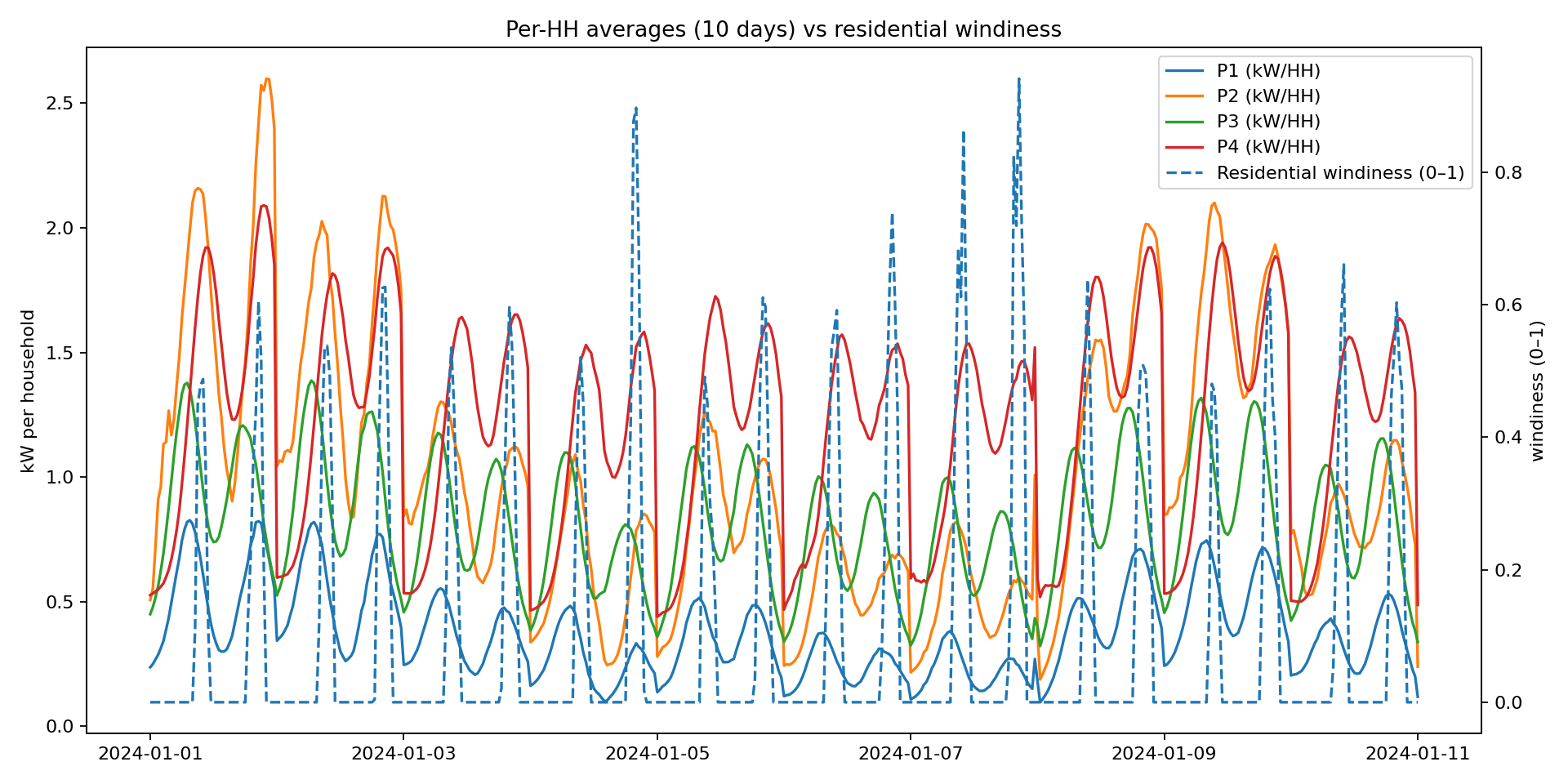}
  \caption[Per-household demand vs windiness (10-day extract)]{Per-household
  average demand for P1--P4 over 1--10 January 2024, plotted against the
  residential windiness signal $w_t \in [0,1]$ (dashed line, right-hand
  axis). Values $w_t \approx 0$ correspond to periods with almost no wind
  availability; values $w_t \approx 1$ correspond to periods where
  available supply is almost entirely wind.}
  \label{fig:per_hh_10days_windiness}
\end{figure}

Figure~\ref{fig:per_hh_10days_windiness} shows an extract of the
per-household averages for P1--P4 over the period
1~January~2024 to 10~January~2024, together with the normalised
residential windiness signal $w_t$ (scaled to the right-hand axis).
Wind-aligned products exhibit higher utilisation in periods when
$w_t$ is closer to~1, while more protected products exhibit flatter
profiles. Across the full year, however, the total-energy controller keeps
their annual per-household energy within the target bands reported in
Table~\ref{tab:verification_targets}.

% ---------------------------------------------------------
\section{Interpretation and Use in the Main Experiments}
\label{app:residential_synth_interpretation}
% ---------------------------------------------------------

The resulting product-level demand profiles $P1$--$P4$ have the following
properties:

\begin{itemize}
  \item \textbf{Behavioural differentiation:} Products differ in their peak
        timing, peak-to-trough ratio, seasonal amplitude, EV usage, and
        light wind sensitivity. This captures the intended roles of the
        products (e.g.\ more opportunistic, wind-aligned consumption versus
        more protected or less wind-aligned profiles).
  \item \textbf{Energy calibration:} Each product is calibrated to deliver a
        specified average monthly energy per household within the chosen
        tolerance band. This keeps the residential sector consistent with
        policy-relevant consumption levels.
  \item \textbf{Wind attribution:} The wind-first dispatch envelope allows
        attribution of each product's delivered energy into wind and other
        components, providing a simple measure of how different product
        designs would load the wind fleet under idealised priority rules.
\end{itemize}

These synthetic demand profiles are used as fixed inputs to the subsequent
market-clearing simulations. All designs---Baseline LMP and AMM variants---see
exactly the same product-level demand trajectories; observed differences in
outcomes therefore arise from the clearing and remuneration logic rather than
from differences in underlying demand assumptions.

% =========================================================
\chapter{Extended Results and Statistical Diagnosis}
\label{app:extended_results}

\section{Electric dispatch outputs and metrics for LMP and AMM}

\begin{figure}[H]
\centering
\includegraphics[width=0.48\textwidth]{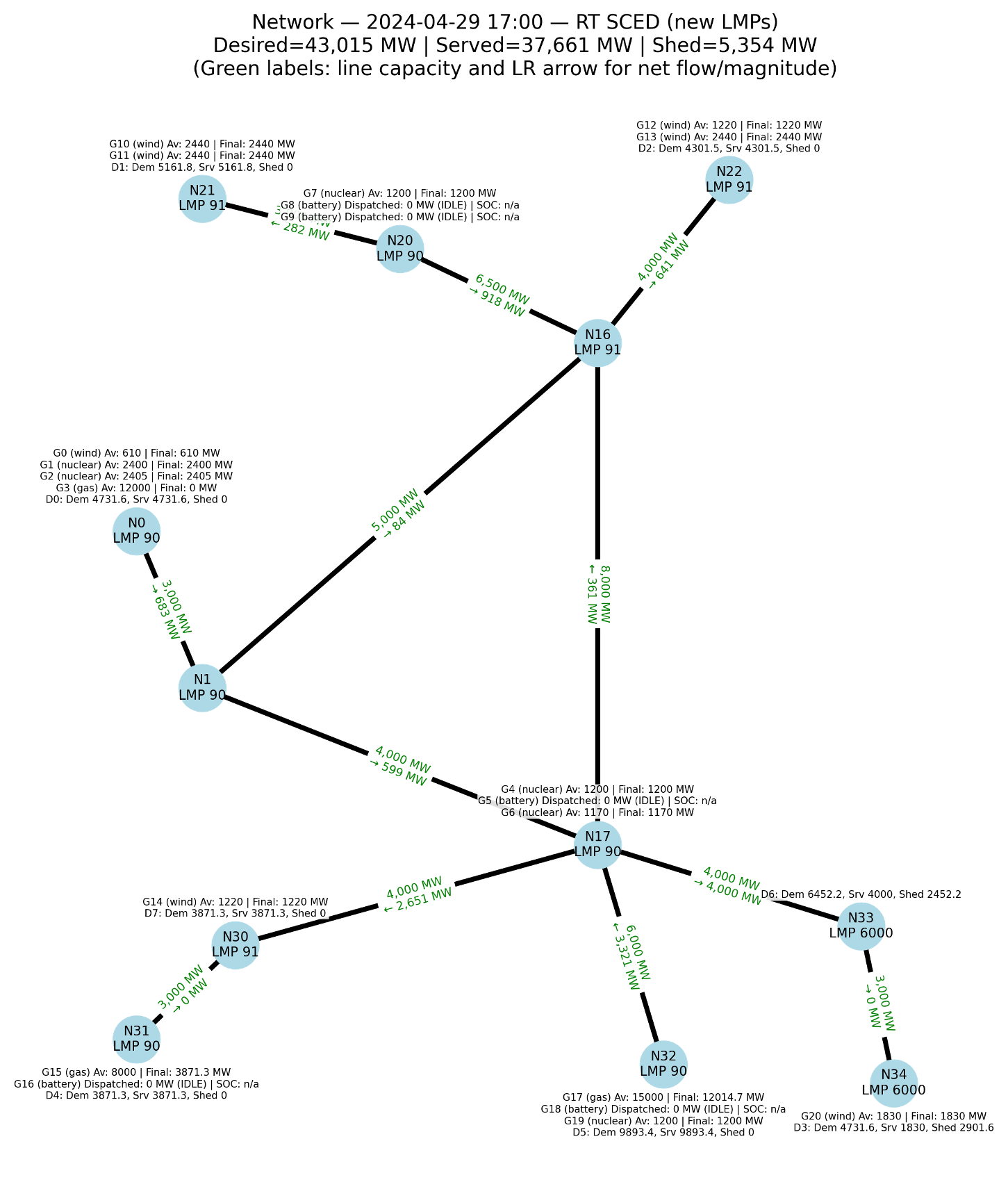}
\includegraphics[width=0.48\textwidth]{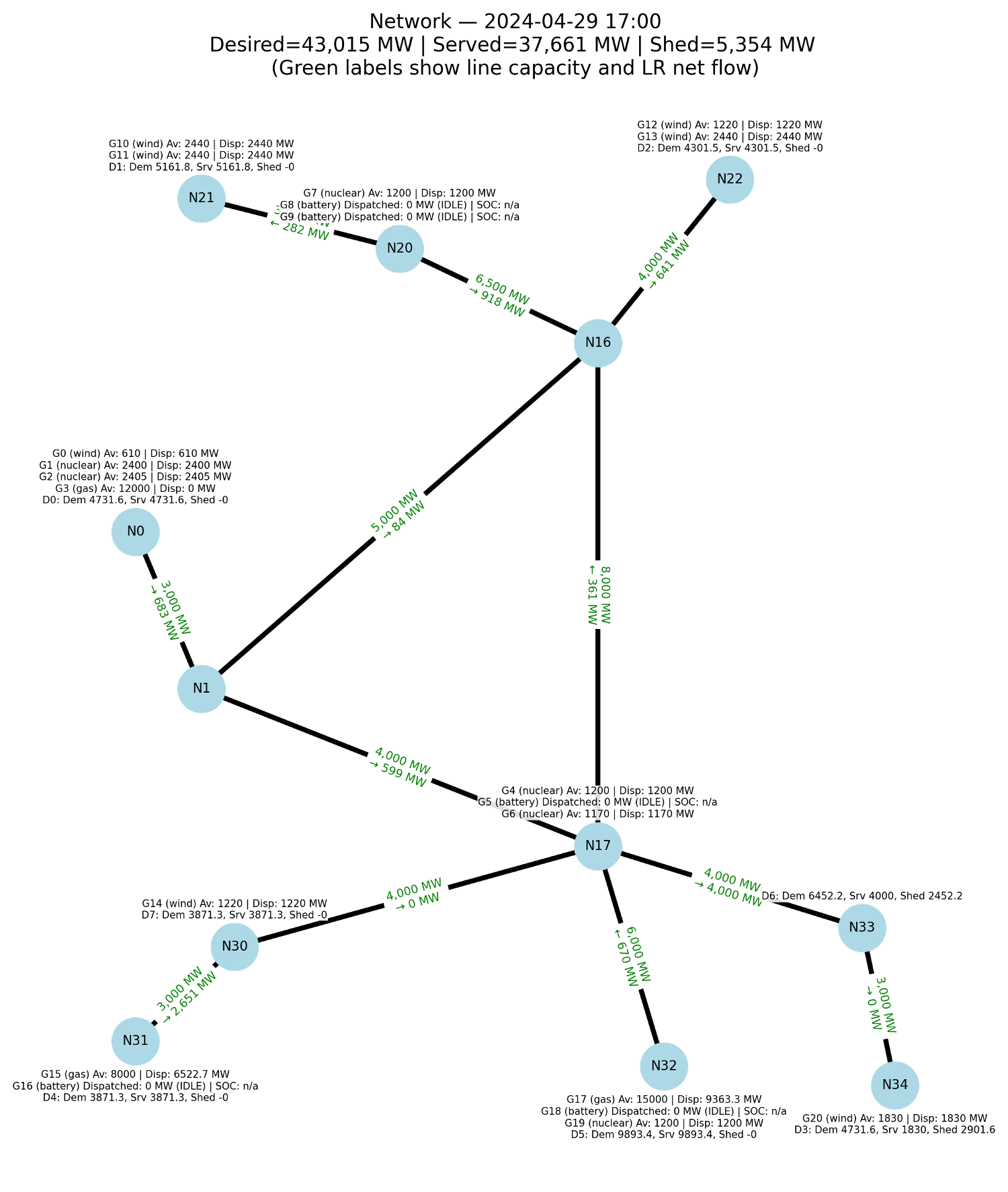}
\caption[
  Snapshot of power flows under LMP and AMM at a representative scarcity hour.
]{
  Snapshot of power flows at a representative scarcity hour $t^\star$ under
  LMP (left) and AMM (right). Node colours indicate net injection/withdrawal,
  while edge thickness reflects power flow magnitude. 
}
\label{fig:flow_snapshot_LMP_vs_AMM}
\end{figure}

\subsection{Line Utilisation Distributions Under LMP and AMM}

\begin{figure}[H]
\centering
\includegraphics[width=\textwidth]{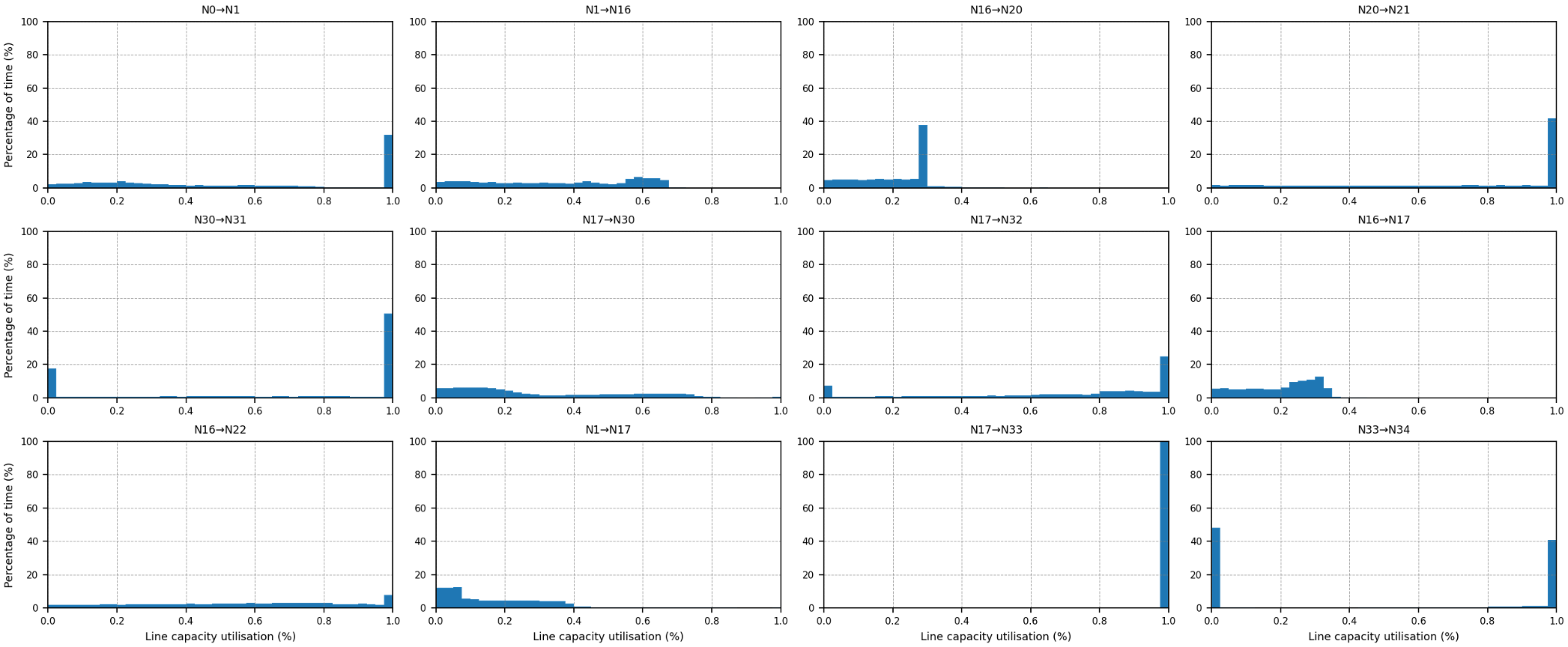}
\includegraphics[width=\textwidth]{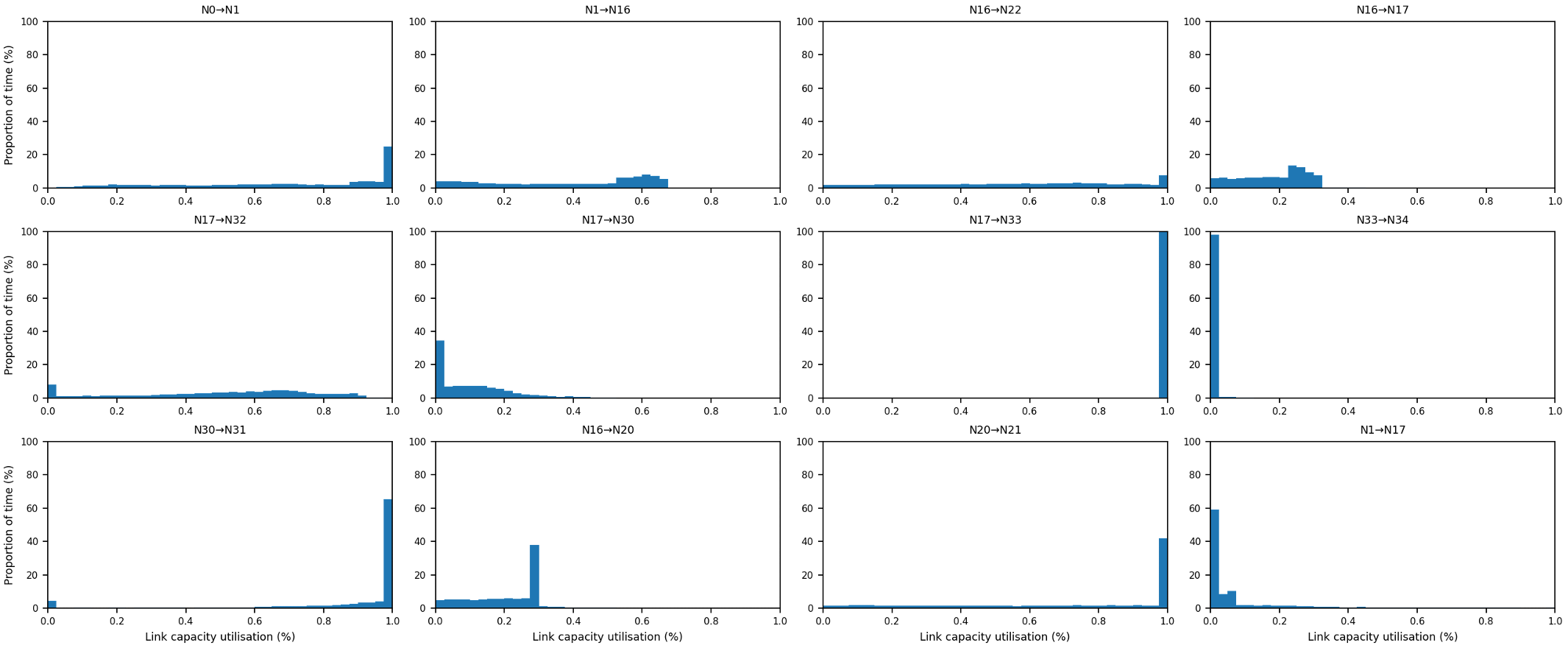}
\caption[
  Line utilisation histograms under LMP and AMM.
]{
  Histograms of normalised line utilisation (flow / thermal limit) under
  LMP (left) and AMM (right) across all timestamps and transmission links.
}
\label{fig:hist_line_utilisation_LMP_AMM}
\end{figure}

\subsection{Energy Supplied by Generation Type}

\begin{figure}[H]
\centering
\includegraphics[width=0.7\textwidth]{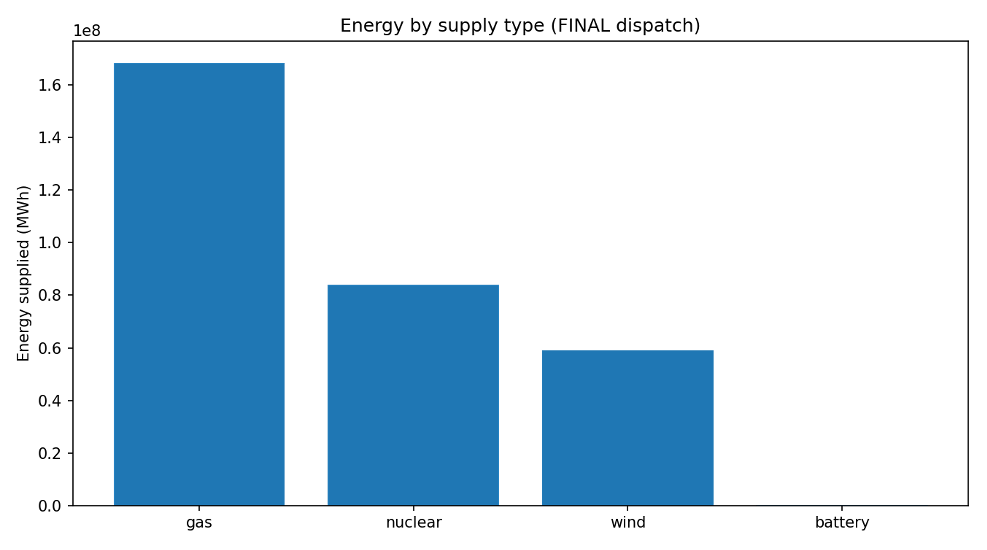}
\caption[
  Energy supplied by generation type under LMP and AMM.
]{
  Annual energy supplied by generation type (wind, nuclear, gas, battery)
  under LMP. Bars show total MWh delivered, with stacked components
  by technology.
}
\label{fig:energy_by_fuel_LMP}
\end{figure}

\begin{figure}[H]
\centering
\includegraphics[width=0.7\textwidth]{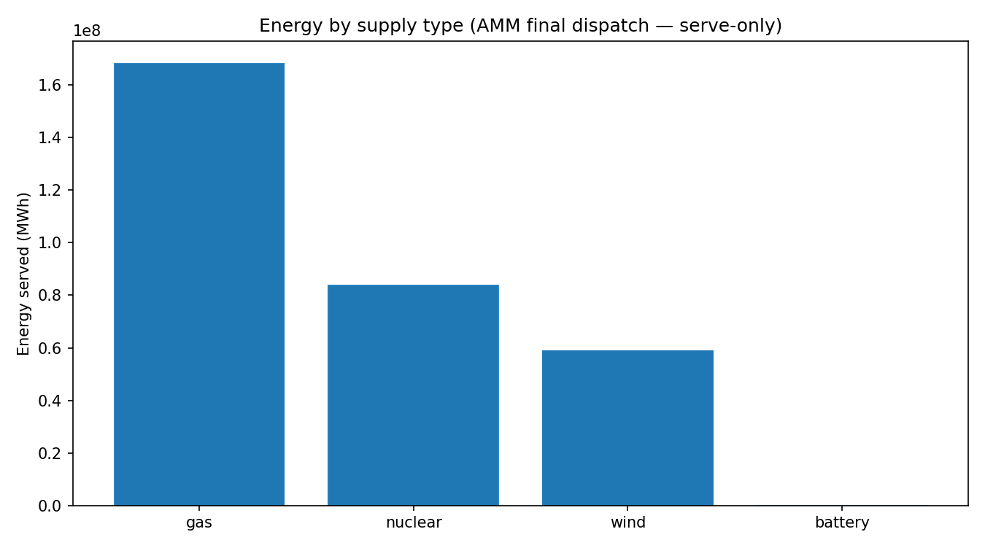}
\caption[
  Energy supplied by generation type under AMM.
]{
  Annual energy supplied by generation type (wind, nuclear, gas, battery)
  under AMM. Bars show total MWh delivered, with stacked components
  by technology.
}
\label{fig:energy_by_fuel_AMM}
\end{figure}

\subsection{Reserves: Required vs Procured and Who Delivers Them}

\begin{figure}[H]
\centering
\includegraphics[width=0.48\textwidth]{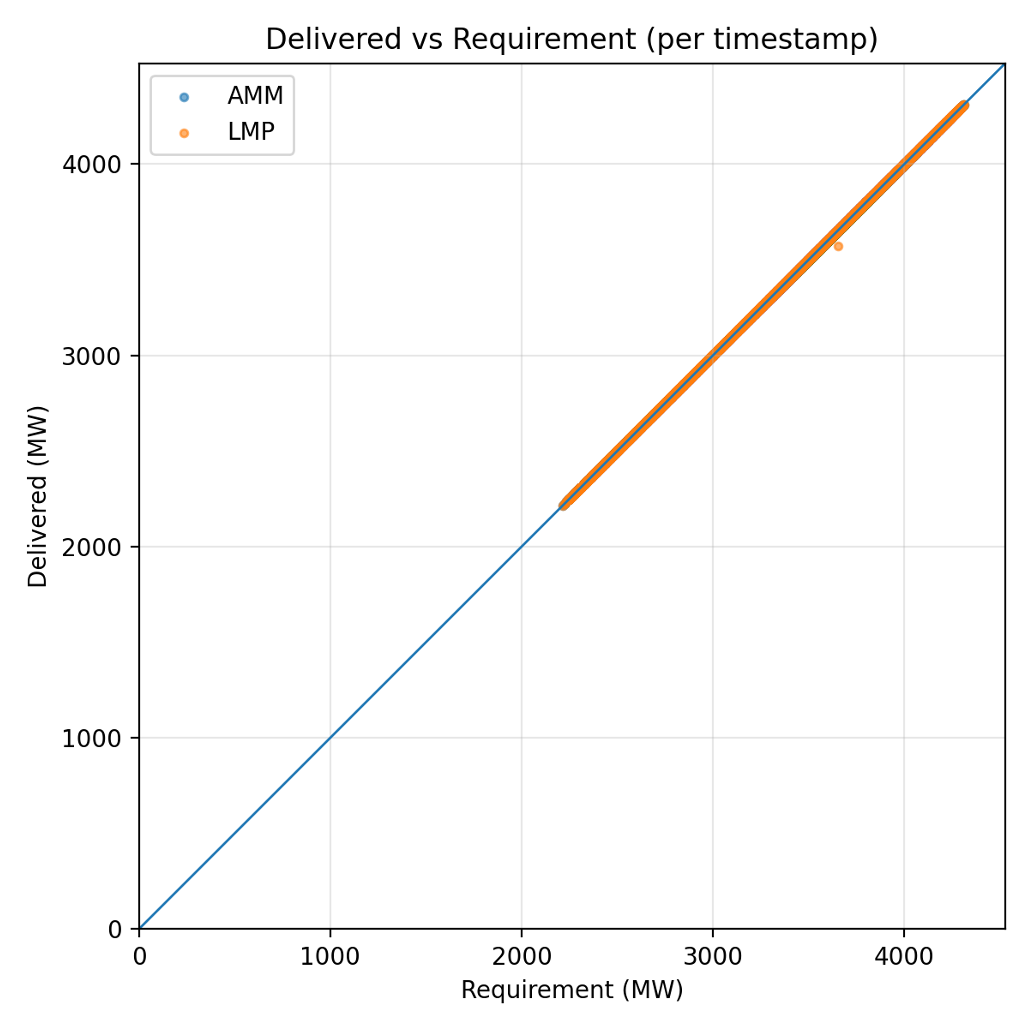}
\caption[
  Reserves required vs procured under LMP and AMM.
]{
  Reserve requirements versus procured reserves under LMP and AMM, aggregated over all timestamps. The figure compares the system-level requirement (solid line) with the realised procured volume
  (dots)
}
\label{fig:reserves_required_vs_procured}
\end{figure}

\begin{figure}[H]
\centering
\includegraphics[width=0.7\textwidth]{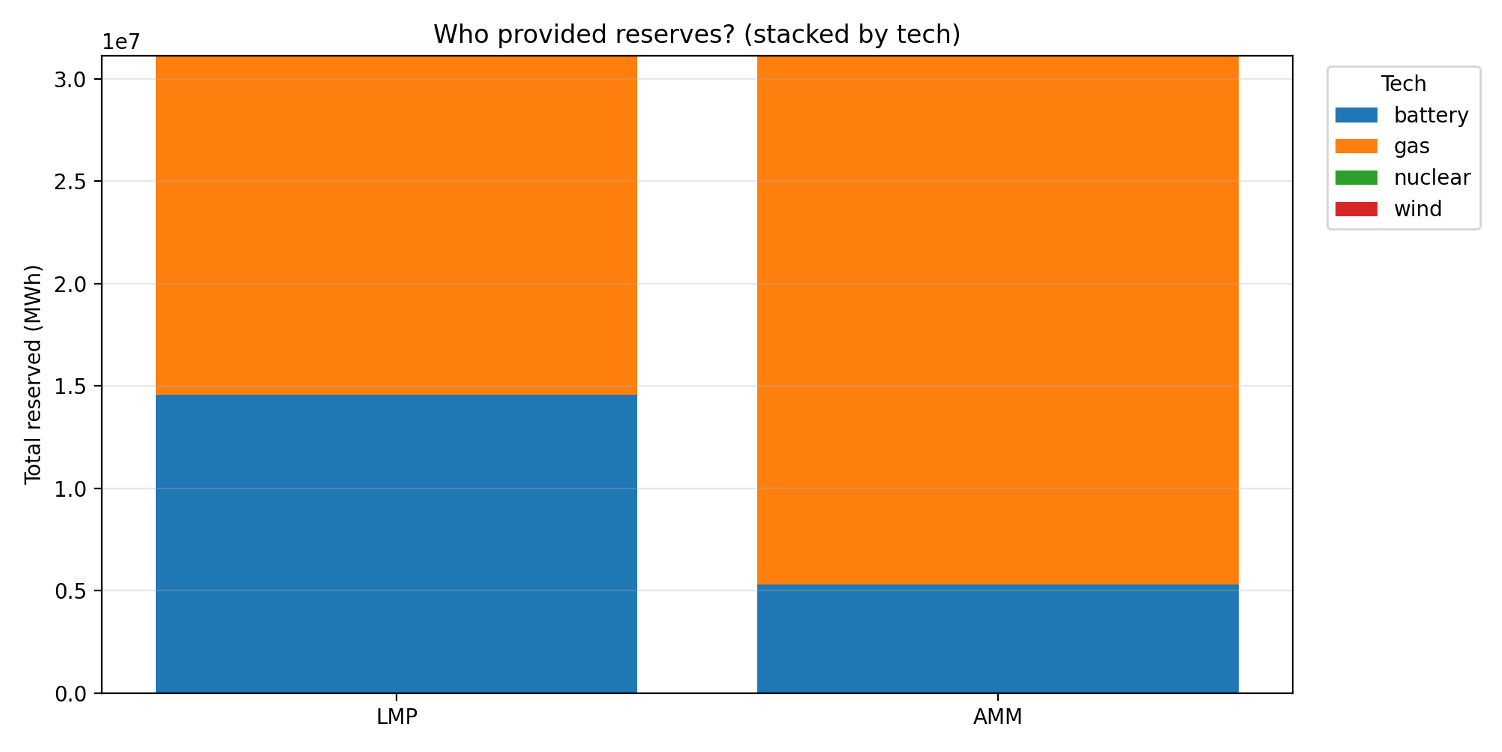}
\caption[
  Reserve contribution by technology and design.
]{
  Share of reserves delivered by each technology under LMP and AMM. Only batteries and gas generators are enabled to provide reserves.
}
\label{fig:reserve_contributors_LMP_vs_AMM}
\end{figure}

\subsection{Demand Served and Curtailed by Node}

\begin{figure}[H]
\centering
\includegraphics[width=0.48\textwidth]{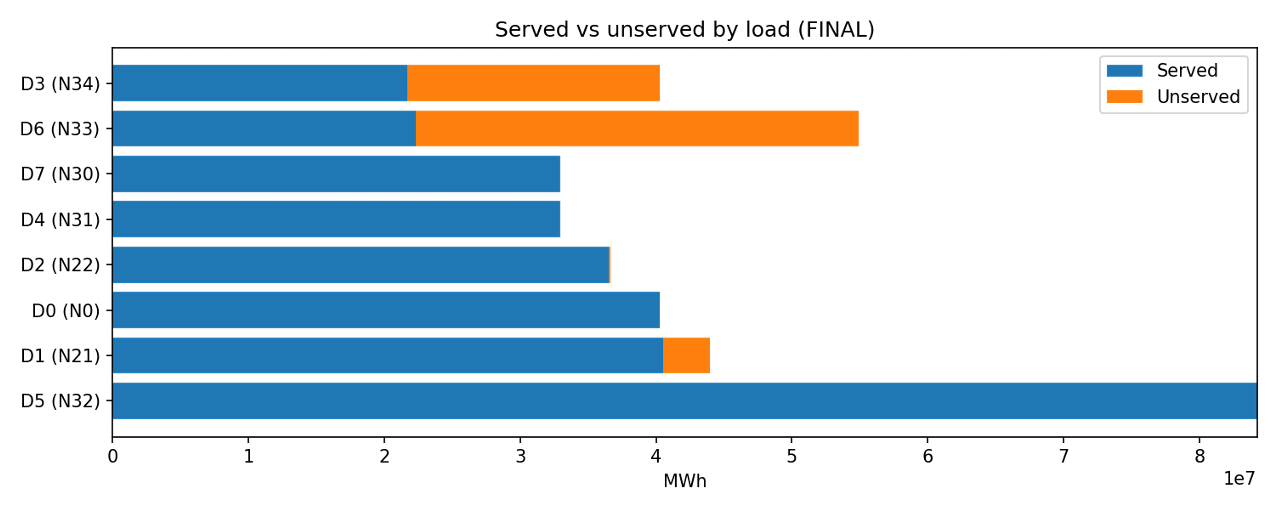}
\includegraphics[width=0.48\textwidth]{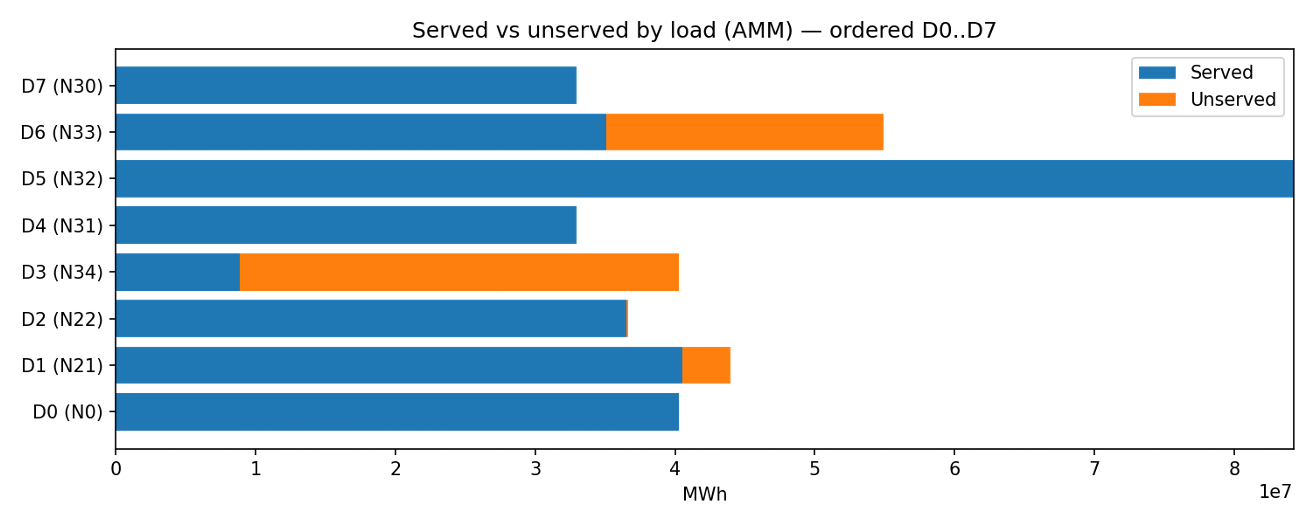}
\caption[
  Demand served and curtailed by node under LMP and AMM.
]{
  Demand served (dark bars) and curtailed (light bars) by node under LMP
  (left) and AMM (right). 
}
\label{fig:demand_served_curtailed_by_node}
\end{figure}

\subsection{Wind Curtailment by Node and Design}

\begin{figure}[H]
\centering
\includegraphics[width=0.48\textwidth]{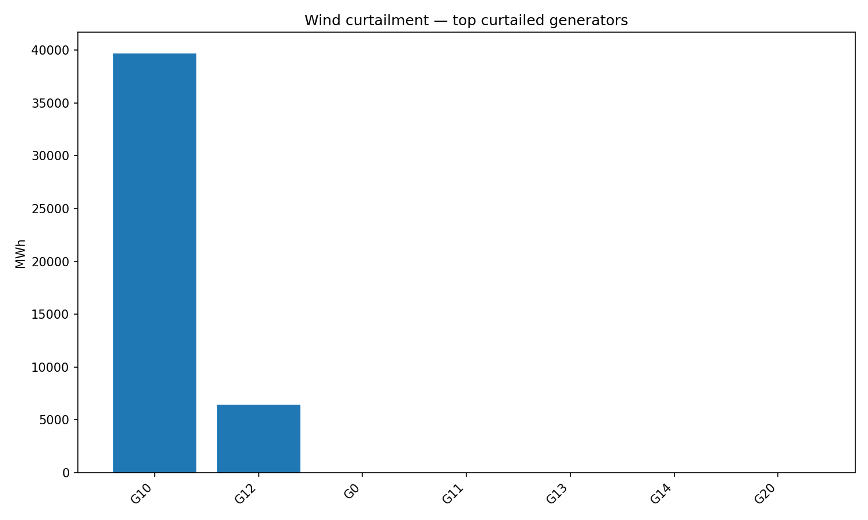}
\includegraphics[width=0.48\textwidth]{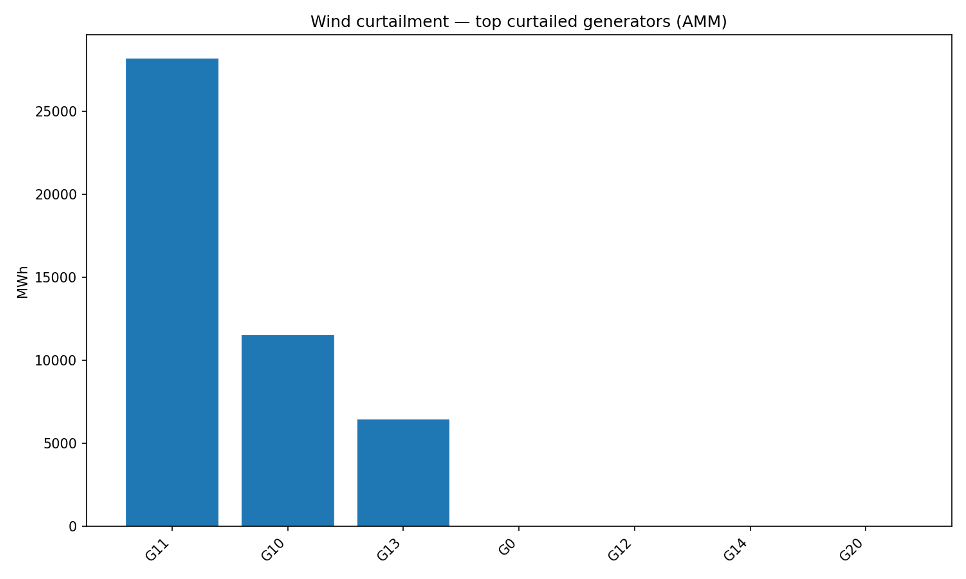}
\caption[
  Wind curtailment by node under LMP and AMM.
]{
  Annual wind curtailment by node under LMP (left) and AMM (right).
}
\label{fig:wind_curtailment_by_node}
\end{figure}

\subsection{Battery Charge/Discharge Profiles Under LMP and AMM}

\begin{figure}[H]
\centering
\includegraphics[width=0.48\textwidth]{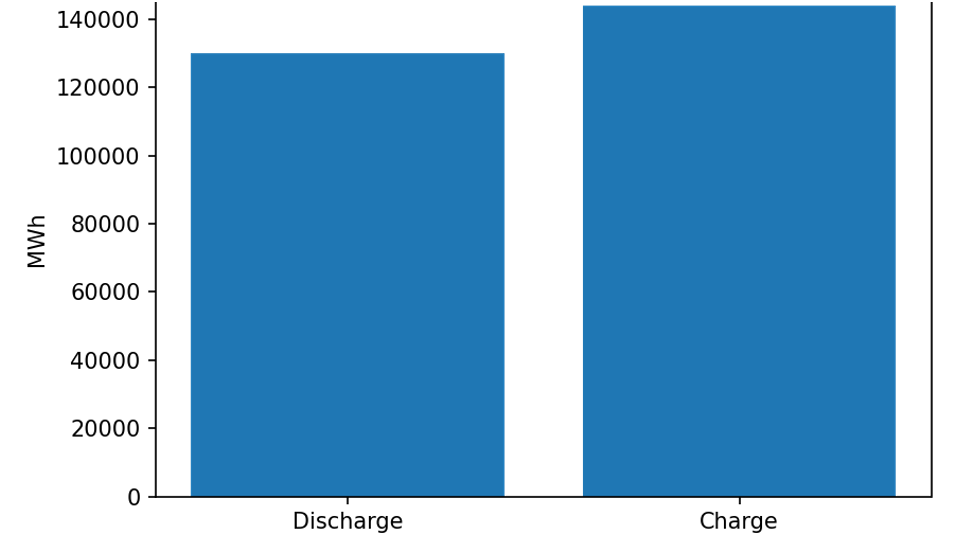}
\includegraphics[width=0.48\textwidth]{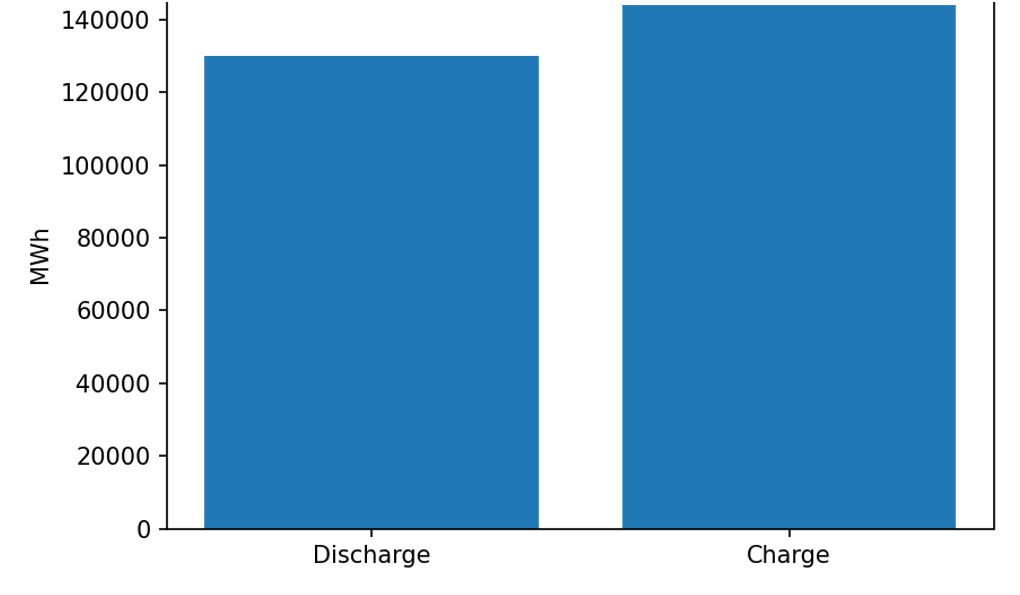}
\caption[
  Battery charge/discharge patterns under LMP and AMM.
]{
  Aggregate battery charge (negative) and discharge (positive) power over time
  under LMP (left) and AMM (right). Under AMM, battery dispatch aligns more
  systematically with tightness peaks and congestion periods, delivering
  higher scarcity relief per MWh cycled, whereas under LMP, dispatch
  primarily tracks short-run price spreads.
}
\label{fig:battery_profiles_LMP_vs_AMM}
\end{figure}

\section{Verification of Shapley alignment}

\subsection{Logic check: intuition}
\label{sec:fairness_logic_check}

Before turning to the formal fairness metrics and Shapley allocations, it is
useful to articulate some intuitive expectations based on the network
topology in Figure \ref{fig:network_topology} and nameplate capacities. 

At a coarse level, the network can be read as a stylised map of the GB
system:

\begin{itemize}[leftmargin=*]

  \item \textbf{(A) Bulk power transfer nodes.}
        Nodes $N1$, $N16$, and $N17$ play the role of bulk regional
        hubs, roughly corresponding to Wales / the western system
        ($N1$), northern England / Scotland ($N16$), and London and the
        South ($N17$). In the results, we will aggregate net injections
        at and ``behind'' each of these hubs (generation minus demand)
        to form three regional time series. These provide a quick visual
        check of when each region is in surplus or deficit and therefore
        when we should expect congestion rents and locational value
        differences to arise.

  \item \textbf{(B) Power-transfer corridors.}
        Three main corridors connect these bulk nodes:
        \begin{itemize}[leftmargin=1.5em]
          \item $N1 \rightarrow N16$ with a 5\,GW limit, providing a
                reasonable amount of north--west transfer capacity;
          \item $N16 \rightarrow N17$ with an 8\,GW limit, representing
                a strong north--to--south transfer path, assuming there
                is spare generation in the north; and
          \item $N1 \rightarrow N17$ with a 4\,GW limit, but with the
                effective export from $N0$ constrained by the 3\,GW
                limit on the $N0$--$N1$ line.
        \end{itemize}
        Intuitively, when northern generators are abundant and southern
        loads are high, we expect these corridors---especially
        $N16$--$N17$ and $N0$--$N1$---to bind. A fair allocation
        mechanism should then reflect higher marginal value for
        generators ``upstream'' of a binding constraint and lower value
        for those sitting behind uncongested capacity.

  \item \textbf{(C) Loads in potentially constrained pockets.}
        Several loads are located in parts of the network that may
        become import-constrained even when the system as a whole is
        well-supplied. In particular, loads $D3$ at $N34$ and $D6$ at
        $N33$ sit behind the $N17$--$N33$ and $N33$--$N34$ interfaces.
        Even though there is substantial generation connected at $N17$,
        the 4\,GW limit on $N17$--$N33$ and the 3\,GW limit on
        $N33$--$N34$ cap how much power can be imported into this
        ``peninsula''. We therefore anticipate:
        \begin{itemize}[leftmargin=1.5em]
          \item higher local scarcity signals and cost shares for
                consumers at $N33$--$N34$ during stressed periods; and
          \item correspondingly higher per-MWh value for generators
                that can directly serve these nodes without transiting
                congested corridors.
        \end{itemize}
        By contrast, load $D1$ at node $N21$ is directly served by
        generators $G10$ and $G11$ and has 3\,GW of import capacity
        from $N20$. On low-wind days when upstream supply at $N20$ is
        tight, we expect $D1$ to experience some scarcity, but in
        general it is better connected than the $N33$--$N34$ pocket.

  \item \textbf{(D) Generators in surplus versus constrained locations.}
        On the supply side, node $N0$ hosts four large units
        ($G0$--$G3$) with a combined nameplate capacity of
        18.3\,GW but only 0.2\,GW of local demand. We therefore expect
        $N0$ to behave as a bulk export node whose generators are often
        competing to supply the rest of the system through a
        3\,GW-limited interface. In fairness terms, this suggests:
        \begin{itemize}[leftmargin=1.5em]
          \item relatively \emph{low} Shapley value per installed MW at
                $N0$, reflecting abundant supply behind a tight
                interface; but
          \item relatively \emph{high} Shapley value per MW for more
                isolated units such as $G20$ at $N34$, which sits close
                to potentially import-constrained loads and faces less
                competition at the margin.
        \end{itemize}
        We therefore expect the fairness metrics to recognise not just
        raw nameplate capacity but \emph{where} that capacity sits
        relative to congestion and demand.

\end{itemize}

These qualitative expectations provide a simple ``logic check'' for the
fairness analysis that follows. If the AMM--Fair Play allocations are
behaving sensibly, we should see:
(i) higher relative rewards for generators in constrained, demand-rich
locations than for over-supplied export nodes; and
(ii) consumer cost shares that increase when they are located behind
binding constraints, but remain bounded and transparent rather than
dominated by arbitrary uplift. In the next sections, we quantify these
patterns and use them to validate the fairness of the design for each
party.

\subsection{Validation: Do Shapley Values Allocate as Expected?}
\label{sec:validation_shapley_expected}

This section provides empirical validation that the Shapley allocation behaves in a manner consistent
with the physical and economic structure of the system. Whereas the formal fairness tests in the main
results chapters evaluate \emph{outcomes} (distributional equity, risk allocation, deprivation reduction),
the diagnostics here evaluate whether the \emph{mechanism itself} allocates value in the right places:
towards generators that contribute marginal value under scarcity, exhibit location-specific importance,
or operate in environments with limited local competition. These checks follow immediately from the
intuition developed in the preceding subsection.

\vspace{1em}
\noindent\textbf{(1) Scarcity responsiveness}

\begin{figure}[H]
  \centering
  \includegraphics[width=0.85\textwidth]{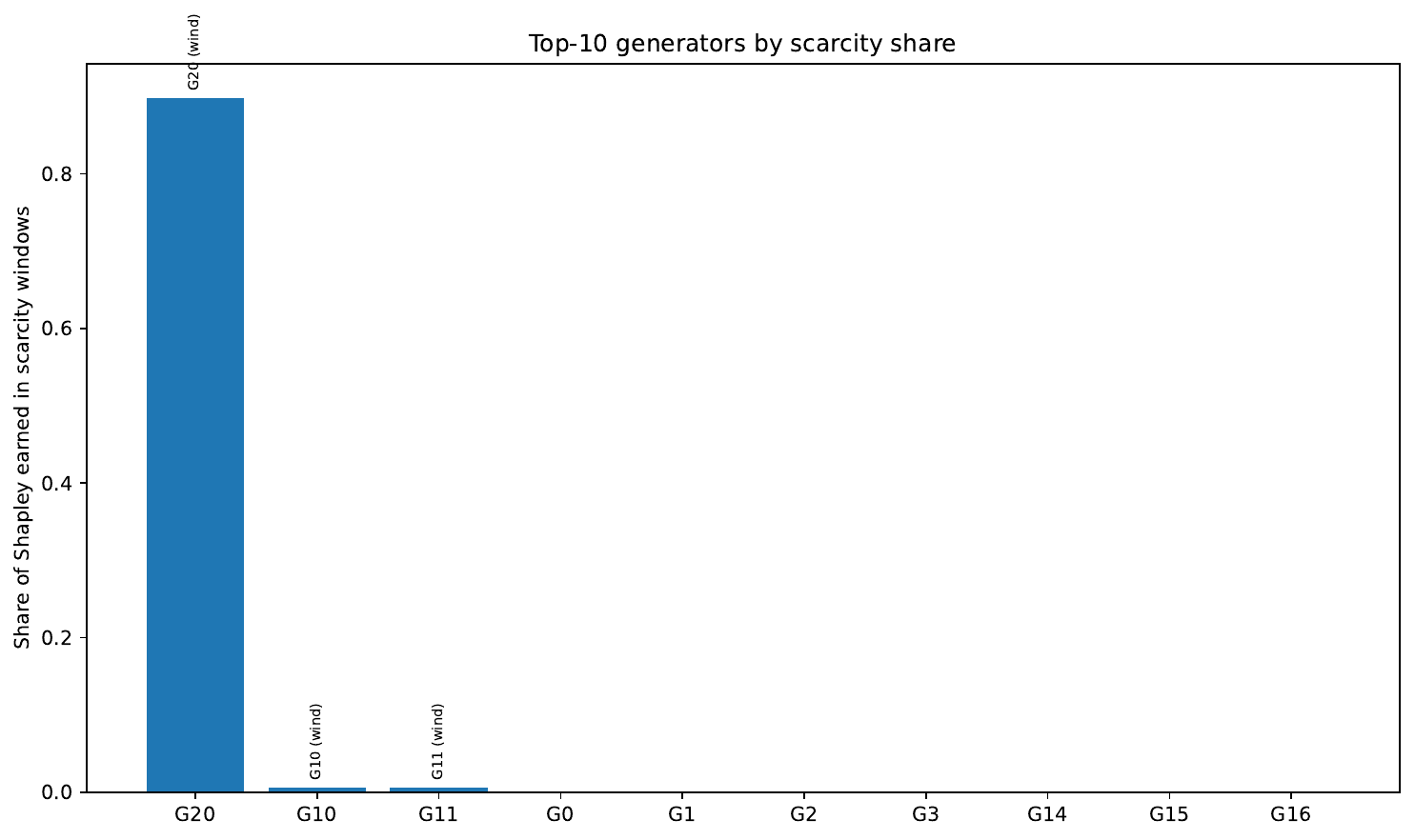}
  \caption{
    Top-10 generators ranked by the share of their total Shapley value earned in scarcity windows.
    See also parsed data from uploaded file.%
    \protect\footnotemark[1]
  }
  \label{fig:scarcity_share_bar}
\end{figure}

A core behavioural requirement is that Shapley value increases when the system becomes tight. To test
this, we compute each generator's share of total Shapley earned specifically during scarcity windows.
Figure~\ref{fig:scarcity_share_bar} illustrates that only a small set of generators earn a large fraction
of their Shapley value during scarcity, and that these generators tend to be located at structurally tight
or weakly connected nodes. This confirms that the AMM--Shapley mechanism correctly identifies marginal
contributors in scarcity periods.

\vspace{1em}
\noindent\textbf{(2) Alignment with nodal scarcity conditions (tightness)}

\begin{figure}[H]
  \centering
  \includegraphics[width=0.85\textwidth]{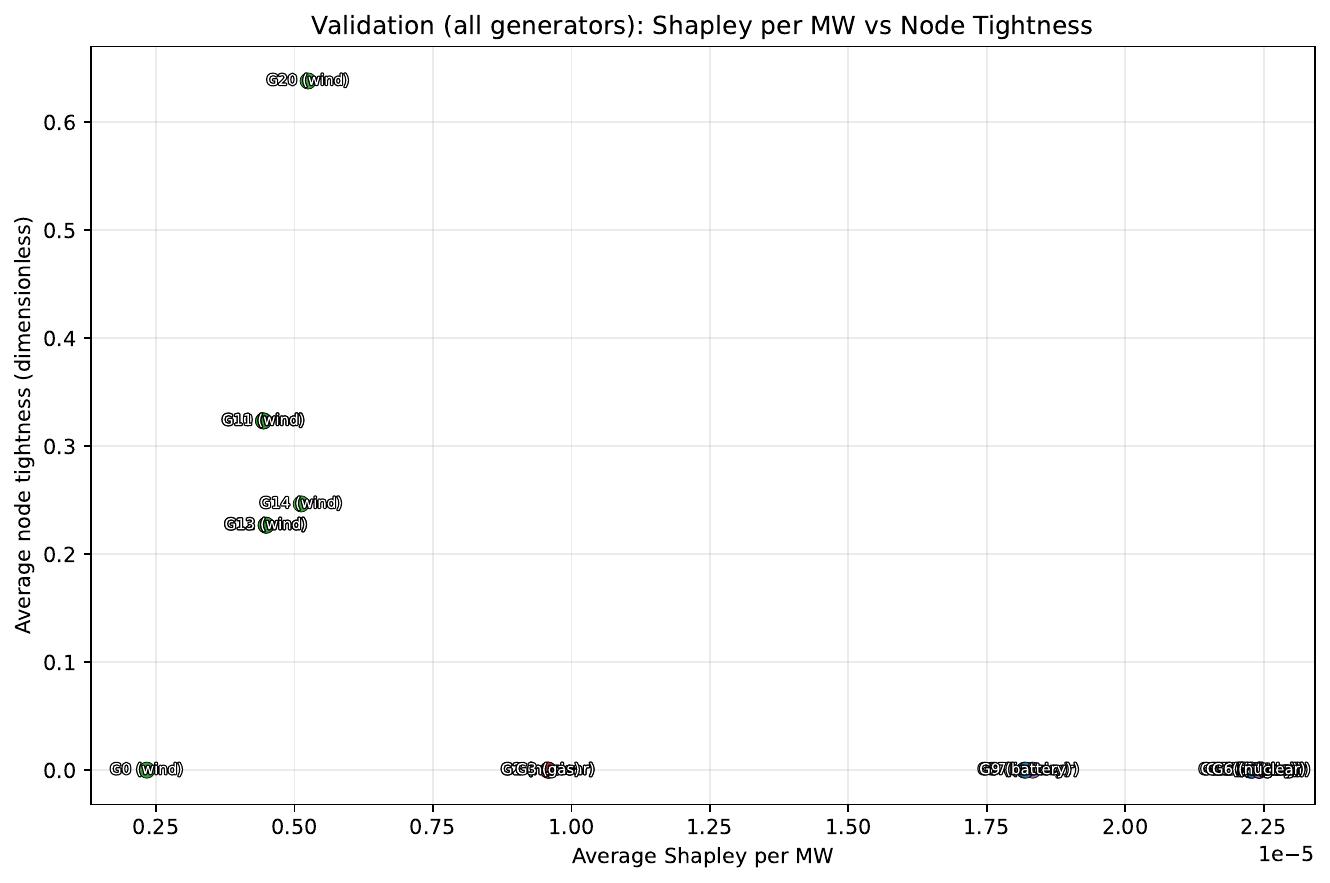}
  \caption{
    Shapley value per MW versus average nodal tightness. The increasing trend confirms that
    generators at tighter nodes receive higher marginal contributions.
  }
  \label{fig:shap_tightness_validation}
\end{figure}

If the Shapley mechanism is behaving correctly, generators situated at tighter nodes should exhibit
higher Shapley-per-MW values. Figure~\ref{fig:shap_tightness_validation} shows a strong monotonic
pattern: average Shapley-per-MW rises with average nodal tightness, with wind units at constrained
nodes receiving substantially higher shares. This provides direct validation that Shapley captures
location-specific marginal value associated with structural scarcity.

\vspace{1em}
\noindent\textbf{(3) Scarcity-only validation against nodal prices}

\begin{figure}[H]
  \centering
  \includegraphics[width=0.85\textwidth]{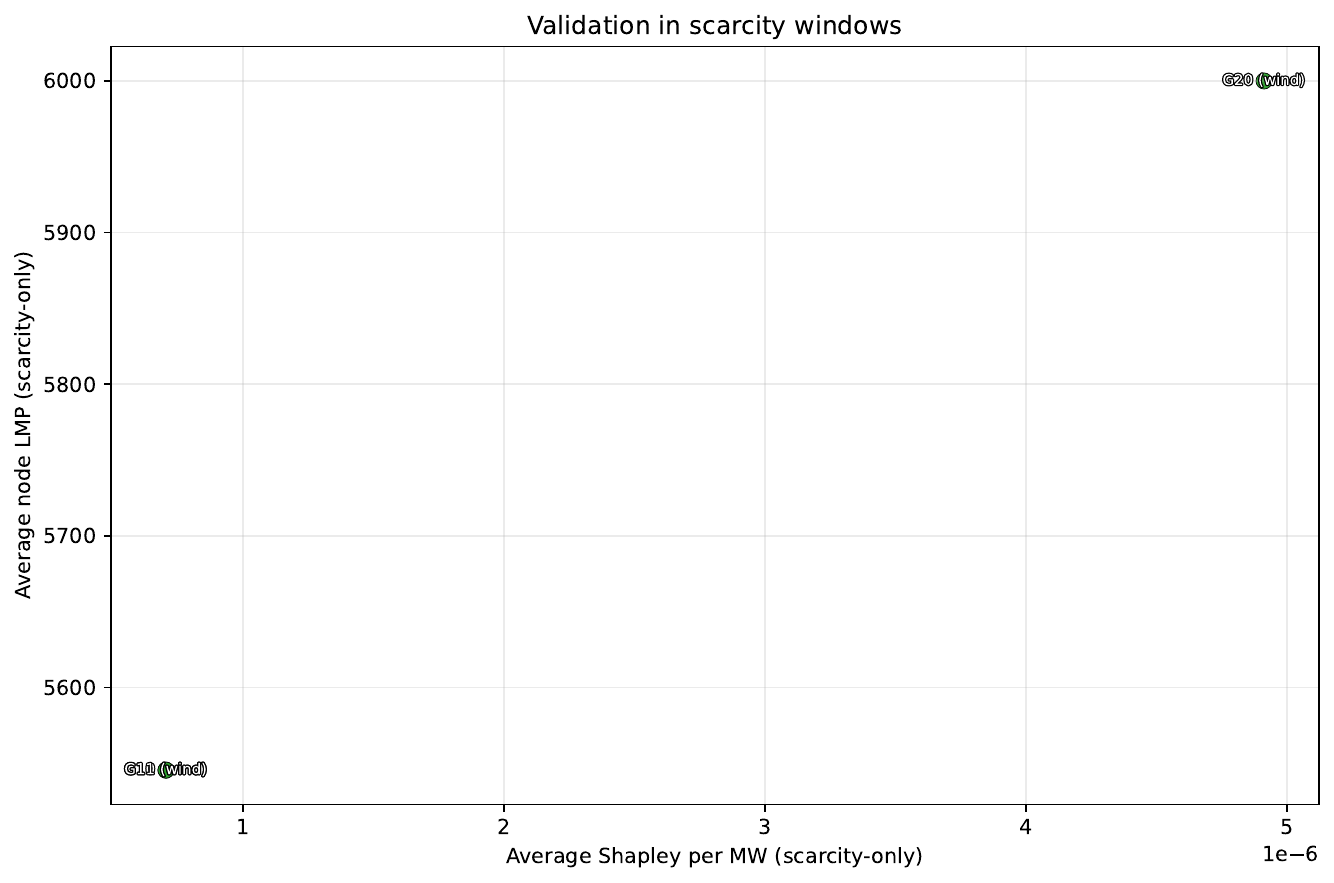}
  \caption{
    Scarcity-only validation: Shapley-per-MW versus average nodal LMP during scarcity windows.
    The positive relationship indicates correct marginal attribution under high-price conditions.
  }
  \label{fig:scarcity_lmp_validation}
\end{figure}

During scarcity windows, marginal generators should align with high nodal prices. Figure~\ref{fig:scarcity_lmp_validation}
confirms that scarcity-period Shapley-per-MW increases with scarcity-period average LMP, validating that the Shapley
mechanism correctly loads value onto generators that matter when prices spike and flexibility is most valuable.

\vspace{1em}
\noindent\textbf{(4) Overall consistency with nodal price levels}

\begin{figure}[H]
  \centering
  \includegraphics[width=0.85\textwidth]{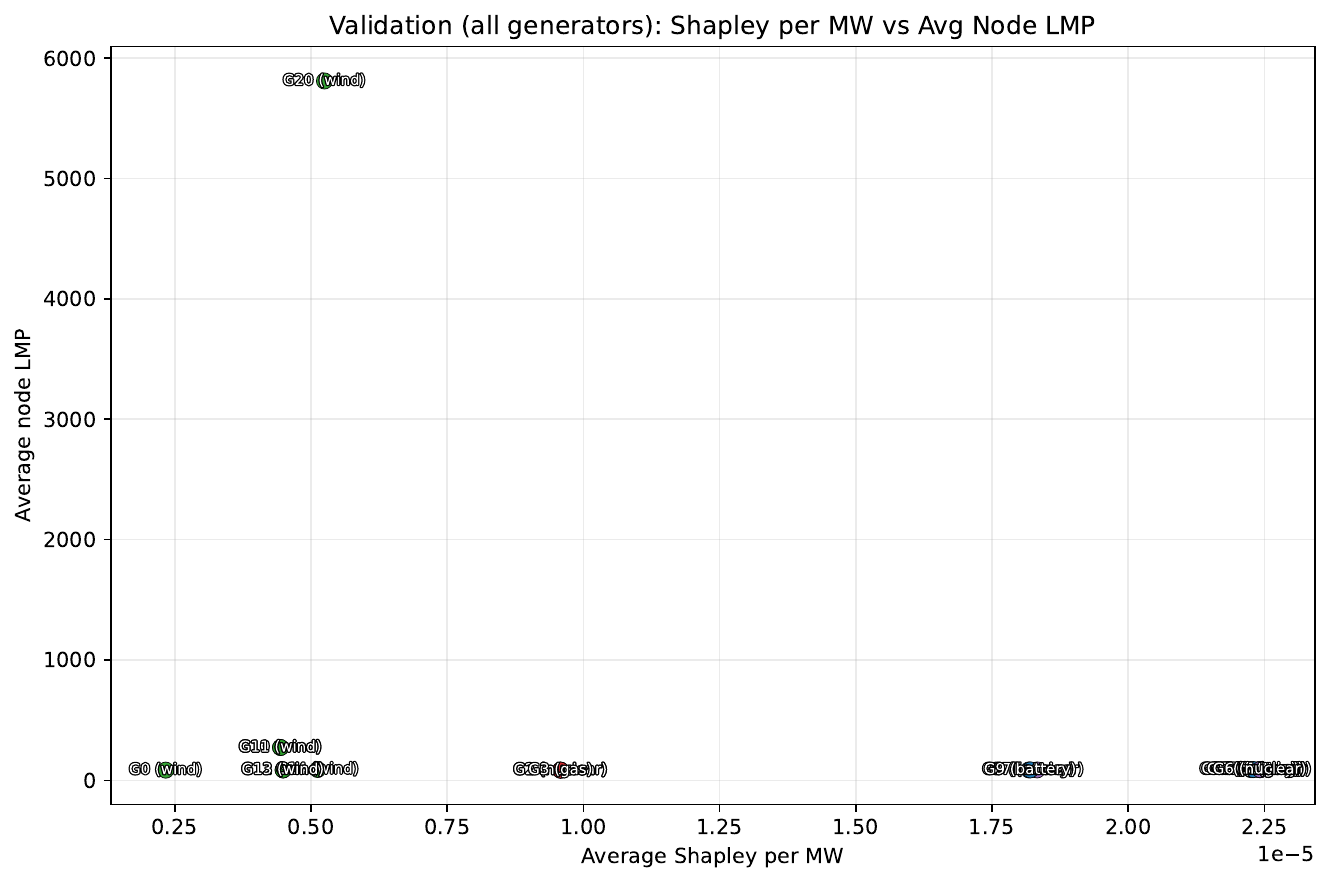}
  \caption{
    Shapley-per-MW versus average nodal LMP across all timestamps.
    Long-run high-value locations correspond to higher marginal Shapley contributions.
  }
  \label{fig:shap_lmp_validation}
\end{figure}

Beyond scarcity, Shapley values should exhibit qualitative alignment with long-run average LMPs.
Figure~\ref{fig:shap_lmp_validation} shows that generators located at persistently high-LMP nodes
receive correspondingly higher Shapley-per-MW, demonstrating that the mechanism correctly internalises
spatial variation in marginal energy value.

\vspace{1em}
\noindent\textbf{(5) Revenue alignment for paid technologies}

\begin{figure}[H]
  \centering
  \includegraphics[width=0.85\textwidth]{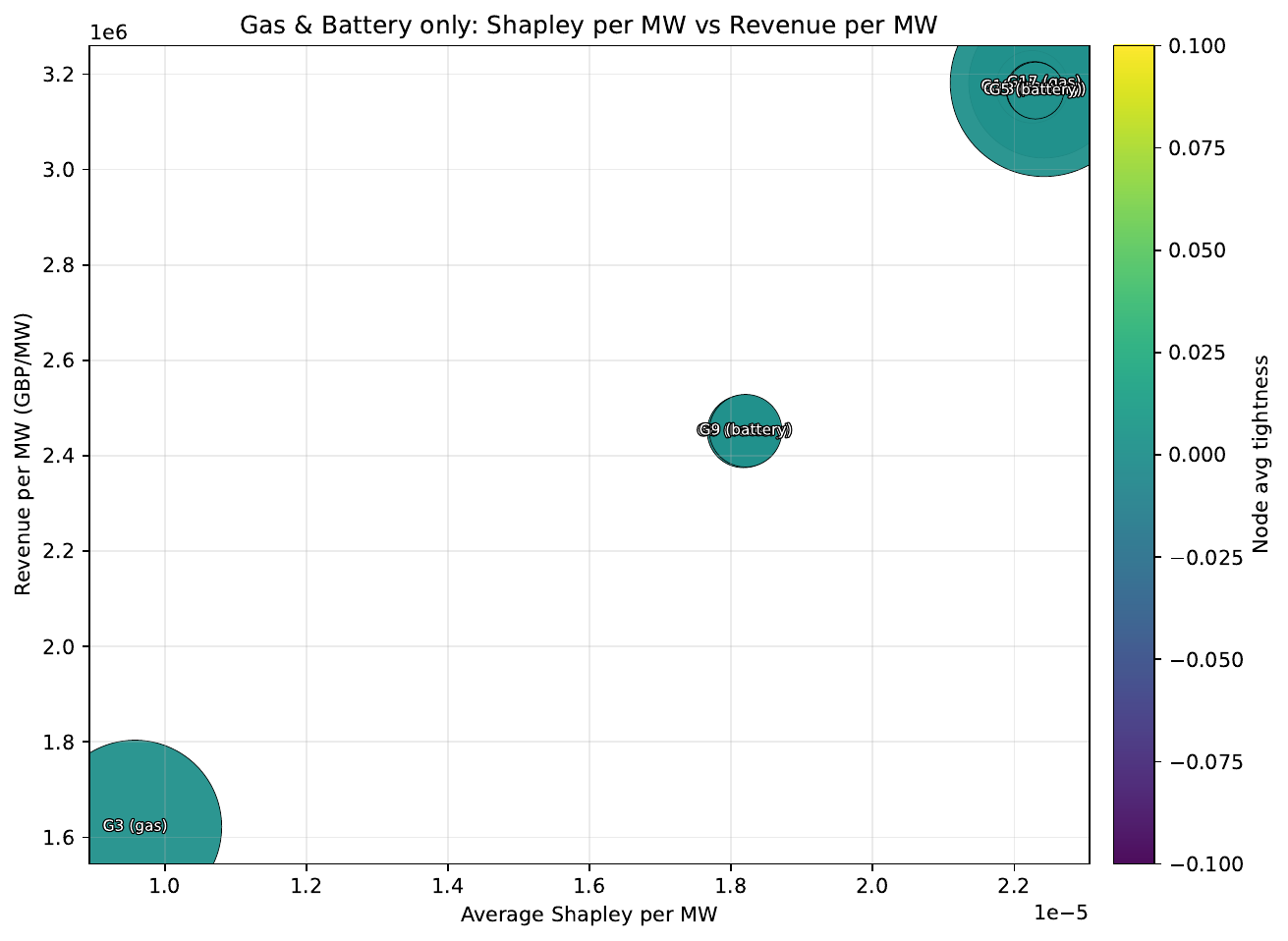}
  \caption{
    Gas \& battery generators: Shapley-per-MW versus revenue-per-MW.
    Alignment confirms that Shapley value tracks economic contribution for paid assets.
  }
  \label{fig:shap_revenue_validation}
\end{figure}

For gas and battery generators---the only technologies directly remunerated in the experiments---we
expect revenue-per-MW to align with Shapley-per-MW. Figure~\ref{fig:shap_revenue_validation} shows
precisely this pattern: Shapley-per-MW is strongly predictive of realised revenue-per-MW. Moreover,
the colour scale reveals that this relationship is mediated by nodal tightness, again demonstrating
mechanistic correctness.

\vspace{1em}
\noindent\textbf{(6) Competition index validation}

\begin{figure}[H]
  \centering
  \includegraphics[width=0.85\textwidth]{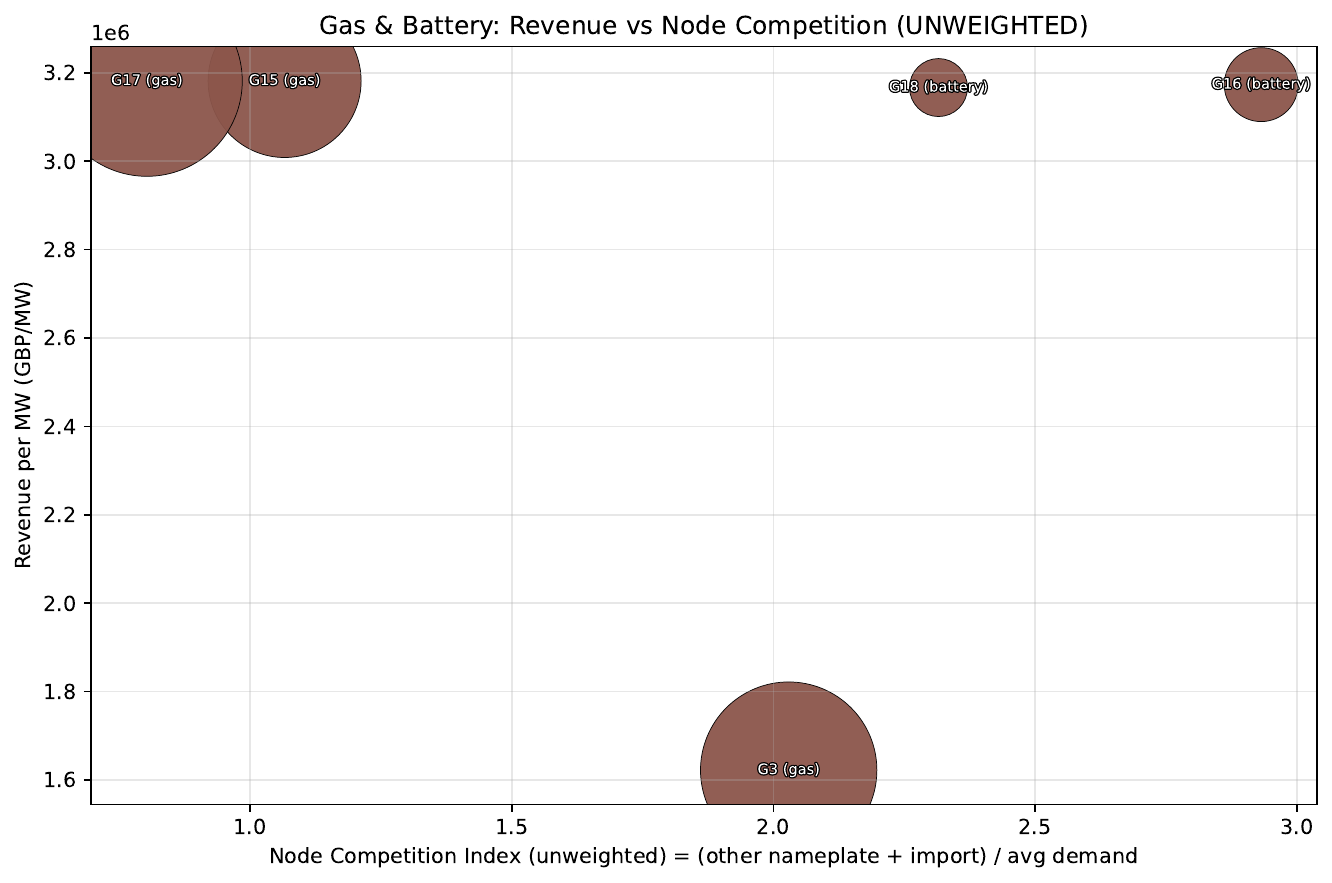}
  \caption{
    Structural Node Competition Index versus revenue-per-MW for gas \& battery generators.
    Higher competition corresponds to lower value.
  }
  \label{fig:structural_nci_validation}
\end{figure}

\begin{figure}[H]
  \centering
  \includegraphics[width=0.85\textwidth]{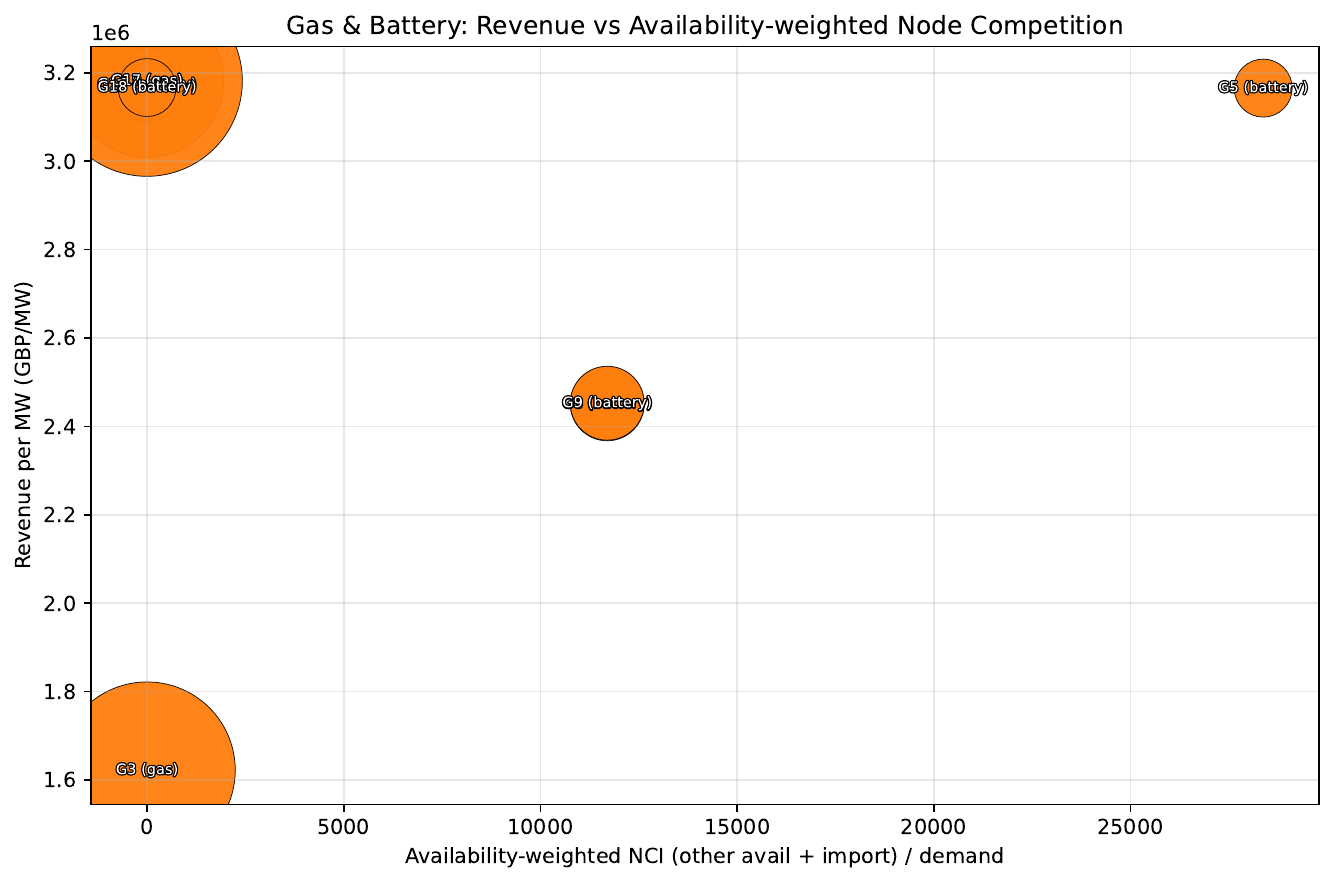}
  \caption{
    Availability-weighted competition index versus revenue-per-MW.
    Time-varying competition further strengthens the expected relationship.
  }
  \label{fig:availability_nci_validation}
\end{figure}

Shapley value should also fall when local competition is strong. This is tested using two competition
indices: (i) a purely structural Node Competition Index (NCI), capturing other available nameplate and
import capacity; and (ii) an availability-weighted NCI incorporating time-varying generator outages.
Figures~\ref{fig:structural_nci_validation} and~\ref{fig:availability_nci_validation} show that both
indices predict lower revenue-per-MW for generators facing greater competition. This confirms that the
AMM--Shapley mechanism assigns less value to generators that are easily substitutable.

\vspace{1em}

Taken together, these diagnostics confirm that the Shapley mechanism allocates value in the correct
places: towards generators that matter during scarcity, are located at high-value or constrained nodes,
and face limited local competition. This establishes the behavioural validity of the mechanism prior
to the formal fairness comparisons with LMP presented in the main results chapters.

% -% ---------------------------------------------------------
\section{Nested--Shapley tractability and empirical validation}
\label{sec:ext_nested_shapley}

Computing exact generator-level Shapley values is combinatorially expensive.
For $|\mathcal{G}|$ generators, the classical Shapley allocation requires
evaluating $2^{|\mathcal{G}|}$ coalitions (up to constant factors), which is
infeasible for realistically sized power systems. To make generator fairness
operational at system scale, this thesis employs a \emph{nested--Shapley}
procedure: generators are first grouped into physically cohesive clusters,
Shapley values are computed over the reduced cluster game, and each cluster-level
value is then disaggregated back to individual generators in proportion to
capacity.

The formal construction, assumptions, and exactness conditions of this approach
are given in Chapter~\ref{ch:mathematics},
Section~\ref{sec:shapley_comp} (Theorem~\ref{thm:nested_shapley_exact}).  
This section provides an empirical validation that the nested--Shapley procedure
reproduces the full generator-level Shapley allocation on the benchmark network
used in the main experiments, and that it yields substantial computational
savings.

\paragraph{Benchmark system and clustering rules.}
We consider a 13-generator test system with an explicit transmission network and
line capacity constraints. Generators are located at different buses and
connected via a set of trunk corridors and branches. Using the clustering
algorithm described in Chapter~\ref{ch:mathematics}, generators are merged into
clusters only when all of the following conditions hold:
\begin{enumerate}[leftmargin=*]
    \item they lie on the same main transmission corridor (common trunk branch);
    \item at least one generator pair across the two groups is within two network
          hops (electrical proximity constraint);
    \item there exists a connecting path whose minimum line capacity exceeds the
          larger of their rated outputs (internal capacity feasibility).
\end{enumerate}
These criteria ensure that generators grouped into the same cluster are
operationally substitutable in the OPF sense and therefore approximate the
within-cluster symmetry and capacity-substitutability assumptions required by
Theorem~\ref{thm:nested_shapley_exact}.

\paragraph{Validation protocol.}
For this benchmark system, we compute:
\begin{itemize}[leftmargin=*]
    \item the \emph{full} generator-level Shapley values $\phi_g$ by evaluating
          the OPF-based characteristic function
          \(
            W(S)
          \)
          for every coalition $S \subseteq \mathcal{G}$; and
    \item the \emph{nested} Shapley values by computing cluster-level Shapley
          values $\Phi_{C_j}$ on the reduced game over
          $\mathcal{C}=\{C_1,\dots,C_m\}$, then disaggregating each $\Phi_{C_j}$
          back to individual generators in proportion to their maximum
          capacities.
\end{itemize}

\paragraph{Results.}
Table~\ref{tab:shapcomp} reports the resulting Shapley values for all 13
generators under both methods. Within numerical precision, the nested--Shapley
procedure exactly reproduces the full Shapley vector:
\[
\phi_g^{\mathrm{nested}} - \phi_g^{\mathrm{exact}} = 0
\qquad \text{for all } g.
\]
This confirms that, for the benchmark network, the clustering rules preserve the
relevant feasible redispatch structure and eliminate infeasible cross-cluster
coalitions without distorting marginal contributions.

\begin{table}[H]
  \centering
  \caption{Shapley value comparison: full generator-level versus nested--Shapley allocation.}
  \label{tab:shapcomp}
  \begin{tabular}{lrrr}
    \toprule
    \textbf{Generator} & $\phi_g$ (full) & $\phi_g$ (nested) & Difference \\
    \midrule
    $G0$  & 0.1034 & 0.1034 & 0.0000 \\
    $G1$  & 0.0690 & 0.0690 & 0.0000 \\
    $G2$  & 0.0862 & 0.0862 & 0.0000 \\
    $G3$  & 0.0690 & 0.0690 & 0.0000 \\
    $G4$  & 0.0862 & 0.0862 & 0.0000 \\
    $G5$  & 0.1034 & 0.1034 & 0.0000 \\
    $G6$  & 0.0517 & 0.0517 & 0.0000 \\
    $G7$  & 0.0690 & 0.0690 & 0.0000 \\
    $G8$  & 0.0690 & 0.0690 & 0.0000 \\
    $G9$  & 0.0690 & 0.0690 & 0.0000 \\
    $G10$ & 0.0690 & 0.0690 & 0.0000 \\
    $G11$ & 0.0862 & 0.0862 & 0.0000 \\
    $G12$ & 0.0690 & 0.0690 & 0.0000 \\
    \bottomrule
  \end{tabular}
\end{table}

\begin{figure}[H]
  \centering
  \includegraphics[width=0.9\textwidth]{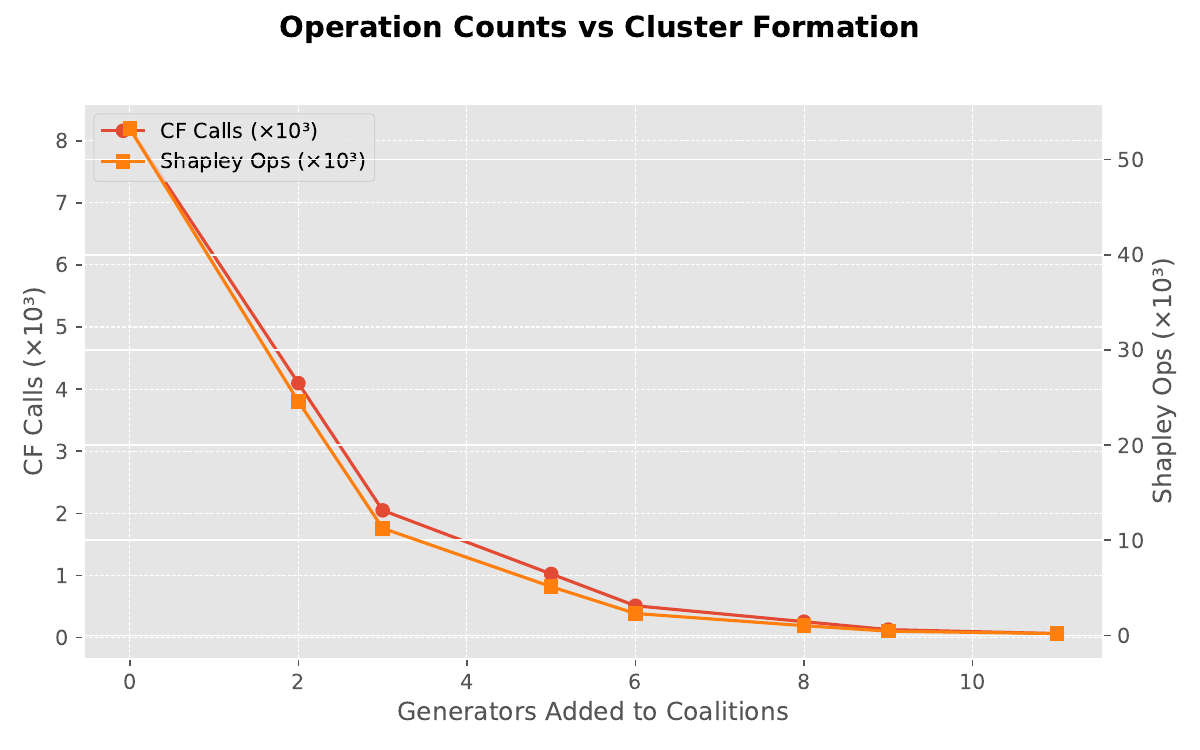}
  \caption[
    OPF evaluations vs number of clusters.
  ]{
    Number of OPF evaluations required to compute Shapley values as a function
    of the number of clusters. Clustering reduces the combinatorial burden
    from exponential in the number of generators to exponential in the number
    of clusters.
  }
  \label{fig:ops_vs_clusters}
\end{figure}

\begin{figure}[H]
  \centering
  \includegraphics[width=0.9\textwidth]{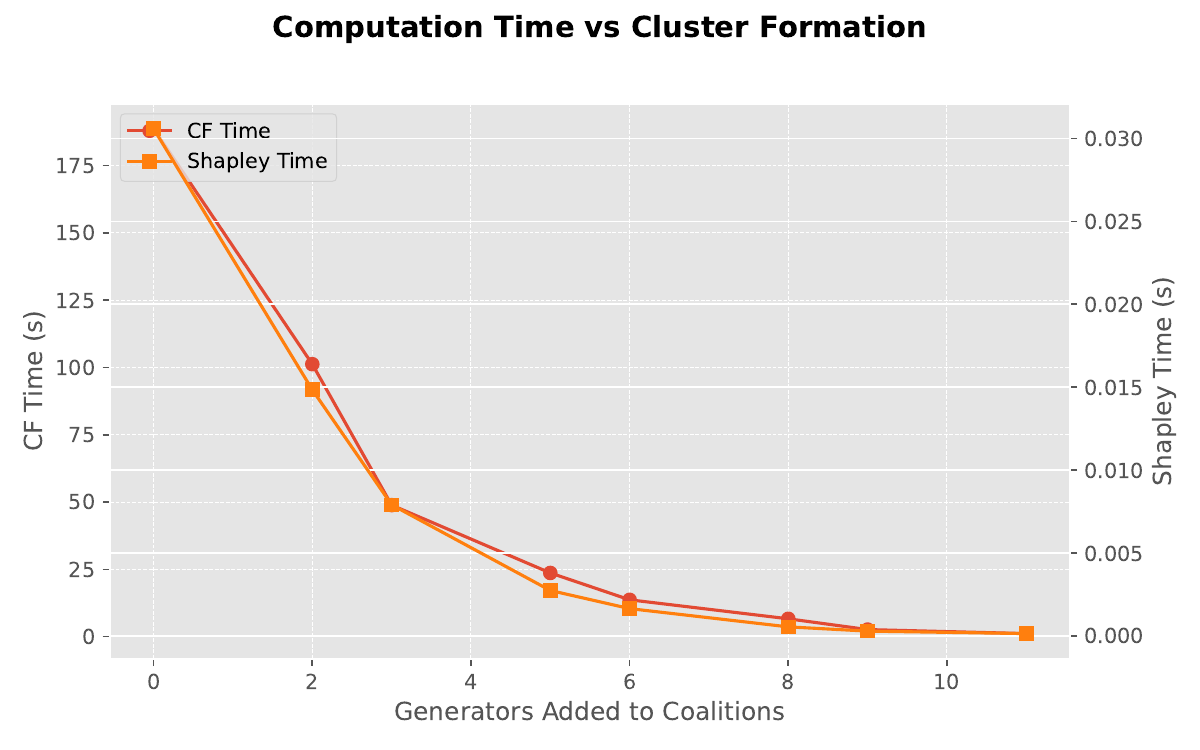}
  \caption[
    Runtime vs number of clusters.
  ]{
    Total computation time for Shapley evaluation as a function of the number
    of clusters. Runtime collapses rapidly as the clustered (nested) game
    replaces the full generator-level coalition enumeration.
  }
  \label{fig:time_vs_clusters}
\end{figure}

\paragraph{Computational tractability.}
Algorithmic profiling results (Figures~\ref{fig:ops_vs_clusters} and
\ref{fig:time_vs_clusters}, reported in Chapter~\ref{ch:mathematics}) show that
the number of OPF evaluations and total runtime grow rapidly with the number of
individual generators under the full Shapley computation, but remain tractable
when Shapley values are computed over clusters. This confirms that nested--Shapley
achieves the intended dimensionality reduction without sacrificing allocation
accuracy on the benchmark network.

Taken together, Theorem~\ref{thm:nested_shapley_exact} and
Table~\ref{tab:shapcomp} justify the use of nested--Shapley in the main AMM
experiments: generator fairness is evaluated using a computationally efficient
procedure that is provably exact under the stated symmetry conditions and
empirically exact for the test system considered here.

% ---------------------------------------------------------
% ---------------------------------------------------------

% ---------------------------------------------------------

% =========================================================
\section{Extended Network and Scarcity Diagnostics}
\label{sec:extended_network_scarcity}

This section reports extended diagnostic results that explain the physical and
temporal mechanisms underlying the fairness outcomes documented in
Section~\ref{sec:results_fairness}. These results are not themselves fairness
tests: rather, they characterise the network bottlenecks, scarcity propagation,
and value formation dynamics that give rise to the observed remuneration
patterns under LMP and AMM. All results are computed on identical dispatch,
demand, and network inputs.

\subsection{Congestion Frequency and Structural Bottlenecks}

\begin{figure}[H]
\centering
\includegraphics[width=0.48\textwidth]{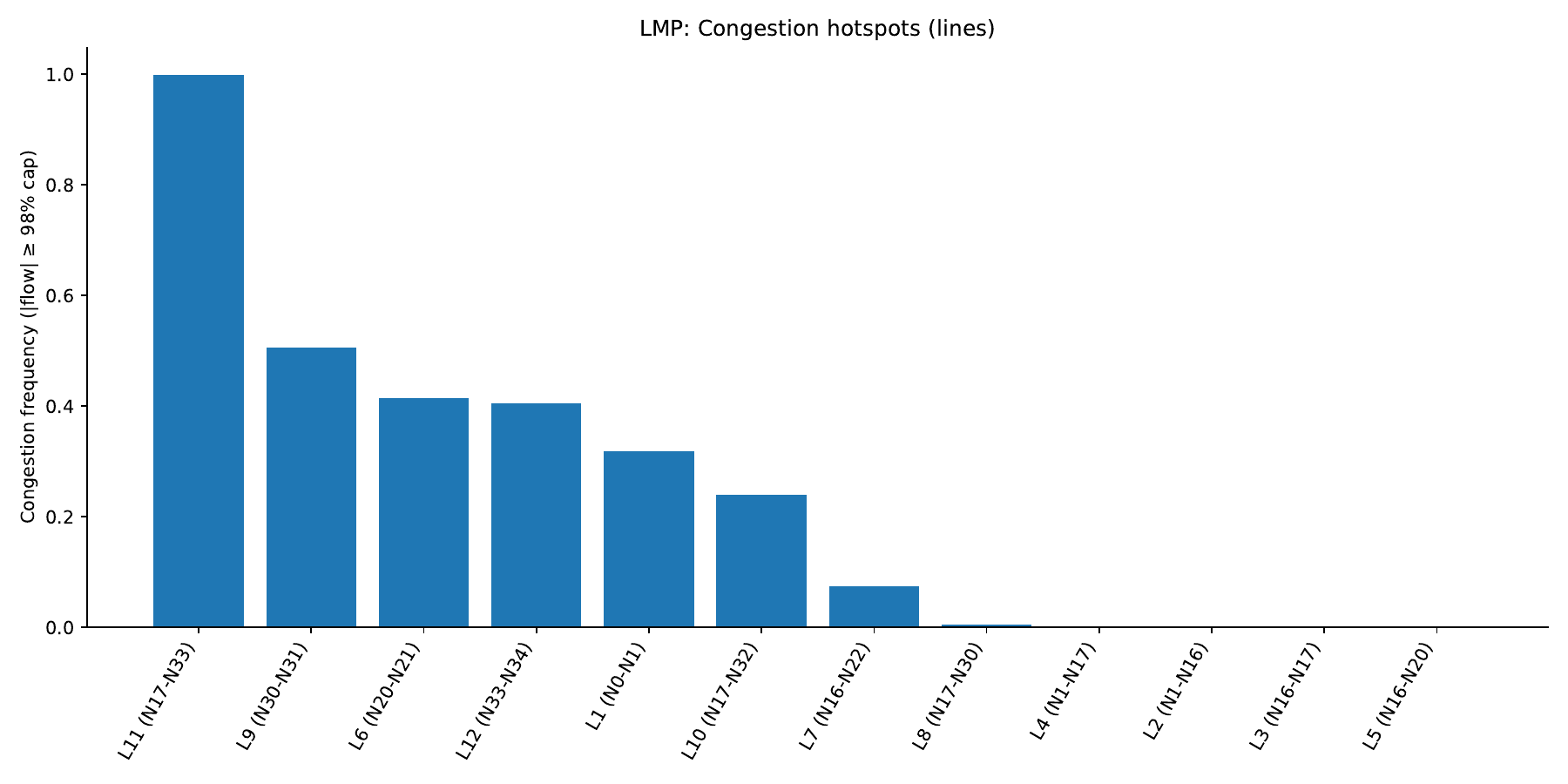}
\includegraphics[width=0.48\textwidth]{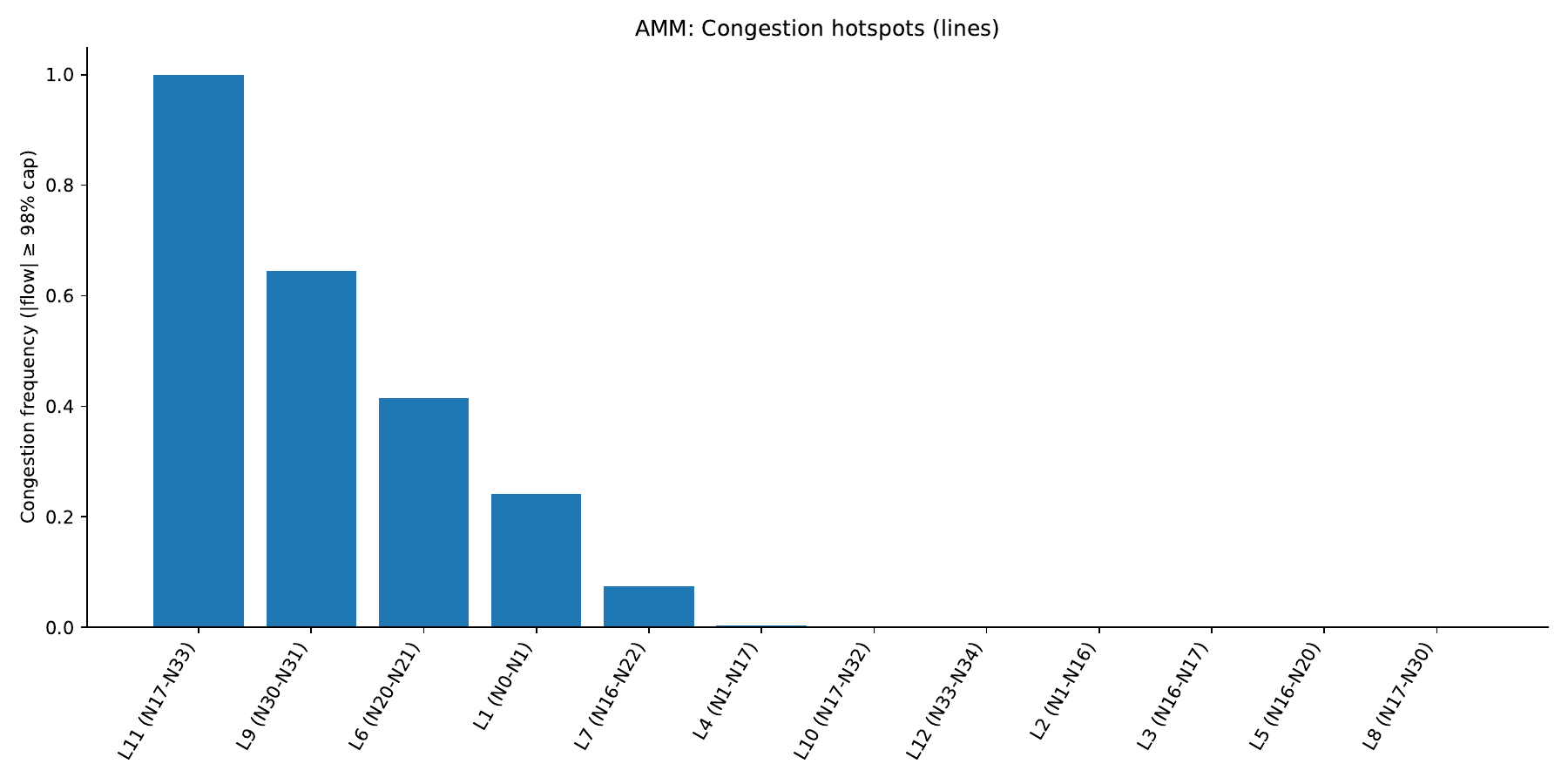}
\caption[
  Congestion frequency by transmission line under LMP and AMM.
]{
  Frequency with which each transmission line operates within 98\% of its
  thermal limit. Both designs identify the same structural bottlenecks, but
  differ in how congestion is economically internalised.
}
\label{fig:congestion_lines_LMP_AMM}
\end{figure}

Figure~\ref{fig:congestion_lines_LMP_AMM} reports the frequency with which each
transmission line operates within 98\% of its thermal limit under LMP and AMM.
Across both designs, congestion is highly concentrated on a small subset of
corridors, notably those connecting the N16--N17--N33 and N30--N31 regions. This
confirms that scarcity is driven by persistent structural bottlenecks rather
than stochastic or evenly distributed stress.

While the identity of congested lines is broadly consistent across designs,
their economic interpretation differs. Under LMP, congestion is resolved
ex post through nodal price separation, with no anticipatory adjustment of
demand or remuneration. Under AMM, the same bottlenecks inform tightness
signals that feed directly into the allocation of scarcity value. As a result,
congestion under LMP manifests primarily as price volatility, whereas under AMM
it acts as an input into controlled, bounded scarcity pricing.

\subsection{Local Adequacy Alignment by Node}

\begin{figure}[H]
\centering
\includegraphics[width=0.48\textwidth]{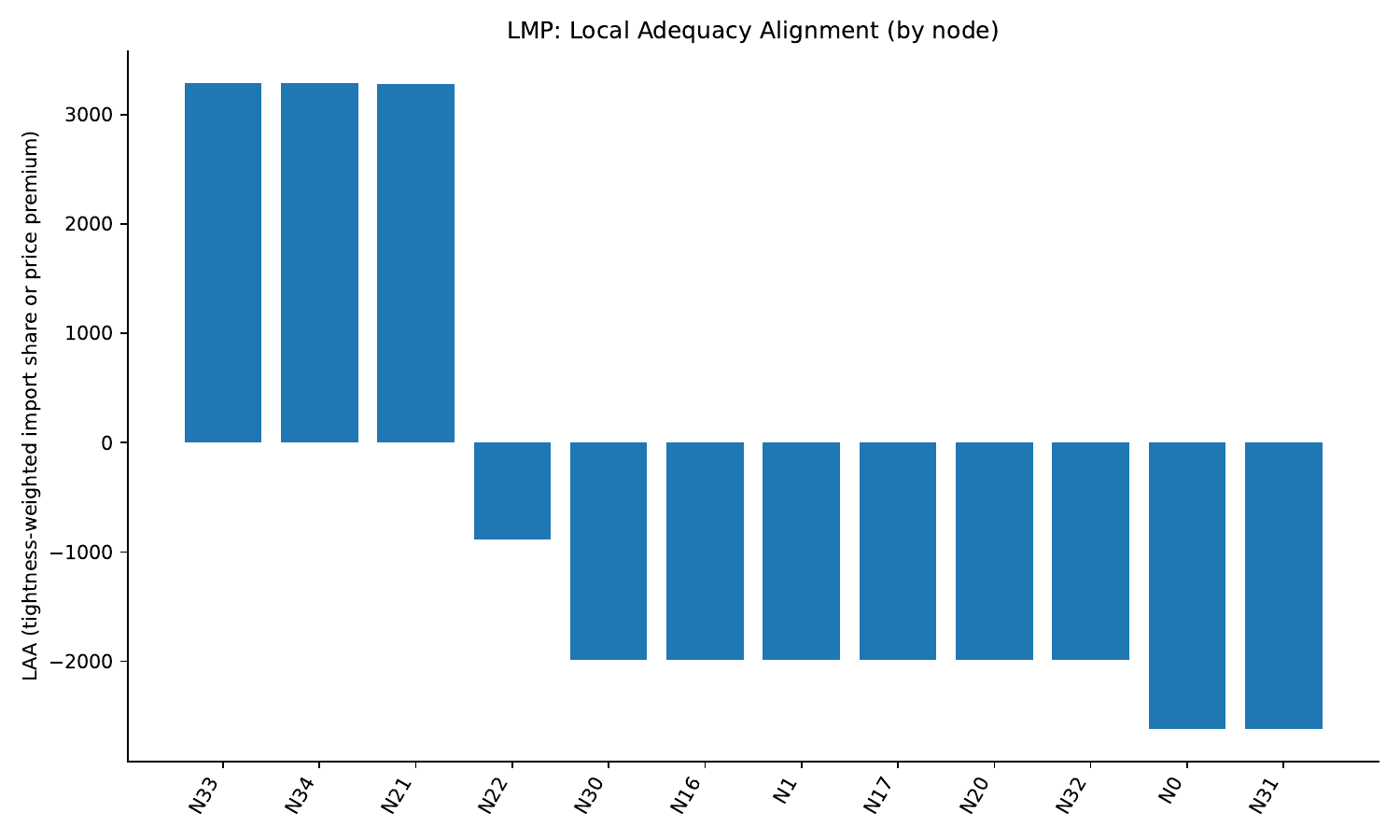}
\includegraphics[width=0.48\textwidth]{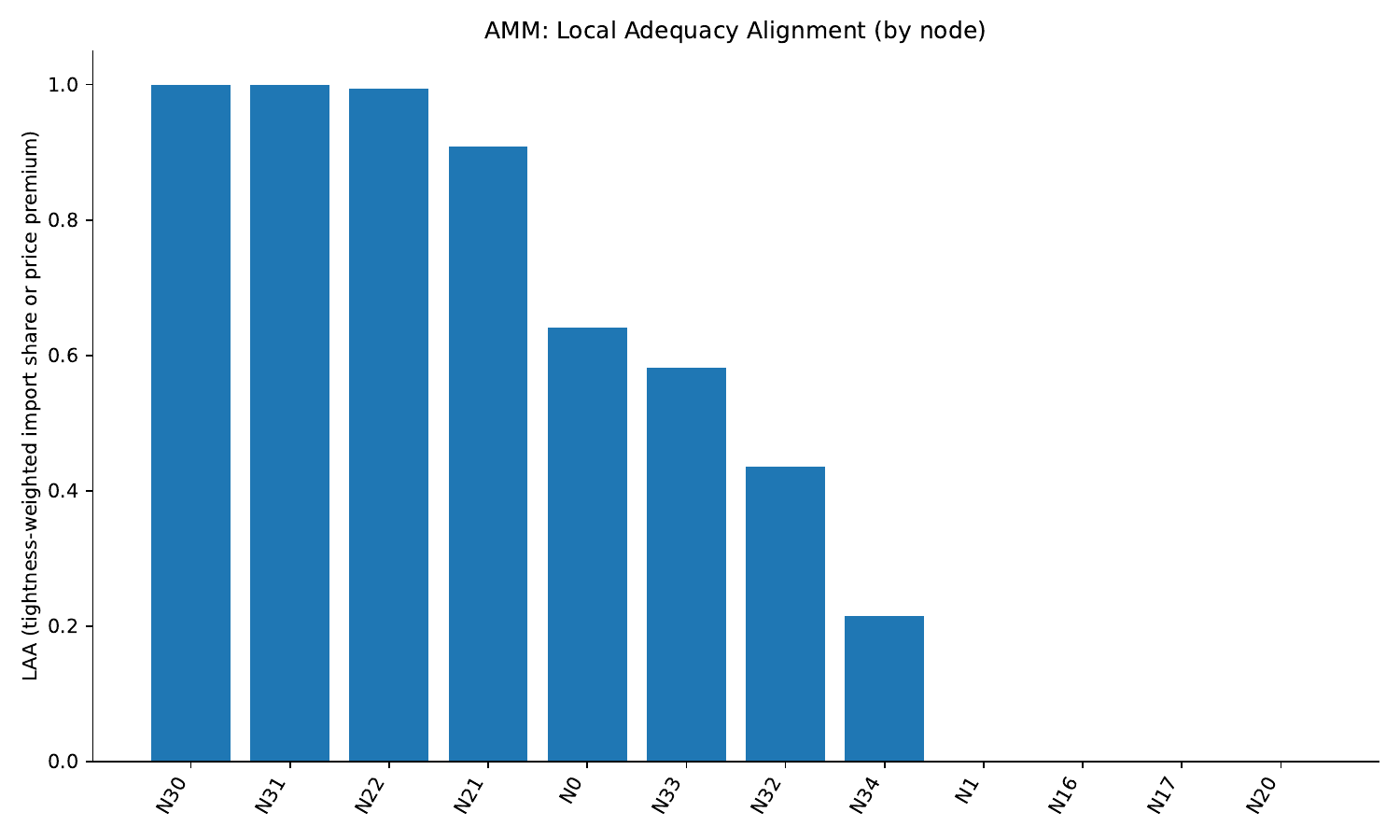}
\caption[
  Local Adequacy Alignment (LAA) by node under LMP and AMM.
]{
  Tightness-weighted import dependence by node. Under AMM, LAA values are
  bounded and interpretable as structural dependence. Under LMP, LAA is
  dominated by tail price behaviour and becomes unbounded.
}
\label{fig:node_LAA_LMP_AMM}
\end{figure}

Figure~\ref{fig:node_LAA_LMP_AMM} reports the Local Adequacy Alignment (LAA) metric
by node, defined as the tightness-weighted share of imports or equivalent price
premium exposure during system stress events. LAA provides a spatial diagnostic
of which locations rely most heavily on the rest of the system when capacity is
scarce.

Under AMM, LAA values are bounded and smoothly distributed across nodes, with
higher values indicating persistent structural dependence on imports during
tight conditions. This yields an interpretable ranking of nodal dependence that
is stable across time.

Under LMP, by contrast, LAA values span multiple orders of magnitude. Nodes
located behind frequently congested interfaces exhibit extreme LAA not because
of persistent inadequacy, but because rare scarcity events produce unbounded
price premia. As a result, LAA under LMP is dominated by tail price behaviour
rather than structural dependence, reducing its usefulness as a diagnostic of
physical causation.

\subsection{System Alignment versus Revenue Alignment (LMP)}

\begin{figure}[H]
\centering
\includegraphics[width=0.65\textwidth]{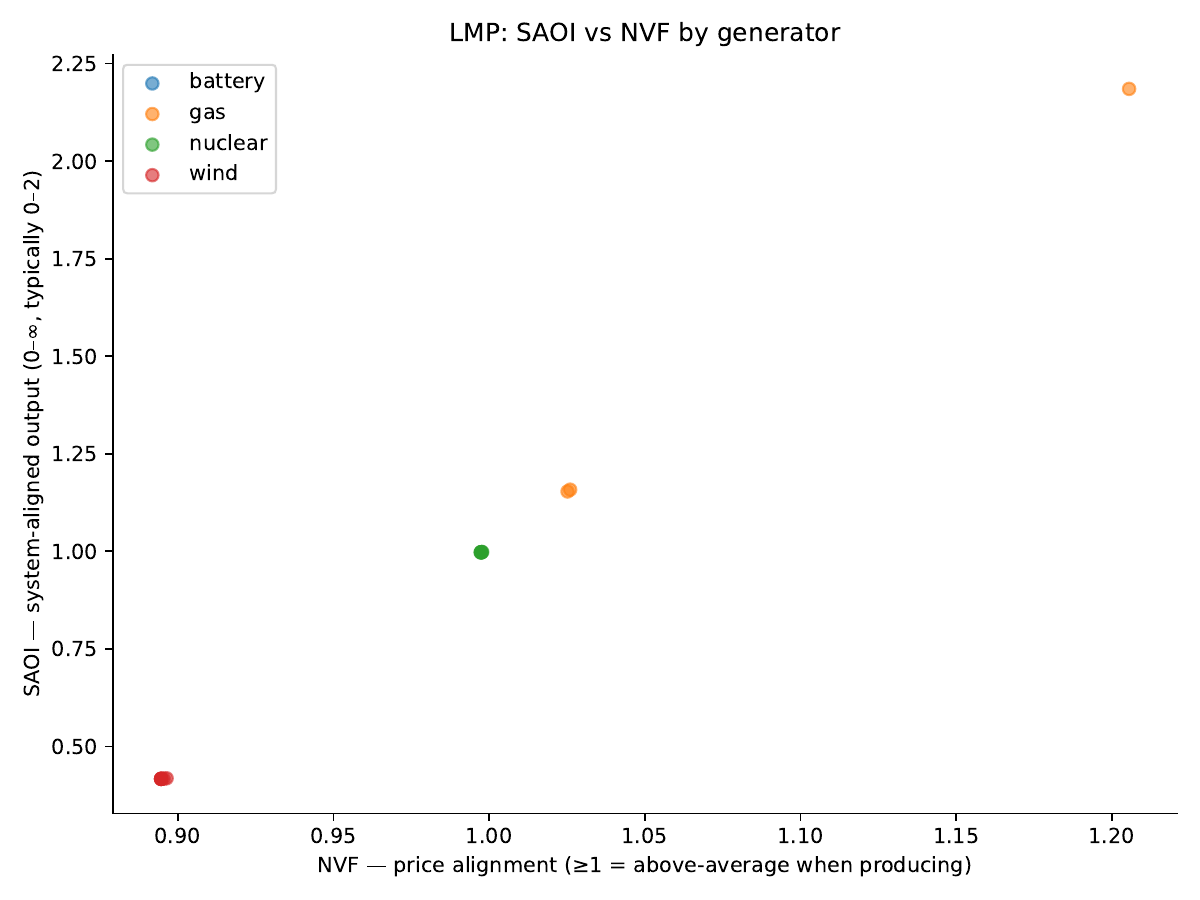}
\caption[
  System-aligned output versus revenue alignment under LMP.
]{
  Scatter of system-aligned output (SAOI) against normalised value factor (NVF)
  for individual generators under LMP. Weak alignment indicates that realised
  revenues reflect scarcity rents and locational effects rather than marginal
  contribution to system adequacy.
}
\label{fig:saoi_vs_nvf}
\end{figure}

Figure~\ref{fig:saoi_vs_nvf} plots, for each generator, system-aligned output
(SAOI) against the Normalised Value Factor (NVF) under LMP. SAOI measures the
extent to which a generator produces during system-tight periods, while NVF
captures whether realised prices when producing exceed the system average.

The resulting scatter shows weak and noisy alignment between contribution and
remuneration. Several generators with high SAOI receive only average or below-
average prices, while others with modest system contribution achieve elevated
NVF due to locational or temporal scarcity rents. Technology clusters are also
clearly separated, with flexible and fast-ramping units exhibiting higher NVF
irrespective of aggregate contribution.

This decoupling illustrates the extent to which LMP remuneration reflects
exposure to scarcity rents rather than proportional contribution to system
adequacy.

\subsection{Temporal Concentration of Value}

\begin{figure}[H]
\centering
\includegraphics[width=0.48\textwidth]{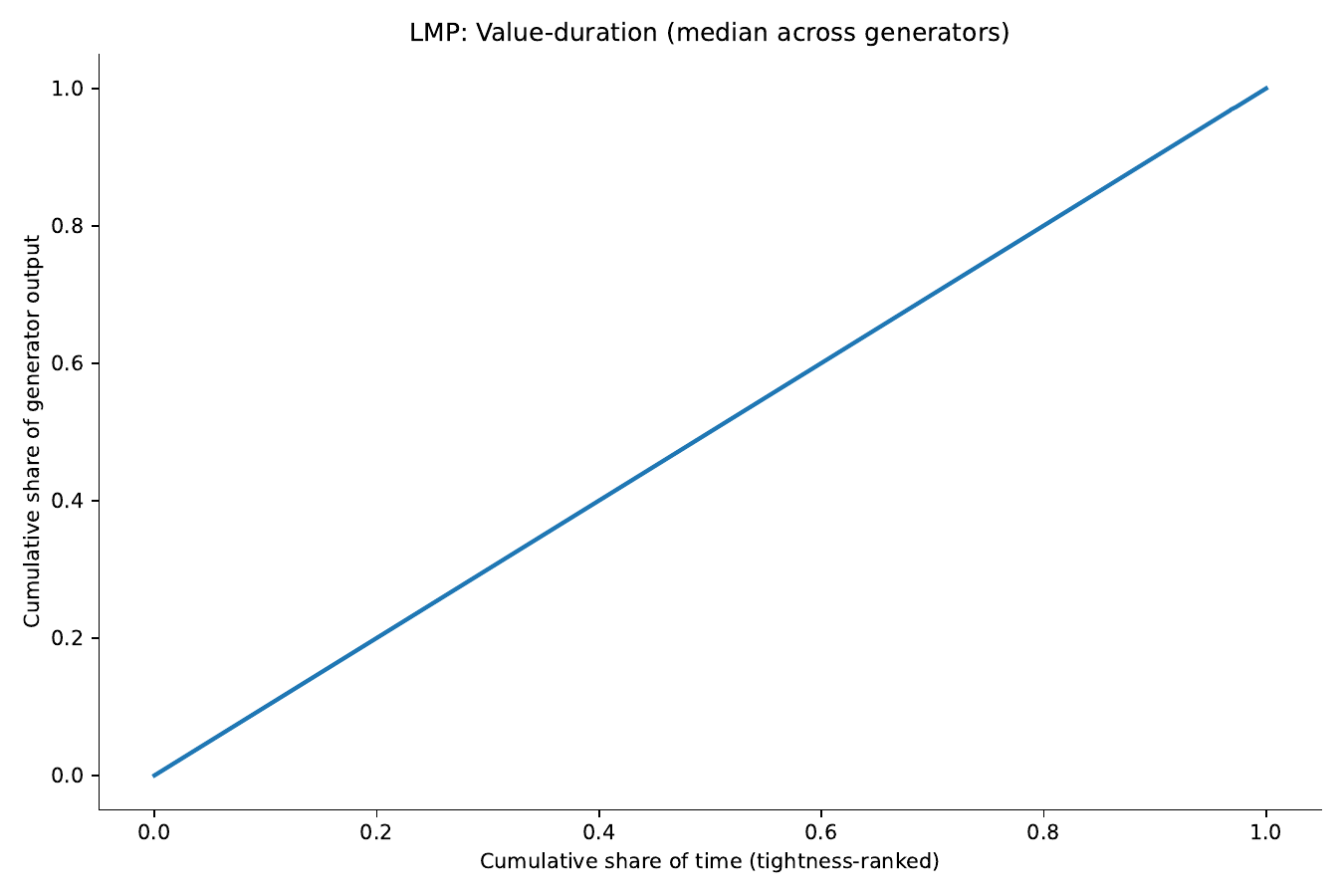}
\includegraphics[width=0.48\textwidth]{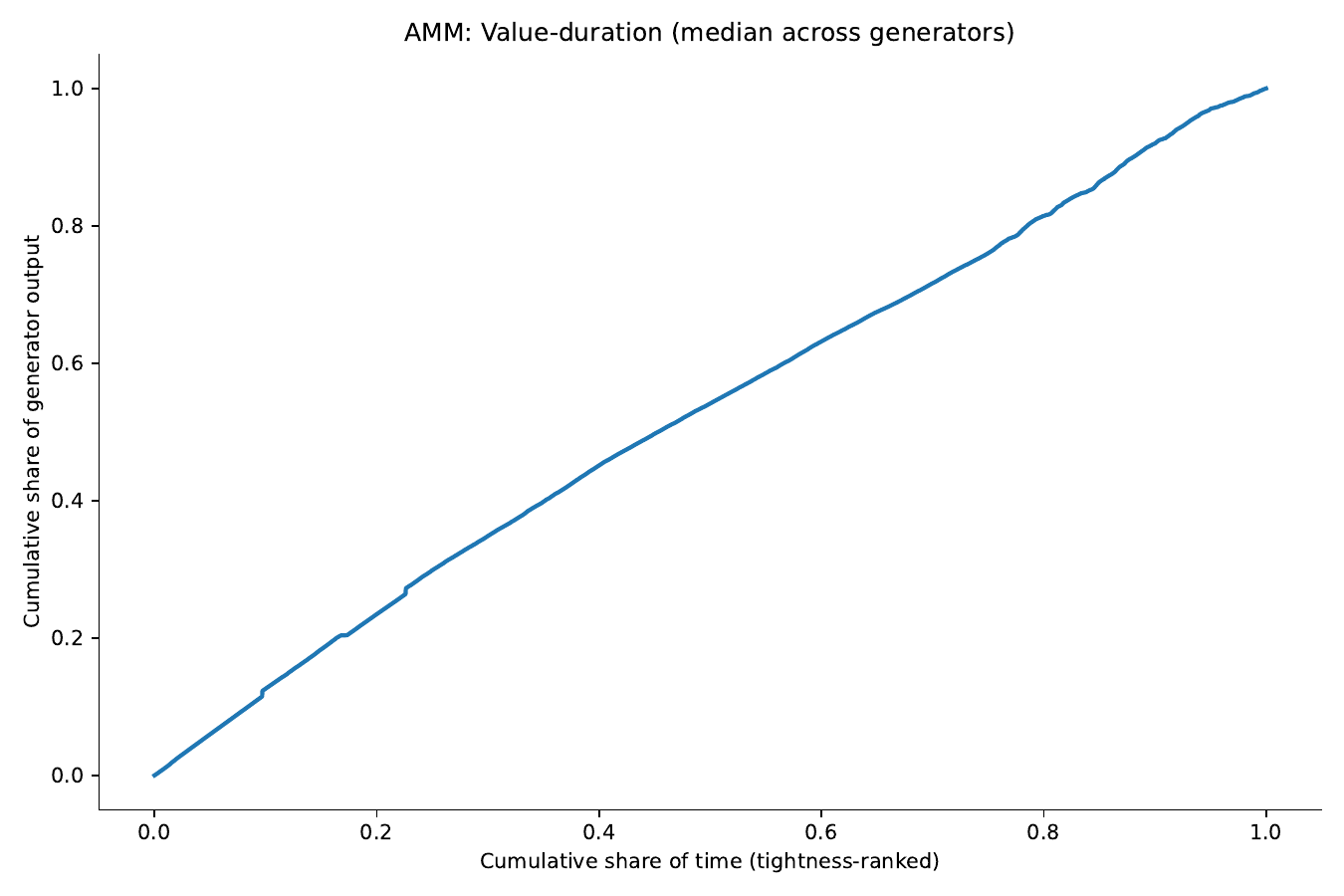}
\caption[
  Value--duration curves under LMP and AMM.
]{
  Cumulative share of generator output as a function of system tightness.
  LMP exhibits strong tail concentration, with a small number of scarcity hours
  accounting for a large share of value. AMM distributes value more smoothly
  across time through bounded tightness pricing and Shapley-based allocation.
}
\label{fig:value_duration_LMP_AMM}
\end{figure}

Figure~\ref{fig:value_duration_LMP_AMM} reports value--duration curves for LMP and
AMM, showing the cumulative share of generator output delivered as system
tightness increases. Under LMP, generator value exhibits a pronounced
``hockey-stick'' profile: a small fraction of tight hours accounts for a
disproportionate share of total revenue. This reflects the reliance of
energy-only pricing on rare scarcity events to recover fixed costs.

Under AMM, value is distributed more smoothly across time. Scarcity rents are
spread over a broader set of hours through bounded tightness pricing and
Shapley-based allocation, reducing the dependence of generator viability on
extreme tail events. This temporal smoothing is a structural consequence of the
AMM design rather than a tuning artefact.

\subsection{Interpretation and Link to Fairness Results}

Taken together, these diagnostics clarify why LMP and AMM produce markedly
different fairness outcomes. Both designs operate on the same physical network
and identify the same structural bottlenecks. However, LMP translates these
bottlenecks into unbounded, tail-driven prices that weakly align remuneration
with contribution and concentrate value into rare events. AMM instead internalises
network tightness into a bounded, anticipatory allocation of scarcity value,
yielding spatially coherent and temporally smooth remuneration.

These mechanisms explain the generator revenue distributions, payback profiles,
and inequality reductions reported in Section~\ref{sec:results_fairness}, while
remaining analytically distinct from the fairness tests themselves.

% =========================================================
\section{Diagnostics: Demand-side subscription construction (BASE vs DELTA; Aggregate vs IndividualTS)}
\label{sec:ext_demand_subscription_diagnostics}
% =========================================================

This section reports diagnostic outputs from the subscription construction script
(availabilityPayments $\rightarrow$ per-product flat subscriptions) and clarifies
the interpretation of:
(i) \textbf{BASE vs.\ DELTA} revenue streams, and
(ii) \textbf{Aggregate vs.\ IndividualTS} non-fuel allocation rules.

\subsection{What the script is doing (conceptual map)}
\label{subsec:diag_script_map}

The script starts from three system objects evaluated on the \emph{served} load:

\begin{enumerate}[leftmargin=*]
  \item \textbf{Served demand decomposition.}
  Using \texttt{served\_breakdown\_D*.csv}, the script constructs
  served residential power by product ($P1$--$P4$) and the served residential
  share of total served demand,
  \[
    f^{\mathrm{served}}_{\mathrm{res}}(t)
    =
    \frac{D^{\mathrm{served}}_{\mathrm{res}}(t)}{D^{\mathrm{served}}_{\mathrm{total}}(t)}.
  \]

  \item \textbf{Uncontrollable vs controllable supply (U/C).}
  Using dispatch and generator technology classes, the script forms
  total served generation by class,
  $U(t)$ (wind-like / uncontrollable) and $C(t)$ (controllable).
  These define global contemporaneous shares
  \[
    \alpha_U(t)=\frac{U(t)}{U(t)+C(t)}, \qquad
    \alpha_C(t)=1-\alpha_U(t).
  \]
  Residential product served MW is split into U/C components by the same global shares:
  $U_p(t)=\alpha_U(t)\,D_p(t)$ and $C_p(t)=\alpha_C(t)\,D_p(t)$ (served basis).

  \item \textbf{Generator revenue ``pots'' by class and approach.}
  Using \texttt{generator\_revenue\_timeseries\_ALL.csv}, the script aggregates
  generator revenues by technology class (U/C) and by \texttt{approach} prefix:
  \texttt{BASE*} and \texttt{DELTA*} (including reserve sub-approaches via prefix match).
\end{enumerate}

The output is therefore a decomposition of residential payments into:
(i) \emph{fuel cost} (from controllable dispatch costs),
(ii) \emph{non-fuel cost recovery} (allocated from BASE or DELTA pots, split into U- and C-attributed components),
and (iii) an optional \emph{uniform reserves adder} per household, computed from total reserves and the residential
share of total served MWh.

\subsection{Meaning of BASE vs.\ DELTA in the diagnostics}
\label{subsec:diag_base_delta_meaning}

The terms \textbf{BASE} and \textbf{DELTA} here are \emph{not} alternative names
for the same money. They refer to two different revenue streams recorded in the
availabilityPayments accounting:

\begin{itemize}[leftmargin=*]
  \item \textbf{BASE} (\texttt{BASE*} rows in the generator revenue timeseries)
  corresponds to the \emph{base} cost-recovery layer (the minimum revenue stream
  the design assigns as a stable subscription-like recovery component).

  \item \textbf{DELTA} (\texttt{DELTA*} rows) corresponds to the additional
  \emph{equalisation / top-up} layer used in the DELTA variant accounting
  (i.e.\ an additional settlement component, conceptually distinct from the base layer).
\end{itemize}

Accordingly, the diagnostic tables below should be read as:
\emph{``what flat subscriptions would be if we recover the non-fuel pot using BASE accounting''}
versus
\emph{``what flat subscriptions would be if we recover the non-fuel pot using DELTA accounting''},
with fuel treated consistently in both.

\subsection{Aggregate vs.\ IndividualTS: why the allocations differ}
\label{subsec:diag_aggregate_vs_individualts}

The script produces two allocation variants for the \textbf{non-fuel} pot:

\begin{enumerate}[leftmargin=*]
  \item \textbf{IndividualTS (per-timestamp) non-fuel allocation:}
  at each timestamp, residential non-fuel pots are allocated to products in proportion to
  \emph{capacity-weighted controllable MW}:
  \[
    \mathrm{share}_p(t)\ \propto\ w_p\,C_p(t),
  \]
  where $w_p$ is a product-specific capacity weight (e.g.\ EV-capable products weighted higher).
  This pushes non-fuel recovery toward products that (i) rely more on controllable supply in tight
  periods and (ii) are designed for higher peak capability.

  \item \textbf{Aggregate (period-level) non-fuel allocation:}
  the residential non-fuel pots are summed over the full period, and then allocated by
  \emph{aggregate} U and C energy shares:
  \[
    \mathrm{share}^{U}_p\ \propto\ \sum_t U_p(t)\Delta t,\qquad
    \mathrm{share}^{C}_p\ \propto\ \sum_t C_p(t)\Delta t.
  \]
  This treats non-fuel recovery as an energy-proportional subscription over the period,
  rather than a peak/availability-driven charge.
\end{enumerate}

Fuel is allocated in the same way in both cases (by per-timestamp controllable energy shares),
so the observed differences between Aggregate and IndividualTS are \emph{specifically} the
effect of the non-fuel rule.

\subsection{Reported results (from summary sheets)}
\label{subsec:diag_reported_results}

\begin{figure}[H]
\centering
\includegraphics[width=0.85\textwidth]{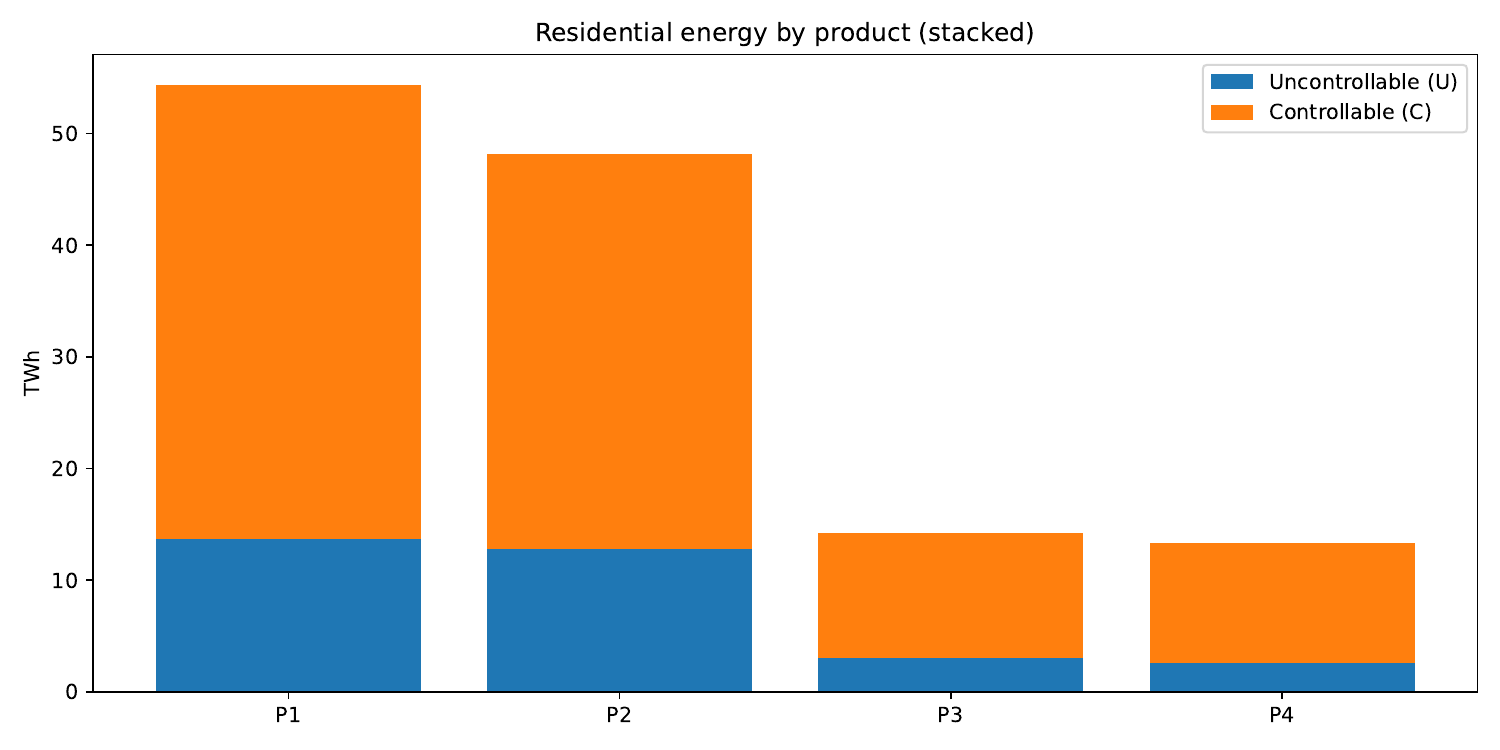}
\caption{Residential served energy split into uncontrollable (U) and controllable (C) by product (served basis).}
\label{fig:diag_uc_split_residential}
\end{figure}

\begin{figure}[H]
\centering
\includegraphics[width=0.92\textwidth]{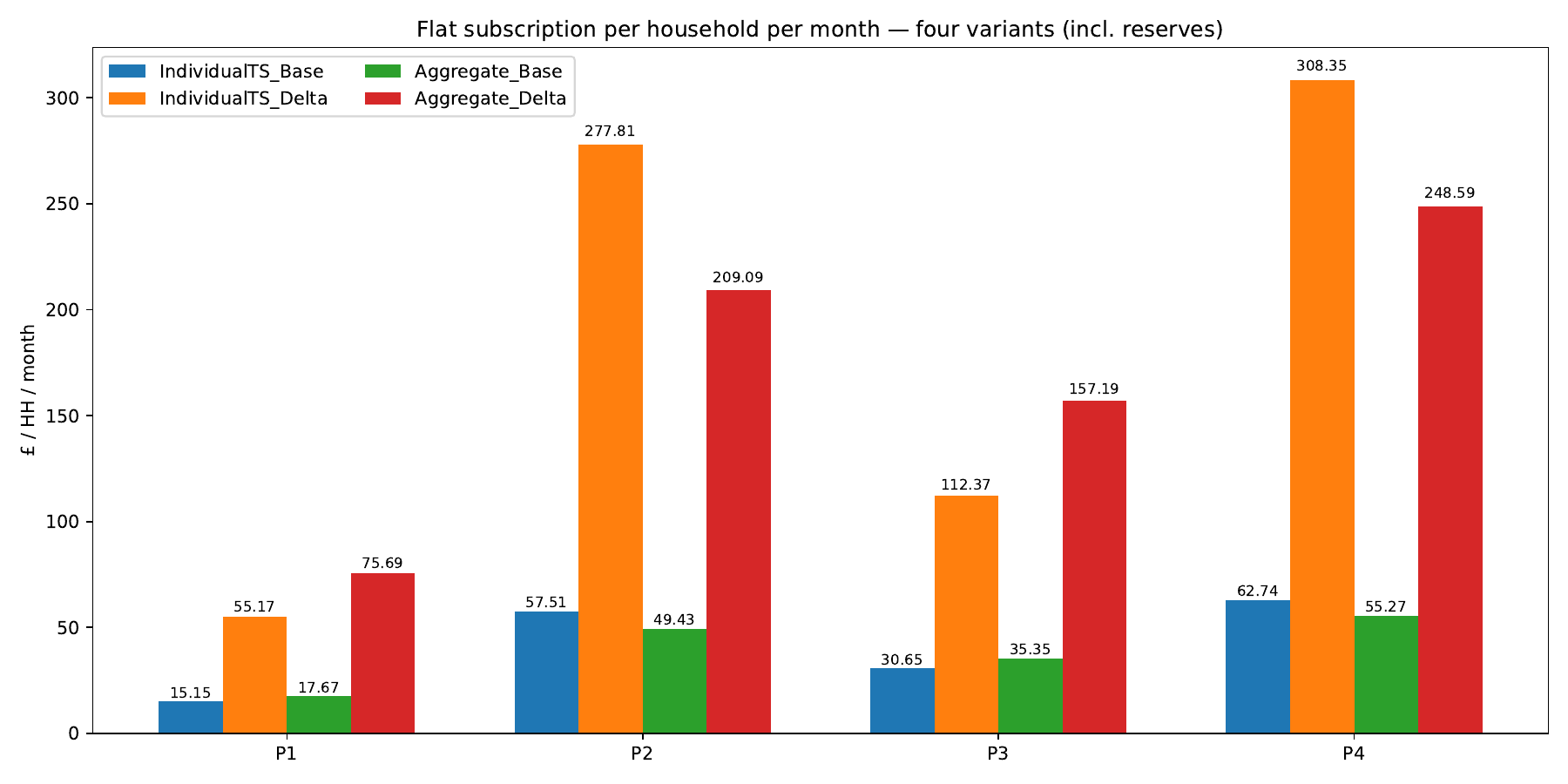}
\caption{Flat subscription per household per month for $P1$--$P4$ under the four diagnostic variants
(IndividualTS vs Aggregate) $\times$ (BASE vs DELTA).}
\label{fig:diag_subscriptions_all_variants}
\end{figure}

\begin{figure}[H]
\centering
\includegraphics[width=0.92\textwidth]{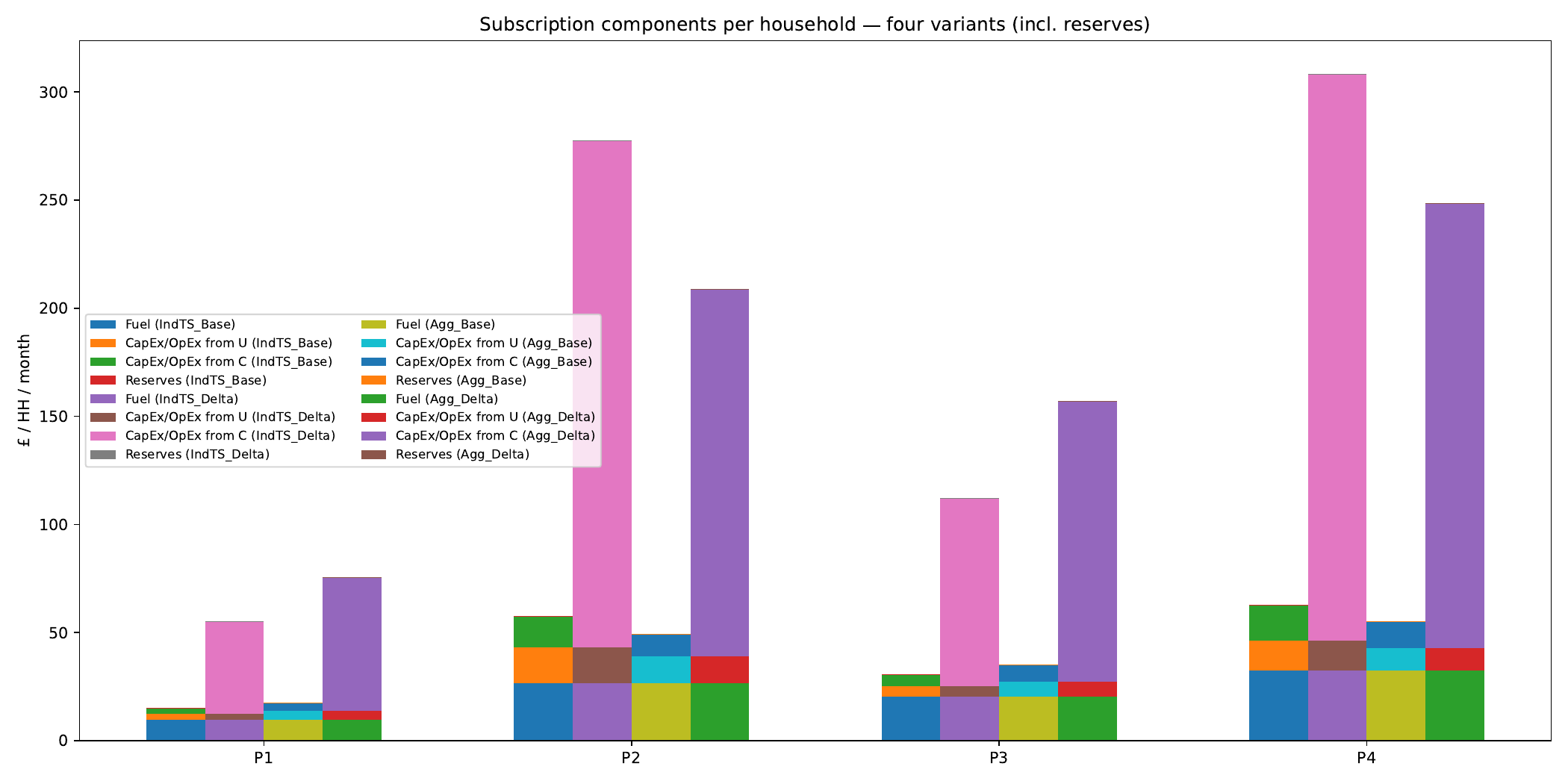}
\caption{Component breakdown of flat subscriptions (fuel, non-fuel U, non-fuel C, reserves adder if enabled)
for the four diagnostic variants.}
\label{fig:diag_subscription_components_all_variants}
\end{figure}

Figure~\ref{fig:diag_uc_split_residential} reports the served residential U/C energy
split by product, and Figures~\ref{fig:diag_subscriptions_all_variants}--\ref{fig:diag_subscription_components_all_variants}
report the derived per-household monthly subscriptions and their component breakdowns.

\paragraph{Per-household monthly subscription levels.}
Table~\ref{tab:diag_subscriptions_levels} reproduces the headline values from the summary CSV outputs.

\begin{table}[H]
\centering
\caption{Per-household flat subscription (\pounds/HH/month) by product under four diagnostic variants.}
\label{tab:diag_subscriptions_levels}
\renewcommand{\arraystretch}{1.15}
\begin{tabular}{lcccc}
\toprule
\textbf{Variant} & \textbf{P1} & \textbf{P2} & \textbf{P3} & \textbf{P4} \\
\midrule
Aggregate\_BASE     & 17.67 & 49.43 & 35.35 & 55.27 \\
IndividualTS\_BASE  & 15.15 & 57.51 & 30.65 & 62.74 \\
Aggregate\_DELTA    & 75.69 & 209.09 & 157.19 & 248.59 \\
IndividualTS\_DELTA & 55.17 & 277.81 & 112.37 & 308.35 \\
\bottomrule
\end{tabular}
\end{table}

\paragraph{What changes when we move from Aggregate to IndividualTS (non-fuel rule only).}
Table~\ref{tab:diag_shift_aggregate_to_individual} reports the percentage change in subscription
when switching the \emph{non-fuel} allocation rule from Aggregate to IndividualTS, holding the pot choice fixed.

\begin{table}[H]
\centering
\caption{Change in \pounds/HH/month when switching from Aggregate $\rightarrow$ IndividualTS non-fuel allocation (same pot).}
\label{tab:diag_shift_aggregate_to_individual}
\renewcommand{\arraystretch}{1.15}
\begin{tabular}{lcccc}
\toprule
\textbf{Pot} & \textbf{P1} & \textbf{P2} & \textbf{P3} & \textbf{P4} \\
\midrule
BASE  & $-14.26\%$ & $+16.33\%$ & $-13.29\%$ & $+13.50\%$ \\
DELTA & $-27.11\%$ & $+32.87\%$ & $-28.51\%$ & $+24.04\%$ \\
\bottomrule
\end{tabular}
\end{table}

The direction is structurally consistent with the design intent of IndividualTS:
because non-fuel recovery is allocated proportional to capacity-weighted controllable MW,
higher-capability (EV-capable / higher peak) products carry a larger share of the fixed pot,
while lower-capability products carry less. Aggregate allocation, by contrast, behaves as an
energy-proportional recovery over the period, producing a more ``averaged'' distribution.

\subsection{How this links back to demand-side fairness (H2) without making the wrong claim}
\label{subsec:diag_link_back_to_fairness}

These diagnostics are \emph{not} claiming that ``flexibility is punished'' under LMP, nor that
``costs are flat'' under AMM. The point is narrower and cleaner:

\begin{quote}
\textbf{Subscription levels are aligned with the product definitions and the chosen recovery rule.}
Moving between Aggregate and IndividualTS changes the incidence of \emph{non-fuel} recovery in an
interpretable way (energy-proportional vs peak/capability-weighted), while preserving a consistent
fuel allocation logic. This confirms the accounting pipeline does what it is designed to do, and
provides traceable levers for policy choice about how fixed-cost recovery should fall across product tiers.
\end{quote}

% ---------------------------------------------------------
\subsection{Additional diagnostic: served controllable power burden (mean/peak; system and per-HH)}
\label{subsec:diag_power_burden}
% ---------------------------------------------------------

\begin{figure}[H]
\centering
\includegraphics[width=0.85\textwidth]{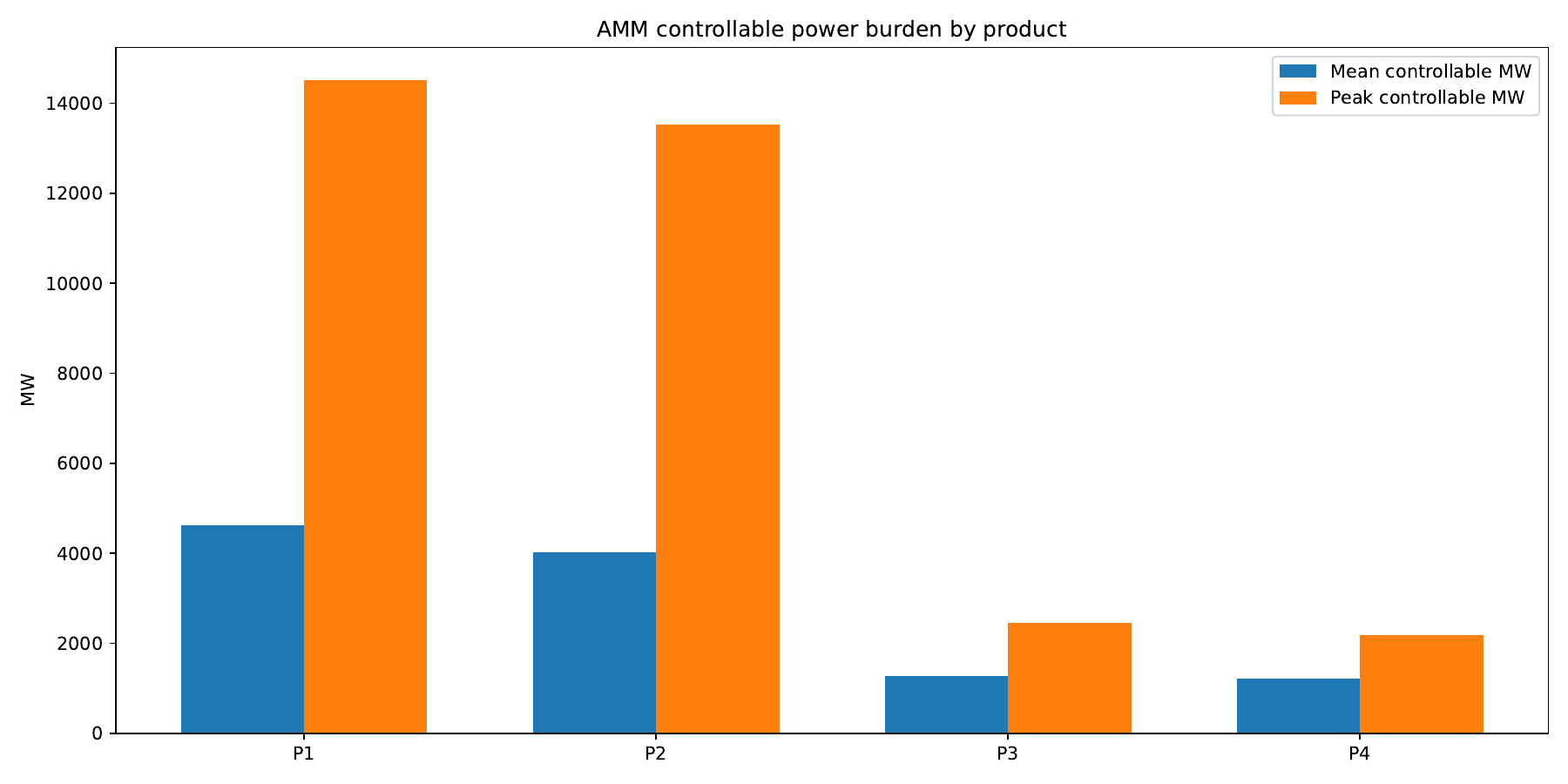}
\caption{Served controllable power by product under AMM: mean and peak over time (MW).}
\label{fig:amm_power_burden}
\end{figure}

\begin{figure}[H]
\centering
\includegraphics[width=0.85\textwidth]{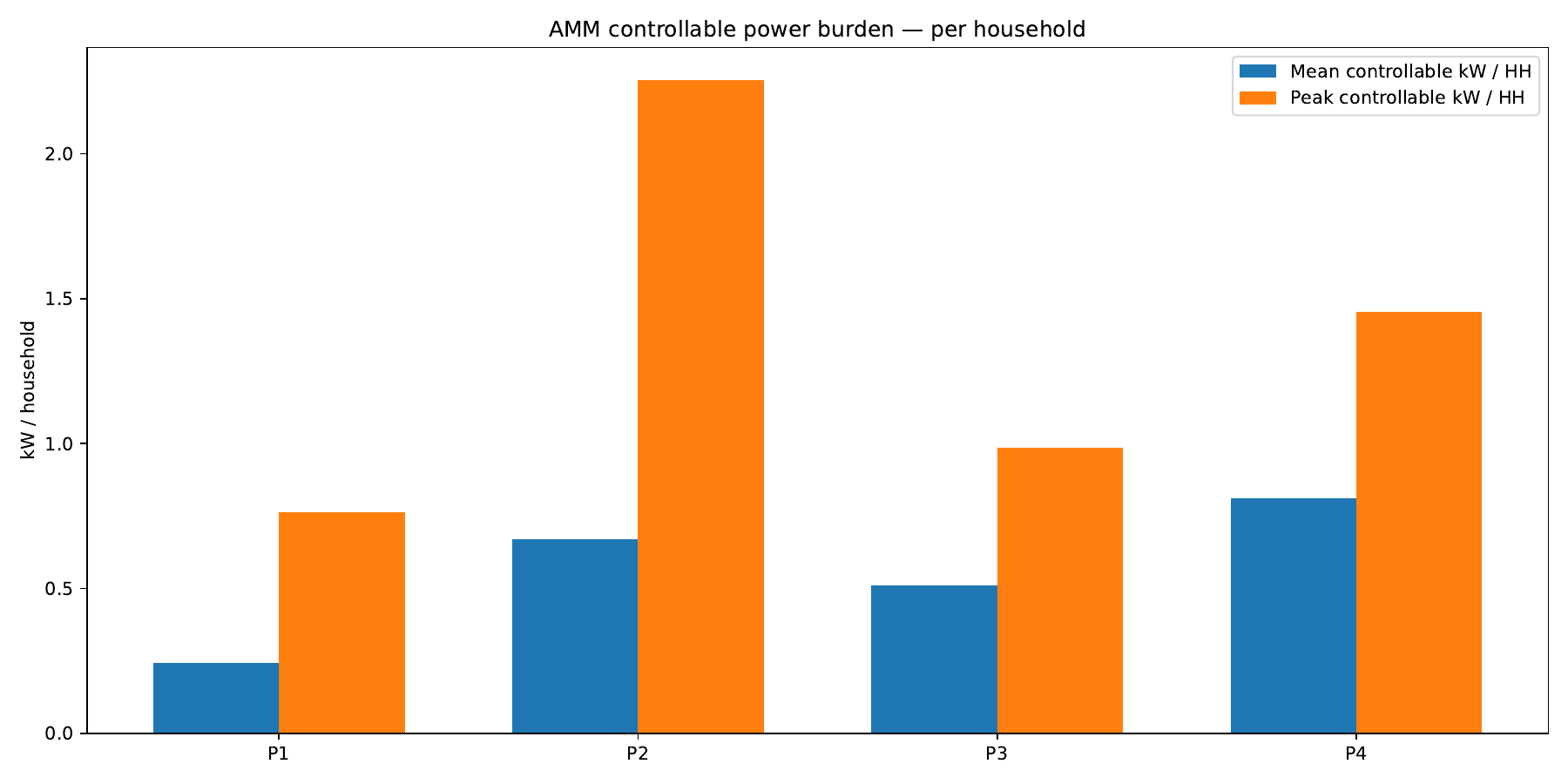}
\caption{Served controllable power per household by product under AMM: mean and peak over time (kW/HH), computed by dividing system-level MW by the assumed household count per product tier.}
\label{fig:amm_power_burden_perHH}
\end{figure}

As an additional diagnostic, Figures~\ref{fig:amm_power_burden} and
\ref{fig:amm_power_burden_perHH} summarise the \emph{served controllable power
burden} implied by each product tier. These plots are generated from the optional
file \texttt{per\_product\_UC\_power\_timeseries\_RES\_served.csv} (if present),
and therefore describe the time-series $C_p(t)$ used by the subscription
construction pipeline on a \emph{served basis}.

Concretely, the script computes, for each product $p \in \{P1,\dots,P4\}$:
(i) the mean and peak of the served controllable power time series, and
(ii) a system-average per-household normalisation obtained by dividing the
system-level MW quantities by the assumed household count in that product tier.
These diagnostics are \emph{burden-facing}: they indicate how strongly each
product relies on controllable supply over time, and therefore provide an
interpretability check on the direction of the \textbf{IndividualTS} non-fuel
allocation rule (which is proportional to capacity-weighted $C_p(t)$).

\paragraph{Interpretation.}
These figures do \emph{not} constitute a fairness claim on their own (they do not
measure dispersion, exposure, or ``jackpots''). Their role is narrower: they
confirm that the controllable-capacity burden is ordered and interpretable across
tiers, so that shifts observed when switching Aggregate $\rightarrow$ IndividualTS
(Table~\ref{tab:diag_shift_aggregate_to_individual}) can be traced back to a
transparent physical driver (served controllable power, with optional weighting
by $w_p$ in the non-fuel rule).

This is the appropriate ``results-facing'' bridge back into the main fairness narrative:
\textbf{the mechanism produces predictable, diagnosable, and design-consistent household charges},
rather than accidental or exposure-driven outcomes.

\paragraph{Verification via product--metric alignment.}
A separate verification check is whether \emph{subscription levels themselves}
are ordered by the \emph{metrics that define each product}: typical power
capability, target energy, and implied controllable burden.

This alignment is not assumed in the write-up: it is checked on the realised run
outputs. End-to-end verification is provided in Appendix~\ref{app:residential_synth}
(controller verification and dispatch consistency) and Appendix~\ref{app:price_allocation}
(pricing and allocation logic). Table~\ref{tab:product_metric_alignment} summarises
the observed ordering: products with higher capability and higher controllable burden
are assigned higher subscriptions, while lower-capability products are priced lower.

This delivers the intended ``you pay for what you buy'' property: differences in
bills are explained by product choice and service level, not by postcode,
incidental scarcity coincidence, or exposure to extreme spot prices.

\begin{sidewaystable}[p]
\centering
\caption{Observed alignment between product subscription levels and product-defining burden metrics (AMM run outputs).}
\label{tab:product_metric_alignment}
\renewcommand{\arraystretch}{1.15}
\begin{tabular}{lrrrr}
\toprule
Product &
\makecell{AMM subscription\\(\pounds/HH$\cdot$yr)} &
\makecell{Controllable energy\\(kWh/HH$\cdot$yr)} &
\makecell{Mean controllable power\\(kW/HH)} &
\makecell{Peak controllable power\\(kW/HH)} \\
\midrule
P1 & 269.53 & 2126.31 & 0.244 & 0.623 \\
P3 & 368.05 & 4495.09 & 0.512 & 1.296 \\
P2 & 690.34 & 5890.55 & 0.671 & 1.745 \\
P4 & 975.83 & 6731.53 & 0.812 & 2.075 \\
\bottomrule
\end{tabular}
\end{sidewaystable}

\chapter{How Generator Revenues Are Determined}
\label{app:amm_allocation}

This appendix documents the algorithm used to allocate revenues to generators
under the AMM in each scenario.  The implementation is identical for AMM~1
(cost-recovery) and AMM~2 (LMP-equivalent), differing only in how the annual
revenue pots are defined.

\section{Inputs}

\begin{itemize}[leftmargin=*]
  \item Half-hourly Shapley values:
        \[
        \phi_{g,t} \;\; \text{for all generators } g \text{ and timestamps } t.
        \]
  \item Generator cost data:
        non-fuel OpEx, CapEx (annual or total with payback years), and fuel cost.
  \item Tech classification identifying wind and nuclear (fixed-class) units.
  \item Annual revenue pots:
        \begin{itemize}
            \item BASE (controllable OpEx+CapEx),
            \item DELTA (LMP--AMM reconciliation),
            \item TARGET (user-defined multiple of BASE).
        \end{itemize}
\end{itemize}

\subsection{Step 1: Identify Fixed-Class Generators}

Wind and nuclear units are excluded from Shapley-sharing and instead receive a
fixed annual payment:
\[
F_g = \text{OpEx}_g^{\text{nonfuel}} + \text{CapEx}_g^{\text{per-year}}.
\]

If only total CapEx is known, annualisation uses a default payback horizon.

\subsection{Step 2: Shape Fixed Revenues by Each Generator's Own Shapley Profile}

For each fixed-class generator $g$:

\[
w_{g,t} = 
\begin{cases}
\displaystyle
\frac{\max\{\phi_{g,t},0\}}{\sum_\tau \max\{\phi_{g,\tau},0\}} &
    \text{if } \exists t : \phi_{g,t} > 0, \\[1.25em]
\displaystyle
\frac{1}{T} & \text{otherwise.}
\end{cases}
\]

Its half-hourly revenue is:

\[
R_{g,t}^{\text{fixed}} = F_g \cdot w_{g,t}.
\]

\subsection{Step 3: Compute Scarcity-Based Time Weights}

Total scarcity per timestamp:

\[
S_t = \sum_g \max\{\phi_{g,t},0\}.
\]

Normalised to obtain time weights:

\[
w_t = \frac{S_t}{\sum_\tau S_\tau}.
\]

Each annual pot (BASE, DELTA, TARGET) is distributed over time as:

\[
P_t^{(k)} = w_t \cdot P^{(k)},
\]
where $k \in \{\text{BASE},\,\text{DELTA},\,\text{TARGET}\}$.

\subsection{Step 4: Allocate Scarcity-Driven Pots Among Controllable Generators}

Let $\mathcal{G}_{\mathrm{elig}}$ denote controllable (non fixed-class) generators.

Per timestamp:

\[
R_{g,t}^{(k)} =
P_t^{(k)} \cdot
\frac{\max\{\phi_{g,t},0\}}
     {\sum_{h \in \mathcal{G}_{\mathrm{elig}}} \max\{\phi_{h,t},0\}}.
\]

\subsection{Step 5: Aggregate Revenues}

Total revenue per generator under scenario $k$ is:

\[
R_g^{(k)}
=
\sum_t \left(
R_{g,t}^{\text{fixed}} + R_{g,t}^{(k)}
\right).
\]

The full half-hourly series is exported for settlement and downstream analysis.

\subsection{Validation}

The implementation enforces:
\begin{itemize}[leftmargin=*]
  \item exact equality of fixed-class annual revenues and their cost requirements;
  \item conservation of annual pot sizes:
        \[
        \sum_g R_{g}^{(k)} = P^{(k)} + \sum_{g \in \text{fixed}} F_g;
        \]
  \item non-negativity of all timestamp allocations;
  \item strict timestamp alignment across all inputs.
\end{itemize}
.

\subsection{Rationale for Treating Wind and Nuclear as Fixed-Class Resources}
\label{app:fixed_class_rationale}

Wind and nuclear generators are deliberately excluded from the competitive
Shapley-based allocation and placed on a regulated cost-recovery footing. This
choice reflects both their physical system role and their investment and risk
profiles, and ensures that the Shapley mechanism focuses on genuinely
dispatchable, marginal scarcity response.

\paragraph{Wind (uncontrollable generation).}

Wind farms are weather-driven and non-dispatchable. Once built, their
short-run operational decisions have limited influence on real-time scarcity
relief: output is determined by meteorology, not strategic behaviour.
Under a marginal-contribution metric such as the Shapley value,
\emph{deliverable} scarcity relief dominates, meaning wind units would
naturally earn very low scarcity payments even when the system planner wishes
to remunerate them for decarbonisation and diversification benefits.

Accordingly:
\begin{itemize}[leftmargin=*]
    \item wind’s scarcity contribution is reflected in energy-market revenues;
    \item non-fuel OpEx and capital costs are recovered via a fixed annual
          payment $F_g$ rather than via the Shapley pot.
\end{itemize}

This avoids artefacts in which capital-intensive renewable capacity appears
uneconomic solely because it is structurally mismatched to a dispatchability-
based scarcity metric.

\paragraph{Nuclear (security-of-supply backbone).}

Nuclear stations exhibit:
\begin{itemize}[leftmargin=*]
    \item extremely high up-front capital costs,
    \item slow ramping and technical minimums,
    \item safety and maintenance constraints, 
    \item and a long-lived baseload contribution to adequacy.
\end{itemize}

If remunerated purely via Shapley-based “marginal deliverability”, their
long-run payback would be unrealistically long, despite their crucial role in
system security. In practice, nuclear capacity is typically underwritten by
long-term contracts or regulated asset-base models.

In line with that reality, nuclear units are treated here as:
\begin{itemize}[leftmargin=*]
    \item \emph{must-pay backbone plant}, recovering OpEx and annualised CapEx
          on a regulated basis;
    \item excluded from the competitive scarcity pot to keep marginal-flexibility
          signals undistorted.
\end{itemize}

\paragraph{Purpose of the separation.}

This fixed-class treatment ensures that:
\begin{enumerate}[leftmargin=*]
    \item the Shapley pot focuses on technologies with genuine
          dispatchable, marginal scarcity response (gas, batteries);
    \item long-lived, capital-intensive baseload plant are not penalised for
          being structurally uncorrelated with real-time scarcity;
    \item decarbonisation-critical zero-carbon resources receive stable,
          investment-compatible cost recovery.
\end{enumerate}

Together, the rules in Sections~\ref{app:amm_allocation}–\ref{app:fixed_class_rationale}
ensure that the AMM revenue mechanism:
\begin{itemize}[leftmargin=*]
    \item preserves Shapley-based fairness for responsive generators,
    \item maintains financial viability of essential zero-carbon capacity,
    \item and avoids artefacts arising from mismatch between physical roles and 
          scarcity-based remuneration.
\end{itemize}

% =========================================================
\chapter{How suppliers are charged in the wholesale market and how retail pricing works}
\label{app:price_allocation}
% =========================================================

This appendix specifies the \emph{wholesale charging basis faced by suppliers}
under the AMM--Fair Play architecture evaluated in this thesis, and clarifies
how those wholesale charges can be mapped into retail-facing tariffs.

\medskip

\noindent
\textbf{Interpretation in the two-axis model (this thesis).}
In the evaluated implementation, suppliers are charged through a
\emph{product-indexed wholesale charging framework}. The monthly amounts derived
below are \emph{per-enrolled-household wholesale charges} that a supplier faces
\emph{because of the assumed usage and flexibility characteristics of its
customer base}, conditional on how many customers are enrolled in each product
category (P1--P4). These charges therefore represent the supplier’s wholesale
\emph{liability} for serving a portfolio of households with a given product
composition.

Suppliers are not themselves charged ``subscriptions'' as contractual objects.
Rather, the AMM procures energy, reserves, and adequacy at system level, and
suppliers are charged for the implied demand liabilities of their customers
under the product framework. This removes exposure to nodal wholesale price
volatility while preserving cost reflectivity at the level of aggregate customer
behaviour.

\medskip

\noindent
\textbf{Scope limitation: ``all demand treated as essential'' during transition.}
The charging construction in this appendix corresponds to the \emph{two-axis}
version of the architecture. For the purposes of the LMP comparison, residential
demand is treated as \emph{essential service} and recovered via flat
product-indexed wholesale charges. This is a deliberate transitional
simplification: it matches the constrained AMM configuration used in the main
experiments and avoids assuming direct device-level enrolment or automated
flexibility control.

\medskip

\noindent
\textbf{Third axis (future work).}
In the full holarchic deployment (the \emph{third axis}), customer devices are
directly enrolled for flexibility and reliability services. Extending the
charging methodology to that setting requires an explicit framework for
(i) pricing reliability rights and (ii) settling enrolled flexibility and
performance, and a corresponding extension of how suppliers are charged for
those services. This extension is left as future work and does not affect the
interpretation of the two-axis results reported in the main text.

\medskip

\noindent
\textbf{What this appendix computes.}
Starting from system-level AMM payments to generators, the appendix derives:
\begin{itemize}[leftmargin=*]
  \item \textbf{Residential product-indexed wholesale charges} for P1--P4,
        expressed as \pounds/household/month and interpreted as the supplier’s
        per-enrolled-household wholesale liability for that product category; and
  \item \textbf{An aggregate non-residential wholesale cost total}
        (no product subdivision in the evaluated implementation).
\end{itemize}

The procedure starts from the generator-level revenue time series and dispatch
outcomes described in Appendix~\ref{app:amm_allocation}, and from the
residential/non-residential served-demand breakdown produced by the AMM run. It
allocates fuel, CapEx/OpEx recovery, and reserves costs across the residential
and non-residential segments, and then across residential products.

The absolute number of households in each residential product is taken from the
product-classification exercise based on synthetic residential demand profiles
in Appendix~\ref{app:residential_synth}. In particular, the script fixes
\[
H_{\text{P1}} = 19\,\text{M},\quad
H_{\text{P2}} = 6\,\text{M},\quad
H_{\text{P3}} = 2.5\,\text{M},\quad
H_{\text{P4}} = 1.5\,\text{M},
\]
so that per-household charges can be computed for each product.

\section{Inputs and Overall Structure}

The allocation script takes as inputs:
\begin{itemize}[leftmargin=*]
  \item \emph{Served demand by load and product:}
        files \texttt{served\_breakdown\_D*.csv} from the AMM run, produced
        by \texttt{compute\_served\_by\_product.py}. Each file corresponds
        to a demand node $D_k$ and contains, for every time step,
        \begin{itemize}
          \item total served demand $D^{\text{served}}_{k}(t)$,
          \item served residential demand by product
                $P1_{k}(t),\dots,P4_{k}(t)$, and
          \item served non-residential demand
                $D^{\text{nonres}}_{k}(t)$.
        \end{itemize}
  \item \emph{Generator dispatch time series:} a tree of
        \texttt{dispatch.csv} files under the AMM output directory,
        containing, at least, generator output $p_{g}(t)$ and energy prices
        or costs.\footnote{These are the same dispatch outcomes that are
        used in Appendix~\ref{app:amm_allocation} to compute generator
        revenues.}
  \item \emph{Static generator metadata:} \texttt{gens\_static.csv}, which
        provides a technology label for each generator.
  \item \emph{Generator revenue time series:}
        \texttt{generator\_revenue\_timeseries\_ALL.csv}, which holds the
        time-resolved AMM payments to each generator under different
        approaches (e.g.\ \texttt{BASE}, \texttt{DELTA}), as described in
        Appendix~\ref{app:amm_allocation}.
  \item \emph{Reserves payments:} a separate file
        \texttt{reserves\_by\_generator.csv} containing the total monetary
        amount paid for reserve services.
\end{itemize}

All time series are first reindexed to a common step length
$\Delta t = 30$ minutes (by default) and aligned on a shared set of
timestamps. The remainder of this appendix describes, step-by-step, how
the script uses these inputs to derive residential and non-residential
charges.

\section{Splitting Served Demand into Residential and Non-residential}

For each demand node $D_k$, the script reads
\texttt{served\_breakdown\_Dk.csv} and constructs two derived data sets:

\begin{enumerate}[label=(\alph*), leftmargin=*]
  \item \textbf{Per-load residential served demand.} For every timestamp
        $t$ and demand node $k$:
        \[
        P1_k(t),\; P2_k(t),\; P3_k(t),\; P4_k(t),
        \]
        together with
        \[
        D^{\text{res}}_k(t)
          = P1_k(t) + P2_k(t) + P3_k(t) + P4_k(t).
        \]
        These are stored as a long-format table with columns
        \texttt{timestamp}, \texttt{demand\_id}, \texttt{P1--P4} and a
        computed \texttt{total\_res\_served\_MW}.
  \item \textbf{System-level served totals.} Summing across all demand
        nodes, the script computes, for each time $t$,
        \[
        D^{\text{total}}(t)
          = \sum_k D^{\text{served}}_k(t), \quad
        D^{\text{res}}(t)
          = \sum_k D^{\text{res}}_k(t), \quad
        D^{\text{nonres}}(t)
          = \sum_k D^{\text{nonres}}_k(t).
        \]
        The residential share of served demand is then
        \[
        f_{\text{res}}(t)
          = \begin{cases}
              \dfrac{D^{\text{res}}(t)}{D^{\text{total}}(t)}, & 
                D^{\text{total}}(t) > 0,\\[0.3cm]
              0, & \text{otherwise}.
            \end{cases}
        \]
        This time series $(f_{\text{res}}(t))_t$ is central: it is used to
        split generator revenues and fuel costs between residential and
        non-residential segments.
\end{enumerate}

In parallel, the script aggregates the per-load residential data into a
system-level residential product time series:
\[
Pp(t) = \sum_k Pp_k(t), \quad p \in \{\text{P1},\dots,\text{P4}\},
\]
and total residential demand
$D^{\text{res}}(t) = \sum_{p} Pp(t)$. This is the basis for all subsequent
residential allocations.

\section{Classifying Generation into Uncontrollable and Controllable}

As in Appendix~\ref{app:amm_allocation}, each generator is classified
into either:
\begin{itemize}[leftmargin=*]
  \item class U (``uncontrollable'') for wind and other zero-marginal-cost
        renewables; or
  \item class C (``controllable'') for thermal, storage and other flexible
        assets.
\end{itemize}
This classification is derived from the \texttt{tech} field in
\texttt{gens\_static.csv}.

The dispatch tree is then scanned and all \texttt{dispatch.csv} files are
stacked into a single time series of generator outputs $p_g(t)$. For each
timestamp, effective generation is defined as
\[
p^{\text{eff}}_g(t) = \max\{p_g(t), 0\},
\]
and aggregated by class:
\[
U(t) = \sum_{g \in \mathcal{G}_U} p^{\text{eff}}_g(t), \qquad
C(t) = \sum_{g \in \mathcal{G}_C} p^{\text{eff}}_g(t).
\]

To ensure consistency with served demand, total generation is capped at
total served load. Defining
\[
\widehat{P}(t) = U(t) + C(t), \qquad
\widehat{D}(t) = D^{\text{total}}(t),
\]
the script sets a scaling factor
\[
\alpha(t) =
  \begin{cases}
    \min\left\{1, \dfrac{\widehat{D}(t)}{\widehat{P}(t)} \right\}, 
      & \widehat{P}(t) > 0,\\[0.3cm]
    1, & \text{otherwise},
  \end{cases}
\]
and defines capped class outputs
\[
U^{\ast}(t) = \alpha(t)\, U(t), \qquad
C^{\ast}(t) = \alpha(t)\, C(t).
\]

These form global uncontrollable and controllable supply time series.

\section{Splitting Residential Demand into U and C Components}

The script does not attempt to match individual residential loads to
specific generators. Instead, it applies the global U/C shares at each
timestamp to each product:
\[
s_U(t) =
  \frac{U^{\ast}(t)}{U^{\ast}(t) + C^{\ast}(t)}, \qquad
s_C(t) = 1 - s_U(t),
\]
with the convention $s_U(t) = s_C(t) = 0$ if
$U^{\ast}(t) + C^{\ast}(t) = 0$.

For each product $p \in \{\text{P1},\dots,\text{P4}\}$, the script sets:
\[
U_p(t) = s_U(t)\, Pp(t), \qquad
C_p(t) = s_C(t)\, Pp(t).
\]
By construction, $U_p(t) + C_p(t) = Pp(t)$ for all $t$ and each product.

These per-timestamp power allocations are then integrated over time to
obtain energy:
\[
U_p^{\text{MWh}}(t) = U_p(t) \Delta t, \qquad
C_p^{\text{MWh}}(t) = C_p(t) \Delta t,
\]
where $\Delta t$ is expressed in hours.

The same global U/C shares can also be applied to the non-residential
served demand $D^{\text{nonres}}(t)$ for diagnostic purposes, yielding
time series $U^{\text{nonres}}(t)$ and $C^{\text{nonres}}(t)$.

\section{Fuel Costs from Dispatch}

Fuel costs are constructed consistently with the dispatch and AMM
configuration described in Appendix~\ref{app:amm_allocation}. For each
generator $g$ and timestamp $t$, the script either:
\begin{itemize}[leftmargin=*]
  \item uses a direct field \texttt{energy\_cost\_gbp}(t) if available, or
  \item reconstructs the fuel cost as
        \[
        \text{fuel\_cost}_{g}(t) =
          p^{\text{eff}}_g(t)\, \Delta t \, c_g(t),
        \]
        where $c_g(t)$ is the energy cost or bid price in \pounds/MWh.
\end{itemize}

Fuel costs are only accrued for controllable generators,
$g \in \mathcal{G}_C$. The total controllable fuel cost is
\[
F^{\text{total}}(t) =
  \sum_{g \in \mathcal{G}_C} \text{fuel\_cost}_{g}(t).
\]

\section{Generator Revenue Pots for U and C}

Generator revenues are taken from the time series constructed in
Appendix~\ref{app:amm_allocation}. The script reads
\texttt{generator\_revenue\_timeseries\_ALL.csv} and, for a given AMM
approach (e.g.\ \texttt{BASE} for AMM1, \texttt{DELTA} for AMM2), selects
all rows whose \texttt{approach} string starts with the corresponding
prefix. This ensures that any sub-services (including reserve-related
revenue tagged as \texttt{BASE\_RESERVE}, \texttt{DELTA\_RESERVE}, etc.)
are included in the same aggregate.

For each timestamp $t$, generator revenues are aggregated by class:
\[
\text{pot}_U(t) = \sum_{g \in \mathcal{G}_U} R_g(t), \qquad
\text{pot}_C(t) = \sum_{g \in \mathcal{G}_C} R_g(t),
\]
where $R_g(t)$ is the total revenue to generator $g$ under the chosen
approach at time $t$. These time series represent the total monetary
flows into uncontrollable and controllable generators respectively.

\section{Splitting Pots and Fuel Between Residential and Non-residential}

The next step is to split both the class-level revenue pots and the fuel
costs into residential and non-residential components. This uses the
time-varying residential share of served demand $f_{\text{res}}(t)$.

For the revenue pots,
\[
\text{pot}_U^{\text{res}}(t)
  = f_{\text{res}}(t)\, \text{pot}_U(t), \qquad
\text{pot}_C^{\text{res}}(t)
  = f_{\text{res}}(t)\, \text{pot}_C(t),
\]
and the remainder is implicitly non-residential:
\[
\text{pot}_U^{\text{nonres}}(t)
  = \text{pot}_U(t) - \text{pot}_U^{\text{res}}(t), \qquad
\text{pot}_C^{\text{nonres}}(t)
  = \text{pot}_C(t) - \text{pot}_C^{\text{res}}(t).
\]

Similarly, for fuel costs,
\[
F^{\text{res}}(t) = f_{\text{res}}(t)\, F^{\text{total}}(t),
\]
with $F^{\text{nonres}}(t) = F^{\text{total}}(t) - F^{\text{res}}(t)$.

This ensures that, at every timestamp, the fraction of total pots and
fuel attributed to the residential segment matches the fraction of
served energy that is residential.

\section{Allocation of Residential Costs by Product}

Given the residential-scaled pots and fuel timeseries, the script
allocates these between products P1–P4. Two alternative methods are
implemented:
\begin{enumerate}[leftmargin=*]
  \item \emph{Per-timestamp non-fuel allocation} 
        (\texttt{NonFuelOpexCapExIndividualTS}): non-fuel pots are
        allocated at every timestamp using controllable \emph{power}
        shares; fuel is allocated using controllable \emph{energy}
        shares.
  \item \emph{Aggregate non-fuel allocation}
        (\texttt{NonFuelOpexCapExAggregate}): residential-scaled pots are
        summed over the entire period, and the totals are allocated using
        aggregate controllable energy shares; fuel is still allocated
        per timestamp.
\end{enumerate}

\subsection{Per-timestamp Non-fuel Allocation (IndividualTS)}

For each timestamp $t$, product $p$ and controllable allocation $C_p(t)$,
the script computes a capacity-weighted share:
\[
w^{C}_p(t)
  = \frac{w_p \, C_p(t)}
         {\sum_{q} w_q\, C_q(t)},
\]
where $w_p$ is a product-specific weighting factor (currently
$w_{\text{P1}} = w_{\text{P3}} = 1$ and
$w_{\text{P2}} = w_{\text{P4}} = 2$). These weights allow the allocation
to reflect differences in typical peak demand between products.

The residential pots at time $t$ are then split as:
\[
\text{pot}_{U,p}(t) = w^{C}_p(t)\, \text{pot}_U^{\text{res}}(t), \qquad
\text{pot}_{C,p}(t) = w^{C}_p(t)\, \text{pot}_C^{\text{res}}(t).
\]

Fuel is allocated using controllable energy shares. For each $t$, set
\[
\widetilde{C}_p^{\text{MWh}}(t)
  = C_p^{\text{MWh}}(t), \quad
\widetilde{C}^{\text{MWh}}_{\text{tot}}(t)
  = \sum_q C_q^{\text{MWh}}(t),
\]
and define fuel shares
\[
\phi_p(t)
  = \frac{\widetilde{C}_p^{\text{MWh}}(t)}
         {\widetilde{C}^{\text{MWh}}_{\text{tot}}(t)}
   \quad \text{(with } \phi_p(t) = 0 \text{ if denominator is 0)}.
\]
Then residential fuel costs at time $t$ are split as
\[
F_p(t) = \phi_p(t)\, F^{\text{res}}(t).
\]

Summing over all timestamps yields the annual costs per product:
\[
\text{CapEx/OpEx}_{U,p}
  = \sum_t \text{pot}_{U,p}(t), \quad
\text{CapEx/OpEx}_{C,p}
  = \sum_t \text{pot}_{C,p}(t), \quad
\text{Fuel}_p = \sum_t F_p(t).
\]
The base annual cost per product is then
\[
\text{Cost}_p^{\text{base}}
  = \text{CapEx/OpEx}_{U,p}
  + \text{CapEx/OpEx}_{C,p}
  + \text{Fuel}_p.
\]

\subsection{Aggregate Non-fuel Allocation (Aggregate)}

In the aggregate variant, the residential-scaled pots are first summed
over the entire period:
\[
\text{pot}_U^{\text{res, tot}}
  = \sum_t \text{pot}_U^{\text{res}}(t), \qquad
\text{pot}_C^{\text{res, tot}}
  = \sum_t \text{pot}_C^{\text{res}}(t).
\]

These totals are allocated using aggregate uncontrollable and
controllable energy shares:
\[
U_p^{\text{MWh, tot}} = \sum_t U_p^{\text{MWh}}(t), \qquad
C_p^{\text{MWh, tot}} = \sum_t C_p^{\text{MWh}}(t),
\]
and
\[
s^{U}_p = \frac{U_p^{\text{MWh, tot}}}
               {\sum_q U_q^{\text{MWh, tot}}}, \qquad
s^{C}_p = \frac{C_p^{\text{MWh, tot}}}
               {\sum_q C_q^{\text{MWh, tot}}}.
\]

The non-fuel components are then
\[
\text{CapEx/OpEx}_{U,p}
  = s^U_p \, \text{pot}_U^{\text{res, tot}}, \qquad
\text{CapEx/OpEx}_{C,p}
  = s^C_p \, \text{pot}_C^{\text{res, tot}}.
\]

Fuel is still allocated per timestamp, exactly as in the
\texttt{IndividualTS} variant, and summed to obtain $\text{Fuel}_p$.
The resulting base annual cost per product,
$\text{Cost}_p^{\text{base}}$, is defined as before.

\section{Reserves Allocation and Per-household Subscription Charges}

Reserves payments are handled separately from the generator revenue pots
above. The script reads \texttt{reserves\_by\_generator.csv} and
identifies the relevant monetary column (e.g.\ \texttt{reserve\_revenue\_gbp}
or \texttt{revenue\_gbp}). Summing across generators yields a total
reserves payment
\[
R^{\text{tot}}_{\text{reserves}}.
\]

To split this between residential and non-residential segments, the
script uses the \emph{total} served energy over the entire period:
\[
E^{\text{res}}_{\text{tot}} = \sum_t D^{\text{res}}(t)\, \Delta t, \qquad
E^{\text{nonres}}_{\text{tot}} =
  \sum_t D^{\text{nonres}}(t)\, \Delta t,
\]
and
\[
E^{\text{tot}} = E^{\text{res}}_{\text{tot}} + E^{\text{nonres}}_{\text{tot}}.
\]
The residential share of reserves is then
\[
\gamma_{\text{res}} =
  \begin{cases}
    \dfrac{E^{\text{res}}_{\text{tot}}}{E^{\text{tot}}}, & E^{\text{tot}} > 0,\\[0.3cm]
    0, & \text{otherwise},
  \end{cases}
\]
so that
\[
R^{\text{res}}_{\text{reserves}}
  = \gamma_{\text{res}} R^{\text{tot}}_{\text{reserves}}, \qquad
R^{\text{nonres}}_{\text{reserves}}
  = (1 - \gamma_{\text{res}}) R^{\text{tot}}_{\text{reserves}}.
\]

The residential reserves amount is then spread \emph{uniformly} across
all residential households:
\[
H_{\text{tot}} = H_{\text{P1}} + H_{\text{P2}} + H_{\text{P3}} + H_{\text{P4}},
\]
\[
r_{\text{HH}}^{\text{year}}
  = \frac{R^{\text{res}}_{\text{reserves}}}{H_{\text{tot}}}, \qquad
r_{\text{HH}}^{\text{month}}
  = \frac{r_{\text{HH}}^{\text{year}}}{12}.
\]
This monthly amount $r_{\text{HH}}^{\text{month}}$ is added as a uniform
per-household ``reserves'' line item for each product.

Given the base annual costs $\text{Cost}_p^{\text{base}}$ and the product
household counts $H_p$, the script computes:
\begin{align*}
\text{Base subscription per HH per month:}\quad
  s_p^{\text{base}} &=
    \frac{\text{Cost}_p^{\text{base}}}{H_p \cdot 12},\\[0.2cm]
\text{Component breakdown:}\quad
  s_{p,\text{fuel}} &=
    \frac{\text{Fuel}_p}{H_p \cdot 12},\\
  s_{p,\text{CapEx/OpEx-U}} &=
    \frac{\text{CapEx/OpEx}_{U,p}}{H_p \cdot 12},\\
  s_{p,\text{CapEx/OpEx-C}} &=
    \frac{\text{CapEx/OpEx}_{C,p}}{H_p \cdot 12},
\end{align*}
and a uniform reserves component
\[
s_{\text{reserves}} = r_{\text{HH}}^{\text{month}}, \quad
\text{independent of } p.
\]

The final flat subscription for product $p$ is therefore
\[
s_p^{\text{total}}
  = s_p^{\text{base}} + s_{\text{reserves}}.
\]

\section{Non-residential Totals and Consistency Checks}

The non-residential component is not subdivided by product. Instead, the
script computes non-residential totals by subtracting residential totals
from system-wide totals. For each variant and AMM approach, it reports:
\begin{align*}
\text{CapEx/OpEx}_{U}^{\text{nonres}} &=
  \sum_t \text{pot}_U(t) - \sum_t \text{pot}_U^{\text{res}}(t), \\
\text{CapEx/OpEx}_{C}^{\text{nonres}} &=
  \sum_t \text{pot}_C(t) - \sum_t \text{pot}_C^{\text{res}}(t), \\
\text{Fuel}^{\text{nonres}} &=
  \sum_t F^{\text{total}}(t) -
  \sum_t F^{\text{res}}(t),\\
\text{Reserves}^{\text{nonres}} &=
  R^{\text{nonres}}_{\text{reserves}},
\end{align*}
and the corresponding totals.

Several verification steps are included:
\begin{itemize}[leftmargin=*]
  \item \emph{Energy balance per timestamp and product:} checks that
        $U_p(t) + C_p(t) = Pp(t)$ (within numerical tolerance) for all
        $t$ and $p$.
  \item \emph{Cost balance:} checks that the sum of allocated CapEx/OpEx
        across all products (plus the non-residential portion) matches the
        total pots integrated over time, and similarly for fuel costs.
  \item \emph{Subscription vs.\ allocated costs:} the script reports the
        difference between the sum of subscription revenue (including
        reserves) and the underlying allocated costs.
\end{itemize}

Finally, a high-level comparison table is written which summarises, for
each AMM configuration (e.g.\ AMM1--\texttt{BASE},
AMM2--\texttt{DELTA}), the total services cost recovered from the
residential and non-residential segments. This links the generator-side
revenue allocations in Appendix~\ref{app:amm_allocation} to the
customer-side pricing structure used in the main numerical experiments.

\subsection{Two-sided market interpretation and the supplier's retail choice}
\label{app:price_allocation:two_sided}

The AMM--Fair Play architecture is designed to restore a \emph{two-sided
marketplace} structure.

\medskip

\noindent
\textbf{Supply side.}
At the system layer, generators sell physical services (energy, reserves, and
adequacy) into a clearing process whose investment-facing remuneration is
allocated explicitly via the Fair Play pots (Appendix~\ref{app:amm_allocation}).
This makes the allocation of scarcity rents and fixed-cost recovery a
first-class design object, rather than an emergent artefact of scarcity prices.

\medskip

\noindent
\textbf{Demand/supplier side.}
Suppliers act as retail market-makers: they purchase standardised wholesale
\emph{liabilities} (product-indexed service bundles) and then decide how to
package and price these bundles for end customers. In the two-axis model used in
this thesis, these liabilities take the form of flat residential product
subscriptions (P1--P4) plus non-residential aggregate charges. These wholesale
charges are therefore not themselves ``the retail price''; they are the
supplier-facing cost base from which retail offerings are constructed.

\medskip

\noindent
\textbf{Why this matters for competition.}
In legacy price-capped retail architectures, suppliers often behave as residual
warehouses for wholesale volatility and tail risk they cannot control, so entry
and innovation are suppressed and competition collapses into a thin margin game.
By contrast, once the structural risk--volume separation problems identified in
Chapter~\ref{ch:fairness_definition} are removed (and wholesale volatility is
managed at the system layer), suppliers compete primarily on dimensions that are
\emph{within their control}: product design, customer service, portfolio
management, hedging strategy against predictable exposure, and the quality of
behavioural and flexibility support offered to customers.

\medskip

\noindent
\textbf{Retail price formation (conceptual).}
Let $s_p^{\text{total}}$ denote the wholesale subscription charge per household
per month for product $p$ computed in this appendix. A supplier $i$ with
$H^{(i)}_p$ households enrolled in product $p$ faces an annual wholesale charge
\[
C_i^{\text{wholesale}}
  = 12 \sum_{p \in \{\mathrm{P1},\dots,\mathrm{P4}\}}
      H^{(i)}_p \, s_p^{\text{total}}
  \;+\; C_{i,\mathrm{nonres}}^{\text{wholesale}},
\]
where $C_{i,\mathrm{nonres}}^{\text{wholesale}}$ is the supplier's non-residential
charge (if applicable).

The supplier then chooses a retail tariff structure---for example a flat
subscription, a hybrid (subscription + usage), or differentiated bundles---by
adding its operational cost, risk premium for \emph{controllable} risks, desired
margin, and any competitive discounts:
\[
\text{RetailPrice}_{i,p}
  = s_p^{\text{total}}
    + \underbrace{m_{i,p}}_{\text{supplier margin \& controllable risk}}
    + \underbrace{c_{i,p}}_{\text{service / acquisition / overhead}}
    + \underbrace{\epsilon_{i,p}}_{\text{competitive adjustment}}.
\]
Because the wholesale layer is stable and role-consistent, the key competitive
degrees of freedom lie in $(m_{i,p}, c_{i,p}, \epsilon_{i,p})$, rather than in
attempts to survive rare wholesale tail events.

% =========================================================
\chapter{Fairness Metrics: Definitions and Computation}
\label{app:fairness_metrics}
% =========================================================

This appendix defines the fairness metrics used throughout the thesis and
documents the \emph{exact} outcome objects and conventions used in computation.
We group metrics into:
(i) \emph{distributional inequality diagnostics} (ECDF, Lorenz/Gini, Atkinson,
Theil/GE); (ii) \emph{adequacy and cost-recovery diagnostics} (revenue adequacy
ratios and headcounts); (iii) \emph{payback and jackpot diagnostics} (median
and tail paybacks, ultra-rapid payback shares); (iv) \emph{concentration
metrics} (HHI and top concentration ratios); and (v) \emph{burden--cost
alignment} diagnostics (correlations/slopes between cost and burden variables).

The definitions are generic, but we explicitly record how each metric is
instantiated for: (a) generator outcomes under LMP and AMM variants; and
(b) household/product outcomes in the product-cost comparisons.

\section{Setup and notation}
\label{sec:fairness_metrics_setup}

Let $x=(x_1,\dots,x_n)$ denote a sample of $n$ observed outcomes for a given
population (e.g.\ household annual bills, generator annual payments, unit
cost per kWh, or revenue per MWh). Let $\bar{x}=\frac{1}{n}\sum_{i=1}^n x_i$
denote the sample mean, and let $x_{(1)}\le \dots \le x_{(n)}$ denote the
sorted outcomes.

Unless stated otherwise, distributional inequality indices (Lorenz/Gini,
Atkinson, Theil/GE) are applied to \emph{nonnegative} outcome vectors. When the
underlying economic quantity can be negative (e.g.\ net profit), we use the
\emph{nonnegative component} convention:
\[
x_i^+ \coloneqq \max\{0, x_i\},
\]
and compute inequality indices on $x^+=(x_1^+,\dots,x_n^+)$. This matches the
implementation used for generator \emph{net} outcomes in this thesis.

\paragraph{Weights.}
Unless explicitly stated, all distributional metrics reported here use
\emph{equal weight per agent} (each generator counts once; each household/load
counts once). Capacity scaling (e.g.\ ``per GW'') changes the outcome object
$x$, but does not introduce metric weights.

\section{Empirical cumulative distribution function (ECDF)}
\label{sec:ecdf}

For any real-valued outcome sample $z=(z_1,\dots,z_n)$ (not necessarily
nonnegative), the empirical cumulative distribution function (ECDF) is:
\[
\widehat{F}(t) = \frac{1}{n}\sum_{i=1}^n \mathbb{1}\{z_i \le t\}.
\]
Interpretation: $\widehat{F}(t)$ is the fraction of the population whose
outcome is at most $t$.

\paragraph{Quantiles.}
The $q$-quantile is defined as:
\[
\widehat{Q}(q) = \inf\{t : \widehat{F}(t) \ge q\}.
\]
In results sections we often summarise distributions using
$\widehat{Q}(0.25)$, $\widehat{Q}(0.5)$, $\widehat{Q}(0.75)$, and tail quantiles
when relevant.

\paragraph{Use in this thesis.}
ECDF plots are used both for nonnegative outcome objects (e.g.\ unit costs) and
for real-valued diagnostics such as the \emph{payback differential} (actual
minus expected), which can be negative.

\section{Lorenz curve}
\label{sec:lorenz}

For a nonnegative outcome vector $x\ge 0$ with $\sum_i x_i>0$, define the
Lorenz curve:
\[
L(k/n) = \frac{\sum_{i=1}^{k} x_{(i)}}{\sum_{i=1}^{n} x_{(i)}},
\qquad k = 0,1,\dots,n,
\]
with $L(0)=0$ and $L(1)=1$. The Lorenz curve is the piecewise-linear curve
connecting the points $\bigl(k/n,\, L(k/n)\bigr)$.

\paragraph{Interpretation.}
Perfect equality corresponds to the $45^\circ$ line $L(p)=p$. Greater bowing
below the line indicates greater inequality: more concentration of outcomes in
a small fraction of the population. For bills, ``inequality'' reflects uneven
burden; for revenues, it reflects jackpot concentration.

\section{Gini coefficient}
\label{sec:gini}

The Gini coefficient is twice the area between the Lorenz curve and the line of
equality:
\[
G = 1 - 2\int_{0}^{1} L(p)\,dp,
\qquad 0 \le G \le 1.
\]
For a finite sample, a common discrete formula is:
\[
G
=
\frac{2\sum_{i=1}^{n} i\,x_{(i)}}{n\sum_{i=1}^{n} x_{(i)}} - \frac{n+1}{n}.
\]
We report $G$ as a headline inequality statistic because it is scale-invariant,
bounded in $[0,1]$, and directly interpretable as concentration relative to
perfect equality.

\section{Atkinson index}
\label{sec:atkinson}

The Atkinson index introduces an explicit inequality-aversion parameter.
For $\varepsilon \ge 0$, define:
\[
A_\varepsilon
=
1 - \frac{x^{\mathrm{ede}}_\varepsilon}{\bar{x}},
\]
where $x^{\mathrm{ede}}_\varepsilon$ is the equally distributed equivalent
(EDE) outcome.

For $\varepsilon \neq 1$,
\[
x^{\mathrm{ede}}_\varepsilon
=
\left(
\frac{1}{n}\sum_{i=1}^{n} x_i^{\,1-\varepsilon}
\right)^{\!\frac{1}{1-\varepsilon}},
\]
and for $\varepsilon = 1$ (log case),
\[
x^{\mathrm{ede}}_{1}
=
\exp\!\left(
\frac{1}{|\mathcal{I}_{+}|}\sum_{i\in\mathcal{I}_{+}} \ln x_i
\right),
\qquad
\mathcal{I}_{+}\coloneqq\{i: x_i>0\}.
\]
That is, in computation we restrict the log term to strictly positive outcomes
($x_i>0$) rather than adding a numerical offset; if $x_i=0$ for all $i$, we
set $A_1=1$.

\paragraph{Interpretation.}
$A_\varepsilon \in [0,1]$ is the fraction of mean outcome $\bar{x}$ that would
be forgone to achieve equality at the same welfare level. Larger $\varepsilon$
places more weight on the lower tail. We report $A_{0.5}$ and $A_{1}$ in this
thesis.

\section{Generalised entropy (GE) family and Theil indices}
\label{sec:ge_theil}

The Generalised Entropy (GE) family provides inequality measures with
different tail sensitivities. For $\alpha\neq 0,1$,
\[
GE(\alpha)
=
\frac{1}{\alpha(\alpha-1)}
\left[
\frac{1}{n}\sum_{i=1}^{n}
\left(\frac{x_i}{\bar{x}}\right)^{\alpha} - 1
\right].
\]

\paragraph{Theil T (GE(1)).}
The Theil-$T$ index (the $\alpha\to 1$ limit) is:
\[
T
=
\frac{1}{n}\sum_{i=1}^{n}
\left(\frac{x_i}{\bar{x}}\right)\ln\!\left(\frac{x_i}{\bar{x}}\right),
\]
computed in practice on the strictly positive subset $\{i:x_i>0\}$.

\paragraph{Theil L / Mean log deviation (GE(0), optional).}
The Theil-$L$ (mean log deviation) is:
\[
L
=
\frac{1}{n}\sum_{i=1}^{n}
\ln\!\left(\frac{\bar{x}}{x_i}\right),
\]
which requires $x_i>0$; when used, it is computed on $\{i:x_i>0\}$.

\paragraph{Interpretation.}
GE measures are additively decomposable across groups, which can be useful for
within- vs between-group reporting (e.g.\ region, node, or product group).

\section{Tail diagnostics: quantiles, top shares, and jackpot indicators}
\label{sec:tail_diagnostics}

Market failures often manifest as \emph{jackpots} (extreme upper tail for
revenues or profits) or \emph{deprivation} (lower-tail exposure for households).
We therefore report additional tail diagnostics.

\paragraph{Quantiles.}
For any outcome distribution, we report selected quantiles such as
$\widehat{Q}(0.25)$, $\widehat{Q}(0.5)$, $\widehat{Q}(0.75)$, and high-tail
quantiles where relevant (e.g.\ $0.9$).

\paragraph{Top-$p$ share.}
For $p\in(0,1)$ and $m=\lceil pn\rceil$, the top-$p$ share is:
\[
S_{\mathrm{top}}(p)
=
\frac{\sum_{i=n-m+1}^{n} x_{(i)}}{\sum_{i=1}^{n} x_{(i)}}.
\]

\paragraph{Palma ratio (optional).}
The Palma ratio compares the top 10\% share to the bottom 40\% share:
\[
\mathrm{Palma}
=
\frac{\sum_{i=\lceil 0.9n\rceil}^{n} x_{(i)}}
{\sum_{i=1}^{\lfloor 0.4n\rfloor} x_{(i)}}.
\]

\paragraph{Ultra-rapid payback (jackpot) share.}
In generator results we additionally report the fraction of generators whose
implied payback time is below a short threshold (e.g.\ 1 year, and 0.2 years
$\approx$ 60--75 days), as a direct measure of extreme ``jackpot'' outcomes.

\section{Adequacy, cost recovery, and payback diagnostics}
\label{sec:adequacy_payback}

In addition to distributional inequality indices, we report operational and
investment-alignment diagnostics constructed from annualised revenues and
costs.

\paragraph{Modelled costs and annual revenue.}
For each generator $i$, we define annual modelled non-fuel costs as:
\[
C_i \coloneqq \mathrm{OpEx}^{\mathrm{nonfuel}}_i + \mathrm{CapEx}^{\mathrm{annual}}_i.
\]
Let $R_i^{(m)}$ denote total annual revenue under market design $m$.

\paragraph{Adequacy ratio.}
The adequacy ratio is:
\[
\mathrm{Adeq}_i^{(m)} \coloneqq \frac{R_i^{(m)}}{C_i},
\]
with summary statistics reported across generators (e.g.\ mean, $p25$, $p75$).

\paragraph{Net (non-fuel) margin and cost-recovery headcount.}
Define the annual net margin against non-fuel OpEx:
\[
\mathrm{Net}_i^{(m)} \coloneqq R_i^{(m)} - \mathrm{OpEx}^{\mathrm{nonfuel}}_i.
\]
The \emph{cost-recovery headcount} is the share of generators with
$\mathrm{Net}_i^{(m)}\ge 0$, and we also report the corresponding count.

\paragraph{Implied payback time.}
Let $Y_i^{\mathrm{exp}}$ denote the expected payback horizon from the cost
calibration, and define a total capex proxy
\[
\mathrm{CapEx}^{\mathrm{total}}_i \coloneqq \mathrm{CapEx}^{\mathrm{annual}}_i \, Y_i^{\mathrm{exp}}.
\]
The implied payback time under market design $m$ is computed as:
\[
\mathrm{PB}_i^{(m)} \coloneqq
\begin{cases}
\displaystyle \frac{\mathrm{CapEx}^{\mathrm{total}}_i}{\mathrm{Net}_i^{(m)}} & \text{if }\mathrm{Net}_i^{(m)} > 0,\\[1em]
+\infty & \text{otherwise.}
\end{cases}
\]
We report the median payback, a high-tail payback (e.g.\ 90th percentile), and
the ultra-rapid payback shares $\mathbb{P}(\mathrm{PB}\le 1)$ and
$\mathbb{P}(\mathrm{PB}\le 0.2)$.

\paragraph{Payback differential.}
We define the payback differential as:
\[
\Delta \mathrm{PB}_i^{(m)} \coloneqq \mathrm{PB}_i^{(m)} - Y_i^{\mathrm{exp}}.
\]
ECDF plots of $\Delta \mathrm{PB}$ summarise whether paybacks are typically
faster or slower than the expected horizon.

\section{Revenue concentration: HHI and concentration ratios}
\label{sec:concentration}

To quantify concentration of \emph{total} revenue across generators (distinct
from per-GW or net inequality), we report standard concentration metrics.

Let $R_i^{(m)}\ge 0$ denote nonnegative annual revenue under design $m$, and let
$S_i^{(m)} = R_i^{(m)}/\sum_j R_j^{(m)}$ denote the revenue share.

\paragraph{Herfindahl--Hirschman Index (HHI).}
\[
\mathrm{HHI}^{(m)} \coloneqq \sum_{i=1}^{n} \left(S_i^{(m)}\right)^2,
\]
with $\mathrm{HHI}\in[1/n,1]$; larger values indicate higher concentration.

\paragraph{Concentration ratios.}
Let $S_{(1)}^{(m)}\ge \dots \ge S_{(n)}^{(m)}$ be the sorted shares. For $k\in\{4,10\}$,
\[
\mathrm{CR}k^{(m)} \coloneqq \sum_{i=1}^{k} S_{(i)}^{(m)}.
\]
We report $\mathrm{CR4}$ and $\mathrm{CR10}$.

\section{Composite fairness score (reporting convenience)}
\label{sec:composite_fairness}

For compact comparison in summary tables, we report a transparent composite
fairness score built from four components computed at the approach level:
\begin{itemize}[leftmargin=*]
  \item \textbf{Inequality component:} $1-G$, where $G$ is the Gini of the
        nonnegative per-GW net outcome.
  \item \textbf{Adequacy component:} the cost-recovery headcount share
        $\mathbb{P}(\mathrm{Net}\ge 0)$.
  \item \textbf{Median-payback component:} $1/(1+\mathrm{median}(\mathrm{PB}))$.
  \item \textbf{Anti-jackpot component:} $1-\mathbb{P}(\mathrm{PB}\le 0.2)$.
\end{itemize}
Each component is min--max normalised across the compared approaches, and the
composite score is the simple average of the normalised components. This
composite is \emph{not} a new axiom; it is a reporting convenience that
summarises inequality, adequacy, investment alignment, and jackpot risk.

\section{Outcome objects used in computation}
\label{sec:outcome_objects}

This section records the exact $x$ used for each metric family in the scripts.

\subsection{Generator distributional fairness (annual)}
\label{subsec:generator_distributional_fairness}

Let $\mathrm{GW}_i$ denote generator nameplate capacity in GW. For market design
$m$, define annual total revenue $R_i^{(m)}$ and net against non-fuel OpEx:
$\mathrm{Net}_i^{(m)} = R_i^{(m)} - \mathrm{OpEx}^{\mathrm{nonfuel}}_i$.

\paragraph{Per-GW net outcome (inequality object).}
The inequality indices (Lorenz Gini, Atkinson, Theil) are applied to:
\[
x_i^{(m)} \coloneqq \left(\frac{\mathrm{Net}_i^{(m)}}{\mathrm{GW}_i}\right)^{+}
= \max\!\left\{0,\ \frac{\mathrm{Net}_i^{(m)}}{\mathrm{GW}_i}\right\}.
\]
This matches the implementation: per-GW net is computed and then clipped at
zero before calculating inequality indices.

\paragraph{Totals (concentration object).}
HHI and concentration ratios are computed on nonnegative total revenue shares:
\[
S_i^{(m)} \coloneqq \frac{\max\{0, R_i^{(m)}\}}{\sum_j \max\{0, R_j^{(m)}\}}.
\]

\paragraph{Equal weighting.}
All generator-level distributional metrics use equal weight per generator; they
are \emph{not} capacity-weighted.

\subsection{Household/product cost comparisons and geographic dispersion}
\label{subsec:household_product_objects}

For household/product results, the outcome objects include:
\begin{itemize}[leftmargin=*]
  \item \textbf{Per-household annual cost} (e.g.\ \pounds/HH/year) at node level
        (LMP nodal), and at socialised level (flat), and AMM product subscription
        (flat per product).
  \item \textbf{Geographic dispersion objects} such as node-level deltas
        $\Delta = \mathrm{LMP}_{\mathrm{nodal}} - \mathrm{AMM}$ and corresponding
        ECDF/boxplot summaries.
\end{itemize}

\section{Burden--cost alignment diagnostics (Pearson r and slope)}
\label{sec:burden_cost_alignment}

Some fairness claims in this thesis concern \emph{alignment} between a burden
metric (e.g.\ controllable energy per household) and the cost assigned by an
approach. These are not inequality indices; they are proportionality
diagnostics.

Let $b_p$ denote a per-product burden metric (e.g.\ controllable kWh per HH for
product $p$) and let $c_p^{(m)}$ denote the corresponding per-product cost
under approach $m$ (per HH per year, or total). We report:

\paragraph{Pearson correlation.}
\[
r(b,c) = \frac{\sum_p (b_p-\bar{b})(c_p-\bar{c})}
{\sqrt{\sum_p (b_p-\bar{b})^2}\sqrt{\sum_p (c_p-\bar{c})^2}}.
\]

\paragraph{Slope from a fitted line.}
We also report the slope $\hat{\beta}_1$ from the least-squares fit
$c_p = \beta_0 + \beta_1 b_p + \varepsilon_p$, as an effect-size measure.

\paragraph{Small-sample caution.}
When the alignment diagnostic is computed over a small number of products
(e.g.\ $N=4$), correlations and slopes are treated as indicative rather than
conclusive, and interpretation focuses on direction and relative magnitude.

\section{Practical notes for computation and comparability}
\label{sec:fairness_practical_notes}

\paragraph{Scale invariance.}
Gini, Atkinson, and GE measures are scale-invariant: multiplying all outcomes
by a constant leaves the index unchanged. This supports comparison across
scenarios with different absolute levels.

\paragraph{Zeros and logs (implementation convention).}
Atkinson with $\varepsilon=1$ and Theil indices involve logarithms and are
computed on the strictly positive subset $\{i:x_i>0\}$. This avoids introducing
an arbitrary numerical offset; where all outcomes are zero, the relevant
statistic is handled by explicit conventions in the implementation.

\paragraph{Outcome choice matters.}
The same index represents different notions depending on the object:
\begin{itemize}[leftmargin=*]
  \item Bills or unit costs: burden concentration.
  \item Revenues or revenue/MWh: jackpot and market-power concentration.
  \item Scarcity exposure or curtailment incidence: deprivation concentration.
  \item Adequacy ratios and headcounts: solvency/cost-recovery feasibility.
  \item Payback diagnostics: investment alignment and extreme tail risk.
  \item Burden--cost alignment: proportional cost responsibility (F4-type tests).
\end{itemize}

\section*{Summary}
\addcontentsline{toc}{section}{Summary}

We use ECDFs and quantiles for distributional diagnostics (including real-valued
payback differentials), Lorenz/Gini/Atkinson/Theil for inequality on explicitly
nonnegative outcome objects, adequacy and cost-recovery headcounts to test
whether revenues cover calibrated non-fuel cost bases, payback and ultra-rapid
payback shares to quantify investment alignment and jackpot risk, HHI/CR$n$ to
quantify revenue concentration, and Pearson $r$/slope diagnostics to test
burden--cost alignment. All metrics are reported with explicit outcome
definitions and conventions consistent with the implemented scripts.

% =========================================================
\chapter{Notation Table}
\label{app:notation}

% Optional: tighten vertical space before/after longtable
\setlength{\LTpre}{0pt}
\setlength{\LTpost}{0pt}

\begin{longtable}{@{} l p{0.78\textwidth} @{}}
\caption{Key notation used across Chapters~\ref{ch:fairness_definition}--\ref{ch:discussion}.}
\label{tab:notation_all}\\
\toprule
\textbf{Symbol} & \textbf{Description} \\
\midrule
\endfirsthead

\multicolumn{2}{c}%
{{\tablename\ \thetable{} -- continued from previous page}}\\
\toprule
\textbf{Symbol} & \textbf{Description} \\
\midrule
\endhead

\midrule
\multicolumn{2}{r}{{Continued on next page}}\\
\bottomrule
\endfoot

\bottomrule
\endlastfoot

% ---------- Time, network, actors ----------
\multicolumn{2}{l}{\textbf{Time, Network, and Actors}}\\
$t \in \mathcal{T}$ & Discrete time index (e.g.~30-minute settlement interval). \\
$n \in \mathcal{N}$ & Network node (household, feeder, cluster, control zone, ESO). \\
$c \in \mathcal{C}$ & Cluster of generators, loads, or households (nested Shapley cluster). \\
$h \in \mathcal{H}_n$ & Household or load entity at node $n$. \\
$g \in \mathcal{G}_n$ & Generator or controllable asset at node $n$. \\
$i \in \mathcal{I}(n)$ & Flexible device or request associated with node $n$. \\[0.5em]
$r \in \mathcal{R}$ & Request / bid / capability profile (consumer, generator, or device). \\[0.5em]

% ---------- System tightness ----------
\multicolumn{2}{l}{\textbf{System Tightness and Feasibility}}\\
$\alpha_{t,n}$ & Tightness index: ratio of available supply to demand at node $n$ and time $t$. \\
$\tilde{\alpha}_{t,n}$ & AMM-internal tightness after holarchic aggregation and forecasting. \\
$\Delta_{t,n}$ & Local deficit $= D_{t,n} - S_{t,n}$. \\
$W_{t,n}$ & Curtailment or wasted energy: $\max(0,\,S^{\mathrm{avail}}_{t,n} - G^{\mathrm{used}}_{t,n})$. \\
$\Gamma$ & Physical/network constraint set (flow, thermal, voltage). \\
$\mathcal{A}$ & Set of allocatively feasible dispatch outcomes. \\[0.5em]

% ---------- Demand, supply, prices ----------
\multicolumn{2}{l}{\textbf{Electricity Demand, Supply, and Prices}}\\
$D_{t,n}$ & Total demand at node $n$ and time $t$. \\
$S_{t,n}$ & Maximum physically secure supply or import capacity. \\
$G^{\mathrm{used}}_{t,n}$ & Supply actually allocated/served at time $t$. \\
$p_{t,n}$ & Real-time scarcity-aware price signal from AMM at node $n$. \\
$p^{\mathrm{base}}_{t,n}$ & Non-scarcity base price component at node $n$. \\
$p^{\mathrm{tight}}_{t,n}$ & Scarcity/tightness-driven price component at node $n$. \\
$v_{t,n}$ & Measured local voltage at node $n$ (physical shadow price of local scarcity). \\[0.5em]

% ---------- Contract / products ----------
\multicolumn{2}{l}{\textbf{Contract Representation and Products}}\\
$p \in \mathcal{P}$ & Contracted QoS product/class (P1--P4). \\
$\rho^{\mathrm{QoS}}(p)$ & Reliability entitlement: probability of delivery in scarcity for product $p$. \\
$\pi^{\mathrm{sub}}(p)$ & Subscription price for service class $p$. \\
$w(p)$ & Priority weight used for product $p$ within Fair Play sampling. \\[0.5em]
$\Gamma^{\mathrm{contract}}_r$ & Contract attribute vector for request $r$ (magnitude, timing, reliability). \\
$E_r$ & Requested or offered energy volume in request $r$. \\
$\bar{P}_r$ & Maximum power rate associated with request $r$. \\
$[t^{\mathrm{start}}_r, t^{\mathrm{end}}_r]$ & Allowable delivery window for request $r$. \\
$\sigma^r$ & Time-shifting or flexibility tolerance parameter for request $r$. \\[0.5em]

% ---------- Fair Play / entitlement ----------
\multicolumn{2}{l}{\textbf{Fair Play Allocation and Entitlement}}\\
$F_n(T)$ & Fairness ratio for participant $n$: delivered/desired flexible energy over horizon $T$. \\
$\delta_i$ & Fairness deficit for flexible request $i$. \\
$\Pr_i$ & Selection probability of request $i$ under Fair Play. \\
$q^{\mathrm{ess}}_{h}$ & Essential protected block of energy for household $h$. \\
$U_{h,t}$ & Uplift cost attributed to household $h$ at time $t$ (F4 proportional responsibility). \\[0.5em]
$m_s$ & Service-level priority weight for tier $s$ (e.g.\ premium vs standard) under scarcity. \\[0.5em]

% ---------- Shapley and nested aggregation ----------
\multicolumn{2}{l}{\textbf{Shapley Value and Nested Aggregation}}\\
$W(S)$ & System value with coalition $S$ of generators, clusters, or flexibility participants. \\
$\phi_g$ & Shapley value allocated to generator, household, or agent $g$. \\
$\Phi_c$ & Aggregated Shapley value allocated to cluster $c$. \\
$\mathcal{C}_k$ & Set of clusters in level $k$ of the nested Shapley aggregation. \\
$v_g$ & Value-attribute vector (e.g.\ $(E_g,F_g,R_g,K_g,S_g)$). \\[0.5em]

% ---------- Cost / waste ----------
\multicolumn{2}{l}{\textbf{Cost, Waste, and Architectural Terms}}\\
$B$ & Total annual household bill. \\
$B_{\mathrm{phys}}$ & Physical system component of the bill (energy, network, capacity). \\
$B_{\mathrm{policy}}$ & Policy component (CfDs, ECO, carbon, capacity mechanisms). \\
$B_{\mathrm{arch}}$ & Architecture-induced bill component (uplift, bailouts, risk premia). \\
$\Phi_{\mathrm{waste}}$ & Cost of infeasible dispatch, wrong-sided curtailment, or misallocated energy. \\
$\Lambda_{\mathrm{risk}}$ & Risk premium from price-cap hedging and liquidity buffers. \\
$\Gamma_{\mathrm{intervention}}$ & Pass-through cost of bailouts, crisis schemes, and emergency support. \\
$\Xi_{\mathrm{inefficiency}}$ & Settlement and tariff inefficiency (standing charges, misaligned TOU, etc.). \\
$\mathrm{Saving\%}$ & Proportional reduction of $B_{\mathrm{arch}}$ under AMM--Fair Play. \\[0.5em]

% ---------- Mechanism / digital governance ----------
\multicolumn{2}{l}{\textbf{Market Mechanism and Digital Governance}}\\
$\mathcal{M}$ & Market mechanism mapping state $\mathcal{S}_t$ to allocation (e.g.\ AMM, LMP, Fair Play). \\
$\mathcal{S}_t$ & State of knowledge at time $t$ (prices, $\alpha$, congestion, histories, entitlements). \\
$XR$ & Explainability record associated with a particular allocation or curtailment decision. \\

\end{longtable}

\chapter{Estimated Cost Impact of the AMM Architecture}
\label{app:cost_impact}

This appendix provides the detailed methodology, assumptions, and numerical
estimates underlying the comparison of customer-facing costs between the legacy
price-capped retail architecture and the AMM--Fair Play design. It supplements,
but does not interrupt, the qualitative policy discussion in
Chapter~\ref{ch:discussion}.

The previous Chapters argued that legacy price-capped retail architectures
separate \emph{who chooses volume} from \emph{who bears tail risk}, and that
this risk--volume separation makes insolvency cascades and structural waste
(avoidable curtailment, default costs, and risk premia) effectively
unavoidable (Lemma~\ref{lem:price_cap_insolvency},
Lemma~\ref{lem:risk_volume_instability}). The AMM architecture, by contrast,
co-locates volume decisions and risk-bearing at the market-making layer and
implements a zero-waste, fairness-aware allocation of scarcity.
a
This section translates that structural difference into an \emph{estimated}
cost impact, expressed as a percentage saving in the total customer-facing
energy bill, conditional on external bill breakdown data.

\section{Bill decomposition and comparison metric}

Following the cost accounting framework in
Chapter~\ref{ch:market_scenarios}, we decompose the retail price in each
regime $k \in \{\mathrm{cap}, \mathrm{AMM}\}$ as:
\[
P_R^{k}(t)
=
P_{\mathrm{phys}}(t)
+
P_{\mathrm{pol}}^{k}(t)
+
P_{\mathrm{arch}}^{k}(t),
\]
where:
\begin{itemize}[leftmargin=*]
    \item $P_{\mathrm{phys}}(t)$ captures physical system costs (fuel, losses,
          short-run network Opex, and the amortised component of CapEx);
    \item $P_{\mathrm{pol}}^{k}(t)$ consists of policy-driven levies and
          ``stealth taxes'' (e.g.\ carbon funding mechanisms, socialised
          surcharges);
    \item $P_{\mathrm{arch}}^{k}(t)$ captures costs arising from the market
          architecture itself: risk premia, insolvency and restructuring costs,
          inefficient hedging constraints, and the cost of avoidable waste
          (curtailment and involuntary unserved energy).
\end{itemize}

For a given demand path $Q(t)$, the total bill in regime $k$ over horizon
$[0,T]$ is:
\[
\mathcal{B}^{k}
=
\int_0^T P_R^{k}(t) Q(t)\, dt
=
\mathcal{B}^{k}_{\mathrm{phys}}
+
\mathcal{B}^{k}_{\mathrm{pol}}
+
\mathcal{B}^{k}_{\mathrm{arch}}.
\]

To isolate the \emph{architectural} effect, we perform a like-for-like
comparison which:
\begin{enumerate}[leftmargin=*]
    \item holds the physical system and demand trajectory fixed, i.e.\
    $P_{\mathrm{phys}}(t)$ and $Q(t)$ are the same under both regimes;
    \item treats policy costs $P_{\mathrm{pol}}^{k}(t)$ as either identical
    across regimes or accounted for separately from the energy bill; and
    \item focuses on the difference in $\mathcal{B}^{k}_{\mathrm{arch}}$.
\end{enumerate}

The primary comparison metric is the percentage reduction in the
\emph{physical-plus-architectural} bill:
\[
\mathrm{Saving\%}
=
100 \times
\frac{
\mathcal{B}^{\mathrm{cap}}_{\mathrm{phys}+\mathrm{arch}}
-
\mathcal{B}^{\mathrm{AMM}}_{\mathrm{phys}+\mathrm{arch}}
}{
\mathcal{B}^{\mathrm{cap}}_{\mathrm{phys}+\mathrm{arch}}
},
\]
where
\[
\mathcal{B}^{k}_{\mathrm{phys}+\mathrm{arch}}
=
\mathcal{B}^{k}_{\mathrm{phys}}
+
\mathcal{B}^{k}_{\mathrm{arch}}.
\]

For reporting at the household level, we also define the average unit price
(in \pounds/kWh) in regime $k$:
\[
\bar{P}^{k}
=
\frac{
\int_0^T P_R^{k}(t) Q(t)\, dt
}{
\int_0^T Q(t)\, dt
}
\]
and the corresponding unit saving:
\[
\Delta \bar{P}
=
\bar{P}^{\mathrm{cap}} - \bar{P}^{\mathrm{AMM}}.
\]

\subsection{Mapping experimental outcomes to architectural costs}

The experiments in Chapter~\ref{ch:experiments} simulate paired markets
under identical physical conditions and demand scenarios, varying only the
market architecture (Baseline vs.\ AMM/subscription). For each experiment and
regime $k$ we track:
\begin{itemize}[leftmargin=*]
    \item total served demand $E^{k}_{\mathrm{served}}$;
    \item curtailed or stranded energy $E^{k}_{\mathrm{curt}}$;
    \item involuntary unserved energy $E^{k}_{\mathrm{unserved}}$;
    \item the time series of marginal prices and scarcity signals.
\end{itemize}

To translate these into architectural costs, we adopt the following stylised
mapping:
\begin{align*}
\mathcal{B}^{k}_{\mathrm{arch}}
&=
\underbrace{
  v_{\mathrm{curt}} \, E^{k}_{\mathrm{curt}}
}_{\text{avoidable procurement and curtailment}}
+
\underbrace{
  v_{\mathrm{lost}} \, E^{k}_{\mathrm{unserved}}
}_{\text{value-of-lost-load penalties}}
+
\underbrace{
  C^{k}_{\mathrm{risk}}
}_{\text{risk premia, default, restructuring}},
\end{align*}
where:
\begin{itemize}[leftmargin=*]
    \item $v_{\mathrm{curt}}$ represents the effective cost of energy that is
          procured and then curtailed or stranded (e.g.\ strike price or
          marginal procurement cost);
    \item $v_{\mathrm{lost}}$ is a value-of-lost-load (VOLL) proxy used to
          monetise unserved energy (e.g.\ regulatory benchmarks or scenario
          values);
    \item $C^{k}_{\mathrm{risk}}$ aggregates architecture-induced risk costs
          (default, restructuring, and risk premia on contracts). In the
          absence of detailed balance-sheet data, this can be calibrated
          from external bill breakdowns (e.g.\ the observed fraction of bills
          attributed to supplier failures and risk premia in the legacy
          system).
\end{itemize}

Under the zero-waste AMM design, the experiments are constructed such that:
\[
E^{\mathrm{AMM}}_{\mathrm{curt}} \approx 0,
\qquad
E^{\mathrm{AMM}}_{\mathrm{unserved}}
\ \text{is minimal and explicitly allocated via Fair Play},
\]
whereas in the Baseline price-capped architecture we typically observe
$E^{\mathrm{cap}}_{\mathrm{curt}} > 0$ and, in stressed scenarios, higher
$E^{\mathrm{cap}}_{\mathrm{unserved}}$ or implicit rationing.

Thus, for any fixed choice of $(v_{\mathrm{curt}}, v_{\mathrm{lost}})$ and
externally calibrated $(C^{\mathrm{cap}}_{\mathrm{risk}},
C^{\mathrm{AMM}}_{\mathrm{risk}})$, the experiments yield an empirical estimate
of $\mathcal{B}^{k}_{\mathrm{arch}}$ and therefore of $\mathrm{Saving\%}$.

\subsection{Embedding external bill breakdowns}

Let $\theta_{\mathrm{phys}}$, $\theta_{\mathrm{pol}}$, and
$\theta_{\mathrm{arch}}$ denote the observed shares of an average customer
bill attributed to physical system costs, policy levies, and architectural
costs, respectively, in a given jurisdiction:
\[
\theta_{\mathrm{phys}} + \theta_{\mathrm{pol}} + \theta_{\mathrm{arch}} = 1.
\]

Let $B_{\mathrm{avg}}$ denote the average annual bill per household. Then:
\[
\mathcal{B}^{\mathrm{cap}}_{\mathrm{phys}+\mathrm{arch}}
=
(\theta_{\mathrm{phys}} + \theta_{\mathrm{arch}}) \, B_{\mathrm{avg}},
\]
and the absolute annual saving per household implied by the AMM architecture is:
\[
\Delta B_{\mathrm{annual}}
=
\mathrm{Saving\%} \times
\frac{
\mathcal{B}^{\mathrm{cap}}_{\mathrm{phys}+\mathrm{arch}}
}{
100
}
=
\mathrm{Saving\%} \times
\frac{
(\theta_{\mathrm{phys}} + \theta_{\mathrm{arch}}) \, B_{\mathrm{avg}}
}{
100
}.
\]

In practice, the procedure is:
\begin{enumerate}[leftmargin=*]
    \item Obtain external estimates of $(\theta_{\mathrm{phys}},
    \theta_{\mathrm{pol}}, \theta_{\mathrm{arch}})$ and $B_{\mathrm{avg}}$
    from bill breakdown or regulatory reports.
    \item Use the paired market simulations to compute
    $(E^{k}_{\mathrm{curt}}, E^{k}_{\mathrm{unserved}})$ and a calibrated
    $(C^{k}_{\mathrm{risk}})$ for each regime.
    \item Compute $\mathcal{B}^{k}_{\mathrm{arch}}$ for
    $k \in \{\mathrm{cap}, \mathrm{AMM}\}$ via the mapping above.
    \item Evaluate $\mathrm{Saving\%}$ and $\Delta B_{\mathrm{annual}}$.
\end{enumerate}

Table~\ref{tab:cost_impact_summary} provides a template for reporting the
resulting estimates.

\begin{table}[H]
\centering
\caption{Illustrative structure for reporting cost impact of AMM vs.\ price-capped regime. External bill breakdown parameters $(\theta_{\mathrm{phys}}, \theta_{\mathrm{pol}}, \theta_{\mathrm{arch}})$ and $B_{\mathrm{avg}}$ are taken from regulatory or industry data; architectural costs are estimated from the paired simulations.}
\label{tab:cost_impact_summary}
\begin{tabular}{lccc}
\toprule
 & Price-capped regime & AMM regime & Difference \\
\midrule
Physical system bill component ($\pounds$/year)    & to be calibrated & to be calibrated & $\Delta \mathcal{B}_{\mathrm{phys}}$ \\
Policy/levies bill component ($\pounds$/year)     & aligned / separate & aligned / separate & -- \\
Architectural bill component ($\pounds$/year)     & from sims + data & from sims + data & $\Delta \mathcal{B}_{\mathrm{arch}}$ \\
\midrule
Total phys+arch bill ($\pounds$/year)             & $\mathcal{B}^{\mathrm{cap}}_{\mathrm{phys}+\mathrm{arch}}$ & $\mathcal{B}^{\mathrm{AMM}}_{\mathrm{phys}+\mathrm{arch}}$ & $\Delta B_{\mathrm{annual}}$ \\
Estimated saving (\%)                              & \multicolumn{3}{c}{$\mathrm{Saving\%}$} \\
\bottomrule
\end{tabular}
\end{table}

\subsection{Interpretation}

This cost impact assessment should be read as follows:

\begin{itemize}[leftmargin=*]
    \item It does \emph{not} assume that the AMM changes the underlying physics
          of the system: $P_{\mathrm{phys}}(t)$ and $Q(t)$ are held fixed.
    \item It explicitly separates policy choices from market architecture:
          $P_{\mathrm{pol}}^{k}(t)$ is treated as exogenous or accounted for
          outside the energy bill.
    \item Any positive estimate of $\mathrm{Saving\%}$ therefore reflects
          \emph{reduced architectural waste and risk}, not cheaper turbines,
          wires, or diminished policy ambition.
\end{itemize}

In other words, for a given physical system and policy stance, the AMM
architecture can be interpreted as \emph{running the same grid, serving the
same demand, with less structural waste and fewer insolvency-induced costs}.
The empirical value of $\mathrm{Saving\%}$ depends on the chosen calibration,
but the direction of the effect is a direct consequence of the zero-waste,
risk-co-locating design demonstrated in the experiments.
% =========================================================
\chapter{Illustrative Legislative Abstraction}
\label{app:legislative_abstraction}
% =========================================================

\section*{Energy System Renewal and Fair Market Design Act (Condensed Form\footnote{A full legislative draft corresponding to this condensed abstraction is available
online for reference \cite{energy_system_renewal_bill_full}.})}

\subsection*{Purpose and Scope}

\begin{enumerate}[label=\arabic*.]
    \item This Act establishes a statutory framework for an electricity market
    designed around system resilience, essential service protection, fairness,
    and long-term investment adequacy.
    \item The Act provides for institutional reform, market design restructuring,
    and digital settlement integration across national and local system levels.
    \item The Act applies to Great Britain.
\end{enumerate}

\subsection*{Key Definitions}

For the purposes of this Act:
\begin{itemize}[leftmargin=*]
    \item \textbf{Essential consumption} means electricity supply required for
    health, safety, and critical public services.
    \item \textbf{Flexible consumption} means electricity demand that may be
    shifted, curtailed, or scheduled in response to system conditions.
    \item \textbf{Market Participation Entity} means any actor authorised to bid
    supply, demand, storage, or flexibility into energy markets.
    \item \textbf{Energy Service Provider} means an entity licensed to provide
    retail-facing energy services, billing, and consumer representation.
    \item \textbf{Market Facilitator} means an operator authorised to run approved
    market clearing and settlement algorithms.
\end{itemize}

\subsection*{Establishment of the Energy System Regulator}

\begin{enumerate}[label=\arabic*.]
    \item An independent statutory regulator is established with responsibility
    for electricity market oversight, system resilience, and consumer protection.
    \item The Regulator shall exercise its functions independently of commercial
    market participants.
\end{enumerate}

\subsection*{Principal Regulatory Duties}

In exercising its functions, the Regulator must prioritise:
\begin{itemize}[leftmargin=*]
    \item continuity of supply for essential consumption;
    \item system adequacy, redundancy, and resilience under stress;
    \item fairness and proportional cost allocation;
    \item facilitation of participation and competition;
    \item provision of stable, bankable investment signals;
    \item alignment with decarbonisation objectives.
\end{itemize}

Where duties conflict, protection of essential supply and system resilience take
precedence over short-term economic efficiency.

\subsection*{Market Structure and Design}

\begin{enumerate}[label=\arabic*.]
    \item Electricity markets shall operate as layered procurement mechanisms,
    including:
    \begin{itemize}
        \item energy delivery,
        \item flexibility and demand response,
        \item strategic reserve and adequacy,
        \item locational congestion relief.
    \end{itemize}
    \item Market clearing and settlement algorithms must be constraint-aware and
    certified by the Regulator.
    \item Essential consumption shall not be exposed to real-time scarcity pricing
    or involuntary disconnection except under formally declared emergency
    conditions.
\end{enumerate}

\subsection*{Licensing Framework}

\begin{enumerate}[label=\arabic*.]
    \item Existing supplier licensing categories are replaced with:
    \begin{itemize}
        \item Energy Service Providers;
        \item Market Participation Entities;
        \item Market Facilitators;
        \item Local Energy Stewards (distribution-level operators).
    \end{itemize}
    \item Licensing requirements must be proportionate to system risk and role.
\end{enumerate}

\subsection*{National and Local System Coordination}

\begin{enumerate}[label=\arabic*.]
    \item A National System Steward shall be responsible for national adequacy,
    interregional coordination, and strategic reserve.
    \item Local Energy Stewards shall manage local congestion, flexibility, and
    protection of critical supply.
    \item Local markets shall clear prior to national markets where feasible.
\end{enumerate}

\subsection*{Digital Settlement and Transparency}

\begin{enumerate}[label=\arabic*.]
    \item Settlement systems shall allocate costs and revenues based on real-time
    system contribution, including energy delivery, flexibility provision, and
    congestion relief.
    \item Market algorithms must be auditable, explainable, and subject to
    regulatory certification.
\end{enumerate}

\subsection*{Enforcement and Emergency Powers}

\begin{enumerate}[label=\arabic*.]
    \item The Regulator may issue compliance orders, impose penalties, suspend
    licences, or direct system operators where necessary to protect essential
    supply or system stability.
    \item Emergency powers may be exercised only to preserve public safety and
    system integrity.
\end{enumerate}

\subsection*{Transitional Provisions}

\begin{enumerate}[label=\arabic*.]
    \item Existing market arrangements shall continue during a defined transition
    period.
    \item Pilot and sandbox markets may operate in parallel with legacy systems
    prior to full implementation.
    \item No lawful contractual rights may be expropriated without due process and
    compensation.
\end{enumerate}

% =========================================================
\chapter{Final Epilogue: From Homogeneity and Control to Diversity and Enablement}

A deep insight underlying this thesis is that many systemic distortions in today’s electricity markets are not caused by physical limits alone, but by conceptual limits inherited from 20th-century economics. These models assumed homogeneity—of consumers, preferences, technologies, behaviours, and value. In reality, humans, devices, and energy needs have never been homogeneous. They only appeared so because our information systems were coarse and our institutions were built around aggregate abstractions. 

The communications revolution—smart meters, digital twins, machine learning, and increasingly AI-driven interpretation—has made visible what always existed: diversity of priority, diversity of needs, diversity of capability, diversity of contribution. What were once “market failures” or “model errors” are often simply manifestations of differences that could not previously be recognised, expressed, or valued.

The thesis therefore makes a broader argument: fairness, resilience, and participation must be reconceived not as corrective interventions, but as design principles that embrace non-homogeneity rather than suppress it. The role of the AMM + Fair Play architecture is precisely to map diversity into system coordination—without forcing conformity, without arbitrary rules, and without masking variety under uniform price or identical tariff designs.

The future electricity system (and, by extension, the future economic system) is not one of centralised control or unmanaged chaos. It is one of structured enablement, where individuals, households, and technologies can express their roles, priorities, and contributions—and where the system can respond intelligently, transparently, and fairly.

\section{Redefining Growth for a Digital, Electrified Society}

The GDP paradigm—originally designed to count factory outputs—was never intended to measure societal wellbeing, resilience, or knowledge creation. Yet it continues to dominate economic policy and fiscal priorities, often crowding out investment in human development, digital infrastructure, and public trust.

Digital and cyber–physical economies, by contrast, reward not the size of transactions, but the quality of interactions. True growth in the 21st century should be measured through:

\begin{itemize}[leftmargin=2em]
    \item Growth in learning, knowledge, insight, and truth-seeking.
    \item Growth in kindness, civility, tolerance, and humanity.
    \item Growth in resource efficiency, technological capability, and system resilience.
    \item Growth in the reduction of poverty, inequality, waste, corruption, and involuntary exclusion.
    \item Growth in transparency, diversity of thought, participation, and democratic legitimacy.
\end{itemize}

These are not philosophical sentiments; they are design requirements for modern infrastructure. Digital markets, including the AMM framework, explicitly reward contribution, stabilise risk, expose underused capacity, and allocate fairly—not because fairness is moral, but because it is efficient, persistent, and legitimacy-preserving.

\section{Economics, Democracy, and Freedom in a Post-GDP World}

Capitalism, in its purest sense, is the right to deploy one’s capability freely. Communism, in its purest sense, is the collective provision of certain essential protections. Both degenerate in practice when constrained by centralised information systems that assume uniformity.

The future requires neither ideological polarity nor forced convergence. It requires systems that can recognise individual roles, value diverse contributions, and support legitimate prioritisation—not control behaviour, but enable choice.

Markets must therefore evolve from price-only arbitrators to digitally governed allocation systems that embed:
\begin{itemize}[leftmargin=2em]
    \item Priority for essential access,
    \item Opportunity for contribution,
    \item Clarity of entitlement, and
    \item Recognition of diversity of needs and abilities.
\end{itemize}

This is the essence of democracy in cyber–physical infrastructures: not just voting every few years, but continuous, traceable, algorithmic representation of roles and rights.

\section{Finance, Debt, Inflation, and the Transition Path}

The transition outlined in this thesis is not only a technical or conceptual transformation; it has a financial dimension. National economies today spend significant portions of tax revenue—often £1 in every £10—on interest payments rather than productive investment. High debt, misallocated subsidies, and poorly designed incentives erode both resilience and legitimacy.

In algorithmically governed markets, defaults are not simply failures; they are breaches of trust and transparency. Fair, data-driven contracting means:
\begin{itemize}[leftmargin=2em]
    \item Cost recovery must be traceable and proportionate,
    \item Investment signals should reduce uncertainty rather than amplify it,
    \item Inflation should be studied as redistribution of risk, not just price mechanics,
    \item Interest rates should reflect future productive capability, not short-term scarcity or speculation.
\end{itemize}

A digitally regulated energy system—with explicit contract tiers, transparent risk allocation, and Shapley-derived value—provides a robust foundation for national financial governance. It reduces volatility, improves bankability, supports sovereign credibility, and creates options for debt restructuring and strategic investment without eroding fairness or trust.

Fairness, therefore, becomes not only ethically desirable—but fiscally stabilising.

\section{Toward a Democratically Governed, Digitally Regulated Economy}

This thesis concludes that the future of energy markets—and arguably economic governance more broadly—lies neither in central control nor in pure abstraction, but in structured digital enablement.

We now have the technology: data platforms, cyber–physical systems, agent-based digital clearing engines, and explainable allocation mechanisms like the AMM + Fair Play.  
We now have the mathematics: cooperative game theory, graph theory, nested aggregation, and bounded digital scarcity control.  
We now have the information: peri-second physical measurement, settlement-grade data, and machine learning for pattern recognition at scale.

What remains is design.

Design is how technology becomes trust.  
Design is how fairness becomes enforceable.  
Design is how democracy becomes continuous, not occasional.

\section{Final Reflection}

Markets are not just for trading.  
They are agreements—on how we allocate what matters.  
This thesis shows that agreements can now be digitally precise, physically grounded, mathematically fair, and democratically governable.

If designed intentionally, the next version of our markets will not only allocate electricity.  
They will allocate dignity, resilience, and agency—  
in a world where diversity is no longer noise,  
but signal.

\end{document}